\documentclass[11pt]{cernrep}
\pdfoutput=1
\usepackage{graphicx,epsfig}
\usepackage{hyperref}
\usepackage{amsmath,amssymb,amsfonts}
\usepackage{url}
\usepackage{fixmath}
\usepackage{subfig}
\usepackage{xspace}
\usepackage{placeins}
\usepackage{units}
\usepackage{slashed}
\usepackage{booktabs}

\newcommand{\tms}{t_{\textnormal{\tiny{MS}}}}

\newcommand{\W}{\mathrm{W}}
\newcommand{\Wp}{\ensuremath{\W^+}}

\newcommand{\mz}{\mathrm{M}_\text{Z}}
\newcommand{\mw}{\mathrm{M}_\text{W}}
\newcommand{\Wjets}{\W+\mathrm{jets}}
\newcommand{\Wpjets}{\Wp+\mathrm{jets}}

\begin{document}

\title{\centering{THE SM AND NLO MULTILEG AND SM MC WORKING GROUPS:\\
          \textbf{Summary Report}}}

\author{\underline{Convenors}:
J.~Alcaraz Maestre$^{1}$,
G.~Heinrich$^{2}$,
J.~Huston$^{3}$, 
F.~Krauss$^{4}$,
D.~Ma\^itre,$^{4,5}$, 
E.~Nurse,$^{6}$,
R.~Pittau$^{7}$\\
\underline{Contributing authors}:
S.~Alioli$^{8}$, 
J.~R.~Andersen$^{9}$, 
R.~D.~Ball$^{10}$,
A.~Buckley$^{11}$,
M.~Cacciari,$^{12,13}$,   
F.~Campanario$^{14}$, 
N.~Chanon$^{15}$,  
G.~Chachamis$^{16}$,
V.~Ciulli$^{17}$, 
F.~Cossutti$^{18}$,
G.~Cullen$^{19}$,
A.~Denner$^{20}$, 
S.~Dittmaier$^{21}$, 
J.~Fleischer$^{22}$, 
R.~Frederix$^{23}$, 
S.~Frixione$^{5,24,25}$,
J.~Gao$^{26}$,
L.~Garren$^{27}$, 
S.~Gascon-Shotkin$^{28}$, 
N.~Greiner$^{2}$,
J.~P.~Guillet$^{29}$, 
T.~Hapola$^{9}$,
N.~P.~Hartland$^{10}$,
G.~Heinrich$^{2}$,
G.~Hesketh$^{6}$,
V.~Hirschi$^{24}$,
H.~Hoeth$^{4}$,
J.~Huston$^{30}$, 
T.~Je\v{z}o$^{31}$, 
S.~Kallweit$^{23}$,
K.~Kova\v{r}\'{\i}k$^{14}$,
F.~Krauss$^{4}$,
A.~Kusina$^{26}$,
Z.~Liang$^{26}$,
P.~Lenzi$^{1}$,
L.~L{\"o}nnblad$^{32}$, 
J.~J.~Lopez-Villarejo$^{33,5}$,
G.~Luisoni$^{4}$,
D.~Ma\^itre$^{4,5}$,
F.~Maltoni$^{34}$,
P.~Mastrolia$^{2,35}$, 
P.~M.~Nadolsky$^{26}$,
E.~Nurse,$^{6}$,
C.~Oleari$^{36}$,
F.~I.~Olness$^{26}$,
G.~Ossola$^{37,38}$, 
E.~Pilon$^{29}$,
R.~Pittau$^{7}$,
S.~Pl\"atzer$^{39}$,
S.~Pozzorini$^{23}$,
S.~Prestel$^{32}$,
E.~Re$^{4}$,
T.~Reiter$^{2}$, 
T.~Riemann$^{19}$,
J.~Rojo$^{5}$,
G.~P.~Salam$^{5,39,12}$, 
S.~Sapeta$^{4,12}$,
I.~Schienbein$^{31}$,
M.~Sch{\"o}nherr$^{4}$,
H.~Schulz$^{41}$, 
M.~Schulze$^{42}$, 
M.~Schwoerer$^{43}$,  
P.~Skands$^{5}$,
J.~M.~Smillie$^{10}$, 
G.~Somogyi$^{19}$,
G.~Soyez$^{44}$, 
T.~Stavreva$^{31}$,
I.~W.~Stewart$^{45}$,
M.~Stockton$^{46}$, 
Z.~Sz\H or$^{47}$,
F.~J.~Tackmann$^{39}$, 
P.~Torrielli$^{24}$,
F.~Tramontano$^{5}$,
M.~Tripiana$^{48}$, 
Z.~Tr\'ocs\'anyi$^{47}$,
M.~Ubiali$^{49}$,
V.~Yundin$^{50}$, 
S.~Weinzierl$^{51}$,
J.~Winter$^{5}$,
J.~Y.~Yu$^{31,26}$,
K.~Zapp$^{4}$
\mbox{} }
\institute{\centering{\small
$^{1}$ CERN, CH-1211 Geneva 23, Switzerland \\
$^{2}$ Max-Planck Insitut f\"ur Physik,  F\"ohringer Ring, 6, D-80805 M\"unchen, Germany\\
$^{3}$ Physics and Astronomy Department, Michigan State University, East Lansing, MI, 48824 USA\\
$^{4}$ IPPP, University of Durham, Science Laboratories, South Rd, Durham DH1 3LE, UK\\
$^{5}$ PH-TH Department, CERN, CH-1211 Geneva 23, Switzerland\\
$^{6}$ Universtiy College London, Gower Street, WC1E 6BT, London, UK\\ 
$^{7}$ Departamento de F\'{\i}sica Te\'orica y del Cosmos, Universidad de Granada\\
$^{8}$ Ernest Orlando LBNL, University of California, Berkeley, CA 94720, USA\\
$^{9}$ CP$^3$-Origins, University of Southern Denmark, Campusvej 55, DK-5230 Odense M, Denmark\\
$^{10}$ Tait Institute, School of Physics and Astronomy, University of Edinburgh,
EH9 3JZ, UK\\
$^{11}$ PPE Group, School of Physics, University of Edinburgh, EH9 3JZ, UK\\
$^{12}$ LPTHE, UPMC Univ.~Paris 6 and CNRS UMR 7589, Paris, France\\  
$^{13}$ Universit\'e Paris Diderot, Paris, France\\
$^{14}$ Institut f\"ur Theoretische Physik, Universit\"at Karlsruhe, KIT, 76128 Karlsruhe, Germany\\
$^{15}$ Institute for particle physics, ETH Z\"urich, Switzerland\\
$^{16}$ Paul Scherrer Institut, CH-5232 Villigen PSI, Switzerland\\
$^{17}$ Universit\`a di Firenze and INFN, Sezione di Firenze, via G. Sansone 1, 50019 Sesto F. (FI), Italy \\
$^{18}$ INFN, Sezione di Trieste, Via Valerio 2, 34127 Trieste, Italia\\
$^{19}$ Deutsches Elektronen-Synchrotron, DESY, Platanenallee 6, 15738 Zeuthen, Germany \\
$^{20}$ Universit\"at W\"urzburg, Institut f\"ur Theoretische Physik und Astrophysik, 97074 W\"urzburg, Germany \\
$^{21}$ Albert-Ludwigs-Universit\"at Freiburg, Physikalisches Institut, 79104 Freiburg, Germany \\
$^{22}$ Fakult\"at f\"ur Physik, Universit\"at Bielefeld, Universit\"atsstr. 25, 33615 Bielefeld, Germany\\
$^{23}$ Institut f\"ur Theoretische Physik, Universit\"at Z\"urich, Winterthurerstrasse 190, CH-8057 Z\"urich, Switzerland\\
$^{24}$ ITPP, EPFL, CH-1015 Lausanne, Switzerland\\
$^{25}$ On leave of absence from INFN, Sezione di Genova, Italy \\
$^{26}$ Department of Physics, Southern Methodist University, Dallas, TX 75275-0175, USA\\
$^{27}$ Fermi National Accelerator Laboratory, Batavia, USA\\
$^{28}$ Universit\'e de Lyon, Universit\'e Claude Bernard Lyon 1/Institut de Physique Nucl\'eaire de Lyon IN2P3-CNRS\\
$^{29}$ LAPTH, 9 Chemin de Bellevue, B.P. 110, Annecy-le-Vieux 74951, France\\
$^{30}$ Department of Physics and Astronomy, Michigan State University, United States of America\\
$^{31}$ Laboratoire de Physique Subatomique et de Cosmologie, Universit\'e Joseph Fourier/CNRS-IN2P3/INPG,53 Avenue des Martyrs, 38026 Grenoble, France\\
$^{32}$ Department of Astronomy and Theoretical Physics, Lund University,  S\"{o}lvegatan 14A, SE-22362 Lund, Sweden\\
$^{33}$  Madrid, Autonoma U. \& IFT, 28049 Cantoblanco, Spain \\
$^{34}$ Centre for Cosmology, Particle Physics and Phenomenology (CP3) Universit\'e catholique de Louvain, B-1348 Louvain-la-Neuve, Belgium\\
$^{35}$ Dipartimento di Fisica, Universita di Padova, Italy \\
$^{36}$ Universit\`a di Milano-Bicocca and INFN, Sezione di Milano-Bicocca, 20126 Milan, Italy\\
$^{37}$ Physics Department, New York City College of Technology, City University of New York, USA \\ 
$^{38}$ The Graduate School and University Center, City University of New York, USA \\
$^{39}$ Theory Group, Deutsches Elektronen-Synchrotron (DESY), D-22607 Hamburg, Germany\\
$^{40}$ Department of Physics, Princeton University, Princeton, NJ 08544, USA\\
$^{41}$ Physics Dept., Berlin Humboldt University, Germany\\
$^{42}$ High Energy Physics Division, Argonne National Laboratory, Argonne, IL 60439, USA\\
$^{43}$ LAPP, 9 Chemin de Bellevue, B.P. 110, Annecy-le-Vieux 74951, France\\
$^{44}$ Institut de Physique Th\'eorique, CEA Saclay, France\\
$^{45}$ Center for Theoretical Physics, Massachusetts Institute of Technology, Cambridge, MA 02139, USA\\
$^{46}$ Department of Physics, McGill University, Montreal QC, Canada\\
$^{47}$ University of Debrecen and Institute of Nuclear Research of HAS, H-4001 P.O.Box 51, Hungary\\
$^{48}$ Instituto de F\'isica de La Plata, UNLP-CONICET, La Plata, Argentina\\
$^{49}$ Institut f\"ur Theoretische Teilchenphysik und Kosmologie, RWTH Aachen University, D-52056 Aachen, Germany\\
$^{50}$ Niels Bohr International Academy and Discovery Center, Niels Bohr Institute, University of Copenhagen, Blegdamsvej 17, DK-2100, Copenhagen, Denmark\\ 
$^{51}$ Institut f{\"u}r Physik, Universit{\"a}t Mainz, D -55099 Mainz, Germany\\
}}

\maketitle

\begin{abstract}
The 2011 Les Houches workshop was the first to confront LHC data. In the two years since the previous workshop there have been significant advances in both soft and hard QCD, particularly in the areas of multi-leg NLO calculations, the inclusion of those NLO calculations into parton shower Monte Carlos, and the tuning of the non-perturbative parameters of those Monte Carlos. These proceedings describe the theoretical advances that have taken place, the impact of the early LHC data, and the areas for future development.
\end{abstract}

 \begin{center}
   \textit{Report of the SM and NLO Multileg and SM MC Working Groups 
     for the Workshop ``Physics at TeV
     Colliders'', Les Houches, France, 31 May--8 June, 2011.  }
\end{center}

\newpage
  
\setcounter{tocdepth}{1}
\tableofcontents
\setcounter{footnote}{0}


\part[INTRODUCTION]{INTRODUCTION}


                                                                                 
The workshop in 2011 was the first for which the long-awaited LHC data (at 7     
TeV) was available                                                               
for analysis and comparison to theory. Even though of limited statistical        
power  compared to the ultimate goals of the LHC, this data accesses             
a very wide kinematic range, and probes regions where multiple scales are        
important. The presence of large scales for some processes, on the TeV level,    
points to the importance of electroweak corrections, which have been             
calculated only for some of the important processes. The first hints of a        
Higgs boson have now been observed. In order to search for signs of              
New Physics, as well as to completely understand the Standard Model at the       
LHC, it is important to                                                          
understand the {\it perturbative framework} at the LHC. The data taken so far    
provides many                                                                    
challenges for perturbative QCD predictions; and it is clear that New            
Physics, if it is present in current data, is hiding well.                              
                                                                                 
On the theoretical side, there has been a great deal of                          
productivity in the area of multi-particle calculations at next-to-leading         
order  (NLO)  and  next-to-next-to-leading order (NNLO).                                            
NLO is the first order at which the normalization,                               
and in some cases the shape, of perturbative cross sections can be considered    
reliable~\cite{Campbell:2006wx}.  A full                                         
understanding for both Standard Model and beyond the Standard Model physics      
at the LHC requires the development of fast, reliable programs for the           
calculation of multi-parton final states at NLO. There have been many            
advances in                                                                      
the development of NLO techniques, especially in the area of                     
automation~\cite{Agrawal:2011tm,Belanger:2003sd,Berger:2008sj,Badger:2010nx,Mastrolia:2010nb,Ellis:2011cr,Hirschi:2011pa,Campbell:2011cu,Bevilacqua:2011xh,Reina:2011mb,Cullen:2011ac}. 

Some of these approaches also allow for
relatively easy~\cite{Alioli:2010xd,Garzelli:2011vp,Cullen:2011ac} and/or
automatic~\cite{Frederix:2011zi} inclusion of the NLO matrix elements into
parton shower Monte Carlo programs.
For more details we refer to the individual contributions in these proceedings.

A prioritized list of NLO (and some NNLO) cross sections was assembled at Les Houches in         
2005~\cite{Buttar:2006zd} and added to in 2007~\cite{Bern:2008ef} and             
2009~\cite{Binoth:2010ra}. This list                                                          
includes cross sections which are experimentally important, and which are        
theoretically feasible (if difficult) to calculate. As we stand now,             
basically all NLO                                                                    
$2\rightarrow 3$  and $2\rightarrow 4$ cross sections of interest have been        
calculated, see Tables~\ref{tab:wishlist1},\ref{tab:wishlist2} below, 
and even some processes which were not on the 2009 wishlist  
are available at NLO, see e.g.~\cite{Bern:2011ep,Ita:2011wn,Campanario:2011ud,Berger:2010zx,Bevilacqua:2010qb,Dittmaier:2011ti,Denner:2010jp}.
The success of                                                                               
automation techniques means that future NLO calculations of similar              
complexity can be completed                                                      
without the man-years of labor previously required. Thus, we do not add to       
the NLO wish list in 2011.                                                       
Instead, we comment on calculations needed at NNLO, and processes at NLO for     
which it is important to                                                         
calculate the impact of electroweak corrections, and/or the influence of         
interference effects with                                                        
other processes with the same final state.                                       
                                                                                 
For many of the processes calculated at the LHC (such as for Higgs               
production), it is important either to apply a veto for the production of        
extra jets, or to bin the analysis results according to the jet multiplicity.    
While such cuts are useful for dealing with the experimental backgrounds, the    
exclusivity of the cross sections results in increases to the theoretical        
uncertainties obtained for the corresponding inclusive results, see e.g.~\cite{Stewart:2011cf}. 
The impact of    
such cuts is explored in the contribution of Stewart and Tackmann in these proceedings.               
                                                                                                                                                                  
\begin{table}                                                                    
  \begin{center}                                                                 
     \begin{tabular}{|l|l|}                                                      
\hline \hline                                                                    
Process ($V\in\{Z,W,\gamma\}$) & Comments\\                                      
\hline                                                                           
Calculations completed since Les Houches 2005&\\                                 
\hline                                                                           
&\\                                                                              
1. $pp\to VV$\,jet & $WW$\,jet completed by                                          
Dittmaier/Kallweit/Uwer~\cite{Dittmaier:2007th,Dittmaier:2009un};\\              
 &                                                                               
Campbell/Ellis/Zanderighi~\cite{Campbell:2007ev}.\\                              
 &                                                                               
$ZZ$\,jet completed by \\                                                          
&                                                                                
Binoth/Gleisberg/Karg/Kauer/Sanguinetti~\cite{Binoth:2009wk}\\                   
& $WZ$\,jet,$W\gamma$\,jet  completed by  Campanario et al.\,\cite{Campanario:2009um,Campanario:2010hp}     \\                
2. $pp \to$ Higgs+2\,jets & NLO QCD to the $gg$ channel \\                         
& completed by Campbell/Ellis/Zanderighi~\cite{Campbell:2006xx};\\               
& NLO QCD+EW to the VBF channel\\                                                
& completed by                                                                   
Ciccolini/Denner/Dittmaier~\cite{Ciccolini:2007jr,Ciccolini:2007ec} \\ 
& Interference QCD-EW in VBF channel~\cite{Andersen:2007mp,Bredenstein:2008tm} \\          
3. $pp\to V\,V\,V$ & $ZZZ$ completed                                             
by Lazopoulos/Melnikov/Petriello~\cite{Lazopoulos:2007ix}                        
 \\                                                                              
 & and $WWZ$ by Hankele/Zeppenfeld~\cite{Hankele:2007sb},\\                       
 & see also Binoth/Ossola/Papadopoulos/Pittau~\cite{Binoth:2008kt}  \\ 
 & VBFNLO~\cite{Arnold:2011wj,Arnold:2008rz} meanwhile also contains  \\
 & $WWW,ZZW,ZZZ,WW\gamma,ZZ\gamma,WZ\gamma,W\gamma\gamma,Z\gamma\gamma,$\\    
&$\gamma\gamma\gamma, W\gamma \gamma j $~\cite{Bozzi:2011wwa, Bozzi:2011en,Bozzi:2010sj,Bozzi:2009ig,Campanario:2008yg,Campanario:2011ud}\\                                                                              
&\\                                                                              
4. $pp\to t\bar{t}\,b\bar{b}$ &  relevant for $t\bar{t}H$, computed by\\          
 &                                                                               
Bredenstein/Denner/Dittmaier/Pozzorini~\cite{Bredenstein:2009aj,Bredenstein:2010rs} \\                                                                        
 & and Bevilacqua/Czakon/Papadopoulos/Pittau/Worek~\cite{Bevilacqua:2009zn}      
\\                                                                               
5. $pp \to V$+3\,jets & $W$+3\,jets calculated by the Blackhat/Sherpa~\cite{Berger:2009ep}     
\\                                                                               
 & and Rocket~\cite{Ellis:2009zw} collaborations\\ 
 &$Z$+3jets  by Blackhat/Sherpa~\cite{Berger:2010vm}  \\                                                                                                   
 \hline                                                                          
Calculations remaining from Les Houches 2005&\\                                  
\hline                                                                           
&\\                                                                              
6. $pp\to t\bar{t}$+2jets & relevant for $t\bar{t}H$, computed by  \\             
& Bevilacqua/Czakon/Papadopoulos/Worek                                           
~\cite{Bevilacqua:2010ve,Bevilacqua:2011aa}                                                        
\\                                                                               
7. $pp\to VV\,b\bar{b}$, & Pozzorini et al.\cite{Denner:2010jp},Bevilacqua et
al.\cite{Bevilacqua:2010qb}  \\
8. $pp\to VV$+2jets  & $W^+W^+$+2jets~\cite{Melia:2010bm},
$W^+W^-$+2jets~\cite{Melia:2011dw,Greiner:2012im},  \\
& VBF contributions calculated by \\                                             
& (Bozzi/)J\"ager/Oleari/Zeppenfeld~\cite{Jager:2006zc,Jager:2006cp,Bozzi:2007ur}                                                                               
\\                                                                               
\hline                                                                           
NLO calculations added to list in 2007&\\                                        
\hline                                                                           
&\\                                                                              
9. $pp\to b\bar{b}b\bar{b}$ &Binoth et al.~\cite{Binoth:2009rv,Greiner:2011mp} \\                                            
&\\                                                                              
\hline                                                                           
NLO calculations added to list in 2009&\\                                        
\hline                                                                           
&\\                                                                              
10. $pp \to V$\,+\,4 jets & top pair production, various new physics signatures\\ 
& Blackhat/Sherpa: $W$+4jets~\cite{Berger:2010zx},  $Z$+4jets~\cite{Ita:2011wn} \\
& see also HEJ~\cite{Andersen:2011hs} for $W+n$jets\\
11. $pp \to W b \bar{b}j$ & top, new physics signatures,  
 Reina/Schutzmeier~\cite{Reina:2011mb}\\                      
12. $pp \to t\bar{t}t\bar{t}$ & various new physics signatures \\
&\\   
\hline
also completed: &\\
$pp\to  W\,\gamma\gamma$ \,jet &Campanario/Englert/Rauch/Zeppenfeld\,\cite{Campanario:2011ud}\\
$pp\to $ 4\,jets &   Blackhat/Sherpa~\cite{Bern:2011ep}  \\ 
 &\\       
\hline 
\end{tabular}
\end{center}
\caption{The updated experimenter's wishlist for LHC processes                   
\label{tab:wishlist1}}                                                            
\end{table} 

\begin{table}                                                                    
  \begin{center}                                                                 
     \begin{tabular}{|l|l|}                                                      
\hline                                                                   
Calculations beyond NLO added in 2007&\\                                         
\hline                                                                           
&\\                                                                              
13. $gg\to W^*W^*$ ${\cal O}(\alpha^2\alpha_s^3)$& backgrounds to Higgs\\        
14. NNLO $pp\to t\bar{t}$ & normalization of a benchmark process\\               
15. NNLO to VBF and $Z/\gamma$+jet  & Higgs couplings and SM benchmark\\ 
&\\                                                                                     
\hline                                                                           
Calculations including electroweak effects&\\                                    
\hline                                                                           
&\\                                                                              
16. NNLO QCD+NLO EW for $W/Z$ & precision calculation of a SM benchmark\\
NLO EW to $W/Z$ &\cite{Kuhn:2007cv,Hollik:2007sq}\\
NLO EW to $W/Z$+jet&\cite{Denner:2009gj,Denner:2011vu}\\ 
NLO EW to $WH/ZH$ &\cite{Denner:2011id}\\       
&\\                                                                              
\hline                                                                           
\hline                                                                           
\end{tabular}                                                                    
\end{center}                                                                     
\caption{The updated experimenter's wishlist for LHC processes continued                  
\label{tab:wishlist2}}                                                            
\end{table}                                                                      
                                                                                  
Much of the complexity for multi-parton NLO processes consists of the            
calculation of the non-leading color contributions. Such contributions           
typically contribute only at the level of a few percent 
and approximations to the non-leading color contributions should be accurate     
within a percent or so~\cite{Berger:2009zg,Melnikov:2009wh,Berger:2009ep}. 
So it may be more time-prudent for groups carrying out such calculations to estimate the           
non-leading color effects before carrying out the full calculation.       
                                                                                 
To reach full utility, the codes for any of these                                
complex NLO calculations should be made public and/or the authors should         
generate \textsc{ROOT} ntuples                                                            
providing the parton level event information from which experimentalists         
can assemble any cross sections of interest. Where possible, decays (with        
spin correlations) should be included. A \textsc{ROOT} output option is especially        
useful where the creation of a user-friendly NLO program may be very             
time-consuming.                                                                  
We now have some experience with the use of \textsc{ROOT} ntuples with both MCFM and      
Blackhat+Sherpa calculations.                                                    
 The latter, in particular, does not exist as a public program, while \textsc{ROOT}       
tuples have been made available for NLO W/Z + n jet multiplicities (with n up    
to 4) for W/Z + jets, and (also for n up to 4) for inclusive jet production.     
The estimation of the {\it correct} scale for use in multi-parton NLO            
calculations, and the proper evaluation of the uncertainty on this scale, is     
more complex than for simpler calculations. The use of \textsc{ROOT} ntuples can make     
these evaluations easier to carry out.                                           
A contribution describing their use has been included in these proceedings.      
                                                                                 
While NLO is sufficient for most predictions,  it is also crucial to             
understand certain critical cross sections at NNLO. To date, NNLO                
calculations have been carried out primarily for  
processes in $e^+e^-$ annihilation~\cite{GehrmannDeRidder:2007hr,GehrmannDeRidder:2008ug,Weinzierl:2009ms}, 
and in hadronic collisions for $2\rightarrow 1$ processes,        
with the exception of VH~\cite{Brein:2003wg,Brein:2011vx,Ferrera:2011bk} 
and $\gamma\gamma$ production~\cite{Catani:2011qz}. 

To calculate a $2\to 2$ scattering process at NNLO, 
the divergent contributions arising from the              
tree-level $2\rightarrow 4$, the one-loop $2\rightarrow 3$            
and the two-loop $2\rightarrow 2$ subprocesses have to be 
properly subtracted and cancelled, such that the finite remainders 
can be combined into a parton-level event generator. 
To combine the three contributions, an infra-red subtraction scheme 
for unresolved real radiation is required.   
Several approaches have been used and are being further developed:
antenna subtraction~\cite{GehrmannDeRidder:2005cm}, 
which currently is extended to hadronic and semi-hadronic 
initial
states~\cite{Daleo:2009yj,Gehrmann:2009vu,Gehrmann:2011wi,Abelof:2011jv,Boughezal:2010mc}, 
a method based on sector decomposition appplied to 
real radiation~\cite{Heinrich:2002rc,Anastasiou:2003gr,Binoth:2004jv} 
where the decomposition is guided by the physical 
singularity structure~\cite{Czakon:2010td,Boughezal:2011jf}, 
$q_T$-subtraction~\cite{Catani:2007vq}, which is very elegant but appplicable only to 
colourless final states, and the one of \cite{Bolzoni:2010bt} described in these proceedings.

Further, two-loop amplitudes are interesting in     
their own right from a field theory point of view, 
for example to study asymptotic behaviour, or to 
gain insights into the all-order infared structure of massless 
field theories. 

Below we construct a table of calculations needed at the LHC, and which are      
feasible within the next few years.                                              
Certainly, results for  inclusive cross sections at NNLO will be 
easier to achieve than  differential distributions, 
but most groups are working towards a partonic Monte Carlo program 
capable of producing fully differential distributions for measured observables.
                                                                                 
\begin{itemize}    
\item $t\bar{t}$ production:\\  
needed for accurate background estimates, top mass measurement, 
top quark asymmetry  (which is zero at tree level, so NLO is the 
leading non-vanishing order for this observable, and a discrepancy 
of theory predictions with Tevatron data 
needs to be understood).  Several groups are already well on the way to 
complete NNLO results for $t\bar{t}$
production~\cite{Bonciani:2010mn,Abelof:2011ap,Czakon:2011ve,Czakon:2011xx}.

\item $W^+W^-$ production:\\
importand background to Higgs search. 
At the LHC, $gg\to WW$ is the dominant subprocess, but
$gg\to WW$ is a loop-induced process, such that two loops 
need to be calculated to get a reliable estimate of the cross section. 
Advances towards the full two-loop result are 
reported in~\cite{Chachamis:2010zz,Chachamis:2008yb}.                                                                       

\item inclusive jet/dijet production:\\
NNLO parton distribution function (PDF) fits are starting to become the norm     
for predictions and comparisons at the LHC. Paramount in these global fits is the    
use of inclusive jet production to tie down the behavior of the gluon            
distribution, especially at high $x$. However, while the other essential         
processes used in the global fitting are known to NNLO, the inclusive jet        
production cross section is only known at NLO. Thus, it is crucial for           
precision predictions for the LHC for the NNLO corrections for this process      
to be calculated, and to be available for inclusion in the global PDF fits. 
First results for the real-virtual and double real corrections to 
gluon scattering 
can be found in~\cite{GehrmannDeRidder:2011aa,Pires:2010jv}.     
                                                                                 
\item V+1 jet production:\\  
$W/Z/\gamma$ + jet production form the signal channels (and backgrounds) for     
many key physics processes, for both SM and BSM. In addition, they also serve    
as calibration tools for the jet energy scale and for the crucial                
understanding of the missing transverse energy resolution.  
The two-loop amplitudes for this process are
known~\cite{Garland:2001tf,Garland:2002ak}, 
therefore it can be calculated once the parts involving unresolved real
radiation are available.

\item V+$\gamma$ production:\\
important signal/background processes for Higgs and New Physics searches.
The two-loop helicity amplitudes for 
$q\bar q \to W^\pm \gamma$ and $q\bar q \to Z^0 \gamma$  
recently have become available~\cite{Gehrmann:2011ab}.                   
                                                                                 
\item Higgs+1 jet production:\\  
As mentioned previously, events in many of the experimental Higgs analyses       
are separated by the number of additional jets accompanying the Higgs boson.     
In many searches, the Higgs + 0 jet and Higgs + 1 jet bins contribute            
approximately equally to the sensitivity. It is thus necessary to have the       
same theoretical accuracy for the Higgs + 1 jet cross section as already         
exists for the inclusive Higgs cross section, i.e. NNLO.  
The two-Loop QCD Corrections to the Helicity Amplitudes for
$H \to$ 3 partons are already available~\cite{Gehrmann:2011aa}.
\end{itemize}                     
                                                                                 
The contributions in this document are arranged as follows.
In section~\ref{part:nlo}, various developments concerning techniques for NLO and NNLO calculations are described, in particular in view of
providing automated tools for NLO corrections.
In section~\ref{part:pdf}, issues related to parton distribution functions are discussed. Section ~\ref{part:pheno} contains phenomenological studies
of observables and uncertainties, based on theory input where
higher order corrections obtained by different approaches are available.
Section~\ref{part:corr} includes phenomenological studies on the definition of experimental observables and corrections applied to data.
Finally Section~\ref{part:mc} discusses issues related to the tuning of Monte Carlos and standardised Monte Carlo output formats.




\part[NLO AUTOMATION AND (N)NLO TECHNIQUES]{NLO AUTOMATION AND (N)NLO TECHNIQUES}
\label{part:nlo}
\section[PJFry -- a C++ package for tensor reduction of one-loop Feynman integrals]
{PJFRY -- A C++ PACKAGE FOR TENSOR REDUCTION OF ONE-LOOP FEYNMAN INTEGRALS \protect\footnote{Contributed by: J.~Fleischer, T.~Riemann, V.~Yundin}}
{\graphicspath{{pjfry-lesHouches2011-riemann-1/}}
%
%

\newcommand{\bea}{\begin{eqnarray}}
\newcommand{\eea}{\end{eqnarray}}


\title{%
PJFry -- a C++ package for tensor reduction of one-loop Feynman integrals 
}

\author{%
J. Fleischer$^1$, T. Riemann$^2$, V. Yundin$^3$
}
\institute{%
$^1$Fakult\"at f\"ur Physik, Universit\"at Bielefeld, Universit\"atsstr. 25, 
33615
Bielefeld, Germany
\\
$^2$Deutsches Elektronen-Synchrotron, DESY, Platanenallee
  6, 15738 Zeuthen, Germany
\\
$^3$Niels Bohr International Academy and Discovery Center, Niels Bohr
Institute, University of Copenhagen, Blegdamsvej 17, DK-2100, Copenhagen,
Denmark
}


\begin{abstract}
The C++ package PJFry 1.0.0 \cite{pjfry:2011aa,Fleischer:2010sq} -- a one loop
tensor integral library -- is introduced.
We use an algebraic approach to tensor reduction.
As a result, the tensor integrals are presented in terms of scalar one- to
four-point functions,  which have to be provided by an external library, e.g.
QCDLoop/FF 
or OneLOop
or LoopTools/FF.
The reduction is implemented until five-point functions of rank five.
A numerical example is shown, including a special treatment for small or
vanishing inverse four-point Gram determinants.
Future modules of PJFry might cover the treatment of $n$-point functions with
$n \geq 6$; the corresponding formulae are worked out.
Further, an extremely efficient approach to tensor reduction relies on 
evaluations of complete contractions of the tensor integrals with external
momenta. For this, we worked out an algorithm for the analytical
evaluation of sums over products of signed minors with scalar products of
chords, i.e. differences of external momenta. As a result, the usual multiple
sums over tensor coefficients are replaced for the numerical evaluation by
compact combinations of the basic scalar functions.  
\end{abstract}


\subsection{PJFry\label{pjfry}}
The goal of the C++ package PJFry is a stable and fast open-source
implementation of one-loop tensor reduction of Feynman integrals 
\begin{eqnarray}\label{definition}
 I_n^{\mu_1\cdots\mu_R} &=&  ~~
C(\epsilon)
~
\int \frac{d^d
k}{i\pi^{d/2}}~~\frac{\prod_{r=1}^{R} k^{\mu_r}}{\prod_{j=1}^{n}
(k-q_j)^2-m_j^2 +i\epsilon 
},
\end{eqnarray}
suitable for any physically relevant
kinematics.\footnote{An extended description of notations and of the formalism
may be found in
\cite{Fleischer:2010sq,Fleischer:2011nt,Fleischer:2011hc,Fleischer:2011bi}.
The normalization of PJFry follows that chosen in the scalar library. For
QCDLoop,
$C(\epsilon)
= \Gamma(1-2\epsilon)/[\Gamma(1+\epsilon)\Gamma^2(1-\epsilon)]$. }
The algorithm was invented in \cite{Fleischer:2010sq}. PJFry performs the
reduction of 5-point 1-loop tensor integrals up to rank 5. The
4- and 3-point tensor integrals are obtained as a by-product.
Main features are:
\begin{itemize}
 \item Any combination of internal or external masses
\item Automatic selection of optimal formula for each coefficient
\item Leading $()_5$ are eliminated in the reduction
\item Small $()_4$ are avoided using asymptotic expansions where appropriate
\item Cache system for tensor coefficients and signed minors
\item Interfaces for C, C++, FORTRAN and Mathematica
\item Uses QCDLoop \cite{Ellis:2007qk,vanOldenborgh:1990yc} or OneLOop
\cite{vanHameren:2010cp} for 4-dim scalar integrals
\item Available from the project webpage \url{https://github.com/Vayu/PJFry/}
\cite{pjfry:2011aa,Fleischer:2010sq}
\end{itemize}
The installation of PJFry may be performed following the instructions given at
the project webpage. 
\\
The project subdirectories are\\
./src       - the library source code\\
./mlink     - the MathLink interface\\
./examples  - the FORTRAN examples of library use, built with make check

\bigskip

A build on Unix/Linux and similar systems is done in a standard way by
sequential performing
   ./configure, make, make install.
See the INSTALL file for a detailed description of the ./configure options.

The functions for tensor coefficients for up to rank $R=5$
pentagon integrals are declared in the Mathematica interface:
\begin{verbatim}
In:= Names["PJFry`*"]

Out= {A0v0, B0v0, B0v1, B0v2, C0v0, C0v1, C0v2, C0v3, \
D0v0, D0v1, D0v2, D0v3, D0v4, E0v0, E0v1, E0v2, \
E0v3, E0v4, E0v5, GetMu2, SetMu2}
\end{verbatim}
The C++ and Fortran interface syntax is very close to that of e.g. LoopTools/FF:
{\small{
\begin{verbatim}
E0v3[i,j,k,p1s,p2s,p3s,p4s,p5s,s12,s23,s34,s45,s15,m1s,m2s,m3s,m4s,m5s,ep=0]
\end{verbatim}
}}
where:\footnote{One has to carefully control accuracies; e.g. the on-shell
conditions for massless particles have to be fulfilled with a numerical
precision expected by the scalar functions library in use; for QCDLoop this
means on default at least 10 digits.} 
\\
{\tt i,j,k} are indices of the tensor coefficient $(0 < i \leq  j \leq  k <
n)$, 
\\
{\tt p1s,p2s,...} are squared external masses $p_i^2$, 
\\
{\tt s12,s23,...} are Mandelstam invariants $(p_i + p_j)^2$,
\\
{\tt m1s,m2s,...} are squared internal masses $m_i^2$,
\\
{\tt  ep=0,-1,-2} selects the coefficient of the $\epsilon$-expansion.

The average evaluation time per phase-space point on a 2 GHz Core 2 
laptop for the evaluation of all 81 rank 5 tensor form-factors amounts to 2 ms.

A numerical example is shown, for a configuration as in figure
\ref{pjfry_fig3}, in figures \ref{pjfry_fig1} and \ref{pjfry_fig2} for a
five-point rank $R=4$ tensor coefficient in a region, where one of the 4-point
sub-Gram determinants vanishes [at $x=0$]:

$E_{3333}(0,0,-6{\times}10^4(x+1),0,0,%
10^4,-3.5{\times}10^4,2{\times}10^4,-4{\times}10^4,1.5{\times}10^4,%
             0,6550,0,0,8315)$

\noindent
The red curve is produced with standard PJFry, and the blue one with 
Passarino-Veltman [PV] reduction \cite{Passarino:1978jh}; we mention that for
the case treated here ($x \to 0$), the PV reduction is no standard option.
Our expansion in terms of higher dimensional scalar 3-point functions in case
of vanishing 4-point sub-Gram determinants uses only functions $I_3^{d+2l}$
\cite{Fleischer:2010sq}. These are tensor coefficients of the
pure $g^{\mu\nu}$ type \cite{Davydychev:1991va}, and so our method is different
from others with a mixed
numerical approach \cite{Binoth:2005ff} or  with use of additional tensor
coefficients \cite{Denner:2005nn}.

Tensor reduction by  PJFry is used as one option of the GoSam package
\cite{Cullen:2011ac}.
An older version of the  algorithm, as described in 
\cite{Diakonidis:2008ij}, has been implemented independently in
\cite{Reina:2011mb}.

\begin{figure}[t]
\begin{center}
\vspace*{5mm}\includegraphics[width=0.6\textwidth]{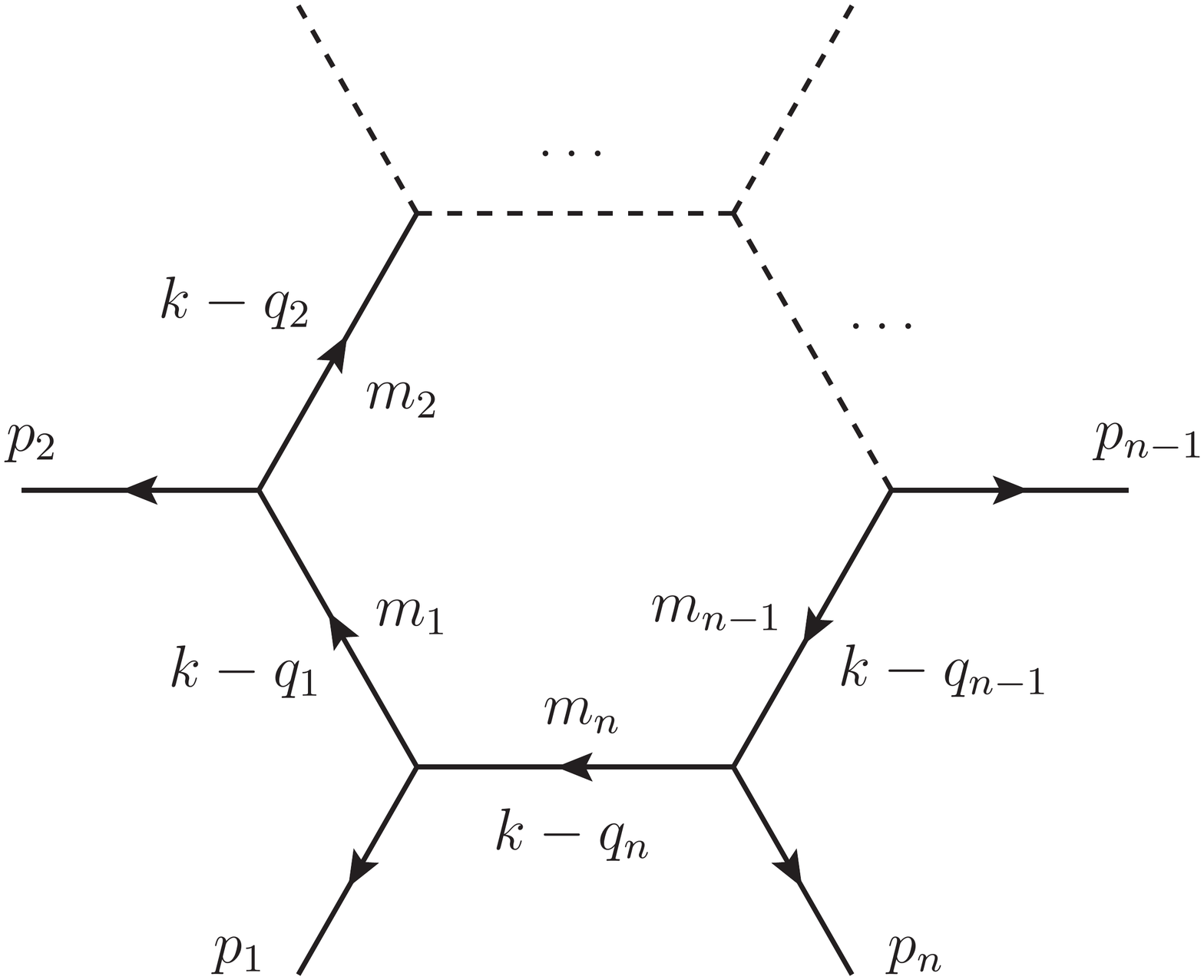}
 \caption{Momenta definitions for PJFry.
\label{pjfry_fig3}}
\end{center}
\end{figure}
\begin{figure}[h]
\begin{center}
\includegraphics[width=0.6\textwidth]{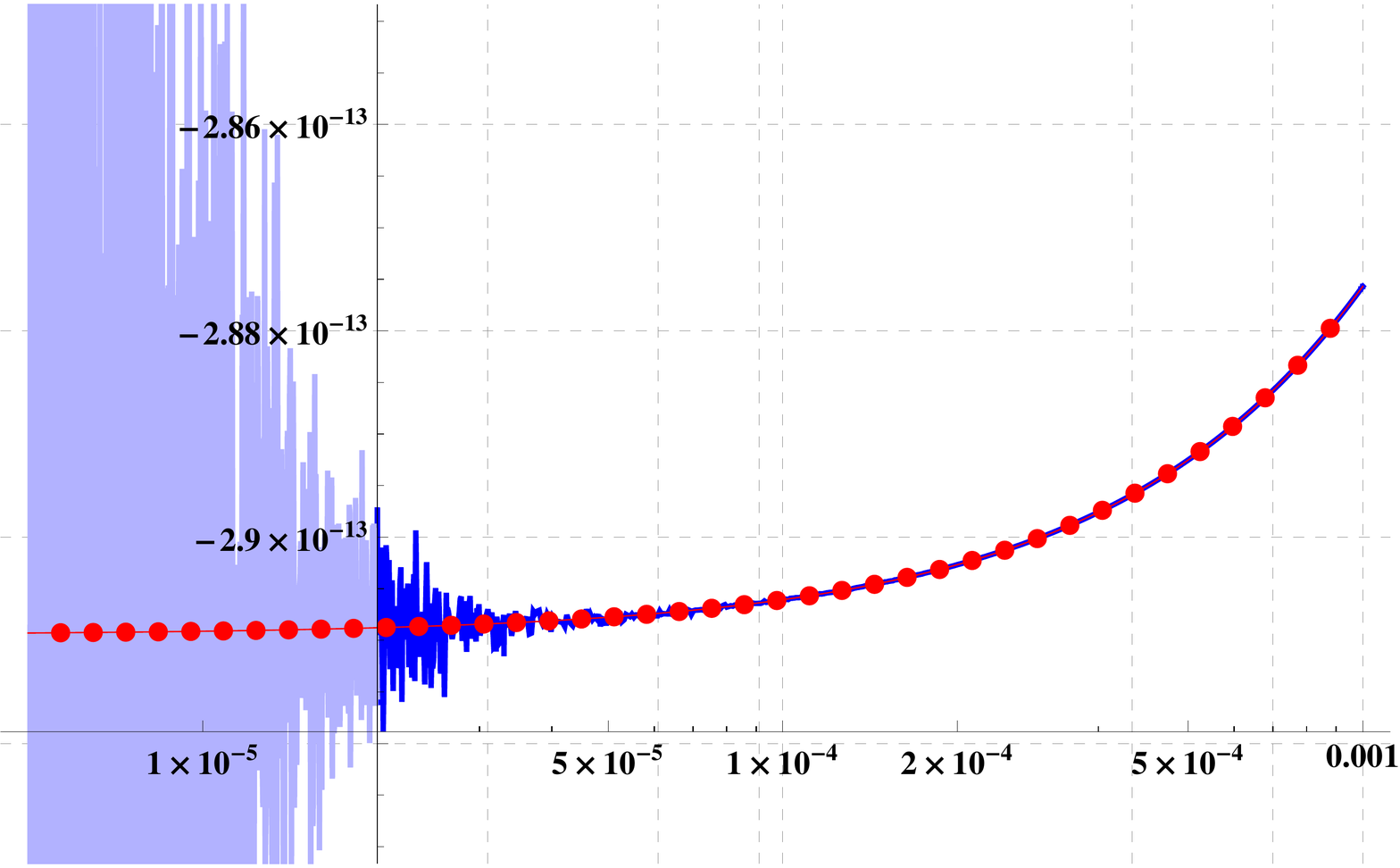}
\caption{Absolute accuracy of $E_{3333}$ in  the region of vanishing sub-Gram
determinant.
Blue curve: conventional Passarino-Veltman reduction, red curve:
PJFry.
\label{pjfry_fig1}}
\end{center}
\end{figure}

\begin{figure}[h]
\begin{center}
\includegraphics[width=0.6\textwidth]{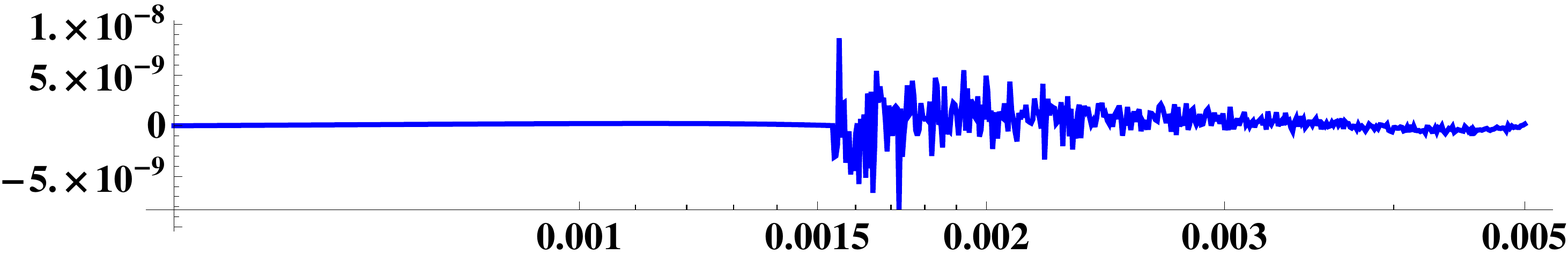}
\caption{Relative accuracy of $E_{3333}$ in  the region of vanishing sub-Gram
determinant.
At $x \sim 0.0015$,  PJFry switched to the asymptotic expansion.
\label{pjfry_fig2}}
\end{center}
\end{figure}

\subsection{POTENTIAL UPGRADES}
\subsubsection{Tensor reduction for higher-point functions}
So far, PJFry is foreseen for 5-point functions and simpler ones.
The extension to 6-point functions is known from e.g.
\cite{Binoth:2005ff,Denner:2005nn,Diakonidis:2008ij}.
In \cite{Fleischer:2011hc} we solve analytically generalized recursions for
$n\geq 6$, derived in \cite{Binoth:2005ff}:
\bea
I_n^{{\mu}_1 {\mu}_2 \dots {\mu}_R}= -\sum_{r=1}^n 
C_r^{{\mu}_1}(n) 
I_{n-1}^{{\mu}_2 \cdots {\mu}_R,r},
\label{npoint}
\eea
where in $I_{n-1}^{\mu,\cdots,r}$ the line $r$ is scratched.
The coefficients for 6-point functions are:
\bea
C_r^{{s},{\mu}}(6)=
\sum_{i=1}^5  \frac{1}{{0\choose s}_6}  {0r\choose {s}i}_6 ~ q_i^{\mu_1}
,~~{s}=0 \dots 6 ,
\label{6point}
\eea 
where the $q_i$ are chords, and ${0r\choose{s}i}_6$ etc. are signed minors with
arbitrary $s$. For
the 7-point and 8-point functions, we found several representations, among them
\bea
C_r^{st,\mu}(7)=\sum_{i=1}^{6} \frac{1}{{st\choose st}_7} {sti\choose str}_7
 ~ q_i^{\mu}
\label{C71}
\eea
and
\bea
C_r^{stu,\mu}(8)=\sum_{i=1}^{7} \frac{1}{{stu\choose stu}_8} {stui\choose
stur}_8  ~ q_i^{\mu}
\label{C81}
\eea
The upper indices $s,t$ and $u$ stand for the redundancy of the solutions and
can be freely chosen. 

\subsubsection{Evaluation of contracted tensor integrals using sums over signed
minors}
The contraction of a tensor integral with chords may be written as a sum over
basic scalar integrals (at a stage where they are free of tensor coefficient
indices),
multiplied by (multiple) sums over chords times signed minors.
If one may perform these sums algebraically, the method becomes very efficient.
And this has been systematically worked out in \cite{Fleischer:2011nt}, see also
\cite{Fleischer:2011bi}.

We reproduce here two 7-point examples.

The rank $R=2, 3$ integrals become by contraction
\bea
q_{a,\mu}q_{b,\nu}~I_7^{\mu \nu}&=&
\sum_{r,t=1}^{7} K^{ab,rt} I_5^{rt},
\label{tens2}
\\
q_{a,\mu}q_{b,\nu}q_{c,\lambda}~I_7^{\mu \nu \lambda}&=&
\sum_{r,t,u=1}^{7} K^{abc,rtu} I_4^{rtu},
\label{tens3}
\eea
where $I_5^{rt}$ and $I_4^{rtu}$ are scalar 5- and 4-point functions, arising
from the 7-point function by scratching lines $r,t,\ldots$ 
In the general case, we have at this stage higher-dimensional integrals
$I_n^{d+2l},
n=2,\ldots,5$, to be further reduced following the known scheme, if needed.
Here, the $I_5^{rt}$ have to be expressed by 4-point functions.

The expansion coefficients are factorizing here,
\bea
K^{ab,rt} &=& K^{a,r}K^{b,rt},
\\
K^{abc,rtu} &=& - K^{a,r}K^{b,rt}K^{c,rtu},
\label{k}
\eea
and the sums over signed minors have been  performed analytically:
\bea
K^{a,r}=\frac{1}{2}\left( {\delta}_{ar}-{\delta}_{7r} \right),
\eea

\bea
\label{appearance}
K^{b,rt}
&=&
\sum_{j=1}^{6} (q_b q_j) \frac{{rst\choose rsj}_7}{{rs\choose
rs}_7}
~~\equiv~~  \frac{\Sigma^{1,stu}_{b}} {{rs\choose rs}_7}
~~=~~ \frac{1}{2} \left( {\delta}_{bt}-{\delta}_{7t} \right)
-\frac{1}{2} \frac{{rs\choose ts}_7}{{rs\choose rs}_7} \left(
{\delta}_{br}-{\delta}_{7r} \right),
\eea

\bea
K^{a,stu}
&=&
\label{eq-wa-87}
\sum_{i=1}^{6}(q_a q_i)  {0stu\choose 0sti}_7
~~\equiv~~
\Sigma^{2,stu}_{a}
 \\ \nonumber
&=&
\frac{1}{2} 
 \Biggl\{{stu\choose st0}_7  \left(Y_{a7}-Y_{77} \right)
+{0st\choose 0st}_7\left({\delta}_{au}-{\delta}_{7u} \right)
-{0st\choose 0su}_7\left({\delta}_{at}-{\delta}_{7t} \right)
-{0ts\choose 0tu}_7\left({\delta}_{as}-{\delta}_{7s} \right) 
\Biggr\},
\eea
with
\bea
\label{yjk}
Y_{jk}=-(q_j-q_k)^2+m_j^2+m_k^2. 
\eea
Conventionally, $q_7=0$.

The sums may be found in eqns. (A.15) and (A.16) of
\cite{Fleischer:2011nt}.
The $s$ is redundant and fulfils $s  \neq r,b,7$ in $K^{b,rt}$. In
$K_0^{a,stu}$ it is $s,t,u = 1,\ldots 7$ with $s \neq u, t\neq u$.

\subsection*{ACKNOWLEDGEMENTS}
J.F. thanks DESY for kind hospitality.
Work is supported in part by Sonderforschungsbereich/Trans\-re\-gio SFB/TRR 9 of
DFG
``Com\-pu\-ter\-ge\-st\"utz\-te Theoretische Teil\-chen\-phy\-sik" and
European Initial Training Network LHCPHENOnet PITN-GA-2010-264564.


}

\section[The GoSam approach to Automated One-Loop Calculations]
{THE GOSAM APPROACH TO AUTOMATED ONE-LOOP CALCULATIONS \protect\footnote{Contributed by: G.~Cullen,
N.~Greiner,
G.~Heinrich,
G.~Luisoni,
P.~Mastrolia,
G.~Ossola,
T.~Reiter,
F.~Tramontano}}
{\graphicspath{{GoSam_Houches/}}

\title{The GoSam approach to Automated One-Loop Calculations}

\author{G.~Cullen$^1$,
N.~Greiner$^2$,
G.~Heinrich$^2$,
G.~Luisoni$^3$,
P.~Mastrolia$^{2,4}$,
G.~Ossola$^{5,6}$,
T.~Reiter$^{2}$,
F.~Tramontano$^{7}$}

\institute{
$^1$ Deutsches Elektronen-Synchrotron DESY, Zeuthen, Germany \\
$^2$ Max-Planck Insitut f\"ur Physik,  
M\"unchen, Germany\\
$^3$	IPPP, %
	University of Durham, %
UK\\
$^4$ Dipartimento di Fisica, Universita di Padova, Italy \\
$^5$ Physics Department, New York City College of Technology,
	City University of New York, 
	USA \\
$^6$ The Graduate School and University Center, City University of New York,
USA \\
$^7$ Theory Group, Physics Department, CERN, 
Switzerland
}


\newcommand{\GOSAM}{{\textsc{GoSam}}}
\newcommand{\SHERPA}{{\textsc{SHERPA}}}
\newcommand{\POWHEG}{{\texttt{POWHEG}}}
\newcommand{\QGRAF}{{\texttt{QGRAF}}}
\newcommand{\FORM}{{\texttt{FORM}}}
\newcommand{\SPINNEY}{{\texttt{spinney}}}
\newcommand{\HAGGIES}{{\texttt{haggies}}}
\newcommand{\SAMURAI}{{\textsc{SAMURAI}}}
\newcommand{\GOLEMVC}{{\texttt{Golem95C}}}
\newcommand{\PJFRY}{{\texttt{PJFRY}}}
\newcommand{\FORTRAN}{{\texttt{Fortran}}}
\newcommand{\PYTHON}{{\texttt{Python}}}
\newcommand{\UFO}{{\texttt{UFO}}}
\newcommand{\LANHEP}{{LanHEP}}
\newcommand{\CUTTOOLS}{{CutTools}}

\begin{abstract}
We describe the \GOSAM{} framework for the automated computation of multi-particle scattering 
amplitudes at the one-loop level. The amplitudes are generated explicitly in terms of
Feynman diagrams, and can be evaluated using either $d$-dimensional reduction at the integrand level or
tensor decomposition. \GOSAM{} can be used to compute one-loop QCD and EW corrections to
Standard Model processes, and it is ready to link generic model files for theories Beyond the Standard Model.
\end{abstract}

\subsection{Introduction and General Motivations}

In the last few years we observed major advances in the direction of constructing
packages for fully automated one-loop calculations, which profited from the new developments in the
field of NLO QCD calculations~\cite{Bern:2008ef,Binoth:2010ra}. The continuous improvement of the 
techniques for one-loop computations led to important new results for processes with many particles~\cite{Berger:2009ep,Berger:2009zg,KeithEllis:2009bu,Melnikov:2009wh,Berger:2010vm,
Bredenstein:2009aj,Bredenstein:2010rs,Bevilacqua:2009zn,Bevilacqua:2010ve,
Binoth:2009rv,Greiner:2011mp,Bevilacqua:2010qb,Bevilacqua:2011hy,Denner:2010jp,Melia:2010bm,Melia:2011dw,
Campanario:2011ud,Berger:2010zx,Ita:2011wn,Bern:2011ep}.

Very advanced calculations have been performed with improved algebraic reduction methods based on Feynman-diagrammatic algorithms, as well as with new numerical techniques based on the idea of reconstructing one-loop amplitudes from their unitarity cuts. These theoretical developments found an ideal counterpart in the integrand-level reduction algorithm, that allows for the reduction of any scattering amplitudes to scalar master integrals, simply by evaluating numerically the integrand at given fixed values of the integration momentum.
In both scenarios, to tackle the increase
in the complexity and in the number of diagrams that contribute to the amplitudes, 
automation becomes indispensable for processes with many external legs. 

The purpose of the present document is to illustrate the main features of \GOSAM{}~\cite{Cullen:2011ac}, a new framework which allows the automated calculation of one-loop scattering amplitudes for multi-particle processes. This approach combines the automated algebraic generation of $d$-dimensional unintegrated amplitudes obtained via Feynman diagrams, with the numerical integrand-level reduction provided by the $d$-dimensional extension~\cite{Ellis:2008ir, Mastrolia:2008jb, Mastrolia:2010nb} of the OPP integrand-level reduction method~\cite{Ossola:2006us, Ossola:2007bb, Ossola:2008xq} and improved tensorial techniques~\cite{Heinrich:2010ax, Reiter:2010md}. 

The integrands of the one-loop amplitudes are generated via Feynman diagrams, using \QGRAF~\cite{Nogueira:1991ex}, \FORM~\cite{Vermaseren:2000nd}, \SPINNEY~\cite{Cullen:2010jv} and \HAGGIES~\cite{Reiter:2009ts}. 
The only task required from the user is the preparation of an ``input card'' to start the generation of the source code and its compilation, without having to worry about internal details of the code generation. 
The individual program tasks are efficiently managed by python scripts. 
Concerning the reduction, the program offers the possibility to use either the $d$-dimensional extension of
the OPP method, as implemented in \SAMURAI~\cite{Mastrolia:2010nb}, or tensor reduction as implemented in
\GOLEMVC~\cite{Binoth:2008uq,Cullen:2011kv} interfaced through tensorial reconstruction at the integrand level~\cite{Heinrich:2010ax}.

\subsection{Algebraic approach to Automation}  

There are several approaches to the automated computation of multi-particle scattering amplitudes at the one-loop
level, which provide different recipes for the construction of multi-purpose tools. The goal of such tools is the evaluation of one-loop scattering amplitudes for any choice of particles in the initial and final states, in a fully automated manner. 

In the algebraic approach to multi-purpose automation, amplitudes can be generated from Feynman diagrams by employing 
tools for algebraic manipulation: already some time ago, the idea of automating NLO calculations has been pursued by public programs 
like FeynArts\,\cite{Hahn:2000kx} and \QGRAF~\cite{Nogueira:1991ex} 
for diagram generation and FormCalc/LoopTools~\cite{Hahn:1998yk} and 
{\small GRACE}~\cite{Belanger:2003sd} for the automated calculation of NLO corrections, primarily in the electroweak sector.

When we combine the algebraic generation with the integrand-level reduction, the set of algebraic operations required are quite different with respect to a traditional tensorial reduction. Since the target is to provide the numerical value of the numerator function at given values of integration momentum, we should aim at expressions for the unintegrated numerator that are easily evaluated numerically. To achieve this task, for example, expressions in terms of spinor products are particularly convenient.

We briefly list here some of the advantages of the "algebraic approach": i) the algebraic generation is executed separately from the numerical reduction, 
therefore algebraic manipulations  are possible before starting the numerical
integration; CPU-time can be spent, once for all at the beginning of the calculation, 
to optimize and reduce the size of the integrands that will be evaluated numerically 
several times later on during the reduction; ii) the algebraic method allows us to 
group sets of diagrams and cache all factors that do not depend on the integration 
momentum; iii) easy access to sub-parts of the computation; subsets of diagrams 
can be easily added or removed from the final results, simplifying comparisons and 
tests; iv) computer algebra can be performed in dimension $d$ using alternative 
regularization schemes; v) the choice between different reduction algorithms can 
be performed at run-time, providing flexibility and internal cross-checks.
In the next section we will briefly illustrate how these properties are used 
within \GOSAM{}. 

Important progress in a similar direction has been also recently 
achieved by means of FeynArts, FormCalc and 
LoopTools~\cite{Hahn:2010zi, Agrawal:2011tm} to provide 
amplitudes that can be processed using the integrand-level reduction 
provided by \CUTTOOLS~\cite{Ossola:2007ax} and/or \SAMURAI~\cite{Mastrolia:2010nb} 
or with the traditional Passarino-Veltman reduction~\cite{Passarino:1978jh}.

\subsection{A brief introduction to \GOSAM}  

\GOSAM{} produces in a fully automated way all the code required to perform the calculation of 
virtual one-loop amplitudes. The only task left to the user is the preparation of an ``input card'' which contains all the information related to the particular process namely initial and final particles, model, helicities, 
selection rules to exclude particular sets of diagrams, regularization scheme. The card also contains flags to
select the preferred reduction methods and some optimization flags to adapt the diagram generation
to the needs of the user.
 
There are three main steps that \GOSAM{} follows in order to prepare the code for the 
calculation: 
the generation of diagrams that contribute to the process, the optimization
and algebraic manipulation to simplify their expressions, and the writing of a FORTRAN code ready to be
used within a phase-space integration. It is important to remember that these steps 
will only be performed once.  
After the code is generated, the reduction of unintegrated amplitudes to 
linear combinations of scalar (master) integrals is fully embedded in the process 
and can be performed with different options, all available at run-time. 
Only the last part, namely the reduction and evaluation of master integrals, 
will be repeated for all the different phase-space points that contribute to
the cross-section.

\subsubsection{Diagram Generation} For the diagram generation both at tree level 
and one-loop level
we employ \QGRAF~\cite{Nogueira:1991ex} which we complemented by adding 
another filter over diagrams
implemented in \PYTHON{}. 
 This gives several advantages since it increases the ability of the code to
distinguish certain classes of diagrams and group them according to the sets of 
their propagators, in order to fully optimize the reduction.

At this stage \GOSAM{} generates three sets
of output files: an expression for each diagram for
\FORM~\cite{Vermaseren:2000nd}, \PYTHON{} code for drawing each diagram,
and \PYTHON{} code for computing the properties of the diagram. 
Information about the model is either read from the built-in Standard Model of \QGRAF{} or 
can be defined by the user by means of \LANHEP~\cite{Semenov:2010qt}
or an Universal FeynRules Output (\UFO) file~\cite{Degrande:2011ua} .

The \PYTHON{} program automatically performs several operations:
diagrams whose color factor turns out to be zero are dropped;
the number of propagators containing the loop momentum, the tensor rank 
and the kinematic invariants of the associated loop integral are computed;
diagrams with a vanishing loop integral associated are detected and flagged for the diagram selection;
all propagators and vertices are classified for the diagram selection;
diagrams containing massive quark self-energy insertions or closed massless quark loops are specially flagged.

During this phase, \GOSAM{} also generates a \LaTeX{} file which contains, among other useful information of the generated process, 
the drawings of all contributing diagrams. To achieve this task, we use our own
implementation of the algorithms described in Ref.~\cite{Ohl:1995kr} and 
Axodraw~\cite{Vermaseren:1994je} to actually draw the diagrams.

\subsubsection{Lorentz Algebra}

Concerning the algebraic operations performed by \GOSAM{} 
to render the integral suitable for efficient numerical evaluation,
one of the primary goals is to split the $(4-2\varepsilon)$ dimensional
algebra into strictly four-dimensional objects and symbols representing
the higher-dimensional remainder.
%
All external vectors
(momenta and polarisation vectors) are kept in four dimensions;
internal vectors, however, are kept in the $d$-dimensional vector space.

We adopt the conventions used in~\cite{Cullen:2010jv}, where
$\hat{k}$ denotes the four dimensional projection of an in general
$d$~dimensional vector $k$. The $(d-4)$~dimensional orthogonal projection
is denoted as~$\tilde{k}$. For the integration momentum $q$ we introduce
in addition the symbol $\mu^2=-\tilde{q}^2$, such that
\begin{equation}
q^2=\hat{q}^2+\tilde{q}^2=\hat{q}^2-\mu^2.
\end{equation}
We also introduce suitable projectors by splitting the metric tensor
\begin{equation}
g^{\mu\nu}=\hat{g}^{\mu\nu}+\tilde{g}^{\mu\nu},\quad%
\hat{g}^{\mu\nu}\tilde{g}_{\nu\rho}=0,\quad%
\hat{g}^\mu_\mu=4,\quad\tilde{g}^\mu_\mu=d-4.
\end{equation}

Once all propagators and all vertices have been replaced by their corresponding
expressions, according to the model file, all vector-like quantities and
metric tensors are split into their four-dimensional and their orthogonal
part. 
As we use the 't~Hooft algebra, $\gamma_5$ is defined as a purely
four-dimensional object, $\gamma_5=i\epsilon_{\mu\nu\rho\sigma}%
\hat{\gamma}^\mu\hat{\gamma}^\nu\hat{\gamma}^\rho\hat{\gamma}^\sigma$.
By applying the usual anti-commutation relation for Dirac matrices we can
always 
separate the four-dimensional and $(d-4)$-dimensional parts of
Dirac traces.

While the $(d-4)$-dimensional traces are reduced completely to
products of $(d-4)$-dimensional metric tensors, 
the four-dimensional part, which will be reduced numerically,  
is treated such that the number of terms in the resulting expression is kept as small as possible. 


\subsubsection{Treatment of rational terms $R_2$}
Instead of relying on the construction of $R_2$ from specialized Feynman
rules~\cite{Ossola:2008xq,Draggiotis:2009yb,Garzelli:2009is,Garzelli:2010qm,Garzelli:2010fq}, 
we can generate the $R_2$ part along with all other
contribution using automated algebraic manipulations.
The code offers the option between the \emph{implicit} and \emph{explicit} construction of the $R_2$ terms. 
The implicit construction treats the $4-$ and $(d-4)$~dimensional numerators on equal grounds: they are
generated algebraically and reduced numerically.
The explicit construction of $R_2$
is based on the fact that the $(d-4)$~dimensional part of the
numerator function contains expressions for the 
corresponding integrals that are relatively simple and known explicitly.
Therefore, after separating it using the algebraic manipulation described before, 
the $(d-4)$~dimensional part is computed analytically whereas
the purely four-dimensional part is passed to the numerical
reduction. This approach also allows for an efficient calculation of the part $R_2$ alone.

\subsubsection{Reduction to scalar (master) integrals}

\GOSAM{} allows to choose at run-time (i.e. without re-generating the code) the preferred method of reduction.    
Available options include the integral-level $d$-dimensional reduction, 
as implemented in \SAMURAI, 
or traditional tensor reduction as implemented in
\GOLEMVC{} interfaced through tensorial reconstruction at the integrand level,
or a combination of both.
Concerning the scalar (tensorial) integrals, \GOSAM{} allows to choose among a variety of options, including
QCDLoop~\cite{Ellis:2007qk}, OneLoop~\cite{vanHameren:2010cp}, \GOLEMVC~\cite{Binoth:2008uq}, plus the recently added \PJFRY~\cite{Fleischer:2011bi}.
Among these codes, OneLoop and \GOLEMVC{} also fully support complex masses.

\subsection{Installation and Usage}

\GOSAM{} can be used within a standard Linux/Unix environment. In order to work, it requires some programs to be 
installed on the system: these include a recent version of \PYTHON~(version $\ge$ 2.6), Java ($\ge$  1.5), 
a Fortran95 compiler, \FORM~(version $\ge$ 3.3), and \QGRAF.
Further, at least one of the libraries \SAMURAI\ or \GOLEMVC\ needs to be present at the time the code is compiled.

To facilitate this task, we have prepared a package containing \SAMURAI\ and \GOLEMVC\ together with the
libraries for the integrals: OneLOop, QCDLoop, and FF. 
The package, which is called {\tt gosam-contrib-1.0.tar.gz} is available for download from 
\begin{center}
{\tt http://projects.hepforge.org/gosam/}. 
\end{center}
The installation procedure 
is facilitated by the use of {\tt Autotools}.

The user can download the \GOSAM{} code either as a tar-ball or from the subversion repository
at {\tt http://projects.hepforge.org/gosam/}. The installation of \GOSAM{} is controlled by  {\tt Python Distutils} and can be performed by simply running the command
\begin{center}
{\tt python setup.py install}
\end{center}
In order to generate the code for a process, the user needs to prepare an input file (process card) which contains:
\begin{itemize}
\item[-] \emph{process} specific information, such as a list of initial and final state particles, their helicities
(optional) and the order of the coupling constants;
\item[-] \emph{scheme} specific information, such as the regularisation and renormalisation schemes, the underlying model, masses and widths which are set to zero;
\item[-] \emph{system} specific information, such as paths to programs and libraries or compiler options;
\item[-] \emph{optional} information for optimisations within the code generation.
\end{itemize}
Assuming that the process card is called {\tt myprocess.in}, the generation of the code can be started by simply running the command
{\tt gosam.py myprocess.in}. 
All further steps are controlled by {\tt makefile}s which are automatically generated by \GOSAM{}: 
the command {\tt make compile} generates the source code and compiles all files relevant for the production of the matrix element.
The code can be tested with the program {\tt test.f90} (located in the subdirectory {\tt matrix}) which provides, for a random phase-space point, the tree-level LO matrix element and the NLO result for the finite part, single and double poles. Examples of process cards for a selection of benchmark processes are provided with the main distubution.

For more details about the usage and installation of \GOSAM{}, 
we refer the user to a more technical presentation~\cite{Cullen:2011xs} 
or to the original publication~\cite{Cullen:2011ac} and the user manual
which accompanies the code.

\subsection{Examples of Applications}

The BLHA interface~\cite{Binoth:2010xt} allows to link \GOSAM{} to a general 
Monte Carlo event generator, which is responsible for supplying the missing 
ingredients for a complete NLO calculation of a physical cross section. 
Among those, \SHERPA{}~\cite{Gleisberg:2008ta} offers the possibility to 
compute the LO cross section and the real corrections with both the subtraction 
terms and the corresponding integrated counterparts~\cite{Krauss:2001iv}. 
Using the BLHA interface, we linked \GOSAM{} with \SHERPA{} to compute the 
physical cross section for $W^{\pm}+1$-jet at NLO, which is described in Section~\ref{sec:case}.

The codes produced by \GOSAM{} have been tested on several processes of 
increasing complexity, some of which are shown in Table~1.
The full list of processes produced by \GOSAM{} 
and compared to the literature where available 
is given in Ref.~\cite{Cullen:2011ac}.
\begin{table}[h]
\begin{center}
{\small \begin{tabular}{|l|l|}
\hline
$e^+e^-\to u\overline{u}$&\cite{Ellis:1991qj}\\
$e^+e^-\to t\overline{t}$&\cite{Jersak:1981sp,Catani:2002hc}, own analytic calculation\\
$u\overline{u}\to d\overline{d}$&\cite{Ellis:1985er,Hirschi:2011pa}\\
$g g \to gg$&\cite{Binoth:2006hk}\\
$g g \to gZ$&\cite{vanderBij:1988ac}\\
$b g \to H\,b$&\cite{Campbell:2002zm,Hirschi:2011pa}\\
$\gamma \gamma \to \gamma \gamma $ (W loop) &\cite{Gounaris:1999gh}\\
$\gamma \gamma \to \gamma \gamma \gamma \gamma $ (fermion loop) &\cite{Bernicot:2008nd}\\
$pp \to t\overline{t}$&\cite{Hirschi:2011pa}, MCFM
\cite{Campbell:1999ah,Campbell:2011bn}\\
$pp\to W^\pm\,j$ (QCD corr.) &\cite{Campbell:1999ah,Campbell:2011bn} \\
$pp\to W^\pm\,j$ (EW corr.) &for IR poles: \cite{Kuhn:2007cv,Gehrmann:2010ry} \\
$pp\to W^\pm\,t$ &\cite{Campbell:1999ah,Campbell:2011bn} \\
$pp\to W^\pm\,jj$ &\cite{Campbell:1999ah,Campbell:2011bn} \\
$pp\to W^\pm b\bar{b}$ (massive b) &\cite{Campbell:1999ah,Campbell:2011bn} \\
$e^+e^-\to e^+e^-\gamma$ (QED)&\cite{Actis:2009uq}\\
$pp \to H\,t\overline{t}$&\cite{Hirschi:2011pa}\\
$pp \to Z\,t\overline{t}$&\cite{Bevilacqua:2011xh}\\
$pp\to W^+W^+jj$&\cite[v3]{Melia:2010bm}\\
$pp\to b\overline{b} b\overline{b}$ &\cite{Binoth:2009rv,Greiner:2011mp}\\
$pp\to W^+W^- b\overline{b}$ &\cite{Hirschi:2011pa,vanHameren:2009dr}\\
$pp \to t\overline{t}b\overline{b}$&\cite{Hirschi:2011pa,vanHameren:2009dr}\\
$u\overline{d} \to W^+ ggg$&\cite{vanHameren:2009dr}\\
\hline
\end{tabular} }
\end{center}
\caption{Some of the processes computed and checked with \GOSAM{}}
\end{table}

As an example of the usage of \GOSAM{} with a model file different from the 
Standard Model, we calculated the QCD corrections to neutralino pair production
in the MSSM. The model file has been imported using the \UFO{} interface. 
In this calculation, we combined the one-loop amplitude with the 
real radiation corrections to obtain results for differential cross sections.
For the infrared subtraction terms we employed {\tt MadDipole}
\cite{Frederix:2008hu}, while the real emission part is 
calculated using MadGraph/MadEvent \cite{Alwall:2007st}.
The virtual matrix element is renormalized in the $\overline{MS}$ scheme, 
while massive particles are treated in the on-shell scheme. 
The renormalization terms specific to the massive MSSM particles have been added manually.
In Fig.\ref{fig:qqNNm12JVfull} we show the differential cross section for the 
$m_{\chi_{1}^{0} \chi_{1}^{0}}$ invariant  mass,
where we employed a jet veto to suppress large contributions from 
the channel $qg \rightarrow \chi_{1}^{0} \chi_{1}^{0} q$ which opens up 
at order $\alpha^2 \alpha_s$, but for large $p_{T}^{jet}$ belongs to
the distinct process of neutralino pair
plus one hard jet production at leading order.
\begin{figure}[h!]
\centering
\includegraphics[height=2.5in]{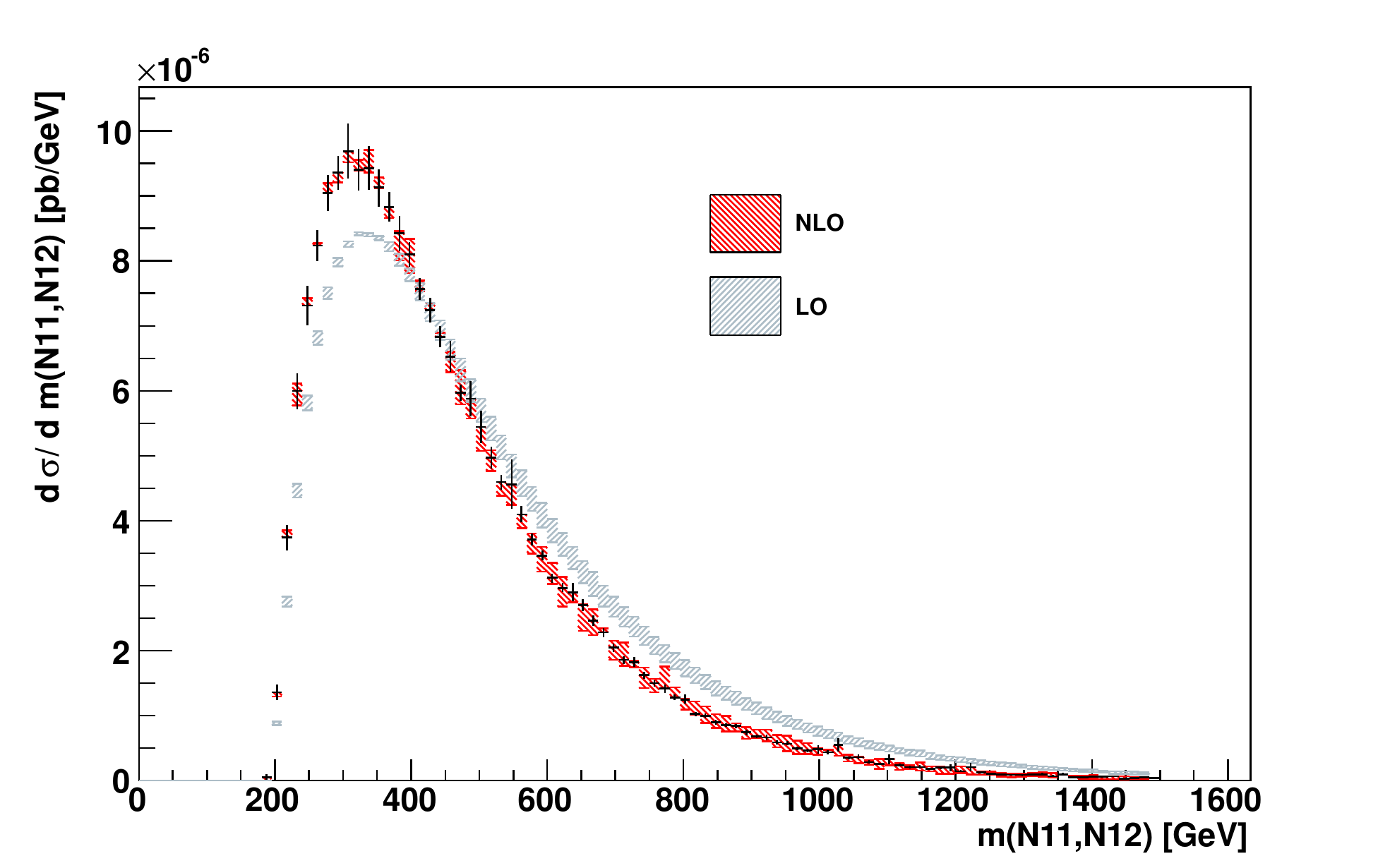}
\caption
{Comparison of the NLO and LO $m_{\chi_{1}^{0} \chi_{1}^{0}}$ distributions for the process $pp \rightarrow \chi_{1}^{0} \chi_{1}^{0}$
with a jet veto on jets with $p^{jet}_{T} > 20$ GeV and $\eta<4.5$.
The band gives the dependence of the result on $\mu = \mu_{F} = \mu_{R}$ between
$\mu_{0}/2$ and $2 \mu_{0}$. We choose $\mu_{0} = m_{Z}$. }
\label{fig:qqNNm12JVfull}
\end{figure}

\subsection*{Conclusions and Outlook}  
Several groups are currently working at the development of automated multi-purpose tools for one-loop calculations.
For quite a long time, tree-level calculation have been fully automated and 
included in flexible multi-process tools~\cite{Kanaki:2000ey, Stelzer:1994ta}. 
The level of automation achieved by one-loop calculations is suggesting the 
possibility of a similar success also at the next-lo-leading order. 
The target is to build efficient and flexible NLO programs which can be used to 
tackle the increasing need 
of precision required by the experimental collaborations.

\GOSAM{} is a flexible and broadly applicable tool for the fully automated evaluation of one-loop scattering amplitudes. 
In this approach, scattering amplitudes are generated in terms of Feynman 
diagrams and their reduction to master integrals can be performed in several ways, which can be selected at run-time. 
\GOSAM{} can be used to calculate one-loop corrections both in QCD and electro-weak 
theory and offers the flexibility to link general model files for theories Beyond the Standard Model. 
The code performed well in reproducing a wide range of examples and we are 
looking forward to tackle more challenging calculations and interfacing with 
other existing tools in the near future.

\subsection*{Acknowledgments} 
The work of G.C. was supported by DFG Sonderforschungsbereich Transregio 9, Computergest\"utzte
Theoretische Teilchenphysik. N.G. was supported in part by the U.S. Department of Energy under
contract No. DE-FG02-91ER40677. P.M. and T.R. were supported by the Alexander von
Humboldt Foundation, in the framework of the Sofja Kovaleskaja Award Project “Advanced Mathematical
Methods for Particle Physics”, endowed by the German FederalMinistry of Education and
Research. The work of G.O. was supported in part by the National Science Foundation under Grant
PHY-0855489. G.O. also acknowledges the support of the Center for Theoretical Physics at City Tech.  
The research of F.T. is supported by Marie-Curie-IEF, project:
“SAMURAI-Apps”. We also acknowledge the support of the Research Executive Agency (REA) of
the European Union under the Grant Agreement number PITN-GA-2010-264564 (LHCPhenoNet).


}

\section[Automation and numerical loop integration]
{AUTOMATION AND NUMERICAL LOOP INTEGRATION \protect\footnote{Contributed by: S.~Weinzierl}}
{\graphicspath{{weinzierl_submit/}}

\title{Automation and numerical loop integration}

\author{S. Weinzierl}
\institute{Institut f{\"u}r Physik, Universit{\"a}t Mainz, D - 55099 Mainz, Germany}


\subsection{INTRODUCTION}

Numerical methods are nowadays routinely used in fully differential fixed-order perturbative calculations 
for the integration over the phase-space of the final-state particles.
The use of numerical methods for the phase-space integration
allows the flexibility to compute any infrared-safe observable for a given process within a single
numerical program.
It is thus natural to investigate if numerical methods can also be applied for the loop integration in the virtual 
corrections
\cite{Becker:2011vg,Becker:2010ng,Assadsolimani:2010ka,Assadsolimani:2009cz,Gong:2008ww,Anastasiou:2007qb,Nagy:2006xy,Nagy:2003qn,Soper:2001hu,Soper:1999xk,Soper:1998ye}. 
A major breakthrough was achieved recently by showing that the numerical method is compatible in efficency with
the commonly used approaches
based on cut techniques and generalised unitarity or on Feynman graphs
\cite{Berger:2009zg,Berger:2009ep,Berger:2010vm,Berger:2010zx,Ita:2011wn,Ellis:2009zw,KeithEllis:2009bu,Melia:2010bm,Bevilacqua:2010ve,Bevilacqua:2009zn,Bredenstein:2009aj,Frederix:2010ne,vanHameren:2010cp,Badger:2010nx,Cascioli:2011va,Hirschi:2011pa,Cullen:2011ac}.
The implementation of the numerical method for the loop integration is process-independent and offers therefore the flexibility to
compute several processes within one numerical program.
We discuss the main principles of the numerical method for the loop integration at one-loop.
In addition we give an outlook towards higher loops.

\subsection{THE SUBTRACTION METHOD FOR THE LOOP INTEGRATION}

The contributions to an infrared-safe $n$-jet observable observable $O$ at next-to-leading order are given by
\begin{eqnarray}
 \langle O \rangle^{NLO} &=& \int\limits_{n+1}O_{n+1}d\sigma^{R}+\int\limits_{n}O_{n}d\sigma^{V}+\int\limits_{n}O_{n}d\sigma^{C}.
\end{eqnarray}
Here a rather condensed notation is used. $d\sigma^{R}$ denotes the real emission contribution, 
whose matrix elements are given by the square of the Born amplitudes with $(n+3)$ partons $|A_{n+3}^{(0)}|^{2}$. 
$ d\sigma^{V}$ denotes the virtual contribution, whose matrix elements are given by the interference term of the one-loop 
and Born amplitude $\mbox{Re}\;(A_{n+2}^{(0)^{*}}A_{n+2}^{(1)})$ and $d\sigma^{C}$ denotes a collinear subtraction term, 
which subtracts the initial state collinear singularities. 
Each term is separately divergent and only their sum is finite. 

The subtraction method is widely used to render the real emission part of a NLO calculation suitable for a numerical Monte Carlo integration. 
One adds and subtracts a suitably chosen piece to be able to perform the phase-space integrations by Monte Carlo methods:
\begin{eqnarray}
 \langle O \rangle^{NLO} &=& \int\limits_{n+1}\left(O_{n+1}d\sigma^{R}-O_{n}d\sigma^{A}\right)+\int\limits_{n}\left(O_{n}d\sigma^{V}+O_{n}d\sigma^{C}+O_{n}\int\limits_{1} d\sigma^{A}\right).
\end{eqnarray} 
The first term $\left(O_{n+1}d\sigma^{R}-O_{n}d\sigma^{A}\right)$ is by construction integrable over the $\left(n+1\right)$-particle phase-space and can be evaluated numerically. 
The result of the integration 
of the subtraction term over the unresolved one-parton phase-space 
is written in a compact notation as
\begin{eqnarray}
 d\sigma^C + \int\limits_1 d\sigma^A  
 & = &
 \left( {\bf I} + {\bf K} + {\bf P} \right) \otimes d\sigma^B.
\end{eqnarray}
The notation $\otimes$ indicates that colour correlations due to the colour charge operators
${\bf T}_i$ still remain.
The terms with the insertion operators ${\bf K}$ and ${\bf P}$ pose no problem for a numerical evaluation.
The term ${\bf I} \otimes d\sigma^B$ lives on the phase-space of the $n$-parton configuration and has the appropriate
singularity structure to cancel the infrared divergences coming from the one-loop amplitude.
Therefore $d\sigma^V + {\bf I} \otimes d\sigma^B$ is infrared finite.

We extend this subtraction method to the virtual part such that we can evaluate the one-loop integral of the one-loop amplitude numerically. 
The renormalised one-loop amplitude $\mathcal{A}^{(1)}$ is related to the bare amplitude $\mathcal{A}_{\mathrm{bare}}^{(1)}$ by
$\mathcal{A}^{(1)} = \mathcal{A}_{\mathrm{bare}}^{(1)}+\mathcal{A}_{\mathrm{CT}}^{(1)}$,
where $\mathcal{A}_{\mathrm{CT}}^{(1)}$ denotes the ultraviolet counterterm from renormalisation.
The bare amplitude involves the loop integration
\begin{eqnarray}
\mathcal{A}_{\mathrm{bare}}^{(1)}&=&\int\frac{d^{D}k}{(2\pi)^{D}}\mathcal{G}_{\mathrm{bare}}^{(1)}.
\end{eqnarray}  
where $\mathcal{G}_{\mathrm{bare}}^{(1)}$ denotes the integrand of the bare one-loop amplitude.
We introduce subtraction terms which match locally the singular behaviour of the bare integrand:
\begin{eqnarray}
\mathcal{A}_{\mathrm{bare}}^{(1)}+\mathcal{A}_{\mathrm{CT}}^{(1)} 
 = 
 \int\frac{d^{D}k}{(2\pi)^{D}}\left(\mathcal{G}_{\mathrm{bare}}^{(1)}-\mathcal{G}_{\mathrm{soft}}^{(1)}-\mathcal{G}_{\mathrm{coll}}^{(1)}
                                  -\mathcal{G}_{\mathrm{UV}}^{(1)}\right)
 +\left(\mathcal{A}_{\mathrm{CT}}^{(1)}+\mathcal{A}_{\mathrm{soft}}^{(1)}+\mathcal{A}_{\mathrm{coll}}^{(1)}+\mathcal{A}_{\mathrm{UV}}^{(1)}\right).
\end{eqnarray}
Analogous to $\mathcal{G}_{\mathrm{bare}}^{(1)}$, the integrands of the subtraction terms $\mathcal{A}_{x}^{(1)}$ are denoted by $\mathcal{G}_{x}^{(1)}$, 
where $x$ is equal to $\mathrm{soft}$, $\mathrm{coll}$ or $\mathrm{UV}$. 
The expression in the first bracket is finite and can therefore be integrated numerically in four dimensions.
The integrated subtraction terms in the second bracket are easily calculated analytically in $D$ dimensions.
The result can be written as
\begin{eqnarray}
 2 \;\mbox{Re}\; {\cal A}^{(0)}
   \left( {\cal A}_{\mathrm{CT}}^{(1)} + {\cal A}_{\mathrm{soft}}^{(1)} + {\cal A}_{\mathrm{coll}}^{(1)} + {\cal A}_{\mathrm{UV}}^{(1)}\right)^\ast 
                O_n d\phi_n
 & = & 
 {\bf L} \otimes d\sigma^B.
\end{eqnarray}
The insertion operator ${\bf L}$ contains the explicit poles in the dimensional regularisation parameter related
to the infrared singularities of the one-loop amplitude.
These poles cancel when combined with the insertion operator ${\bf I}$:
\begin{eqnarray}
 \left( {\bf I} + {\bf L} \right) \otimes d\sigma^B & = &
 \mbox{finite}.
\end{eqnarray}
The operator ${\bf L}$ contains, as does the operator ${\bf I}$, colour correlations due to soft gluons.
In analogy to the one-loop amplitude we can write $d\sigma^{V}=d\sigma_{\mathrm{CT}}+\int \frac{d^Dk}{(2\pi)^D}d\sigma_{\mathrm{bare}}^{V}$ and then the NLO contributions reads
\begin{eqnarray}
\lefteqn{
\langle O \rangle^{NLO}
 = } & & \\
 & &
 \int\limits_{n+1}\left(O_{n+1}d\sigma^{R}-O_{n}d\sigma^{A}\right)
 +\int\limits_{n+\mathrm{loop}} O_{n} \left(d\sigma_{\mathrm{bare}}^{V}-d\sigma^{A'}\right)
 +\int\limits_{n} O_{n} \left( {\bf I} + {\bf L} + {\bf K} + {\bf P} \right) \otimes d\sigma^B.
 \nonumber
\end{eqnarray}
In a condensed notation this reads
\begin{eqnarray}
\langle O \rangle^{NLO}&=& \langle O \rangle_{\mathrm{real}}^{NLO}+\langle O \rangle_{\mathrm{virtual}}^{NLO}+\langle O \rangle_{\mathrm{insertion}}^{NLO}.
\end{eqnarray}
Every single term is finite and can be evaluated numerically. 

\subsection{THE SUBTRACTION TERMS}

Amplitudes in QCD may be decomposed into group-theoretical factors (carrying the colour structures) multiplied by kinematic factors 
called partial amplitudes.
At the loop level partial amplitudes may further be decomposed into primitive amplitudes.
It is simpler to work with primitive one-loop amplitudes instead of a full one-loop amplitude.
Our method exploits the fact that primitive one-loop amplitudes have a fixed cyclic ordering of the
external legs and that they are gauge-invariant.
The first point ensures that there are at maximum $n$ different loop propagators in the problem, where $n$ is the 
number of external legs, while the second property of gauge invariance is crucial for the proof of the
method.
We therefore consider in the following just a single primitive one-loop amplitude,
which we denote by $A^{(1)}$, while keeping in mind that the full one-loop amplitude is just
the sum of several primitive amplitudes multiplied by colour structures.
We label the external momenta clockwise by $p_1$, $p_2$, ..., $p_n$ 
and define $q_i=p_1+p_2+...+p_i$, $k_i=k-q_i$.
We can write the bare primitive one-loop amplitude in Feynman gauge as
\begin{eqnarray}
\label{starting_point}
 A^{(1)}_{bare} = \int \frac{d^Dk}{(2\pi)^D} 
 G^{(1)}_{bare},
 & &
 G^{(1)}_{bare} = 
 P(k) \prod\limits_{i=1}^n \frac{1}{k_i^2 - m_i^2 + i \delta}.
\end{eqnarray}
$G^{(1)}_{bare}$ is the integrand of the bare one-loop amplitude.
$P(k)$ is a polynomial in the loop momentum $k$.
The $+i\delta$-prescription instructs us to deform -- if possible -- the integration contour into the complex plane
to avoid the poles at $k_i^2=m_i^2$.
If a deformation close to a pole is not possible, we say that the contour is pinched.
If we restrict ourselves to non-exceptional external momenta,
then the divergences of the one-loop amplitude related to a pinched contour are either due to soft
or collinear partons in the loop.
These divergences are regulated within dimensional regularisation by setting the number of space-time dimensions
equal to $D=4-2\varepsilon$.
A primitive amplitude which has soft or collinear divergences must have at least one loop propagator which corresponds
to a gluon. An amplitude which just consists of a closed fermion loop does not have any infrared divergences.
We denote by $I_g$ the set of indices $i$, for which the 
propagator $i$ in the loop corresponds to a gluon.
The soft and collinear subtraction terms for massless QCD read \cite{Assadsolimani:2009cz}
\begin{eqnarray}
G_{\mathrm{soft}}^{(1)} & = & 16 \pi \alpha_s i\sum\limits_{j\in I_g} \frac{p_{j}.p_{j+1}}{k_{j-1}^2k_{j}^2k_{j+1}^2} A_{j}^{(0)} \;, \nonumber \\
G_{\mathrm{coll}}^{(1)}
& = & -8 \pi \alpha_s i\sum\limits_{j\in I_g} \bigg[ \frac{S_{j}g_{\mathrm{UV}}(k_{j-1}^2,k_{j}^2)}{k_{j-1}^2k_{j}^2}
                                    + \frac{S_{j+1}g_{\mathrm{UV}}(k_{j}^2,k_{j+1}^2)}{k_{j}^2k_{j+1}^2} \bigg] A_{j}^{(0)} \;,
\end{eqnarray}
where $S_{j}=1$ if the external line $j$ corresponds to a quark 
and $S_{j}=1/2$ if it corresponds to a gluon. 
The function $g_{\mathrm{UV}}$ ensures that the integration over the loop momentum is ultraviolet finite.
Integrating the soft and the collinear part we obtain
\begin{eqnarray}
S_{\varepsilon}^{-1}\mu_s^{2\varepsilon}\int \frac{d^Dk}{(2\pi)^D} \, G_{\mathrm{soft}}^{(1)} & = &
-\frac{\alpha_s}{4\pi}\frac{e^{\varepsilon\gamma_E}}{\Gamma(1-\varepsilon)}\sum\limits_{j\in I_g} \frac{2}{\varepsilon^2} \Big( \frac{-2p_{j}p_{j+1}}{\mu_s^2}\Big)^{-\varepsilon} A_j^{(0)} \; + {\cal O}(\varepsilon),
 \nonumber \\
S_{\varepsilon}^{-1}\mu_s^{2\varepsilon}\int \frac{d^Dk}{(2\pi)^D} \, G_{\mathrm{coll}}^{(1)} & = &
-\frac{\alpha_s}{4\pi}\frac{e^{\varepsilon\gamma_E}}{\Gamma(1-\varepsilon)}\sum\limits_{j\in I_g} (S_j+S_{j+1}) \frac{2}{\varepsilon} \Big( \frac{\mu_{uv}^2}{\mu_s^2}\Big)^{-\varepsilon} A_j^{(0)} \; + {\cal O}(\varepsilon),
\end{eqnarray}
$S_\varepsilon = (4\pi)^\varepsilon e^{-\varepsilon\gamma_E}$ is the 
typical volume factor of dimensional regularisation, $\gamma_E$ is Euler's constant  and $\mu$ is the renormalisation scale.

The ultraviolet subtraction terms correspond to propagator and vertex corrections.
The subtraction terms are obtained by expanding the relevant loop propagators
around a new ultraviolet propagator $(\bar{k}^2-\mu_{\mathrm{UV}}^2)^{-1}$, where $\bar{k} = k - Q$:
For a single propagator we have
\begin{eqnarray}
 \frac{1}{\left(k-p\right)^2}
 & = &
 \frac{1}{\bar{k}^2-\mu_{\mathrm{UV}}^2}
       + \frac{2\bar{k}\cdot\left(p-Q\right)}{\left(\bar{k}^2-\mu_{\mathrm{UV}}^2\right)^2}
 - \frac{\left(p-Q\right)^2+\mu_{\mathrm{UV}}^2}{\left(\bar{k}^2-\mu_{\mathrm{UV}}^2\right)^2}
 + \frac{\left[ 2\bar{k}\cdot\left(p-Q\right)\right]^2}{\left(\bar{k}^2-\mu_{\mathrm{UV}}^2\right)^3}
 + {\cal O}\left(\frac{1}{|\bar{k}|^5}\right).
 \nonumber
\end{eqnarray}
We can always add finite terms to the subtraction terms.
For the ultraviolet subtraction terms we choose the finite terms such that
the finite parts of the 
integrated ultraviolet subtraction terms
are independent of $Q$ and proportional to the pole part, 
with the same constant of proportionality for all ultraviolet subtraction terms.
This ensures that the sum of all integrated UV subtraction terms is again proportional to a tree-level amplitude \cite{Becker:2010ng}.

\subsection{CONTOUR DEFORMATION}

Having a complete list of ultraviolet and infrared subtraction terms at hand, we can ensure that the integration
over the loop momentum gives a finite result and can therefore be performed in four dimensions.
However, this does not yet imply that we can safely integrate each of the four components of the loop momentum $k^\mu$
from minus infinity to plus infinity along the real axis.
There is still the possibility that some of the loop propagators go on-shell for real values of the loop momentum.
If the contour is not pinched this is harmless, as we may escape into the complex plane in a direction indicated by
Feynman's $+i\delta$-prescription.
However, it implies that the integration should be done over a region of real dimension $4$ in the complex space
${\mathbb C}^4$.
Let us consider an integral corresponding to a primitive one-loop amplitude with $n$ propagators minus the appropriate
IR- and UV-subtraction terms:
\begin{eqnarray}
\int\frac{d^{4}\tilde{k}}{(2\pi)^{4}}\left(\mathcal{G}_{\mathrm{bare}}^{(1)}-\mathcal{G}_{\mathrm{soft}}^{(1)}-\mathcal{G}_{\mathrm{coll}}^{(1)}-\mathcal{G}_{\mathrm{UV}}^{(1)}\right) &=& 
\int\frac{d^{4}\tilde{k}}{(2\pi)^{4}}P(\tilde{k})\prod\limits_{j=1}^{n}\frac{1}{\tilde{k}_{j}^{2}-m_{j}^{2}+i\delta}
\end{eqnarray}
where $P(\tilde{k})$ is a polynomial of the loop momentum $\tilde{k}^{\mu}$ and the integration is over a complex contour in order to avoid 
whenever possible the poles of the propagators. 
We set 
$\tilde{k}^\mu =  k^\mu+i\kappa^\mu(k)$,
where $k^{\mu}$ is real \cite{Gong:2008ww}. After this deformation our integral equals
\begin{eqnarray}
&&\int\frac{d^{4}k}{(2\pi)^{4}}\left|\frac{\partial \tilde{k}^{\mu}}{\partial k^{\nu}}\right|P(\tilde{k}(k))\prod\limits_{j=1}^{n}\frac{1}{k_{j}^{2}-m_{j}^{2}-\kappa^{2}+2i k_{j}\cdot\kappa}.
\end{eqnarray}
To match Feynman's $+i\delta$-prescription
we have to construct the deformation vector $\kappa$ such that
\begin{eqnarray}
k_{j}^{2}-m_{j}^{2} & = & 0 \quad \rightarrow \quad k_{j}\cdot \kappa \geq 0. 
\end{eqnarray}
We remark that the numerical stability of the Monte Carlo integration depends strongly on the definition of the deformation vector $\kappa$. 

\subsection{NLO results for n-jets in electron-positron annihilation}

We have calculated results for jet observables in electron-positron annihilation, 
where the jets are defined by the Durham jet algorithm \cite{Becker:2011vg}. 
The cross section for $n$ jets normalised to the LO cross section for $e^{+}e^{-}\rightarrow$ hadrons reads
\begin{eqnarray}
\frac{\sigma_{n-jet}(\mu)}{\sigma_{0}(\mu)}&=& \left(\frac{\alpha_{s}(\mu)}{2\pi}\right)^{n-2}A_{n}(\mu)+\left(\frac{\alpha_{s}(\mu)}{2\pi}\right)^{n-1}B_{n}(\mu)+\mathcal{O}(\alpha_{s}^{n}).
\end{eqnarray}
One can expand the perturbative coefficient $A_{n}$ and $B_{n}$ in $1/N_{c}$:
\begin{eqnarray}
A_{n}&=& N_{c}\left(\frac{N_{c}}{2}\right)^{n-2}\left[A_{n,\mathrm{lc}}+\mathcal{O}\left(\frac{1}{N_{c}}\right)\right],\qquad B_{n}\ = \ N_{c}\left(\frac{N_{c}}{2}\right)^{n-1}\left[B_{n,\mathrm{lc}}+\mathcal{O}\left(\frac{1}{N_{c}}\right)\right].
\nonumber
\end{eqnarray}
We calculate the leading order coefficient $A_{n,\mathrm{lc}}$ and the next-to-leading order coefficient $B_{n,\mathrm{lc}}$ for $n\leq 7$ 
at the renormalisation scale $\mu$ equal to the centre-of-mass energy. 
The centre-of-mass energy is taken to be equal to the mass of the $Z$-boson. 
The scale variation can be restored from the renormalisation group equation. The calculation is done with five massless quark flavours.
\begin{figure}[ht]
\begin{center}
\includegraphics[width=0.32\textwidth]{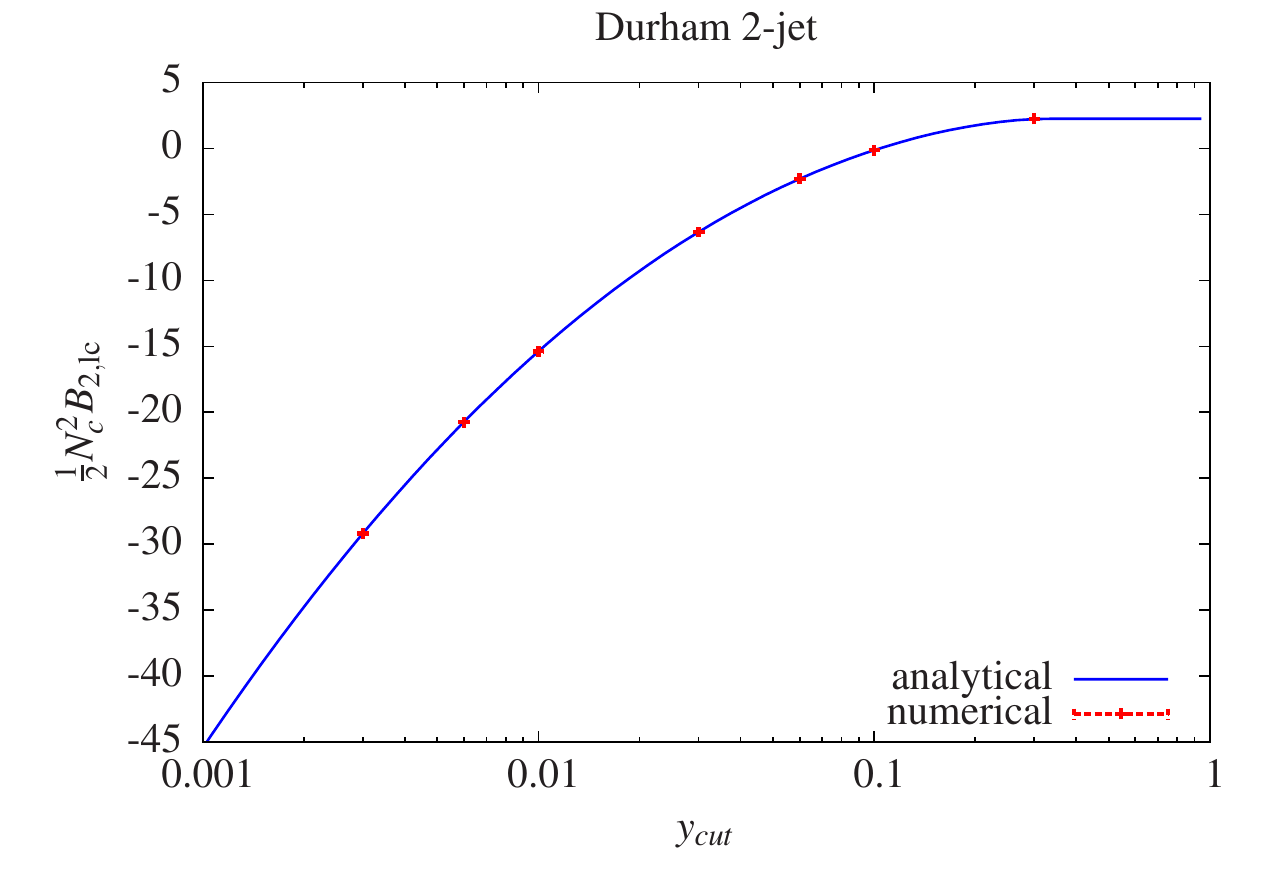}
\includegraphics[width=0.32\textwidth]{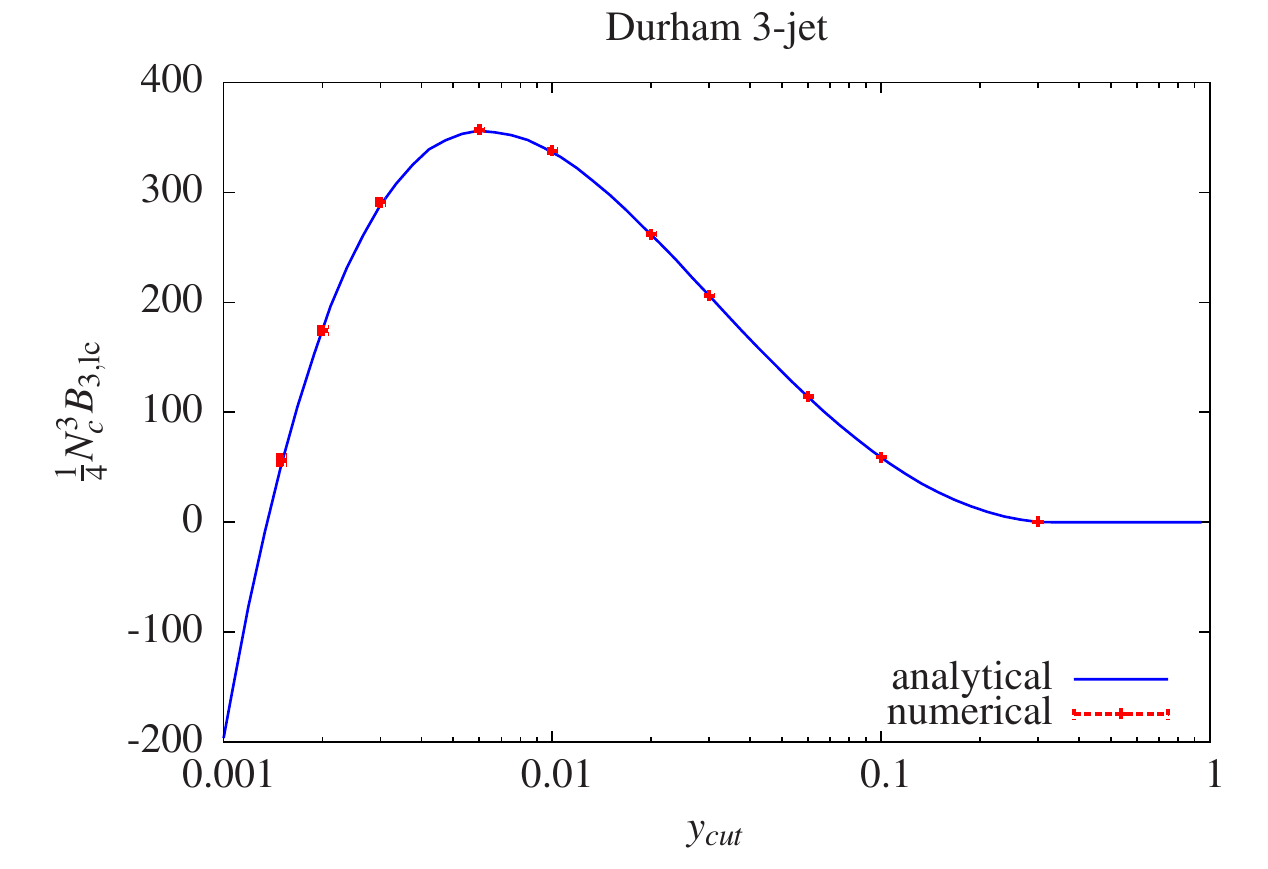}
\includegraphics[width=0.32\textwidth]{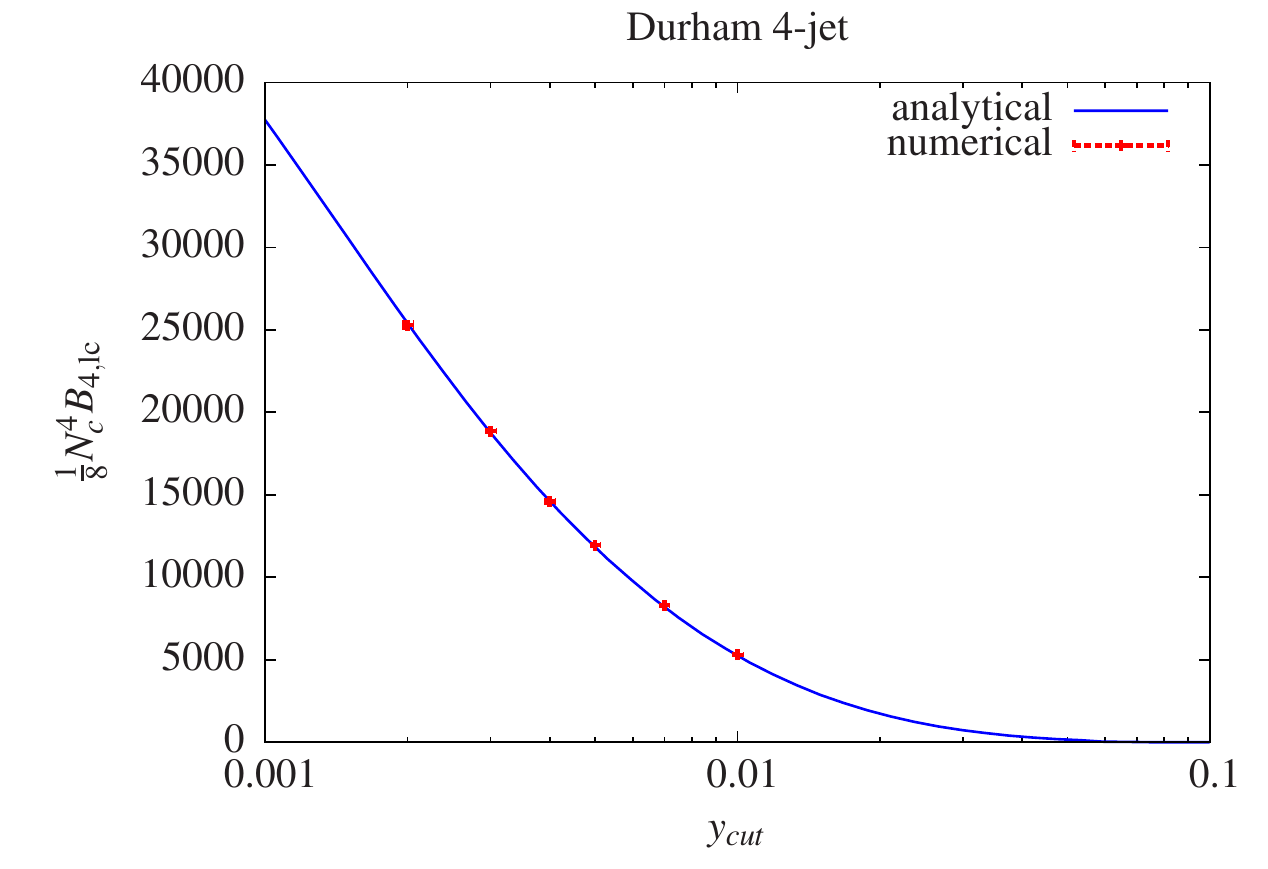}
\end{center}
\caption{
Comparison of the NLO corrections to the two-, three- and four-jet rate between the numerical calculation and an analytic calculation.
The error bars from the Monte Carlo integration are shown and are almost invisible.
}
\label{fig_jetrates}
\end{figure}
Fig.~\ref{fig_jetrates} shows the comparison of our numerical approach with the well-known results for two, three and four jets \cite{Weinzierl:1999yf,Weinzierl:2010cw,Weinzierl:2005dd}. 
We observe an excellent agreement. 
The results for five, six and seven jets for the jet parameter $y_{cut}=0.0006$ are
\begin{align}
 & 
 \frac{N_c^4}{8}  A_{5,\mathrm{lc}} = \left( 2.4764 \pm 0.0002 \right) \cdot 10^{4},
 & &
 \frac{N_c^5}{16} B_{5,\mathrm{lc}} = \left( 1.84 \pm 0.15 \right) \cdot 10^{6},
 \nonumber \\
 &
 \frac{N_c^5}{16} A_{6,\mathrm{lc}} = ( 2.874 \pm 0.002 ) \cdot 10^{5},
 & &
 \frac{N_c^6}{32} B_{6,\mathrm{lc}} = ( 3.88 \pm 0.18 ) \cdot 10^{7},
 \nonumber \\
 &
 \frac{N_c^6}{32} A_{7,\mathrm{lc}} = \left( 2.49 \pm 0.08 \right) \cdot 10^{6},
 & &
 \frac{N_c^7}{64} B_{7,\mathrm{lc}} = ( 5.4 \pm 0.3 ) \cdot 10^{8}.
\end{align}

\subsection{FIRST STEPS TOWARDS NNLO}

An NNLO calculation requires among other things also the calculation of the one-loop amplitude squared.
The expansion in the dimensional regularisation parameter $\varepsilon$ of the one-loop amplitude starts at order $(-2)$
one would naively expect that up to order $\varepsilon^0$ the 
$\cal{O}(\varepsilon)$- and ${\cal O}(\varepsilon^2)$-terms of the one-loop amplitude are needed for an NNLO calculation.
However, it is by no means obvious how the approaches for one-loop amplitudes based on unitarity or the numerical method
can be extended to include the higher-order terms in the $\varepsilon$-expansion.
It turns out that the computation of these higher-order terms can be avoided, provided a method is known to compute the finite
two-loop remainder function.
The one- and two-loop amplitudes can be written as \cite{Catani:1998bh}
\begin{eqnarray}
{\cal A}^{(1)} & = & 
 {\bf Z}^{(1)} {\cal A}^{(0)} 
 + {\cal F}^{(1)}_{\mathrm{minimal}},
 \nonumber \\
{\cal A}^{(2)} & = & 
 \left( {\bf Z}^{(2)} - {\bf Z}^{(1)} {\bf Z}^{(1)} \right) {\cal A}^{(0)} 
 + {\bf Z}^{(1)} {\cal A}^{(1)} 
 + {\cal F}^{(2)}_{\mathrm{minimal}},
\end{eqnarray}
where the operators ${\bf Z}^{(1)}$ and ${\bf Z}^{(2)}$ contain all the infrared poles and
${\cal F}^{(1)}_{\mathrm{minimal}}$ and ${\cal F}^{(2)}_{\mathrm{minimal}}$ are finite remainders.
Here we used the convention that the operators ${\bf Z}^{(1)}$ and ${\bf Z}^{(2)}$ contain only pole terms, but no terms
of order $\varepsilon^k$ with $k \ge 0$. This corresponds to a minimal scheme.
The operators ${\bf Z}^{(1)}$ and ${\bf Z}^{(2)}$ are well-known.
At NNLO it is sufficient to know the $\varepsilon^0$-terms of ${\cal F}^{(1)}_{\mathrm{minimal}}$ and ${\cal F}^{(2)}_{\mathrm{minimal}}$,
the $\varepsilon^1$- or $\varepsilon^2$-terms of ${\cal A}^{(1)}$ or ${\cal F}^{(1)}_{\mathrm{minimal}}$ are not required \cite{Weinzierl:2011uz}.

\subsubsection*{ACKNOWLEDGEMENTS}

I would like to thank Simon Pl\"atzer for fruitful discussions during the workshop.


}

\section[Towards the automation of one-loop amplitudes]
{TOWARDS THE AUTOMATION OF ONE-LOOP AMPLITUDES \protect\footnote{Contributed by: F.~Campanario}}
{\graphicspath{{CampanarioLesHouches/}}

\title{Towards the automation of one-loop amplitudes}

\author{F.Campanario$^1$}
\institute{$^1$Institut f\"ur Theoretische Physik, 
    Universit\"at Karlsruhe, KIT, 76128 Karlsruhe, 
    Germany}


\begin{abstract}
A program is presented that computes one-loop amplitudes automatically 
for processes with up to 6 external particles based on the Feynman-diagram approach.
Additionally, universal one-loop building blocks, which can be used to compute 
several processes at NLO QCD are calculated.
\end{abstract}

\subsection{INTRODUCTION}
The calculation of processes with multi-particle final states beyond
the leading order approximation has been an active field of research
during the last years as a consequence of the demand of high accuracy
for signal and background processes at the LHC. 
A next-to-leading~(NLO) calculation consists of virtual and
real radiation processes which are infrared divergent~(IR) separately
and can be computed numerically only after extracting the divergences of the real radiation contributions.
The one-loop virtual calculation for multiple particles poseses a
challenge of complexity not only due to the large number of contributing
diagrams, but also concerning the stability of the numerical code to evaluate them. 
%
In the last years, an enormous progress has been achieved 
applying new techniques and using traditional
Feynman-diagram approach, leading to new NLO predictions. 

Due to the
large number of processes of potential interest at the LHC, the scientific
community has worked in the automation of the NLO calculations.
The automation of the real contributions including their infrared
subtraction terms has been successfully implemented in several
packages and the automation of the virtual corrections, which is a harder problem, is
currently being achieved in several programs~(see~\cite{Gehrmann:2010rj} and references therein).

In Ref.~\cite{Campanario:2011cs}, the early stage of a program, in
the framework of Mathematica~\cite{Wolfran} and FeynCalc~\cite{Mertig:1990an}, to compute automatically one-loop amplitudes 
based on traditional Feynman-diagram techniques and involving up to $2\to4$ processes was presented. 
This program will become publicly available in the future. 
The method used is described in Section~\ref{s:2;qcdsm:camp}. In Section~\ref{s:3;qcdsm:camp}, 
we present a set of universal one-loop building blocks that has been used to compute recently several processes included in the 
VBFNLO package~\cite{Arnold:2008rz,Arnold:2011wj}.
\subsection{TOWARDS AN AUTOMATIC ONE-LOOP AMPLITUDE GENERATOR}
\label{s:2;qcdsm:camp}
The program above mentioned automatically simplifies a set of amplitudes up to Hexagons
of rank 5. The result is given in terms of scalar and tensor integrals following the
Passarino-Veltman convention~\cite{Passarino:1978jh,Campanario:2011cs}, spinor chains, polarization vectors and model parameters. 
The simplified expression is written automatically to FORTRAN routines. For massless propagators, the
amplitudes can be evaluated also in Mathematica with unlimited precision, which is used for testing purposes. 
To achieve that, the scalar integrals, the tensor
reduction formalism to extract the tensor coefficient integrals, and also the helicity
method described in Ref.~\cite{Hagiwara:1985yu,Hagiwara:1988pp} to
compute the spinor products have been implemented at the FORTRAN and
Mathematica level. 
For the determination of the tensor integrals up to
the box level, the Passarino-Veltman 
tensor reduction formalism~\cite{Passarino:1978jh} is used 
applying the LU decomposition method to avoid the
explicit calculation of inverse Gram matrices by solving a system of
linear equations, which is a more stable procedure close to singular
points. Finally, for singular Gram determinants, special
tensor reduction routines following Ref.~\cite{Denner:2005nn} have been
implemented, however, the external momenta convention~(Passarino-like) was used.  
The impact of these methods is discussed in detail in Ref.~\cite{Campanario:2011cs}. 
For pentagons, in addition to the Passarino-Veltman formalism, the method proposed by Denner and
Dittmaier~\cite{Denner:2005nn,Denner:2002ii}, applied also to hexagons, has been
implemented. For that, the recursion relations of
Ref.~\cite{Denner:2005nn} in terms of the Passarino-Veltman external
momenta convention have been re-derived. This last method is used for the numerical
implementation at the FORTRAN level. 

The Mathematica function does several algebraic manipulations that are summarized as follows:
\begin{itemize}
\setlength{\itemsep}{-1pt}
\item Simultaneous extraction of rational terms based on Dirac algebra manipulations
  and cancelation of scalar products against propagators.
\item Reduction to a minimal basis of tensor and scalar integrals.
\item Reduction to a minimal basis of spinor chains.
\item The use of Chisholm identities, which are only valid in 4 dimensions, for the contraction of Lorentz indices among different
  spinor chains is applied, if selected.
\item Factorization of loop dependent and independent factors~(Useful to
 perform gauge tests, Ward identities or the re-evaluation of the
 amplitudes for different helicity polarization of gluons and fermions
 at a lower CPU cost).
\end{itemize}

As an example of the notation used, the following Hexagon diagram is used. This is written as follows:\\

\begin{minipage}[h]{1\linewidth}
\begin{minipage}[h]{0.30\linewidth}
\hspace*{1.3cm}
\includegraphics[scale=0.795]{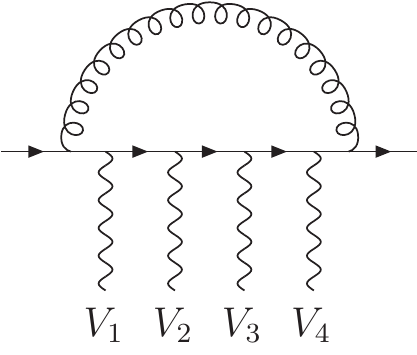}
\end{minipage}
\begin{minipage}[h]{0.65\linewidth}
\vspace*{-0.85cm}
\hspace*{-3.3cm}
\begin{equation}
 = {\cal M
}_{V_1V_2V_3V_4,\tau} = \mbox{{\small $ g^{V_1f}_{\tau}g^{V_2f}_{\tau}g^{V_3f}_{\tau}g^{V_4f}_{\tau}$}} \frac{g_0^2}{(4 \pi)^2}\mbox{{\small ${\cal
  C}^{V_1 V_2 V_3 V_4}_{ij}$}}  {\cal
  M}^{ij}_{\tau},
\label{fig:hex;qcdsm:camp}
\end{equation}
\end{minipage}
\end{minipage}

\noindent where $g_0$ is the strong unrenormalized coupling, ${\cal C}^{V_1 V_2 V_3 V_4}_{ij}$ is a color diagram dependent factor,
e.g, ${\cal C}^{\gamma \gamma g \gamma}_{ij}=(T_a)_{ij} (C_F-1/2 C_A)$. $g^{V_i f}_{\tau}$
are electroweak couplings and ${\cal M}_{\tau}^{ij}$ represents the amplitude considering generic off-shell vector bosons 
with color indices $ij$ for a given helicity $\tau$. The amplitude ${\cal M}^{ij}_{\tau}$, omitting color indices, is written in terms of
\begin{equation}
{\cal M}_\tau
={\cal M}^{D=4}_\tau  + (D-4) {\cal M}^{DR}_\tau,
\label{DR;qcdsm:camp}
\end{equation}
where ${\cal M}^{D=4}_\tau$ is the amplitude that one would obtain performing the Dirac algebra manipulation 
in four dimensions, $D=4$, and ${\cal M}^{DR}_\tau$ contains
the rational terms and vanishes in Dimensional Reduction ($DR$). These functions are decomposed in the form:
\begin{equation}
{\cal M}^{(D=4,DR)} = \sum_{i,j} \mbox{SM}_{i,\tau}~\mbox{F1}_j,
\label{eq:helmethod;qcdsm:camp}
\end{equation}
where SM$_{i,\tau}$ is a basis of Standard Matrix elements corresponding to
spinor products describing the quark line of Eq.~(\ref{fig:hex;qcdsm:camp}) which are computed
following the helicity method~\cite{Hagiwara:1985yu,Hagiwara:1988pp}  
with a defined helicity, $\tau$. F1$_j$ are complex
functions which are further decomposed into dependent and
independent loop integral parts,
\begin{equation}
\mbox{F1}_j= \sum_{l,k} \mbox{F}_l T_k\Big(  \epsilon(p_n) \cdot p_m
  ;\epsilon(p_i) \cdot \epsilon(p_r) \Big).
\label{eq:F1;qcdsm:camp} 
\end{equation}
$T_k$ is a monomial 
function at most for each polarization vector $\epsilon(p_x)$, i.e., $\epsilon(p_x)^0$ or $\epsilon(p_x)^1$. 
The first possibility, $\epsilon(p_x)^0$, implies that the  
polarization vector appears in the set of Standard Matrix elements SM$_{i,\tau}$. %
$F_l$ contains kinematic variables~($p_i\cdot p_j$), the scalar integrals ($B_0, C_0, D_0$), and the tensor integral coefficients
 $(B_{ij},C_{ij},D_{ij},E_{ij},F_{ij})$.
Then, the full result is obtained from ${\cal M}^{D=4}_\tau$ and
${\cal M}^{DR}_\tau$ using the finite and the coefficients of the $1/\epsilon^n$ poles of
the scalar and tensor coefficient integrals:
\begin{equation}
 {\cal M}^{D=4}_v= \widetilde{{\cal M}}_v
 +\frac{{\cal M}^1_v
}{\epsilon} +\frac{{\cal M}^2_v
}{\epsilon^2}, \qquad \qquad (D-4) {\cal M}^{DR}_v= \widetilde{{\cal N}_v} 
+\frac{{\cal N}_v^1
}{\epsilon},
\label{D4terms;qcdsm:camp}
\end{equation}
where, e.g.,  $ \widetilde{{\cal M}}_v $ is the finite contribution obtained
using the finite pieces of the scalar and tensor coefficient integrals including the finite contributions from rational terms
arising in ultraviolet tensor coefficient integrals.
\subsection{UNIVERSAL BUILDING BLOCKS}
\label{s:3;qcdsm:camp}
Based on the observation that the same one-loop virtual amplitudes appear in many processes (Fig.~\ref{fig:boxcont;qcdsm:camp}), we are aiming to 
collect a basis of universal building blocks, which can be used to compute all of the $2 \to 4$ processes at LHC at the QCD one-loop level~%
(Similar to the philosophy of older versions of MADGRAPH~\cite{Alwall:2011uj}  calling the HELAS~\cite{Murayama:1992gi} routines). 
This methodology of collecting topologies in groups has been proved very successful in the program VBFNLO, 
where for example a boxline routine, first line of Fig.~\ref{fig:boxcont;qcdsm:camp}, is computed and applied to 
$pp\to VV$, $pp\to VVV$, $pp\to VV j$ and EW production of 
$pp\to  V j j$ and $pp\to H V j j$.
\begin{figure}[ht!]
\begin{center}
\includegraphics[scale=0.67]{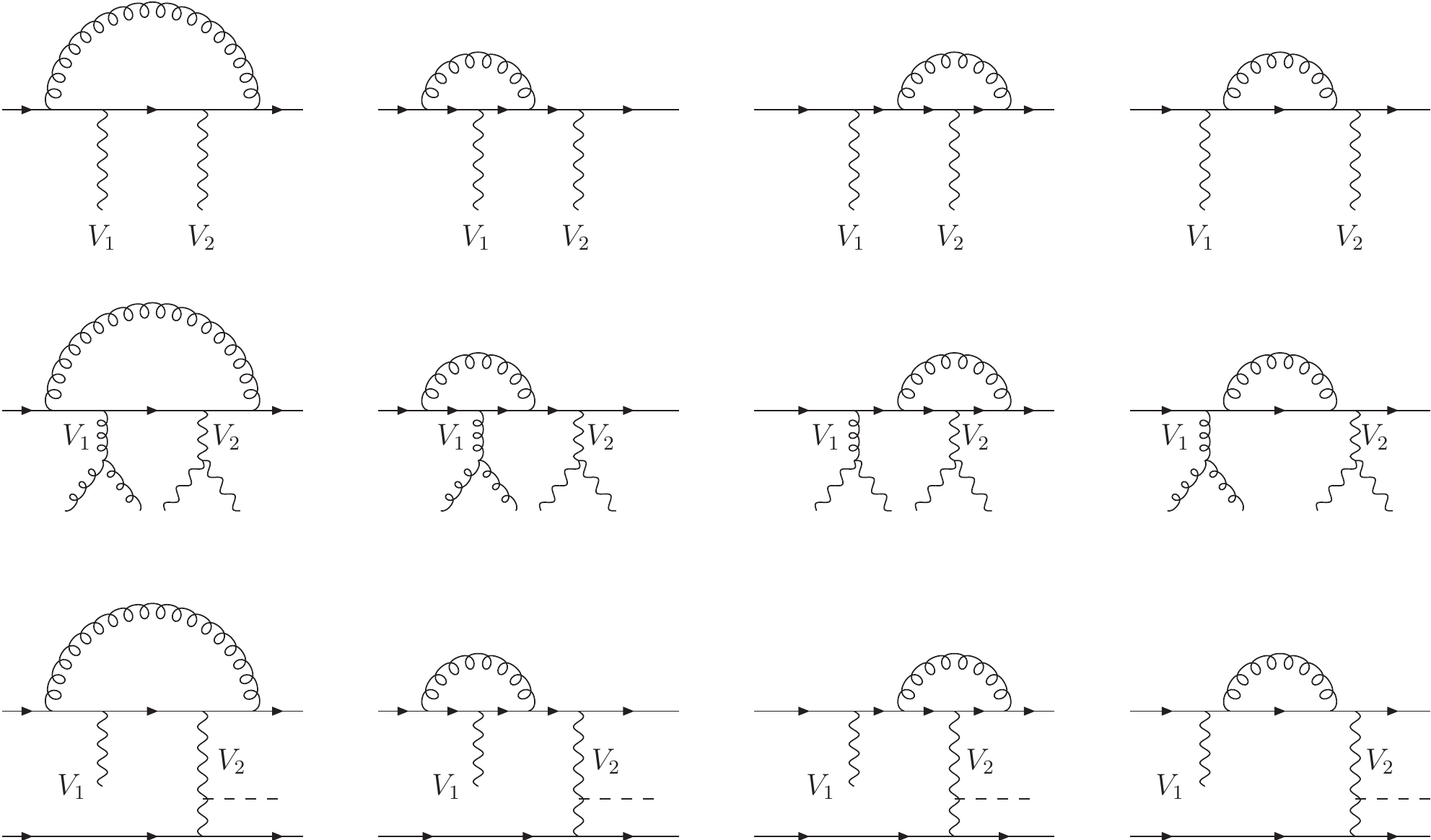}
\end{center}
\caption{Boxline contributions appearing in different processes.}
\label{fig:boxcont;qcdsm:camp}
\end{figure}

To do that, we use the effective current approach described and applied in
Refs.~\cite{Hankele:2007sb,Campanario:2008yg,Campanario:2009um,Campanario:2010mi,Campanario:2011ud}.
As illustration, the first diagram of the second raw of Fig.~\ref{fig:boxcont;qcdsm:camp} is used. This can be
written as, 
\begin{equation}
A_{V_1V_2V_3V_4,\tau}= J^{\mu_1}_{V_1^*} J^{\mu_2}_{V_2^*} {\cal
  M}_{\mu_1\mu_2,\tau}\equiv {\cal
  M}_{V_1^*V_2^*,\tau},
\end{equation}
where the color indices have been omitted. Here, $J^{\mu_1}_{V_1^*} $
and $J^{\mu_2}_{V_2^*}$ represent effective polarization vectors
in the unitarity gauge for the EW sector including finite width effects in the scheme of Refs.~\cite{Denner:1999gp,Oleari:2003tc} and
propagator factors, e.g., 
\begin{equation}
J^{\mu_1}_{V_1^*}(q_1) = \frac{-i}{q_1^2-M_{V_1^*}^2 -i \,M_{V_1^*}
  \Gamma_{V_1^*}} \left( g^{\mu_1}_{\mu} -
  \frac{q_1^{\mu_1}q_{1 \mu}}{q_1^2-M_{V_1^*}^2- i \,M_{V_1^*}\Gamma_{V_1^*}}\right) \Gamma^{\mu}_{V_1^*V_1 V_3},
\end{equation} 
with $\Gamma_{V_1^*}$, the width of the $V_1^*$ vector boson, and
$\Gamma^{\mu}_{V_1^*V_1 V_3}$, the triple vertex, which can also contain the leptonic decay of the EW vector bosons
including all off-shell effects or BSM physics.
In this manner, 
we can then concentrate in computing, instead of
$A_{V_1V_2V_3V_4,\tau}$, the virtual correction to
two massive vector bosons attached to the quark line, ${\cal  M}_{V_1^*V_2^*,\tau}$, or equivalently  ${\cal
  M}_{\mu_1\mu_2,\tau}$, where the polarization vectors or effective
currents have been factored out. In our approach, this basic building block is the so-called Boxline, 
which is computed only once and re-used in different processes.
%

We plan to do a classification of all the topologies that appear at 1 loop level for up to $2\to 4$ processes and install 
a library with all the basic one-loop building blocks already computed and simplified. 
This would be an advantage since, for example for $qq\to VVVV$ production,  up to 24 hexagons for a single subprocess would appear, corresponding 
to the permutations of the vector bosons on the hexagon of Eq.~\ref{fig:hex;qcdsm:camp}. In this approach, the amplitude is obtained 
by calling the same one-loop amplitude 24 times with the corresponding ordering of momenta and polarization vectors. We aim towards an automation 
of this procedure, which will result into a faster and shorter final FORTRAN code generation.
The specific building blocks are collected into groups with specific gauge and IR factorization properties, e.g, 
factorization of the IR divergences against the corresponding born, known behavior under Ward identity checks.

In Fig.~\ref{fig:top;qcdsm:camp}, we present the topologies that have been computed and tested. 
In the first line, corrections to a quark line with the emission of 
V$_n$ vector bosons in a fixed order are represented for 4 different topologies.
(The first 2 were explained in detail in Ref.~\cite{Campanario:2011cs}, including their stability behavior).
 We have only depicted the virtual amplitude with the higher complexity for a giving building block, e.g. the boxline of Fig.\ref{fig:boxcont;qcdsm:camp} 
is obtained from the first diagram with two vector bosons attached, i.e., $n=2$ in $V_n$.
The first two topologies of the second line are collected by putting together all possible Feynman-diagrams 
with a fixed order of the vector bosons and attaching it to the quark lines in all possible ways. 
The crossing of the fermion lines are treated as independent building blocks and are not depicted.
Finally, the fermion-loop corrections for a fixed order of vector bosons, $V_n$, are computed in the last diagram of the second line
 
\begin{figure}[ht!]
\begin{center}
\includegraphics[scale=0.8]{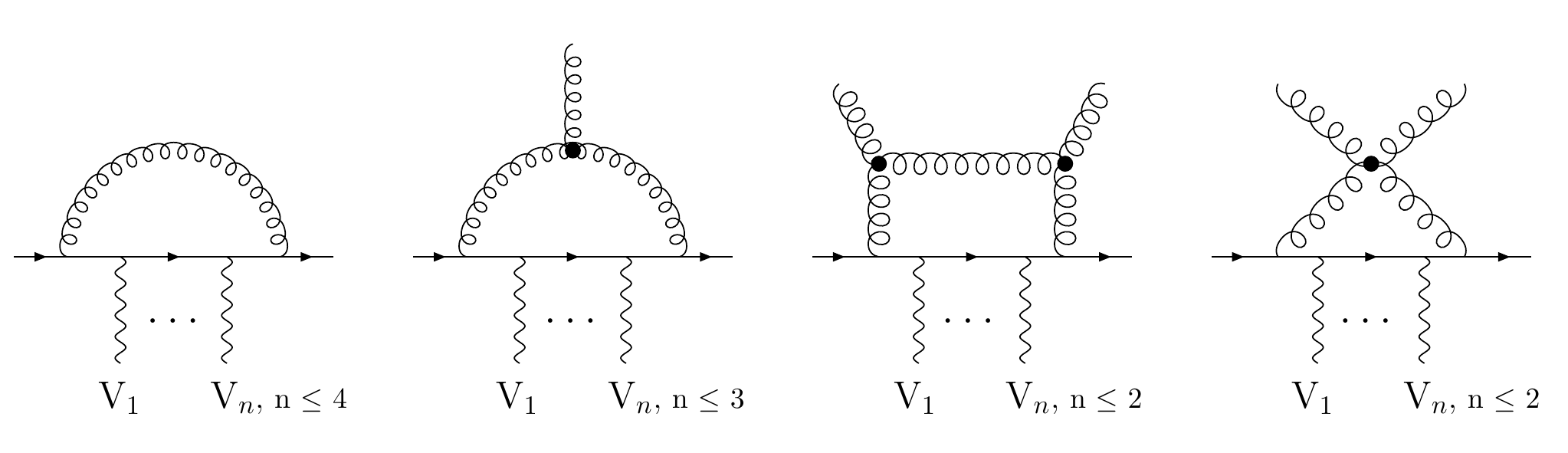}
%
%
\includegraphics[scale=0.85]{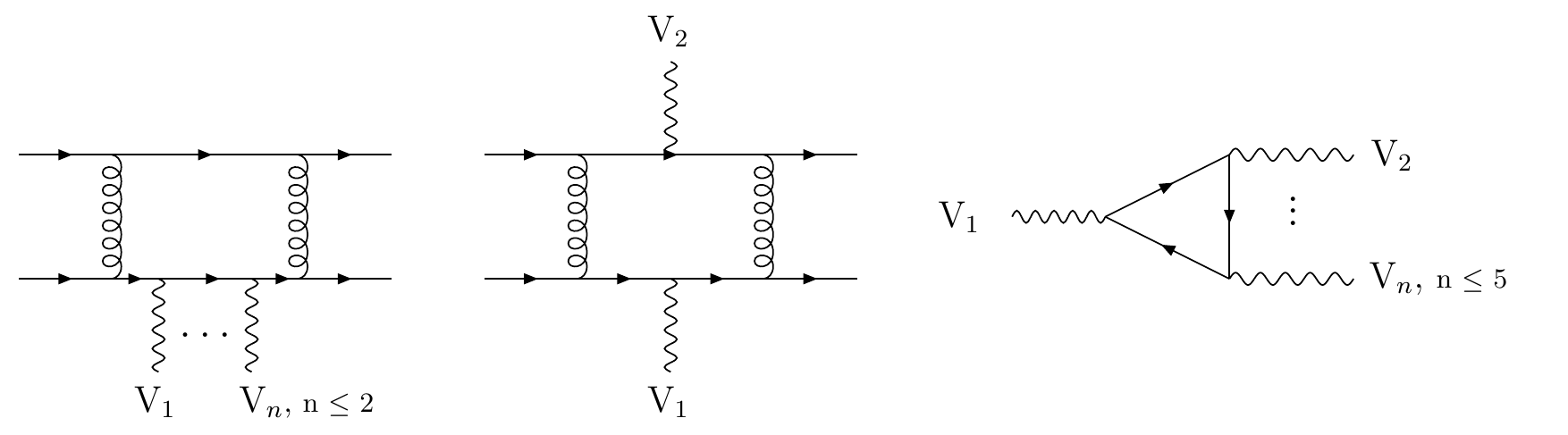}
\end{center}
\caption{Topologies of universal building one-loop blocks. Only the most complicated diagram of each topology is depicted, 
e.g, the boxline of Fig.\ref{fig:boxcont;qcdsm:camp} 
is obtained from the first diagram with two vector bosons attached, i.e., $V_n, n=2$.}
\label{fig:top;qcdsm:camp}
\end{figure}
The use of modular structure routines, as the above presented, has been
proved to be an advantage in the program {\it VBFNLO}~\cite{Arnold:2008rz,Arnold:2011wj} since once a
structure is computed and checked it can be re-used for different
processes. For example, using the building blocks of the first and second topology together with the fermion-loop diagrams,
results at NLO QCD for all $VVV$~\cite{Hankele:2007sb,Campanario:2008yg,Bozzi:2009ig,Bozzi:2010sj,Bozzi:2011wwa,Bozzi:2011en}, several 
$VVj$~\cite{Campanario:2009um,Campanario:2010hp,Campanario:2010xn,Campanario:2010hv}, $H \gamma jj$~\cite{Arnold:2010dx}  and 
$W\gamma \gamma j$~\cite{Campanario:2011ud} production channels have been computed recently. The last one representing the first 
calculation at this accuracy falling in the category of $VVV+j$ production. 
Up to the pentagon level, these building blocks are publicly available as part of the
{\it VBFNLO}~\cite{Arnold:2008rz,Arnold:2011wj} package together with the tensor reduction routines, excluding the routines 
for small Gram determinants which will become available in the future, in addition to the other building blocks. 
\subsection{CONCLUSIONS}
\label{sec:concl;qcdsm:camp}
A program which automatically evaluates one-loop amplitudes for up to $2 \to 4$ processes has been presented based on the traditional 
Feynman-diagram approach. The program has been developed in the framework of Mathematica and FeynCalc and writes down automatically 
the simplified expression to FORTRAN.
Up to the pentagon level and for massless propagators, the code can be evaluated numerically inside Mathematica with unlimited 
precision which can be used for testing purposes.
For the reduction of tensor integrals, we have developed a library that includes expansion for small Gram determinants.
Using the leptonic tensor formalism, we are building a library of universal one-loop building blocks, which can be used to compute 
several processes at NLO QCD.  Recently, following this strategy, we have reported results for all $VVV$
~\cite{Hankele:2007sb,Campanario:2008yg,Bozzi:2009ig,Bozzi:2010sj,Bozzi:2011wwa,Bozzi:2011en}, several $VVj$~\cite{
Campanario:2009um,Campanario:2010hp,Campanario:2010xn,Campanario:2010hv}, $H \gamma jj$~\cite{Arnold:2010dx}  and 
$W\gamma \gamma j$~\cite{Campanario:2011ud} production channels inside the VBFNLO collaboration.
The ultimate goal is to generalize the library to compute all of the $2 \to 4$ processes at LHC at the QCD one-loop level, %
similar to the philosophy of older versions of MADGRAPH~\cite{Alwall:2011uj} calling the HELAS~\cite{Murayama:1992gi} routines,
and deliver a Mathematica package compatible with FeynArts~\cite{Hahn:2000kx}, which can be used to compute full one-loop amplitudes 
automatically using the universal building blocks, resulting into a faster and shorter code generation.

\subsection*{ACKNOWLEDGEMENTS}
F.C acknowledges fruitful discussions with the members of the VBFNLO collaboration and  partial support of a postdoctoral fellowship of the 
Generalitat Valenciana, Spain (Beca Postdoctoral d`Excel$\cdot$l\`encia), by the European FEDER and Spanish MICINN
under the grant FPA2008-02878 and by the Deutsche Forschungsgemeinschaft via the Sonderforschungsbereich/Transregio  SFB/TR-9 ``Computational Particle
Physics''. The Feynman diagrams in this paper were drawn using Axodraw~\cite{Vermaseren:1994je}.


}

\section[The two-loop QCD virtual amplitude for W pair production with full mass dependence] 
{THE TWO-LOOP QCD VIRTUAL AMPLITUDE FOR W PAIR PRODUCTION 
WITH FULL MASS DEPENDENCE  \protect\footnote{Contributed by: G.~Chachamis}}
{\graphicspath{{chachamis_LesHouches2011/}}



\title{THE TWO-LOOP QCD VIRTUAL AMPLITUDE FOR W PAIR PRODUCTION 
WITH FULL MASS DEPENDENCE}

\author{G. Chachamis$^1$}
\institute{$^1$Paul Scherrer Institut, CH-5232 Villigen PSI, Switzerland}


\subsection{INTRODUCTION}

One of the main aims in the Large Hadron Collider (LHC) physics program
is undoubtedly the discovery (or the exclusion) of the Higgs boson which
is responsible for the fermion and gauge boson masses and also part
of the mechanism of dynamical breaking of
the Electroweak (EW) symmetry.
Another important goal for the LHC is the precise measurement of the
hadronic production of gauge boson pairs, 
$W W$, $W Z$, $Z Z$, $W \gamma$, $Z \gamma$, 
this in connection to the investigation
of the non-Abelian gauge structure of the SM.
W pair production,
\begin{equation}
q {\bar q} \rightarrow W^+ \, W^- \, ,
\label{chachamis_qq_channel}
\end{equation}
plays an essential role as it serves as a signal process 
in the search for New Physics and also is the dominant irreducible
background to the Higgs discovery channel 
$p p \rightarrow H \rightarrow W^* W^* \rightarrow l {\bar \nu} {\bar l}' \nu' $, 
in the intermediate Higgs mass range~\cite{Dittmar:1996ss}.
Both ATLAS and CMS collaborations have released first
values for the $WW$ cross section~\cite{Yang:2011db, VVCMS}.

The process~(\ref{chachamis_qq_channel}) is currently known at next-to-leading order (NLO) 
accuracy~\cite{Brown:1978mq, Ohnemus:1991kk, Frixione:1993yp, Dixon:1998py, 
Dixon:1999di, Campbell:1999ah, Grazzini:2005vw }.
The NLO corrections were proven to be large
enhancing the tree-level result by
almost 70\% which falls to a (still) large 30\% after 
imposing a jet veto. 
Therefore, if a theoretical estimate for  the
W pair production is to be compared against 
experimental measurements at the LHC, one is bound to 
go one order higher in the perturbative expansion, namely,
to the next-to-next-to-leading order (NNLO). This would 
allow, in principle, an accuracy of around 10\%.

High accuracy for the W pair production is also needed
when the process is studied as background to Higgs production
in order to match accuracies between signal and background.
The signal process for the Higgs discovery via
gluon fusion,
$g g \rightarrow H$, 
as well as the process
$H \rightarrow W W \rightarrow l {\bar \nu} {\bar l}' \nu'$
are known at 
NNLO~\cite{Spira:1995rr,Dawson:1990zj,Harlander:2002wh,Anastasiou:2002yz,
Ravindran:2003um,Catani:2001cr,Davatz:2004zg,Anastasiou:2004xq, 
Anastasiou:2007mz,Grazzini:2008tf},
whereas the EW corrections are known
beyond NLO~\cite{Bredenstein:2006rh}.
Another process that needs to be included in the background
is the W pair production
in the loop induced gluon fusion channel, 
\begin{equation}
g g \rightarrow W^+ W^- \, .
\label{chachamis_gg_channel}
\end{equation}
The latter
contributes at $\mathcal{O}(\alpha_s^2)$ relative to the 
quark-anti-quark-annihilation channel but is 
nevertheless enhanced due to the large gluon flux
at the LHC~\cite{Binoth:2005ua, Binoth:2006mf}.

The first main difficulty in studying the NNLO QCD
corrections for W pair production is the calculation
of the two-loop virtual amplitude since it is
a $2 \rightarrow 2$ process with massive external particles.
We have already computed the virtual corrections
at the high energy 
limit~\cite{Chachamis:2007cy, Chachamis:2008yb, Chachamis:2008xu}.
However, this is not enough as it cannot
cover the kinematical region close to threshold. Therefore,
in order to cover all kinematical regions we proceed as follows.
We perform a deep expansion in the W mass around the
high energy limit which in combination with
the method
of numerical integration of differential
equations~\cite{Caffo:1998du, Boughezal:2007ny, Czakon:2007qi} 
allows us the numerical computation of the two-loop amplitude
with full mass dependence over the whole phase space.

\subsection{THE HIGH ENERGY LIMIT}

The methodology for obtaining the massive amplitude in the
high energy limit, namely the limit where all the invariants are
much larger than the W mass,
is similar to the one followed in Refs.~\cite{Czakon:2007ej, Czakon:2007wk}.
The amplitude is reduced to an expression
that only contains a small number of integrals (master integrals) 
with the help of the Laporta algorithm~\cite{Laporta:2001dd}. 
In the calculation for the two-loop amplitude there 
are 71 master integrals.  Next step is the construction, in a fully 
automatised way, of the Mellin-Barnes (MB) 
representations~\cite{Smirnov:1999gc, Tausk:1999vh} 
of all the master integrals by using the {\bf MBrepresentation} 
package~\cite{MBrepresentation}. 
The representations are then analytically continued in the number 
of space-time dimensions by means
of the {\bf MB} 
package~\cite{Czakon:2005rk}, thus revealing the full singularity 
structure. An asymptotic expansion in
the mass parameter (W mass) is performed by closing contours and the 
integrals are finally resummed, either
with the help of {\bf XSummer}~\cite{Moch:2005uc} or the 
{\bf PSLQ} algorithm~\cite{pslqAlg}.
The result is expressed in terms of harmonic polylogarithms.

\subsection{POWER CORRECTIONS AND NUMERICAL EVALUATION}

The high energy limit by itself is not
enough, as was mentioned before. 
The next step, following the methods applied in Ref.~\cite{Czakon:2008zk},
is to compute power corrections in the W mass. 
Power corrections are good enough to
cover most of the
phase space, apart from the region near 
threshold as well as the regions corresponding
to small angle scattering.

We recapitulate here some of the notation of Ref.~\cite{Chachamis:2008xu}
for completeness.
The charged vector-boson 
production in the leading partonic scattering process
corresponds to
\begin{equation}
\label{chachamis_qqWW}
q(p_1) + {\overline q}(p_2) 
\:\:\rightarrow\:\: W^-(p_3,m) + W^+(p_4,m) \, ,
\end{equation}
where $p_i$ denote 
the quark and W momenta and $m$ is the mass of the W boson.

We have chosen to express the amplitude in terms
of the kinematic variables $x$ and $m_s$ which are defined to be
\begin{equation}
  x = -\frac{t}{s}, \;\; m_s = \frac{m^2}{s},
\end{equation}
where
\begin{equation}
s = (p_1+p_2)^2 \;\; {\rm and} \;\;t = (p_1-p_3)^2-m^2\,.
\end{equation}
The variation then of
$x$ within the range $ [ 1/2(1-\beta), 1/2(1+\beta) ] $, where
$\beta=\sqrt{1-4m^2/s}$ is the velocity, corresponds to angular variation
between the forward and backward scattering.

It should be evident that any master integral $M_i$ can be written then as
\begin{equation}
M_i = M_i \left( m_s, x, \epsilon \right) = \sum_{j=k}^l \epsilon^j {I_i}_j(m_s, x),
\label{chachamis_epReveal}
\end{equation}
where $\epsilon$ is the usual regulator in dimensional regularization ($d = 4 - 2 \epsilon$)
and the lowest power of $\epsilon$ in the sum can be $-4$.

The crucial point now is that
the derivative of any Feynman integral with
respect to any kinematical variable is 
again a Feynman integral with possibly
higher powers of denominators or numerators which can also be
reduced anew in terms of the initial set of master integrals.
This means that one can construct a
partially triangular system of differential equations in the mass,
which can subsequently be solved in the form of a power series expansion,
with the expansion parameter in our case being $m_s$ following
the conventions above.

Let us differentiate with respect to $m_s$ and $x$, we will then have
respectively
\begin{equation}
m_s \frac{d}{dm_s} M_i(m_s,x,\epsilon) = 
\sum_j C_{i j}(m_s,x,\epsilon) ~M_j(m_s,x,\epsilon)
\label{chachamis_dms}
\end{equation}
and
\begin{equation}
x \frac{d}{dx} M_i(m_s,x,\epsilon) = 
\sum_j C_{i j}\sp{\prime}(m_s,x,\epsilon) ~M_j(m_s,x,\epsilon)\, .
\label{chachamis_dx}
\end{equation}
We use Eq.~(\ref{chachamis_dms}) to obtain the 
mass corrections for the master integrals  
calculating the power series expansion
up to order $m_s^{11}$ (see also Ref.~\cite{Czakon:2008zk} for more
details).
This deep expansion in $m_s$ should be sufficient for most
of the phase space but still not enough to
cover the whole allowed kinematical region. The way to
proceed from this point is to numerically integrate
the system of differential equations.

In particular, we choose to work with
the master integrals in the form of Eq.~(\ref{chachamis_epReveal}), where the
$\epsilon$ dependence is explicit. We can then work with the coefficients
of the $\epsilon$ terms and accordingly have
\begin{equation}
m_s \frac{d}{dm_s} I_i(m_s,x) = 
\sum_j J^M_{i j}(m_s,x) ~I_j(m_s,x)
\label{chachamis_I_dms}
\end{equation}
and
\begin{equation}
x \frac{d}{dx} I_i(m_s,x) = 
\sum_j J^X_{i j}(m_s,x) ~I_j(m_s,x),
\label{chachamis_I_dx}
\end{equation}
where the Jacobian matrices $J^M$ and $J^X$ have rational
function elements.

By using this last system of differential equations, one can obtain a full
numerical solution to the problem. 
What we are essentially dealing now with is an initial
value problem and the main requirement is 
to have the initial conditions to proper accuracy. 
The initial conditions, namely the values of the master integrals
at a proper kinematical point which we call initial
point, are provided by the power series expansion.
The initial point has to be chosen somewhere
in the high energy limit region, where
$m_s$ is small and therefore, the values obtained by the power series
are very accurate.
Starting from there, one can evolve to any other point
of the phase space by numerically integrating the system of differential
equations Eq.~(\ref{chachamis_I_dms}) and Eq.~(\ref{chachamis_I_dx}).

We parametrise with a suitable grid of points the region close
to threshold and then we calculate the master integrals for all points of the
grid by evolving as described previously.
Given that the master integrals have to be
very smooth (we remain above all thresholds) one can use,
after having the values for the grid points,
interpolation to get the values at any point of the region.
We use 1600 points for the grid and take as initial conditions
the values of the master integrals at the
point $m_s = 5 \times 10^{-3}$, $x = 1/4$. The
relative errors at that point
were estimated not to exceed $10^{-18}$.

The numerical integration is performed by using
one of the most advanced
software packages implementing the variable coefficient multistep method
(ODEPACK)~\cite{ODEPACK}. We use quadruple precision to maximise
accuracy. 
The values at any single grid  point can be obtained
in about 15 minutes in average (with a typical 2GHz Intel Core 2 Duo system)
after compilation with the Intel Fortran compiler.
The accuracy is around 10 digits for most of the points
of the grid.
It is also worth noting that in order to perform the numerical
integration one needs to deform the contour in the complex plane
away from the real axis. This is due to the fact
that along the real axis there are spurious
singularities. 
We use an elliptic contour and 
we achieve a  better estimate of the final global error
by calculating more than once for each point of the grid, 
using each time different eccentricities.
Grids of solutions can actually be constructed, which
will be subsequently interpolated
when implemented as part of a Monte Carlo program.

One very stringent test we use to cross-check the
correctness and also the accuracy of our calculation is to compare the
infrared pole structure of our two-loop result against the one predicted
by Catani~\cite{Catani:1998bh} (see also 
Refs.~\cite{Catani:1996vz,Giele:1991vf,Kunszt:1994mc}).
According to Catani, the infrared poles of the interference of the tree 
and the two-loop amplitudes follow a generic formula which in our case,
since we work with the rescaled variables $m_s$ and $x$,
can be cast into the following form:
\begin{equation}
\mathcal{C}_{atani}^{(0 \times 2)}(m_s, x, \frac{s}{\mu})=2 {\rm Re} \left\{ 
\rm{I}^{(1)}(\epsilon) \langle M^{(0)}|M^{(1)} \rangle 
+ \rm{I}^{(2)}(\epsilon) \langle M^{(0)}|M^{(0)}\rangle \right\},
\label{chachamis_catani}
\end{equation}  
where  ${\rm M}^{(0)}$ and ${\rm M}^{(1)}$ are the tree level and 
one-loop amplitudes respectively and
$\mu$ is the renormalization scale. The operators $\rm{I}^{(1)}(\epsilon)$ 
and $\rm{I}^{(2)}(\epsilon)$
encode the information for the infrared pole structure and their exact expressions 
can be found in Ref.~\cite{Chachamis:2008yb}.

The way to perform the test is straightforward. For each point of the grid
with coordinates $(m_{s(i)}, x_{(i)})$, we compute the numerical value of the
two-loop amplitude ($\rm{M}^{(2)}$) interfered with the tree level amplitude
\begin{equation}
\mathcal{A}^{(0 \times 2)}(m_{s(i)}, x_i, \frac{s}{\mu}) = \langle \rm{M}^{(0)}|\rm{M}^{(2)} \rangle + 
\langle \rm{M}^{(2)}|\rm{M}^{(0)} \rangle 
\end{equation}
by numerically integrating the differential equations as described
previously and we also
calculate the numerical value of the quantity
$\mathcal{C}_{atani}^{(0 \times 2)}(m_{s(i)}, x_{(i)}, \frac{s}{\mu})$ by
using Eq.~(\ref{chachamis_catani}).
Then, all we need to make sure is that the infrared singularities
of the quantity $\left\{\mathcal{A}^{(0 \times 2)}(m_{s(i)}, x_i, \frac{s}{\mu}) - 
\mathcal{C}_{atani}^{(0 \times 2)}(m_{s(i)}, x_i, \frac{s}{\mu})\right\}$
cancel numerically for every point $(m_{s(i)}, x_{(i)})$ of the grid (ultraviolet divergencies
have been removed by renormalization).
We will not present here any numbers since the aim was to describe the 
general methods. The details and the results of the study will be  presented in a future 
publication~\cite{chachamis_czakon}. 

\subsection{CONCLUSIONS}
W pair production via quark-anti-quark-annihilation
is an important signal process in the search for
New Physics as well as the dominant irreducible
background for one of the main Higgs discovery channels:
$H \rightarrow W W \rightarrow 4$ leptons. 
Therefore, the accurate knowledge
of this process is essential for the LHC. 
After having calculated
the two-loop and the one-loop-squared virtual QCD corrections
to the W boson pair production in the high energy limit 
we proceed to the next step. Namely, we use
a combination of a deep expansion in the W mass around the
high energy limit and of numerical integration of differential
equations to compute the two-loop amplitude
with full mass dependence over the whole phase space. A strigent
cross-check of our calculation is to verify
that the infrared structure of our result agrees with the 
prediction of the Catani formalism for the infrared structure of QCD amplitudes.
\\
\\
\hspace{-1.5cm}\large {ACKNOWLEDGEMENTS}

\hspace{-1cm}The author wishes to thank the Les Houches Workshop organizers
for the friendly and stimulating atmosphere. 


}

\section[Computation of integrated subtraction terms numerically]
{COMPUTATION OF INTEGRATED SUBTRACTION TERMS NUMERICALLY \protect\footnote{Contributed by:
G.~Somogyi, Z.~Sz\H or, Z.~Tr\'ocs\'anyi }}
{\graphicspath{{nnlo_subtract/}}

\title{Computation of integrated subtraction terms numerically}

\author{G. Somogyi$^1$, Z. Sz\H or$^2$, Z. Tr\'ocs\'anyi$^2$}
\institute{$^1$DESY, Platanenallee 6, D-15738 Zeuthen, Germany
\\$^2$University of Debrecen and Institute of Nuclear Research of
HAS, H-4001 P.O.Box 51, Hungary
}


\begin{abstract}
We report on a numerical representation of the integrated subtraction
terms of the NNLO subtraction scheme defined in 
Refs.~\cite{Somogyi:2005xz,Somogyi:2006cz,Somogyi:2006da,Somogyi:2006db}.
The integrated approximate cross sections themselves can be written as
products of insertion operators (in colour space) times the Born,
or the one-loop cross section. The insertion operator is constructed from
the numerical representation of the integrated subtraction terms. We
give selected results for the integrated doubly-collinear subtraction term.
\end{abstract}

\subsection{INTRODUCTION}

We consider the NNLO correction to a generic $m$-jet observable,
\begin{align}
\sigma^{\mathrm{NNLO}} &=
\int_{m+2}\!\mathrm{d}\sigma^{{\rm RR}}_{m+2} J_{m+2}
+ \int_{m+1}\!\mathrm{d}\sigma^{{\rm RV}}_{m+1} J_{m+1}
+ \int_m\!\mathrm{d}\sigma^{{\rm VV}}_m J_m\,.
\label{eq:sigmaNNLO}
\end{align}
The three contributions on the right hand side are separately divergent in
$d=4$ dimensions, but their sum is finite for IR safe observables. To
obtain the finite NNLO correction, we first continue analytically all
integrals to $d=4-2\epsilon$ dimensions and then rewrite
Eqn.~(\ref{eq:sigmaNNLO}) as
\begin{equation}
\sigma^{\mathrm{NNLO}} =
\int_{m+2}\!\mathrm{d}\sigma^{{\rm NNLO}}_{m+2}
+ \int_{m+1}\!\mathrm{d}\sigma^{{\rm NNLO}}_{m+1}
+ \int_m\!\mathrm{d}\sigma^{{\rm NNLO}}_m\,,
\label{eq:sigmaNNLOfin}
\end{equation}
that is a sum of three integrals where the integrands,
\begin{equation}
\mathrm{d}\sigma^{{\rm NNLO}}_{m+2} =
\Big\{\mathrm{d}\sigma^{{\rm RR}}_{m+2} J_{m+2}
- \mathrm{d}\sigma^{{\rm RR,A_2}}_{m+2} J_{m}
- \Big[\mathrm{d}\sigma^{{\rm RR,A_1}}_{m+2} J_{m+1}
- \mathrm{d}\sigma^{{\rm RR,A_{12}}}_{m+2} J_{m}\Big]
\Big\}_{\epsilon=0}\,,
\label{eq:sigmaNNLOm+2}
\end{equation}
\begin{equation}
\mathrm{d}\sigma^{{\rm NNLO}}_{m+1} =
\Big\{\Big[\mathrm{d}\sigma^{{\rm RV}}_{m+1}
+ \int_1\mathrm{d}\sigma^{{\rm RR,A_1}}_{m+2}\Big] J_{m+1}
- \Big[\mathrm{d}\sigma^{{\rm RV,A_1}}_{m+1}
+ \Big(\int_1\mathrm{d}\sigma^{{\rm RR,A_1}}_{m+2}\Big)
  \strut^{{\rm A}_{\scriptscriptstyle 1}}
\Big] J_{m} \Big\}_{\epsilon=0}\,,
\label{eq:sigmaNNLOm+1}
\end{equation}
and
\begin{equation}
\mathrm{d}\sigma^{{\rm NNLO}}_{m} =
\Big\{\mathrm{d}\sigma^{{\rm VV}}_m
+ \int_2\Big[\mathrm{d}\sigma^{{\rm RR,A_2}}_{m+2}
- \mathrm{d}\sigma^{{\rm RR,A_{12}}}_{m+2}\Big]
+ \int_1\Big[\mathrm{d}\sigma^{{\rm RV,A_1}}_{m+1}
+ \Big(\int_1\mathrm{d}\sigma^{{\rm RR,A_1}}_{m+2}\Big)
  \strut^{{\rm A}_{\scriptscriptstyle 1}}\Big]\Big\}_{\epsilon=0} J_{m}\,,
\label{eq:sigmaNNLOm}
\end{equation}
are integrable in four dimensions by construction.
The approximate cross sections $\mathrm{d}\sigma^{{\rm RR,A_2}}_{m+2}$ and
$\mathrm{d}\sigma^{{\rm RR,A_1}}_{m+2}$ regularise the doubly- and
singly-unresolved limits of the real-emission contribution,
$\mathrm{d}\sigma^{{\rm RR}}_{m+2}$ respectively. The double subtraction due
to the \ overlap of these two terms is compensated by
$\mathrm{d}\sigma^{{\rm RR,A_{12}}}_{m+2}$.
These terms are given explicitly in Ref.~\cite{Somogyi:2006da}.
Finally, $\mathrm{d}\sigma^{{\rm RV,A_1}}_{m+1}$ and
$\Big(\int_1\mathrm{d}\sigma^{{\rm RR,A_1}}_{m+2}\Big)\strut^{{\rm A}_{\scriptscriptstyle 1}}$ regularise
the singly-unresolved limits of $\mathrm{d}\sigma^{{\rm RV}}_{m+1}$ and
$\int_1\mathrm{d}\sigma^{{\rm RR,A_1}}_{m+2}$
respectively. They are given explicitly in Ref.~\cite{Somogyi:2006db}.

The construction of each approximate cross section in
Eqns.~(\ref{eq:sigmaNNLOm+2}--\ref{eq:sigmaNNLOm}) is based on the
known and universal IR limits of tree level and one-loop squared matrix
elements, and proceeds in two steps. First, the IR factorisation
formulae are written in such a way that their complicated overlap
structure can be disentangled (``matching of limits'')
\cite{Somogyi:2005xz,Nagy:2007mn}.
Second, we define ``extensions'' of the formulae, so that
they are unambiguously defined away from the strict IR limits
\cite{Somogyi:2006cz,Somogyi:2006da,Somogyi:2006db}.
These extensions are defined by the use of various momentum mappings
that map a set of $m+1$ or $m+2$ momenta into a set of $m$ momenta,
\begin{equation}
\{p\}_{m+1} \longrightarrow \{\tilde{p}\}_{m}
\quad\mbox{and}\quad
\{p\}_{m+2} \longrightarrow \{\tilde{p}\}_{m}\,,
\end{equation}
such that (i) the delicate structure of cancellations among the matched
limit formulae in various limits is respected (ii) exact momentum conservation
is implemented, and (iii) the original $m+1$ or $m+2$ particle phase space
factorises exactly into the product of an $m$ particle phase space and a
one- or two-particle phase space measure,
\begin{align}
\mathrm{d}\phi_{m+r}(\{p\}_{m+r};Q) &=
\mathrm{d}\phi_{m}(\{\tilde{p}\}_{m};Q) [\mathrm{d} p_{r,m}]
\,,\qquad r = 1,2
\,.
\label{eq:dphim+r}
\end{align}
To finish the definition of the scheme, one must compute once and for all
the one- and two-particle integrals, denoted formally as $\int_1$ and
$\int_2$, appearing in Eqns.~(\ref{eq:sigmaNNLOm+1}--\ref{eq:sigmaNNLOm}).

In general the integrated subtraction terms are integrals of extensions
over the whole phase space of combinations of the QCD splitting
functions and squared soft currents. In this proceedings we discuss two
examples: 
(i) the singly-collinear subtractions
${\cal C}_{ir}^{(\ell,0)}$ and
(ii) the doubly-collinear subtractions
${\cal C}_{ir,js}^{(0,0)}$, which are
part of $\mathrm{d}\sigma^{{\rm RR,A_1}}_{m+2}$ and
$\mathrm{d}\sigma^{{\rm RR,A_2}}_{m+2}$ in Eqn.~(\ref{eq:sigmaNNLOm+2}),
respectively. The precise definitions of these terms can be found in
Ref.~\cite{Somogyi:2006da}. The meaning of the superscript is
irrelevant for our present purpose (also explained in
Ref.~\cite{Somogyi:2006da}).

Denoting a generic subtraction term by ${\cal X}^{(\ell,k)}$ (such as
${\cal C}_{ir}^{(\ell,0)}$) the integrated counterterms can be written
in the following general form: 
\begin{equation}
\int_r {\cal X}^{(\ell,k)} =
\left[\frac{\alpha_{{\rm s}}}{2\pi}S_\epsilon
\left(\frac{\mu^2}{Q^2}\right)^\epsilon\right]^{r+\ell}
N_X(\epsilon) X^{(\ell)}(x,\ldots)
{\rm Re} \langle{\cal M}_m^{(0)}(\{\tilde{p}\})|
\mbox{{\boldmath $T$}}_i\cdot \mbox{{\boldmath $T$}}_j
\ldots |{\cal M}_m^{(k)}(\{\tilde{p}\})\rangle
\,,
\label{eq:intX}
\end{equation}
where
$ 
S_\epsilon =
(4\pi)^\epsilon/\Gamma(1-\epsilon)\,,
$ 
and $X^{(\ell)}(x,\ldots)$ represents a function that depends on
kinematical invariants of the factorized $m$-parton
phase space. It results in the integration of the subtraction term
${\cal X}^{(\ell,k)}$ over the factorized phase spaces
$[\mathrm{d} p_{r,m}]$ in Eqn.~(\ref{eq:dphim+r}).
In a NNLO computation the possible cases are $r+\ell+k=1$ with
$\ell+k=0$ or 1, and $r=2$ with $\ell+k=0$.  We use the colour- and
spin-state notation of Ref.~\cite{Catani:1996vz}, when the amplitude
for a scattering process involving $m$ final-state momenta, $|{\cal
M}_m^{(k)}\rangle$, is an abstract vector in colour and spin space; $k$
denotes the number of loops.  Colour interactions at QCD vertices are
represented by associating colour charges $\mbox{{\boldmath $T$}}_i$
with the emission of a gluon from each parton $i$. There are $2r$ such
colour charges. Then the functions $X^{(\ell)}$ are dimensionless in
colour-space. For certain subtraction terms, universal, possibly
$\epsilon$-dependent numerical factors, $N_X(\epsilon)$ appear
naturally, which can be factored out. Our purpose is to compute all
functions $X^{(\ell)}$, which we discuss next.

\subsection{INTEGRATING THE COUNTERTERMS}

The actual computation of the integrated counterterms leads to a large
number of multi-dimensional integrals. The ultimate goal is to find the
analytical form of the coefficients of a Laurent expansion (in $\epsilon$)
of these integrals, which turns out to be a rather tedious job. In
order to compute these coefficients as efficiently as possible, we have
explored several methods.

First, it is possible to extend the method of integration-by-parts identities
and solving of differential equations, developed for computing multi-loop
Feynman integrals \cite{Kotikov:1991pm,Remiddi:1997ny},
to the relevant phase space integrations \cite{Aglietti:2008fe}. This method
yields $\epsilon$-expansions with fully analytical coefficients, with the final
results being expressed in terms of two-dimensional harmonic polylogarithms
(after a suitable basis extension, see Ref.~\cite{Aglietti:2008fe} for details).
This approach was used successfully to compute a class of singly-unresolved
integrals \cite{Aglietti:2008fe}.

Second, the phase space integrals that arise can be computed via the method
of Mellin--Barnes (MB) representations \cite{Smirnov:1999gc,Tausk:1999vh,Smirnov:2004ym}.
Here we obtain the $\epsilon$-expansion
coefficients in terms of complex contour integrals over $\Gamma$-functions.
Performing these integrals by the use of the residue theorem, a
representation in terms of harmonic sums is obtained. In many cases, the
sums can be evaluated in a closed form, yielding an analytical result. In
some instances however, we find multi-dimensional MB integrals that are very
difficult to compute fully analytically. Nevertheless, in these situations
a direct numerical evaluation of the appropriate MB representations provides
a fast and reliable way to obtain final results with small numerical
uncertainties. We stress that for phenomenological applications, this is
all that is required, since the numerical uncertainty of the complete
computation is dominated by the phase space integrations.  We
have used the MB method to compute all singly-unresolved integrals
\cite{Bolzoni:2009ye}, and all two-particle integrals appearing in
$\int_2 \mathrm{d}\sigma^{RR,A_{12}}_{m+2}$ as well \cite{Bolzoni:2010bt}.

Finally, the method of iterated sector decomposition
\cite{Heinrich:2008si} can also be used to calculate the integrals we
encounter \cite{Somogyi:2008fc}. Sector decomposition produces a
representation of the $\epsilon$-expansion where the coefficients are given
in terms of (mostly quite cumbersome) finite integrals over the unit
hypercube. The analytical evaluation of these integrals is not feasible
except for the simplest cases. Nevertheless, this method is simple to
implement and can be automated to a large extent. In fact there are
several computer programs that use various implementations of sector
decomposition to provide numerical values of coefficients of the powers
of $\epsilon$ in the Laurent expansion of dimensionally regulated
integrals \cite{Bogner:2007cr,Smirnov:2008py,Carter:2010hi}. 
We found the program {\tt SecDec} powerful and
flexible to generate sufficiently precise values of our integrated
subtraction terms.

Choosing the Cuhre integrator implemented in {\tt SecDec}, we can
easily reach $10^{-7}$ relative precision for the integration. Such
precision is sufficient for our purposes: (i) to demonstrate the
cancellation of the $\epsilon$ poles numerically, and (ii) to compute
the finite integrals in Eqns.~(\ref{eq:sigmaNNLOm+1}) and
(\ref{eq:sigmaNNLOm}). As the numerical uncertainty of the second item
is limited more by the Monte Carlo integration over the $m+1$ and $m$
particle phase spaces, for item (ii) much lower (not better than
$10^{-3}$) precision is sufficient. This looser requirement on the
precision for the O(1) terms and the fact that the integrated
subtraction terms are smooth functions of their parameters,
with logarithmic behaviour for asymptotically small values of the
parameters, makes possible that we find sufficient approximations to
the integrated subtraction terms.

\subsection{APPROXIMATE INTEGRATED SUBTRACTION TERMS}

The computation of the integrated subtraction terms at any given values
of the kinematical parameters, as required in the Monte Carlo integration
over the phase space, is not feasible. In order to demonstrate the
cancellation of the $\epsilon$ poles numerically we can choose several
randomly selected phase space points and evaluate the necessary integrals
with high precision. The cancellation cannot depend on the particular
phase space point. In the case of the finite remainders, in order to
compute the phase space integrals in Eqns.~(\ref{eq:sigmaNNLOm+1}) and
(\ref{eq:sigmaNNLOm}), we are able to find sufficiently precise
approximations to the integrated subtraction terms using a procedure
that can be automated to high degree. The latter point is also
important as there are several hundred integrals to compute. In the
following, we outline our procedure for two cases:
(i) an example with integrals depending on one kinematical parameter,
$(\int_1{\cal X}^{(\ell,k)} = \int_1{\cal C}_{ir}^{(\ell,0)})$ and
(ii) another example with integrals depending on two kinematical parameters,
$(\int_2{\cal X}^{(\ell,k)} = \int_2{\cal C}_{ir,js}^{(0,0)})$.

In order to compute $\int_1{\cal C}_{ir}^{(\ell,0)}$, we have to
integrate the azimuthally averaged Altarelli-Parisi splitting functions
$P_{f_i f_r}^{(\ell)}(z_{i,r},z_{r,i};\epsilon)$ in $4-2\epsilon$
dimensions for the splitting process $f_{ir} \to f_i + f_r$, with
$z_i$ being the momentum fraction of parton $f_i$. It was discussed in
Ref.~\cite{Bolzoni:2009ye} that the corresponding functions
$C_{ir}^{(\ell)}$ can be expressed as combinations of the integrals
(we changed the notation from ${\cal I}$ to ${\cal I}_C$)
\begin{equation}
{\cal I}_C(x;\epsilon,\alpha_0,d_0,\kappa,k,\delta,g_I^{(\pm)}) =
\frac{16\pi^2}{S_\epsilon}Q^{2\epsilon}
\int_1 [\mathrm{d} p^{(ir)}_{1,m+1}]
\frac{z_r^{k+\delta \epsilon}}{s_{ir}^{1+\kappa\epsilon}}g_I^{(\pm)}(z_r)
f(\alpha_0,\alpha_{ir},d(m,\epsilon))
\,.
\label{eq:cI-def}
\end{equation}
In terms of explicit integration variables these collinear integrals
have the general form \cite{Bolzoni:2009ye}
\begin{eqnarray}
&&
{\cal I}_C(x;\epsilon,\alpha_0,d_0;\kappa,k,\delta,g_I^{(\pm)}) =
x
\int_0^{\alpha_0}\! \mathrm{d} \alpha\, \alpha^{-1-(1+\kappa)\epsilon}
\,(1-\alpha)^{2d_0-1}\,[\alpha+(1-\alpha)x]^{-1-(1+\kappa)\epsilon}
\nonumber\\[2mm]&&\qquad
\times
\int_0^1\! \mathrm{d} v
[v\,(1-v)]^{-\epsilon}
\left(\frac{\alpha+(1-\alpha)xv}{2\alpha+(1-\alpha)x}\right)^{k+\delta\epsilon}
\,g_I^{(\pm)}\left(\frac{\alpha+(1-\alpha)xv}{2\alpha+(1-\alpha)x}\right)
\,.
\label{eq:Iint}
\end{eqnarray}
The necessary functions $g_I^{(\pm)}$ are listed in
Ref.~\cite{Bolzoni:2009ye}, where analytic results of these integrals for 
$\alpha_0 = 1$ and $d_0 = 3$ are also presented.

Our present goal is to provide sufficiently precise numerical
approximations to the functions ${\cal I}_C(x)$ in a simple way. The
motivation is that often it is difficult to perform the analytic
computation with arbitrary values of the parameters. For instance, the
derivation with $\alpha_0 = 1$ is rather different from a derivation with
$\alpha_0 < 1$. Also, the choice for $d_0$ is to some extent arbitrary,
and a new choice requires a completely new analytic computation.  Thus,
for the sake of flexibility we propose a fully numerical approach here.

First we used the program {\tt SecDec}, modified such that it can
compute the value of the integral at multiple values of the parameter
$x$ in a single run. For simplicity, we call the O(1) terms of the
integral `measurements'. Then, inspired by the analytic results in
Ref.~\cite{Bolzoni:2009ye}, we fitted these measurements by
combinations of logarithms and polynomials in $x$ of the form 
\begin{equation}
{\cal F}_C(x;\kappa=0,k,\delta=0,g_I^{(\pm)}=1) =
\sum_{n=0}^{n_{\max}} P_n^{(m)}(x,k) \log^n(x)
\,,\qquad
P_n^{(m)}(x,k) = \sum_{n=0}^{m} a_n^{(k)} x^n
\label{eq:cFC}
\end{equation}
where the upper limit $n_{\max}$ is determined by the power $-n_{\max}$ of
the leading pole in the Laurent-expansion (in $\epsilon$) of the integral.
As for the degree of the polynomials we tried several simple choices
($m=1,2,3$). We found that splitting the region of the parameter space
into an asymptotic ($0< x \le 10^{-4}$) and a non-asymptotic  ($10^{-4}
< x \le 1$) region, we could provide a fit with $m=2$ that approximates
the analytic result within relative difference few times $10^{-4}$.
The loss of relative precision is associated with phase space points
where the function changes sign, and its numerical value is close to zero
(around $x=0.2$).

In Fig.~\ref{fig:cFC} we show the approximate function
${\cal F}_C(x;\alpha_0,d_0,0,-1,0,1)$ together with the `measurements',
which coincide with the known exact analytic result to at least six
digit accuracy.  We find very good agreement, which is characterized by
the ratio of the two values in the lower panels.  In
Fig.~\ref{fig:cFC}b we show the approximate function for $\alpha_0=0.1$
and $d_0=3-3\epsilon$ together with the corresponding `measurements'. In
this case the analytic results are not available.
\begin{figure}
\begin{center}
\includegraphics[width=0.49\textwidth]{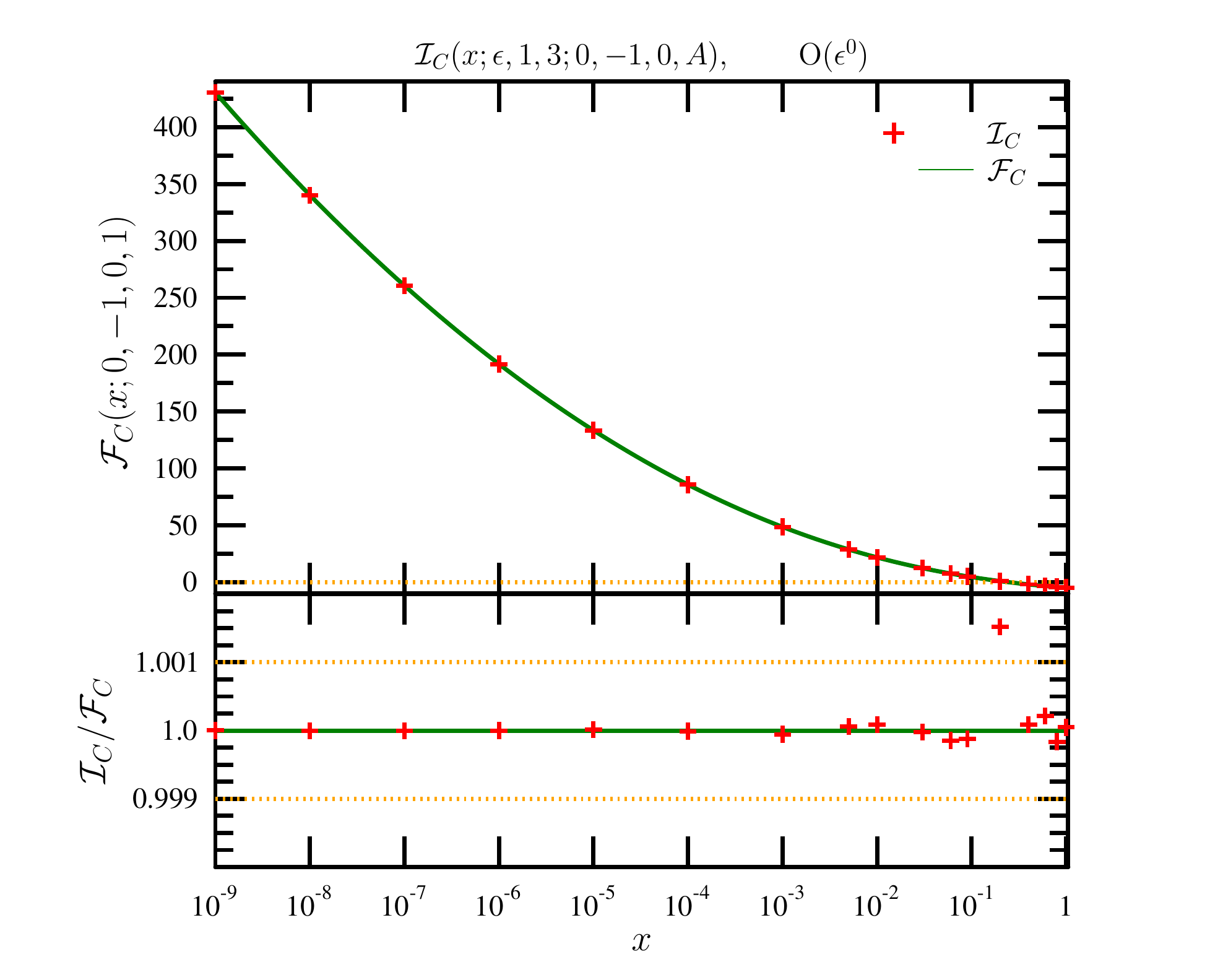}
~
\includegraphics[width=0.49\textwidth]{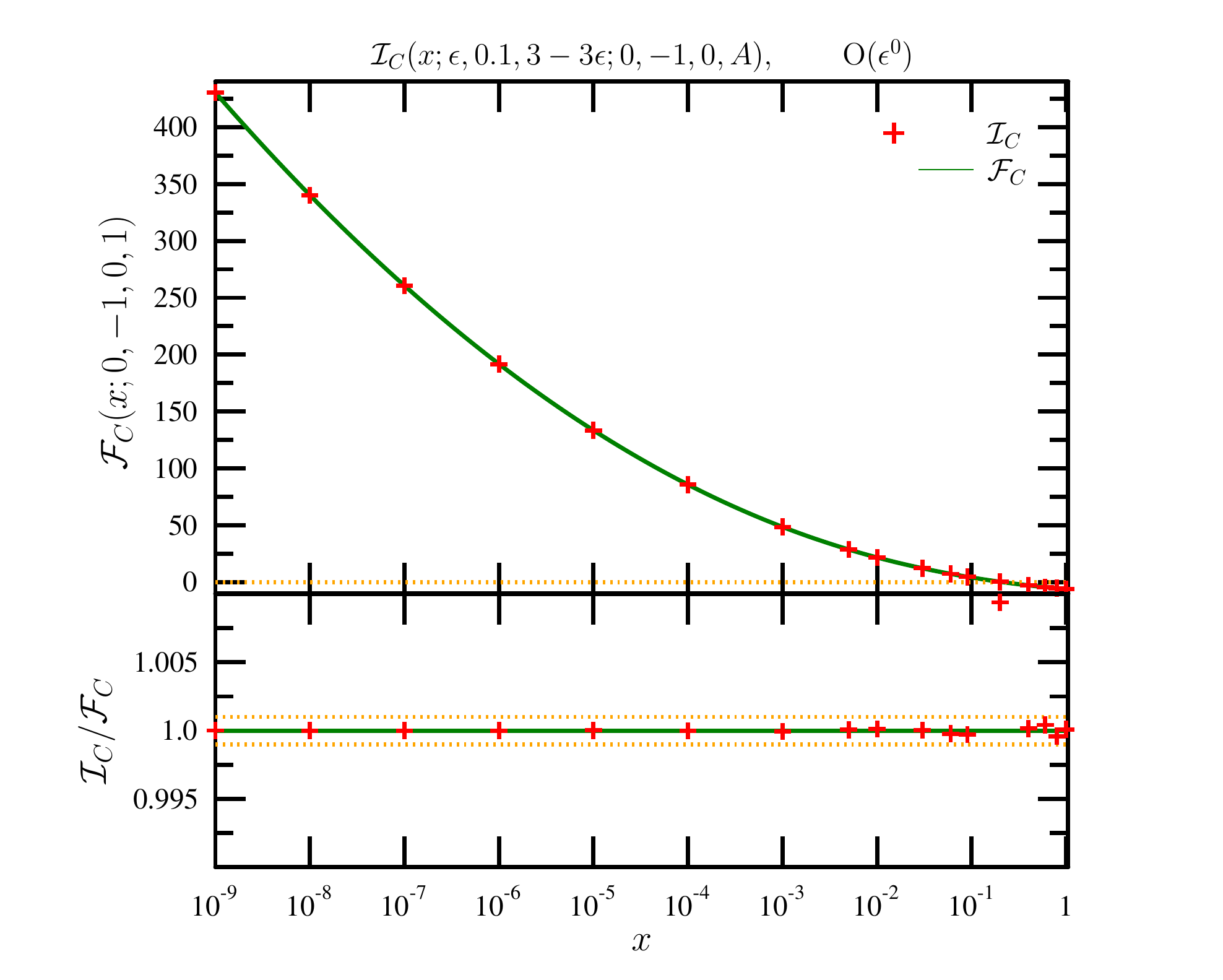}
 \caption{The fitted function ${\cal F}_C(x;0,-1,0,1)$ compared to the
integral ${\cal I}(x;0,-1,0,1)$ at a) $\alpha_0 = 1$ and $d_0 = 3$,
b) $\alpha_0 = 0.1$ and $d_0 = 3 - 3\epsilon$.}
\label{fig:cFC}
\end{center}
\end{figure}

Building on the experience gained in studying the one-parameter case, we
worked out a similar strategy for the integrated subtraction term
$\int_2{\cal C}_{ir,js}^{(0,0)}$. 
The corresponding functions $C_{ir,js}^{(0,0)}$ can be expressed as 
combination of the integrals
\begin{equation}
\begin{split}
{\cal I}_{2C}&(x_i,x_j;\epsilon,\alpha_0,d_0;k,l) = x_i\,x_j
\int_{0}^1\!\mathrm{d}\alpha\,\int_{0}^1\!\mathrm{d}\beta\,
\Theta(\alpha_0-\alpha-\beta)
\\[2mm] &\times
(1-\alpha-\beta)^{2d_0-2(1-\epsilon)}
\alpha^{-1-\epsilon} \beta^{-1-\epsilon}
\left( \alpha+ (1-\alpha-\beta) x_i \right)^{-1-\epsilon}
\left( \beta + (1-\alpha-\beta) x_j \right)^{-1-\epsilon}
\\[2mm] &\times
\int_0^1\!\mathrm{d} v\, v^{-\epsilon} (1 - v)^{-\epsilon}
\int_0^1\!\mathrm{d} u\, u^{-\epsilon} (1 - u)^{-\epsilon}
\left(\frac{\alpha+(1-\alpha-\beta)x_i v}{2\alpha+(1-\alpha-\beta)x_i}\right)^k
\left(\frac{\beta +(1-\alpha-\beta)x_j u}{2\beta +(1-\alpha-\beta)x_j}\right)^l
\,.
\label{eq:I2C}
\end{split}
\end{equation}
We again run {\tt SecDec} with $\alpha_0=0.1$ and $d_0 = 3-3\epsilon$
at several hundred different values of the kinematic parameters to
obtain the `measurements'. To reach $10^{-7}$ relative precision for
all such `measurements' takes several hours on a single CPU. Then we
fitted these `measurements' with the function
\begin{equation}
{\cal F}_{2C}(x_i,x_j;k,l) =
\sum_{n_i=0}^{n_{\max}}\sum_{n_j=0}^{n_{\max}-n_i}
P_{n_i}^{(m)}(x_i,k,l) P_{n_j}^{(m)}(x_j,k,l) \log^{n_i}(x_i) \log^{n_j}(x_j)
\,.
\label{eq:cF2C}
\end{equation}
We divide the parameter space $0< x_i, x_j \le 1$ into four regions:
(i) $0< x_i, x_j \le 10^{-4}$,
(ii) $0< x_i \le 10^{-2}$ and $10^{-4} < x_j \le 1$,
(iii) $0< x_j \le 10^{-2}$ and $10^{-4} < x_i \le 1$,
(iv) $10^{-2}< x_i, x_j \le 1$.
Using $m=2$, we are able to fit the original function ${\cal I}_{2C}$ to
per mille precision almost everywhere. The ratio of the fitted
function ${\cal F}_{2C}$ to the numerical evaluation of ${\cal I}_{2C}$
is shown in Fig.~\ref{fig:cF2C} together whith the fitted function
${\cal F}_{2C}$ itself.

\begin{figure}
\begin{center}
\includegraphics[width=0.49\textwidth]{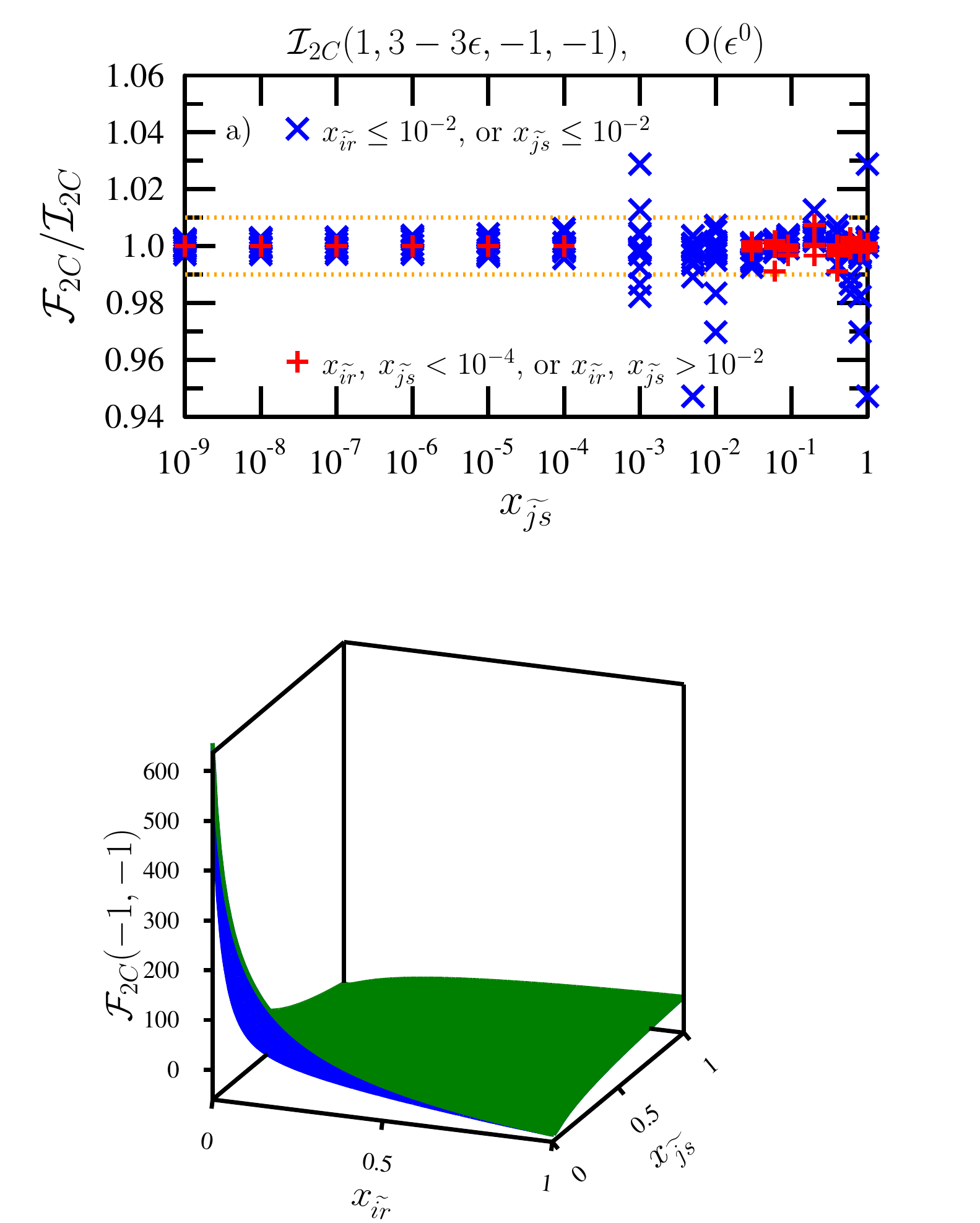}
~
\includegraphics[width=0.49\textwidth]{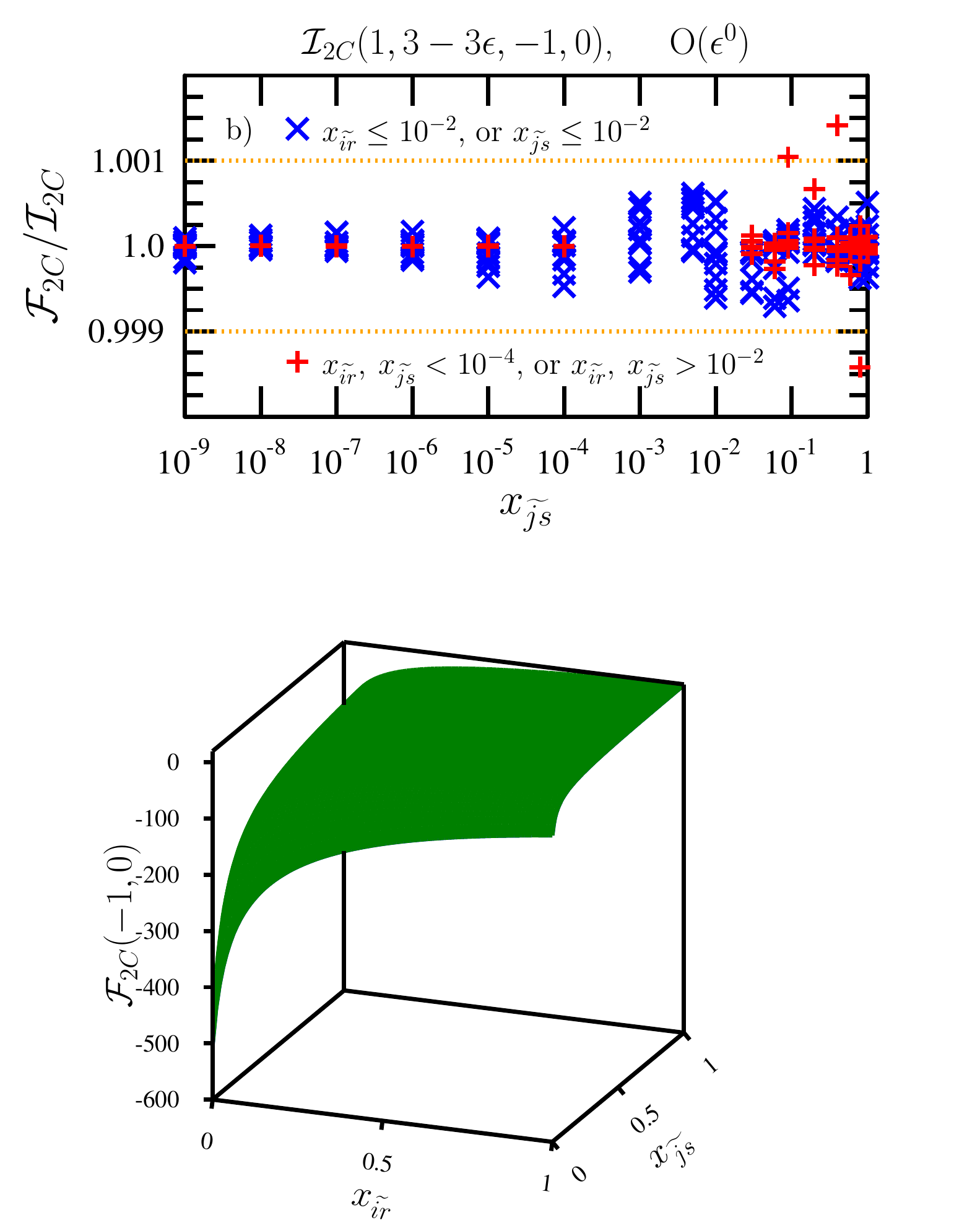}
\caption{The ratio of the fitted function ${\cal F}_{2C}(x_i,x_j;-1,l)$
to the integral ${\cal I}_{2C}(x_i,x_j;-1,l)$. a) $l=-1$ b) $l=0$ 
Also shown the fitted function ${\cal F}_{2C}(x_i,x_j;-1,-l)$.}
\label{fig:cF2C}
\end{center}
\end{figure}

\subsection*{CONCLUSIONS}
We have worked out a numerical procedure for providing simple
approximations of the integrated subtraction terms of the NNLO
subtraction scheme defined in 
Refs.~\cite{Somogyi:2005xz,Somogyi:2006cz,Somogyi:2006da,Somogyi:2006db}.
We use the publicly available program {\tt SecDec} to compute the
coefficients of the Laurent expansion of the necessary integrals to high
numerical precision.  We found that the integrals that depend on one or
two kinematical invariants can be approximated with simple combinations
of polynomials and logarithms. The precision of these approximations is
usually at per mille or better. 

\subsection*{ACKNOWLEDGEMENTS}
We are grateful to G. Heinrich for her help in modifying the {\tt SecDec}
package to our needs. This research was supported in part by the 
the LHCPhenoNet network PITN-GA-2010-264564.

}

\part[PARTON DISTRIBUTION FUNCTIONS]{PARTON DISTRIBUTION FUNCTIONS}
\label{part:pdf}
\section[Which experiments constrain the gluon PDF in a global QCD fit?]
{WHICH EXPERIMENTS CONSTRAIN THE GLUON PDF IN A GLOBAL QCD FIT? \protect\footnote{Contributed by: Z.~Liang, P.~M.~Nadolsky}}
\label{sec:gluonpdf}
{\graphicspath{{nadolsky1/}}



\newcommand{\Dzero}{D\O\xspace}


\title{Which experiments constrain the gluon PDF in a global QCD fit?}

\author{Zhihua Liang, Pavel M. Nadolsky}
\institute{Department of Physics, Southern Methodist University, 
Dallas, TX 75275, USA}


\begin{abstract}
Based on computation of PDF-induced correlations, 
we identify the experiments in CTEQ and MSTW global QCD analyses
that are sensitive to the gluon parton density in the
proton. The Tevatron 
inclusive jet production at large momentum fractions $x$
and DIS charm quark production at moderately small $x$ 
show the strongest correlation with the gluon PDF. 
The strength of the PDF-induced correlation between the gluon PDF
and inclusive  (di)jet production data 
is different in the CTEQ and MSTW analyses. 
\end{abstract}


\subsection{Introduction}
\label{sec:introduction}
\begin{figure}
\centering
\includegraphics[width=0.9\textwidth]{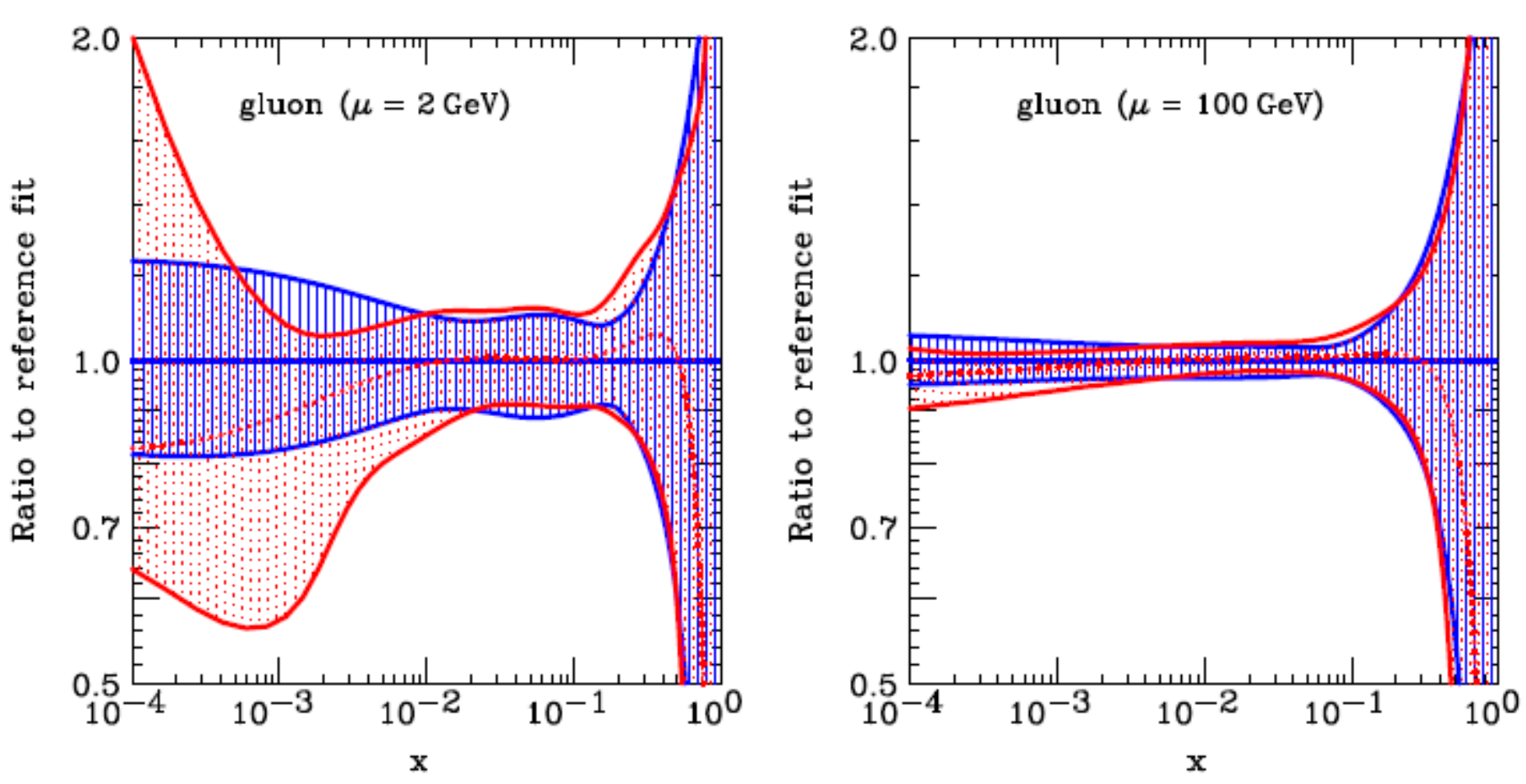}
\caption{CT10 and CTEQ 6.6 PDF uncertainty bands at $\mu = 2$ GeV
  (left) and $100$ GeV (right), taken from Ref.~\cite{Lai:2010vv}. The CTEQ 6.6 best-fit PDFs and uncertainties are indicated by solid curves and hatched bands, while those of CT10 are indicated by dashed curves and dotted bands.}
\label{fig:gluondisct10}
\end{figure}

The parton distribution function (PDF) of gluons in a proton, $g(x,\mu)$, 
plays an important role in hadron
collider phenomenology. It arises in cross sections for production of hadronic final states, massive scalar bosons, and hypothetical elementary particles, often in a combination
with an overall normalization prefactor proportional to $\alpha_s$.  
The gluon distributions from CT10 \cite{Lai:2010vv} and CTEQ 6.6 \cite{Nadolsky:2008zw} PDF
sets are shown in Fig.\ref{fig:gluondisct10}. The figure shows 
that the gluon PDF is
constrained well by fitted experiments 
at the intermediate momentum fractions $x$, but the uncertainty grows
in the region $x > 0.1$. We may ask which experiments in the global fit
impose the most significant constraints on the the gluon PDF. It is
often said that the precise neutral-current DIS data
provides the tightest constraints on the gluon PDF 
at $x$ of order $10^{-3}$, while inclusive jet production at the
Tevatron plays the key role in constraining the gluon at
$x>0.1$. However, the net PDF uncertainty reflects 
subtle interplay of numerous constraints imposed by
QCD theory and multiple experiments, as well as various correlated
uncertainties in experimental measurements.
In this contribution, we identify the 
experiments with the strongest sensitivity
to the gluon PDF by using a method of
PDF-induced correlations that was developed in
Refs.~\cite{Nadolsky:2008zw,Pumplin:2001ct,Nadolsky:2001yg}. The
analysis of correlations provides a systematic way to identify such 
experiments and also to establish specific ranges
of $x$ and $Q$ where the correlations of the experimental data sets with  
the gluon PDF are the most pronounced. 


\subsection{Log-likelihood $\chi^2$ and PDF-induced correlations}
\label{sec:chi_PDF_corr}
The quality of theory description of an experimental data
set can be quantified by the log-likelihood function $\chi^2$. 
Many 
high-energy physics experiments 
publish three kinds of measurement errors for each data point $i$: 
the statistical error $\sigma_i$, uncorrelated systematic error $u_i$, 
and correlated systematic errors 
$\{\beta_{1i},\beta_{2i},\beta_{3i} .... \beta_{Ki}\}$ of $K$ different types. 
To compare a theory prediction $T_i$ to the data value $D_i$ for a
data point $i$,  while accounting for all types of errors, 
the $\chi^2$ function can be constructed as 
 \cite{Stump:2001gu, Pumplin:2002vw}
\begin{equation}
\label{eq:chi2}
\chi^2 = {\displaystyle\sum_{expt.}} \left[ \displaystyle{\sum_{i=1}^{N_e} \left( \frac{  D_i - T_i(a) - \displaystyle{\sum_{k=1}^{K}} r_k \beta_{ki}  }{\alpha_i^2} \right)^2 + \displaystyle{\sum_{k=1}^{K}} r_k^2 }  \right],
\end{equation}
where $\alpha_i^2 = \sigma_i^2 + u_i^2 $ is the combined uncorrelated
error; $r_k$ are random parameters describing each of $K$ correlated
errors (each distributed according to the standard normal
distribution); $N_e$ is the number of the data points; and $K$ is the number of the
sources of the correlated systematic errors.

Analytic minimization of the function (\ref{eq:chi2}) with respect to the correlated systematic parameters $r_k$
renders the following result \cite{Pumplin:2001ct, Stump:2001gu}:
 \begin{equation}
    \label{eq:bestfit}
    \left. r_k\right|_{\rm best\ fit} = \displaystyle{\sum_{k^\prime = 1}^{K} } A^{-1}_{kk^\prime}B_{k^\prime},
 \end{equation}
 where $A_{kk^\prime}$ and $B_k$ are given by
 \begin{equation}
    A_{kk^\prime} = \delta_{kk^\prime} + \displaystyle{\sum_{i=1}^{N_e}} \frac{\beta_{ki}\beta_{k^\prime i}}{\alpha_i^2},
    \quad \text{ and } \quad
    B_k = \displaystyle{ \sum_{i=1}^{N_e} } \frac{\beta_{ki} (D_i - T_i) }{\alpha_i^2}.
 \end{equation}
 Substituting Eq.~(\ref{eq:bestfit}) into Eq.~(\ref{eq:chi2}), we obtain a reduced $\chi^2$ function \cite{Pumplin:2001ct,Stump:2001gu}, 
 \begin{equation}
 \label{eq:chi2anly}
    \chi^2 = \displaystyle{ \sum_{expt.} } \left[ \displaystyle{ \sum_{i=1}^{N_e} } \frac{ (D_i - T_i)^2 }{\alpha_i^2} - \displaystyle{ \sum_{k,k^\prime = 1}^{K}} B_k A_{kk^\prime}^{-1} B_{k^\prime}  \right].
 \end{equation}
In this function, the information about the systematic shifts in $r_k$ is included implicitly. Often, the influence of the correlated shifts on the PDFs
is substantial. 


Next, we wish to discuss correlations between PDF uncertainties of two
variables, $ X(\vec{a}) $ and $ Y(\vec{a}) $, where  $\vec{a}=\{a_1,
a_2, ..., a_N\}$ is the vector of $N$ PDF parameters. The correlations
can be computed either in the Hessian
\cite{Nadolsky:2008zw,Pumplin:2001ct,Nadolsky:2001yg} 
or Monte-Carlo \cite{Ball:2008by} approaches. 
In this note we will adopt the Hessian approach.

A symmetric PDF uncertainty $\Delta X$ corresponds to the maximal
variation of $X$ for all combinations of PDF parameters that lie within the tolerance
hypersphere $\Delta\chi^2 \leq T^2$. This uncertainty
is given by
\begin{equation}
    \Delta X = \frac{1}{2} \sqrt{\sum^N_{i=1} [X^+_i - X^-_i]^2 }
\label{eq:masterEQ}
\end{equation}
in terms of the value $X_0$ of $X$ obtained with the central PDF set,
and values $X_i^+$ and $X_i^-$ of $X$ obtained for maximal positive and negative
displacements of each orthonormal PDF parameter $a_i$ within the
tolerance hypersphere. The same ``master equation'' defines 
$\Delta Y$, the PDF uncertainty of the variable $Y$.

In the linear approximation, the pairs of values of $X$ and $Y$ that are
allowed within the PDF uncertainty correspond to the points inside an
ellipse in the $X$-$Y$ plane. The boundary of the ellipse is
parametrically described by  
\begin{eqnarray}
    &&X = X_0 + \Delta X \cos \theta, \\
    &&Y = Y_0 + \Delta Y \cos( \theta + \varphi ),
\end{eqnarray}
where the parameter $ \theta$ varies between 0 and $2\pi$, and
the relative phase angle $\varphi$ is 
a function of $X^{\pm}_i$ and $Y^\pm_i$. The PDF
uncertainties $\Delta X$ and $\Delta Y$ are calculated according to
Eq.~(\ref{eq:masterEQ}). The angle 
$\varphi$ is included between the gradients $\vec{\nabla} X$ and
$\vec{\nabla} Y$ of $X$ and $Y$ in the PDF parameter space. Its cosine, 
\begin{equation}
    \cos \varphi = \frac{ {\vec{\nabla}X} \cdot {\vec{\nabla} Y}
    }{\Delta X \Delta Y} = \frac{1}{4\Delta X \Delta Y}
    \displaystyle{\sum_{i=1}^{N}} \left (  X_i^{(+)} -  X_i^{(-)}
    \right ) \left (  Y_i^{(+)} -  Y_i^{(-)} \right ),
\label{eq:cosphi}
\end{equation}
quantifies the degree of similarity in the PDF
dependence of $X$ and $Y$. If $X$ and $Y$ are strongly correlated
(corresponding to $\cos \varphi \rightarrow 1$) or anti-correlated ($\cos
\varphi \rightarrow -1$), the PDF uncertainties of $X$ and $Y$ are driven
by essentially the same combinations of PDF parameters. Conversely,
the PDF dependence of $X$ is independent from the PDF dependence of
$Y$ if $\cos \varphi \approx 0$.


\subsection{Which experiments are sensitive to the gluon PDF?}
\label{sec:what_cons_gluon}

If an experimental cross section $\sigma$ strongly constrains a PDF
$f_{a}(x,Q)$ for some combination of $x$ and $Q$, we expect that
Eq.~(\ref{eq:cosphi}) returns $|\cos\varphi|$ close to unity when using
$X=f_{a/A}(x,Q)$ and $Y=\sigma$. If the experimental data set includes
several data points, we can use $Y=\chi^2$. The strength of the constraint
on the PDF from this experiment is determined by $|\cos\varphi|$ and the magnitude of $\chi^2$. In the majority of the fitted experiments, 
$\chi^2/N_e$ is close to 1, so that $|\cos\varphi|$ tends to be more important for distinguishing between the sensitivities of the experiments than the magnitude of $\chi^2$.

Following this approach, we compute
$\cos\varphi$ between the NLO gluon PDF $g(x,Q)$ in various $x$ ranges
(for $Q^2 = 10\mbox{ GeV}^2$),
and $\chi^2$ for typical experimental data sets that are used in the
PDF analysis. In this study, we compute $\cos\varphi$ for the
experiments from the CT10 analysis 
that are listed in Table~\ref{tab:EXP_bin_ID}. In the figures, we
refer to each experiment by its numerical ID that is shown in the left column
of  Table~\ref{tab:EXP_bin_ID}.

The $\cos\varphi$ values between the gluon PDF at a given $x$ value
and $\chi^2$ for each experiment are plotted as two-dimensional contour plots for CT10
NLO PDFs \cite{Lai:2010vv} in the left panel of
Fig.~\ref{fig:corr_exp_PDF}, and for MSTW'08 NLO PDFs
\cite{Martin:2009iq} in the right panel. The horizontal axis indicates
the range of $x$ in $g(x,Q)$.
The vertical axis indicates the ID of
the experiment.  At the bottom of the figure, we show the color legend adopted to draw the contour plots. The color legend is chosen so as to emphasize 
only cells with large
correlation ($\cos\varphi > 0.5$, dark yellow-red colors) 
or large anticorrelation ($\cos\varphi < -0.5$, blue colors). 
The regions with
$|\cos\varphi|<0.5$ are filled with a light-yellow color. The $\chi^2$
values for each data set are computed according to
Eqs. (\ref{eq:chi2}) and (\ref{eq:chi2anly}) 
using the CTEQ fitting code for both CT10 and MSTW PDF sets.

\begin{figure}
\centering
\subfloat{ \includegraphics[width=0.44\textwidth]{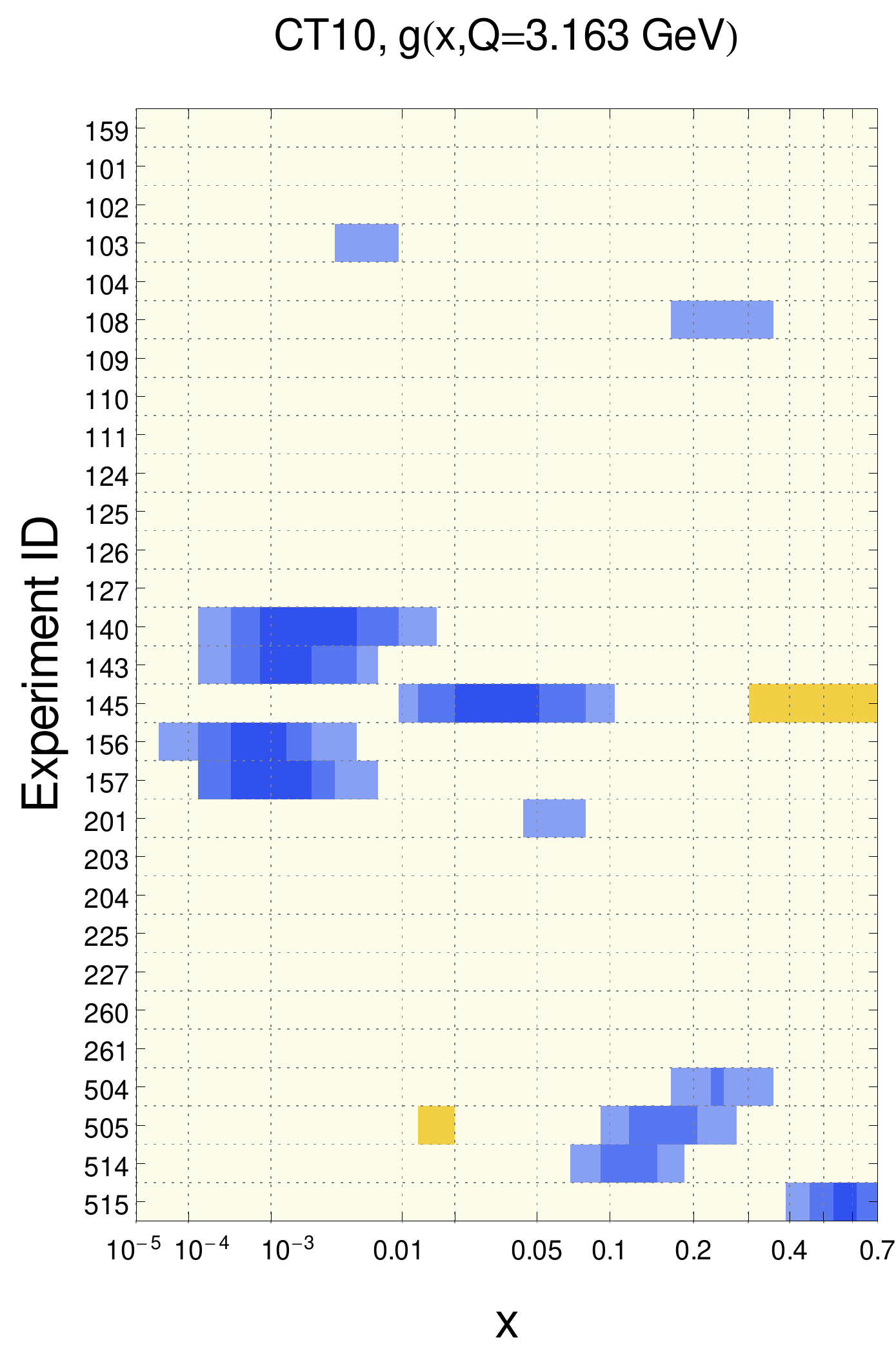} }
\subfloat{
  \includegraphics[width=0.44\textwidth]{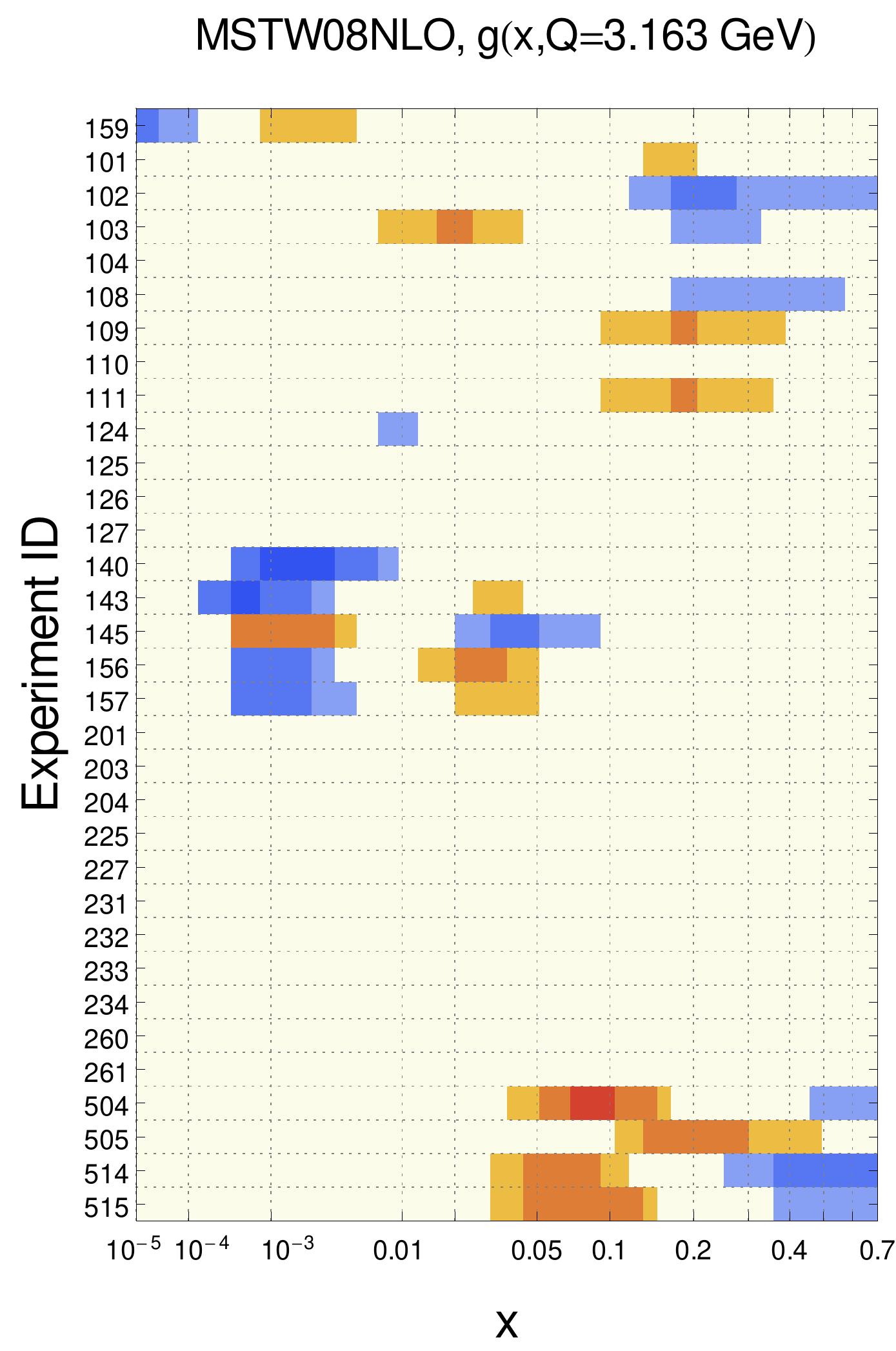}
}\\ \vspace{12pt}
\includegraphics[width=0.5\textwidth]{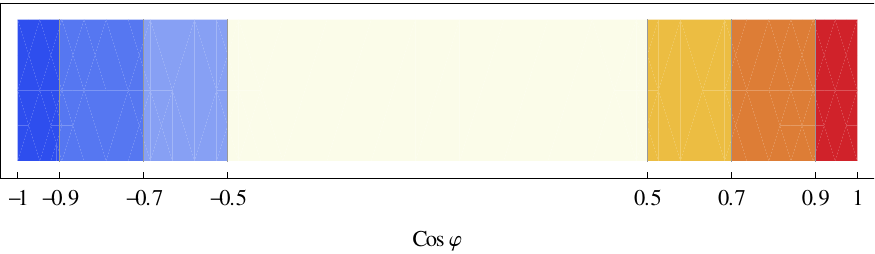}
\caption{ Correlation between the gluon
  distribution from CT10 NLO (left) and MSTW2008 NLO (right) PDF
  sets and $\chi^2$ for the experiments used in the CT10 global QCD analysis.
  The color of each cell indicates the value of $\cos\varphi$
  according to the included legend. The ID's of individual
  experiments are listed in Table~\protect{\ref{tab:EXP_bin_ID}}.}
\label{fig:corr_exp_PDF}
\end{figure}

\begin{table}[p]
\centering
\begin{tabular}{|l|l|}
  \hline
  \hline
  ID  & Experimental data set\\
  \hline
  159 & Combined HERA1 NC and CC DIS                \cite{:2009wt}\\
  101 & BCDMS $F_2^p$                                                  \cite{Benvenuti:1989rh}\\
  102 & BCDMS $F_2^d$                                                  \cite{Benvenuti:1989fm}\\
  103 & NMC $F_2^p$                                                      \cite{Arneodo:1996qe}\\
  104 & NMC $F_2^d/F_2^p$                                              \cite{Arneodo:1996qe}\\
  108 & CDHSW $F_2^p$                                                 \cite{Berge:1989hr}\\
  109 & CDHSW $F_3^p$                                                 \cite{Berge:1989hr}\\
  110 & CCFR $F_2^p$                                                  \cite{Yang:2000ju}\\
  111 & CCFR $x F_3^p$                                                 \cite{Seligman:1997mc} \\
  124 & NuTeV neutrino dimuon SIDIS                            \cite{Mason:2006qa}\\
  125 & NuTeV antineutrino dimuon SIDIS                       \cite{Mason:2006qa}\\
  126 & CCFR neutrino dimuon SIDIS                             \cite{Goncharov:2001qe}\\
  127 & CCFR antineutrino dimuon SIDIS                        \cite{Goncharov:2001qe}\\
  140 & H1 $F_2^c$                                       \cite{Adloff:2001zj} \\
  143 & H1 $\sigma_r^c$ for $c\bar{c}$                           \cite{Aktas:2004az,Aktas:2005iw}\\
  145 & H1 $\sigma_r^b$ for $b\bar{b}$                         \cite{Aktas:2004az,Aktas:2005iw}\\
  156 & ZEUS $F_2^c$                                     \cite{Breitweg:1999ad}\\
  157 & ZEUS $F_2^c$                                             \cite{Chekanov:2003rb} \\
  201 & E605 Drell-Yan process, $\sigma(pA)$                                          \cite{Moreno:1990sf}\\
  203 & E866 Drell Yan process, $\sigma(pd)/(2\sigma(pp))$                                                     \cite{Towell:2001nh}\\
  204 & E866 Drell-Yan process, $\sigma(pp)$                                                      \cite{Webb:2003ps}\\
  225 & CDF Run-1 $W$ charge asymmetry                                          \cite{Abe:1996us}\\
  227 & CDF Run-2 $W$ charge asymmetry                                          \cite{Acosta:2005ud}\\
  231-234 & \Dzero Run-2 $W$ charge asymmetry    \cite{Abazov:2008qv} \\
  260 & \Dzero Run-2 Z rapidity distribution                              \cite{Abazov:2006gs}\\
  261 & CDF Run-2 Z rapidity distribution                     \cite{Aaltonen:2010zza} \\
  504 & CDF Run-2 inclusive jet production                             \cite{Aaltonen:2008eq}\\
  505 & CDF Run-1 inclusive central jet production                   \cite{Affolder:2001fa}\\
  514 & \Dzero Run-2 inclusive jet production                          \cite{:2008hua}\\
  515 & \Dzero Run-1 inclusive jet production                       \cite{Abbott:2000ew}\\
  \hline
\end{tabular}
\caption{Experimental data sets examined in this analysis.}
\label{tab:EXP_bin_ID}
\end{table}

Visual inspection of two panels of Fig.~\ref{fig:corr_exp_PDF}
reveals both similarities and differences in the pattern of
correlations of the gluon PDF in the CT10 and MSTW PDF sets. 
In the case of the CT10 PDF (left panel), the gluon PDF has a
pronounced anti-correlation (blue spots) with HERA charm and
bottom SIDIS production data sets (experiments 140, 143, 145,
156, 157) at $x<0.1$, as well as with
Tevatron inclusive jet production data sets (experiments
504, 505, 514, and 515) at $x>0.05$. Some correlations (brown and red spots) are
also observed, but they are not as pronounced as the anti-correlations. Weaker
(anti-)correlations can be noticed with the NMC $F_2^p$, CDHSW
$F_2^p$, and E605 $pp$ Drell-Yan process data, corresponding to
experiments 103, 108, and 201. 

While the gluon PDF of the MSTW'08 set (right panel of
Fig.~\ref{fig:corr_exp_PDF}) also shows an (anti-)correlation with the
heavy-quark DIS and jet production data, the overall pattern of the
correlations is somewhat different from the CT10 case. 
Here, the gluon PDF is mostly
correlated with high-$x$ jet production (experiments 504, 505, 514,
and 515), while it is either correlated or anti-correlated with
heavy-quark DIS experiments (experiments 140, 143, 145,
156, 157). In addition, we observe significant (anti-)correlations
with the combined HERA DIS data set (ID=159) and fixed-target DIS
experiments (ID=101-124) that are not seen in the CT10 panel.

\begin{figure}
\centering
\includegraphics[width=0.38\textwidth]{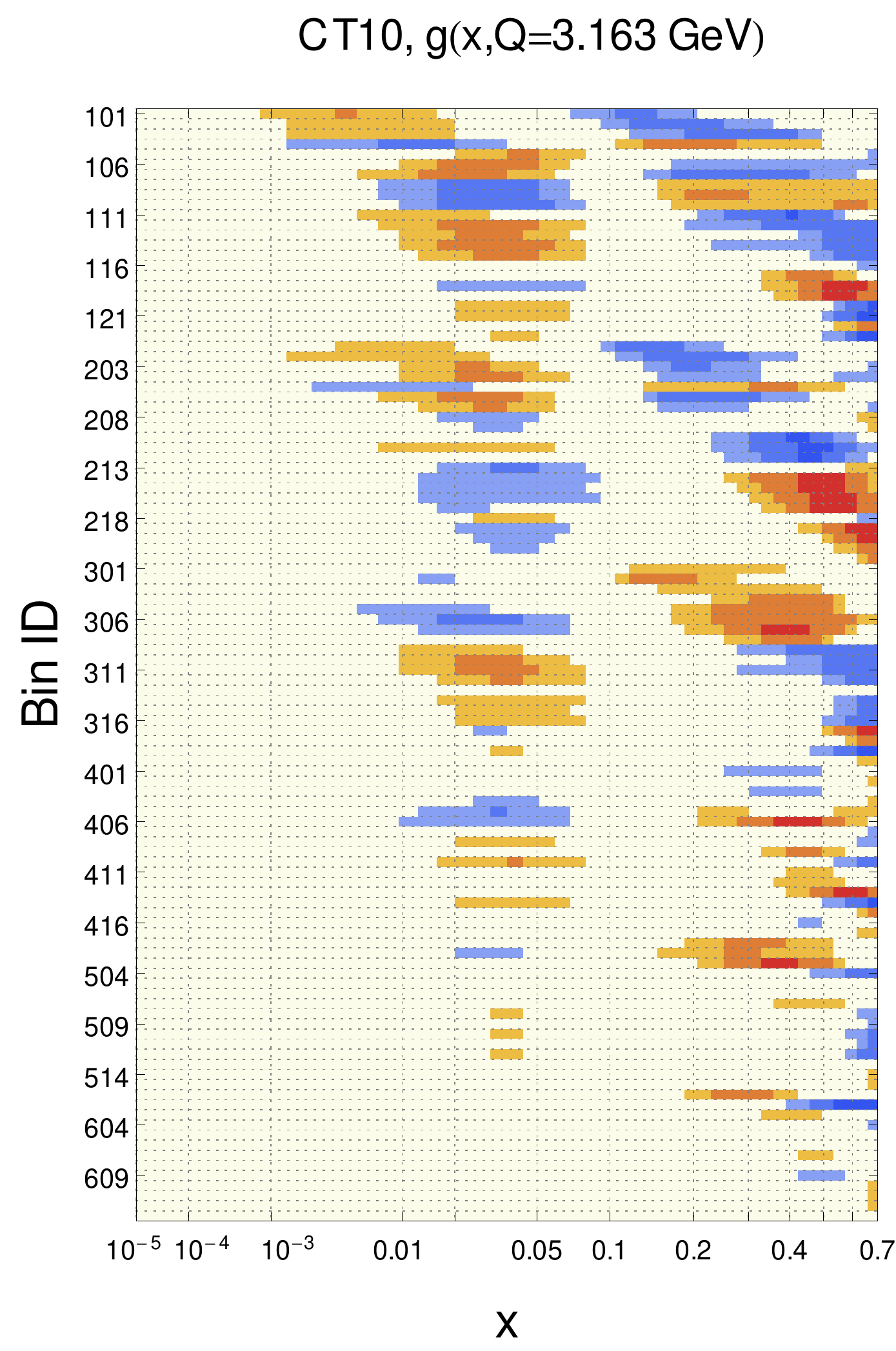} 
\includegraphics[width=0.38\textwidth]{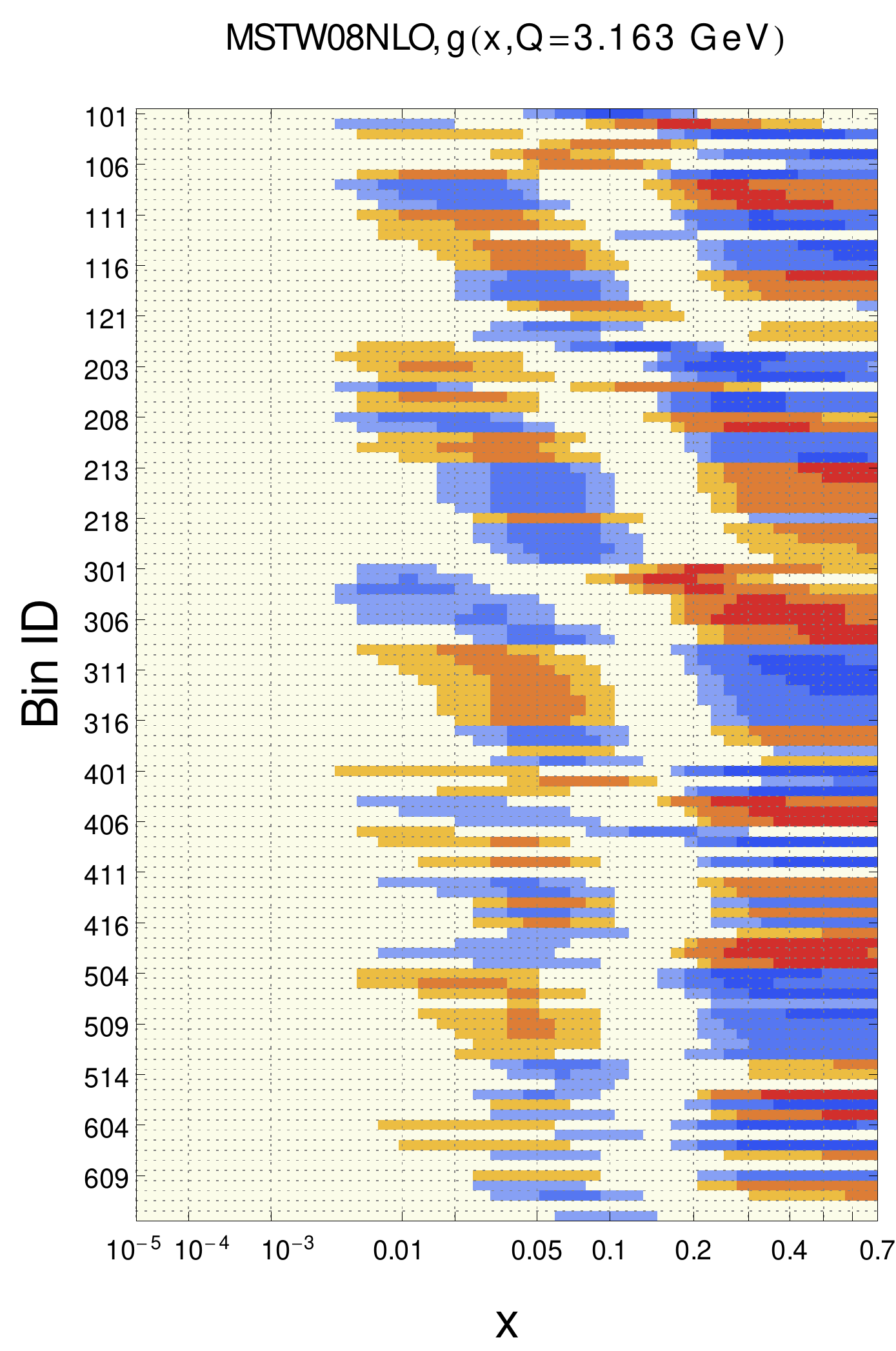} \\
\includegraphics[width=0.38\textwidth]{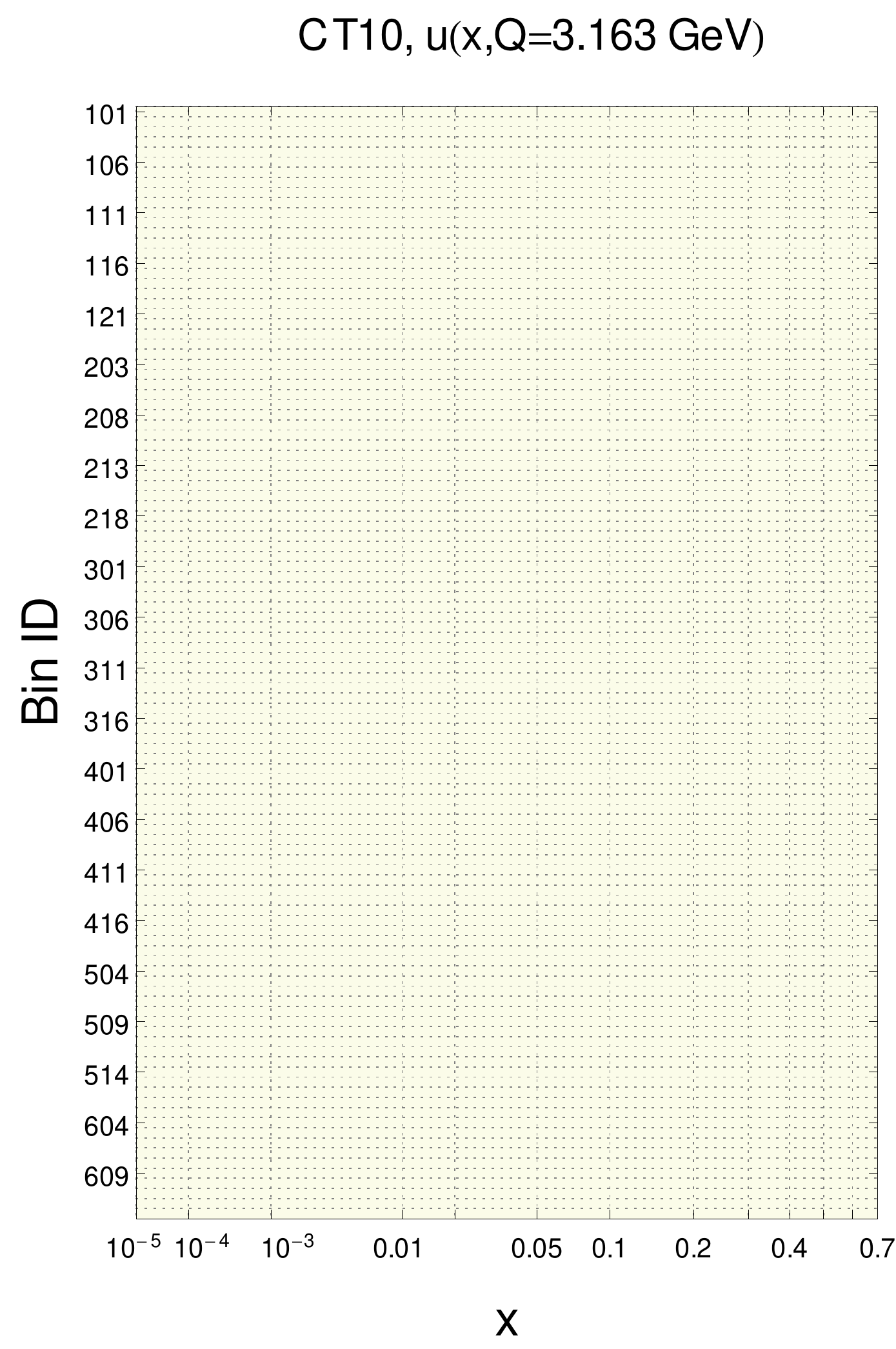} 
\includegraphics[width=0.38\textwidth]{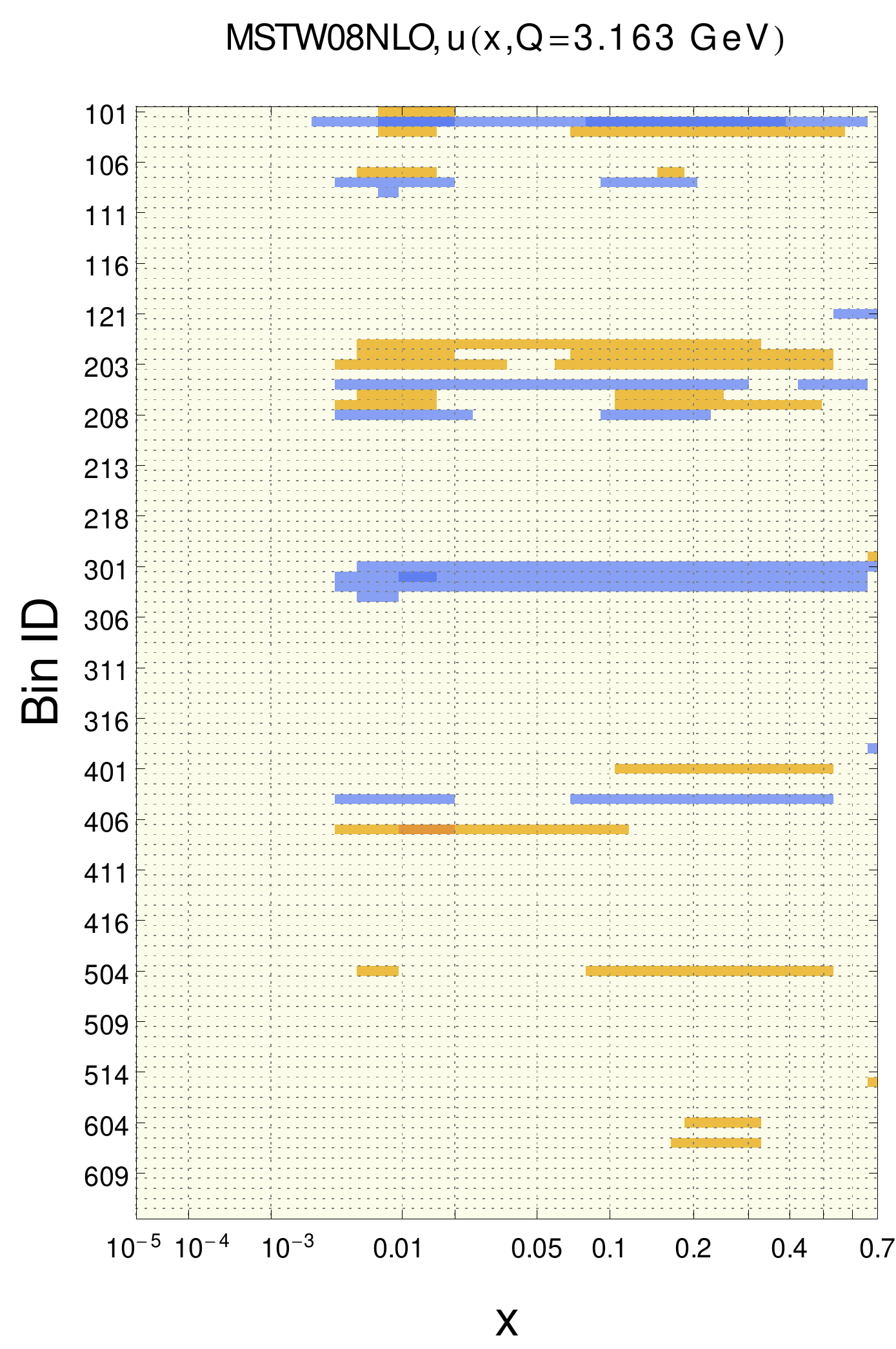} 
\caption{Correlation cosine between $\chi^2_i$ 
in each $p_T$ bin from \Dzero Run-2 inclusive jet production 
and gluon and $u$ quark distributions from CT10 and MSTW 2008 NLO
sets. The horizontal axis refers to the $x$ value in the PDF. 
The vertical axis indicates the numerical ID of the
experimental bin for which $\chi^2$ is computed. The ID for each bin is
indicated as $100\, i_y + i_{p_T}$, where $i_y=1,...6$ and $i_{p_T}$
are the ID's of the corresponding rapidity interval and the $p_T$
interval, respectively.}
\label{fig:corr_sys_shift_ptbins_PDF}
\end{figure}

\begin{figure}
\centering
\includegraphics[width=0.38\textwidth]{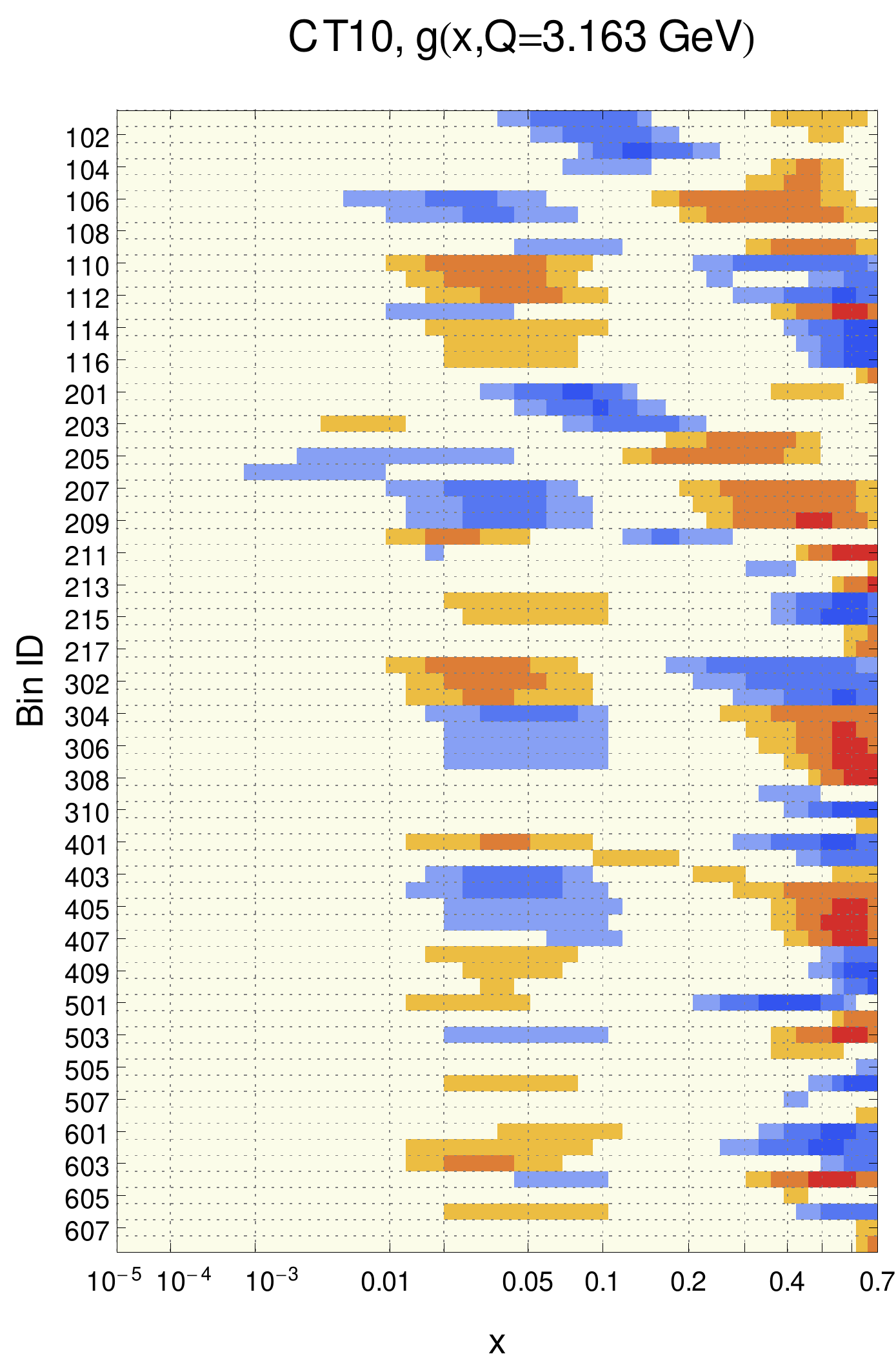} 
\includegraphics[width=0.38\textwidth]{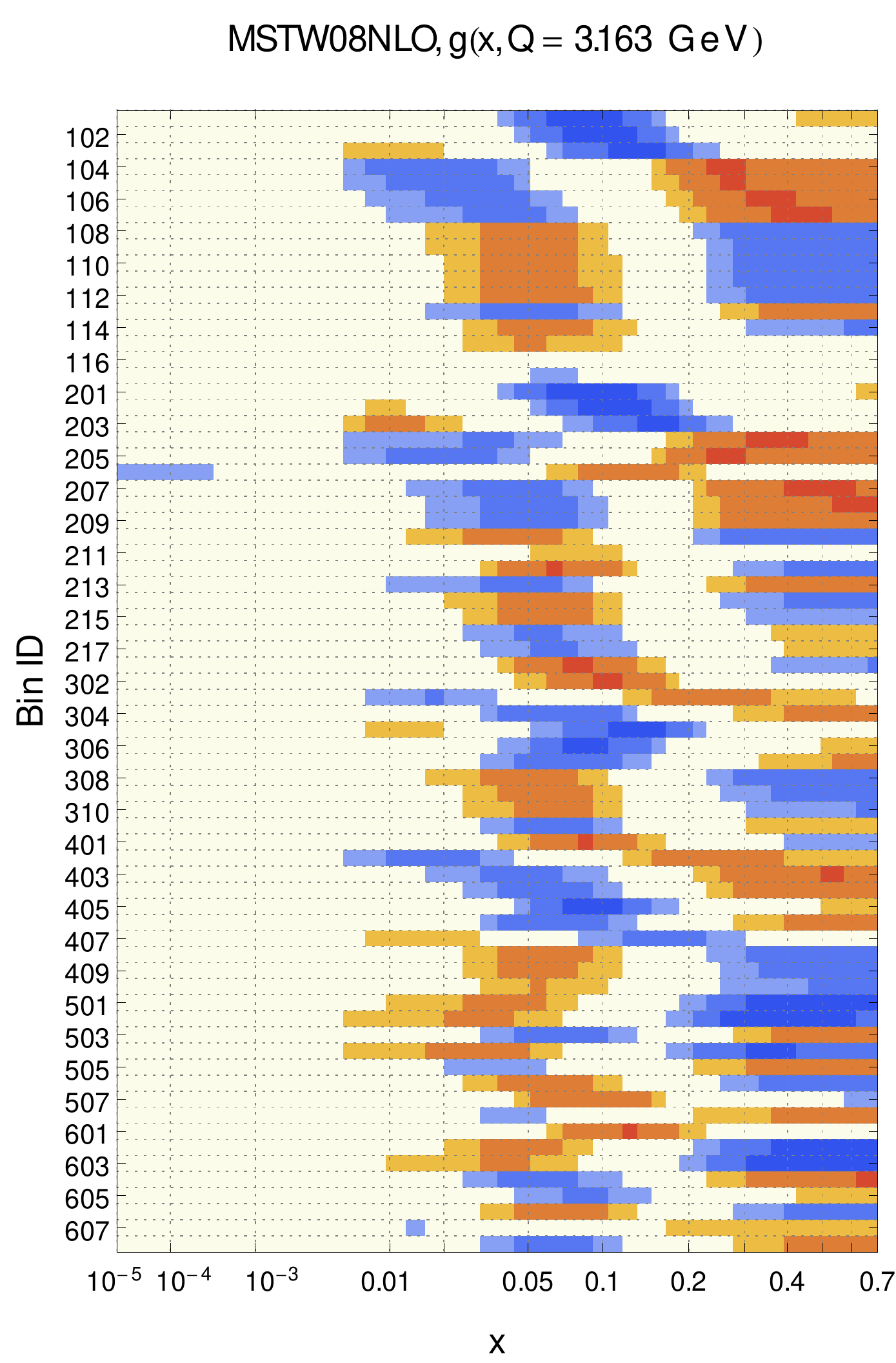} \\
\includegraphics[width=0.38\textwidth]{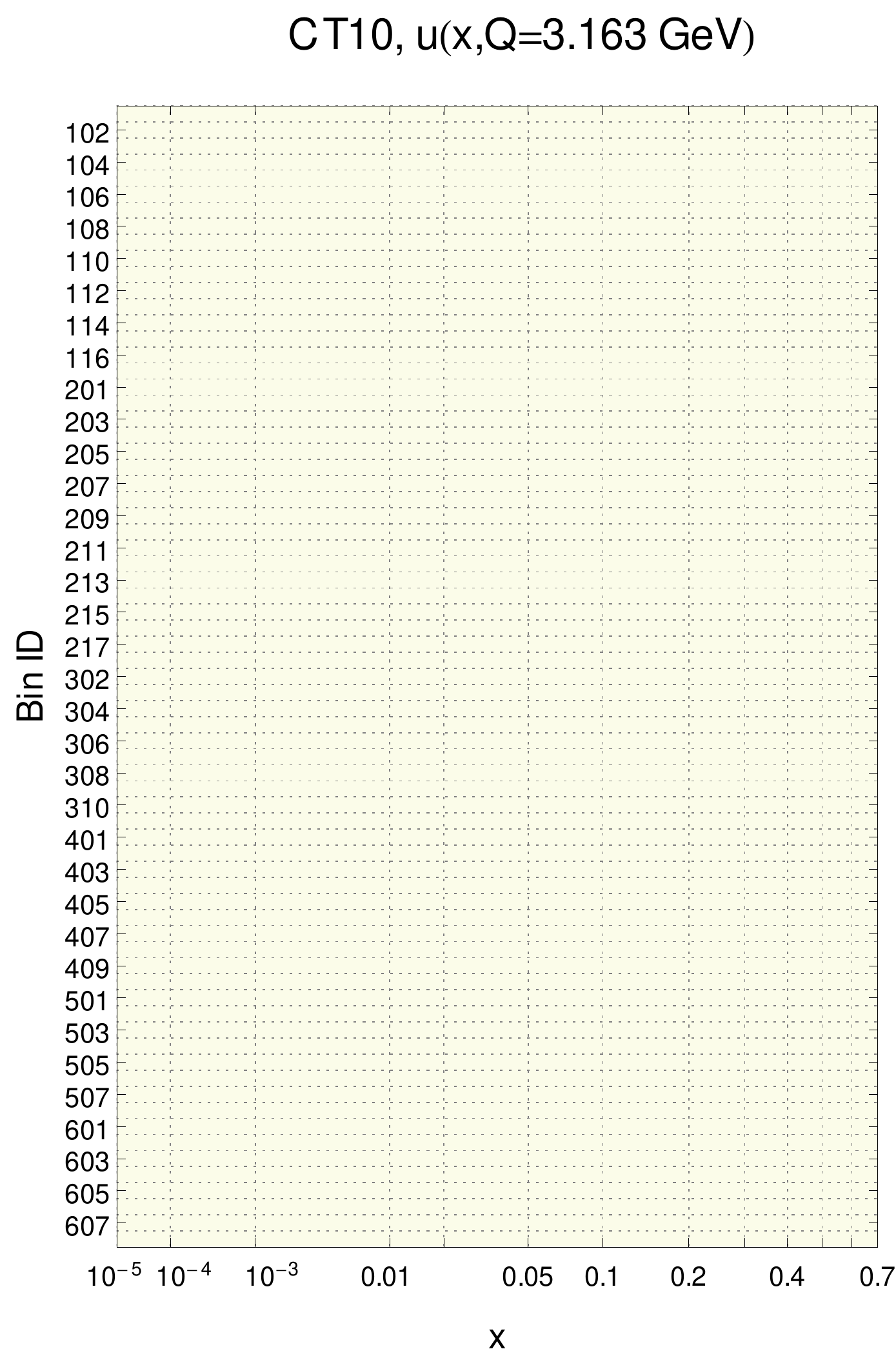}
\includegraphics[width=0.38\textwidth]{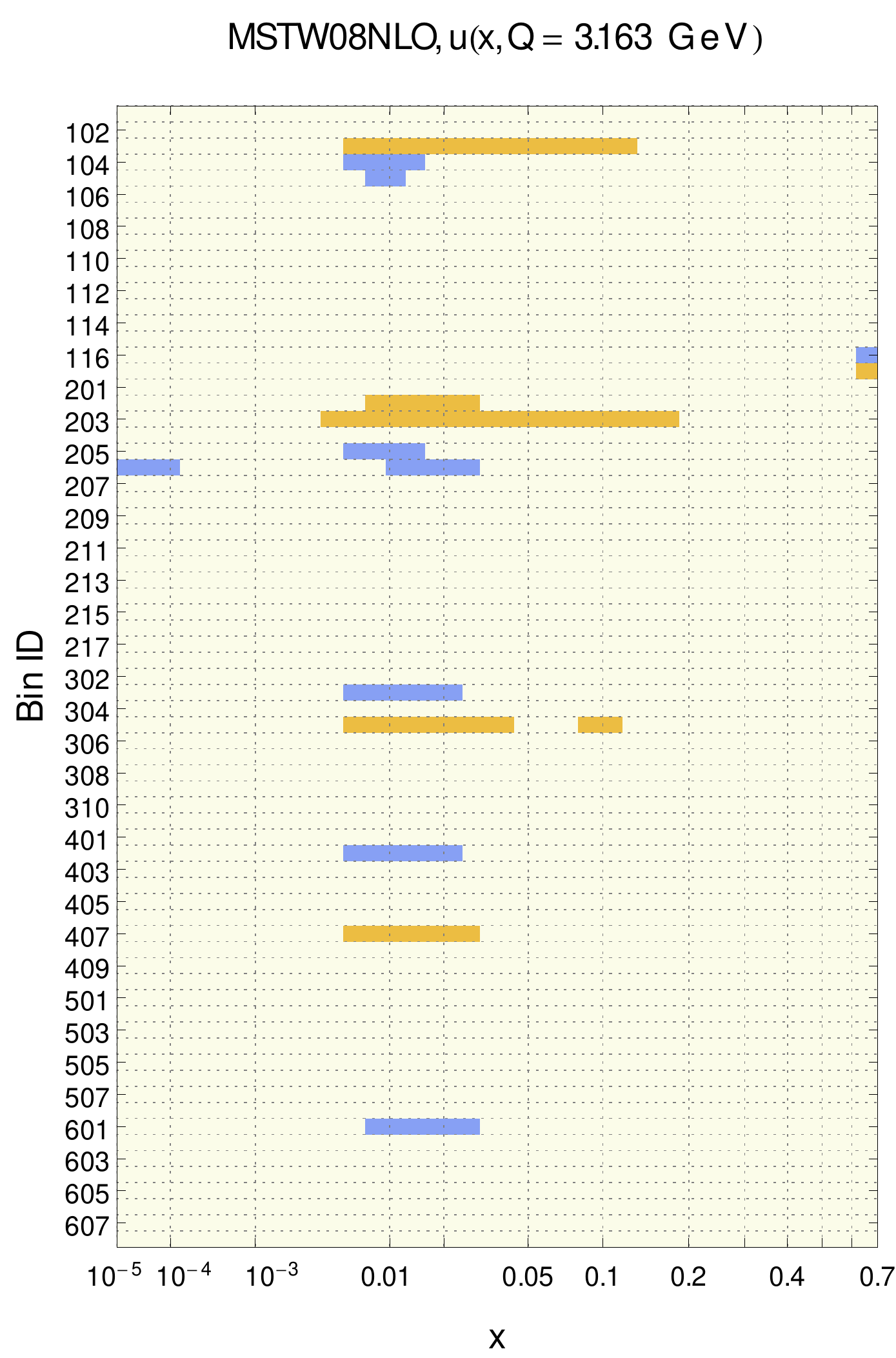} 
\caption{Correlation cosine between $\chi^2_i$ 
in each $m_{jj}$ bin from \Dzero Run-2 dijet production 
and gluon and $u$ quark distributions from CT10 and MSTW 2008 NLO
sets. The horizontal axis refers to the $x$ value in the PDF. The vertical axis indicates the numerical ID of the
experimental bin for which $\chi^2$ is computed. The ID for each bin is
indicated as $100\, i_{y_{max}} + i_{m_{jj}}$, where $i_{y_{max}}=1,..,6$ and $i_{m_{jj}}$
are the ID's of the corresponding intervals in $y_{max}$ and $m_{jj}$,
respectively.}
\label{fig:corr_sys_shift_mjjbins_PDF}
\end{figure}

We now turn to  the correlations of the gluon and $u$-quark PDFs 
with $\chi^2$ values in individual bins of Tevatron inclusive jet and
dijet production data. For this purpose, 
we represent $\chi^2$  for one experimental data set in
Eq.~(\ref{eq:chi2anly}) as a sum of contributions $\chi^2_i$ from
individual data points $i$: 
\begin{equation}
\chi^2 = \sum_{i=1}^{N_e} \chi^2_i, 
\end{equation}
where 
\begin{equation}
\chi^2_i = \frac{D_i-T_i}{\alpha_i} \sum_{j=1}^{N_e}
\left\{\delta_{ij}-
\frac{D_j-T_j}{\alpha_j}
\sum_{k,k'=1}^K\frac{\beta_{ki}}{\alpha_i}
A^{-1}_{kk^\prime}\frac{\beta_{k^\prime j}}{\alpha_j} \right\}.
\label{eq:chi2i}
\end{equation}
 Each contribution $\chi^2_i$ accounts for the effect of
correlated systematic shifts through the term that includes
$A_{kk^\prime}^{-1}$ on the right-hand side of Eq.~(\ref{eq:chi2i}).
Again, the constraining power of each point is determined both by the value of $|\cos\varphi|$ and the magnitude of $\chi^2_i$, with the latter being comparable to unity for the majority of the data points. 

For \Dzero Run-2 single-inclusive jet cross sections \cite{:2008hua},
we plot the $\cos\varphi$ values for the gluon and $u$-quark PDFs,
with $\chi^2_i$ computed for each bin of the jet's transverse momentum $p_T$ and
rapidity $y$. The resulting contour plots are shown 
in Fig.~\ref{fig:corr_sys_shift_ptbins_PDF}. Similarly, for \Dzero
Run-2 dijet cross sections \cite{Abazov:2010fr},
Fig.~\ref{fig:corr_sys_shift_mjjbins_PDF} shows the contour plots of
$\cos\varphi$ for $\chi^2_i$ in the bins of 
of dijet invariant mass $m_{jj}$ and maximal absolute rapidity
$|y|=\max (|y_1|,|y_2|)$ of the dijets. In both figures, theory cross
sections are computed at NLO (without threshold resummation
corrections) with the FASTNLO code~\cite{Kluge:2006xs,ftnlo:2010xy}, using the settings described in Section~\ref{sec:incljet}. The same color legend 
as in Fig.~\ref{fig:corr_exp_PDF} is used. Similar patterns of correlations were found with the CDF Run-2 inclusive jet data (not shown).

The upper panels in both figures show $\cos\varphi$ for CT10 NLO and
MSTW'08 NLO gluon PDFs.  The correlated experimental errors 
modify the correlations by smearing the $\cos\varphi$ distribution.
The pattern of $\cos\varphi$ indicates clearly that the (di)jet data
are very sensitive to the gluon  at $x$ above 0.01. However, the
correlation is weaker for the CT10 gluon PDF (left panel) 
then for MSTW'08 PDF (right panel), suggesting that the importance of
the constraints on the gluon PDF from the jet data is not the same
in two fits. In addition, the MSTW'08 $u$-quark PDF
shows mild (anti-)correlation with both single-inclusive jet data and
dijet data, as can be observed in the right lower panels
 in Figs.~\ref{fig:corr_sys_shift_ptbins_PDF} and
\ref{fig:corr_sys_shift_mjjbins_PDF}. No pronounced
(anti-)correlations with the $u$-quark PDF or other quark PDFs of physical flavors are
observed for the CT10 set, shown in the lower left panels. 

The contour plots confirm the expectation
that the inclusive jet data play an important role in constraining
the gluon PDF. While the constraints are strongest at $x > 0.1$,
they extend down to $x$ as low as 0.05 for both CT10 and MSTW sets,
as can be observed in  Figs.~\ref{fig:corr_sys_shift_ptbins_PDF} and
\ref{fig:corr_sys_shift_mjjbins_PDF}. The gluon PDF is sensitive 
to constraints from heavy-quark
semi-inclusive DIS production at even lower $x$ values, 
cf. Fig.~\ref{fig:corr_exp_PDF}. As the HERA data on
heavy-quark DIS production continue to improve, it will play an increasingly
important role in constraining the low-$x$ gluon density.

While the patterns of PDF-induced correlations are visually similar for
the CT10 and MSTW'08 sets, they are not completely identical. 
 Constraints on the gluon PDF
from Tevatron jet production may not be as strong in the CT10 fit as in
the MSTW'08 fit, according to Figs.~\ref{fig:corr_sys_shift_ptbins_PDF} and
\ref{fig:corr_sys_shift_mjjbins_PDF}. It remains to be investigated what causes the observed
differences between CT10
and MSTW sets
in the correlations involving the gluon PDF. Several
features are different in these fits, including
different heavy-quark DIS schemes, choice of experimental data sets, PDF
parametrizations, and radiative contributions in theoretical cross
sections. A combination of these effects may indirectly affect the strength of the
constraints imposed on the gluon density by the collider jet data.

\subsection*{ACKNOWLEDGMENTS}
This work
was supported by the U.S. DOE Early Career Research Reward
DE-SC0003870 and by Lightner-Sams Foundation. We thank M. Guzzi, 
J. Gao, J. Huston, J. Pumplin, D. Stump, and C.-P. Yuan for related
discussions.

}

\section[PDF constraints from Electroweak Vector Boson production at the LHC]
{PDF CONSTRAINTS FROM ELECTROWEAK VECTOR BOSON PRODUCTION AT THE LHC \protect\footnote{Contributed by: R.~D.~Ball, N.~P.~Hartland, J.~Rojo and M.~Ubiali}}
{\graphicspath{{NNPDFewk/}}


\def\smallfrac#1#2{\hbox{${{#1}\over {#2}}$}}
\newcommand{\be}{\begin{equation}}
\newcommand{\ee}{\end{equation}}
\newcommand{\bea}{\begin{eqnarray}}
\newcommand{\eea}{\end{eqnarray}}
\newcommand{\bi}{\begin{itemize}}
\newcommand{\ei}{\end{itemize}}
\newcommand{\ben}{\begin{enumerate}}
\newcommand{\een}{\end{enumerate}}
\newcommand{\la}{\left\langle}
\newcommand{\ra}{\right\rangle}
\newcommand{\lc}{\left[}
\newcommand{\rc}{\right]}
\newcommand{\lp}{\left(}
\newcommand{\rp}{\right)}
\newcommand{\as}{\alpha_s}
\newcommand{\aq}{\alpha_s\left( Q^2 \right)}
\newcommand{\amz}{\alpha_s\left( M_Z^2 \right)}
\newcommand{\aqq}{\alpha_s \left( Q^2_0 \right)}
\newcommand{\aqz}{\alpha_s \left( Q^2_0 \right)}
\def\toinf#1{\mathrel{\mathop{\sim}\limits_{\scriptscriptstyle
{#1\rightarrow\infty }}}}
\def\tozero#1{\mathrel{\mathop{\sim}\limits_{\scriptscriptstyle
{#1\rightarrow0 }}}}
\def\toone#1{\mathrel{\mathop{\sim}\limits_{\scriptscriptstyle
{#1\rightarrow1 }}}}
\def\frac#1#2{{{#1}\over {#2}}}
\def\gsim{\mathrel{\rlap{\lower4pt\hbox{\hskip1pt$\sim$}}
    \raise1pt\hbox{$>$}}}         
\def\lsim{\mathrel{\rlap{\lower4pt\hbox{\hskip1pt$\sim$}}
    \raise1pt\hbox{$<$}}}         
\newcommand{\mrexp}{\mathrm{exp}}
\newcommand{\dat}{\mathrm{dat}}
\newcommand{\one}{\mathrm{(1)}}
\newcommand{\two}{\mathrm{(2)}}
\newcommand{\art}{\mathrm{art}} 
\newcommand{\rep}{\mathrm{rep}}
\newcommand{\net}{\mathrm{net}}
\newcommand{\stopp}{\mathrm{stop}}
\newcommand{\sys}{\mathrm{sys}}
\newcommand{\stat}{\mathrm{stat}}
\newcommand{\diag}{\mathrm{diag}}
\newcommand{\pdf}{\mathrm{pdf}}
\newcommand{\tot}{\mathrm{tot}}
\newcommand{\minn}{\mathrm{min}}
\newcommand{\mut}{\mathrm{mut}}
\newcommand{\partt}{\mathrm{part}}
\newcommand{\dof}{\mathrm{dof}}
\newcommand{\NS}{\mathrm{NS}}
\newcommand{\cov}{\mathrm{cov}}
\newcommand{\gen}{\mathrm{gen}}
\newcommand{\cut}{\mathrm{cut}}
\newcommand{\parr}{\mathrm{par}}
\newcommand{\val}{\mathrm{val}}
\newcommand{\tr}{\mathrm{tr}}
\newcommand{\checkk}{\mathrm{check}}
\newcommand{\reff}{\mathrm{ref}}
\newcommand{\extra}{\mathrm{extra}}
\newcommand{\draft}[1]{}
\newcommand{\comment}[1]{{\bf \it  #1}}
\def\beq{\begin{equation}}  
\def\eeq{\end{equation}}  

\def\bgamma{\boldsymbol{\gamma}}
\def\nn{\nonumber}
\def \so{\sigma_I^{DIS}(x_I,Q^2_I)}
\def \sh{\frac{d\sigma^{hh}}{dX}}
\def\sdy{\frac{d\sigma^{\mathrm{DY}}}{dQ_I^2dY_I}}
\def \npdf{N_{\mathrm{pdf}}}
\def \gtilda{\tilde\Gamma_J^{\mathrm{OBS}}}
\def \n0{N_j^{(0)}}
\def \a{\alpha}
\def \b{\beta}
\def \g{\gamma}
\def \c{\xi}
\def \z{\zeta}
\def\lapprox{\lower .7ex\hbox{$\;\stackrel{\textstyle <}{\sim}\;$}}
\def\gapprox{\lower .7ex\hbox{$\;\stackrel{\textstyle >}{\sim}\;$}}
\def\half{\smallfrac{1}{2}}
\def\GeV{{\rm GeV}}
\def\TeV{{\rm TeV}}
\def\ap{{a'}}
\def\vp{{v'}}
\def\e{\epsilon}
\def\d{{\rm d}}
\def\calN{{\cal N}}
\def\shat{\hat{s}}
\def\barq{\bar{q}}
\def\qq{q \bar q}
\def\uu{u \bar u}
\def\dd{d \bar d}
\def\pp{p \bar p}
\def\xa{x_{1}}
\def\xb{x_{2}}
\def\xaa{x_{1}^{0}}
\def\xbb{x_{2}^{0}}
\def\smx{\stackrel{x\to 0}{\longrightarrow}}
\def\Li{{\rm Li}}


\title{PDF constraints from Electroweak Vector Boson production at the LHC}

\author{R. D. Ball$^1$, N. P. Hartland$^1$, J. Rojo$^2$ and M. Ubiali$^3$}
\institute{\it ~$^1$ Tait Institute, University of Edinburgh,
JCMB, KB, Mayfield Rd, Edinburgh EH9 3JZ, Scotland\\
~$^2$ PH Department, TH Unit, CERN, CH-1211 Geneva 23, Switzerland\\
~$^3$ Institut f\"ur Theoretische Teilchenphysik und Kosmologie, RWTH 
Aachen University, D-52056 Aachen, Germany\\}


\begin{abstract}
We present a study of the impact of the recent $W$ and $Z$ measurements from ATLAS, CMS and 
LHCb on parton distribution functions. We show that the NNPDF2.1 NNLO predictions are
consistent with all the new data, but that these provide significant further constraints on 
the light quarks and antiquarks at medium and small-$x$. We conclude that these data 
already have the potential to play a useful role in future global PDF analyses.
\end{abstract}

\subsection{LHC measurements sensitive to PDFs}

The LHC has already provided an impressive set of
measurements which are sensitive to parton
distributions: inclusive jet and dijet data~\cite{CMS:2011ab,Chatrchyan:2011qta,Aad:2011fc},
electroweak vector boson production~\cite{Aad:2011dm,Aad:2011yna,CMS234,
Chatrchyan:2011jz,Chatrchyan:2011wt,LHCb} 
(both inclusive and in association with heavy quarks~\cite{CMSWc}) and direct
photon production~\cite{Aad:2011tw,Chatrchyan:2011ue}. The purpose
of this contribution is to quantify the impact on PDFs of a subset of
these data, the $W$ and $Z$ inclusive production measurements.
In this first section  we will review the status of LHC data relevant
for PDF determination and then in the next section we will
study how the $W$, $Z$ data impact on the NNPDF analysis.

Let's begin this short review of LHC
data with electroweak vector boson production.
ATLAS has measured the $W$ lepton and $Z$ rapidity distributions
using the 2010 data (36 pb$^{-1}$) and determined the full  
covariance matrix of correlated experimental uncertainties~\cite{Aad:2011dm}. 
This measurement supersedes the original
muon asymmetry measurement from $W$ decays~\cite{Aad:2011yna}, 
for which the covariance
matrix was not available.
The CMS collaboration has presented
a preliminary measurement of the muon asymmetry with 
2011 data (234 pb$^{-1}$)~\cite{CMS234}
which supersedes the 2010 data~\cite{Chatrchyan:2011jz}. In addition 
it has presented
a measurement of the normalized $Z$ rapidity distribution using 2010 
data~\cite{Chatrchyan:2011wt}.
In neither of these two measurements has the full  
covariance matrix been made available.
Finally, the LHCb Collaboration has presented preliminary
results for the $Z$ rapidity distribution, $W$ lepton asymmetry and
W lepton charge ratio using 2010 data~\cite{LHCb}.

  \begin{table}[h]                                                     
 \footnotesize
 \centering
 \begin{tabular}{|c||c|c|c||c|c|c|}
\hline
 Data Set & Ref.  & $N_{\rm dat}$ &  
$\lc\eta_{\rm min},\eta_{\rm max}\rc$ &
$\la \sigma_{\rm stat}\ra$ (\%) &
   $\la \sigma_{\rm sys}\ra$  (\%) & $\la \sigma_{\rm norm}\ra$  (\%)
 \\ \hline
\hline
ATLAS W,Z 36 pb$^{-1}$      & \cite{Aad:2011dm} & 30 &$\lc 0,3.2\rc$  &  1.9 & 1.7    &  3.4    \\  
ATLAS $W^+$ 36 pb$^{-1}$      & \cite{Aad:2011dm} & 11 & $\lc 0,2.4\rc$ & 1.4 
 & 1.3    & 3.4      \\  
ATLAS $W^-$ 36 pb$^{-1}$      & \cite{Aad:2011dm} & 11 & $\lc 0,2.4\rc$  & 1.6  & 1.4 & 3.4       \\  
ATLAS $Z$ 36 pb$^{-1}$       & \cite{Aad:2011dm}  & 8 & $\lc 0,3.2\rc$ & 2.8
  &  2.4   & 3.4      \\  
\hline
CMS $Z$ rapidity 36 pb$^{-1}$       & \cite{Chatrchyan:2011wt}  &35  &
$\lc 0,3.6\rc$ & 12.3  &  -   & 0    \\
CMS muon asymmetry 234 pb$^{-1}$       &   \cite{CMS234}  & 11 & $\lc 0,2.4\rc$& 1.7  &  3.1
      & 0   \\  
\hline
LHCb $Z$ rapidity 36 pb$^{-1}$      & \cite{LHCb} & 5  & $\lc 2,4.5\rc$ & 20  &   5  &  3.4   \\  
LHCb  $W$ lepton asymmetry 36 pb$^{-1}$      & \cite{LHCb}  & 5 & $\lc 2,4.5\rc$ & 16  &  21   & 0     \\  
\hline
 \end{tabular}
\caption{\small \label{tab:exp-sets-errors} The number of data points,
kinematical coverage and average
statistical, systematic and normalization percentage uncertainties for
each of the experimental LHC $W$ and $Z$ datasets considered in
the present analysis. For the CMS $Z$ rapidity data, 
the systematic uncertainty is included in the statistical uncertainty: there is no normalization uncertainty because these data are normalised to the total cross-section.
}
 \end{table}

The kinematical coverage of each of the various LHC
$W$ and $Z$ dataset with the corresponding
average experimental uncertainties
for each dataset are summarized in Table~\ref{tab:exp-sets-errors}.
As we can see the LHC electroweak data span a large range in rapidity up
to $\eta=$4.5. Each of the three processes considered,
$W^+$, $W^-$ and $Z$ is sensitive to different partonic
subprocesses.

There are other LHC datasets potentially sensitive to PDFs.
Jet production from the Tevatron has been a very important measurement not
only to constrain the gluon at high $x$, but in determining the strong coupling
from a global PDF analysis~\cite{Ball:2011us,Lionetti:2011pw}.
Similar constraints are expected 
from the LHC jet data, extended into a wider kinematical range. 
From the 2010 (36 pb$^{-1}$) dataset inclusive jet and dijet production has been
measured by both CMS~\cite{CMS:2011ab,Chatrchyan:2011qta} and
ATLAS~\cite{Aad:2011fc}, however only for ATLAS is the
full experimental covariance matrix available.
The LHC inclusive jet data can be treated within a global
analysis framework using tools like FastNLO or APPLgrid~\cite{Carli:2010rw}. 
Since the full NNLO corrections to the inclusive jet production
are unknown, jet data in a NNLO analysis can be included
only within some approximation: for example with NNLO
PDF evolution and coupling running but with NLO matrix elements, 
or else with NLO matrix elements supplemented with Sudakov 
estimates of the NNLO corrections.
Another LHC  measurement that has the potential
to constrain the gluon PDFs is prompt photon production from 
ATLAS~\cite{Aad:2011tw} and CMS~\cite{Chatrchyan:2011ue}: its consistency
with NLO QCD and their impact on the NNPDF2.1 PDFs will
be discussed in detail in Ref.~\cite{d'Enterria:2012yj}.

\subsection{PDF constraints from LHC $W$ and $Z$}

Until recently all available NNPDF
sets~\cite{DelDebbio:2004qj,DelDebbio:2007ee,
Ball:2008by,Ball:2009mk,Ball:2010de,Ball:2011mu,Ball:2011uy}
were based on non-LHC data. NNPDF2.2~\cite{Ball:2011gg} was the first set
to include LHC data, the $W$ lepton asymmetry from ATLAS
and CMS~\cite{Chatrchyan:2011jz,Aad:2011yna}. 
However now these two datasets are outdated,
the first because now the full correlation matrix
of the $W$ and $Z$ lepton distributions is available, and
the second because data from higher luminosities is also
available. So we have chosen to continue to use as our baseline
the NNPDF2.1 NNLO set. 

We now study the impact of the latest LHC $W$ and $Z$ data
on the NNPDF parton distributions. All our 
theoretical NNLO predictions will be computed with DYNNLO~\cite{Catani:2009sm}
with the same cuts and settings as in the respective measurements. 
The impact of the new data will be quantified using the 
reweighting method of Refs.~\cite{Ball:2010gb,Ball:2011gg}
applied to the $N_{\rm rep}=1000$ replicas of the NNPDF2.1 NNLO set.

To begin with, we have computed the $\chi^2$ for each of the
datasets in Table~\ref{tab:exp-sets-errors} 
for the most recent NNLO PDF sets currently available on LHAPDF:
NNPDF2.1, MSTW08~\cite{Martin:2009iq}, 
ABKM09~\cite{Alekhin:2009ni}, HERAPDF1.5~\cite{herapdf15} and JR09~\cite{JimenezDelgado:2009tv}.
When available, we use the full experimental covariance matrix.
Normalization uncertainties are included using the
$t_0$ method~\cite{Ball:2009qv}. This is important specially
for the treatment of the ATLAS differential
distributions where normalization
uncertainties are comparable to the statistical
and systematic uncertainties (See Table~\ref{tab:exp-sets-errors}). 

The results are summarized in Table~\ref{tab:summary-chi2}.
For the ATLAS $W$ and $Z$ lepton distributions we show the
results both for the total dataset and the individual
subsets, where in the latter case cross-correlations
between subsets have been neglected. In all cases
the theoretical NNLO predictions have been obtained
with DYNNLO as discussed above. We can see that none of these PDF sets describes
the ATLAS and CMS data perfectly, although NNPDF2.1 and HERAPDF1.5 give probably the best
description, while ABKM09 and JR09 are significantly worse. All five sets 
give a reasonable description of 
the LHCb data within their large uncertainties.

\begin{table}[h]
\footnotesize
\begin{center}
\begin{tabular}{|c|c|c|c|c|c|}
\hline
Dataset    &  $\chi^2$ NNPDF2.1  &  $\chi^2$ MSTW08 & $\chi^2$ ABKM09  
& $\chi^2$ JR09 & $\chi^2$ HERAPDF1.5      \\  
\hline
\hline
ATLAS     & 2.7    & 3.6 & 3.6 & 5.0 & 2.0\\
\hline 
ATLAS $W^+$ 36 pb$^{-1}$     & 5.7 & 6.5 & 11.4& 5.4& 5.3\\ 
ATLAS $W^-$ 36 pb$^{-1}$      & 2.5  & 4.1& 5.4& 8.0& 6.4\\ 
ATLAS $Z$ 36 pb$^{-1}$      &  1.8 & 3.7 & 4.2 & 6.5 & 2.9\\ 
\hline
\hline
CMS  & 2.0   & 3.0 & 2.8 & 3.6& 2.8\\  
\hline
CMS $Z$ rapidity 36 pb$^{-1}$  &  1.9  & 2.9& 2.7& 2.0& 3.0 \\  
CMS muon asymmetry 234 pb$^{-1}$      & 2.0  & 3.4& 3.0& 8.7& 2.1\\ 
\hline
\hline
LHCb     & 0.8   & 0.7 & 1.2 & 0.4& 0.6\\
\hline
LHCb $Z$ rapidity 36 pb$^{-1}$      &  1.1   & 0.7 & 0.8 & 0.6& 0.8\\ 
LHCb  $W$ lepton asymmetry 36 pb$^{-1}$ & 0.5 & 0.6 & 1.6 & 0.2& 0.5\\ 
\hline
\end{tabular}
\end{center}
\caption{\small Comparison between LHC $W$ and $Z$ data
and the most recent NNLO PDFs. For each PDF set
we provide the $\chi^2/dof$ between data and theory predictions, computed using the $t_0$-method.
  \label{tab:summary-chi2}}
\end{table}

For the ATLAS data, we would like to emphasize the importance of
properly taking into account the correlations between datasets, specially
the normalization: the description of the individual $W^+$, $W^-$ and
$Z$ datasets is always worse than the overall description because of these
cross-correlations. For the CMS $Z$ rapidity distribution we find
that the fixed order NNLO description seems rather worse than the NLO+LL prediction 
implemented in POWHEG~\cite{Chatrchyan:2011wt}: the origin of this difference should 
be investigated in future studies.

We now discuss the impact of these
LHC EW data into the NNPDF2.1 NNLO PDFs~\cite{Ball:2011uy}.
In Table~\ref{tab:summary} we summarize  the 
initial $\chi^2$ for each dataset,  the $\chi^2$ after
reweighting,  $\chi^2_{\rm rw} $. We find excellent
agreement with all the LHC electroweak measurements after reweighting.
Some comparisons between data and theory for a selected observables are shown
in Fig.~\ref{fig:datatheo}. From top to bottom we show
the comparison with ATLAS, CMS and LHCb data. In each case
we have included all the most updated electroweak datasets from
each collaboration.

In Table~\ref{tab:summary} we also show the effective number of replicas left 
after the reweighting, defined as in Ref.~\cite{Ball:2010gb} using the Shannon entropy,
\begin{equation}
N_{\rm eff}\equiv \exp \{\smallfrac{1}{N_{\rm rep}}\sum_{k=1}^{N_{\rm rep}} 
 w_k \ln (N_{\rm rep}/w_k)\} \ .
\label{eq:effective}
\end{equation}
In each case we have performed the reweighting separately for each of the experimental 
datasets individually, for the combined datasets from each experiment, and finally 
with all three combined together.

\begin{table}[h]
\small
\begin{center}
\begin{tabular}{|c|c|c|c|c|}
\hline
Dataset  &  $\chi^2$  &  $\chi^2_{\rm rw} $ & $N_{\rm eff}$  \\
\hline
\hline
ATLAS   &  2.7  & 1.2   &  16   \\ 
\hline 
ATLAS $W^+$ 36 pb$^{-1}$     & 5.7    & 1.5  & 17   \\  
ATLAS $W^-$ 36 pb$^{-1}$      &  2.5  & 1.0  &  205  \\  
ATLAS $Z$ 36 pb$^{-1}$      &  1.8  & 1.1  & 581  \\  
\hline
\hline
CMS  & 2.0       & 1.2   &  56   \\  
\hline
CMS $Z$ rapidity 36 pb$^{-1}$  &    1.9    & 1.4  &  223  \\  
CMS muon asymmetry 234 pb$^{-1}$      &  2.0   & 0.4   & 200   \\  
\hline
\hline
LHCb   &   0.8 & 0.8   & 972    \\ 
\hline
LHCb $Z$ rapidity 36 pb$^{-1}$      &  1.1  & 1.0  & 962   \\  
LHCb  $W$ lepton asymmetry 36 pb$^{-1}$      & 0.8   & 0.5  & 961   \\  
\hline
\hline
All data combined &   2.1 & 1.2   & 4    \\ 
\hline
\end{tabular}
\end{center}
\caption{\small The impact of LHC electroweak measurements
on the NNPDF2.1 NNLO PDFs. For each dataset
we show the initial $\chi^2$, the $\chi^2$ after
reweighting these particular dataset and
the effective number of replicas $N_{\rm eff}$  in this case.
We show both the results for individual datasets
as well as for the combined impact of all datasets
within the same experiment. All the results have been computed
starting with $N_{\rm rep}=1000$ replicas.
  \label{tab:summary}}
\end{table}

\begin{figure}[h]
\begin{center}
\includegraphics[width=0.40\textwidth]{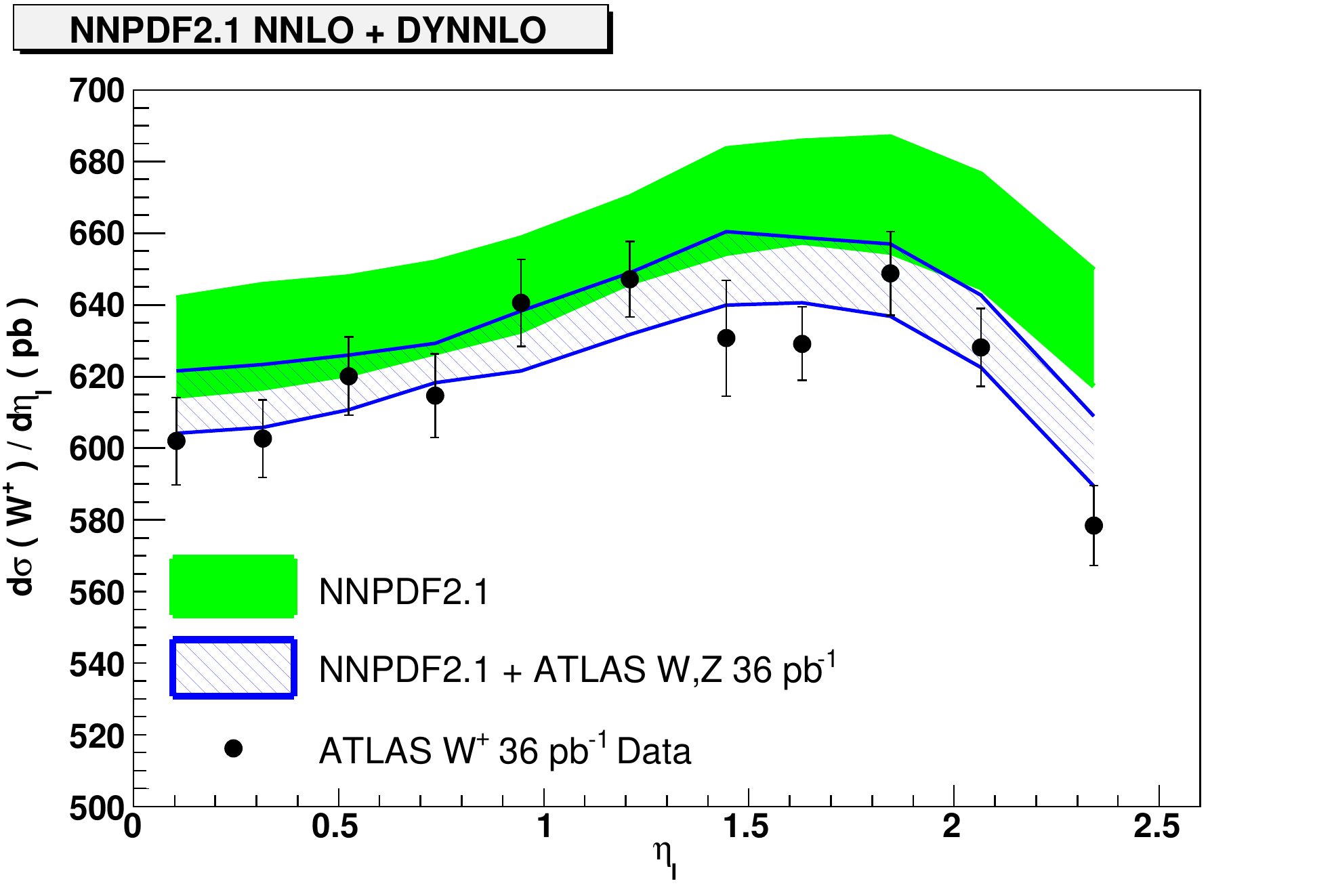}
\includegraphics[width=0.40\textwidth]{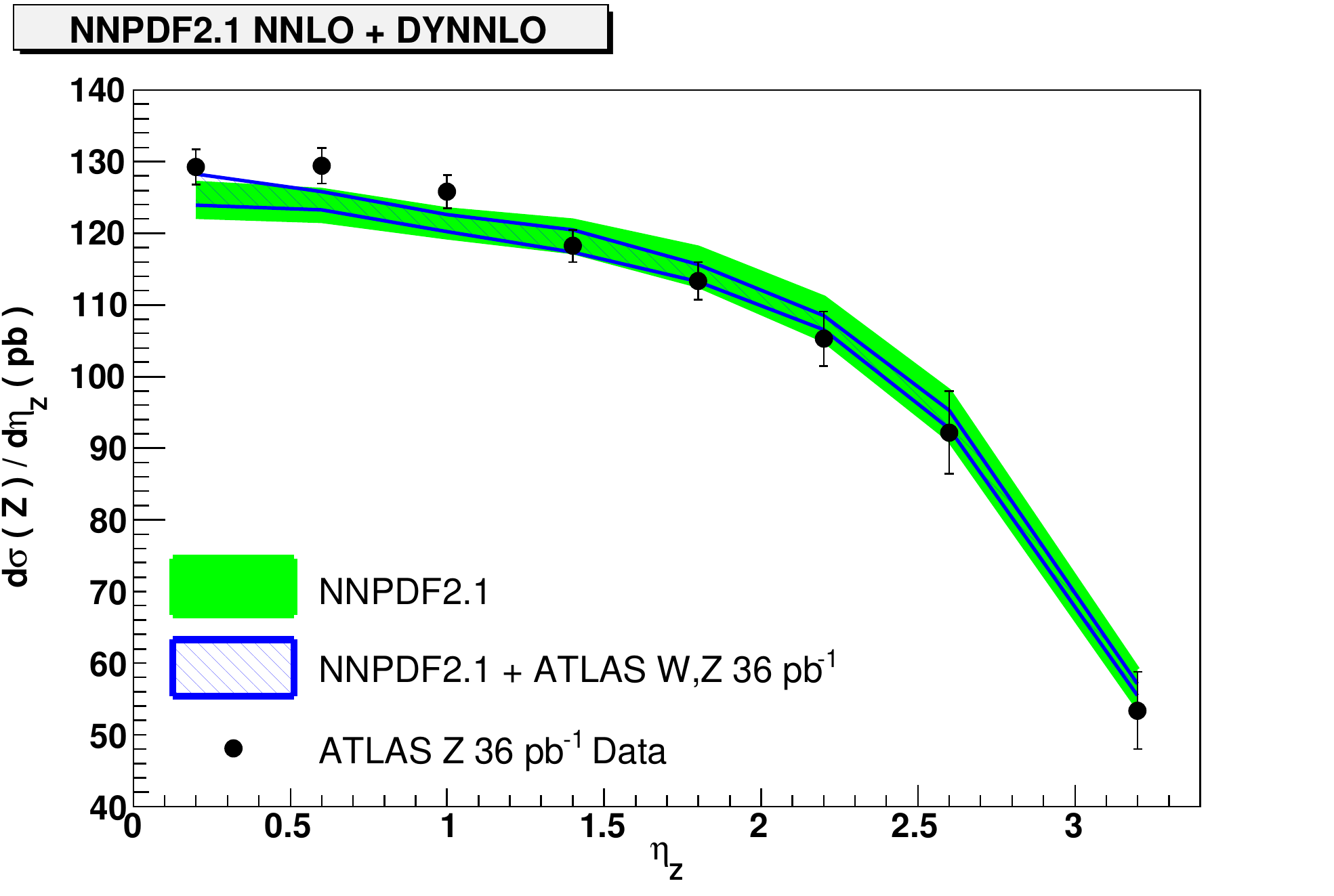}\\
\includegraphics[width=0.40\textwidth]{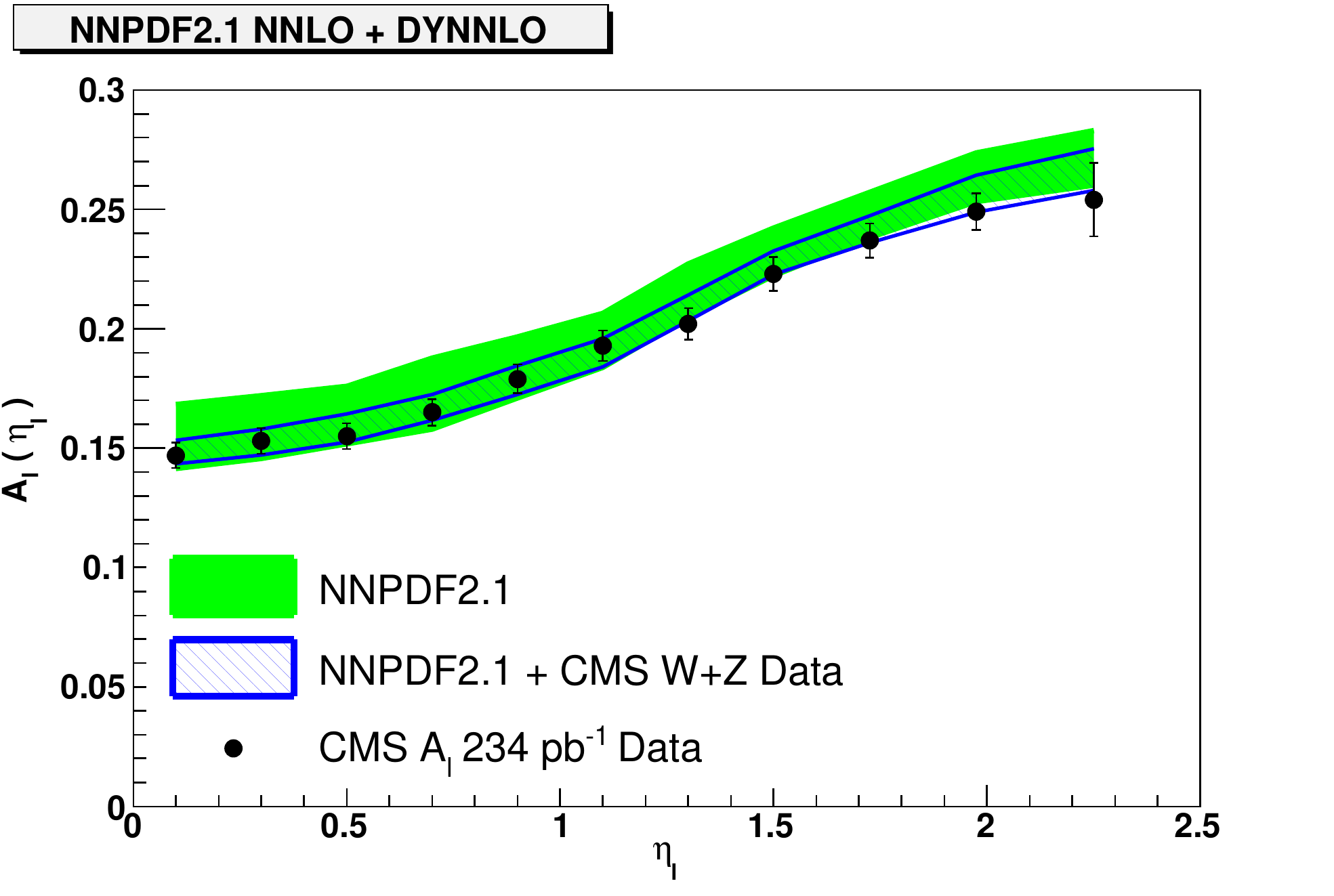}
\includegraphics[width=0.40\textwidth]{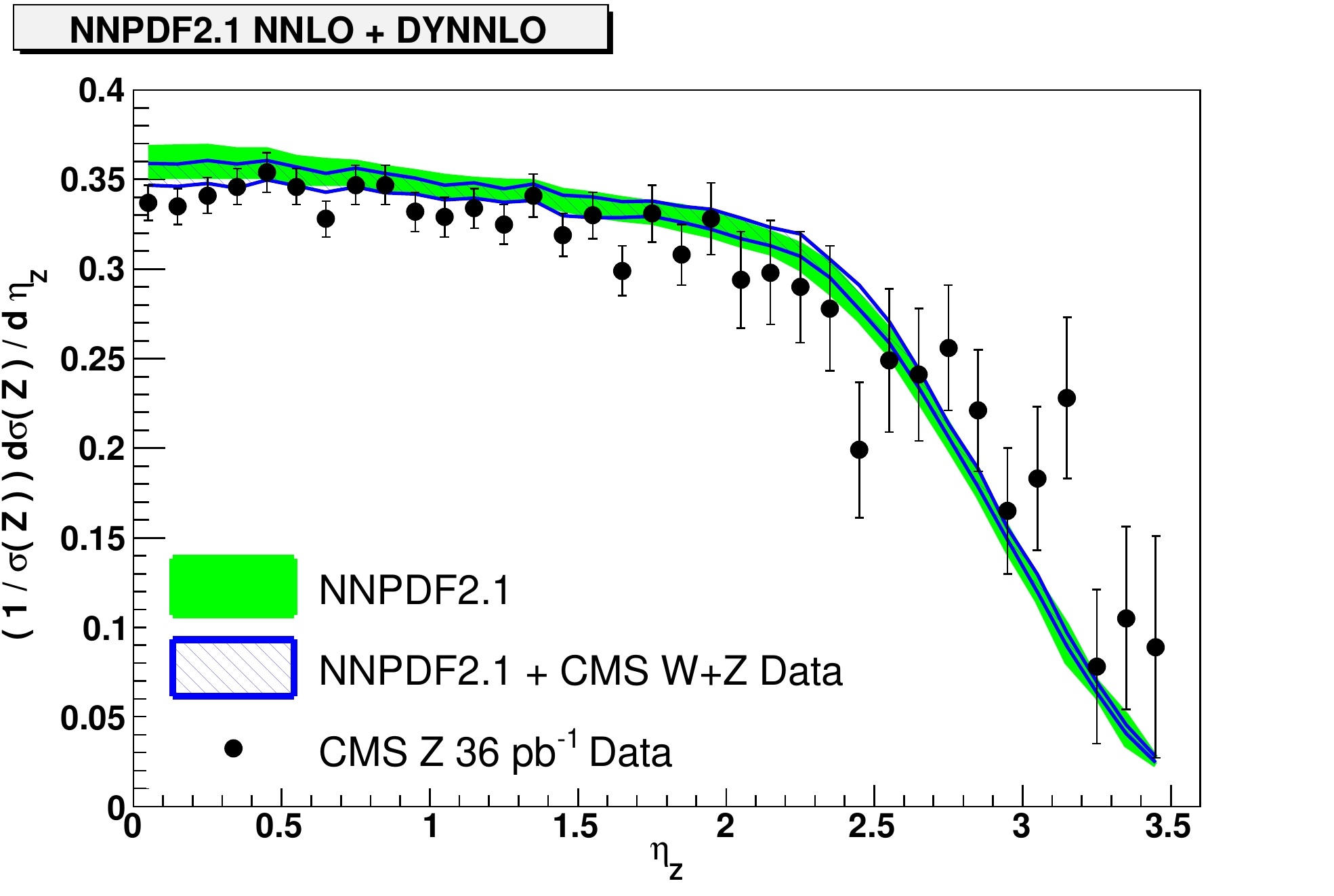}\\
\includegraphics[width=0.40\textwidth]{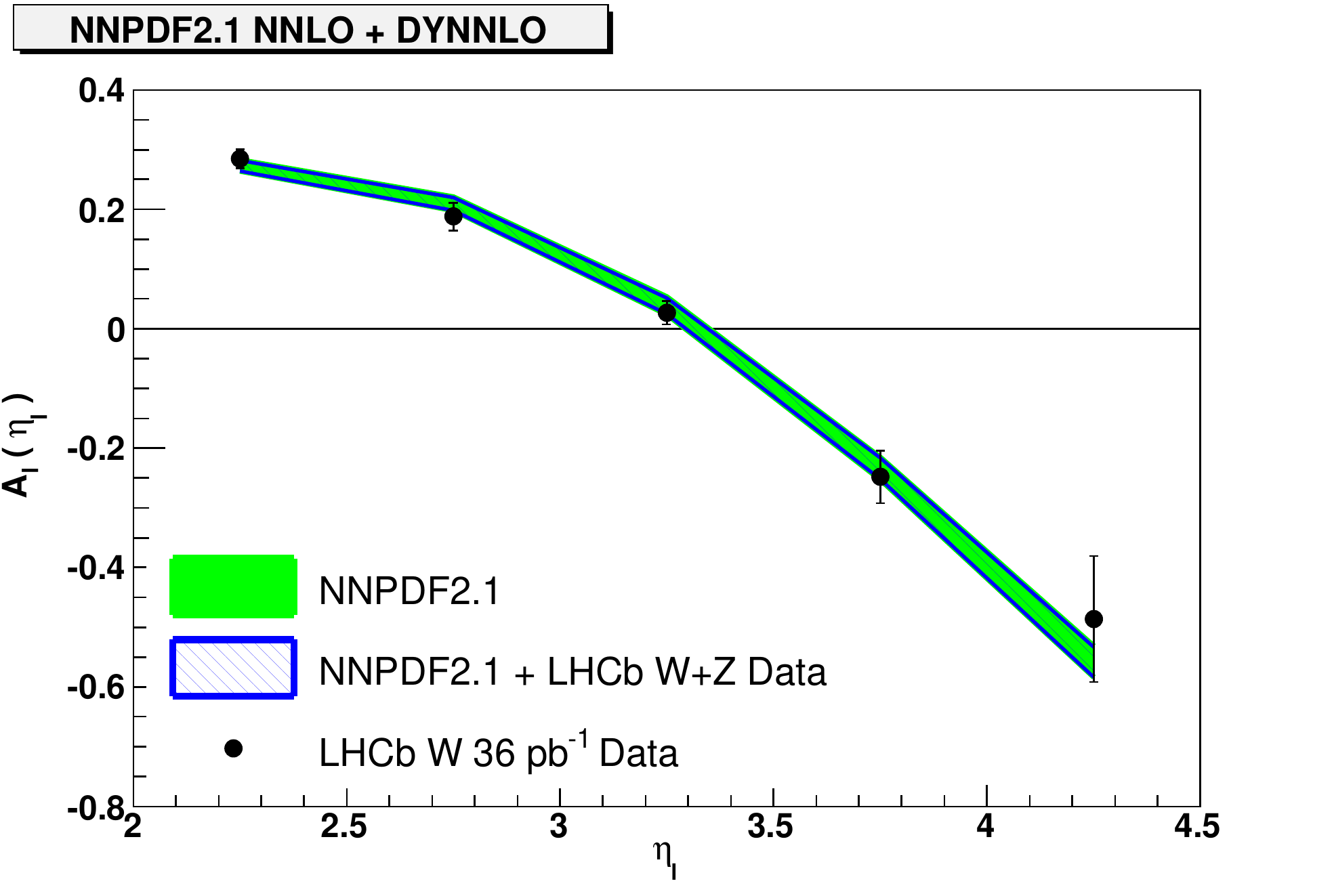}
\includegraphics[width=0.40\textwidth]{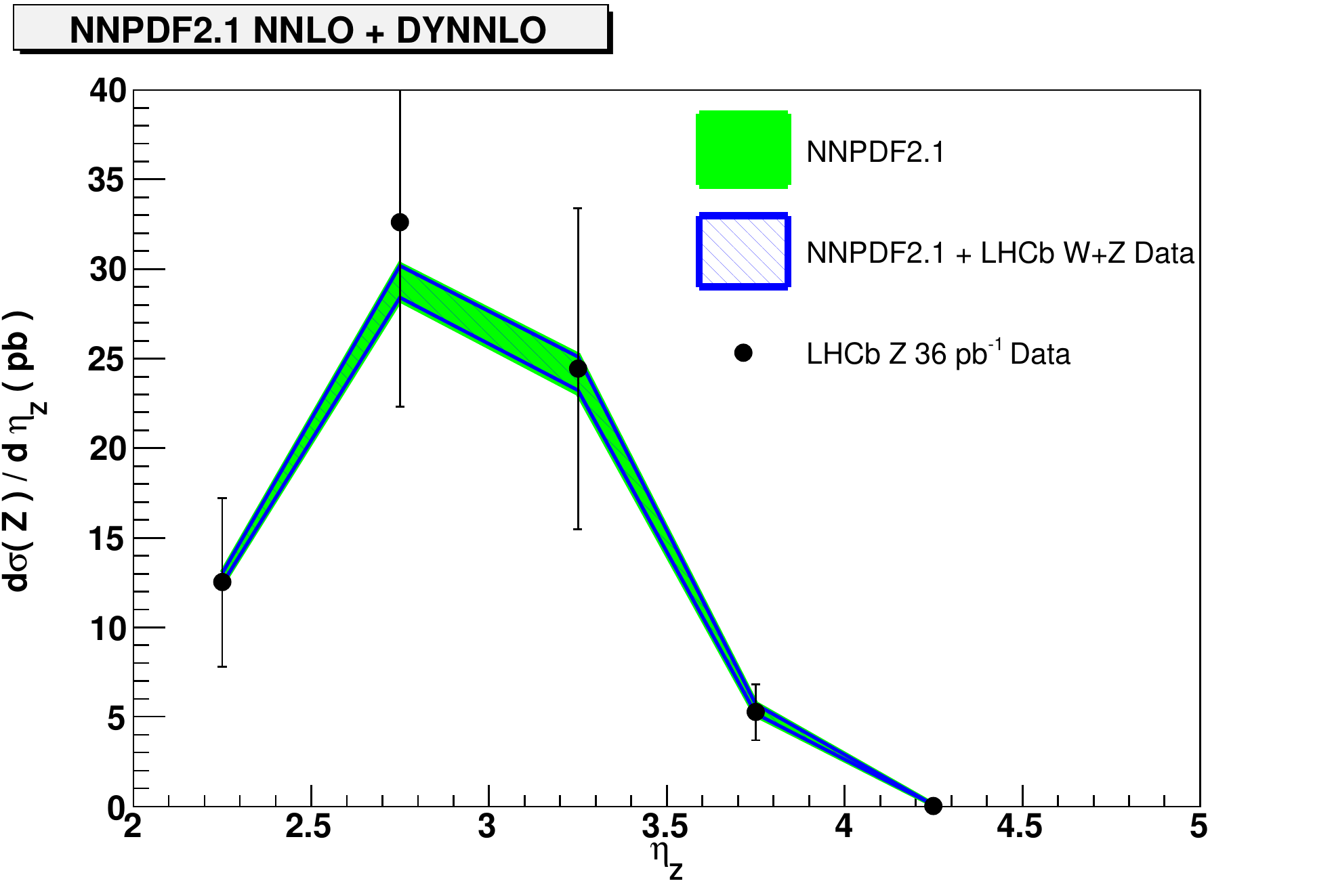}
 \caption{\small Comparison between data and theory before
and after reweighting for NNPDF2.1 NNLO compared to the various LHC
EW datasets considered. From top to bottom we show comparisons
with ATLAS ($W^+$ lepton $Z$ rapidity distributions), CMS
($W$ lepton asymmetry and $Z$ rapidity distribution) and LHCb data
(same as CMS). For the ATLAS data the error bars include statistical
and systematic uncertaintes, but not the normalization
uncertainties.}
\label{fig:datatheo}
\end{center}
\end{figure}

When all the datasets are taken together, the initial $\chi^2=2.1$,
already quite reasonable
is reduced down to $\chi^2_{\rm rw}=1.2$, thus obtaining a very
good overall description of all the most recent LHC electroweak data. 
The effective number of replicas
for all combined datasets is only $N_{\rm eff}=4$ however: from this we conclude 
that to determine the combined impact of these data on PDFs would require many 
more replicas (around 25,000 in fact, to obtain reasonable statistical accuracy), 
or, more practically, a new fit. Note that the fact that the
total effective number of replicas for the whole dataset is rather smaller than
that of any individual subset confirms their mutual compatibility and the
lack of any appreciable tension. Comparing the effective number of replicas 
for the individual datasets, the most constraining data are the ATLAS $W$ and $Z$
distributions, specially the very precise $W^+$ data. On the other hand the
LHCb data have a rather small impact.

\begin{figure}[h]
\begin{center}
\includegraphics[width=0.40\textwidth]{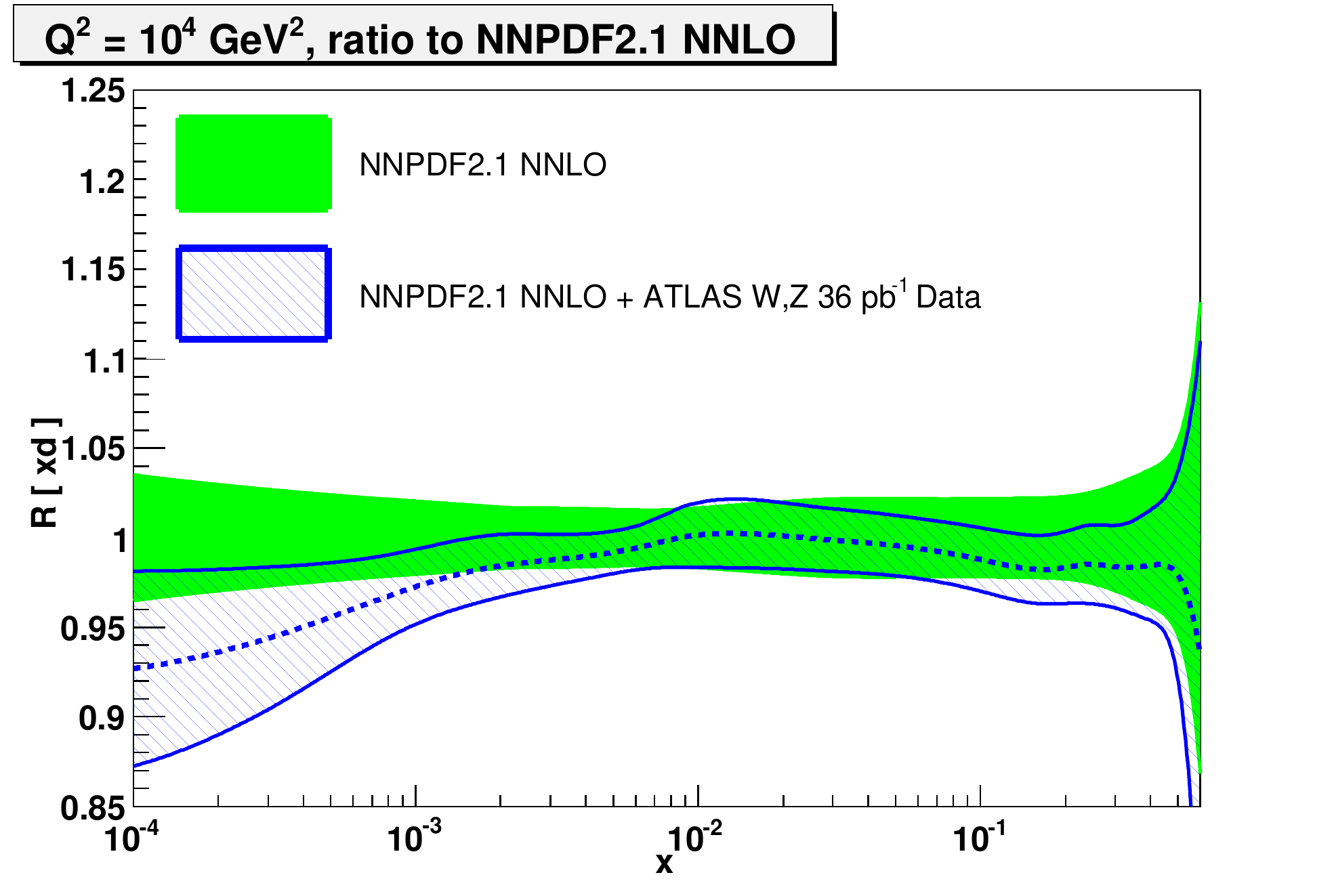}
\includegraphics[width=0.40\textwidth]{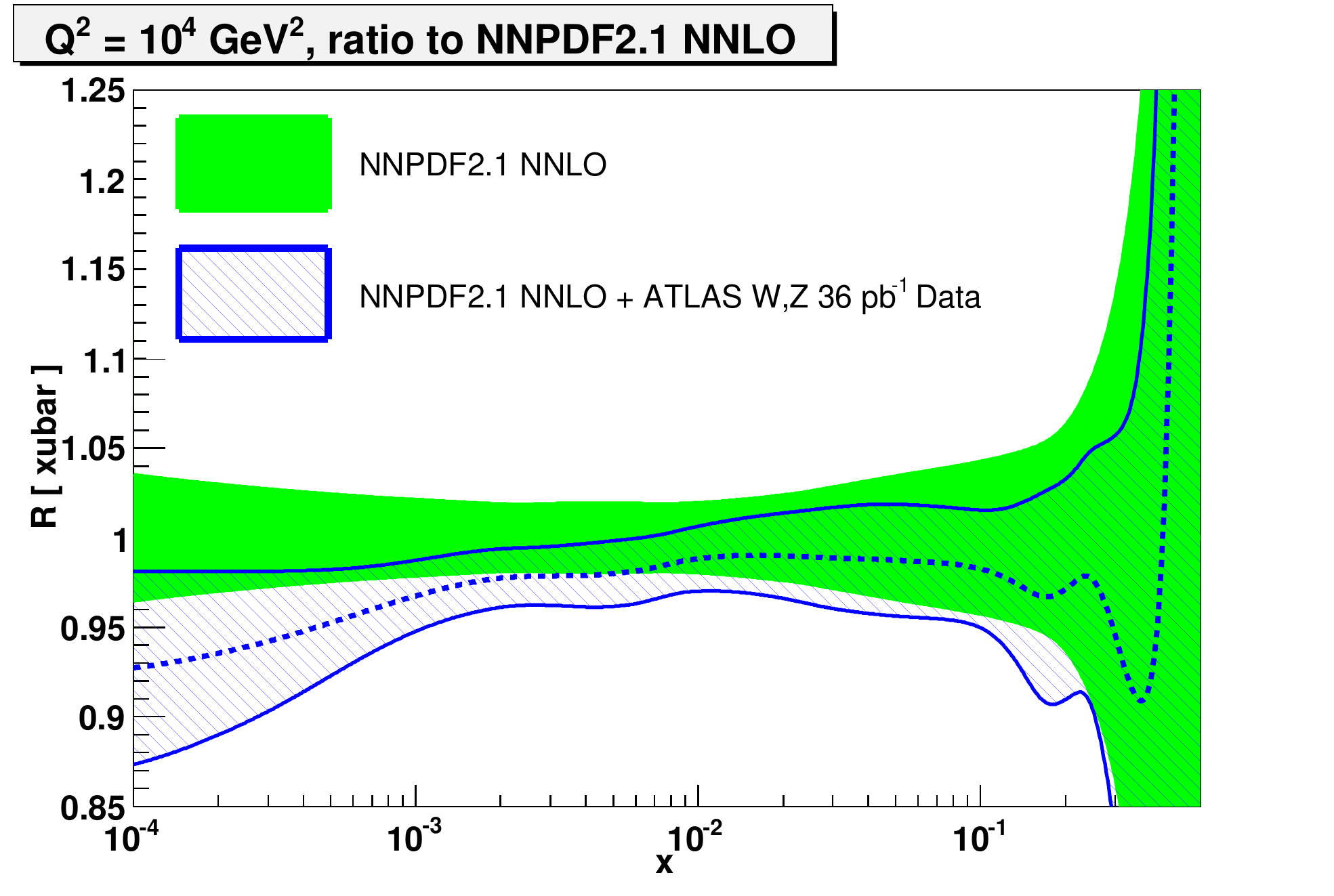}
\includegraphics[width=0.40\textwidth]{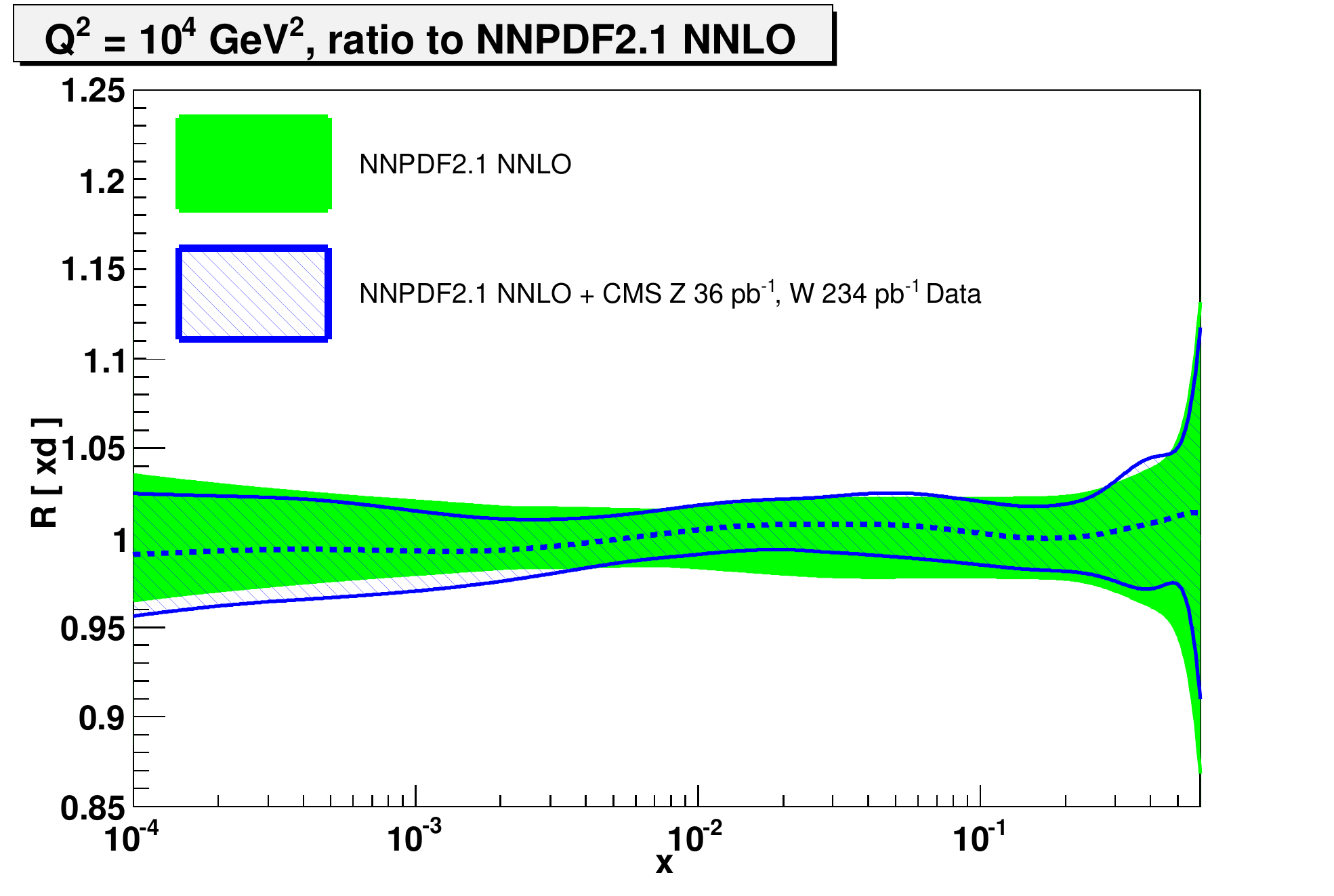}
\includegraphics[width=0.40\textwidth]{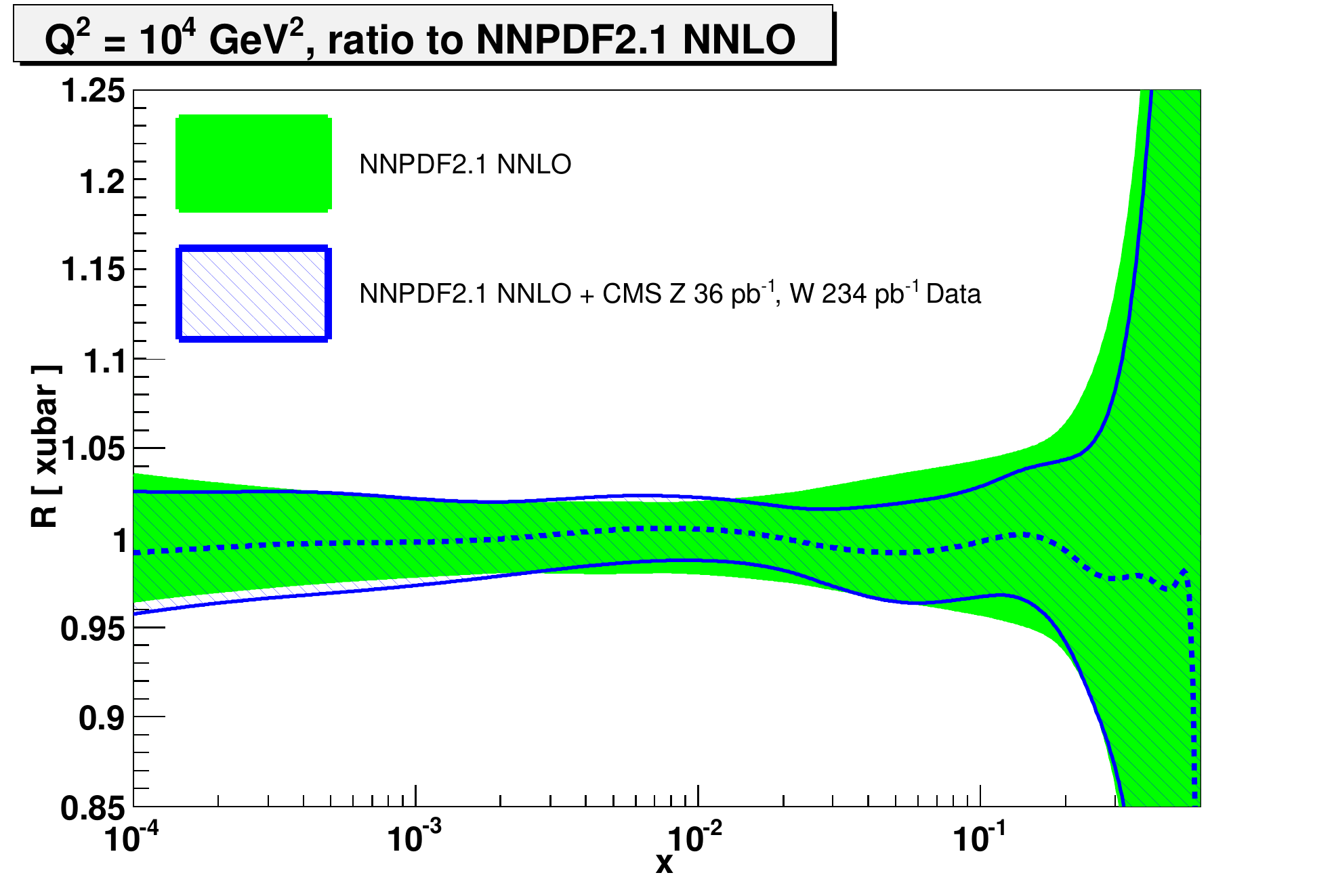}
\includegraphics[width=0.40\textwidth]{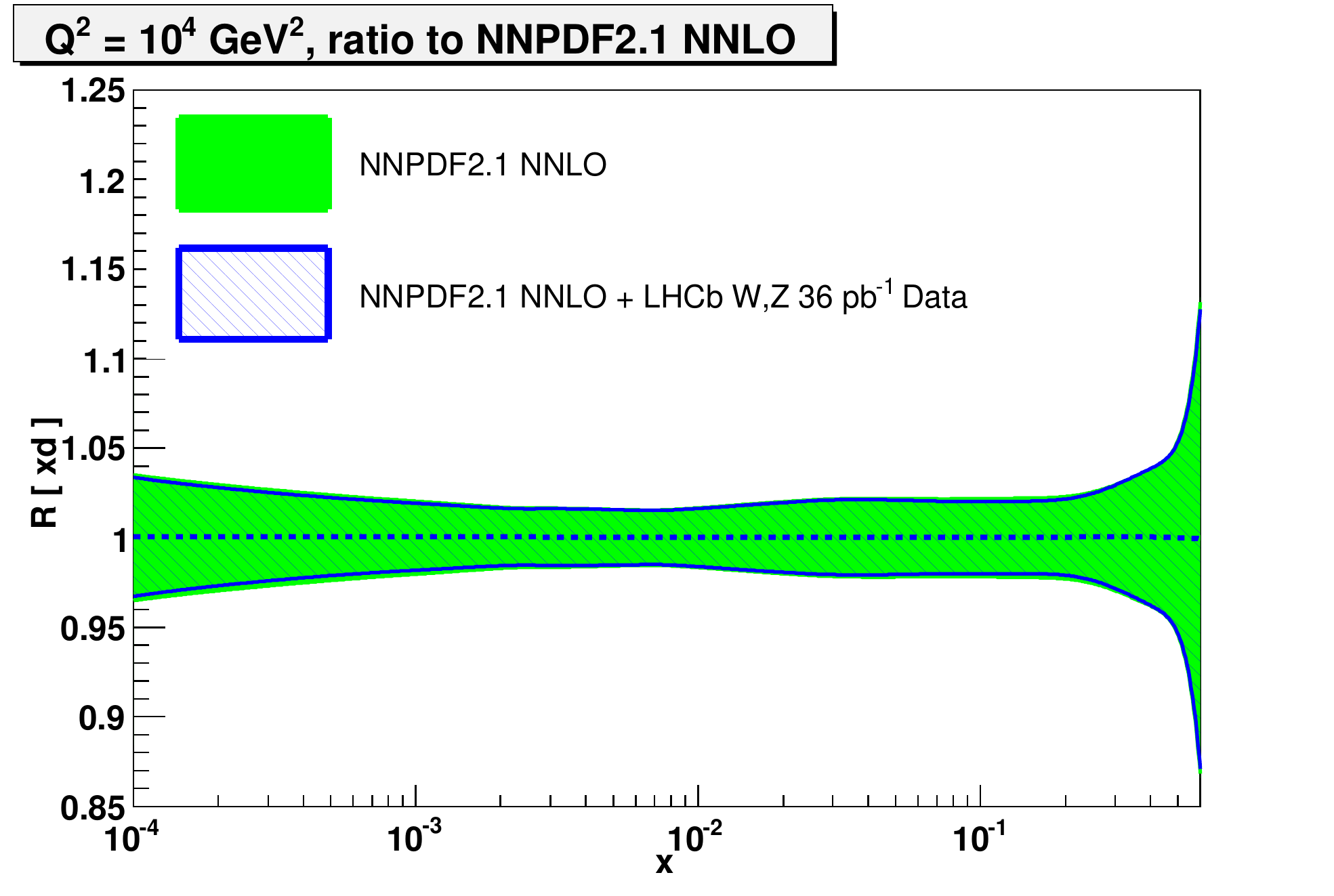}
\includegraphics[width=0.40\textwidth]{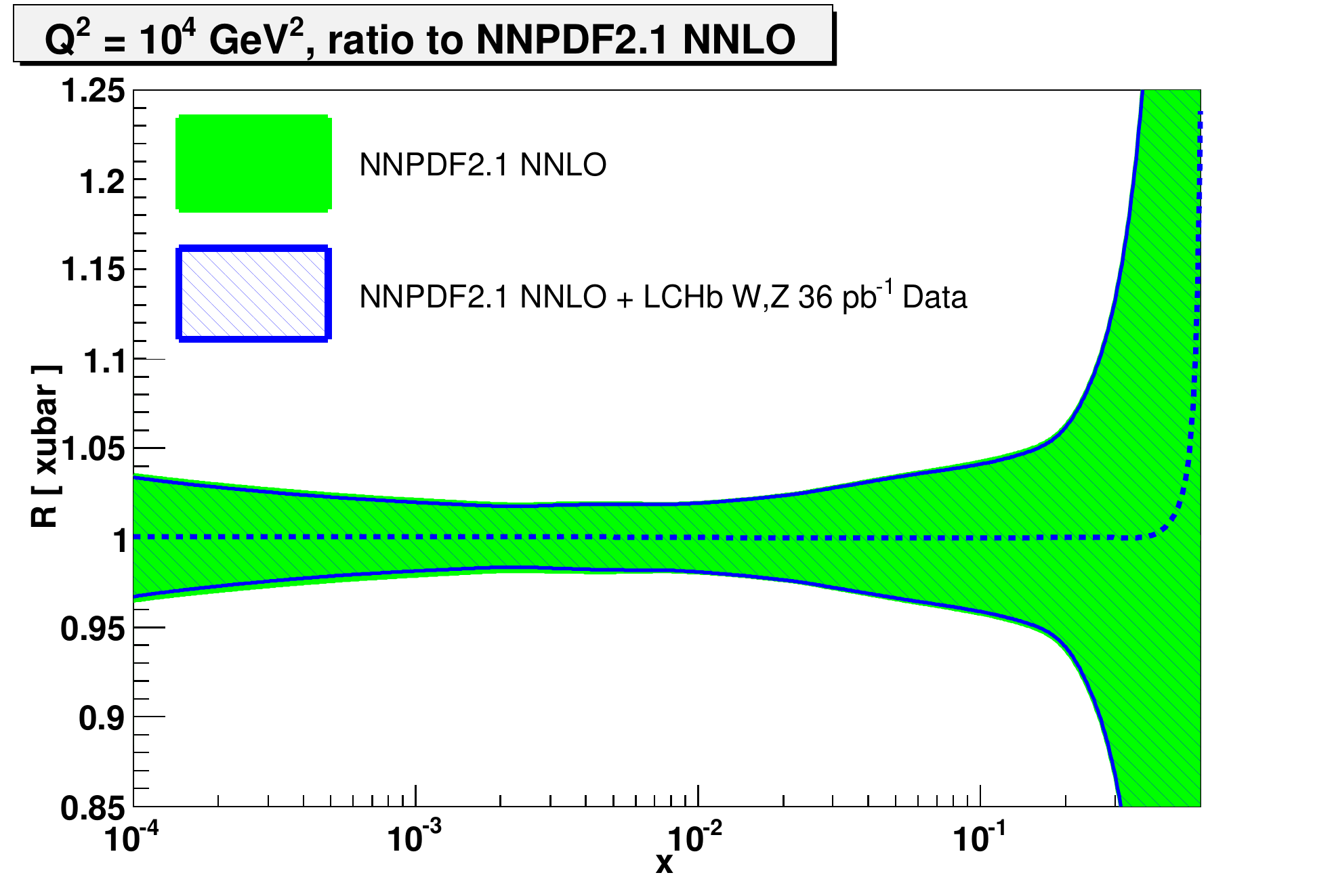}
 \caption{\small Comparison between the NNPDF2.1 NNLO
parton distributions before
and after
reweighting with the various LHC
EW datasets considered. From top to bottom: ATLAS, CMS and LHCb data,
where the left column we show the ratio of the $d$ quark
PDF and in the right column the ratio of the $\bar{u}$ quark
PDF to their central values before reweighting. }
\label{pdfcomp}
\end{center}
\end{figure}

Let us now examine how various 
PDFs change when new experiments are added.
In particular we show in Fig.~\ref{pdfcomp} the NNPDF2.1 NNLO
$d(x,Q^2)$ and $\bar{u}(x,Q^2)$ PDFs at $Q^2=M_W^2$ as ratios
to the central value before including the new data. As described 
above, we put together all 
the data from a given experiment.
As can be seen, the ATLAS data give a moderate reduction in 
PDF uncertainties, and a somewhat softer small-$x$
sea quarks, although the old and new PDFs agree at the
1--sigma level. For CMS the central values for the old and new PDFs
are unchanged with a moderate error reduction at
medium-$x$. Finally, for LHCb the PDF uncertainties
are almost unaffected, due to the low constraining power
of these datasets.

\subsection{Conclusions}
In this contribution we have quantified the impact
of the most updated LHC electroweak data on the NNPDF2.1 NNLO 
parton distributions. NNPDF2.1 provides a reasonable description 
of all these datasets even before their impact on the PDFs is included. 
We find that all the datasets are mutually consistent, with no obvious 
tensions. The PDF uncertainties for the light
quarks and antiquarks at medium and small-$x$ are moderately reduced.
The ATLAS $W,Z$ data seem to prefer a softer small-$x$ sea.
It is clear from our results that the LHC $W$ and $Z$ data should play an 
important part in any future PDF global fit.

\subsection*{Acknowledgments}
The research of J.~R. has been supported by a Marie Curie Intra--European Fellowship
of the European Community 7th Framework Programme under contract
number PIEF-GA-2010-272515. M.U. is supported by the Bundesministerium f\"ur Bildung and Forschung
(BmBF) of the Federal Republic of Germany (project code 05H09PAE).

\clearpage


}

\section[Heavy Quark Production in the ACOT Scheme at NNLO and N$^{3}$LO]
{HEAVY QUARK PRODUCTION IN THE ACOT SCHEME AT NNLO AND N$^{3}$LO \protect\footnote{Contributed by: T.~Stavreva, 
I.~Schienbein,
F.~I.~Olness, 
T.~Je\v{z}o, 
K.~Kova\v{r}\'{\i}k,
A.~Kusina, 
J.~Y.~Yu}}
{\graphicspath{{olness/}}

\def\msbar{\overline{MS}}
\def\lsim{\mathrel{\rlap{\lower4pt\hbox{\hskip1pt$\sim$}} \raise1pt\hbox{$<$}}}
\def\gsim{\mathrel{\rlap{\lower4pt\hbox{\hskip1pt$\sim$}} \raise1pt\hbox{$>$}}}



\title{Heavy Quark Production in the ACOT Scheme at NNLO and N$^{3}$LO}

\author{%
T.~Stavreva$^1$, 
I.~Schienbein$^1$,
F.~I.~Olness$^2$, 
T.~Je\v{z}o$^1$, 
K.~Kova\v{r}\'{\i}k$^3$,
A.~Kusina$^2$, 
J.~Y.~Yu$^{1,2}$}
\institute{$^1$Laboratoire de Physique Subatomique et de Cosmologie,
Universit\'e Joseph Fourier/CNRS-IN2P3/INPG,
53 Avenue des Martyrs, 38026 Grenoble, France
\\$^2$Southern Methodist University, Dallas, TX 75275, USA
\\$^3$Institute for Theoretical Physics,
Karlsruhe Institute of Technology, Karlsruhe, D-76128, Germany}


\begin{abstract}
We extend the ACOT scheme for heavy quark production to NNLO and
N$^{3}$LO for the structure functions $F_{2}$ and $F_{L}$ in 
deep-inelastic scattering (DIS).
We use the fully massive ACOT scheme up to NLO, and estimate the
dominant heavy quark mass effects at the higher orders using the
massless Wilson coefficients together with a generalized
slow-rescaling prescription.
We present results for $F_2$ and $F_L$ showing the effect of the
higher orders and the contributions from the heavy flavors.

\end{abstract}

\subsection{INTRODUCTION}


\begin{figure}[t]
\begin{center}
\includegraphics[clip,width=0.50\textwidth]{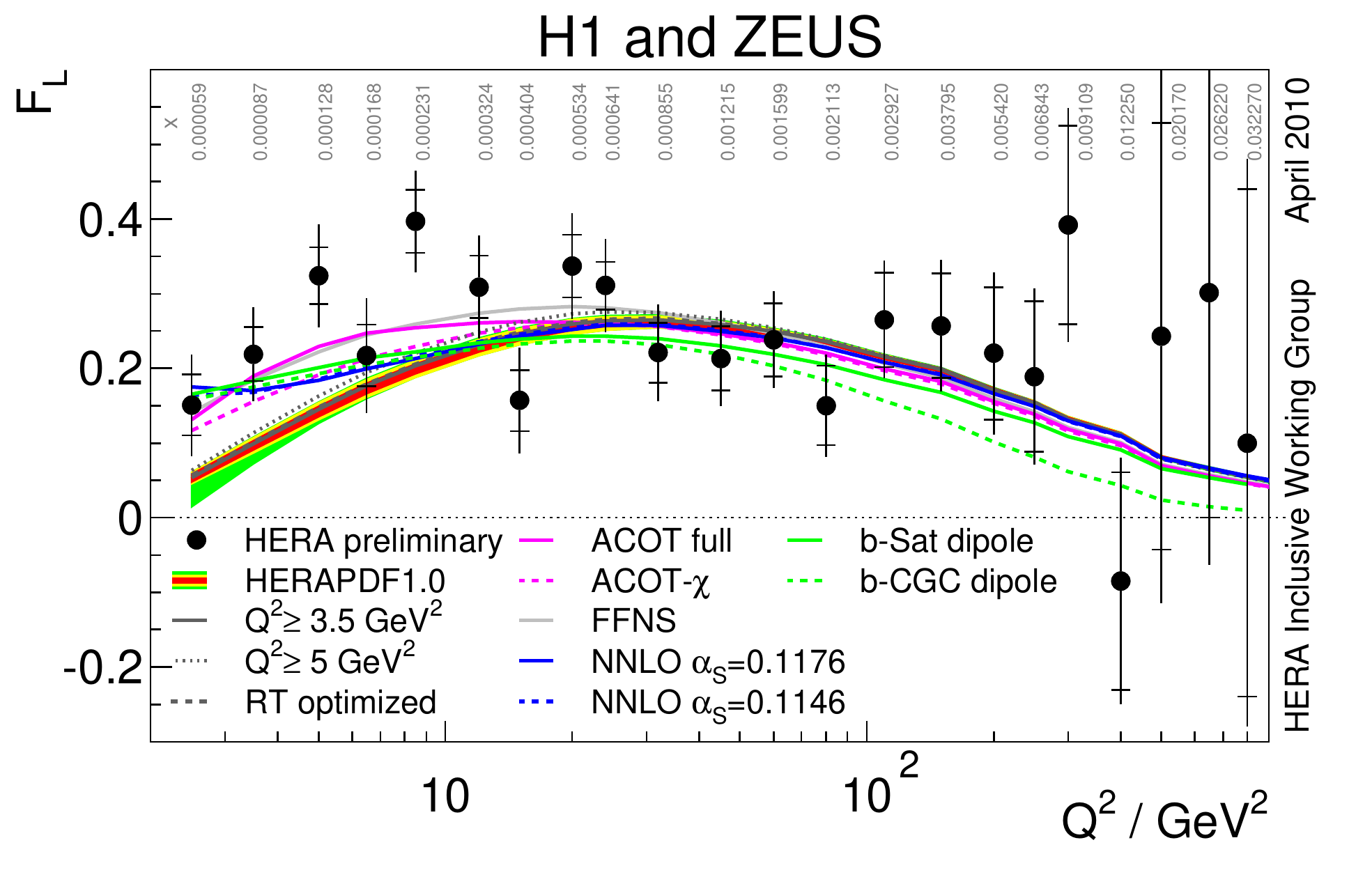}
\caption{$F_{L}$ vs. $Q^2$ for the HERA combined measurements from H1 and ZEUS~\cite{hera:FL}. 
}
\label{olness_label1}
\end{center}
\end{figure}

The production of heavy quarks in high energy processes has become
an increasingly important subject of study both theoretically and
experimentally. The theory of heavy quark production in perturbative
Quantum Chromodynamics (pQCD) is more challenging than that of light
parton (jet) production because of the additional heavy quark mass scale.
The correct theory
must properly take into account the changing role of the heavy quark
over the full kinematic range of the relevant process from the threshold
region (where the quark behaves like a typical {}``heavy particle'')
to the asymptotic region (where the same quark behaves effectively
like a parton, similar to the well known light quarks $\{u,d,s\}$).

With the ever-increasing precision of experimental data and
the progression of theoretical calculations and parton distribution function (PDF) 
evolution to next-to-next-to-leading order (NNLO) of QCD, there is a clear
need to implement the heavy quark schemes at
this order and beyond.
The most important case is arguably the heavy quark treatment
in inclusive deep-inelastic scattering (DIS) since the very precise
HERA data for DIS structure functions and cross sections form the backbone
of any modern global analysis of PDFs. Here, the heavy quark structure functions
contribute up to 30\% or 40\% to the inclusive structure functions at small momentum
fractions $x$.
Extending the heavy quark schemes to higher orders is 
relevant for extracting precision PDFs, and hence for accurate predictions
of observables at the LHC.

An example where higher order corrections are particularly important 
is the longitudinal structure function $F_L$ in DIS.
The leading order ${\cal O}(\alpha_s^0)$ contributions to this structure function
vanishes for massless quarks due to helicity conservation (Callan-Gross relation).
Since the first
unsuppressed contribution to $F_L$ is at next-to-leading order, the NNLO
and N$^3$LO corrections are more important than for $F_2$.
In Fig.~\ref{olness_label1} we show the preliminary results for the $F_L$ measurement 
from the H1 and ZEUS experiments~\cite{hera:FL}.
In Fig.~\ref{olness_label2} displays sample Feynman diagrams at the various orders. 
Producing an accurate prediction for $F_L$ is a challenge, particularly in the 
region of low $Q^2$ and small $x$.

In this paper, we will briefly outline the method we used to incorporate
the higher order terms,  the key elements of the ACOT scheme, 
and the treatment of the heavy quark masses.
We then present results for  the 
$F_2$ and $F_L$ neutral current DIS structure functions.

\subsection{THE ACOT SCHEME AND ITS EXTENSION BEYOND NLO}


\begin{figure*}[t]
\begin{center}
\includegraphics[clip,width=0.17\textwidth]{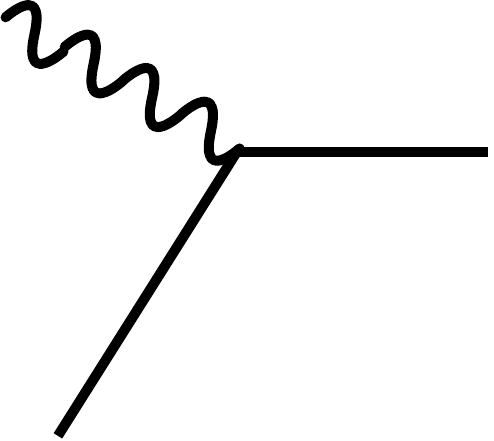}\qquad{}
\includegraphics[clip,width=0.14\textwidth]{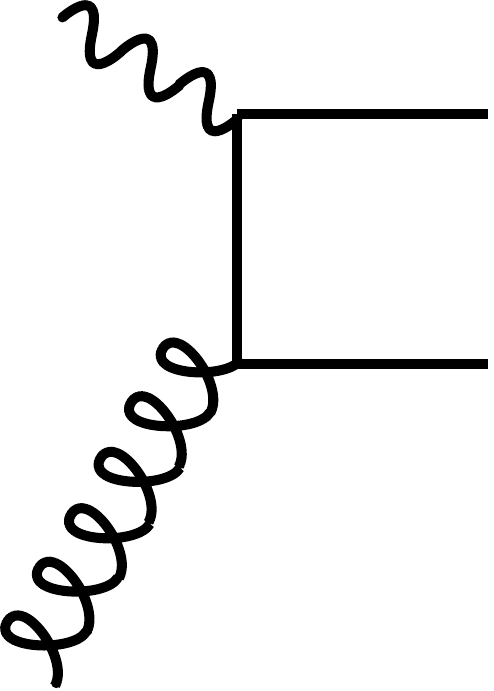}\qquad{}
\includegraphics[clip,width=0.14\textwidth]{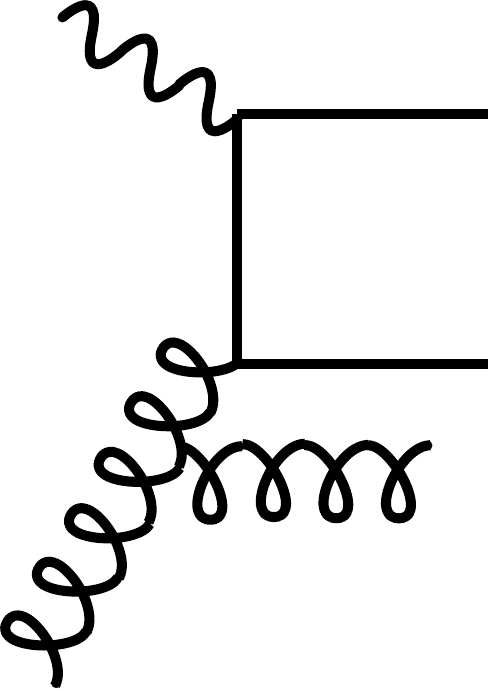}\qquad{}
\includegraphics[clip,width=0.14\textwidth]{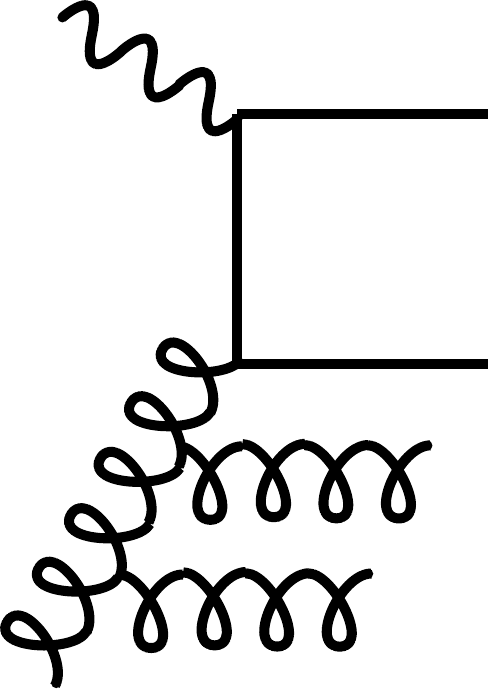}
\caption{Example Feynman diagrams contributing to DIS heavy quark production
(from left): LO ${\cal O}(\alpha_{S}^{0})$ quark-boson scattering
$QV\to Q$, NLO ${\cal O}(\alpha_{S}^{1})$ boson-gluon scattering
$gV\to Q\bar{Q}$, NNLO ${\cal O}(\alpha_{S}^{2})$ boson-gluon
scattering $gV\to gQ\bar{Q}$ and N$^{3}$LO ${\cal O}(\alpha_{S}^{3})$
boson-gluon scattering $gV\to ggQ\bar{Q}$.}
\label{olness_label2}
\end{center}
\end{figure*}

\begin{figure}[ht]
\begin{center}
\includegraphics[width=0.4\textwidth]{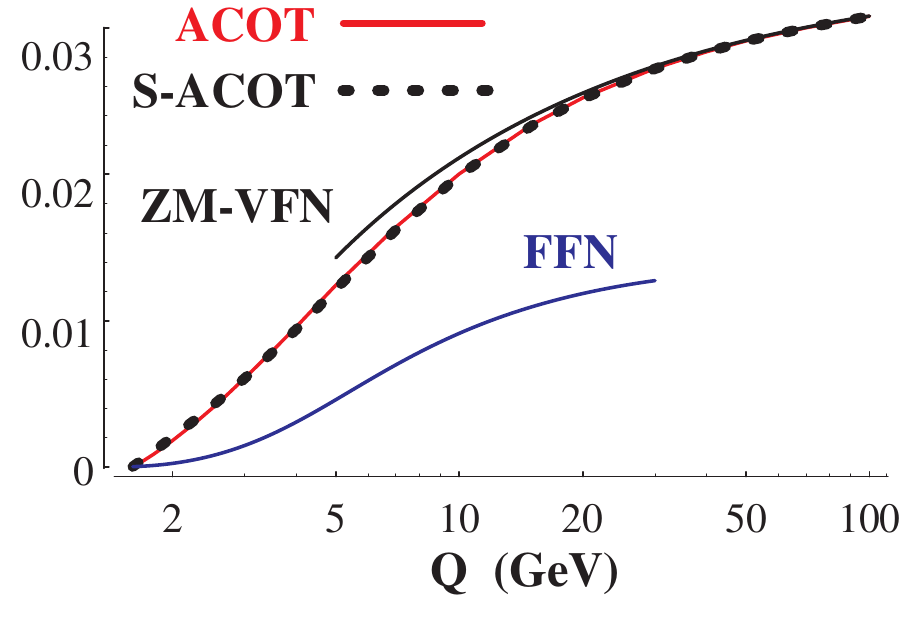}
\caption{Comparison of schemes for $F_2^c$ at $x=0.1$ for NLO DIS heavy quark production as a function
of $Q$. We display calculations using the ACOT, S-ACOT, Fixed-Flavor
Number Scheme (FFNS), and Zero-Mass Variable-Flavor-Number-Scheme
(ZM-VFNS). The ACOT and S-ACOT results are virtually identical.}
\label{olness_label3}
\end{center}
\end{figure}

The ACOT scheme~\cite{Aivazis:1993kh,Aivazis:1993pi} 
 is based upon 
the factorization theorem for heavy quarks\cite{Collins:1998rz}; hence,  it is valid 
at any order of perturbation theory. 
The factorization proof ensures that the ACOT scheme can be applied throughout the full 
kinematic regime, and that there is a smooth transition from 
a massless result ($m=0$) to the heavy-mass decoupling limit ($m\to \infty)$.

In the limit where the quark $Q$ of mass $m$ is relatively heavy compared
to the characteristic energy scale $(\mu\lsim m)$, the ACOT result
naturally reduces to the Fixed-Flavor-Number-Scheme (FFNS). In
the FFNS, the heavy quark is treated as being extrinsic to the hadron,
and there is no corresponding heavy quark PDF, $f_{Q}(x,\mu) = 0$.
Conversely, in the limit where the quark mass is relatively light
$(\mu\gsim m)$, the ACOT result reduces to the $\msbar$\ Zero-Mass
Variable-Flavor-Number-Scheme (ZM-VFNS) exactly--{\it without} any
finite renormalizations. In this limit, the quark mass $m$ no longer
plays any dynamical role; it serves purely as a regulator.
This feature  is presented in Fig.~\ref{olness_label3} 
where we can see that the ACOT scheme precisely 
matches the results of the FFNS and ZM-VFNS schemes in their respective limits.

Additionally Fig.~\ref{olness_label3} shows  the results obtained
within the Simplified-ACOT scheme (S-ACOT)~\cite{Kramer:2000hn}.
The S-ACOT scheme drops the heavy quark mass dependence for the 
hard-scattering processes with incoming heavy quarks or with internal
on-shell cuts on a heavy quark line.
The S-ACOT scheme is {\it not} an approximation;
it is an exact renormalization scheme, extensible to all orders.
Note, the ACOT and S-ACOT results agree throughout the kinematic region.

\subsubsection{Beyond NLO}

\begin{figure*}
\begin{center}
\includegraphics[width=0.3\textwidth]{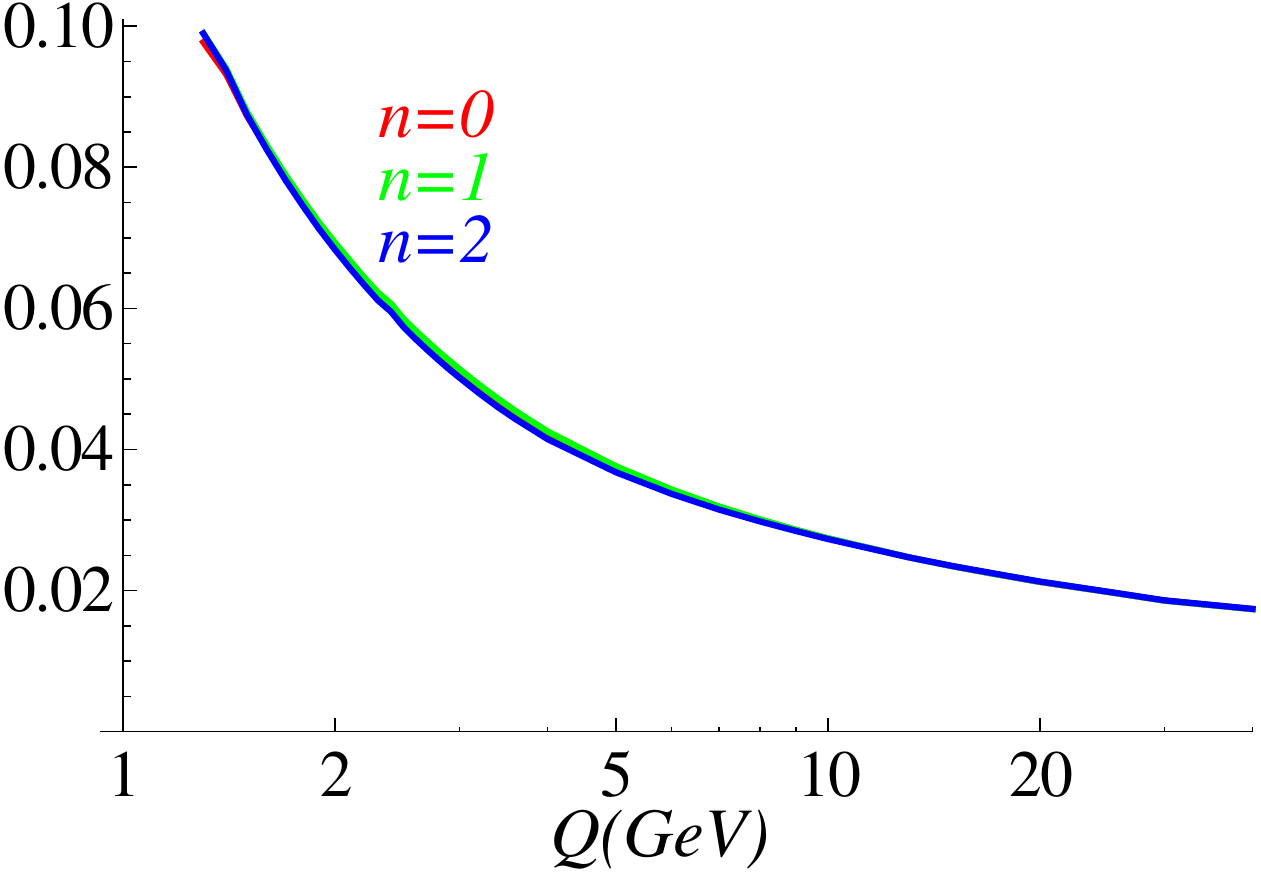}
\quad
\includegraphics[width=0.3\textwidth]{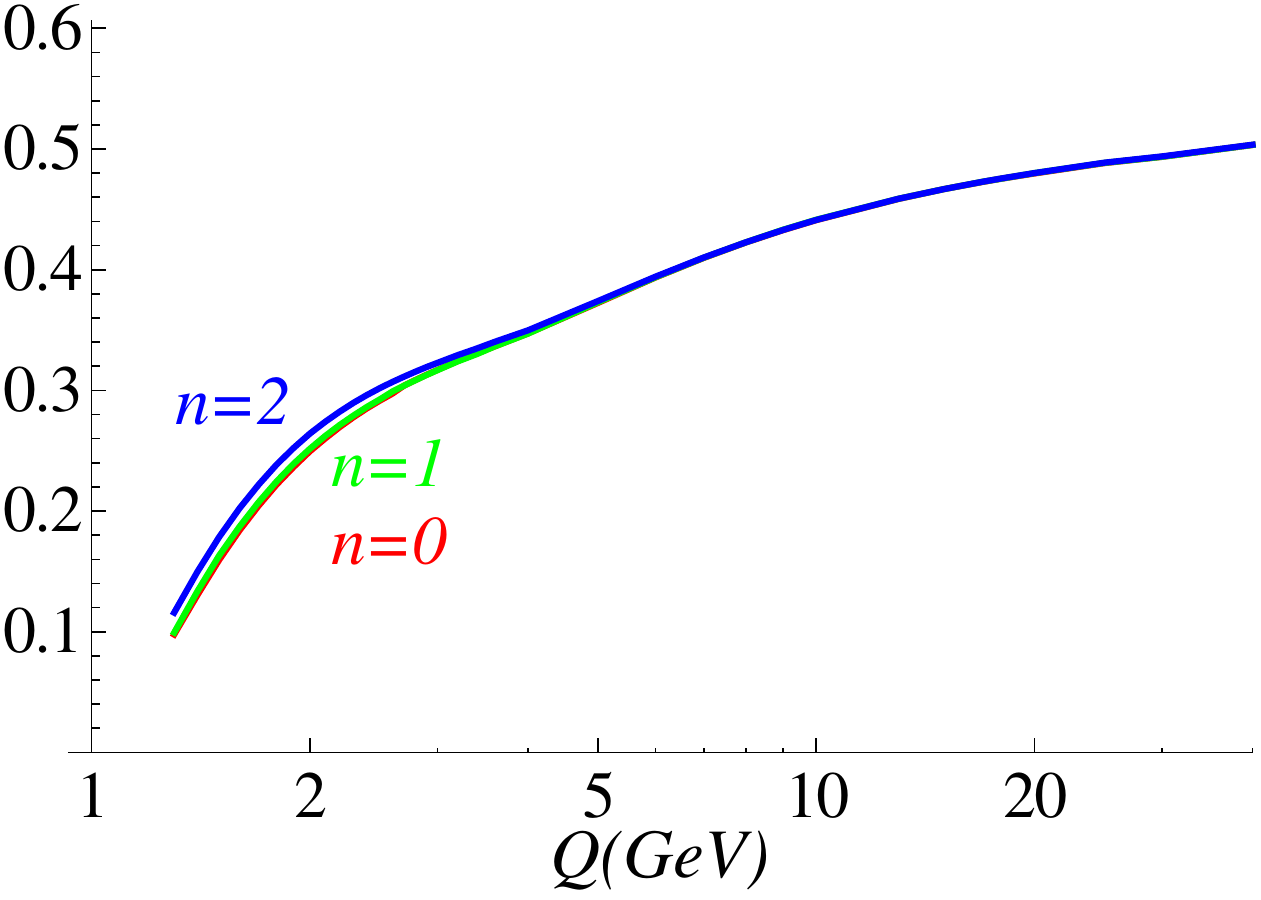}
\quad
\includegraphics[width=0.3\textwidth]{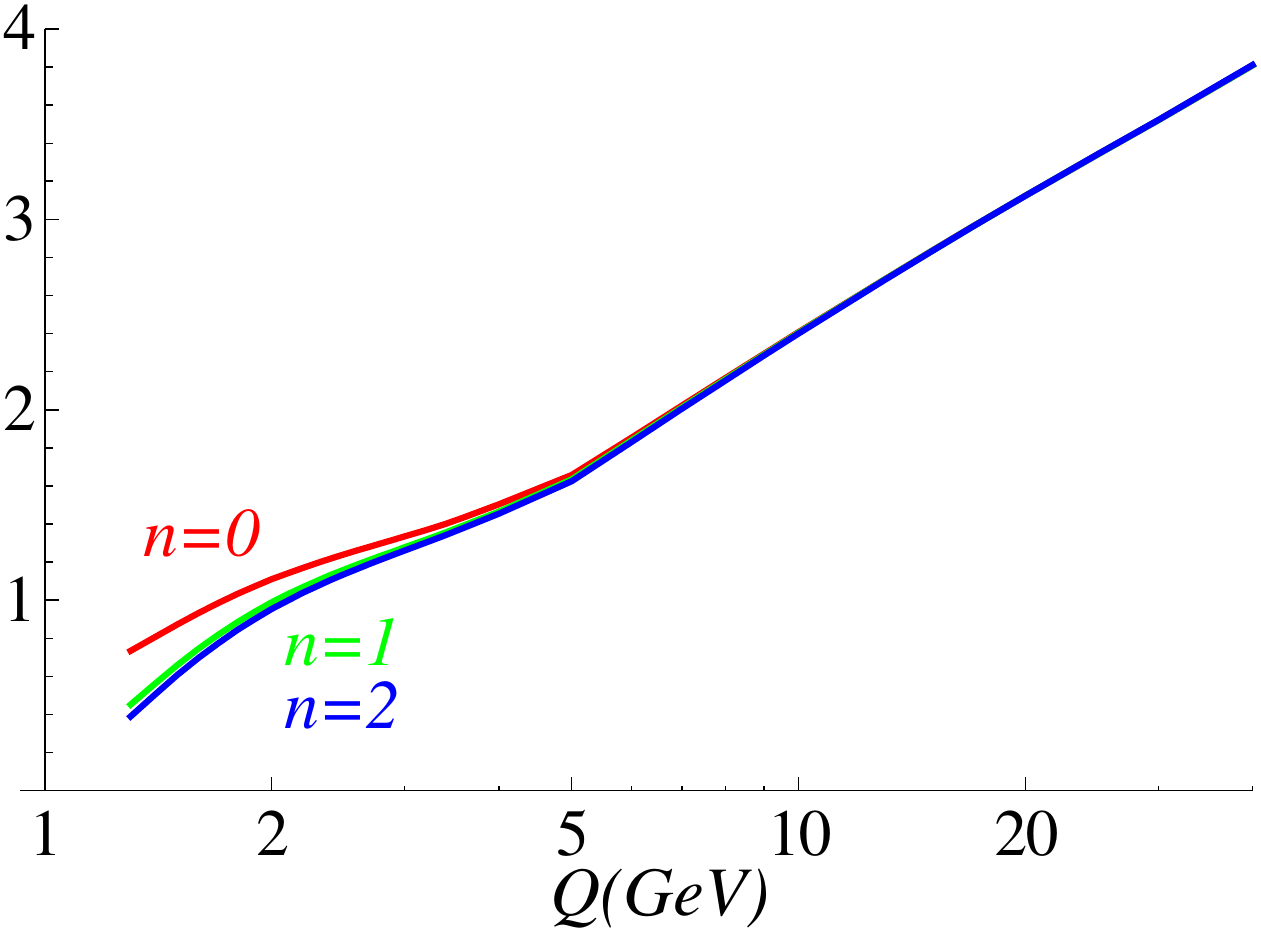}
\caption{$F_{L}$ vs. $Q$ at N$^3$LO for fixed
$x=\{10^{-1},10^{-3},10^{-5}\}$, (left to right).  
The three lines show the mass effects via the scaling variable: 
$n=\{0,1,2\}$, (Red, Green, Blue). 
We observe the effect of the $n$-scaling is negligible except for 
small  $x$ and $Q$ values.}
\label{olness_label4}
\end{center}
\end{figure*}

While there is no conceptual difficulty with extending the 
ACOT scheme beyond NLO, the fully massive Wilson coefficients 
have yet to be computed.\footnote{There has been a calculation of neutral current electroproduction
of heavy quarks ${\cal O}(\alpha_s^2)$ in the FFNS~\cite{Laenen:1992zk}.
however, extra contributions are still required for a VFNS calculation~\cite{Guzzi:2011ew}.}
However massless calculations of  NNLO and even N$^3$LO for
$F_2$ and $F_L$ structure functions are
available.\footnote{See Ref.~\cite{Moch:2004xu} and references therein.}
%
%

The question is: can we use these results, together with the knowledge
that ACOT reduces to 
the massless $\msbar$\ (ZM-VFNS) for $m\to 0$, to estimate mass effects at
NNLO and N$^3$LO?
Obviously we cannot restore the fully massive ACOT result from the
massless limit, but we can try to extract the dominant higher order
contributions.
There are two ways in which mass effects enter the calculation.
The first is ``dynamically'' through the mass dependent Wilson coefficients.
The second is  ``kinematically'' via the restricted phase space.
Comparisons using the fully massive results at NLO suggest that the 
kinematic mass effects are dominant, and that much of this dependence 
can be obtained with a rescaling of the Bjorken $x$ variable. 
We introduce a generalized rescaling 
$x \to x[1+(n \, m/Q)^2]$
where $n=0$ is the massless result, $n=1$ is the original Barnett\cite{Barnett:1976ak} rescaling, 
and $n=2$ is the $\chi$-rescaling~\cite{Amundson:1998zk}.

Thus, our strategy is as follows. 
We use the fully massive ACOT result to NLO~\cite{Kretzer:1998ju},
 and add to this the 
massless 
NNLO and N${}^3$LO contributions using the generalized rescaling prescription. 
By varying $n$, we can investigate the influence of the kinematic mass in our results. 
We argue that the massless Wilson coefficients at NNLO and N${}^3$LO,
together with the generalized rescaling prescription provide a good
approximation of the exact result. At worst, the error is of order
$\alpha\, \alpha_{S}^{2}\times[m^{2}/Q^{2}]$, and comparative studies
at NLO suggest the error is less.%
\footnote{Details will be presented in a forthcoming publication.}
For example, in Fig.~\ref{olness_label4} we display the results of
 $F_L$ for $n=\{0,1,2\}$. 
The effects of the detailed mass dependence is most noticeable for low
$Q^2$ and small $x$.  While the massless scaling result ($n=0$) does deviate
from the other curves, comparing the $n=1$ and $n=2$ curves we observe 
the details of the mass rescaling  are 
relatively small.  While this is not a proof,\footnote{Of course,
once the massive higher order Wilson coefficients have been computed,
it is straightforward to incorporate these results into our
calculations.}  this result does give us confidence that the mass
effects are under control.

\subsection{RESULTS}

\begin{figure*}[t]
\begin{center}
\subfloat[$F_{2}$ vs. $Q$.]{
\includegraphics[width=0.3\textwidth]{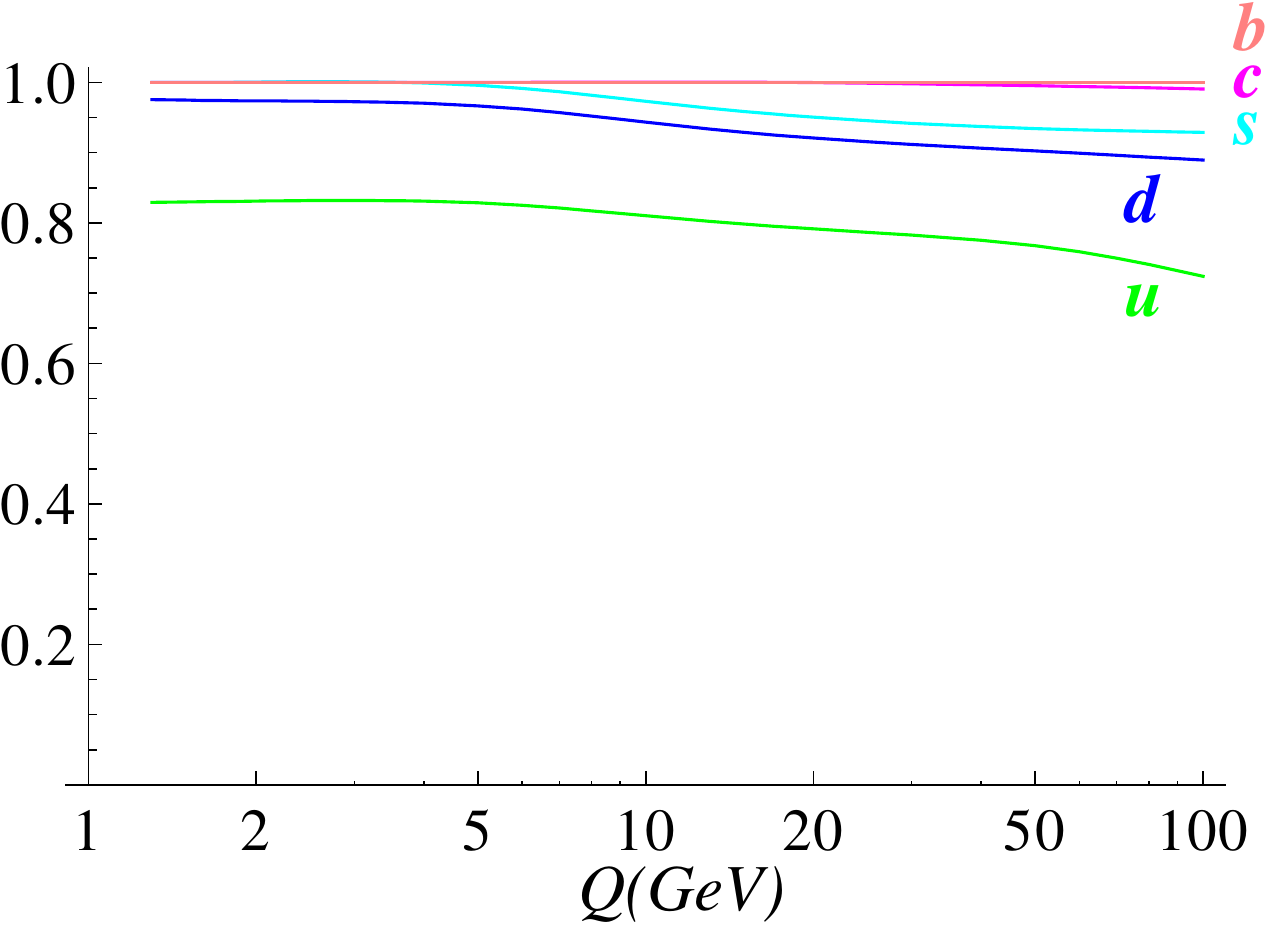}
\quad
\includegraphics[width=0.3\textwidth]{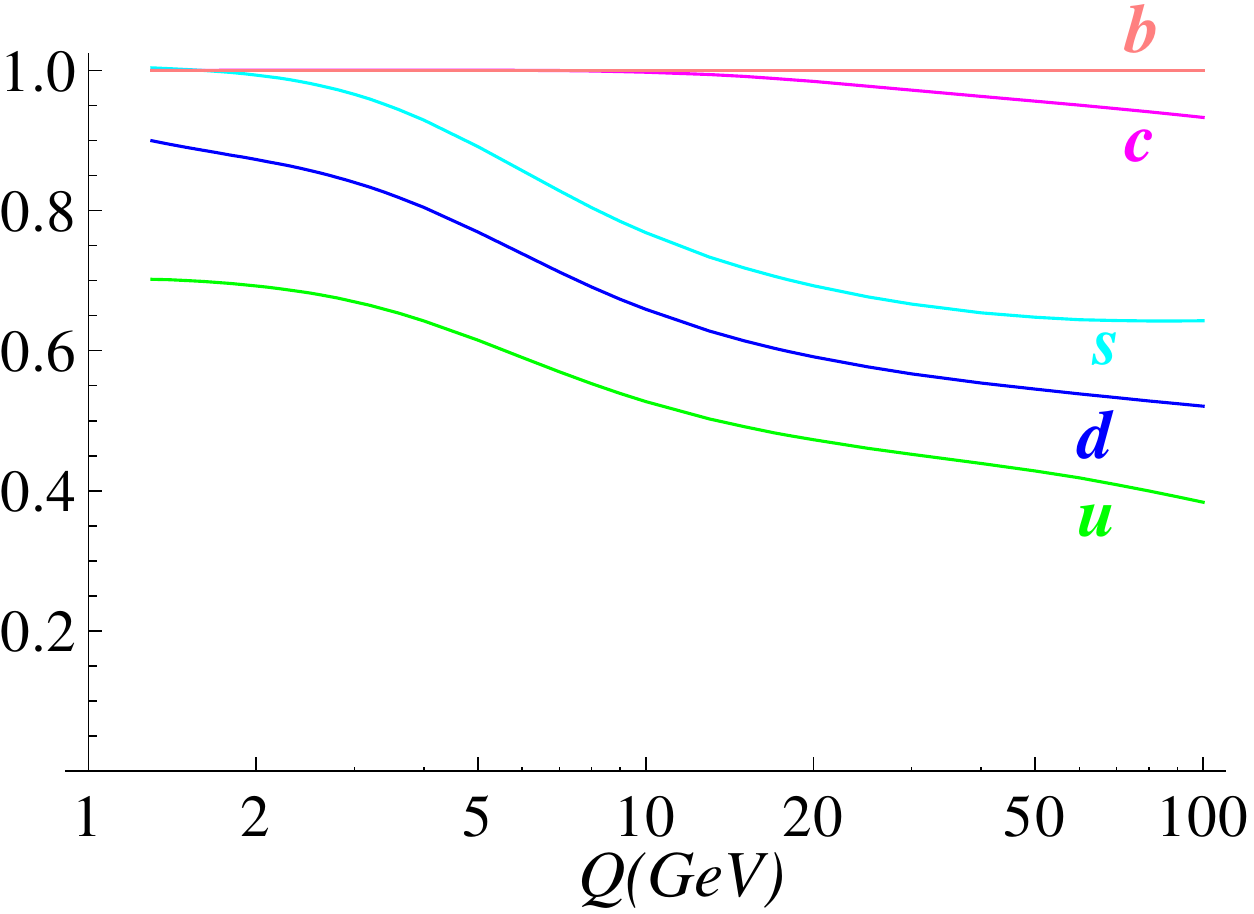}
\quad
\includegraphics[width=0.3\textwidth]{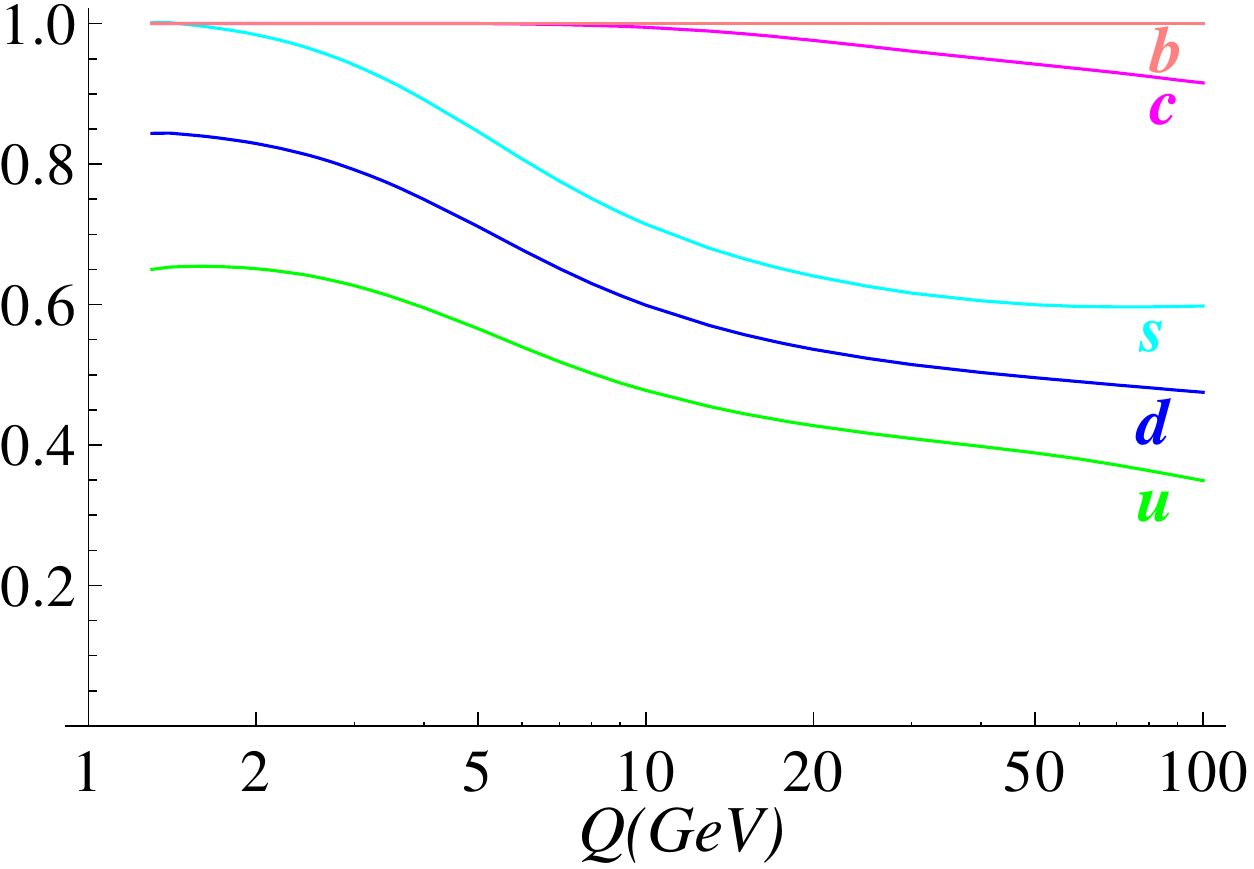}
}
\\
\subfloat[$F_{L}$ vs. $Q$.]{
\includegraphics[width=0.3\textwidth]{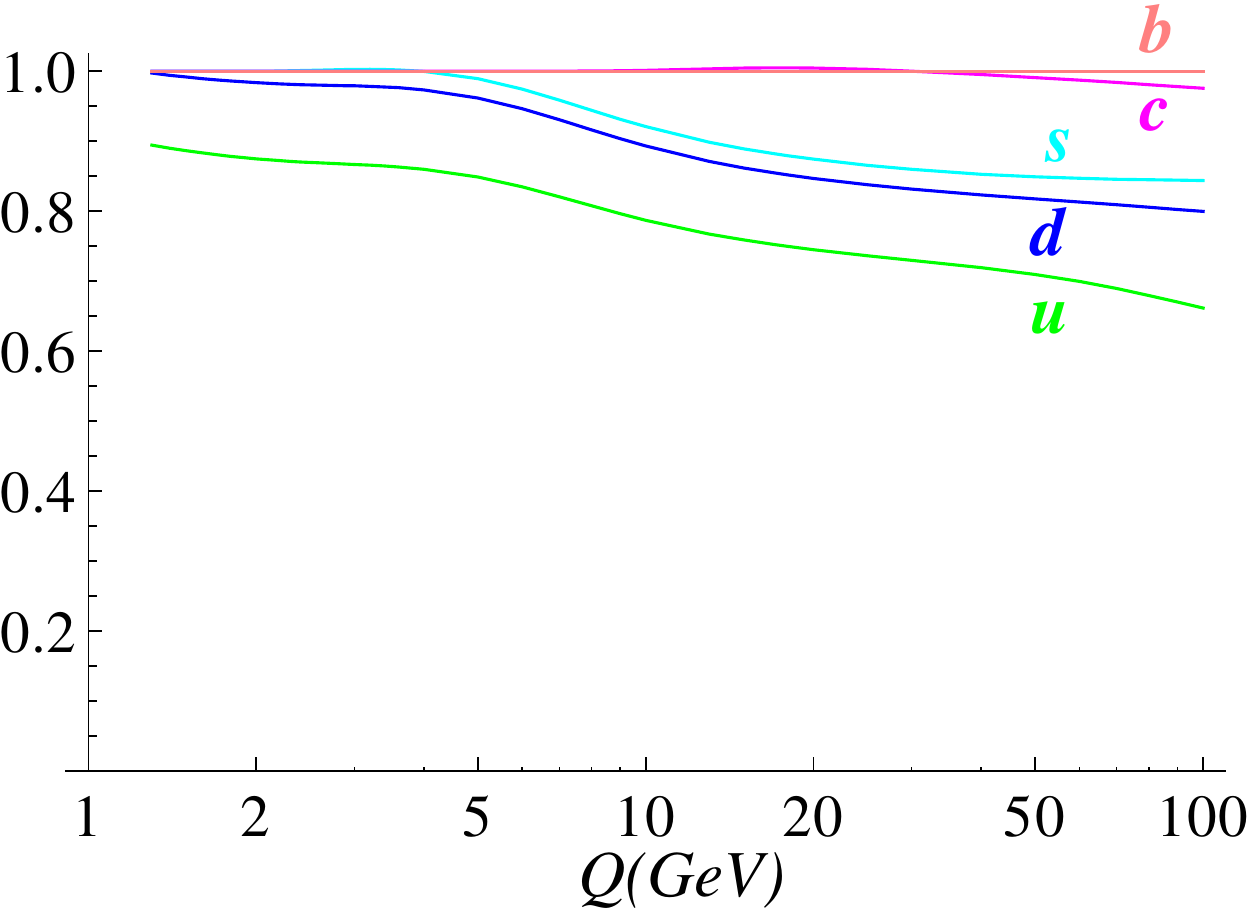}
\quad
\includegraphics[width=0.3\textwidth]{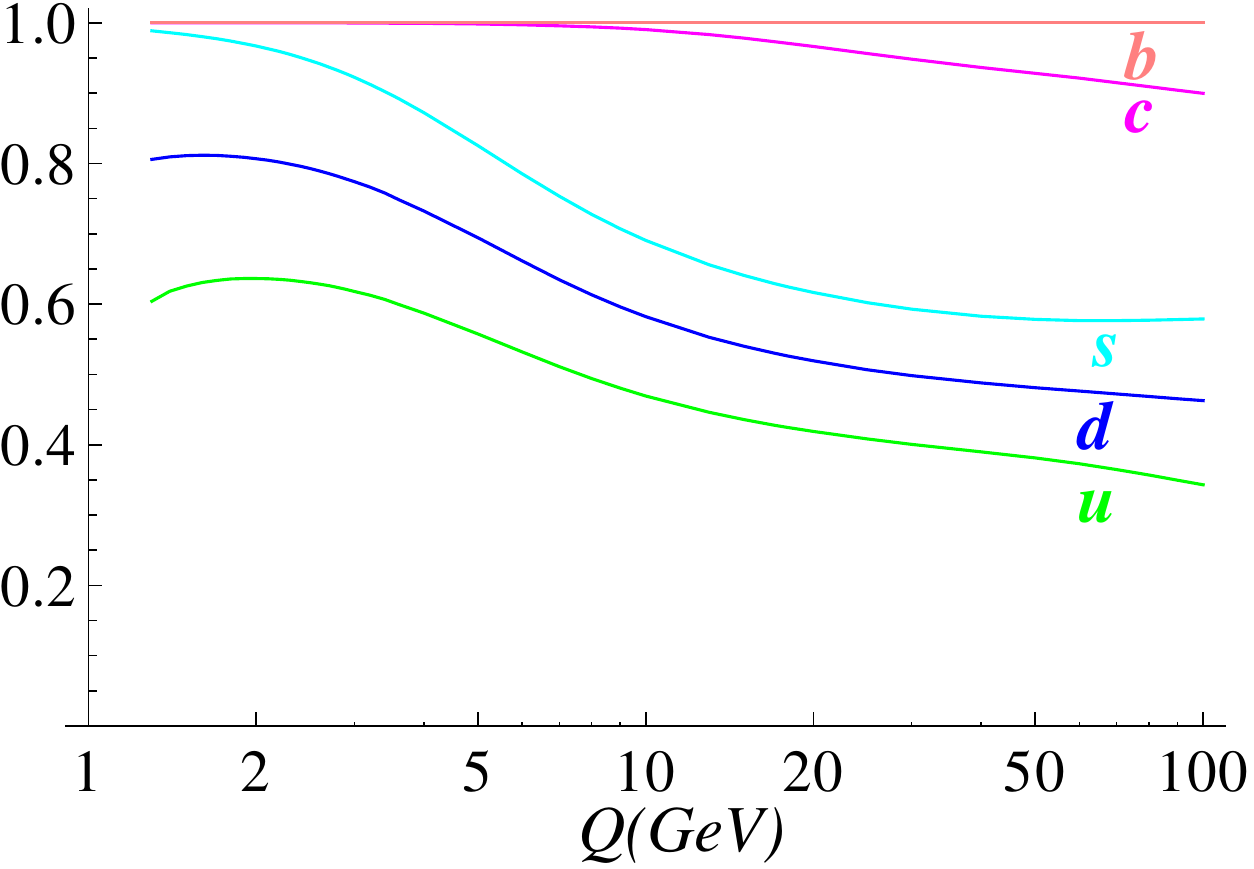}
\quad
\includegraphics[width=0.3\textwidth]{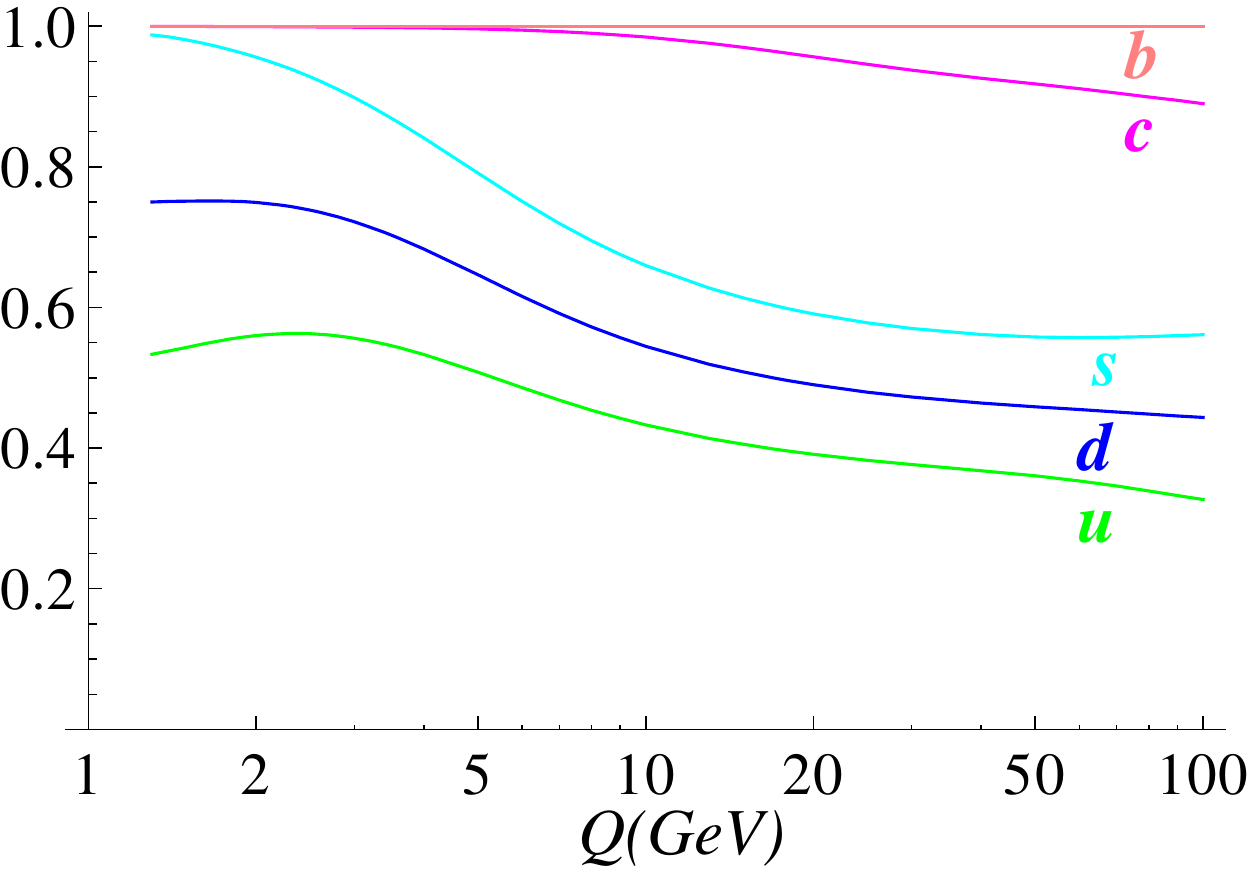}
}
\caption{Fractional contribution for each quark structure function $F_{2,L}^i$ 
for each flavor $i=\{u,d,s,c,b\}$ for (a) $F_{2}^i$
and (b) $F_L^i$ vs. $Q$ at N$^3$LO for fixed $x=\{10^{-1},10^{-3},10^{-5}\}$
(left to right) for the  ACOT-$\chi$ scheme. 
%
}
\label{olness_label5}
\end{center}
\end{figure*}

In Fig.~\ref{olness_label5}  we display
the fractional contributions to the structure functions $F_{2}$ and
$F_{L}$.
At larger values of $x$ and low $Q$, we observe that the heavy flavor
contributions are minimal. For example, for $x=10^{-1}$, we see that
the $u$-quark structure function $F^u$ comprises $\sim80\%$ of the 
total structure
function. In contrast, at $x=10^{-5}$ and large $Q$ we see that
the contributions of the $u$ and $c$ quarks are comparable
(as they couple with a factor 4/9), and the $d$ and $s$ quarks
contributions are comparable (as they couple with a factor 1/9).

Figure~\ref{olness_label5} also shows how the $\chi$-rescaling
introduces a damping of  the heavy quark contributions as we move from 
large  $Q^2$ values to smaller values. 
The $\chi$-rescaling ensures  the heavy quarks 
($c, b$) are appropriately suppressed  for low  $Q^2$  scales.

In Fig.~\ref{olness_label6} we display the results for $F_{2}$
vs. $Q$ computed at various orders; the ratio to the N$^3$LO result is
displayed in Fig.~\ref{olness_label7}. For large $x$ ({\it c.f.} $x=0.1$)
we find the perturbative calculations are particularly stable. 
We see
that the LO result is within 20\% of the others at small $Q$, and
within 5\% at large $Q$. The NLO is within 2\% at small $Q$, and
indistinguishable from the NNLO and N$^3$LO for $Q$ values above $\sim10$~GeV.
The NNLO and N$^3$LO results are essentially identical throughout the
kinematic range. For smaller $x$ values ($10^{-3}$, $10^{-5}$), the
contributions of the higher order terms are slightly larger. Here, the
NNLO and N$^3$LO coincide for $Q$ values above $\sim5$~GeV, but the NLO
result can differ by $\sim5\%$ for  low  $Q^2$  scales. 

\begin{figure*}[t]
\begin{center}
\subfloat[$F_{2}$ vs. $Q$ at \{LO, NLO, NNLO, N$^3$LO\}
(red, green, blue, cyan) for fixed $x=\{10^{-1},10^{-3},10^{-5}\}$,
(left to right) for ACOT-$\chi$ scheme.
\label{olness_label6}]{
\includegraphics[width=0.3\textwidth]{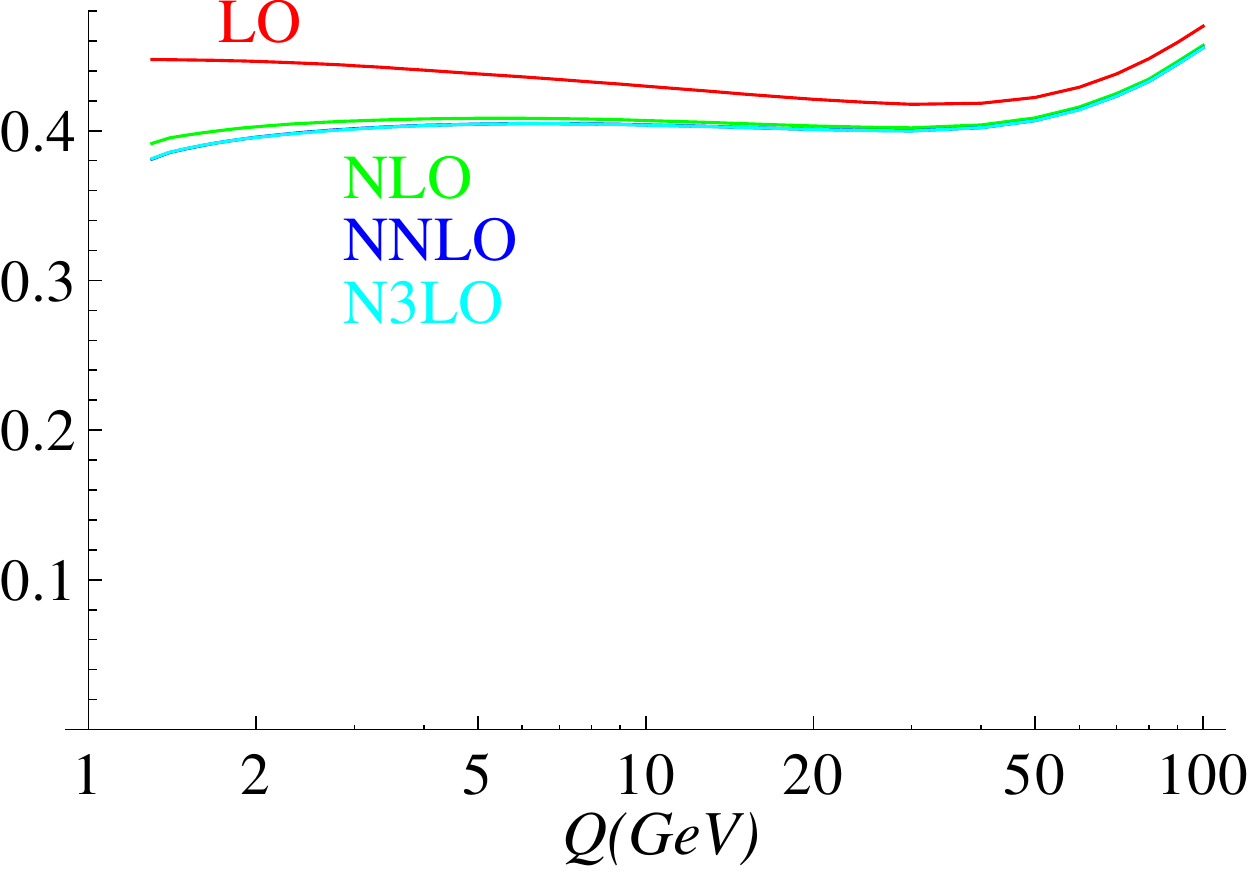}
\quad
\includegraphics[width=0.3\textwidth]{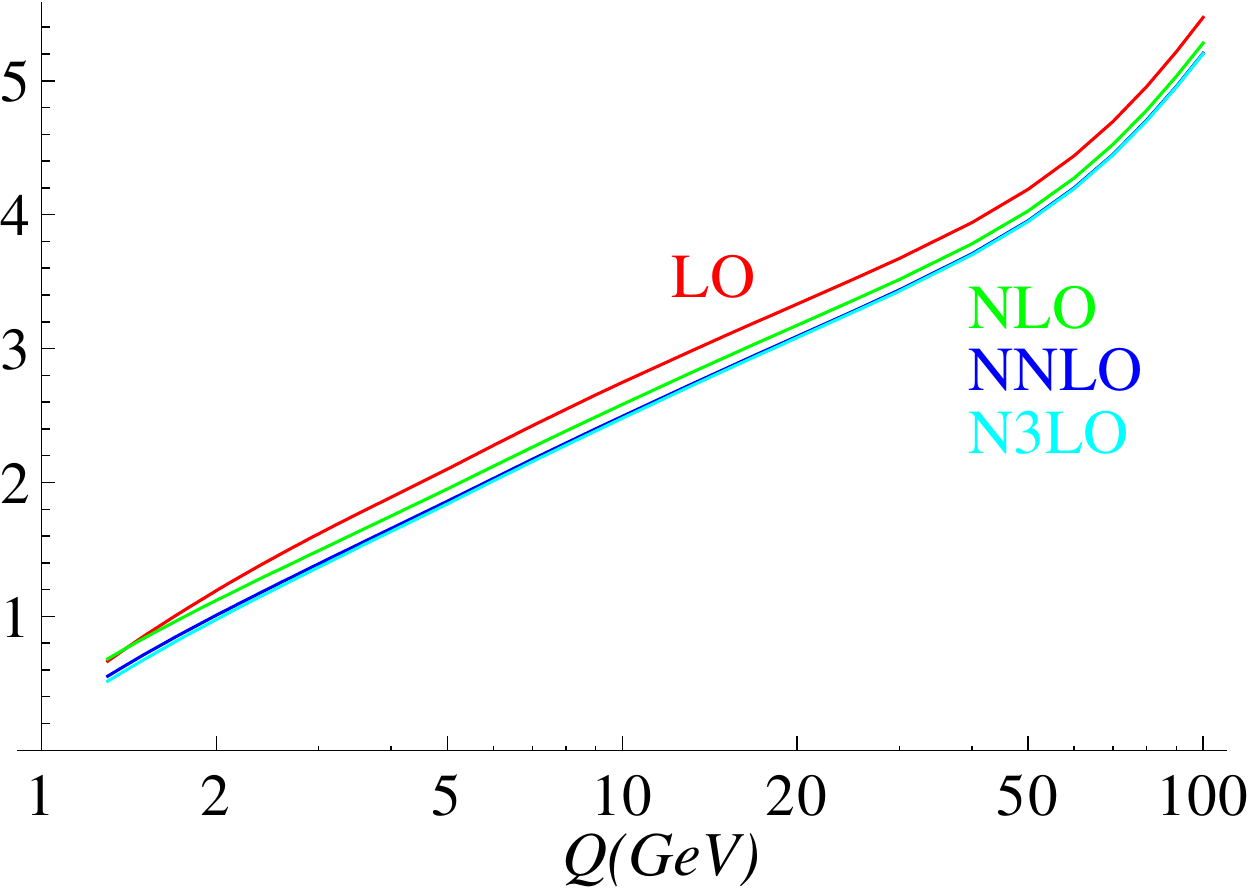}
\quad
\includegraphics[width=0.3\textwidth]{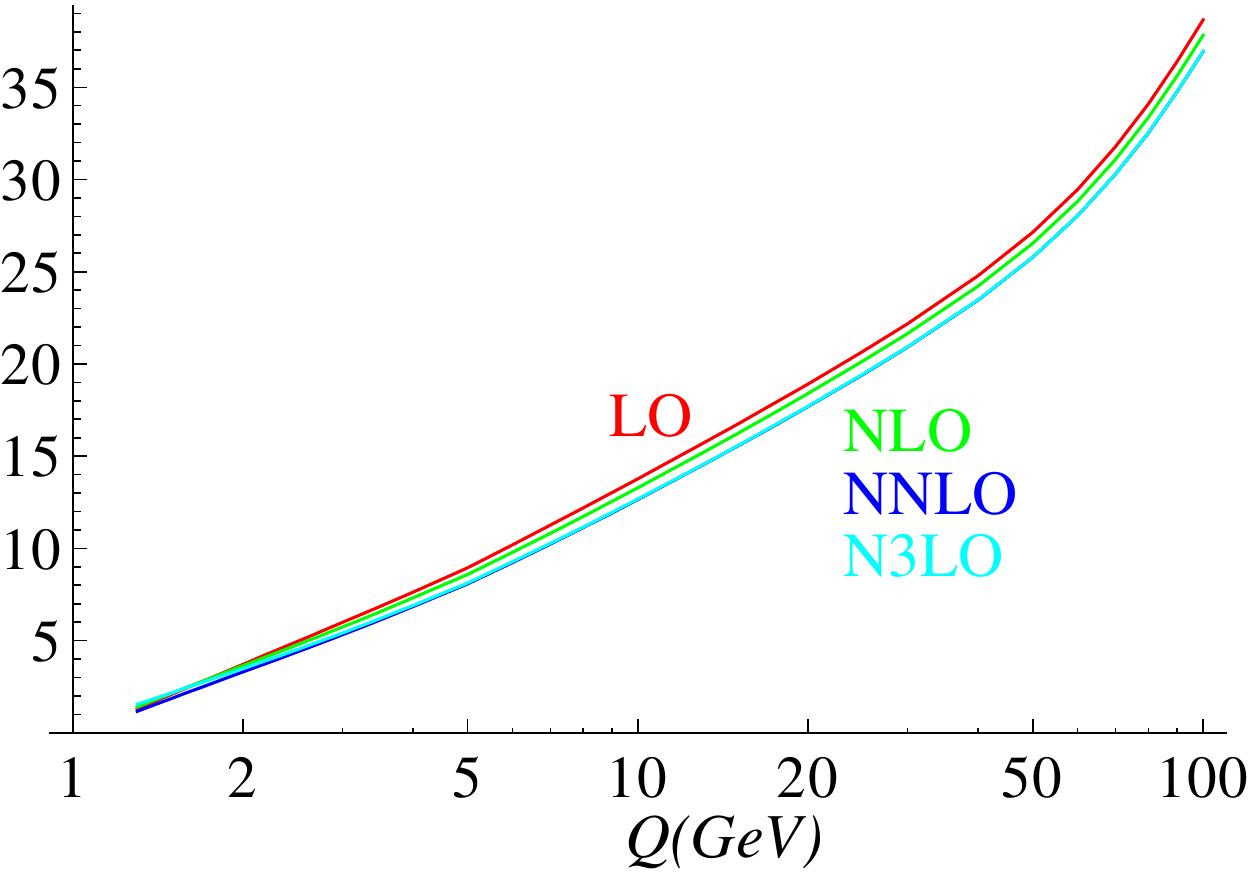}
}
\\
\subfloat[Ratio of $F_{2}$ vs. $Q$ at \{LO, NLO, NNLO, N$^3$LO\}
(red, green, blue, cyan) compared to $F_{2}$ at N$^3$LO for fixed
$x=\{10^{-1},10^{-3},10^{-5}\}$, (left to right) for ACOT-$\chi$ scheme.
\label{olness_label7}]{
\includegraphics[width=0.3\textwidth]{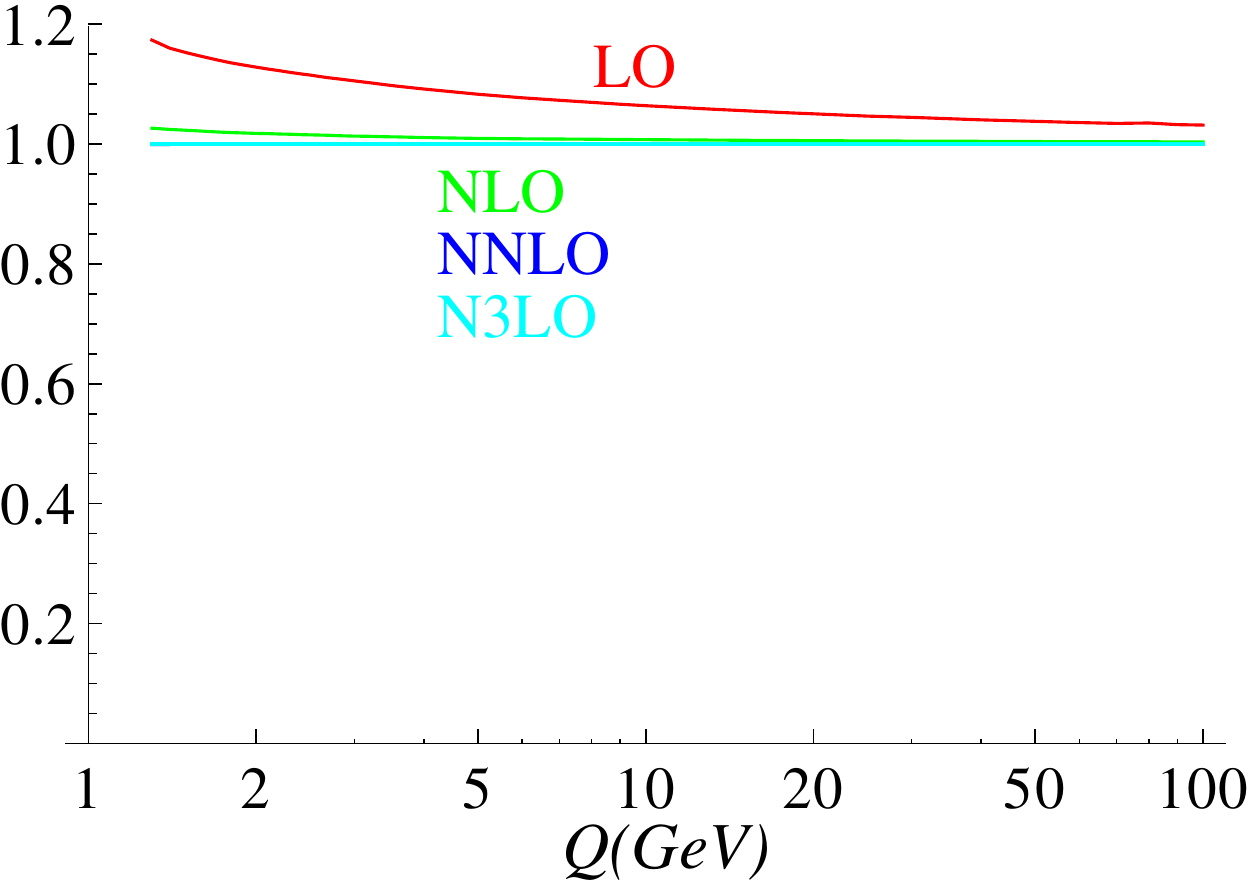}
\quad
\includegraphics[width=0.3\textwidth]{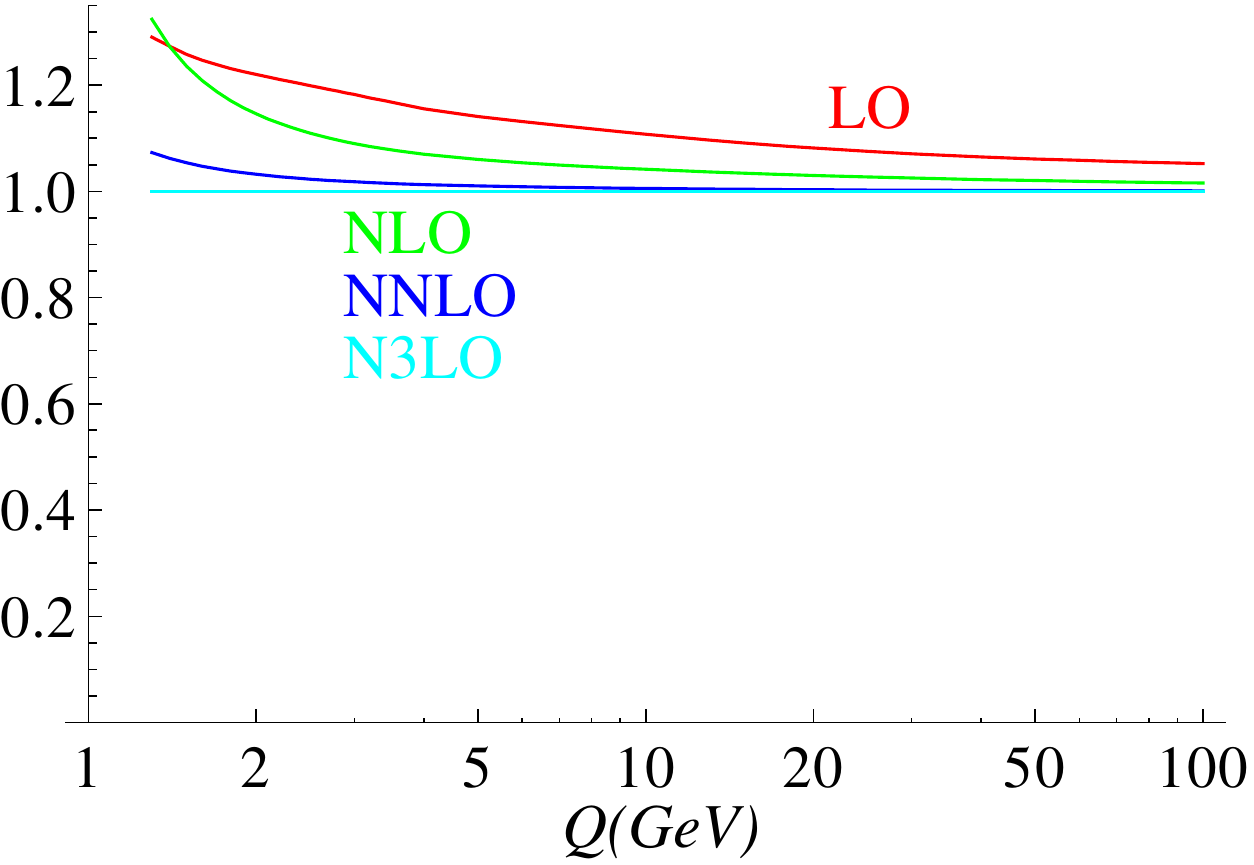}
\quad
\includegraphics[width=0.3\textwidth]{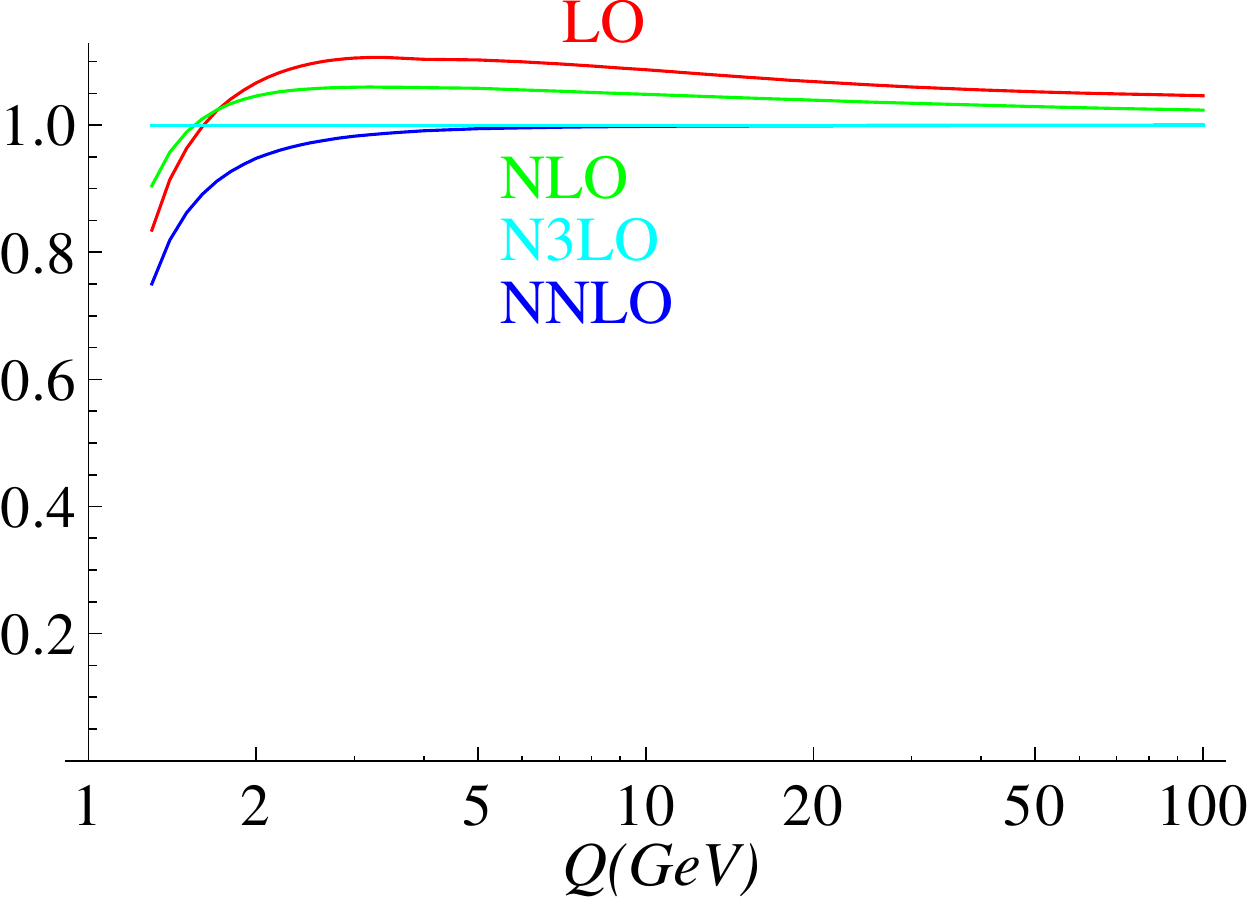}
}
\caption{$F_{2}$ vs. $Q$ at \{LO, NLO, NNLO, N$^3$LO\}.}
\end{center}
\end{figure*}

In Fig.~\ref{olness_label8} we display the results for $F_{L}$
vs. $Q$ computed at various orders. In contrast to $F_{2}$,
we find that NLO corrections are large; this is expected because the LO corrections
to $F_{L}$ (which violate the Callan-Gross relation) are suppressed
by $(m^{2}/Q^{2})$ compared to the dominant gluon contributions which
enter at NLO. Consequently, we observe that the LO result
for $F_{L}$ receives large contributions from the higher order terms.
Essentially, NLO is the first non-trivial order for $F_{L}$, and
the subsequent contributions then converge. 
For example, at large
$x$ ({\it c.f.} $x=0.1$) for $Q\sim10$~GeV we find the NLO results yields
$\sim 70\%$ of the total, the NNLO is a $\sim20\%$ correction,
and the N$^3$LO is a $\sim10\%$ correction. For lower $x$ values
($10^{-3}$, $10^{-5}$) the convergence of the perturbative series
improves, and the NLO results is within $\sim10\%$ of the N$^3$LO result.
Curiously, for $x=10^{-5}$ the NNLO and N$^3$LO roughly compensate
each other so that the NLO and the N$^3$LO match quite closely for
$Q\gsim 2$~GeV.

\begin{figure*}[t]
\begin{center}
\subfloat[$F_{L}$ vs. $Q$ at \{LO, NLO, NNLO, N$^3$LO\}
(red, green, blue, cyan) for fixed $x=\{10^{-1},10^{-3},10^{-5}\}$, 
(left to right) for ACOT-$\chi$ scheme.]{
\includegraphics[width=0.3\textwidth]{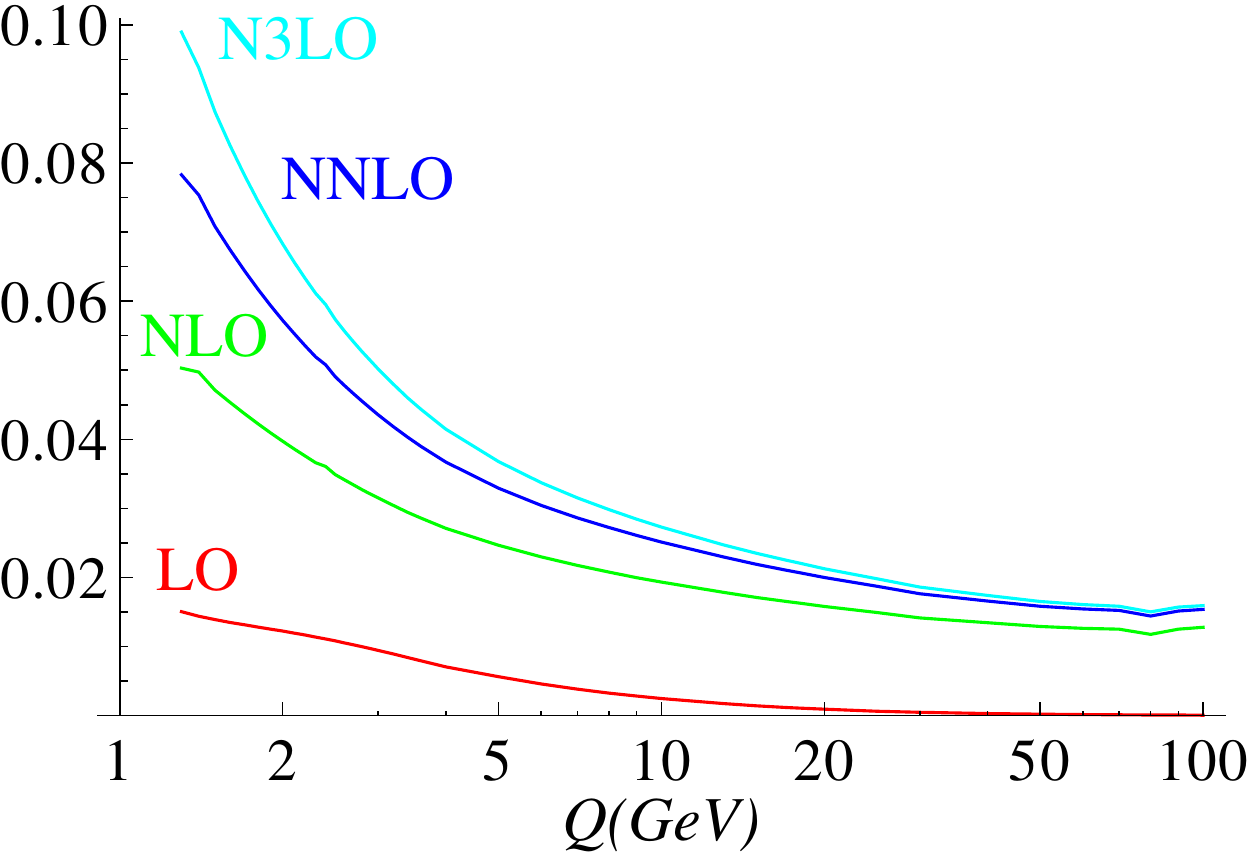}
\quad
\includegraphics[width=0.3\textwidth]{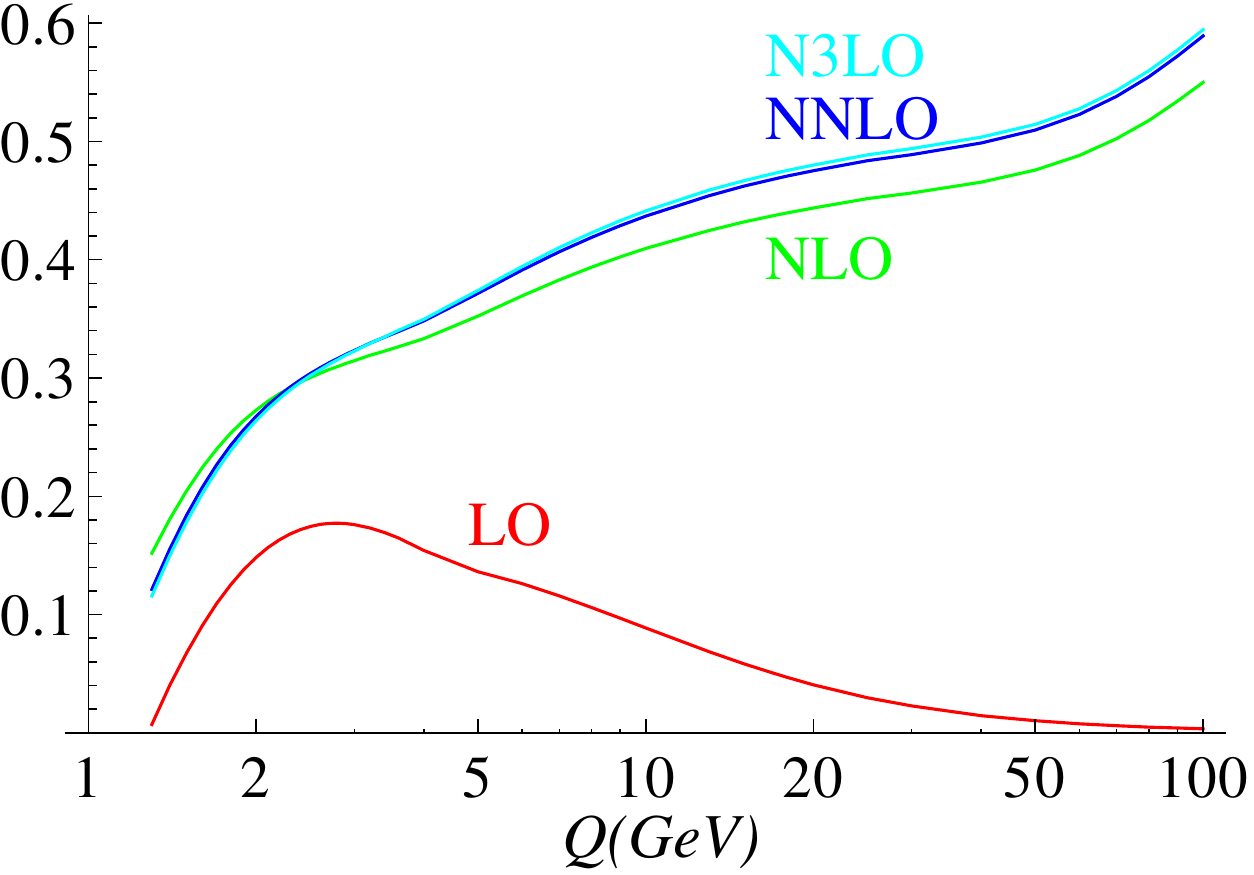}
\quad
\includegraphics[width=0.3\textwidth]{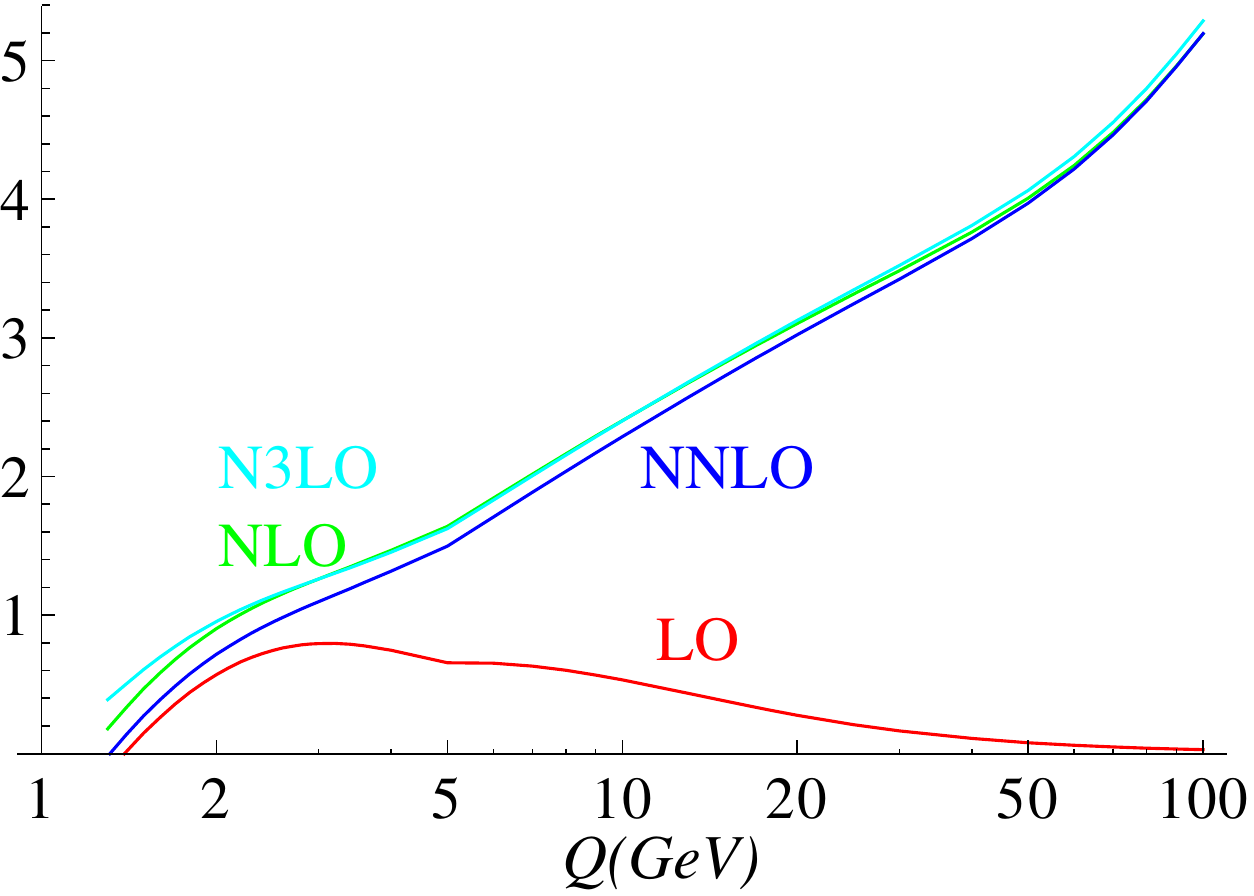}
}
\\
\subfloat[Ratio of $F_{L}$ vs. $Q$ at \{LO, NLO, NNLO, N$^3$LO\}
(red, green, blue, cyan) compared to $F_{L}$ at N$^3$LO for fixed
$x=\{10^{-1},10^{-3},10^{-5}\}$, (left to right) for ACOT-$\chi$ scheme.]{
\includegraphics[width=0.3\textwidth]{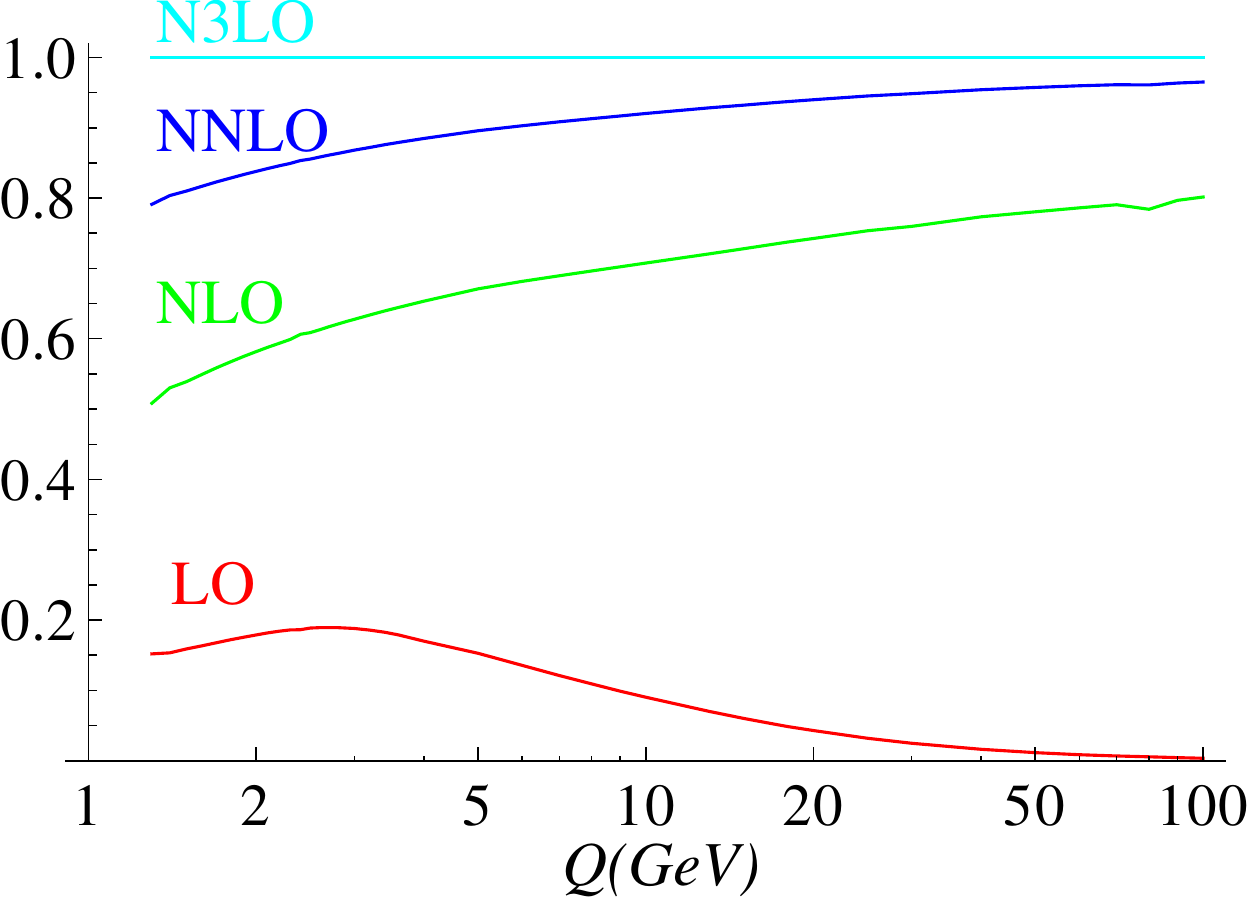}
\quad
\includegraphics[width=0.3\textwidth]{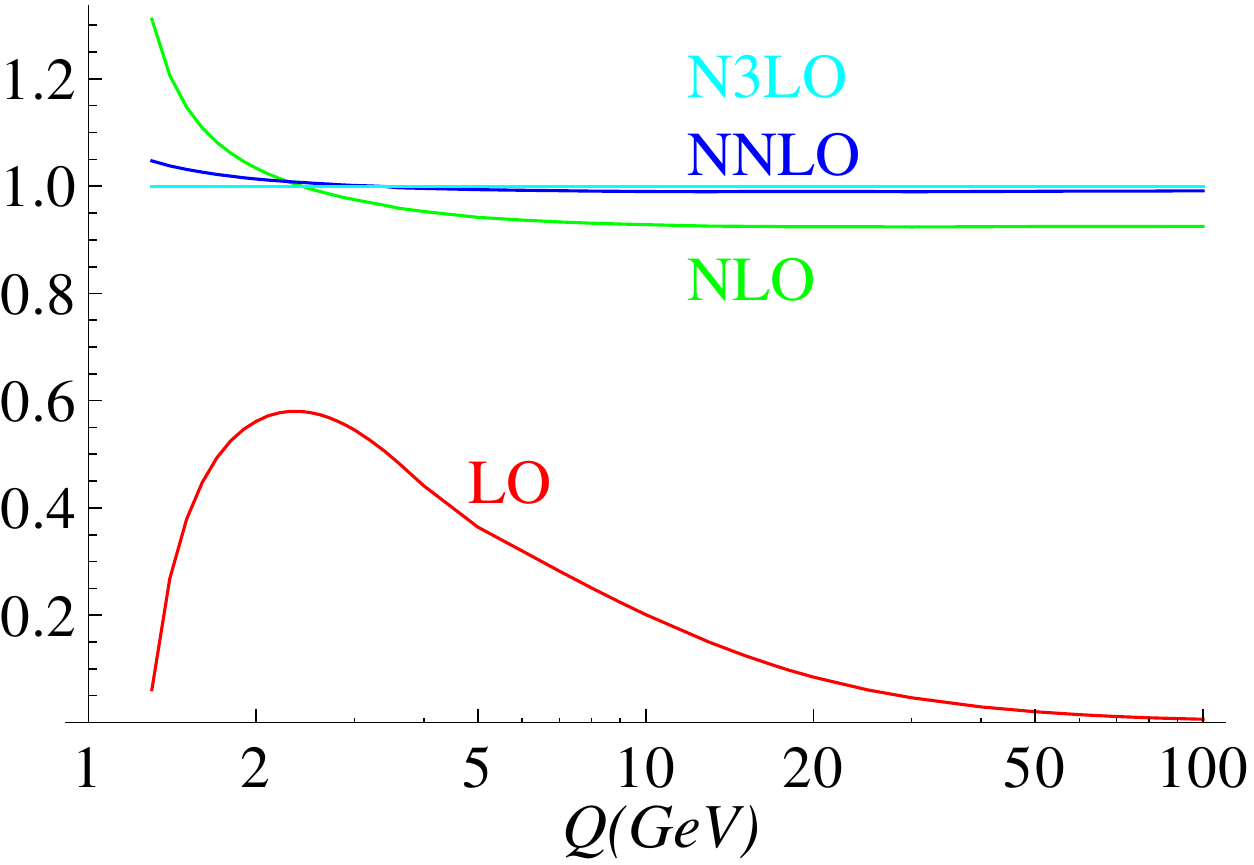}
\quad
\includegraphics[width=0.3\textwidth]{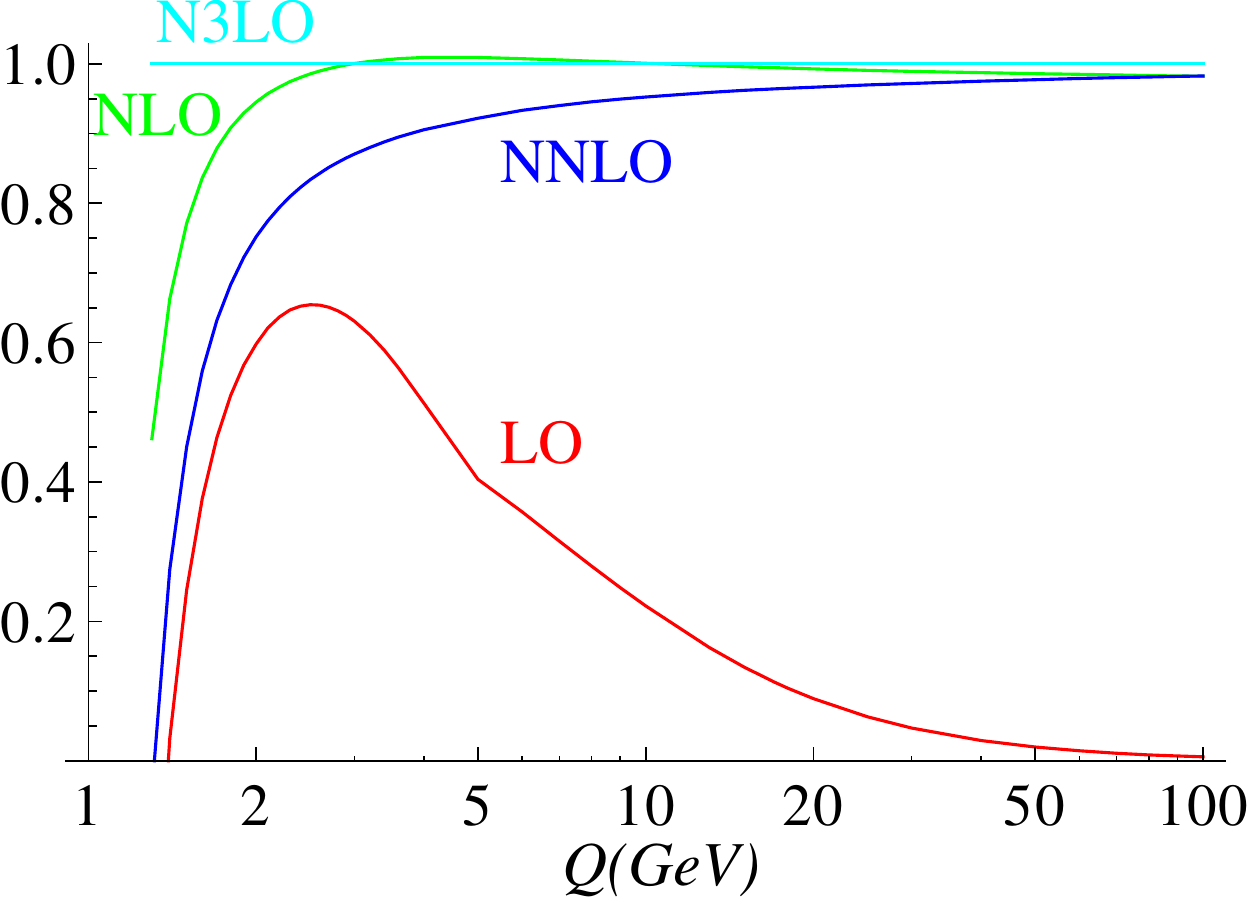}
}
\caption{$F_{L}$ vs. $Q$ at \{LO, NLO, NNLO, N$^3$LO\}.}
\label{olness_label8}
\end{center}
\end{figure*}

\subsection{CONCLUSIONS}

We have computed the  $F_2$ and $F_L$ structure
functions in the ACOT scheme at NNLO
and N$^3$LO.
The full mass dependence is computed to NLO, and 
the dominant mass effects for the higher orders are 
approximated using a generalized rescaling; the details 
of this rescaling are demonstrated to be small. 
This allows us to make detailed predictions throughout 
the kinematic range investigated by HERA, and we obtain a reasonable 
estimate of the uncertainty due to the higher order mass effects. 
Together with the precise HERA data, these calculations 
facilitate accurate determination of the PDFs which are the foundation
of the LHC calculations.

\subsection*{ACKNOWLEDGMENTS}
We thank 
M.~Botje,
A.~M. Cooper-Sakar,
A.~Glazov,
C.~Keppel,
J.~G.~Morf\'{\i}n,
P.~Nadolsky, 
J.~F.~Owens,
V.~A.~Radescu,
for valuable
discussions. 
F.I.O., I.S., and J.Y.Y.\ acknowledge the hospitality of Argonne,
BNL, CERN, Fermilab, and Les Houches where a portion of this work
was performed. 
F.I.O thanks the Galileo Galilei Institute for Theoretical Physics 
and  the INFN for their hospitality and support. 
This work was supported
by the U.S.\ Department of Energy under grant DE-FG02-04ER41299,
and the Lightner-Sams Foundation.
The work of J.~Y.~Yu was supported
by the Deutsche Forschungsgemeinschaft (DFG) through grant No.~YU~118/1-1.
The work of K.~Kova\v{r}\'{\i}k was supported by the ANR projects
ANR-06-JCJC-0038-01 and ToolsDMColl, BLAN07-2-194882.



}

\part[PHENOMENOLOGICAL STUDIES OF OBSERVABLES AND UNCERTAINTIES]{PHENOMENOLOGICAL STUDIES OF OBSERVABLES AND UNCERTAINTIES}
\label{part:pheno}

\section[Finite-width effects in top-quark pair production and decay at the LHC]
{FINITE-WIDTH EFFECTS IN TOP-QUARK PAIR PRODUCTION AND DECAY AT THE LHC \protect\footnote{Contributed by: A.~Denner,
S.~Dittmaier,
S.~Kallweit,
S.~Pozzorini,
M.~Schulze}}
{\graphicspath{{wwbb/}}

\title{Finite-width effects in top-quark pair production and decay at the LHC}

\author{A.~Denner$^1$,
S.~Dittmaier$^2$,
S.~Kallweit$^3$,
S.~Pozzorini$^3$,
M.~Schulze$^{4}$}
\institute{
$^1$ Universit\"at W\"urzburg, Institut f\"ur Theoretische Physik und Astrophysik,
        97074 W\"urzburg, Germany
\\$^2$  Albert-Ludwigs-Universit\"at Freiburg, Physikalisches Institut, 79104 Freiburg, Germany
\\$^3$ Institut f\"ur Theoretische Physik, Universit\"at Z\"urich, 8057 Z\"urich, Switzerland
\\$^4$ High Energy Physics Division, Argonne National Laboratory, Argonne, IL 60439, USA
}


{ 

\newcommand{\Pt}{\mathrm{t}}
\newcommand{\PW}{\mathrm{W}}
\newcommand{\Pb}{\mathrm{b}}
\newcommand{\Pp}{\mathrm{p}}
\newcommand{\Pe}{\mathrm{e}}
\newcommand{\PH}{\mathrm{H}}
\newcommand{\Mt}{m_\Pt}
\newcommand{\rS}{\mathrm{S}}
\newcommand{\GeV}{\mathrm{GeV}}
\newcommand{\MW}{M_\mathrm{W}}
\newcommand{\MZ}{M_\mathrm{Z}}
\newcommand{\MH}{M_\mathrm{H}}
\newcommand{\GF}{G_\mu}
\newcommand{\sw}{s_\mathrm{w}}
\newcommand{\rT}{\mathrm{T}}
\newcommand{\rd}{\mathrm{d}}
\newcommand{\ri}{\mathrm{i}}
\newcommand{\kT}{k_\rT}
\newcommand{\pT}{p_\rT}

\newcommand{\LO}{\mathrm{LO}}
\newcommand{\NLO}{\mathrm{NLO}}
\newcommand{\miss}{\mathrm{miss}}

\newcommand{\tb}{{\bar\Pt}}
\newcommand{\ttb}{{\Pt\bar\Pt}}
\newcommand{\wwbb}{{\PW\PW\Pb\bar\Pb}}
\newcommand{\WWbb}{{\PW^+\PW^-\Pb\bar\Pb}}
\newcommand{\lvlvbb}{{\nu_\Pe\Pe^+\mu^-{\bar\nu}_\mu\Pb\bar\Pb}}

\newcommand{\lsim}
{\mathrel{\raisebox{-.3em}{$\stackrel{\displaystyle <}{\sim}$}}}
\newcommand{\gsim}
{\mathrel{\raisebox{-.3em}{$\stackrel{\displaystyle >}{\sim}$}}}

\newcommand{\refeq}[1]{(\ref{#1})}
\newcommand{\eqintext}{}

\begin{abstract}
We investigate finite-top-width effects in 
top-quark pair production
by comparing  NLO QCD predictions for $\Pp\Pp\to \wwbb$
to corresponding $\Pp\Pp\to \ttb\to\wwbb$ results
in the narrow-top-width limit.
Finite-top-width effects, which result from non-resonant and off-shell contributions, 
are discussed in detail for the case of the inclusive cross section (with experimental cuts)
and for selected differential observables in the di-lepton channel.
\end{abstract}

\subsection{INTRODUCTION}

Top-quark pair production at hadron colliders allows for key tests of the
Standard Model and represents an omnipresent background to  Higgs-boson and new-physics searches.
The very large $\Pt\bar\Pt$ samples from the Tevatron and the
LHC, and the steadily increasing systematic precision
call for a continuous improvement of theory predictions.\footnote{Recent progress in the theoretical description of top-quark pair production at hadron colliders 
is reviewed in Refs.~\cite{Kidonakis:2011ca,Weinzierl:2012tc,Bonciani:2012zt}.
}
In this context, a reliable theoretical description of experimental cuts
and exclusive $\Pt\bar\Pt$ observables, which depend on details of the 
{$\PW^+\PW^-\Pb\bar\Pb$} final state, requires higher-order calculations
for top-pair {\it production and decay}. 
The first NLO QCD predictions for $\Pp\Pp\to
\ttb\to\wwbb+X$~\cite{Bernreuther:2004jv,Melnikov:2009dn, Bernreuther:2010ny}
have been obtained in the narrow-top-width limit, an approximation where the
$2\to 4$ particle process is factorised into 
on-shell $\ttb$ production and (anti)top decays, taking into
account spin correlations.  In this framework, it was shown that NLO QCD effects
in top-quark decays have a significant impact on 
the kinematic properties of final-state leptons and
b-jets~\cite{Bernreuther:2004jv,Melnikov:2009dn, Bernreuther:2010ny},
and play an important role for top-mass measurements at 
the LHC~\cite{Biswas:2010sa}.
More recently, NLO QCD predictions for the complete {$\Pp\Pp\to\PW^+\PW^-\Pb\bar\Pb+X$}
process became available~\cite{Denner:2010jp,Bevilacqua:2010qb}, 
which include all effects related to the finite top-quark width, i.e.~on- and off-shell intermediate top quarks,
non-resonant contributions, and their interference with resonant $\ttb$ production.
Besides new evidence for the importance of NLO corrections to 
$\ttb$ production and decay, 
these studies provided a first quantitative assessment of 
finite-width effects in the inclusive cross section.
Applying a numerical $\Gamma_\Pt\to 0$ extrapolation to the NLO {$\Pp\Pp\to\PW^+\PW^-\Pb\bar\Pb$}
predictions, it was found that 
finite-top-width contributions to the $\wwbb$ cross section  
at the Tevatron and the LHC (7\,TeV) range from 0.2 to 1 percent~\cite{Denner:2010jp,Bevilacqua:2010qb},
which is  perfectly consistent with the expected order of magnitude ($\Gamma_\Pt/\Mt\simeq 0.9\%$) of finite-top-width effects in inclusive
observables.

In this study, we pursue the investigation of finite-top-width effects by means of a 
tuned comparison of the {$\Pp\Pp\to\PW^+\PW^-\Pb\bar\Pb$} NLO calculation of Ref.~\cite{Denner:2010jp}
against the narrow-top-width approximation of Ref.~\cite{Melnikov:2009dn}.
This permits us, for the first time, to investigate $\Gamma_\Pt$-effects in
different phenomenologically interesting regions of the $\wwbb$ phase space,
where large off-shell and non-resonant contributions
cannot be excluded a priori as in the case of inclusive observables.

\subsection{NARROW-TOP-WIDTH APPROXIMATION AND FINITE-WIDTH EFFECTS}
Let us start by recalling the main features of the NLO QCD calculations
of {$\Pp\Pp\to\ttb\to\PW^+\PW^-\Pb\bar\Pb$} in narrow-top-width approximation~\cite{Melnikov:2009dn}
and {$\Pp\Pp\to\PW^+\PW^-\Pb\bar\Pb$} with finite-top-width effects~\cite{Denner:2010jp}.
For brevity, we denote them as $\ttb$ and $\wwbb$ calculations, respectively.
Both calculations implement leptonic W-boson decays in spin-correlated 
narrow-W-width approximation.

In the narrow-top-width limit of Ref.~\cite{Melnikov:2009dn}, top-quark resonances
are approximated by
\begin{equation}
\label{eq:delta}
\lim_{\Gamma_\Pt/\Mt\to 0}\frac{1}{(p_\Pt^2-\Mt^2)^2+\Mt^2\Gamma_\Pt^2} = \frac{\pi}{\Mt\Gamma_\Pt}\delta\big(p_\Pt^2-\Mt^2\big),
\end{equation}
with delta functions that enforce the on-shell conditions, $p_\Pt^2=\Mt^2$, and are accompanied by
$1/\Gamma_\Pt$ factors.  Contributions of $\mathcal{O}(\Gamma_\Pt/\Mt)$, 
i.e.~terms that do not involve two resonant top propagators, are
systematically neglected. The differential $\Pp\Pp\to
\ttb\to\wwbb$ cross section is factorised into the  
$\Pp\Pp\to \ttb$ cross section times  $\Pt\to \PW\Pb$ partial decay widths,
$\rd\sigma =   \left(\rd\sigma_\ttb\,\rd\Gamma_\Pt\rd\Gamma_{\tb}\right)/{\Gamma_\Pt^2}$,
taking into account top-quark spin correlations.
The LO and NLO predictions can be schematically expressed as
\begin{eqnarray}
\label{eq:fact}
\rd\sigma_\LO &=& {\Gamma^{-2}_{\Pt,\LO}} \left(\rd\sigma^{0}_\ttb\,\rd\Gamma_\Pt^{0}\rd\Gamma_{\tb}^{0}\right),\nonumber\\
\rd\sigma_\NLO &=& {\Gamma^{-2}_{\Pt,\NLO}} \left[
\left(\rd\sigma^{0}_\ttb+\rd\sigma^{1}_\ttb\right)
\,\rd\Gamma_\Pt^{0}\rd\Gamma_{\tb}^{0}+
\rd\sigma^{0}_\ttb\,\left(
\rd\Gamma_\Pt^{1}\rd\Gamma_{\tb}^{0}+
\rd\Gamma_\Pt^{0}\rd\Gamma_{\tb}^{1}
\right)
\right],
\end{eqnarray}
where the superscripts 0 and 1 indicate tree-level quantities and NLO corrections, respectively.
The NLO prediction involves three terms, where the corrections are 
applied either to $\rd\sigma_\ttb$ or to one of the decays. 
All ingredients of $\rd\sigma_\LO$ and $\rd\sigma_\NLO$ have to be evaluated 
with input parameters at the corresponding 
perturbative order. In particular,
LO and NLO predictions must be computed using $\Gamma_{\Pt,\LO}$
and $\Gamma_{\Pt,\NLO}$ decay widths, as indicated in~\refeq{eq:fact}.\footnote{
We note that
in the present study the factor $\Gamma^{-2}_{\Pt,\NLO}$ in \refeq{eq:fact}
is not expanded as 
$\left(\Gamma_{\Pt,\LO}+\Gamma^1_{\Pt}\right)^{-2}=
\Gamma^{-2}_{\Pt,\LO}(1-2\,\Gamma^1_{\Pt}/\Gamma_{\Pt,\LO})$, 
like
in Eq.~(6) of Ref.~\cite{Melnikov:2009dn},
since this procedure is not directly applicable to 
the full $\wwbb$ calculation.}
 This guarantees 
that---up to higher-order corrections---the integration over 
the phase space of each top decay in~\refeq{eq:fact}
is consistent with the branching fraction 
\begin{equation}
\label{eq:br}
\frac{\int\rd\Gamma_{\Pt\to \Pb l\nu}}{\Gamma_\Pt}
=
\frac{\Gamma_{\Pt\to \Pb l\nu}}{\Gamma_\Pt}=
\mathrm{BR}(\Pt\to \Pb l\nu).
\end{equation}
In this context, let us point out that a consistent inclusion of finite-W-width 
corrections---both in the scattering amplitudes
and the $\Gamma_\Pt$ input parameters---is expected to lead to doubly-suppressed effects.
This is due to the fact that, in the $\Gamma_\Pt\to 0$ limit, $\mathcal{O}(\Gamma_\PW)$ corrections 
to the numerator and  denominator of the branching fraction~\refeq{eq:br} cancel.
Finite-W-width corrections are thus expected to produce very small effects of
$\mathcal{O}(\frac{\Gamma_\PW\Gamma_\Pt}{\MW\Mt})$ in inclusive observables.
This justifies the use of the narrow-W-width approximation in combination with 
finite-top-width contributions, which is the approach adopted in 
Ref.~\cite{Denner:2010jp}, although in kinematic regions where 
finite-$\Gamma_\Pt$ effects become large
also finite-W-width corrections~\cite{Bevilacqua:2010qb} might become non-negligible.

The calculation of Ref.~\cite{Denner:2010jp}
provides a full description of {$\Pp\Pp\to\PW^+\PW^-\Pb\bar\Pb$} 
at order $\mathcal{O}(\alpha^3_{\mathrm{S}}{\alpha}^2)$. The top-quark width is incorporated into the 
complex top mass, $\mu_\Pt^2=\Mt^2-\ri\Mt\Gamma_\Pt$,
in the complex-mass scheme~\cite{Denner:2005fg}. In this way,
off-shell-top contributions are consistently 
described by Breit--Wigner distributions.
Besides contributions with two intermediate top resonances, also
singly- and non-resonant diagrams  are taken into account, 
including interferences.
A few representative tree diagrams are shown in Fig.\ref{fig:treegraphs}.
The NLO $\wwbb$ predictions involve factorisable corrections
to doubly-resonant diagrams, which provide the off-shell extension 
of NLO corrections in $\ttb$ approximation~\refeq{eq:fact}. In addition, there are
non-factorisable corrections, where $\ttb$ production and decay parts of the 
process are connected via exchange of QCD partons, and
NLO corrections to singly- and non-resonant topologies.
Further technical aspects are discussed in the original publications~\cite{Melnikov:2009dn,Denner:2010jp}.

\begin{figure}[h]
\begin{center}
{\includegraphics[bb=150 650 400 710, width=.68\textwidth]{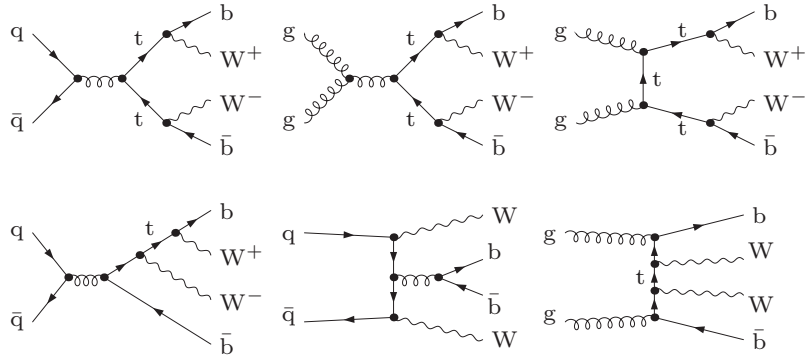}}
\end{center}
\vspace*{4em}
\caption{Representative LO diagrams of doubly-resonant (upper line), 
singly-resonant (first diagram in lower line), and non-resonant type (last two diagrams in lower line).}
\label{fig:treegraphs} 
\end{figure}

\subsection{NUMERICAL RESULTS}

\subsubsection{Input parameters and setup}

In the following we compare $\ttb$ and $\wwbb$  predictions for 
${{\PW^+(\to\nu_\Pe\Pe^+)\PW^-(\to\mu^-{\bar\nu}_\mu)\Pb\bar\Pb}}$
production at the Tevatron 
(\eqintext{$\Pp\bar\Pp$} collisions at 
1.96\,TeV) and the LHC (\eqintext{$\Pp\Pp$} collisions at 7 and 14\,TeV).
These results are based on the same input parameters and cuts as in Ref.~\cite{Denner:2010jp}.
In NLO\,(LO) QCD we employ MSTW2008NLO\,(LO)
parton distributions~\cite{Martin:2009iq} and describe the running of the strong coupling constant $\alpha_\rS$
with two-loop\,(one-loop) accuracy, including five active flavours.
Contributions induced by the strongly suppressed bottom-quark density are neglected.
For the gauge-boson and top-quark masses we use \eqintext{$\Mt=172\,\GeV$},
\eqintext{$\MW=80.399\,\GeV$}, and \eqintext{$\MZ=91.1876\,\GeV$}.  The masses of all
other quarks, including b-quarks, are neglected.  In view of the
negligibly small Higgs-mass dependence we adopt the \eqintext{$\MH\to\infty$} limit,
i.e.~we omit diagrams involving Higgs bosons.  The electroweak couplings
are derived from the Fermi constant
\eqintext{$\GF=1.16637\times10^{-5}\,\GeV^{-2}$} in the $G_\mu$-scheme, where the
sine of the mixing angle and the electromagnetic coupling read
\eqintext{$\sw^2=1-\MW^2/\MZ^2$} and
\eqintext{$\alpha=\sqrt{2}\GF\/\MW^2\sw^2/\pi$}.
For consistency, we perform the LO and NLO calculations using the top-quark
widths \eqintext{$\Gamma_{\Pt,\LO}=1.4655\,\GeV$} and
\eqintext{$\Gamma_{\Pt,\NLO}=1.3376\,\GeV$}~\cite{Jezabek:1988iv}, respectively.
Since the leptonic W-boson decay does not receive NLO QCD corrections
we employ the NLO W-boson width
\eqintext{$\Gamma_\PW=2.0997\,\GeV$} everywhere.

Final-state quarks and gluons with pseudo-rapidity \eqintext{$|\eta|<5$} 
are converted into infrared-safe jets using the anti-$\kT$ algorithm~\cite{Cacciari:2008gp}.
For the Tevatron\,(LHC) we set the jet-algorithm parameter \eqintext{$R=0.4\,(0.5)$} and apply the
transverse-momentum and pseudo-rapidity cuts
\eqintext{$p_{\mathrm{T,b-jet}}>20\,(30)\,\GeV$}, \eqintext{$|\eta_{\mathrm{b-jet}}|<2.5$}. 
Moreover, we require a missing transverse momentum of
\eqintext{$p_{\mathrm{T,miss}}>25\,(20)\,\GeV$} and charged leptons with
\eqintext{$p_{\mathrm{T},l}>20\,\GeV$} and \eqintext{$|\eta_{l}|<2.5$}.

For the renormalisation and factorisation scales we adopt the central value $\mu=\Mt$
and study factor-two  variations of 
{$\mu=\mu_{\mathrm{ren}}=\mu_{\mathrm{fact}}$}, i.e.~we compare predictions at 
$\mu/\Mt=0.5,1,2$. The scale variations are applied also to 
$\Gamma_{\Pt,\NLO}$, but not to $\Gamma_\PW$.

\subsubsection{Integrated cross section}
Results for the integrated $\lvlvbb$ cross sections and scale uncertainties at the Tevatron and the LHC  are 
reported in Table~\ref{tab:integrated}.
While the $\sigma_\wwbb$ results for Tevatron and LHC at 7\,TeV correspond to those of 
Ref.~\cite{Denner:2010jp}\footnote{To be more precise, 
in Ref.~\cite{Denner:2010jp}  the scale dependence was assessed using a fixed
$\Gamma_\Pt$ input, while here we take into account the $\mu$-dependence of $\Gamma_{\Pt,\NLO}$, which 
results into slightly different $\sigma_\wwbb$ variations at NLO.}, 
the ones for LHC at 14\,TeV as well as all $\sigma_\ttb$ predictions are new.
Comparing all $\wwbb$ and  $\ttb$ predictions we find that
finite-top-width effects never exceed one percent, both in LO and NLO.
The statistical precision of the calculations permits 
us to assess the error of the NWA, $\sigma_\ttb/\sigma_\wwbb\,-1$,  with an accuracy of 1--3 permille.
At the Tevatron, the NWA overestimates the $\wwbb$ cross section 
by an amount very close to $\Gamma_\Pt/\Mt\simeq 0.9\%$, both in LO and NLO.
The error of the NWA at the 7(14)\,TeV LHC ranges between 4 and 8 permille.
As shown in the last column of Table~\ref{tab:integrated}, these finite-width effects
are in very good agreement  with the results of the $\Gamma_\Pt\to 0$ extrapolation in Ref.~\cite{Denner:2010jp}.
Similar results can be found also in Ref.~\cite{Bevilacqua:2010qb}.

\begin{table}
$$
\begin{array}{c@{\quad}c@{\quad}c@{\quad}c@{\quad}c@{\quad}c@{\quad}c}
$Collider$ & \sqrt{s}\;$[TeV]$  &$approx.$
                            & \sigma_\ttb \; $[fb]$        &  \sigma_\wwbb \; $[fb]$     & {\sigma_\ttb}/{\sigma_\wwbb} \,-1  & $Ref.~\cite{Denner:2010jp}$ 
\\[2mm] \hline\hline \\[-3mm]
$Tevatron$ & 1.96 & $LO$  &   44.691(8)^{+19.81}_{-12.58}   &  44.310(3)^{+19.68}_{-12.49}   & {}+0.861(19)\% & {}+0.8\%
\\[2mm] \hline \\[-3mm]
%
           &        & $NLO$ &   42.16(3)^{+0.00}_{-2.91}      &  41.75(5)^{+0.00}_{-2.63}     & {}+0.98(14)\% & {}+0.9\%
\\[2mm] \hline\hline \\[-3mm]
%
$LHC\,$    & 7    & $LO$  &   659.5(1)^{+261.8}_{-173.1}    &  662.35(4)^{+263.4}_{-174.1}   & {}-0.431(16)\% & {}-0.4\%
\\[2mm] \hline \\[-3mm]
           &        & $NLO$ &  837(2)^{+42}_{-87}           &  840(2)^{+41}_{-87}           & {}-0.41(31)\% & {}-0.2\%
\\[2mm] \hline\hline \\[-3mm]
%
$LHC$      & 14   & $LO$  & 3306.3(1)^{+1086.8}_{-763.6}    &  3334.6(2)^{+1098.5}_{-771.2}     & {}-0.849(7)\%   & ---
\\[2mm] \hline \\[-3mm]
%
           &        & $NLO$ & 4253(3)^{+282}_{-404}    &  4286(7)^{+283}_{-407}     & {}-0.77(19)\%   & ---
\\[2mm] \hline\hline
%
\end{array}
$$
\caption{
Integrated $\lvlvbb$ 
cross section in narrow-with approximation ($\sigma_\ttb$)
and including finite-top-width effects ($\sigma_\wwbb$).
The relative error of the narrow-width approximation (sixth column) is compared to the 
prediction of Ref.~\cite{Denner:2010jp} (seventh column).
Factor-two scale variations in $\sigma_\ttb$ and $\sigma_\wwbb$
  are shown as sub- and super-scripts, while statistical errors
are given in parenthesis.
}
\label{tab:integrated}
\end{table}

\subsubsection{Differential distributions} 
The small finite-width
corrections to the integrated cross section demonstrate that---in presence
of standard LHC and Tevatron cuts---the NWA provides a fairly accurate
description of inclusive $\wwbb$ production.  It is thus interesting to
investigate to which extent this conclusion applies to the 
various phenomenologically important
regions of the $\wwbb$ phase space.  To this end we have compared $\ttb$ and
$\wwbb$ predictions for a few differential observables that are relevant for
top-pair production, either as signal or as background to Higgs production or new physics.
Note that we refrain from selecting kinematic
variables like the top-quark invariant mass or imposing 
cuts of type $M_{\PW\Pb}>200\,\GeV$,
which would lead to obvious 
enhancements of non-resonant contributions.

In Figs.~\ref{fig:ptmax_lhc7}--\ref{fig:meb_lhc7}
we present predictions for some invariant-mass and transverse-momentum 
distributions, restricting ourselves to the case of the 7\,TeV LHC.
For each observable we display  $\ttb$ (dashed curves) and $\wwbb$ (solid curves) results
in LO (blue) and NLO (red) approximation. Absolute predictions (left plots) are complemented by
the ratios ($\rd\sigma_\LO-\rd\sigma_\NLO)/\rd\sigma_\NLO$ (upper right plots)
and ($\rd\sigma_\ttb-\rd\sigma_\wwbb)/\rd\sigma_\wwbb$ (lower right plots),
which indicate the relative error of LO and narrow-width approximations
w.r.t.~the best predictions, i.e.~NLO and $\wwbb$. 
\begin{figure}[h]
\begin{center}
{\includegraphics[width=1.0\textwidth]
{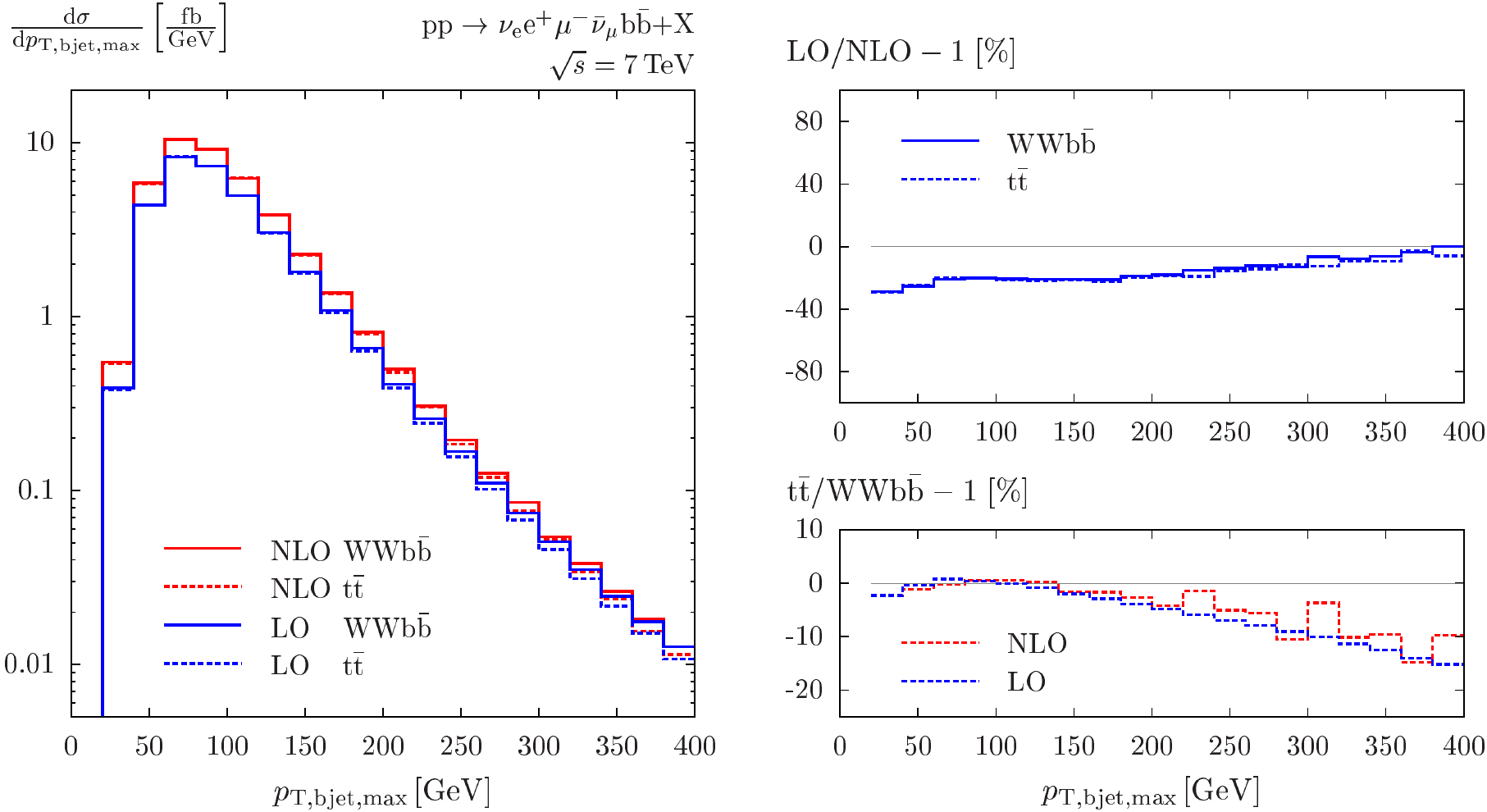}
}
\caption{
Distribution in the transverse momentum of the harder b-jet at the 7\,TeV LHC: 
 LO (blue) and NLO (red) predictions in narrow-width approximation ($\ttb$, dashed) and including finite-top-width effects ($\wwbb$, solid). 
 Plotted are absolute predictions (left) and relative deviations of LO (upper-right) and narrow-width (lower-right) approximations w.r.t.~NLO and $\wwbb$ predictions, respectively.}
\label{fig:ptmax_lhc7}
\end{center}
\end{figure}

The transverse-momentum distribution of the harder b-jet is shown in Fig.~\ref{fig:ptmax_lhc7}.
In the range below 200\,GeV, which contains the bulk of the cross section,
the NLO and finite-width corrections behave similarly as for the integrated cross section:
LO predictions deviate from NLO ones by about $-20\%$, and the error of the NWA
ranges between $+1$ and $-4\%$. Finite-width effects tend to increase with $\pT$ and reach the
10\% level around 300\,GeV. Within the entire $\pT$ range 
the LO/NLO ratios resulting from the $\ttb$ and $\wwbb$ calculations are almost equal. Equivalently, 
 we find the same  $\rd\sigma_\ttb/\rd\sigma_\wwbb$ ratios in LO and NLO.

\begin{figure}
\begin{center}
{\includegraphics[width=1.0\textwidth]
{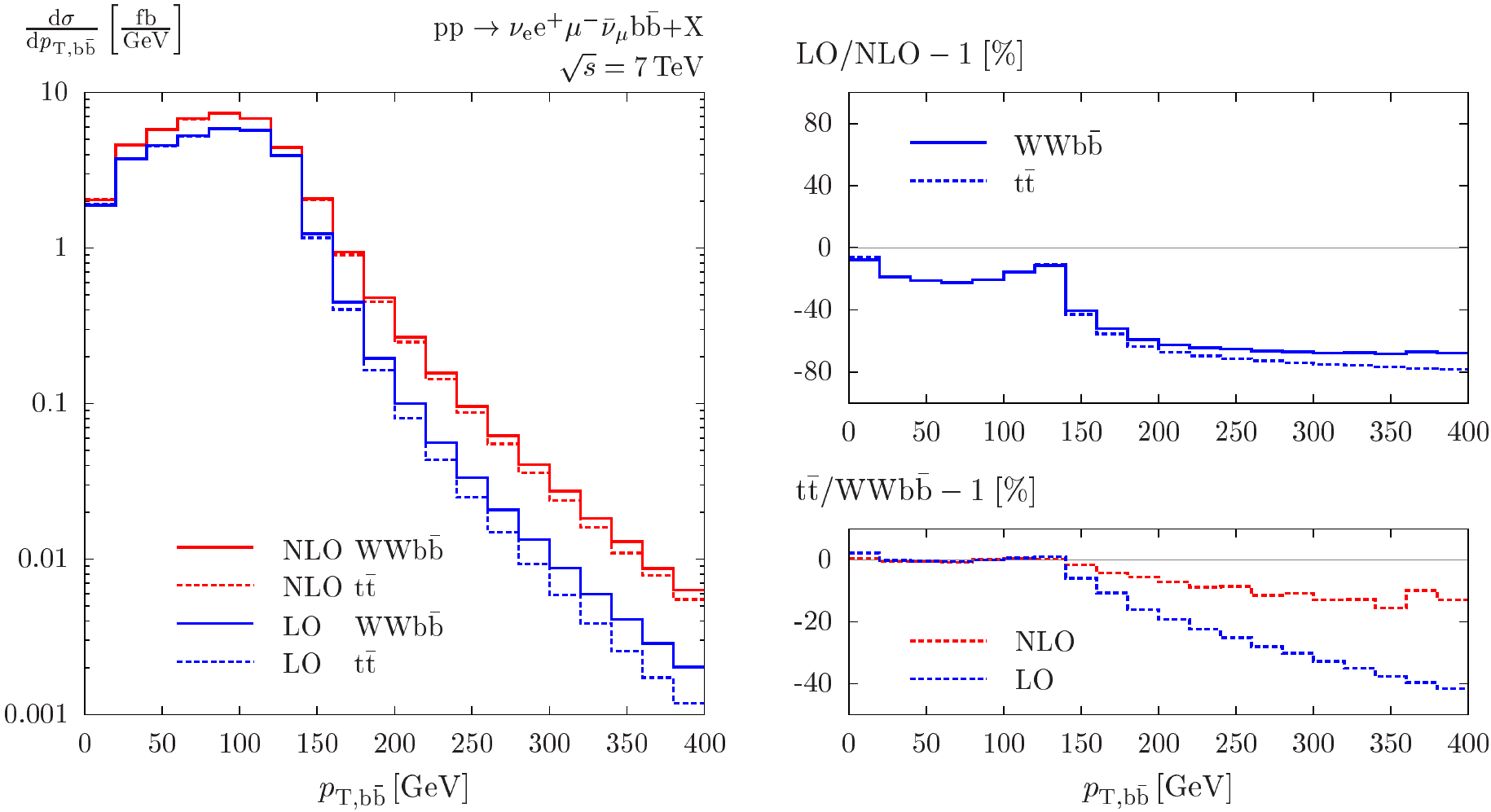}
}
\caption{Distribution in the transverse momentum of the $\Pb\bar\Pb$ di-jet system at the 7\,TeV LHC: 
 LO (blue) and NLO (red) predictions in narrow-width approximation ($\ttb$, dashed) and including finite-top-width effects ($\wwbb$, solid). 
 Plotted are absolute predictions (left) and relative deviations of LO (upper-right) and narrow-width (lower-right) approximations w.r.t.~NLO and $\wwbb$ predictions, respectively.}
\label{fig:ptbb_lhc7}
\end{center}
\end{figure}

In Fig.~\ref{fig:ptbb_lhc7} we show the transverse-momentum distribution of the
$\Pb\bar\Pb$ di-jet system. This kinematic variable plays an important role in boosted-Higgs
searches with a large $\ttb$ background.  In particular, the
strategy proposed in Ref.~\cite{Butterworth:2008iy} to extract a $\Pp\Pp\to
\PH(\to\Pb\bar\Pb)\PW$ signal at the LHC is based on the selection of boosted
$\PH\to \Pb\bar\Pb$ candidates with $p_{\rT,\Pb\bar\Pb}>200\,\GeV$,
which permits to reduce $\ttb$ contamination (and other backgrounds) in a very efficient way.
As can be seen from Fig.~\ref{fig:ptbb_lhc7},  
the suppression of $\ttb$ production is indeed particularly strong 
at $p_{\rT,\Pb\bar\Pb}\gsim 150\,\GeV$.
This is due to kinematic constraints that characterise the LO and narrow-width 
approximations: in order to acquire $p_{\rT,\Pb}> (\Mt^2-\MW^2)/(2\Mt)\simeq 65\,\GeV$
$\Pb$-quarks need to be boosted via the $\pT$ of their parent (anti)top quarks,
and the fact that  top and antitop quarks 
have opposite transverse momenta (at LO) makes it difficult to generate a 
$\Pb\bar\Pb$ system with high $\pT$. The NLO and finite-width corrections undergo less stringent kinematic restrictions,
resulting into a significant enhancement of $\wwbb$ events at large $p_{\rT,\Pb\bar\Pb}$.
This is clearly reflected in the differences between the various curves in the left plot  
of Fig.~\ref{fig:ptbb_lhc7}.
The most pronounced effect comes from the NLO corrections, 
where the $\ttb$ system  can acquire large transverse momentum by recoiling 
against extra jet radiation. As indicated by the right-upper plot,
the NLO correction represents 50--80\% of the cross section at high $\pT$,
corresponding to a huge $K$-factor of 2--5.
Finite-width effects (lower-right plot) lead to a further significant, although less dramatic, enhancement;
for example, non-resonant topologies can lead to direct ${\Pb\bar\Pb}$ production via high-$\pT$  gluons that
recoil against 
$\PW^+\PW^-$ pairs.
For $p_{\rT,\Pb\bar\Pb}>200\,\GeV$, we find that 20--40\% of the LO $\wwbb$ cross section is due to finite-width contributions,
while this fraction decreases to 7--15\% at NLO. 
This reduction is related to the dominance of the jet-emission contribution, 
which we expect to be rather well described by the NWA.
On the other hand, an optimal suppression of the $\ttb$ background will require 
a very tight jet-veto~\cite{Butterworth:2008iy},
and in this case we expect finite-width corrections to the NLO $\ttb$ predictions 
to be as large as in LO.

\begin{figure}
\begin{center}
{\includegraphics[width=1.0\textwidth]
{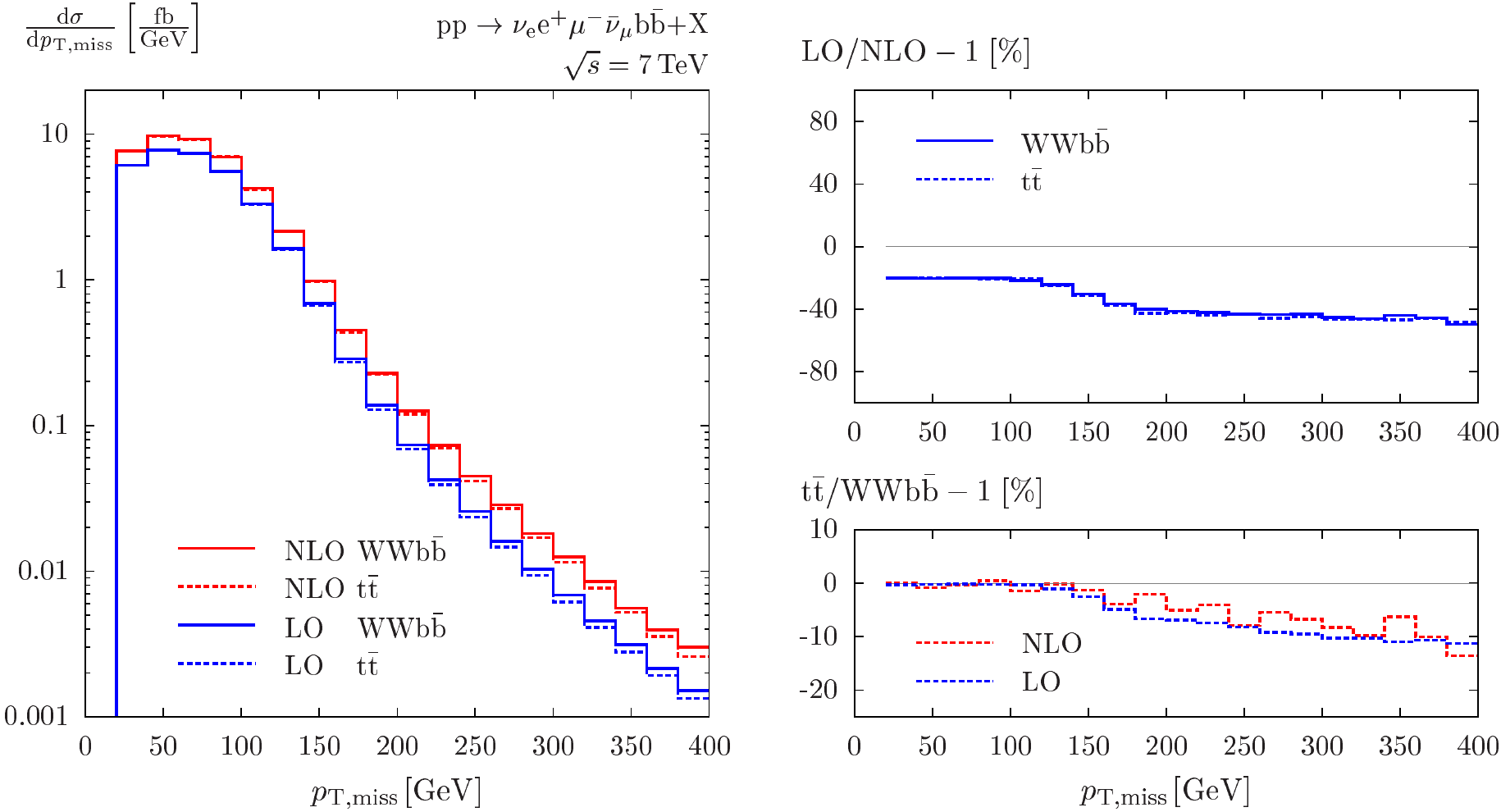}
}
\caption{Distribution in the missing transverse momentum at the 7\,TeV LHC: 
 LO (blue) and NLO (red) predictions in narrow-width approximation ($\ttb$, dashed) and including finite-top-width effects ($\wwbb$, solid). 
 Plotted are absolute predictions (left) and relative deviations of LO (upper-right) and narrow-width (lower-right) approximations w.r.t.~NLO and $\wwbb$ predictions, respectively.}
\label{fig:ptmiss_lhc7}
\end{center}
\end{figure}

The distribution in the  missing transverse momentum, i.e.~the vector sum of the $\nu_e$ and $\bar\nu_\mu$ transverse momenta, is displayed
in  Fig.~\ref{fig:ptmiss_lhc7}. This distribution is relevant for new-physics
searches based on 
missing transverse energy plus jets and leptons. Its tail
features a qualitatively similar behaviour as in the case of $p_{\rT,\Pb\bar\Pb}$,
due to analogous kinematic constraints.
However, in the case of $p_\mathrm{T,miss}$
the corrections are less pronounced:
the NLO correction does not exceed 40--50\% of the full prediction,
and finite-width contributions stay below roughly 10\%.

\begin{figure}[h]
\begin{center}
{\includegraphics[width=1.0\textwidth]
{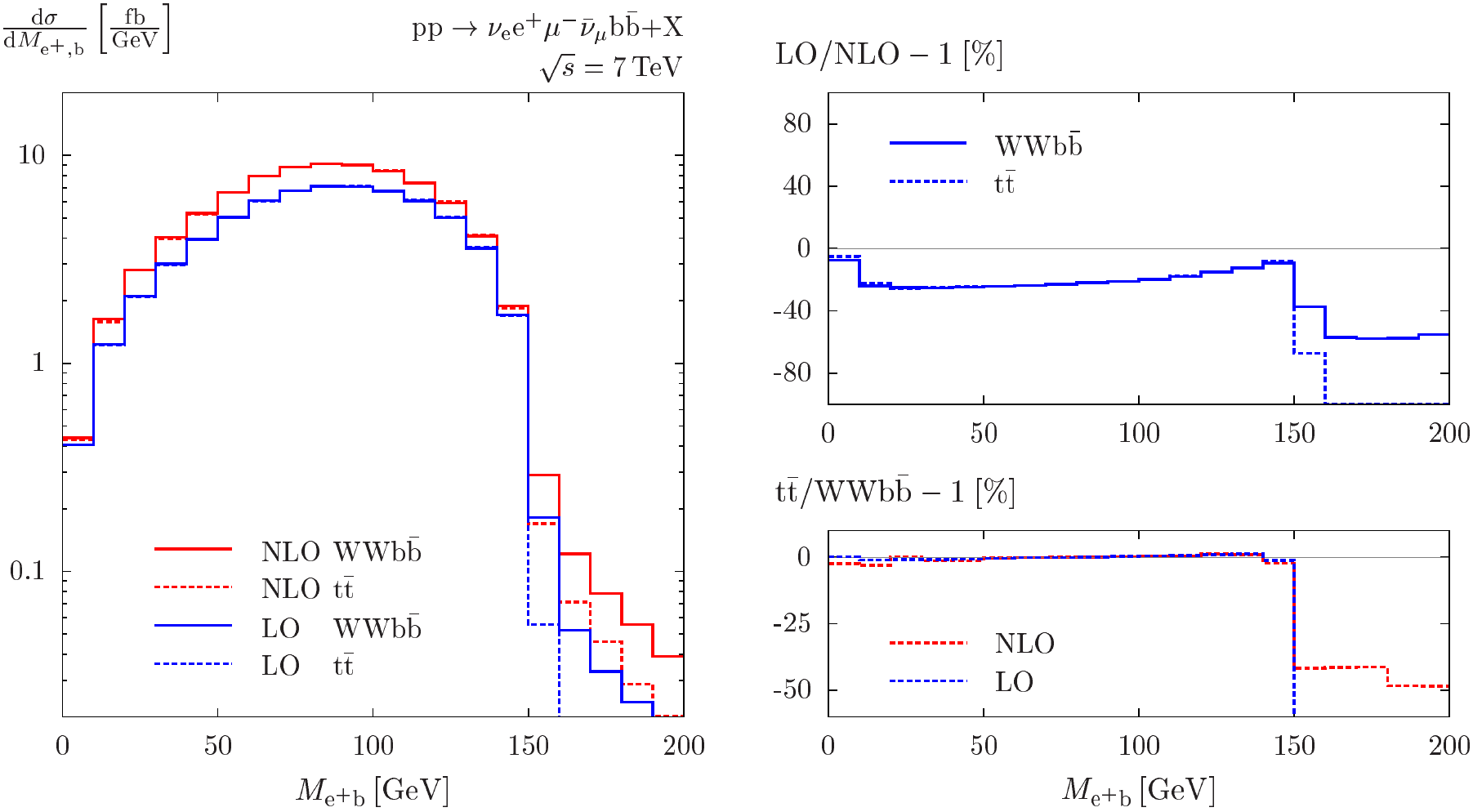}
}
\caption{Distribution in the invariant mass of the positron--b-jet system (as defined in the text) at the 7\,TeV LHC: 
 LO (blue) and NLO (red) predictions in narrow-width approximation ($\ttb$, dashed) and including finite-top-width effects ($\wwbb$, solid). 
 Plotted are absolute predictions (left) and relative deviations of LO (upper-right) and narrow-width (lower-right) approximations w.r.t.~NLO and $\wwbb$ predictions, respectively.}
\label{fig:meb_lhc7}
\end{center}
\end{figure}

Figure~\ref{fig:meb_lhc7} displays 
the distribution in the invariant mass of the positron
and a b-jet, i.e.~the visible products of a top-quark decay.
More precisely, assuming that the charge of the $\Pb$-jet is not known,
the $\Pe^+\Pb$ pair is built by selecting the $\Pb$-jet that yields the smallest invariant mass.\footnote{
Note that the $M_{\Pe^+\Pb}$ distribution in Ref.~\cite{Denner:2010jp}
was defined by selecting (based on the Monte Carlo truth) 
jets that involve negatively charged b-quarks, such that the
${\Pe^+\Pb}$ pairs are consistent with top-quark decays.}
In narrow-width and LO approximation this kinematic quantity is characterised
by a sharp upper bound, \eqintext{$M^2_{\Pe^+\Pb}< \Mt^2-\MW^2\simeq(152\,\GeV)^2$}, which renders it
very sensitive to the top-quark mass.  The value of $\Mt$ can be
extracted with high precision using, for instance,  the invariant-mass distribution of a
positron and a $J/\psi$ from a $B$-meson decay~\cite{Kharchilava:1999yj,Biswas:2010sa}, an
observable that is closely related to $M_{\Pe^+\Pb}$.  
In the region below the kinematic bound, the
NLO corrections to $M_{\Pe^+\Pb}$ vary between 5--30\%, and the impact of the NLO
shape distortion on a precision $\Mt$-measurement is certainly significant.
For $M_{\Pe^+\Pb}<150\,\GeV$, the NWA agrees with the $\wwbb$ predictions at the 1\% level or better. In contrast,
in the vicinity of the kinematic bound the impact of finite-width (and NLO) corrections becomes clearly more important,
giving rise to a tail that extends above $M^2_{\Pe^+\Pb}=\Mt^2-\MW^2$.
The resulting contribution to the total cross section is fairly small,
but the impact of such finite-width effects on the top-mass measurement might be non-negligible,
given the high $\Mt$-sensitivity of the $M^2_{\Pe^+\Pb}\simeq \Mt^2-\MW^2$ region.

\subsection{CONCLUSIONS}

Based on recent NLO QCD calculations,
we have presented a systematic comparison of 
top-pair production 
and decay in narrow-top-width approximation, $\Pp\Pp\to \ttb\to\wwbb$,
against the complete $\Pp\Pp\to \wwbb$ process, which 
involves finite-top-width effects of non-resonant and off-shell type.

At the Tevatron and the LHC (7 and 14\,TeV), finite-top-width contributions
to the integrated cross section  (in the di-lepton channel) 
turn out not to exceed
one percent. This confirms previous estimates based on the
$\Gamma_\Pt\to 0$ extrapolation of $\Pp\Pp\to \wwbb$ predictions.
At the 7\,TeV LHC, we also investigated 
differential observables that are relevant either for 
top-pair production as a signal or as a background in Higgs or new-physics searches.
In the case of the b-jet transverse momentum and $p_{\rT,\miss}$ distributions,
finite-width effects remain very small over a large kinematic range
and reach the 10\% level only around 300\,GeV.
In contrast, the $\pT$-distribution of the $\Pb\bar\Pb$ di-jet system
receives $\Gamma_\Pt$-corrections beyond \mbox{20--30\%} 
for $p_{\rT,\Pb\bar\Pb}\gsim 200\,\GeV$, a kinematic region that plays an important role in 
$\Pp\Pp\to \PH(\to\Pb\bar\Pb)\PW$ searches based on boosted
$\PH\to \Pb\bar\Pb$ candidates.
For the  lepton--b-jet invariant-mass distribution---an observable that provides high sensitivity to the top-quark mass---finite-width corrections
do not exceed
one percent in the range that contains the bulk of the cross section, 
but become more sizable in the region of highest $\Mt$-sensitivity.
This motivates more detailed studies of finite-width effects 
in the context of high-precision $\Mt$-measurements at the LHC.
The results of this investigation of finite-width effects in $\ttb$
production give also useful insights into possible limitations of treating {\it
associated} top-pair production processes in the narrow-width approximation,
since NLO calculations for $\Pp\Pp\to\wwbb\mathrm{j}$ and similar reactions will not be available too soon.

\subsection*{ACKNOWLEDGMENTS}
We thank K.~Melnikov for discussions and comments on the manuscript.
S.P.~would like to acknowledge financial support from the SNSF.
M.S.~is grateful for support from the Director's Fellowship of Argonne National Laboratory under DOE grant DE-AC02-06CD11357.
The presented results of the calculation described in Ref.~\cite{Melnikov:2009dn} were performed on the Homewood High Performance Cluster of Johns Hopkins
University supported by grant NSF-OCI-108849.




}


}

\section[Strong and Smooth Ordering in Antenna Showers]
{Strong and Smooth Ordering in Antenna Showers \protect\footnote{Contributed by: J.~J. ~Lopez-Villarejo,
P.~Skands}}
{\graphicspath{{skands/}}






\begin{abstract}
We comment on strong and smooth ordering in antenna showers, and 
extend the definition of smooth ordering to include the case of $g\to q\bar{q}$
splittings. We define three observables in hadronic $Z$ decays that
can be used to probe the subleading properties of shower models.
\end{abstract}

\subsection{INTRODUCTION}

Traditional parton showers are based on collinear factorization,
and the shower evolution proceeds via $1\rightarrow2$
branchings, on which additional constraints have to be imposed to ensure
momentum conservation and QCD coherence (see \cite{Buckley:2011ms}).
Antenna showers are instead based on momentum-conserving and
intrinsically coherent $2\to 3$ branchings, as pioneered by
Ariadne~\cite{Gustafson:1987rq,Lonnblad:1992tz}. This
note concerns the antenna shower implementation in the 
Vincia code~\cite{Giele:2007di}, a plug-in to Pythia
8~\cite{Sjostrand:2007gs}, though we emphasize that the
notion of smooth ordering  could  be
applied to other shower types as well.

In leading-logarithmic (LL) antenna showers, the fundamental step
is a Lorentz-invariant $2\to 3$ branching process by which two
on-shell ``parent'' partons are replaced by three on-shell ``daughter''
partons.
This $2\to 3$ process makes use of three ingredients~\cite{Giele:2011cb}:
\begin{enumerate}
\item An \emph{antenna function} that captures the leading tree-level
  singularities of QCD matrix elements. 
\item An antenna \emph{phase space} --- an exact, momentum-conserving and 
  Lorentz-invariant factorization of the pre- and post-branching phase
  spaces. 
\item A \emph{kinematics map}, specifying how the global orientation
  of the 
post-branching momenta are related to the pre-branching ones.  
\end{enumerate}

Antenna showers come in two varieties: global and sector. The
two kinds differ in how the  
collinear singularities of gluons are partitioned among neighboring
antennae, see~\cite{Bern:2008ss,LopezVillarejo:2011ap}. Here, we shall 
only be concerned with the global
type~\cite{Gustafson:1987rq, Winter:2007ye, Giele:2007di,
  Giele:2011cb},
in which the gluon-collinear singularity 
is partitioned such that two neigbouring antennae each contain
``half'' of it; their sum reproduces the full singularity.

If each antenna in a global shower is allowed to emit in its full
phase space, the
resulting shower evolution 
amounts to an incoherent addition of  independently radiating
dipoles.  
This tends to overcount regions in which several 
dipole terms contribute at the same level, i.e., in regions where
dipole-dipole interference effects (or, equivalently, multipole
effects) are important \cite{Skands:2009tb,LopezVillarejo:2011ap}. 
The situation is
analogous to, though less severe than, the case of traditional parton
showers with virtuality-ordering~\cite{Bengtsson:1986et}, 
which represent an incoherent addition of independent
monopoles. In parton/monopole showers, multi-parton interference
effects for soft radiation can be taken into account by 
the requirement of angular ordering  \cite{Marchesini:1983bm}, while
in 
dipole/antenna showers, typically a measure of transverse
momentum is used, such as 
\begin{equation}
p_{\perp A}^2 = \frac{s_{ij}s_{jk}}{s_{IK}}~,
\end{equation}
for a branching $IK \to ijk$, with $s_{ab} \equiv 2p_a\cdot p_b =
(p_a+p_b)^2$ for massless partons. Some alternative possibilities are
compared in \cite{Giele:2011cb}.  

\subsection{STRONG AND SMOOTH ORDERING}

In a strongly-ordered shower, each consecutive branching is required
to occur at a lower scale in the evolution variable than that of the previous
one: $Q_{n+1} < Q_{n}$. This can be represented as 
a step function in the evolution variable, multiplying the branching
kernels. In a smoothly-ordered shower~\cite{Giele:2011cb}, 
the step function is replaced by a smooth dampening factor designed to 
leave the soft and collinear limits unchanged while suppressing
radiation at scales above $\sim Q_n$.
Specifically, for evolution in $p_\perp$, we replace the
strong-ordering condition as follows, 
\begin{equation}
\Theta(\hat{p}_\perp - p_\perp) \, P_{LL}
\ \ \to \ \ P_{imp} \, P_{LL} \  \equiv \ \frac{\hat{p}_{\perp}^2}{\hat{p}_{\perp}^2 + p_{\perp}^2} \, P_{LL}~,
\label{eq:unord}
\end{equation}
where $\hat{p}_{\perp}$ characterizes the scale of the previous
branching\footnote{We take $\hat{p}_\perp$ to be the smallest
  $p_{\perp}$ scale among all the color-connected partons in
  the parent configuration, i.e., a 
global measure of the ``current'' $p_{\perp}$ scale of that
topology. This makes the shower a true Markov chain (i.e.,
history-independent) which has beneficial
consequences for matching to matrix elements \cite{Giele:2011cb}.}, 
$p_{\perp}$ is the scale of the emission under
consideration, and $P_{LL}$ is an ordinary LL 
shower kernel, which in our case is represented by a gluon-emission 
antenna function. (We return to the case of $g\to q\bar{q}$ below.) 

Thus, for $p_\perp \ll \hat{p}_\perp$ (the strongly-ordered
limit) the smooth-ordering factor $P_{imp}$ tends to unity, while for 
$p_\perp \sim \hat{p}_\perp$ (the ordering threshold) it tends to
$1/2$, and finally for $p_\perp \gg \hat{p}_\perp$ (highly unordered),
it tends to zero  $\propto \hat{p}_\perp^2/p_\perp^{2}$. Note that, since $P_{LL}$ is
likewise $\propto 1/p_\perp^2$, the
net effect of the suppression factor is to modify the behavior of the
splitting kernel from $1/p_{\perp}^2$ in the strongly-ordered limits
to $1/p_{\perp}^4$ for highly unordered branchings, similar to what
has been studied for initial-state parton showers in
\cite{Corke:2010zj}; above the strong-ordering threshold, the
branching probability is  
explicitly suppressed beyond LL.

For a rigourous interpretation of the $P_{imp}$ factor one would have
to analyze the $2\to 4$ antennae \cite{GehrmannDeRidder:2005cm} 
and check that the combination of two
$2 \to 3$ antennae times 
this factor does indeed reproduce subleading aspects of the full $2\to
4$ function. In the absence of such a study, one may still physically
interpret its purpose in the following way: the LL antenna functions
are derived assuming the outgoing partons/jets
to be massless. This is a good approximation if the virtuality that
they can acquire (through further showering) 
is restricted by the strong-ordering threshold. When allowing
unordered branchings, however, the corresponding Feynman diagrams
contain highly off-shell propagators, which the $P_{imp}$ factor
attempts to mimic by introducing an ``effective mass'' in the
denominator of eq.~(\ref{eq:unord}). 
 
For gluon emissions, it was shown in  \cite{Giele:2011cb} that the
 smooth-ordering condition does lead to a systematic improvement in the
 shower. Since it simultaneously guarantees a complete phase-space coverage
(contrary to the case for strong ordering 
\cite{Andersson:1991he,Giele:2011cb}), it is the default option
in Vincia. 

Antenna showers including $g\to q\bar{q}$ splittings 
were studied in \cite{Ridder:2011dm}, 
in which evolution in $m_{q\bar{q}}^2$ was introduced for such
branchings. This is based on the
observation \cite{Seymour:1994ca} that the scale controlling the
divergences of $g\to q\bar{q}$ splittings is 
 the invariant mass of the pair, not 
 its $p_\perp$. By analogy with the physical interpretation given to
 the $P_{imp}$ factor for gluon emissions above, it therefore seems
 well-motivated to study a 
 ``generalized'' $P_{imp}$ 
factor where each scale depends on whether we are dealing with a
gluon or a quark: 
\begin{equation}
 {P}_{imp}  = \frac{\hat{Q}_E^2}{\hat{Q}_{E}^2 + Q_{E}^2} ,
\label{eq:generalized pimp}
\end{equation}
where $Q_E$ is the evolution variable: $p_\perp$ for gluons and invariant mass for quark-antiquark pairs.

\begin{figure}[t]
\centering
\includegraphics*[scale=0.6]{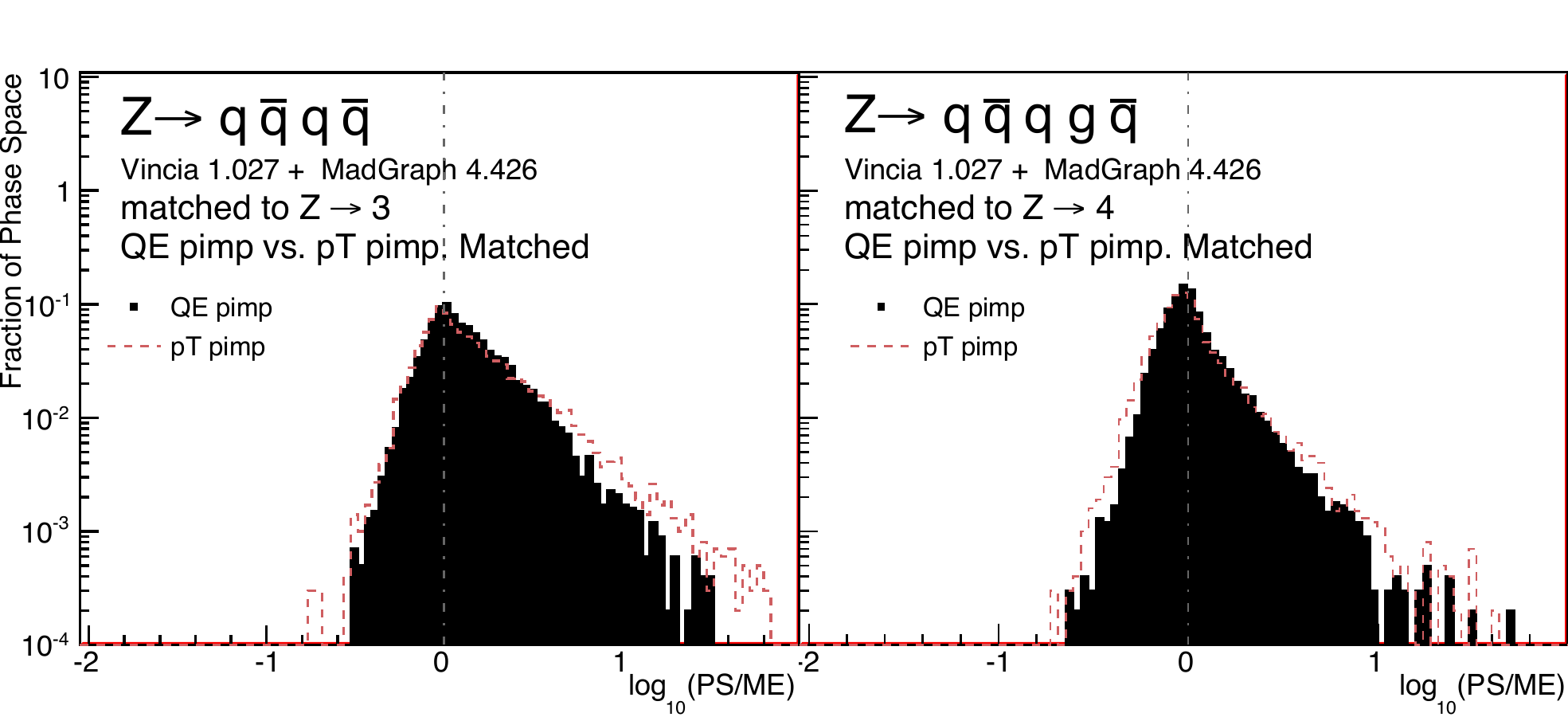}
\caption{Comparison between generalized $\tilde{P}_{imp}$ and ``old" $P_{imp}$ factor in the global shower approximations to LO matrix elements, for 
  processes involving a $g\to q\bar{q}$ splitting. {\sl Left:} $Z\to
  q\bar{q}'q'\bar{q}$. {\sl Right:} $Z\to q\bar{q}'q'g\bar{q}$. 
  In both cases, GKS matching to the LO matrix element for
  the preceding multiplicity ($Z\to 3$ and $Z\to 4$, respectively) 
  has been included, and the Ariadne factor
  was applied to $g\to q\bar{q}$ splittings. }
\label{fig:global_mixedpimpVSoldpimp_matched}
\end{figure}
We can assess the improvement that this produces in the shower by plotting 
the ratio of the shower approximation vs.\ the LO matrix element 
for $Z\to q\bar{q'}q'\bar{q}$ and $Z\to q\bar{q}'q'g\bar{q}$. This is
shown in fig.~\ref{fig:global_mixedpimpVSoldpimp_matched}, where the
histograms represent the distribution of  
  $\log_{10}(\mathrm{PS}/\mathrm{ME})$ in a flat phase-space 
  scan, normalized to unity (i.e., the same type of distributions that were
  shown in \cite{Giele:2011cb,Ridder:2011dm,LopezVillarejo:2011ap}). 
  Points to the left of zero are
  undercounted by the shower approximation, while points to the right
  are overcounted. Although the agreement is by no means perfect, 
we do observe a slight improvement in the shower approximation when
the $P_{imp}$ factor is defined in terms of $Q_E$ (solid black
histogram), as compared to the definition used previously (dashed
histogram). Note that we used the so-called Ariadne factor in the
shower approximation for all cases, see \cite{Ridder:2011dm}, and that
the distributions were made including GKS matching to the preceding
multiplicities \cite{Giele:2011cb}.

\subsection{SENSITIVE OBSERVABLES IN HADRONIC Z DECAYS}
The properties of shower and matrix-element matching 
algorithms are coming under increasing scrutiny, not least due to the
desire of achieving reliable descriptions of jet production and
jet properties, such as jet substructure, for signal and background
estimates at the LHC. 

For final-state radiation, i.e., jet broadening and jet splitting, 
hadronic $Z$ decays are the main
reference, with a large set of 
events shapes and jet resolutions/rates being used to
constrain and tune shower algorithms (see, e.g.,
\cite{Buckley:2010ar,Buckley:2011ms}). However, in the logarithmically dominated
regions, these observables are typically dominated by leading logs,
and are well described by all coherent and reasonably well-tuned
shower algorithms on the market. 
In order to probe the subleading properties in a more dedicated way,
we have found the following three simple observables useful, each
designed to isolate a specific aspect.

We consider hadronic $Z$ events (photon ISR is switched off, and
matching beyond 3 jets is switched off for the strongly-ordered showers)
and use the $k_T$ clustering algorithm \cite{Cacciari:2005hq} to cluster
all events back to two jets. The $3\to 2$ clustering scale is
denoted $y_{23} = k_{T3}^2/m_Z^2$, and so on for 
higher jet numbers. We require all $y_{ij}$ entering in the
observables below to be greater than $0.005$, to 
remove  contamination from $B$ decays and lower scales. 
Since the original topology contains two jets, we
also keep track of which ``side'' each clustering happens on. 
Strong ordering corresponds to $y_{23}\gg y_{34} \gg \ldots$, while
events with, e.g., $y_{34}\sim y_{23}$  should be more
sensitive to the ordering condition and to the effective $1\to 3$
spliting kernels. 

\begin{figure}[t]
\centering
\includegraphics*[scale=0.36]{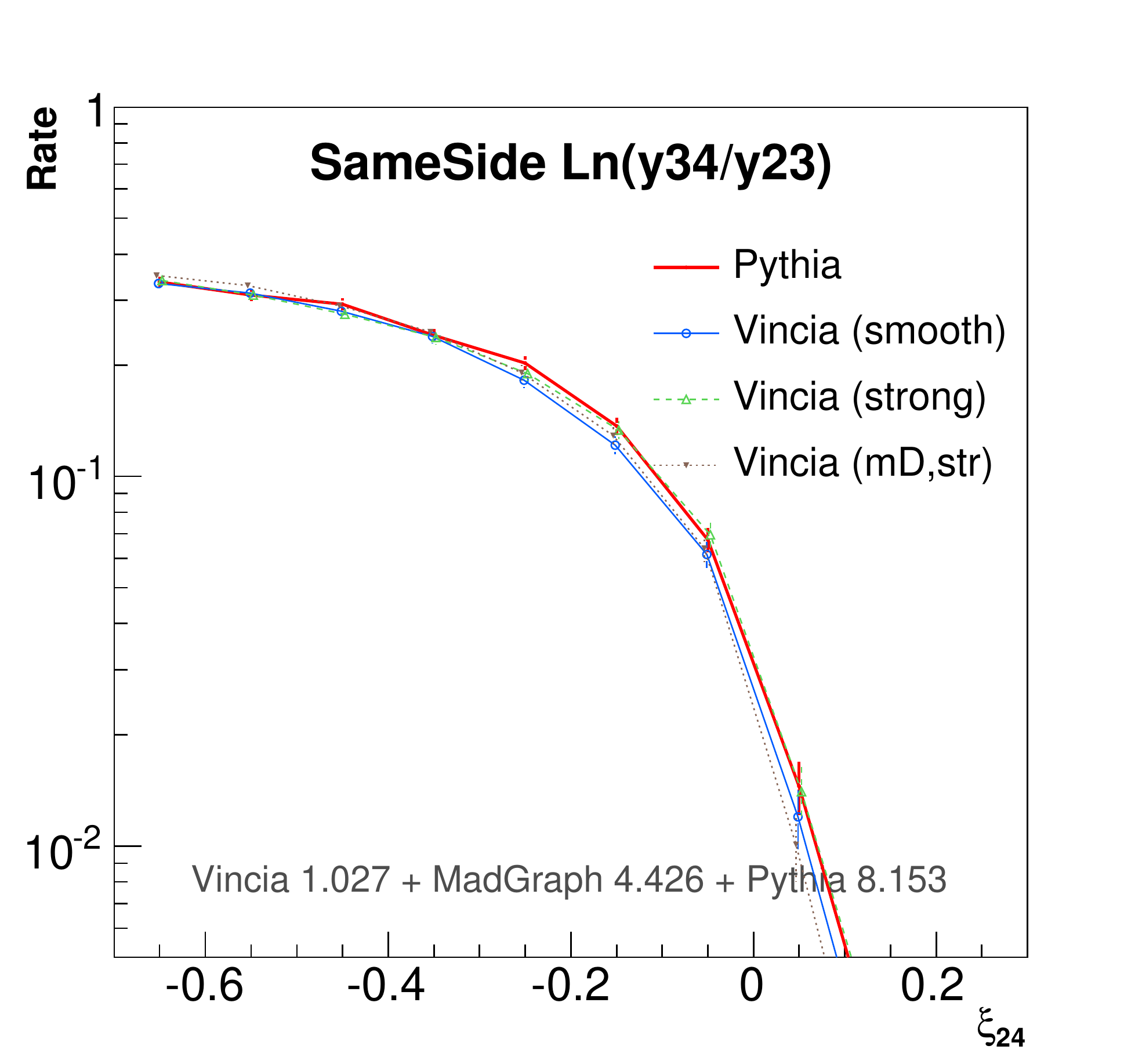}
\includegraphics*[scale=0.36]{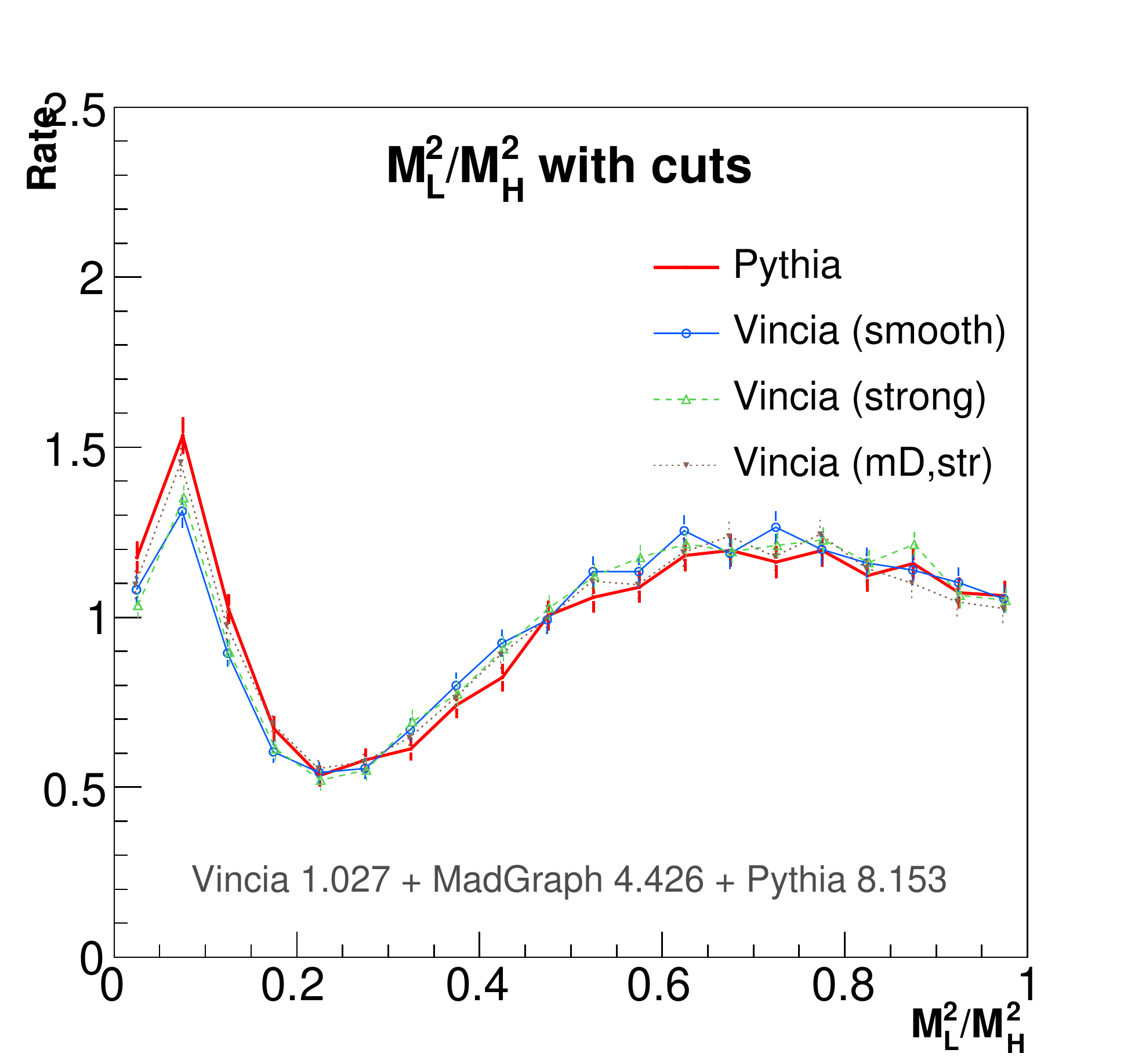}
\caption{{\sl Left:} $\xi_{24} = \ln(y_{34}/y_{23})$ in ``same-side''
  4-jet events.
{\sl Right:} Ratio of jet masses, $m_{L}^2/m_H^2$, in ``compressed''
4-jet events. Error bars indicate expected $1\sigma$ statistical
errors with 400k hadronic $Z$ decays.}
\label{fig:y24}
\end{figure}
The first observable is thus simply the ratio  $y_{34}/y_{23}$, in
events where the $4\to 3$ and $3\to 2$ clusterings happen in the same
jet. This distribution is illustrated in the left-hand pane of
fig.~\ref{fig:y24}, with logarithmic axes. Vertical error bars
  indicate the expected $1\sigma$ statistical error with 400k hadronic
  $Z$ decays. Since
the $k_T$ algorithm allows for unordered clustering scales, the
distribution extends beyond $\xi_{24} = \ln(y_{34}/y_{24}) = 0$. Default
Pythia (thick solid line) 
is compared to three different Vincia
settings: smooth (thin solid) and strong (dashed) ordering in
$p_\perp$ and strong ordering in dipole virtuality, $m_D$
(dotted). Note here that ordering in the variables $p_\perp$ or $m_D$ does not directly imply ordering in $k_T$.
The fact that the $P_{imp}$ factor also suppresses
branchings slightly below the strong-ordering threshold is manifest
in the thin solid line lying below the other ones in the region just
below zero, which should be statistically significant with a sample
size of $\sim 0.5$M events. Note as well that these distributions become
indistinguishable if one does not make the requirement of sameside clustering (not
shown), presumably since opposite-side collinear splittings then
dominate. 

A related observable is shown in the right-hand pane of
fig.~\ref{fig:y24}. To force a ``compressed'' scale hierarchy, 
we impose the cut $y_{34} > 0.5\, y_{23}$,
and plot the ratio $M_L^2/M_H^2$ of the masses of the jets at the end of the
clustering. With four partons at LO, the light jet mass is zero if
both the $4\to3$ and $3\to2$ clusterings happen in the same jet, while
it is non-zero otherwise. Thus, the region close to zero 
isolates events with a $1\to 3$ splitting occurring in one
of the jets, while the region above $\sim 0.25$ is dominated by
opposite-side $1\to 2$ splittings. In Pythia and in
mass-ordered Vincia, the peak at zero is  stronger
than in the $p_{\perp}$-ordered Vincia cases, while there is no
difference between strong and smooth ordering in this variable. 
It thus serves as a useful complement to $\xi_{24}$. 

\begin{figure}[tp]
\centering
\includegraphics*[scale=0.36]{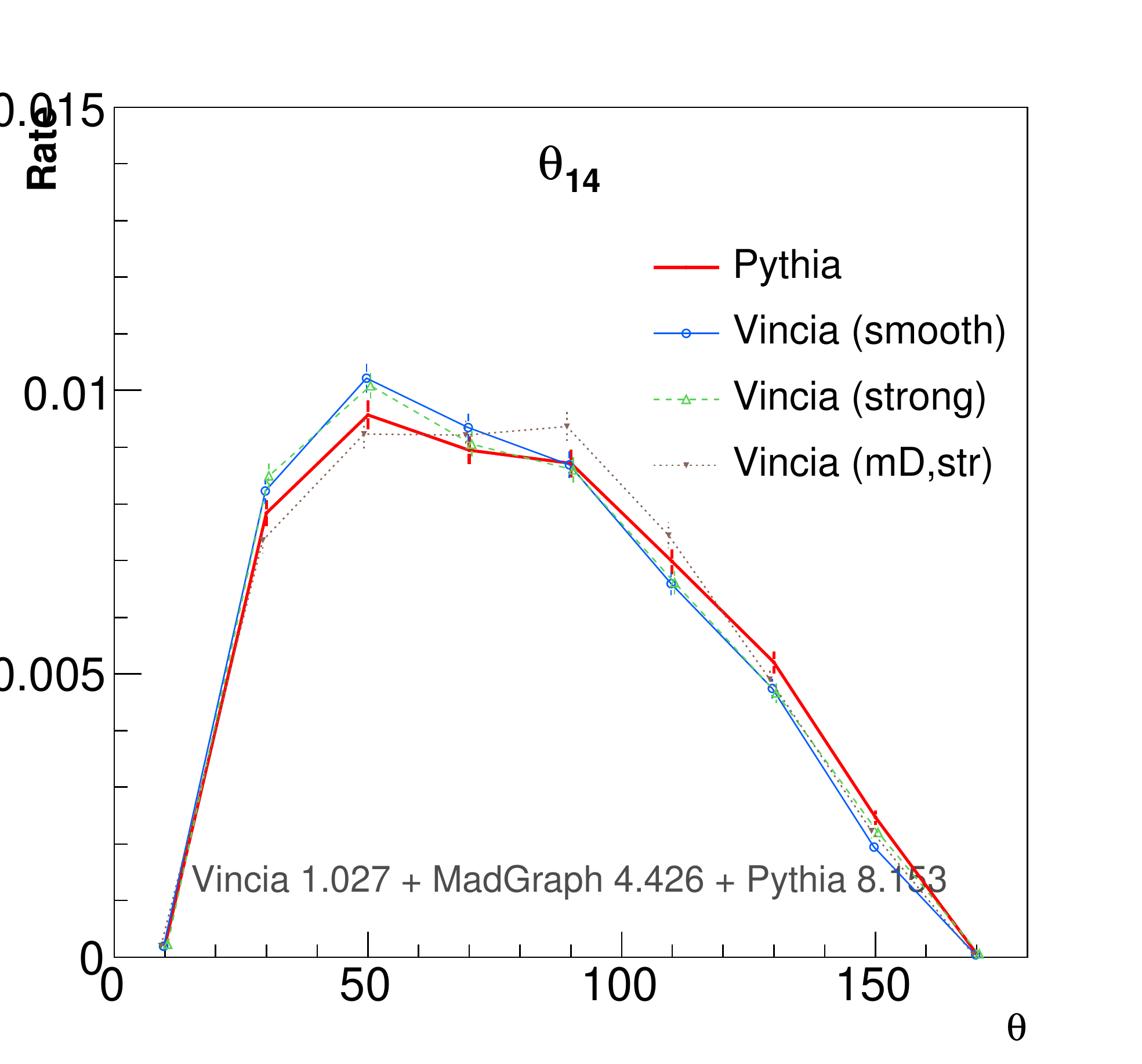}
\caption{Angle between the hardest (1$^\mathrm{st}$) and softest
  (4$^\mathrm{th}$) jets in ``collinear'' 4-jet events.  Error bars indicate expected $1\sigma$ statistical
errors with 400k hadronic $Z$ decays.}
\label{fig:the14} 
\end{figure}
Finally, in fig.~\ref{fig:the14}, we consider 4-jet events in which
the second and third jets (ordered in energy) are nearly collinear and
back-to-back to the hardest jet. Specifically, we impose the cuts
$\theta_{12} > 120^\circ$, $\theta_{13} > 120^\circ$, and $\theta_{23}
< 30^\circ$. We then plot the angle of the fourth (softest) jet with
respect to the hardest one. Again the strong and smooth ordering
options are indistinguishable, but interesting differences with
respect to both Pythia and mass-ordered Vincia 
are visible. Mass-ordering tends to produce a broader distribution,
with more radiation at right angles to the hardest jet (consistent
with mass-ordering prioritizing wide-angle emissions over collinear
ones), and the $p_\perp$-ordered Vincia showers exhibit a
stronger collinear peak than the Pythia one. A similar
observable was proposed in~\cite{Platzer:2009jq}. 

We conclude that, if all three observables could be measured with
an accuracy of $\sim 5-10\%$ or better, 
a useful and multi-dimensional constraint on the
subleading shower aspects would be obtained, including sensitivity
both to the type and shape of the ordering condition, and to the form
of the effective $1\to 3$ probabilities produced by the shower. We emphasize
that we have here restricted our attention to shower models that are 
virtually indistinguishable on all other observables we have
considered. 

\subsection*{ACKNOWLEDGEMENTS}
We thank D.~Kosower, S.~Pl\"atzer, and G.~Salam for discussions
on definitions of sensitive observables. 
This work was supported by the European Commission (HPRN-CT- 200-00148), FPA2009-09017 (DGI del MCyT, Spain) and S2009ESP-1473 (CA Madrid). J.J. L-V is supported by a MEC grant, AP2007-00385, and wants to thank the CERN Theory Division for its hospitality.


}

\section[Perturbative Uncertainties and Resummation for Exclusive Jet Cross Sections]
{PERTURBATIVE UNCERTAINTIES AND RESUMMATION FOR EXCLUSIVE JET CROSS SECTIONS \protect\footnote{Contributed by: Iain W.~Stewart, Frank J.~Tackmann}}
{\graphicspath{{jetvetoLesH/}}


\newcommand{\eq}[1]{Eq.~\eqref{eq:#1}}
\newcommand{\eqs}[2]{Eqs.~\eqref{eq:#1} and \eqref{eq:#2}}
\renewcommand{\sec}[1]{Sec.~\ref{sec:#1}}
\newcommand{\subsec}[1]{Sec.~\ref{subsec:#1}}
\newcommand{\fig}[1]{Fig.~\ref{fig:#1}}
\newcommand{\figs}[2]{Figs.~\ref{fig:#1} and \ref{fig:#2}}
\newcommand{\app}[1]{App.~\ref{app:#1}}

\newcommand{\abs}[1]{\lvert#1\rvert}
\newcommand{\ord}[1]{\mathcal{O}(#1)}
\newcommand{\mae}[3]{\langle#1\lvert#2\rvert#3\rangle}


\newcommand{\e}{\epsilon}
\newcommand{\la}{\lambda}
\newcommand{\w}{\omega}

\newcommand{\nb}{\,\mathrm{nb}}
\newcommand{\pb}{\,\mathrm{pb}}
\newcommand{\GeV}{\,\mathrm{GeV}}
\newcommand{\TeV}{\,\mathrm{TeV}}

\newcommand{\nn}{\nonumber}

\newcommand{\total}{\mathrm{total}}
\newcommand{\cut}{\mathrm{cut}}
\newcommand{\jet}{\mathrm{jet}}
\newcommand{\incl}{\mathrm{incl}}
\newcommand{\excl}{\mathrm{excl}}

\newcommand{\Ecm}{{E_\mathrm{cm}}}
\newcommand{\Tcm}{{\mathcal{T}_\mathrm{cm}}}
\newcommand{\Tcmc}{\mathcal{T}_\mathrm{cm}^\mathrm{cut}}
\newcommand{\Tcmj}{\mathcal{T}_\mathrm{cm}^\mathrm{jet}}

\renewcommand{\arraystretch}{1.3}
\arraycolsep 5pt

\allowdisplaybreaks[4]


\title{Perturbative Uncertainties and Resummation for Exclusive Jet Cross Sections}

\author{B. Nobalma$^1$, N. Schlamoky$^2$, K. Weasel$^2$}
\institute{$^1$LAPTH, 9 Chemin de Bellevue, B.P. 110,
Annecy-le-Vieux 74951, France
\\$^2$Tate Gallery of Fundamental
Research,  Trunka}

\author{Iain W.~Stewart$^1$, Frank J.~Tackmann$^2$}
\institute{$^1$Center for Theoretical Physics, Massachusetts Institute of Technology,
Cambridge, MA 02139, USA
\\$^2$Theory Group, Deutsches Elektronen-Synchrotron (DESY), D-22607 Hamburg, Germany}


\subsection{Introduction} \label{jetbin_intro}

In this writeup we discuss predictions for exclusive jet cross sections, which
have a particular number of jets in the final state. There are several
motivations for analyzing events by dividing the data into exclusive jet bins,
in particular when the relevant backgrounds strongly depend on the number of
jets, or when the sensitivity can be increased by optimizing the analysis for
the individual jet bins.  As our primary example we will consider the Higgs
analysis in the $H\to WW$ channel, which is performed separately in exclusive
$0$-jet, $1$-jet, and $2$-jet
bins~\cite{Benjamin:2011sv,Sung:2011bd,ATLAS:2011aa}. Other examples are
vector-boson fusion analyses, which are typically performed in the exclusive
$2$-jet channel, boosted $H\to b\bar b$ analyses that include a veto on
additional jets, as well as $H\to \tau\tau$ and $H\to \gamma\gamma$ which
benefit from improved sensitivity when the Higgs recoils against a jet. The
importance of the Higgs + $1$ jet channel in $H\to \tau\tau$ and $H\to WW^*$ was
demonstrated explicitly in Refs.~\cite{Mellado:2004tj, Mellado:2007fb}. Another
motivation for studying exclusive jet bins are the $W+$ jets channels, which are
important backgrounds for new physics searches.  We will use the notation
$\sigma_N$ for an \emph{exclusive} $N$-jet cross section (with exactly $N$
jets), and the notation $\sigma_{\ge N}$ for an \emph{inclusive} $N$-jet cross
section (with $N$ or more jets).

To explore the implications of the jet bin restrictions, consider a simple
example where we divide the total cross section, $\sigma_\total$, into an
exclusive $0$-jet bin, $\sigma_0(p^\cut)$, and the remaining
inclusive $(\geq 1)$-jet bin, $\sigma_{\geq 1}(p^\cut)$,
\begin{align} \label{eq:pcut}
\sigma_\total = \int_0^{p^\cut}\!d p\, \frac{d\sigma}{d p} + \int_{p^\cut}\!d p\, \frac{d\sigma}{d p}
\equiv  \sigma_0(p^\cut) + \sigma_{\geq 1}(p^\cut)
\,.\end{align}
Here $p$ denotes the kinematic variable which is used to divide up the cross
section into jet bins.  A typical choice is $p \equiv p_T^\jet$, defined by the
largest $p_T$ of any jet in the event, such that $\sigma_0(p_T^\cut)$ only
contains events with jets having $p_T \leq p_T^\cut$, and $\sigma_{\geq
  1}(p_T^\cut)$ contains events with at least one jet with $p_T \geq p_T^\cut$.
By defining $\sigma_0(p_T^\cut)$ and $\sigma_{\geq 1}(p_T^\cut)$ one has divided
up initial-state radiation from the colliding hard partons and soft radiation in
the event.  This restriction on additional emissions changes the coefficients
appearing in the $\alpha_s$ expansion and leads to the appearance of double and
single logarithms of the form $\alpha_s \ln^2(p^\cut/Q)$ and $\alpha_s
\ln(p^\cut/Q)$ (with higher powers $\alpha_s^n \ln^{m\le 2n}(p^\cut/Q)$
appearing at higher orders in perturbation theory). Here $Q$ is the hard scale
of the process, such as $Q=m_H$ for Higgs production, and most often we have
$p^\cut \ll Q$.  These changes to the perturbation series can modify the
convergence of fixed-order results and make it prudent to consider resummed
cross section predictions that include an all-orders resummation of the large
logarithms.  For $N$ jets the analog of Eq.~(\ref{eq:pcut}) is $\sigma_{\ge N} =
\sigma_N(p_{N+1}^\cut) + \sigma_{\geq N+1}(p_{N+1}^\cut)$ and the same discussion
applies regarding the large logarithms of $p_N^\cut$ that are not present in
$\sigma_{\ge N}$, but are present in each of $\sigma_N$ and $\sigma_{\geq N+1}$.

The definition of $\sigma_0(p^\cut)$ may include dependence on rapidity and on
the grouping of particles. For a jet-based variable like $p_T^\jet$ the former
is induced by only considering jets within the rapidity range $\lvert
\eta^\jet\rvert \leq \eta^\cut$, and the latter enters through the choice of jet
algorithm.  These dependencies make theoretical predictions more difficult. In
Higgs production via gluon fusion the cross section is known to
next-to-next-to-leading order (NNLO)~\cite{Dawson:1990zj, Djouadi:1991tka,
  Spira:1995rr, Harlander:2002wh, Anastasiou:2002yz, Ravindran:2003um,
  Pak:2009dg, Harlander:2009my}, and NNLO results including full kinematic
information are available through FeHiP~\cite{Anastasiou:2004xq,
  Anastasiou:2005qj} and HNNLO~\cite{Catani:2007vq, Grazzini:2008tf} (as well as
by combining the total NNLO cross section with MCFM~\cite{Campbell:1999ah,
Campbell:2010cz} for some distributions).  When the measurements are
performed in exclusive jet bins, the perturbative uncertainties in the
theoretical predictions must also be evaluated separately for each individual
jet bin~\cite{Anastasiou:2009bt}.  When combining channels with different jet
multiplicities, the correlations between the theoretical uncertainties can be
significant and must be taken into account~\cite{Stewart:2011cf}.  The
perturbative predictions can be made more precise by including a resummation of
large $p^\cut$ dependent logarithms on top of the fixed-order predictions. At
the leading logarithmic level this can be achieved with standard parton shower
Monte Carlo programs, regardless of the precise definition of $p^\cut$. So far a
next-to-next-to-leading logarithmic (NNLL) resummed result for a jet-veto variable only
exists for beam thrust~\cite{Stewart:2009yx}, $\Tcm$, which is a rapidity
weighted $E_T$-like inclusive variable. The definitions of the jet-veto
variables we will use are
\begin{align} \label{eq:Tau_def}
 p_T^{\jet} &= \Big| \sum_{k\in \jet} \vec p_{Tk} \Big| \,,
 &\Tcm &= \sum_k | \vec p_{Tk}| e^{-\abs{\eta_k}} 
      = \sum_k (E_k - |p_k^z|) 
\,.\end{align}
For $p_T^\jet$ our jets are defined using anti-$k_T$~\cite{Cacciari:2008gp} with
$R = 0.5$, and we consider jets that satisfy a rapidity cut $|\eta| \le
\eta^\cut$.  For $\Tcm$ the sum is over all objects in the final state except
the Higgs decay products, and can in principle be considered over particles,
topo-clusters, or jets with a small $R$ parameter. In all our results we
consistently use MSTW2008 NNLO PDFs~\cite{Martin:2009iq}.

In this writeup we will explore fixed NNLO and resummed NNLL+NNLO predictions
for $H+$ 0-jet cross sections and compare various methods for evaluating the
uncertainty as a function of cuts on $p_T^\jet$ and $\Tcm$.  The three methods
we will discuss for evaluating the uncertainties in exclusive jet cross sections
are
\begin{itemize}
\item[A)] ``Direct Exclusive Scale Variation''.  Here the uncertainties are
  evaluated by directly varying the renormalization and factorization scales in
  the fixed-order predictions for each exclusive jet cross section $\sigma_N$.
  This implies that the uncertainties are 100\% correlated for different $N$s. 
\item[B)] ``Combined Inclusive Scale Variation'', as proposed in
  Ref.~\cite{Stewart:2011cf} and utilized in
  Refs.~\cite{Benjamin:2011sv,Sung:2011bd,ATLAS:2011aa}.  Here, the perturbative
  uncertainties in the inclusive $N$-jet cross sections, $\sigma_{\geq N}$, are
  treated as the primary uncertainties that can be evaluated by scale variations
  in fixed-order perturbation theory. These uncertainties are treated as
  uncorrelated for different $N$.  The exclusive $N$-jet cross sections are
  obtained using $\sigma_N = \sigma_{\geq N} - \sigma_{\geq N+1}$. The
  uncertainties and correlations follow from standard error propagation,
  including the appropriate anticorrelations between $\sigma_N$ and
  $\sigma_{N\pm 1}$ related to the division into jet bins.
\item[C)] ``Uncertainties from Resummation.''  Resummed
  calculations for exclusive jet cross sections can provide uncertainty
  estimates that allow one to simultaneously include both types of correlated
  and anticorrelated uncertainties as in methods A and B. The magnitude of the
  uncertainties may also be reduced from the resummation of large logarithms.
\end{itemize}
In all three methods, adding the exclusive jet cross sections yields the
expected scale variation in the total cross section.  Method B avoids a
potential underestimate of the uncertainties in individual jet bins due to
strong cancellations that can potentially take place in method A.  Method B
produces realistic perturbative uncertainties for exclusive jet cross sections
when using fixed-order predictions for various processes, since it accounts for
the presence of large logarithms at higher orders caused by the jet binning.  In
Method C one utilizes higher-order resummed predictions for the exclusive jet
cross sections, which allow one to obtain improved central values and further
refined uncertainty estimates.

The basic structure of the large logarithms in the perturbative series is
discussed in Sec.~\ref{sec:jetbin_motivation}.  In Sec.~\ref{sec:jetbin_methods}
we discuss and compare the above three methods to determine the perturbative
uncertainties.  The work discussed here regarding methods A, B, and C builds on
work done in Refs.~\cite{Berger:2010xi,Stewart:2011cf}, was initiated at Les
Houches, and has also been incorporated in the second Higgs Yellow Book
report~\cite{YellowBook2} (Secs.~5.2 and 5.5.) We also review
recent work by others that can be found in~\cite{YellowBook2}(Sec.~5.3).

Note that here we are only discussing the theoretical uncertainties due to
unknown higher-order perturbative corrections, which are commonly estimated
using scale variation. Parametric uncertainties, such as PDF choices and
$\alpha_s(m_Z)$ uncertainties, must be treated appropriately as common sources
for all investigated channels.

\subsection{Theoretical Motivation}
\label{sec:jetbin_motivation}

\subsubsection{Structure of the Perturbative Series}
\label{subsec:jetbin_pertseries}

We begin by discussing the structure of the large logarithms in exclusive jet
cross sections. For Higgs production from gluon fusion with $p_T^\jet\le
p_T^\cut$ the leading double logarithms appearing at $\ord{\alpha_s}$ are
\begin{align} \label{eq:sig0dbleL}
\sigma_0(p_T^\cut) &= \sigma_B \Bigl(1 - \frac{3\alpha_s}{\pi}\, 2\ln^2 \frac{p_T^\cut}{m_H} + \dotsb \Bigr)
\,,\end{align}
where $\sigma_B$ is the Born (tree-level) cross section.

The total cross section only depends on the hard scale $Q=m_H$, which means by
choosing the factorization and renormalization scales $\mu_f\simeq\mu_r \simeq
m_H$, the fixed-order expansion does not contain large logarithms and has the
structure
\begin{equation} \label{eq:sigmatot}
\sigma_\total \simeq \sigma_B\big[ 1 + \alpha_s + \alpha_s^2 + \ord{\alpha_s^3} \big]
\,.\end{equation}
Our expressions for perturbative series such as this one are schematic, showing
the scaling of the terms without the coefficient functions. The convolution with
the parton distribution functions (PDFs) are also not displayed.  For $gg\to H$,
the coefficients of this series can be large, corresponding to the well-known
large K factors.  As usual, varying the scale in $\alpha_s(\mu)$ (and the PDFs)
one obtains an estimate of the size of the missing higher-order terms in this
series, which we denote by $\Delta_\total$.

The inclusive $1$-jet cross section has the perturbative structure
\begin{align} \label{eq:sigma1}
\sigma_{\geq 1}(p^\cut)
\simeq \sigma_B\bigl[ \alpha_s (L^2 + L + 1)
+ \alpha_s^2 (L^4 + L^3 + L^2 + L + 1) + \ord{\alpha_s^3 L^6} \bigr]
\,,\end{align}
where the logarithms $L = \ln(p^\cut/m_H)$. For $p^\cut \ll m_H$ these
logarithms can get large enough to overcome the $\alpha_s$ suppression. In the
limit $\alpha_s L^2 \simeq 1$, the fixed-order perturbative expansion breaks
down and the logarithmic terms must be resummed to all orders in $\alpha_s$ to
obtain a meaningful result. For typical experimental values of $p^\cut$
fixed-order perturbation theory can still be considered, but the logarithms
cause large corrections at each order and dominate the series. 

The exclusive $0$-jet cross section is equal to the difference between
Eqs.~\eqref{eq:sigmatot} and \eqref{eq:sigma1}, and so has the schematic
structure
\begin{align} \label{eq:sigma0}
\sigma_0(p^\cut) &= \sigma_\total - \sigma_{\geq1}(p^\cut)
\nonumber\\
&\simeq \sigma_B \Bigl\{ \bigl[ 1 + \alpha_s + \alpha_s^2 + \ord{\alpha_s^3} \bigr]
- \bigl[\alpha_s (L^2 \!+ L + 1) + \alpha_s^2 (L^4 \!+ L^3 \!+ L^2 \!+ L + 1)
+ \ord{\alpha_s^3 L^6} \bigr] \Bigr\}
\,.\end{align}
In this difference, the large positive corrections in $\sigma_\total$ partly
cancel against the large negative logarithmic corrections in $\sigma_{\geq 1}$.
For example, at $\ord{\alpha_s}$ there is a value of $L$ for which the
$\alpha_s$ terms in Eq.~\eqref{eq:sigma0} cancel exactly. At this $p^\cut$ the NLO $0$-jet
cross section has vanishing scale dependence and is equal to the LO cross
section, $\sigma_0(p^\cut)=\sigma_B$.  Due to this cancellation, a standard use
of scale variation in $\sigma_0(p^\cut)$ does not actually probe the size of the
large logarithms, and does not provide an estimate of $\Delta_\cut$. This issue
impacts the uncertainties in the experimentally relevant region for $p^\cut$.

For example, for $gg\to H$ (with $\sqrt{s} =7\TeV$, $m_H = 165\GeV$, $\mu_f = \mu_r =
m_H/2$), one finds~\cite{Anastasiou:2004xq, Anastasiou:2005qj, Catani:2007vq, Grazzini:2008tf}
\begin{align}
\sigma_\total &= (3.32 \pb) \bigl[1 + 9.5\,\alpha_s + 35\,\alpha_s^2 + \ord{\alpha_s^3} \bigr]
\,,\nonumber\\
\sigma_{\geq 1}\bigl(p_T^\jet \geq 30\GeV, \abs{\eta^\jet} \leq 3.0\bigr)
&= (3.32 \pb) \bigl[4.7\,\alpha_s + 26\,\alpha_s^2 + \ord{\alpha_s^3} \bigr] \,.
\end{align}
In $\sigma_\total$ one can see the impact of the well-known large $K$ factors.
(Using instead $\mu_f = \mu_r = m_H$ the $9.5 \alpha_s$ and $35 \alpha_s^2$
coefficients in $\sigma_\total$ increase to $11\alpha_s$ and $65 \alpha_s^2$.)  In
$\sigma_{\geq 1}$, one can see the impact of the large logarithms on the
perturbative series.  Taking their difference to get $\sigma_0$, one observes a
sizeable numerical cancellation between the two series at each order in
$\alpha_s$.

\subsubsection{Perturbative Series for the Event Fraction}
\label{subsec:jetbin_pertseriesf0}

Experimentally the desired quantity which incorporates the jet-veto cut is the
exclusive $0$-jet event fraction 
\begin{align} \label{eq:f0main}
  f_0(p^\cut) = \frac{\sigma_0(p^\cut)}{\sigma_\total} = 1 - \frac{\sigma_{\ge
      1}(p^\cut)}{\sigma_\total} 
\,.\end{align}
One option for treating $f_0(p^\cut)$ is to consider it as a derived quantity,
given the basic observables $\{\sigma_0,\sigma_\total\}$ or $\{\sigma_{\ge
  1},\sigma_\total\}$.  In this approach, which was utilized in
Ref.~\cite{Stewart:2011cf} and Ref.~\cite{YellowBook2}(Secs.~5.2 and 5.5), one
propagates the uncertainties from the $\sigma_i$s to derive those for
$f_0(p^\cut)$. This approach is natural from the perspective of utilizing
log-resummed computations for $\sigma_0(p^\cut)$. In particular, it maintains
the constraint that for large $p^\cut$ we have monotonic convergence of
$\sigma_0\to \sigma_\total$ and $f_0\to 1$, a property that relies on a phase
space cut reducing the cross section, but does not depend on perturbation theory.

When using fixed-order predictions for the various cross sections, an
alternative to \eq{f0main} considered in Ref.~\cite{YellowBook2}(Sec.~5.3) is to
analyze the perturbation theory for $f_0(p^\cut)$ directly. In this case
different schemes of organizing the perturbation series, by keeping or dropping
various $\ord{\alpha_s^3}$ terms, give a method to estimate the size of the
higher-order perturbative corrections.  Three such schemes were considered in
Ref.~\cite{YellowBook2}(Sec.~5.3) (which we label here by schemes 1,2,3). It is
convenient to define the perturbative corrections to the cross section by
dividing each of them by the Born cross section $\sigma_B$, such that we can
write
\begin{align}
\sigma_\total &= \sigma_B\bigl[1 + \hat \sigma_\total^{(1)} + \hat \sigma_\total^{(2)}
  + \ord{\alpha_s^3} \bigr]
\,,\nn\\
\sigma_{\geq 1}(p^\cut) &= \sigma_B\bigl[ \hat \sigma_{\geq 1}^{(1)}(p^\cut) +
\hat \sigma_{\geq 1}^{(2)}(p^\cut) + \ord{\alpha_s^3} \bigr]
\,.\end{align}
With this notation the result of treating $f_0$ as a derived quantity is
\begin{align} \label{eq:f0derived}
\big[ f_0(p^\cut)\big]^{({\rm scheme\ 1})}
&= 1 - \frac{\hat\sigma_{\geq 1}^{(1)}(p^\cut) + \hat\sigma_{\geq 1}^{(2)}(p^\cut)}{1 +
  \hat\sigma_\total^{(1)} + \hat\sigma_\total^{(2)}} + \ord{\alpha_s^3} \,,
\end{align}
while at the same order in perturbation theory we can also consider the
following expressions for $f_0$:
\begin{align} \label{eq:f0expn}
\big[ f_0(p^\cut)\big]^{(\rm scheme\ 2)}
&= 1 - \frac{\hat\sigma_{\geq 1}^{(1)}(p^\cut) + \hat\sigma_{\geq 1}^{(2)}(p^\cut)}{1 +
  \hat\sigma_\total^{(1)}} + \ord{\alpha_s^3}
\,,\nn\\
\big[ f_0(p^\cut)\big]^{(\rm scheme\ 3)}
&= 1 - \bigl[\hat\sigma_{\geq 1}^{(1)}(p^\cut) + \hat\sigma_{\geq 1}^{(2)}(p^\cut)\bigr]
+ \hat\sigma_{\geq 1}^{(1)}\, \hat\sigma_\total^{(1)} + \ord{\alpha_s^3}
\,.
\end{align}
We will contrast using the expressions in \eq{f0derived} and \eq{f0expn} with
various methods for analyzing the uncertainty in our discussion below.

\subsection{Uncertainty Analysis for Exclusive Jet Bins}
\label{sec:jetbin_methods}

As described in Sec.~\ref{subsec:jetbin_pertseries}, the phase space restriction
defining $\sigma_0$ changes its perturbative structure compared to that of
$\sigma_\total$. In general this gives rise to an additional perturbative
uncertainty due to missing higher-order terms depending on $p^\cut$. We will
call the associated jet-binning uncertainty $\Delta_\cut$.  This can be thought
of as an uncertainty related to the presence of large logarithms of $p^\cut$ at
higher orders in perturbation theory.  In Eq.~\eqref{eq:pcut} both $\sigma_0$
and $\sigma_{\geq 1}$ depend on the phase space cut, $p^\cut$, and by
construction this dependence cancels in $\sigma_0+\sigma_{\geq 1}$.  Hence, the
additional uncertainty $\Delta_\cut$ induced by $p^\cut$ must be 100\%
anticorrelated between $\sigma_0(p^\cut)$ and $\sigma_{\geq 1}(p^\cut)$, such
that it cancels in their sum. For example, using a covariance matrix to model
the uncertainties and correlations, the contribution of $\Delta_\cut$ to the
covariance matrix for $\{\sigma_0, \sigma_{\geq 1}\}$ must be of the form
\begin{equation} \label{eq:cutmatrix}
C_\cut = \begin{pmatrix}
   \Delta_\cut^2 &  - \Delta_\cut^2 \\
   -\Delta_\cut^2 & \Delta_\cut^2
\end{pmatrix}\,.
\end{equation}
The questions then are: (1) How can we estimate $\Delta_\cut$ in a simple way,
and (2) how is the perturbative uncertainty $\Delta_\total$ of $\sigma_\total$
related to the uncertainties of $\sigma_0$ and $\sigma_{\geq 1}$?  

\subsubsection{Perturbative Uncertainties for Method A}

When using method A to estimate the perturbative uncertainties one simply uses a
common scale variation to estimate the uncertainty $\Delta_0$ in $\sigma_0$ and
the uncertainty $\Delta_{\ge 1}$ in $\sigma_{\ge 1}$. By doing so the
uncertainties are 100\% correlated, corresponding to a covariance matrix in
method A for $\{\sigma_0, \sigma_{\geq 1}\}$ given by
\begin{equation} \label{eq:CA}
C_A = \begin{pmatrix}
\Delta_{0}^2 &  \Delta_{0}\,\Delta_{\geq 1}  \\
\Delta_{0}\,\Delta_{\geq 1} & \Delta_{\geq 1}^2 
\end{pmatrix}  \,.
\end{equation}
Here $\Delta_{\total}=\Delta_0 + \Delta_{\ge 1}$ is the scale uncertainty in
$\sigma_{\total}$.  When instead of $\sigma_0$ we directly calculate the 0-jet
event fraction $f_0$ using \eq{f0derived} or one of the expressions in
\eq{f0expn}, we can again determine the method A uncertainty estimate by scale
variation in $f_0$ (we will refer to these results as methods $A_1$, $A_2$, and
$A_3$ respectively).

In this method $\Delta_{\rm cut}$ is not included because, as explained below
\eq{sigma0}, varying the perturbative scale in $\Delta_0$ does not probe the
presence of the higher order large logarithms depending on $p^\cut$. This method
can lead to an underestimate of the perturbative uncertainty in $\sigma_0$ (and
hence $f_0$), since there is a region of $p^\cut$ values where scale variation
is no longer a reasonable estimate of higher order corrections because of the
vanishing of the $\mu$ dependence.

\subsubsection{Perturbative Uncertainties for Method B}

Since the perturbative series for $\sigma_{\ge 1}$ in Eq.~\eqref{eq:sigma1} is
dominated by the large logarithms of $p^\cut$, we can use its scale variation
$\Delta_{\ge 1}$ to get an estimate for their size by taking
$\Delta_\cut=\Delta_{\ge 1}$~\cite{Stewart:2011cf}. Since $\Delta_\cut$ and
$\Delta_\total$ are by definition uncorrelated, by setting $\Delta_\cut =
\Delta_{\geq 1}$ we are effectively treating the perturbative series for
$\sigma_\total$ and $\sigma_{\geq 1}$ as independent with uncorrelated
perturbative uncertainties. That is, considering $\{\sigma_\total, \sigma_{\geq
  1}\}$, the covariance matrix is diagonal,
\begin{equation} \label{eq:diagmatrix}
 \begin{pmatrix}
  \Delta_\total^2 & 0 \\ 0 & \Delta_{\geq1}^2
\end{pmatrix}
\,,\end{equation}
where $\Delta_\total$ and $\Delta_{\geq 1}$ are
evaluated by separate scale variations in the fixed-order predictions for
$\sigma_\total$ and $\sigma_{\geq 1}$.
This is consistent, since for small $p^\cut$ the two series have very different
structures. In particular, there is no reason to believe that the same
cancellations in $\sigma_0$ will persist at every order in perturbation theory
at a given $p^\cut$. It follows that the perturbative uncertainty in
$\sigma_0 = \sigma_\total - \sigma_{\geq 1}$ is given by
$\Delta_\total^2 + \Delta_{\geq 1}^2$, and the resulting
covariance matrix for $\{\sigma_0, \sigma_{\geq 1}\}$ in method B is
\begin{equation} \label{eq:CB}
C_B = \begin{pmatrix}
   \Delta_{\geq 1}^2 + \Delta_\total^2 &  - \Delta_{\geq 1}^2 \\
   -\Delta_{\geq 1}^2 & \Delta_{\geq 1}^2
\end{pmatrix}\,.
\end{equation}
Note that all of $\Delta_\total$ occurs in the uncertainty for $\sigma_0$.
This is reasonable from the point of view that $\sigma_0$ starts at the same
order in $\alpha_s$ as $\sigma_\total$ and contains the same leading virtual
corrections. The method B uncertainty for the event fraction $f_0$ follows most
naturally by error propagation from the cross sections, treating it as a derived
quantity.

The limit $\Delta_\cut = \Delta_{\geq 1}$ that Eq.~\eqref{eq:CB} is
based on is of course an approximation. However, the preceding arguments
show that it is a more reasonable starting point than method A, since the latter
does not account for the additional $p^\cut$ induced uncertainties.

The generalization of the above discussion to more jets and several jet bins is
straightforward.  For the $N$-jet bin we replace $\sigma_\total \to \sigma_{\ge
  N}$, $\sigma_0\to \sigma_N$, and $\sigma_{\ge 1} \to \sigma_{\ge N+1}$. If the
perturbative series for $\sigma_{\ge N}$ exhibits large $\alpha_s$ corrections
due to its logarithmic series or otherwise, then the presence of a different
series of large logarithms in $\sigma_{\ge N+1}$ will again lead to
cancellations when we consider the difference $\sigma_N = \sigma_{\geq N} -
\sigma_{\geq N+1}$. These two cross sections will have different series for
their double logarithms since the number of active partons and their color
structure differ. In this situation $\Delta_{\geq N+1}$ will again give a better
estimate for the extra $\Delta_\cut$ type uncertainty that arises from
separating $\sigma_{\geq N}$ into $\sigma_N$ and $\sigma_{\geq N+1}$.

\subsubsection{Perturbative Uncertainties for Method C}

In method C we assess the perturbative uncertainties using resummed predictions
for variables $p^\cut$ that implement a jet veto, following
Refs.~\cite{Berger:2010xi,Stewart:2011cf}. An advantage of using resummed
predictions is that they contain perturbation theory scale parameters which
allow for an evaluation of two components of the theory error, one which is
100\% correlated with the total cross section (as in method A), and one related
to the presence of the jet-bin cut which is anti-correlated between neighboring
jet bins (as in method B).

The resummed $H+\text{0-jet}$ cross section predictions of
Ref.~\cite{Berger:2010xi} follow from a factorization theorem for the $0$-jet
cross section~\cite{Stewart:2009yx}, $\sigma_0(\Tcmc) = H\, {\cal I}_{gi}\,{\cal
  I}_{gj}\otimes S f_i f_j$, where $H$ contains hard virtual effects, the ${\cal
  I}$s and $S$ describe the veto-restricted collinear and soft radiation, and
the $f$s are standard parton distributions. Fixed-order perturbation theory is
carried out at three scales, a hard scale $\mu_H^2\sim m_H^2$ in $H$, and beam
and soft scales $\mu_B^2\sim m_H\Tcmc$ and $\mu_S^2\sim (\Tcmc)^2$ for ${\cal
  I}$ and $S$, and are then connected by NNLL renormalization group evolution
that sums the jet-veto logarithms, which are encoded in ratios of these scales.
The perturbative uncertainties can be assessed by considering two sources: i) an
overall scale variation that simultaneously varies $\{\mu_H,\mu_B,\mu_S\}$ up
and down by a factor of two which we denote by $\Delta_{H0}$, and ii) individual
variations of $\mu_B$ or $\mu_S$ that each hold the other two scales
fixed~\cite{Berger:2010xi}, whose envelope we denote by the uncertainty
$\Delta_{SB}$. Here $\Delta_{H0}$ is dominated by the same sources of
uncertainty as the total cross section $\sigma_\total$, and hence should be
considered 100\% correlated with its uncertainty $\Delta_\total$. The uncertainty
$\Delta_{SB}$ is only present due to the jet-bin cut, and hence gives the
$\Delta_\cut$ uncertainty that is anti-correlated between neighboring jet bins.

If we simultaneously consider the cross sections $\{\sigma_0, \sigma_{\ge 1}\}$
then the full correlation matrix in method C is
\begin{align} \label{eq:CC}
C_C &=
\begin{pmatrix}
 \Delta_{SB}^2 &  - \Delta_{SB}^2 \\
-\Delta_{SB}^2 & \Delta_{SB}^2
\end{pmatrix}
+
\begin{pmatrix}
\Delta_{H0}^2 &  \Delta_{H0}\,\Delta_{H\geq 1}  \\
\Delta_{H0}\,\Delta_{H\geq 1} & \Delta_{H\geq 1}^2
\end{pmatrix}
,\end{align}
where $\Delta_{H\geq 1} =\Delta_\total - \Delta_{H0}$ encodes the 100\% correlated
component of the uncertainty for the $(\ge 1)$-jet inclusive cross section.
Computing the uncertainty in $\sigma_\total$ gives back $\Delta_\total$.

\begin{figure}[t!]
 \includegraphics[width=0.5\textwidth]{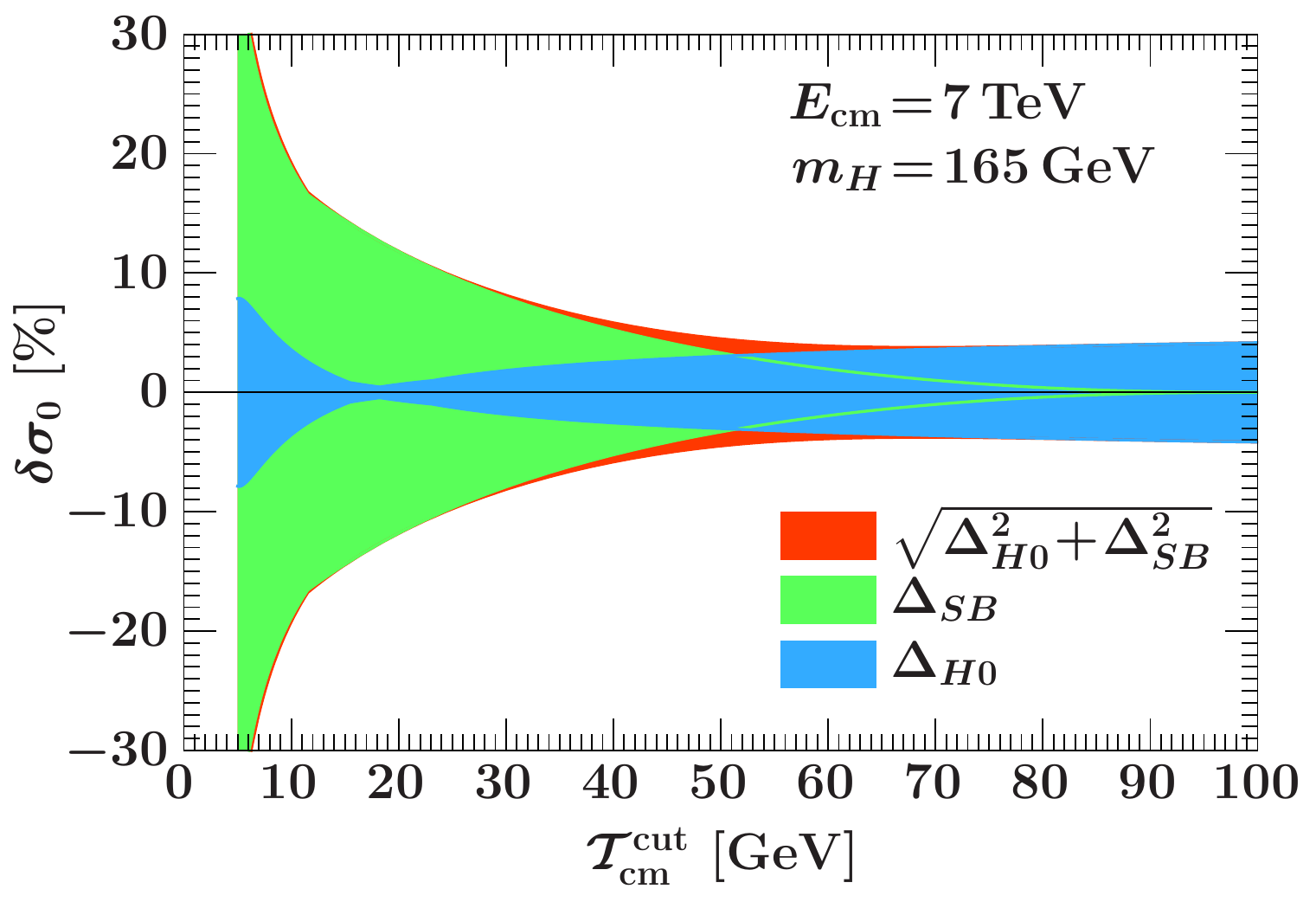}%
 \hfill%
 \includegraphics[width=0.5\textwidth]{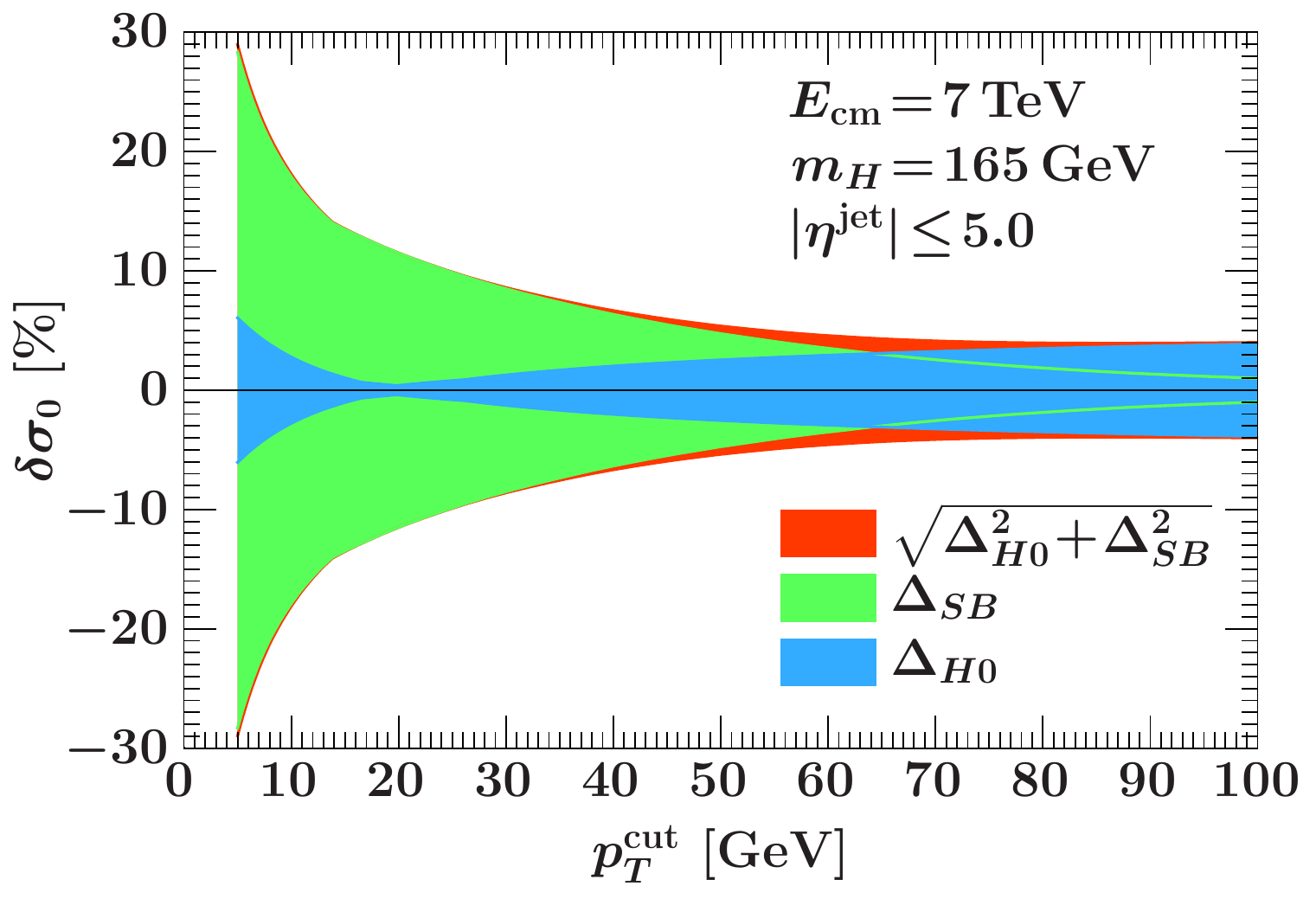}%
 \vspace{-0.5ex}
 \caption{\label{fig:3scales} Relative uncertainties for the 0-jet bin cross
   section from resummation at NNLL+NNLO for beam thrust $\Tcm$ on the left and
   $p_T^{\rm jet}$ on the right.}
\end{figure}

Eq.~\eqref{eq:CC} can be compared to $C_A$ for method A in \eq{CA}, which
corresponds to taking $\Delta_{SB}\to 0$ and obtaining the analog of
$\Delta_{H0}$ by up/down scale variation without resummation
($\mu_H=\mu_B=\mu_S$).  It can also be compared to $C_B$ for method B in
\eq{CB}, which corresponds to taking $\Delta_{SB}\to \Delta_{\ge 1}$ and
$\Delta_{H\ge 1}\to 0$, such that $\Delta_{H0}\to \Delta_\total$. The numerical
dominance of $\Delta_{SB}^2$ over $\Delta_{H0}\Delta_{H\ge 1}$ in the $0$-jet
region is another way to justify the preference for using method B when only
given a choice between methods A and B.  For example, for $p_T^\cut = 30\GeV$
and $|\eta^\jet| \leq 5.0$ we have $\Delta_{SB}^2 = 0.17$ and
$\Delta_{H0}\Delta_{H\ge 1} = 0.02$.

In Fig.~\ref{fig:3scales} we show the uncertainties $\Delta_{SB}$ (light green)
and $\Delta_{H0}$ (medium blue) as a function of the jet-veto variable, as well
as the combined uncertainty adding these components in quadrature (dark orange).
From the figure we see that the $\Delta_{H0}$ dominates at large values where the
veto is turned off and we approach the total cross section, and that the jet-cut
uncertainty $\Delta_{SB}$ dominates for the small cut values that are typical of
experimental analyses with Higgs jet bins. The same pattern is observed in the
left panel which directly uses the NNLL+NNLO predictions for $\Tcmc$, and the
right panel which shows the result from reweighting these predictions to
$p_T^\cut$ as explained in Sec.~\ref{jetbin_compare} below.

\subsubsection{Comparison of Uncertainty Methods}
\label{jetbin_compare}

In Fig.~\ref{fig:ABC} we compare the uncertainties for the 0-jet bin cross
section from methods A (medium green), B (light green), and C (dark orange).  In
the upper panels we use $\Tcmc$ as the jet-veto variable and full results for
the NNLO and NNLL+NNLO cross sections, while in the lower panels we use
$p_T^\cut$ as the jet-veto variable with the full NNLO and the reweighted
NNLL+NNLO results (as explained below).  The upper panels use a cut on beam
thrust, $\Tcmc$ while the lower panels use $p_T^\cut$. The right panels show
the same results as those on the left, but are normalized to the highest-order
result to better show the relative differences and uncertainties.  The
uncertainties in methods A, B, and C are computed from the upper left entry of
the matrices $C_A$, $C_B$, and $C_C$, respectively.

From Fig.~\ref{fig:ABC} we see that in method A (medium green bands) for small
values of $p_T^\cut$ the cancellations that take place in $\sigma_0(p^\cut)$
cause the error bands to shrink and eventually almost vanish at $p_T^\cut\simeq
25\GeV$, where there is an almost exact cancellation between the two series in
Eq.~\eqref{eq:sigma0}.  This is avoided by using method B (light green bands).
For large values of $p_T^\cut$ method B reproduces the method A scale variation,
since $\sigma_{\geq 1}(p^\cut)$ becomes small. On the other hand, for small
values of $p_T^\cut$ the uncertainties estimated using method B are more
realistic, because they explicitly estimate the uncertainties due to the
presence of higher order large logarithmic corrections.  

The features of this plot are quite generic. In particular, the same pattern of
uncertainties is observed for the Tevatron, when using $\mu=m_H$ as our central
scale (with $\mu=2m_H$ and $\mu=m_H/2$ for the range of scale variation),
whether or not we only look at jets at central rapidities, or when considering
the exclusive $1$-jet cross section.  We also note that using independent
variations for $\mu_f$ and $\mu_r$ does not change this picture, in particular
the $\mu_f$ variation for fixed $\mu_r$ is quite small.

\begin{figure}[t!]
\includegraphics[width=0.5\textwidth]{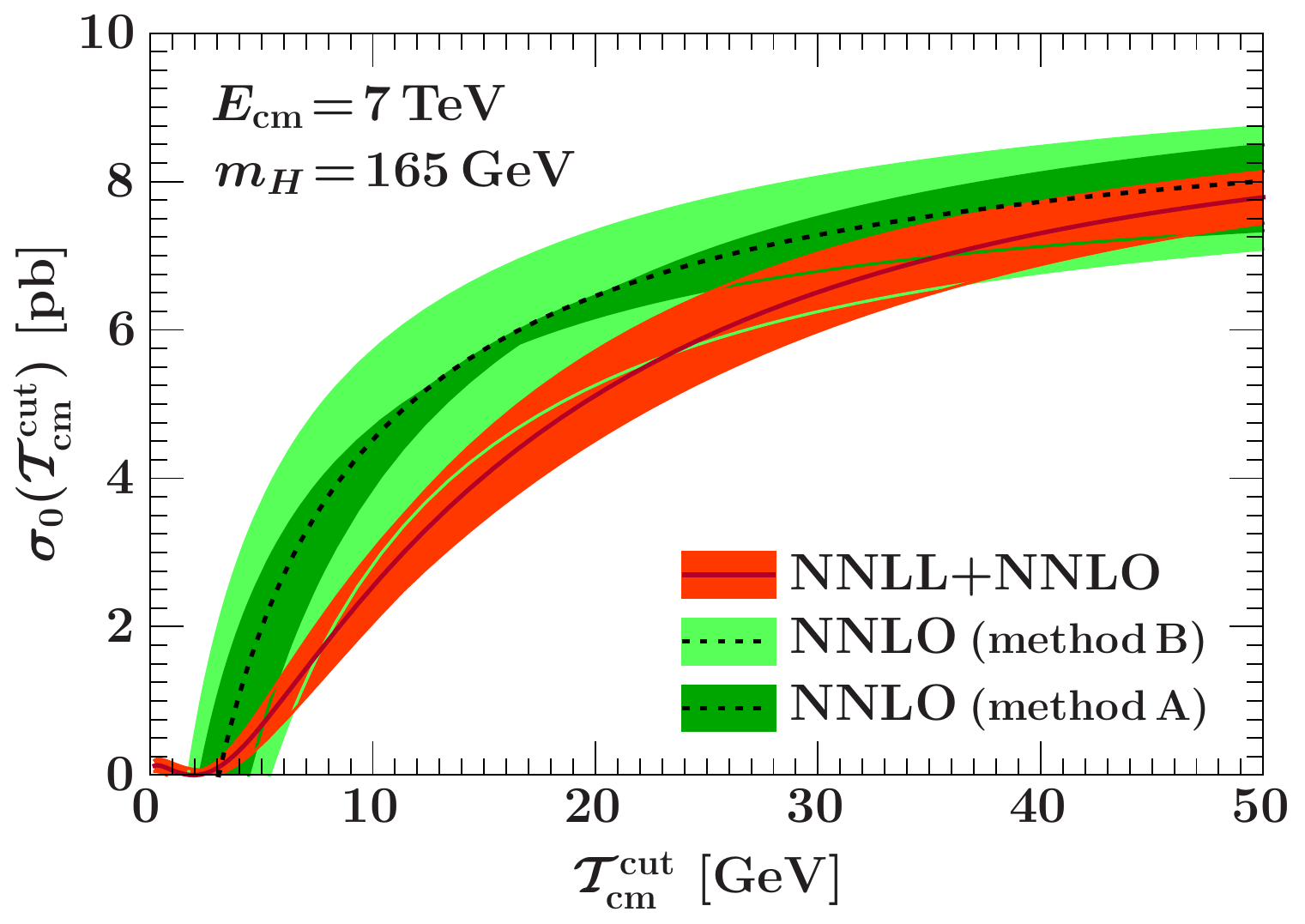}%
\hfill%
\includegraphics[width=0.5\textwidth]{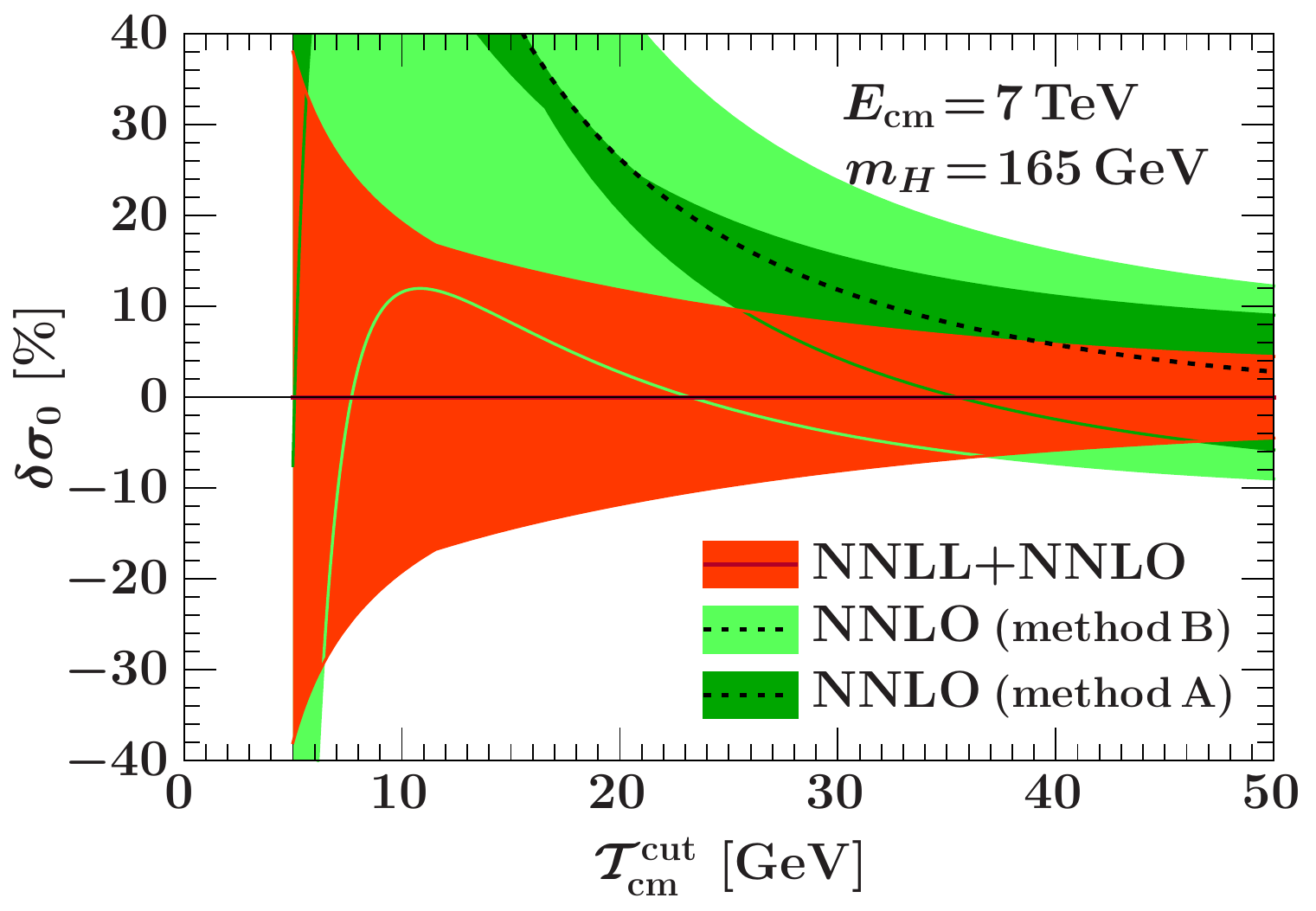}%
\\
\includegraphics[width=0.5\textwidth]{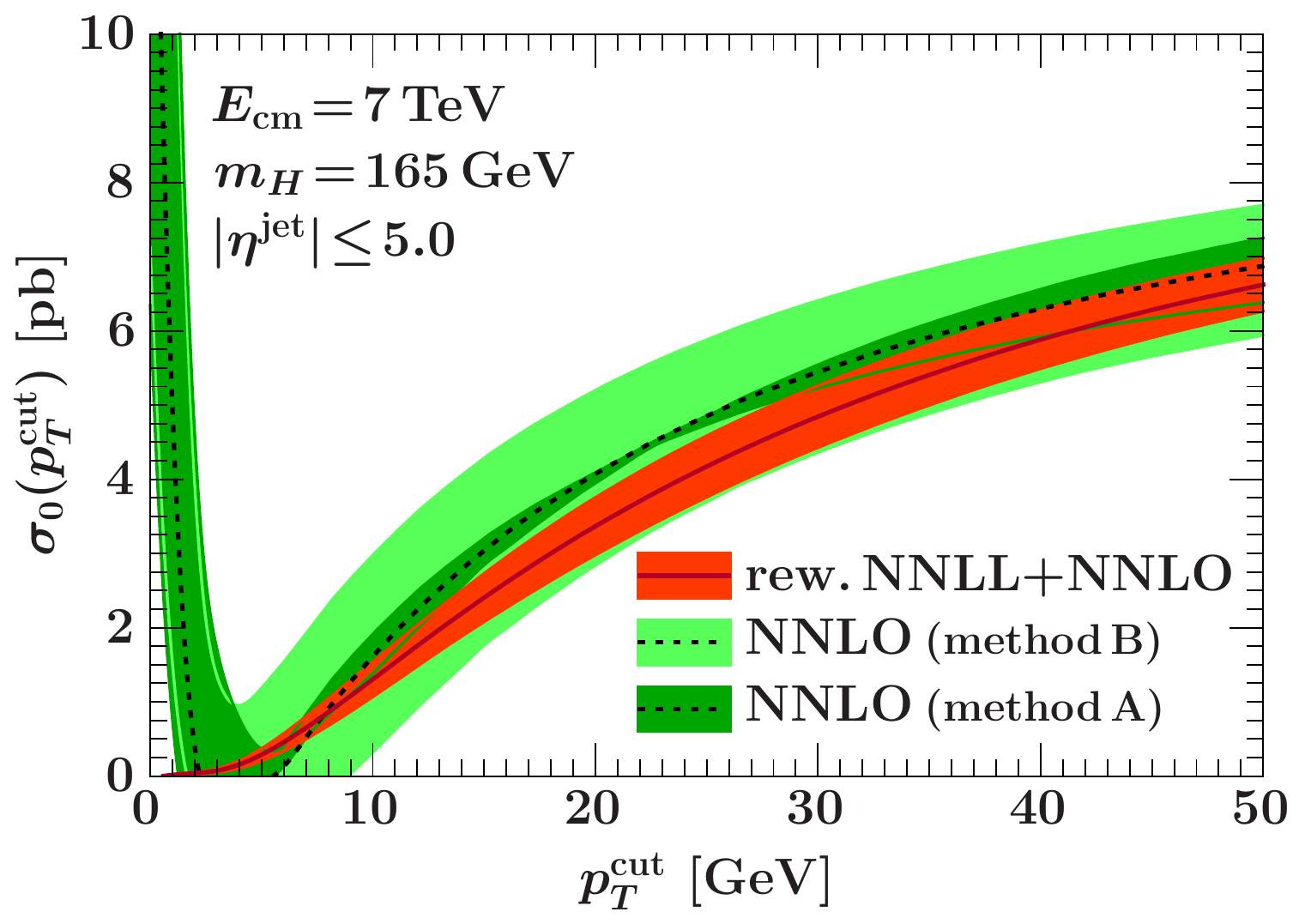}%
\hfill%
\includegraphics[width=0.5\textwidth]{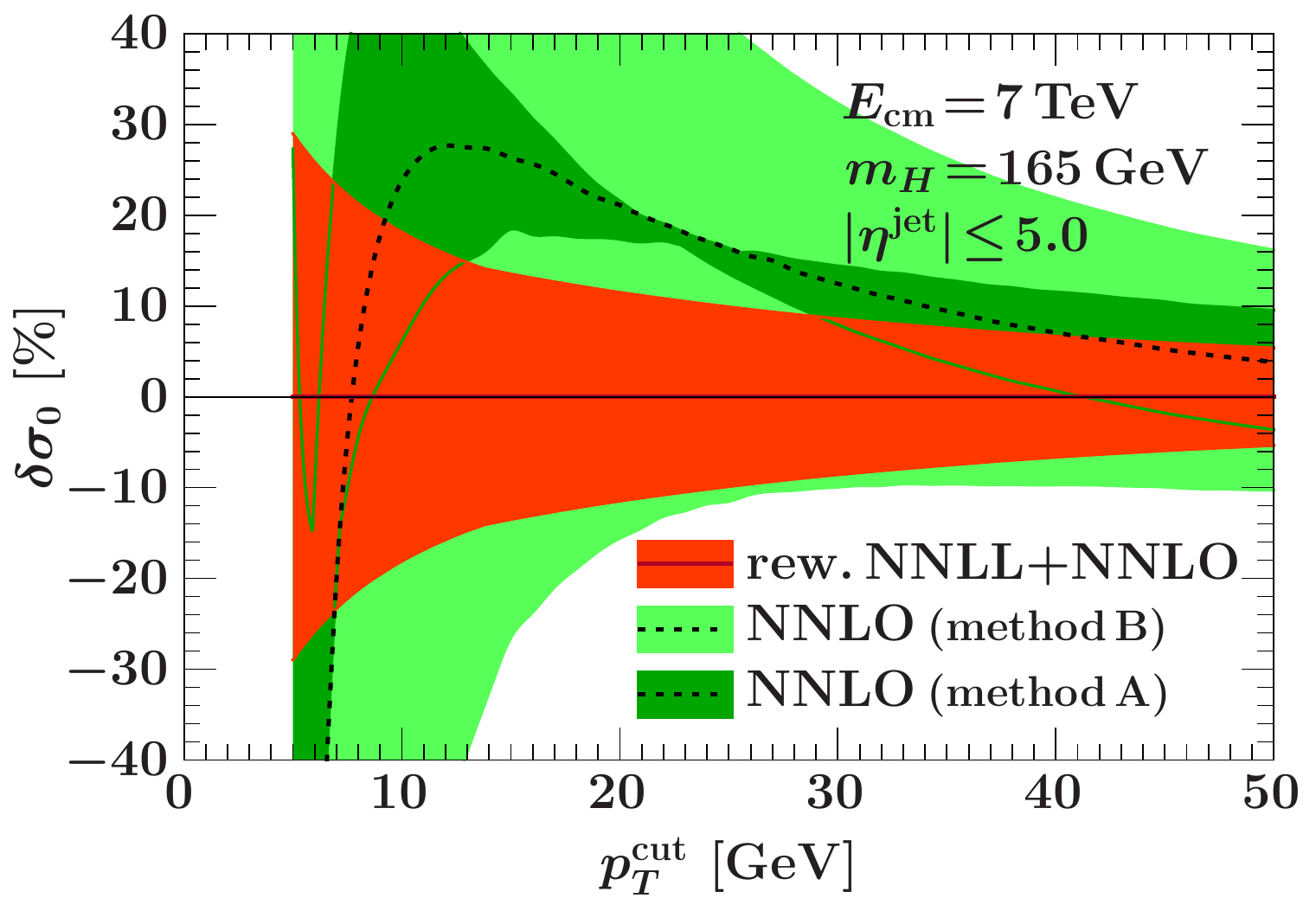}%
\vspace{-0.5ex}
\caption{\label{fig:ABC} Comparison of uncertainties for methods A, B, C for the
  0-jet bin cross section for beam thrust $\Tcm$ (top) and $p_T^{\rm jet}$ (bottom).
  Results are shown at NNLO with uncertainties from methods A and B and for the
  NNLL+NNLO resummed result using method C (reweighted for $p_T^\cut$).
  On the right all curves are normalized relative to the NNLL+NNLO central value.}
\end{figure}

For method C with $\Tcm$ we make use of resummed predictions for $H+0$ jets from
gluon fusion at next-to-next-to-leading logarithmic order (NNLL+NNLO) from
Ref.~\cite{Berger:2010xi}. This includes the correct NNLO fixed-order
corrections for $\sigma_0(\Tcmc)$ for any cut. The resulting cross section
$\sigma_0(\Tcmc)$ has the jet veto implemented by a cut $\Tcm\le \Tcmc$.  This
cross section contains a resummation of large logarithms at two orders beyond
standard LL parton shower programs.  A similar resummation for the case of
$p_T^\jet$ is not available. Instead, we use {\sc MC@NLO} and reweight it to the
resummed predictions in $\Tcm$, doing so for both the central curve as well as
each of the six scale variation curves needed for the uncertainty determination
in method C.\footnote{We thank Fabian St\"ockli for collaboration on this
  NNLL+NNLO reweighting analysis for $p_T^\cut$.} We then use the reweighted
Monte Carlo sample to obtain cross section predictions for the standard jet
veto, $\sigma_0(p_T^\cut)$. We will refer to this as the reweighted NNLL+NNLO
result.  Since the Monte Carlo here is only used to provide a transfer matrix
between $\Tcm$ and $p_T^\jet$, and both variables implement a jet veto, one
expects that most of the improvements from the higher-order resummation are
preserved by the reweighting. However, we caution that this is not equivalent to
a complete NNLL+NNLO result for the $p_T^\cut$ spectrum, since the reweighting
may not fully capture effects associated with the choice of jet algorithm and
other effects that enter at this order for $p_T^\cut$. The dependence on the
Monte Carlo transfer matrix also introduces an additional uncertainty, which
should be studied and is not included in our numerical results.  The transfer
matrix is obtained at the parton level, without hadronization or underlying
event, since we are reweighting a partonic NNLL+NNLO calculation.

From Fig.~\ref{fig:ABC} one observes that the resummation of the large jet-veto
logarithms (dark red central curve) lowers the cross section for both $\Tcmc$
and $p_T^\cut$. Comparing to NNLO for cut values $\gtrsim 25\GeV$ the relative
uncertainties in the resummed result of method C (dark orange bands) and the
reduction in the resummed central value are similar for both jet-veto variables.
Since one expects resummation to decrease the uncertainties, one can also see
that the NNLO uncertainties from method B are more consistent with the higher
order NNLL+NNLO resummed method C results than those in method A. We observe
that the uncertainties in method C are reduced by about a factor of two compared
to those in method B.  Since the zero-jet bin plays a crucial role in the $H\to
WW$ channel for Higgs searches, and these improvements will also be reflected in
uncertainties for the one-jet bin, the improved theoretical precision obtained
with method C has the potential to be quite important.

\begin{figure}[t!]
 \includegraphics[width=0.5\textwidth]{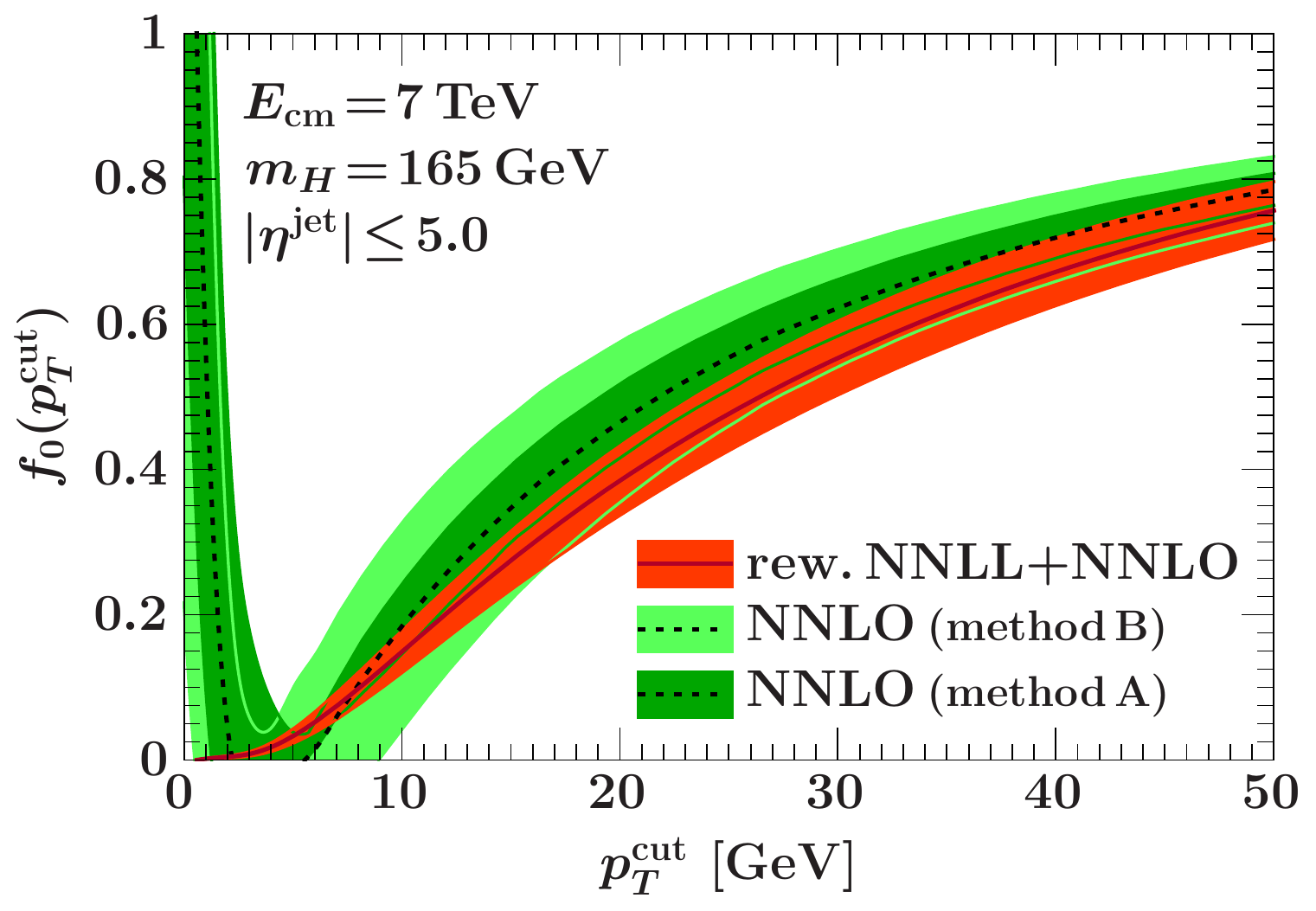}%
 \hfill%
 \includegraphics[width=0.5\textwidth]{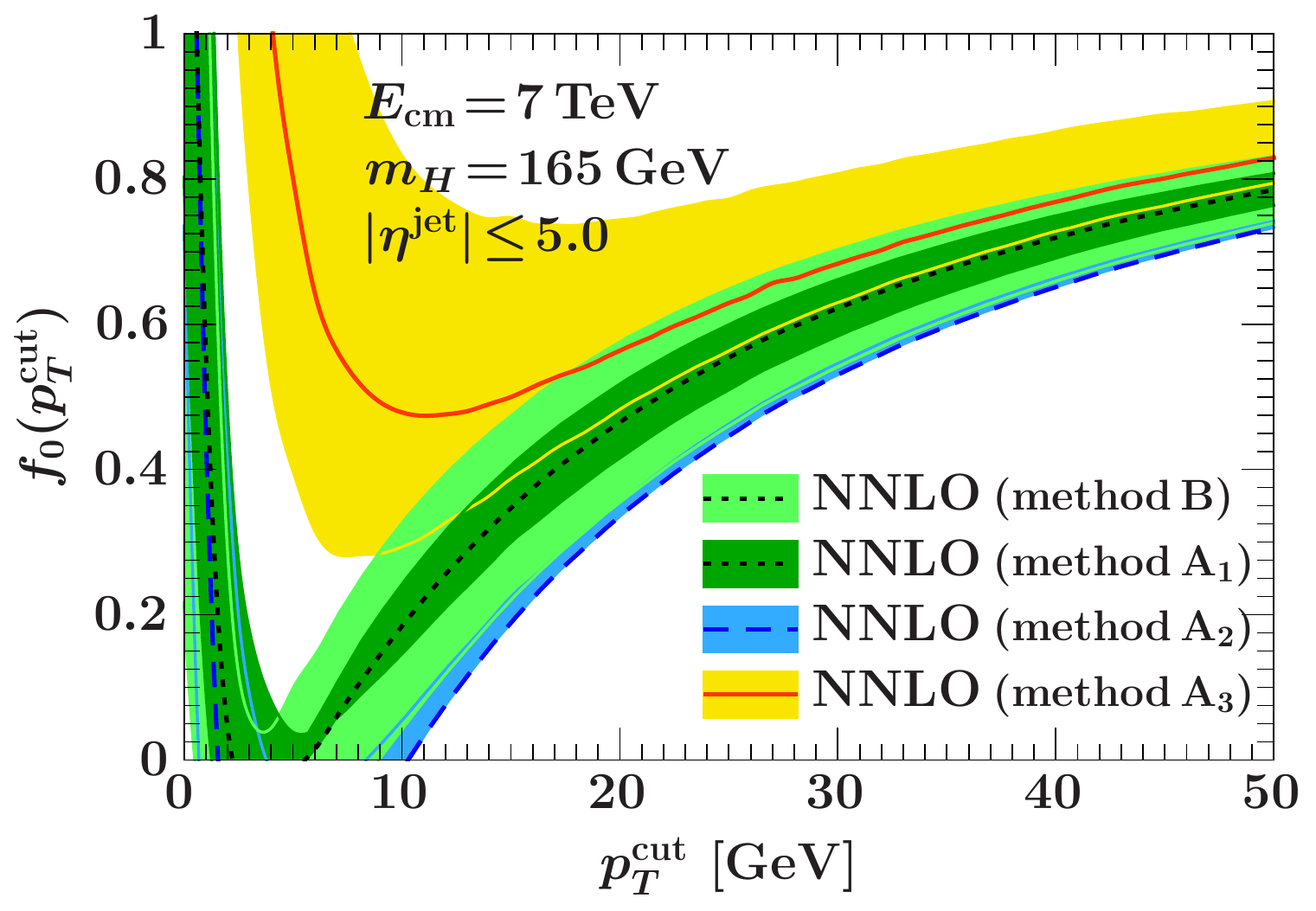}%
 \vspace{-0.5ex}
 \caption{\label{fig:f0} In the left panel we show the same three
   curves as in the bottom-left panel of Fig.~\ref{fig:ABC}, but for the event
   fraction $f_0(p_T^\cut)$ treated as a derived quantity from the jet-bin cross
   sections. In the right panel we contrast the uncertainties obtained using
   Eqs.~(\ref{eq:f0derived}) and (\ref{eq:f0expn}) together with method A, with
   the uncertainty obtained using method B.}
\end{figure}

In Fig.~\ref{fig:f0} we show results for the 0-jet event fraction $f_0$, with
$p_T^\cut$ as the jet-veto variable.  In the left panel we compare the
uncertainties in $f_0(p_T^\cut)$ that result from propagating the uncertainties
from the jet-bin cross sections obtained from methods A (medium green), B (light
green), and C (dark orange). The conclusions are analogous to the corresponding
cross-section results in the bottom-left panel of Fig.~\ref{fig:ABC}, namely
that method B provides a better estimate for the perturbative fixed-order
uncertainties than method A, and that the higher-order logarithmic summation
present in method C leads to a slightly smaller central value together with the
decrease to the uncertainty one expects from incorporating the resummation.  In
the right panel of Fig.~\ref{fig:f0} we show the results of the different
perturbative schemes for $f_0$ defined in Eq.~(\ref{eq:f0derived}) (middle dark
green band) and Eq.~(\ref{eq:f0expn}) (lower narrow blue band and upper wide
yellow band) each at NNLO and in each case obtaining the uncertainties using
method A (direct scale variation)~\cite{YellowBook2}(Sec.~5.3). For comparison,
the middle light green band shows the uncertainties obtained from method B. The
different method A schemes have a wide spread, which demonstrates the large size
of the higher-order perturbative corrections in the total and inclusive $1$-jet
cross sections. The central values of the alternative methods $A_2$ and $A_3$
are not covered by the method $A_1$ uncertainty band, but all three central
curves are covered by the larger uncertainty band from method B (except at small
$p_T^\cut$ where scheme 3 starts to diverge earlier than the other schemes).
This can be taken as a confirmation that method A tends to underestimate the
perturbative uncertainties in the fixed-order
results~\cite{YellowBook2}(Sec.~5.3), while method B produces more realistic
fixed-order uncertainties.

To appreciate the effects of the different methods on the correlation matrix we
consider as an example the results for $p_T^\cut = 30\GeV$ and
$\lvert\eta^\jet\rvert \leq 5.0$.  The inclusive cross sections are $\sigma_\total
= (8.76 \pm 0.80)\,\mathrm{pb}$ at NNLO, and $\sigma_{\geq1} =
(3.31\pm0.64)\,\mathrm{pb}$ at NLO. The relative uncertainties and correlations
at these cuts for the three methods are shown in Table~\ref{tab:corrcoeffs}.
The numbers for the cross sections are also translated into the equivalent
results for the event fractions, $f_0(p_T^\cut)=\sigma_0(p_T^\cut)/\sigma_\total$
and $f_{\ge 1}(p_T^\cut) = \sigma_{\ge 1}(p_T^\cut)/\sigma_\total$. Note that
method A should not be used due to the lack of a contribution corresponding to
$\Delta_\cut$ in this method, and the resulting underestimated $\delta\sigma_0$.
In methods B and C we see, as expected, that $\sigma_0$ and $\sigma_{\geq 1}$
have a substantial anti-correlation due to the jet-bin boundary they share.
%
\begin{table}[t]
\tabcolsep 10pt
  \centering
  \begin{tabular}{c | c c c }
  \hline\hline
  & method A & method B & method C \\
  \hline\hline
  $\delta\sigma_0(p_T^\cut)$         & $3\%$  & $19\%$  & $9\%$ \\
  $\delta\sigma_{\ge 1}(p_T^\cut)$   & $19\%$ & $19\%$  & $14\%$ \\
  $\rho(\sigma_\total,\sigma_0)$       & $1$    & $0.78$  & $0.15$ \\
  $\rho(\sigma_\total,\sigma_{\ge 1})$ & $1$    & $0$     & $0.65$ \\
  $\rho(\sigma_0,\sigma_{\ge 1})$    & $1$    & $-0.63$ & $-0.65$ \\
\hline
  $\delta f_0(p_T^\cut)$             & $6\%$  & $13\%$  & $9\%$ \\
  $\delta f_{\ge 1}(p_T^\cut)$       & $10\%$ & $21\%$  & $11\%$ \\
  $\rho(\sigma_\total,f_0)$            & $-1$   & $0.43$  & $-0.38$ \\
  $\rho(\sigma_\total,f_{\ge 1})$      & $1$    & $-0.43$ & $0.38$ \\
  \hline\hline
  \end{tabular}
  \caption{Example of relative uncertainties $\delta$ and correlations $\rho$ obtained for the
    LHC at $7\,{\rm TeV}$ for $p_T^\cut=30\,{\rm GeV}$ and $\lvert\eta^\jet\rvert \leq 5.0$.
    }
\label{tab:corrcoeffs}
\end{table}

\subsection{Conclusion}
\label{sec:jetbin_conclusion}

To summarize, we have discussed the implications of separating LHC cross
sections into jet bins, using Higgs production from gluon fusion as a concrete
example.  The jet binning induces logarithmic dependences on the jet-bin
boundary which is important to properly take into account when making
predictions and estimating perturbative uncertainties. When using fixed-order
predictions only, the additional logarithms at higher orders in perturbation
theory caused by the jet binning can be taken into account in the perturbative
uncertainty estimate using method B. By resumming the jet-binning logarithms one
can obtain improved predictions with reduced (and more sophisticated)
uncertainties using method C.

Here we have focused our discussion on $\sigma_0$ and $\sigma_{\geq 1}$ and how
to take into account the resulting jet-bin boundary.  To further separate
$\sigma_{\ge 1}$ into a one-jet bin $\sigma_1$ and a $\sigma_{\ge 2}$ one can
use method B for this boundary by treating $\Delta_{\ge 2}$ as uncorrelated with
the total uncertainty for $\sigma_{\ge 1}$ from either methods B or C. Examples
of utilizing method B for this jet bin boundary can be found in
Ref.~\cite{Stewart:2011cf}. Once it becomes available one can also use a resummed
prediction with uncertainties for this boundary with method C.

\section*{Acknowledgments}

This work was supported in part by the Office of Nuclear Physics of the U.S.
Department of Energy under the grant DE-FG02-94ER40818. Preprint: MIT--CTP 4340

}

\section[A NLO benchmark comparison for inclusive jet production at hadron colliders]
{A NLO BENCHMARK COMPARISON FOR INCLUSIVE JET PRODUCTION AT HADRON COLLIDERS \protect\footnote{Contributed by: J.~Gao, Z.~Liang, P.~M.~Nadolsky}}
\label{sec:incljet}
{\graphicspath{{gao_nadolsky1/}}

\title{A NLO benchmark comparison for inclusive jet production at hadron colliders}

\author{Jun Gao, Zhihua Liang, Pavel M. Nadolsky}
\institute{Department of Physics, Southern Methodist University,
Dallas, TX 75275-0175, USA}


\begin{abstract}
We present a benchmark comparison of two
next-to-leading order (NLO) calculations
for inclusive jet and jet pair production
at hadron colliders. A new version of the NLO code EKS is adapted
for computation of  differential cross sections and compared to
an independent calculation based on the FastNLO code. A percent-level
agreement between the two codes is observed for specified settings of
computations at typical transverse momenta and rapidities
of Tevatron and LHC measurements. We identify theoretical
prerequisites for achieving such level of agreement and comment
on the stability of NLO calculations with respect
to the factorization scale choice.
\end{abstract}

\subsection{INTRODUCTION}
Inclusive jet production at hadron colliders provides an excellent
opportunity to test perturbative QCD (PQCD) and look for possible
new physics beyond the Standard Model (SM) over a wide range of
energy scales. Single inclusive jet production in the Tevatron Run-2
has been recently used to determine the QCD coupling
constant~\cite{Abazov:2009nc} and constrain parton distribution
functions (PDF) in the proton in global QCD analyses by several
groups~\cite{Lai:2010vv,Guzzi:2011sv,Martin:2009iq,Ball:2010de}. Jet
production data provide constraints on the gluon PDF at large $x$
values, possibly in a combination with small-$x$ quark
PDFs, as discussed in Section~\ref{sec:gluonpdf}.
Invariant mass distributions of
dijets~\cite{Chatrchyan:2011ns}, angular
distributions~\cite{Khachatryan:2011as,Aad:2011aj}, and other jet
observables at the LHC~\cite{:2010wv,Chatrchyan:2011qta,CMS:2011ab}
provide a unprecedented opportunity to extend searches for quark
compositeness and new particle resonances toward the highest
energies attainable.

In this contribution, we examine agreement between the computer
programs that are available for NLO calculations of jet
production cross sections. NLO QCD predictions for jet
production work remarkably well in a wide kinematical range and
across many orders of magnitude of the cross sections. Nonetheless,
the latest PDF analysis evaluates many scattering processes up to NNLO
in perturbative QCD. Jet production observables are pivotal for
constraining the large-$x$ gluon PDF, but remain known to
NLO only. We identify and document main factors
affecting NLO jet cross sections
at a few-percent level of accuracy and compare the numerical results
for typical collider kinematics. Differences between the programs
used, and choices for the theoretical inputs made, may be responsible
for some differences observed between CT10 and other PDFs, as
explained below. Such NLO benchmark comparison
will be useful for quantifying or reducing the uncertainties on the
resulting PDFs and for the future implementation
of NNLO and higher-order resummed contributions to the jet cross sections.

From the experimental point of view, jet production
has an advantage of very high statistics and a drawback
of sizeable systematical errors associated with complexities of jet
reconstruction. NLO theoretical uncertainties due to the
QCD scale dependence and the fixed-order model
for the jet algorithm are comparable to the experimental errors.
Control of numerical accuracy involves,
in particular, careful tuning of Monte-Carlo integration to handle
steeply falling jet cross sections.

An early numerical code (EKS) for the NLO calculation of
single-inclusive jet and dijet distributions
was developed by S. D. Ellis, Z. Kunszt and D. E. Soper
in 1990's~\cite{Ellis:1992en} based on the subtraction method.
Two other widely used numerical programs are
NLOJET++~\cite{Nagy:2001fj,Nagy:2003tz}
and FastNLO~\cite{Kluge:2006xs,ftnlo:2010xy}. The latter provides
a fast interface to obtain NLO predictions in
kinematical bins of already published experimental jet cross sections by
interpolating table files produced by NLOJET++.
Besides these {\it fixed-order} calculations,
POWHEG combines the NLO jet production cross sections with
{\it leading-logarithm} QCD showering effects~\cite{Alioli:2010xa}.
Some phenomenological studies also include partial
NNLO contributions to jet cross sections obtained by
threshold resummation \cite{Kidonakis:2000gi}.

The agreement between the above NLO numerical programs is not
automatically met, which motivates the present benchmark comparison.
The past CTEQ PDF analyses computed NLO jet cross sections using
NLO K-factor tables produced by the EKS code, while other PDF analysis groups
use FastNLO. Since the CT10 NLO gluon PDF behaves somewhat
differently at large $x$
than the gluon PDF from MSTW'08 or other groups \cite{Martin:2009iq},
one must compare the EKS and FastNLO computations for the same input
values to confirm that these programs do not cause the observed disagreement.

Here we show that the results for the Tevatron ($\sqrt{s}=1.96\mbox{
  TeV}$) and LHC ($\sqrt{s}=7\mbox{ TeV}$)
from EKS and FastNLO agree well when the
computation parameters are chosen as described in the next section.
These settings must be consciously controlled in order to reach acceptable
agreement. As a result of this work, the EKS code
has been revised to improve its stability and
efficiency  and to implement output into new differential cross
sections \cite{EKS2012}.

\subsection{Theoretical setup and inputs}
Several theoretical inputs must be matched exactly between the EKS and
FastNLO programs in order to reach the level of agreement shown in the
figures below.

\begin{itemize}
\item {\bf Jet algorithm.} When
calculating the distribution of jet observables, we need to use the
same jet algorithms as the ones in the experimental measurements. In
this comparison, we utilized the
cone-based Midpoint algorithm~\cite{Blazey:2000qt} for the Tevatron
observables and cluster-based anti-$k_T$ algorithm~\cite{Cacciari:2008gp}
for the LHC. The only difference between the Midpoint algorithm and modified
Snowmass algorithm~\cite{Blazey:2000qt} used in the original EKS
program is that the Midpoint algorithm always starts with the middle point
between the two partons' directions as a seed for a new protojet, no
matter how large their separation is. In the NLO theoretical
calculations for single-jet or dijet production that include at most
three final-state partons, the cluster-based
$k_T$~\cite{Catani:1993hr}, anti-$k_T$, and Cambridge-Aachen
(CA)~\cite{Wobisch:1998wt} algorithms are equivalent.
\item {\bf The recombination scheme} is a procedure for
merging two nearby partons into one jet. For example, the energy
scheme (4D, based on adding the 4-momentum) or $E_T$ scheme (based
on adding the scalar $E_T$, then averaging over the partons'
directions using $E_T$ as the weights) can be employed to find the
momentum of the merged jet~\cite{Salam:2009jx}. Our comparison uses
the energy scheme for both the Tevatron and LHC measurements, as it
is often used by the recent experiments. Different choices of the
recombination scheme can cause differences of up to ten percent in
the NLO distributions, as will be shown later. Note that, with the
energy scheme, the jet could be massive, which means that the jet's
pseudorapidity will not be equal to its rapidity.
\item {\bf The jet trigger} imposes
acceptance conditions on each jet's $p_T$ or rapidity when deciding
if this jet's contribution is included into the jet observable. In
NLO calculations of single-inclusive jet distributions, the jet
trigger conditions have no influence. In dijet production, they may
change the cross sections by small amounts by affecting the
selection of two leading jets in some cases. In our dijet
calculations we choose $p_T>40\,{\rm GeV}$, $|y|<3$ for each jet at
the Tevatron and $p_T>30\,{\rm GeV}$, $|y|<3$ at the LHC.
\item {\bf Renormalization and factorization scales.} The scale
choice is only related to theory and has no correspondence in
experiment. It is conventional to choose the renormalization and
factorization scales to be of order of the typical transverse
momentum $p_T$ of the jet(s): $\mu_R \sim \mu_F \sim p_T$.
In contributions with two resolved jets,
$p_T$  naturally corresponds to the transverse momentum
of either of the final-state jets (which are equal by momentum
conservation). More ambiguity is present in contributions
with three resolved jets, when $p_T$ can correspond
to the transverse momentum of either of the jets in each event
or to a combination of three transverse momenta.
{\it A meaningful comparison must use equivalent definitions
of ``jet $p_T$'' in the renormalization and
factorization scales of both NLO calculations.}

When FastNLO interpolates tables of NLOJET++ cross sections for
single inclusive-jet production, it sets
$\mu_{R}$ and $\mu_F$  proportional to the $p_T$ value at a
fixed point in each $p_T$ bin of the experimental data.
Given the high precision of the latest PDF analyses,
the FastNLO scale convention produces a numerically different
result than the scale proportional to the $p_T$
of the leading jet or the average $p_T$ of two leading jets in each event.
It depends on the binning of the experimental data and is
numerically close to the average $p_T$ in each bin for small enough
bins.

In the EKS calculations for single-jet production, we set the scale
proportional to $p_T$ of each individual jet in any $p_T$ bin, which
means that we repeat the evaluation of the matrix
elements with three resolved
jets (contributing to three $p_T$ bins) by successively
setting $\mu_{R,F}$ to be proportional to the
$p_T$ of each jet in the event. Such matrix elements  are thus
evaluated three times. This event-level scale setting of
EKS turns out to be numerically close to the bin-level scale setting
of FastNLO if the bin sizes are small. However, a few-percent
differences are still observed at the largest rapidities and $p_T$.
For dijet production, FastNLO and EKS choose
the $\mu_{R}$ and $\mu_F$ scales that are proportional to
the average $|p_T|=(|p_{T1}|+|p_{T2}|)/2$ of the two leading jets.

\item{\bf Monte-Carlo integration.} Precision calculations for jet
  production are numerically challenging because of the rapid falloff
  of the cross sections with the jet's $p_T$ and rapidity, and also because of
  large numerical cancellations occurring between some $2\rightarrow 2$
  and $2\rightarrow 3$ contributions.
  Both EKS and NLOJET++ evaluate differential
  cross sections by Monte-Carlo integration, which requires to generate
  of order $10^9$ of sample points to achieve percent-level
  accuracy for the whole kinematical region. The upgraded EKS code
performs the Monte-Carlo integration using the VEGAS method from the CUBA2.1
library~\cite{Hahn:2004fe}. The EKS output is produced in the form of
two-dimensional cross sections ($d^2\sigma/(dp_T\,dy)$,
$d^2\sigma/(dM_{jj}\,dy)$, ...) and stored in finely binned two-dimensional
 histograms. Such output is ``almost fully differential'' in the sense that
the finely grained histograms can be rebinned into any set
of coarse bins of the given experiment at the stage
of the user's final analysis. This format is different from the
FastNLO format, which provides the cross sections in coarse bins
taken from pre-existing experimental publications.

The fine binning in EKS is introduced at the stage of Monte-Carlo
integration in order to improve convergence and to better handle the
NLO cancellations. The Monte Carlo sampling pattern is tuned
automatically to ensure that all fine bins are filled with
comparable numbers of sample points, regardless of the momentum and
scattering angle values associated with each bin.  Then we get
uniform relative errors on the cross sections in all bins without
consuming too much CPU time, and despite the dramatic variation of
cross sections across the bins. Finally, EKS includes a module to allow
for flexible choices of scales $\mu_R$ and $\mu_F$, and another
module for calculating differential cross sections of user-provided
jet observables.
\end{itemize}

\subsection{RESULTS}
Figs.~\ref{jetcom_d0inc}-\ref{jetcom_cmsdim} compare our
representative numerical results with the ones provided by FastNLO
for $p_T$ distributions of single jets, invariant mass distributions
of dijets, and (in the case of D0 Run-2) angular distributions
($\chi$) of dijets. Kinematical bins of the Tevatron
($\sqrt{s}=1.96\mbox{
TeV}$)~\cite{:2008hua,Aaltonen:2008eq,Abazov:2010fr,:2009mh} and LHC
($\sqrt{s}=7\mbox{ TeV}$)~\cite{Chatrchyan:2011qta,CMS:2011ab}
measurements, and CTEQ6.6 PDFs~\cite{Nadolsky:2008zw} were used. The
cone sizes $R$ of the jets are indicated in the figures.

Left panels in the figures show ratios of EKS to FastNLO cross
sections, $\sigma_{\rm EKS}/\sigma_{\rm FastNLO}$, at the LO (red
points) and NLO=LO+NLO correction (blue points), in kinematical bins
provided by the experiments. The horizontal axis indicates the ID of
each bin, which are arranged in the order of increasing jet rapidity
$y$ and then jet's $p_T$ for inclusive jet production, $y$ and then 
$M_{jj}$ for dijet production, and $M_{jj}$ then $\chi$ for dijet angular
dependence. Vertical lines indicate the boundaries of each rapidity
interval for single-jet and dijet distributions, and of each 
dimass interval for the $\chi$ distribution. For example, Fig.~\ref{jetcom_d0inc}
shows $\sigma_{\rm EKS}/\sigma_{\rm FastNLO}$ in 6 bins of jet
rapidity, with bins 1...23 corresponding to the first rapidity bin
($|y|<0.4$), bins 24...45 corresponding to the second rapidity bin
($0.4<|y|<0.8$), and so on.

The left panel includes, from top to bottom, three
plots obtained with the renormalization and factorization scales
equal to 1/2, 1, and 2 times the center scale.
We can see a good overall agreement between EKS and FastNLO both at LO
and NLO. The only
significant discrepancies are found in the highest $p_T$ bins for
both the Tevatron and LHC single inclusive jet production, which
may be due to the difference in the scale choices used in EKS and
FastNLO. [These differences reduce when going to NLO].
In the EKS single-jet calculation, we use the actual
$p_T$ of the partonic jet filled into the bin as the scale input.
FastNLO sets the scale according to a fixed $p_T$ value in
each experimental bin, which tends to be different from the EKS scale
in the highest $p_T$ bins, which have large widths.
The same reason causes a small normalization shift in the other
$p_T$ bins.

For dijet production, we only observe random
fluctuations at highest $M_{jj}$
that are mainly due to numerical integration errors.

In the right panels of Figs.~\ref{jetcom_d0inc}-\ref{jetcom_cmsdim}, we
present plots of the NLO K factor from EKS for each distribution, defined as
the ratio of the NLO differential cross section to the LO
one. The value of the K factor and its stability with
respect to the scale choice may provide an indication of the
magnitude of yet higher-order corrections.

To minimize the potential effect of higher-order terms,
one might opt to choose the renormalization and factorization scales
that bring the K factor close  to unity in most of the kinematical
region. An alternative approach for setting the scale is based on
the minimal sensitivity method, which suggests to
choose the $\mu_R$ and $\mu_F$ values (taken to be equal and designated as
$\mu$ in the following) at the point where the scale
dependence of the NLO cross section is the smallest.

In (di)jet production at central rapidities at the
Tevatron, both requirements ($K\approx 1$\linebreak and
$d\sigma_{NLO}(\mu)/d\mu \approx 0$)
could be satisfied by choosing $\mu \approx 0.5 p_T$;
see, {\it e.g.}, the appendix in Ref.~\cite{Stump:2003yu}.
For this reason, the scale $p_T/2$ was used in the CT10 study.
However, the point of
the minimal sensitivity shifts to higher values (close to $p_T$ or
even higher) at forward rapidities at the Tevatron
or at all rapidities at the LHC.
For such higher scales, however,
it is hard to satisfy the requirement that
$K$ remains close to unity at the same time.

This point is illustrated by our plots of the $K$ factors. At the
central rapidities and $\mu_R =\mu_F=0.5 p_T$ at the Tevatron
(the lowest 3 rapidity bins
in Figs.~\ref{jetcom_d0inc}-\ref{jetcom_d0dic}), $K\approx 1$ and is
relatively independent of $p_T$, as seen in the top
subpanels. However, with this scale choice the K factor deviates
significantly from unity and has strong kinematic dependence
if the rapidity and $p_T$ are large. If one chooses the scale that is equal
to $p_T$ or even $2p_T$ (the middle and bottom figures),
in accord with the minimal sensitivity method for the forward bins,
the kinematical dependence of the K factor reduces,
but its value increases to
1.3-1.6 in most of the bins.

For CMS kinematics (Figs.~\ref{jetcom_cmsinc}-\ref{jetcom_cmsdim}),
the $K$ factor has significant kinematical dependence
for all central scale choices,
however, the choice $\mu_R=\mu_F=p_T$ (the middle subpanels) results in a
comparatively flatter $K$ factor that is also closer to unity.
We can see that it is hard to find a fixed scale (or a scale of the
type $p_T\times (\mbox{a function of\quad} y)$ \cite{Ellis:1992en})
that would simultaneously reduce the magnitude of the NLO correction
and stabilize its scale dependence and
kinematical dependence. The scale $0.5 p_T$ may be
slightly more optimal at the Tevatron, and the scale $p_T$
may be slightly better at the LHC. In the absence of a
clearly superior scale choice, it may be necessary
to vary the scale of jet cross sections
in the global fit in order to estimate its effect on the PDF errors.

In Figs.~\ref{jetcom_sinc} and \ref{jetcom_sdim}, we plot the ratios of
the NLO distributions calculated using different recombination
schemes, where $\sigma_{4D}$ is obtained with the energy scheme, and
$\sigma_{E_T}$ is with the $E_T$ scheme. For
single inclusive jet production at both the Tevatron and LHC,
$\sigma_{E_T}$ is larger then $\sigma_{4D}$. An opposite trend is
observed in dijet production. Differences of the predictions based on the
two schemes are larger with the Midpoint algorithm (used at the
Tevatron) than with the anti-$k_T$ algorithm (used at the LHC).
In an NLO calculation, the Midpoint
algorithm allows a larger maximal angular separation (2$R$)
between the two partons forming a jet, compared to the
anti-$k_T$ algorithm that only allows the angular separation up to $R$.
This produces the shown kinematical differences between the two schemes.

\subsection*{CONCLUSIONS}
Jet production plays an important role at hadron colliders
and is a main background process
in the bulk of new physics searches. A benchmark comparison
of NLO QCD predictions for jet production
from different numerical codes can be useful for both the ongoing
phenomenological studies and upcoming higher-order calculations.
In this work we modify the original EKS program
and compare the single-jet and dijet cross sections that it produces
with the ones from the FastNLO program. We find a good agreement
between two programs, apart from
differences of up to 5-10\% occuring at the highest jet $p_T$'s and rapidities.
We document the exact combination of theoretical settings in EKS that are needed
to reproduce the FastNLO results. Based on the EKS calculation,
we attempted to identify the choice of the renormalization and
factorization scales that could simultaneously reduce the magnitude
of NLO $K$ factors and/or scale dependence of the NLO cross section.
Since we could not easily find such a scale combination,
we propose to vary the factorization and
renormalization scales in future (N)NLO PDF fits
to better estimate theoretical uncertainties in the
resulting PDFs. There is a plan to publish the updated EKS program
in the near future \cite{EKS2012}.

\subsection*{ACKNOWLEDGMENTS}
The work on this preprint 
was supported by the U.S. DOE Early Career Research Reward
DE-SC0003870 and by Lightner-Sams Foundation. We thank our collaborators 
who have contributed to the development of the EKS code, 
Hung-Liang Lai, Zhao Li, Dave Soper, 
and C.-P. Yuan. PMN appreciates helpful
discussions with J. Huston and CTEQ
members. He also benefited from discussing related work with participants
of the Workshop ``High Energy QCD at the start of the LHC''
at the Galileo Galilei Institute of Theoretical
Physics (INFN Florence, September 2011). PMN thanks
the organizers of this workshop for the financial support and hospitality.

\begin{figure}
\begin{center}
\includegraphics[width=1.0\textwidth]{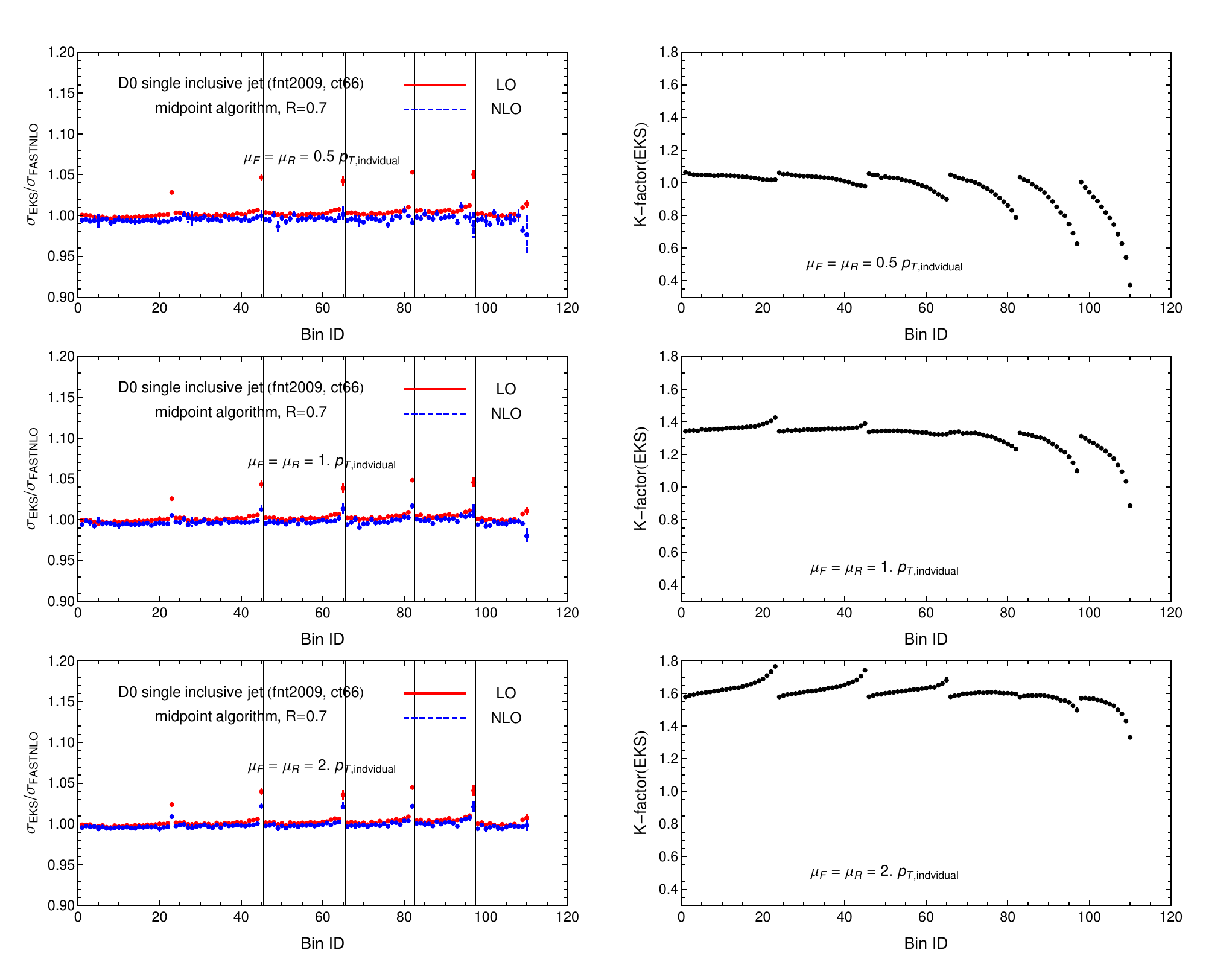}
\end{center}
\caption{\label{jetcom_d0inc} Comparison of $p_T$ distributions
for single inclusive jet production from EKS and FastNLO for D0
Tevatron Run II measurement.\cite{:2008hua}}
\end{figure}
\begin{figure}
\begin{center}
\includegraphics[width=1.0\textwidth]{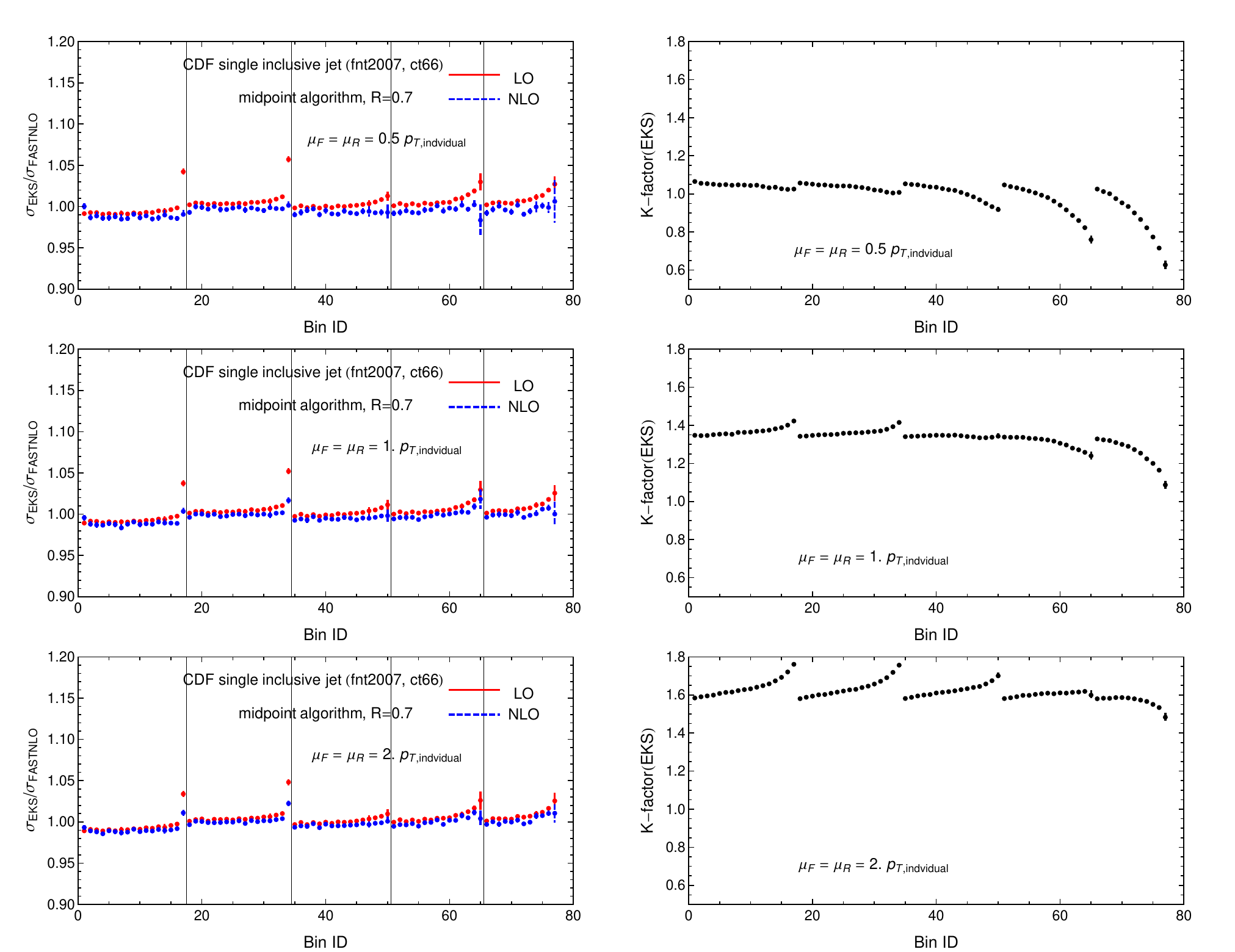}
\end{center}
\caption{\label{jetcom_cdfinc} Comparison of $p_T$ distributions
for single inclusive jet production from EKS and FastNLO for CDF
Tevatron Run II measurement.\cite{Aaltonen:2008eq}}
\end{figure}
\begin{figure}
\begin{center}
\includegraphics[width=1.0\textwidth]{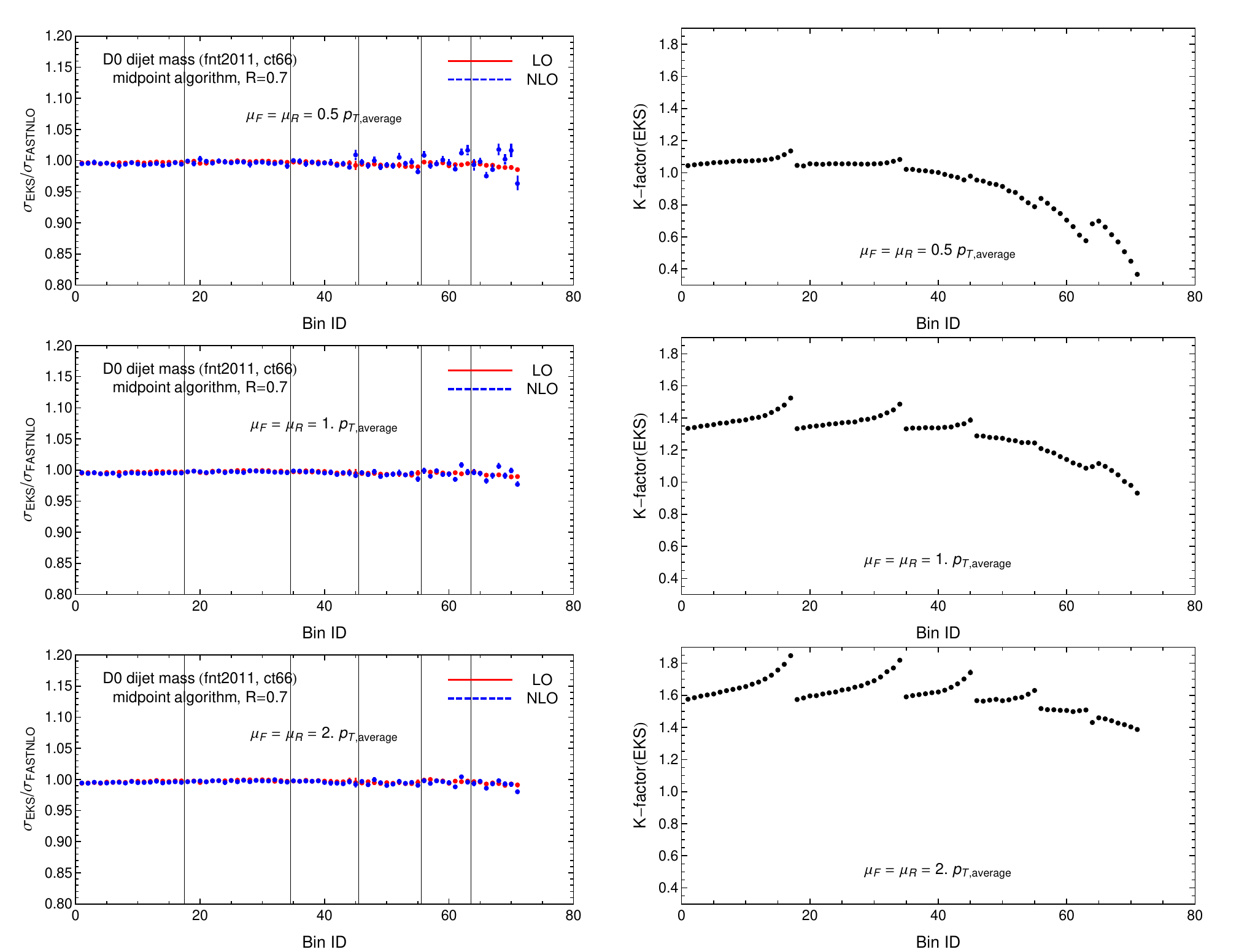}
\end{center}
\caption{\label{jetcom_d0dim} Comparison of invariant mass
distributions for dijet production from EKS and
FastNLO for D0 Tevatron Run II measurement.\cite{Abazov:2010fr}}
\end{figure}
\begin{figure}
\begin{center}
\includegraphics[width=1.0\textwidth]{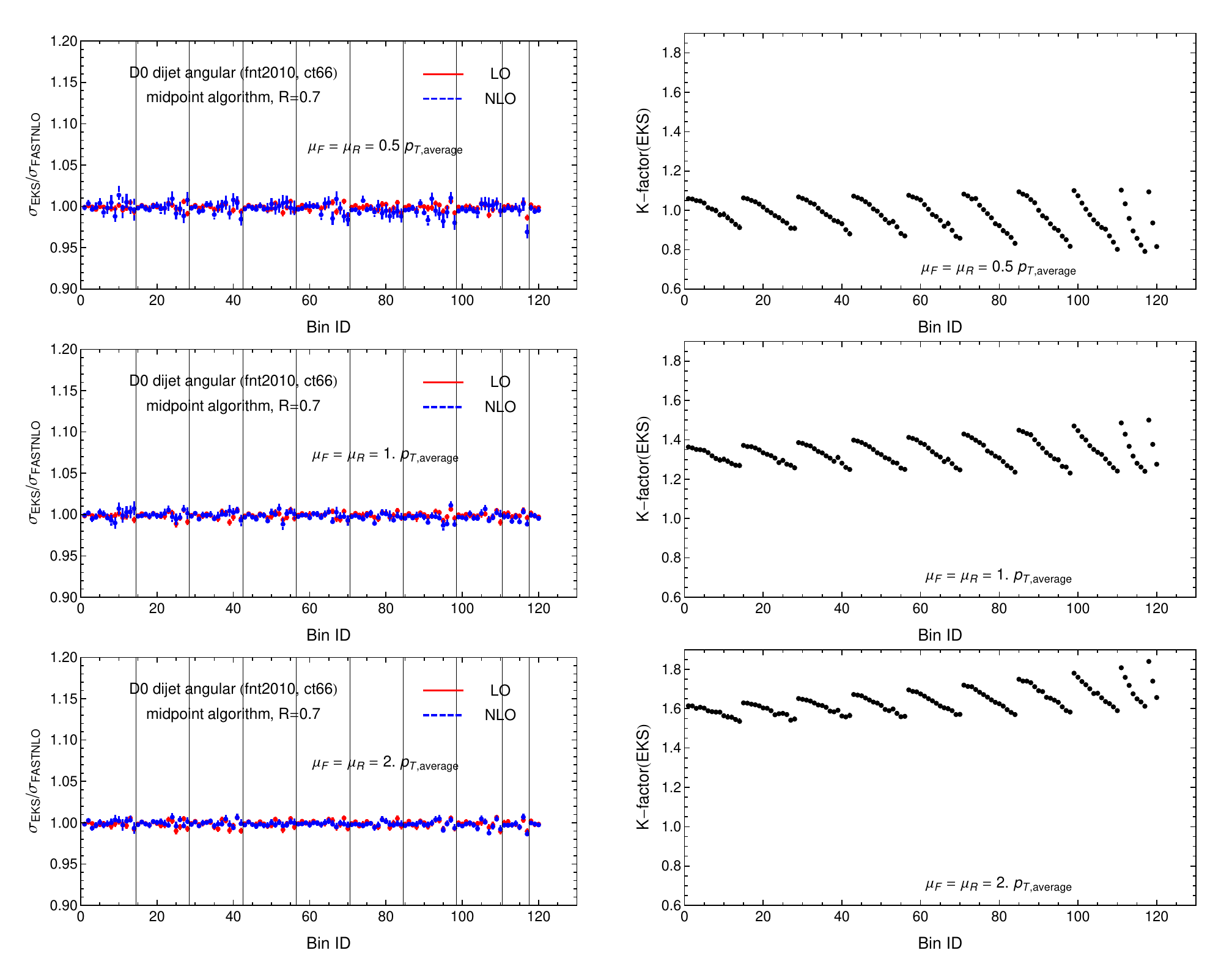}
\end{center}
\caption{\label{jetcom_d0dic} Comparison of angular ($\chi$)
distributions for dijet production from EKS and FastNLO for D0
Tevatron Run II measurement.\cite{:2009mh}}
\end{figure}
\begin{figure}
\begin{center}
\includegraphics[width=1.0\textwidth]{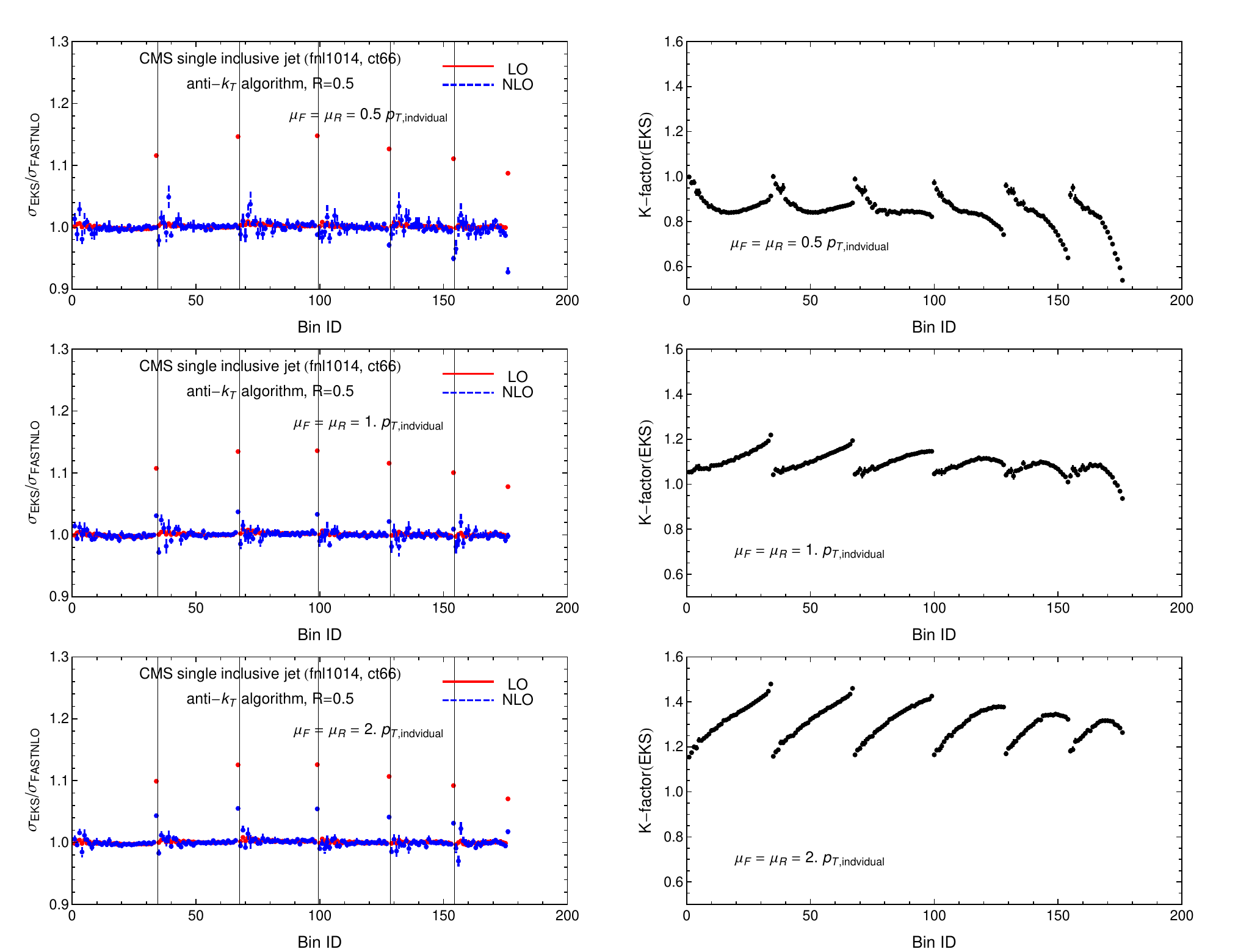}
\end{center}
\caption{\label{jetcom_cmsinc} Comparison of $p_T$ distributions
for single inclusive jet production from EKS and FastNLO for CMS
LHC (7\,TeV) measurement.\cite{CMS:2011ab}}
\end{figure}
\begin{figure}
\begin{center}
\includegraphics[width=1.0\textwidth]{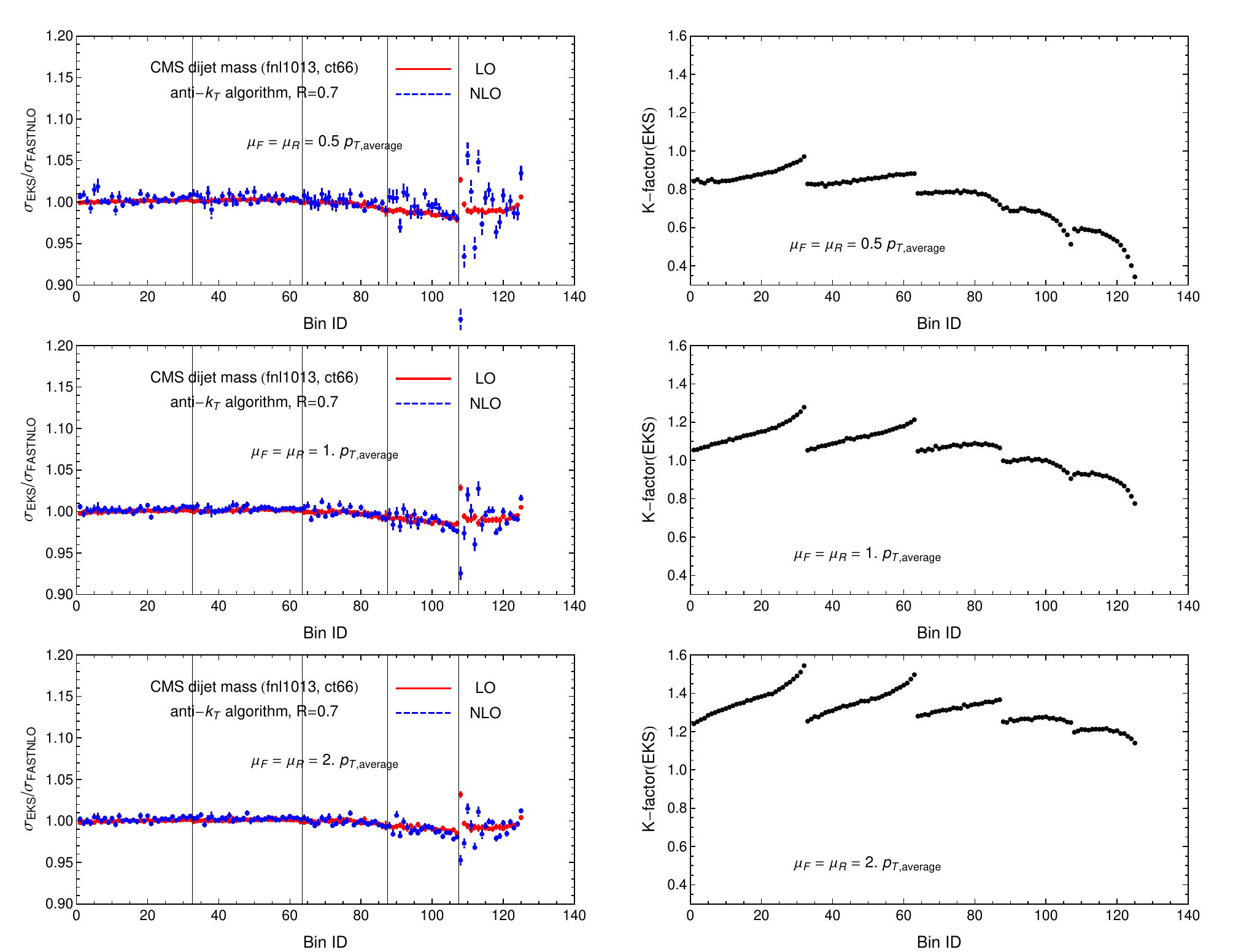}
\end{center}
\caption{\label{jetcom_cmsdim} Comparison of invariant mass
distributions for dijet production from EKS and FastNLO for CMS
LHC (7\,TeV) measurement.\cite{Chatrchyan:2011qta}}
\end{figure}
\begin{figure}
\begin{center}
\includegraphics[width=0.8\textwidth]{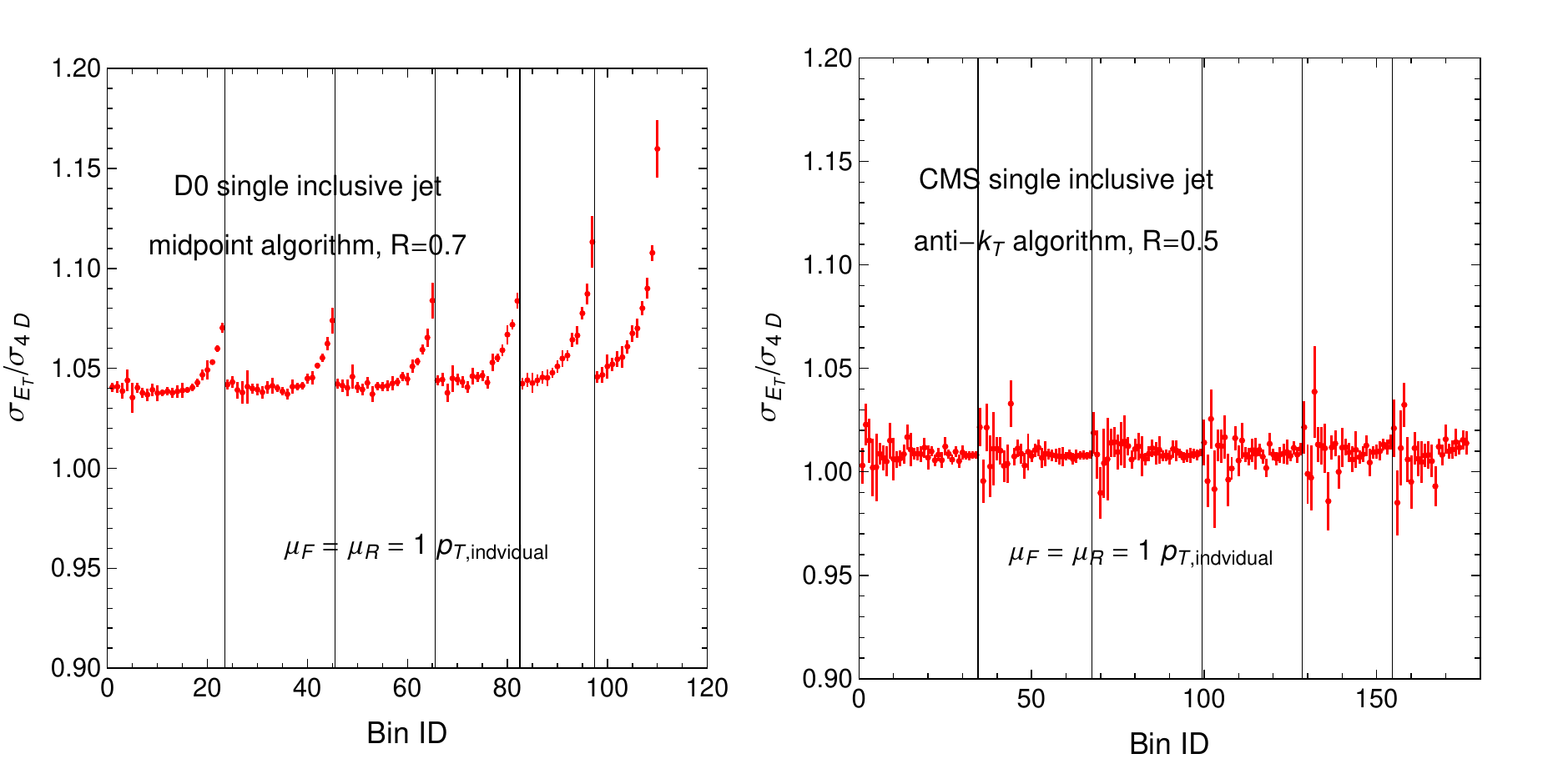}
\end{center}
\caption{\label{jetcom_sinc} Comparison of $p_T$ distributions
for single inclusive jet production using different
recombination schemes.}
\end{figure}
\begin{figure}
\begin{center}
\includegraphics[width=0.8\textwidth]{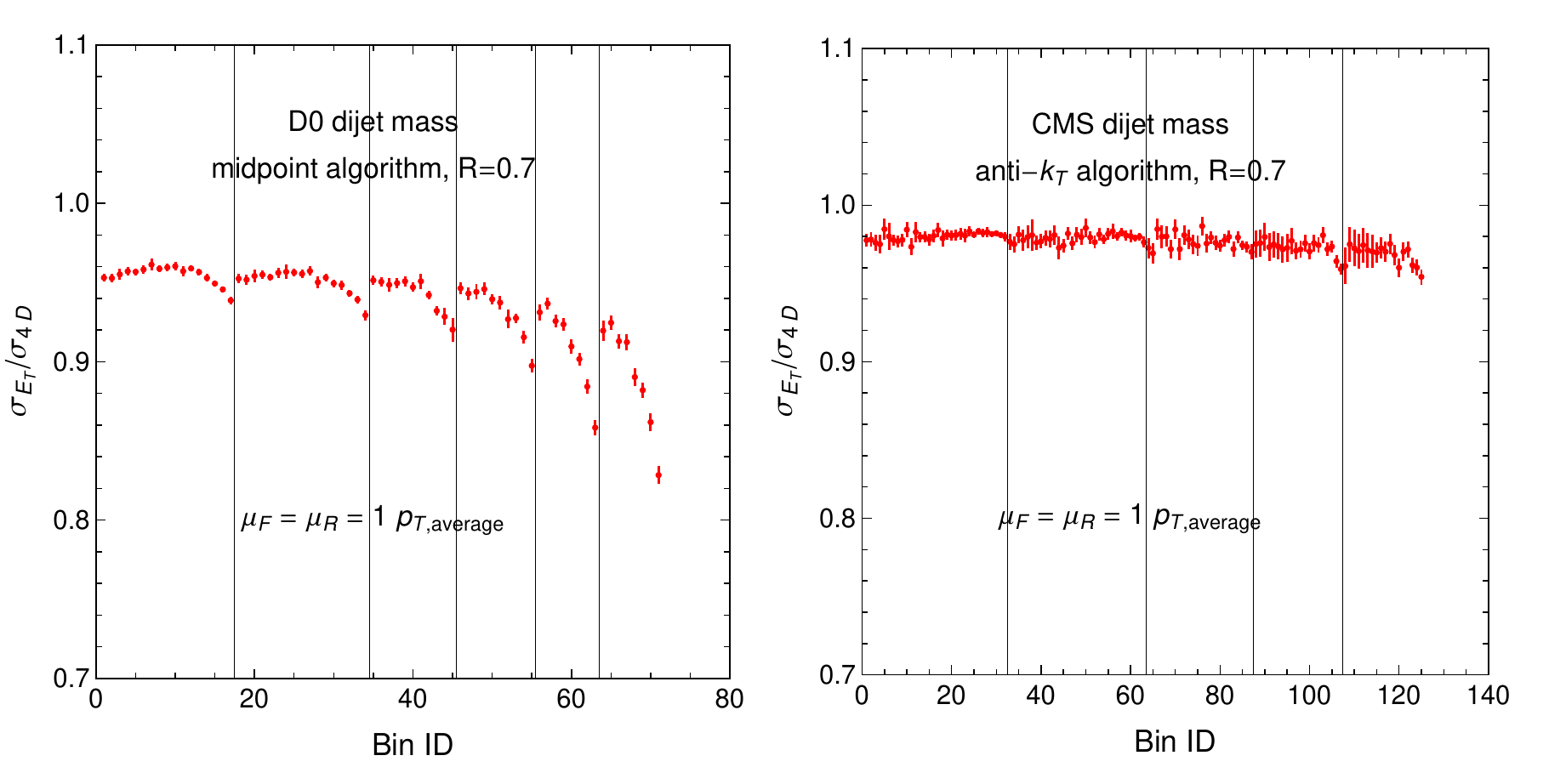}
\end{center}
\caption{\label{jetcom_sdim} Comparison of invariant mass
distributions for the dijet production using different recombination
schemes.}
\end{figure}

}

\clearpage
\section[Phenomenological studies with aMC@NLO]
{PHENOMENOLOGICAL STUDIES WITH AMC@NLO \protect\footnote{Contributed by: 
R.~Frederix,
S.~Frixione,
V.~Hirschi,
F.~Maltoni,
R.~Pittau,
P.~Torrielli}}
{\graphicspath{{amcatnlo/}}

\title{Phenomenological studies with aMC@NLO}

\author{R. Frederix$^1$,
S. Frixione$^{2,3,4}$,
V. Hirschi$^3$,
F. Maltoni$^5$,
R. Pittau$^6$,
P. Torrielli$^3$}

\institute{$^1$ Institut f\"ur Theoretische Physik, 
Universit\"at Z\"urich, Winterthurerstrasse 190,
CH-8057 Z\"urich, Switzerland
\\$^2$PH Department, TH Unit, CERN, CH-1211 Geneva 23, Switzerland
\\$^3$ITPP, EPFL, CH-1015 Lausanne, Switzerland
\\$^4$On leave of absence from INFN, Sezione di Genova, Italy
\\$^5$Centre for Cosmology, Particle Physics and Phenomenology (CP3)
Universit\'e catholique de Louvain, B-1348 Louvain-la-Neuve, Belgium
\\$^6$Departamento de F\'{\i}sica Te\'orica y del Cosmos, Universidad de Granada
}


\begin{abstract}
We present four phenomenological studies of hadron collider processes 
performed within the a{\sc MC@NLO} framework
\end{abstract}

\subsection{Introduction}
a{\sc MC@NLO} (http://amcatnlo.cern.ch) is a fully automated approach to
complete event generation and subsequent parton shower at the NLO accuracy
in QCD, which allows accurate and flexible
simulations for both signals and backgrounds at hadron colliders.  All
calculational aspects in a{\sc MC@NLO} are automated.  One-loop
contributions are evaluated with MadLoop~\cite{Hirschi:2011pa,Hirschi:2011rb},
that uses the OPP integrand reduction method~\cite{Ossola:2006us} as
implemented in CutTools~\cite{Ossola:2007ax}. The other matrix-element
contributions to the cross sections, their phase-space subtractions according
to the FKS formalism~\cite{Frixione:1995ms}, their combinations with the
one-loop results, and their integration are performed by
MadFKS~\cite{Frederix:2009yq} \footnote{The validation of MadLoop and MadFKS
in the context of hadronic collisions has been presented in
Ref.~\cite{Hirschi:2011pa}.}. The matching of the NLO results with {\sc
HERWIG}~\cite{Corcella:2000bw} or {\sc PYTHIA}~\cite{Sjostrand:2006za} parton showers is performed with the MC@NLO
method~\cite{Frixione:2002ik}, and it is also completely automatic.  Finally,
a{\sc MC@NLO} can compute scale and PDF uncertainties at no
extra CPU-time cost with the help of the process-independent reweighting
technique described in~\cite{Frederix:2011ss}.

For all technical details we refer to the original publications. We report
here on the physics results obtained with a{\sc MC@NLO} for observables of
interest at hadron colliders~\cite{Frederix:2011zi,Frederix:2011qg,
Frederix:2011ss,Frederix:2011ig}. We
stress that they are simulated at the hadron level, namely including parton
shower and hadronization effects.  In Sects.~\ref{amcatnlosec1},
~\ref{amcatnlosec2}, and ~\ref{amcatnlosec3} we present results for 
the production of $ttH$, $Vbb$, and four-lepton final states at the
LHC, respectively. Section~\ref{amcatnlosec4} reports on
a study of the $Wjj$ process at Tevatron. Finally, in 
sect.~\ref{sec:concl} we draw our conclusions. The list of the
processes considered here should convince the reader that one
can perform realistic analyses of experimental data, for signals
and backgrounds, entirely within the a{\sc MC@NLO} framework.

\subsection{The $ttH$ process at the LHC \label{amcatnlosec1}}
The production process of a $H$ boson in association 
with a top pair~\cite{Frederix:2011zi} is
a classic mechanism for Higgs production at the 
LHC~\cite{Dittmaier:2011ti,LHCHiggsCrossSectionWorkingGroup:2012vm}, 
where the large $ttH$ Yukawa coupling and the presence of top quarks can be exploited to extract the signal from its QCD multi-jet background.
As an example of the use of  a{\sc MC@NLO} for this process we present, in
Fig.~\ref{amcatnlofig1}, the Higgs transverse momentum distribution 
and the transverse momentum of the $ttH$ or $ttA$ system 
at the $\sqrt{s}$= 7 TeV LHC for a Standard Model (scalar) Higgs with 
$M_H=$ 120 GeV and for a pseudoscalar one with $M_A=$ 120/40 GeV.
The total NLO cross sections in the three cases are 
$\sigma_{\rm NLO}(M_H= 120)=$ 103.4 fb,
$\sigma_{\rm NLO}(M_A= 120)=$ 31.9 fb,
and 
$\sigma_{\rm NLO}(M_A=  40)=$ 77.3 fb, respectively.
\begin{figure}[h]
\begin{center}
\includegraphics[width=0.49\textwidth]{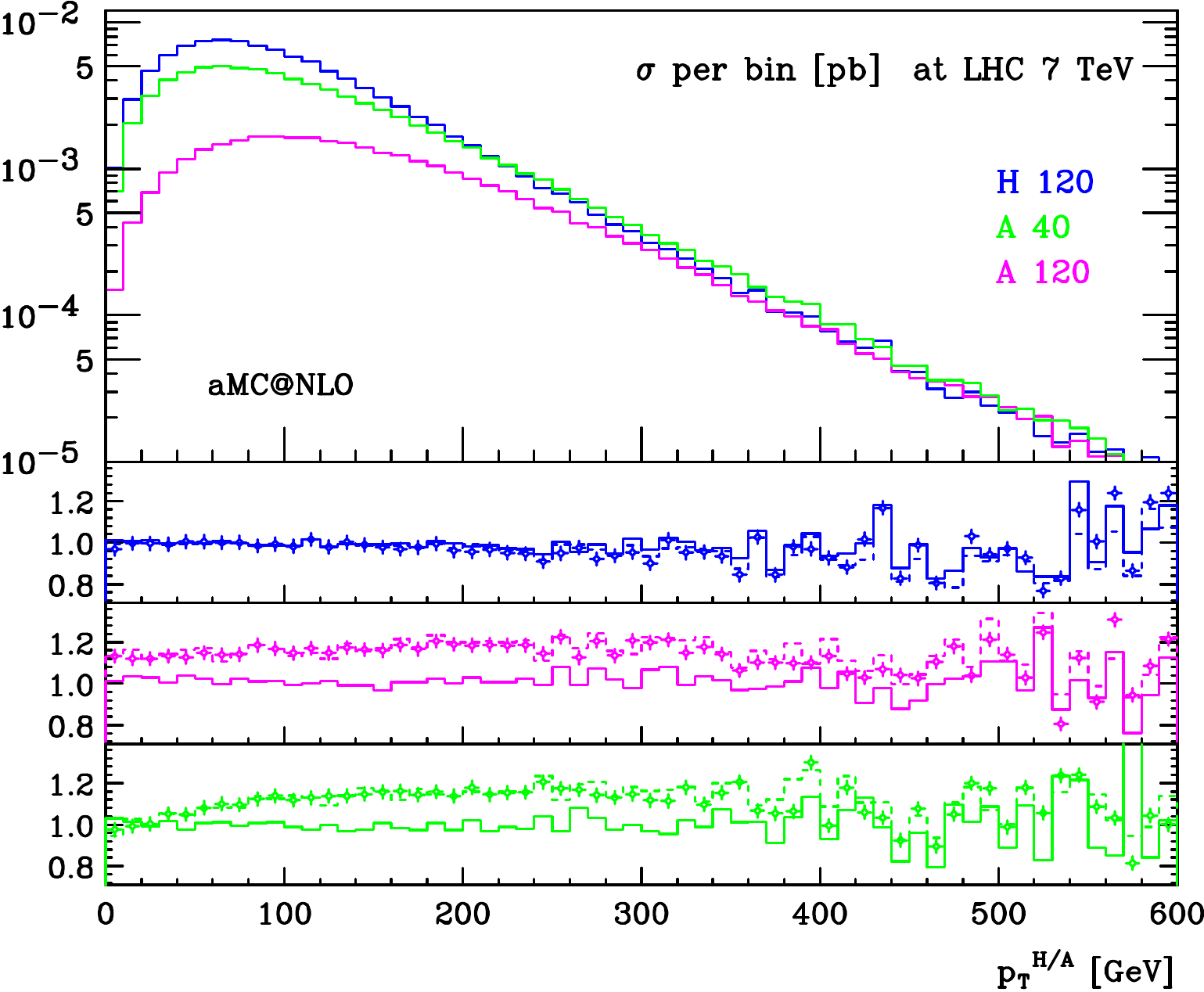}
\includegraphics[width=0.49\textwidth]{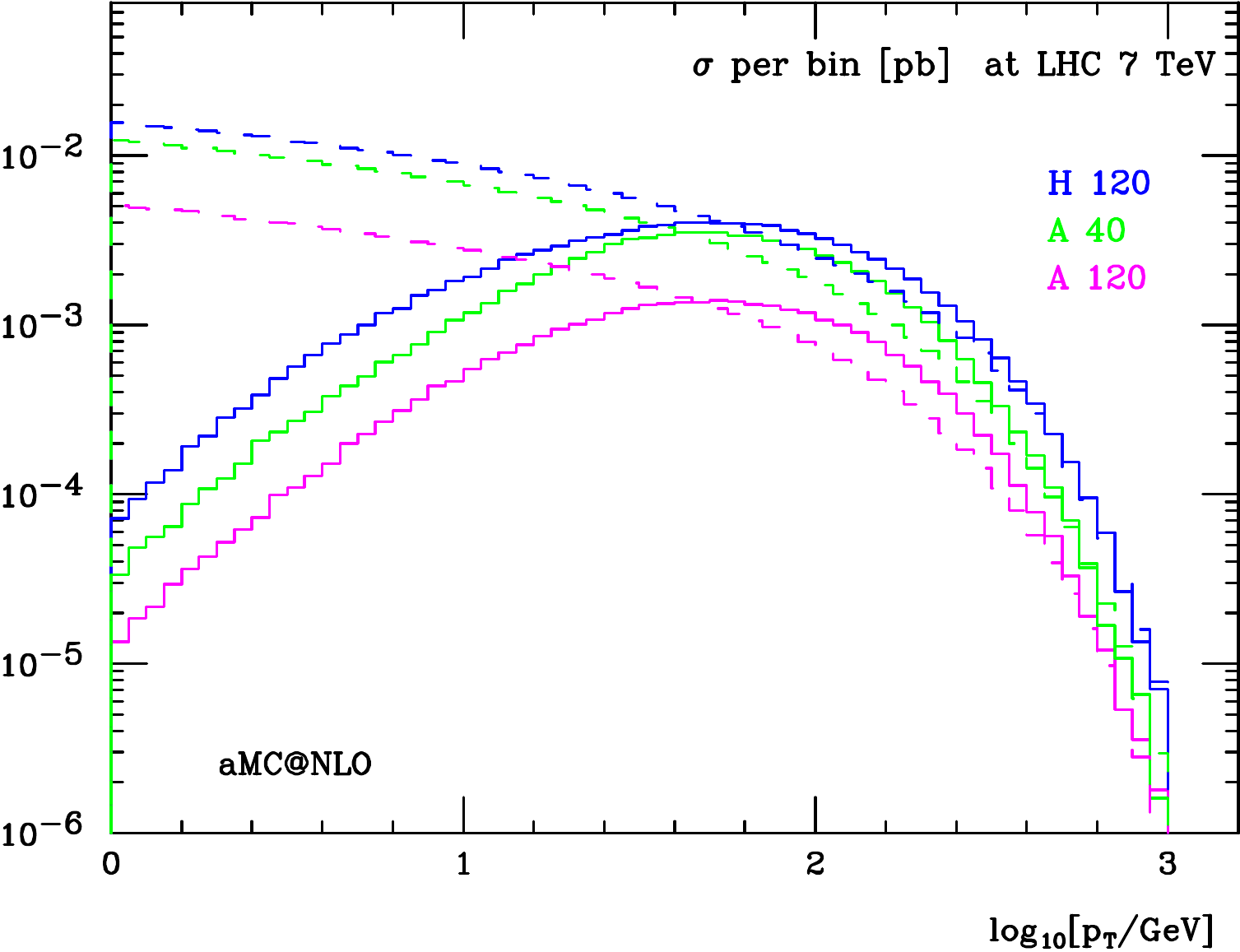}
 \caption{Higgs transverse momentum distributions (left) and 
    transverse momentum of the $ttH$ or $ttA$ system (right)
    in $ttH/ttA$ events at the LHC ($\sqrt{s}$=7 TeV), 
    with a{\sc MC@NLO} in the three cases: Scalar (blue) and pseudoscalar 
    (magenta)
    Higgs with $m_{H/A}=120$ GeV and pseudoscalar (green) with $m_A=40$ GeV. In
    the lower panels of the left part, the ratios of a{\sc MC@NLO} over LO 
    (dashed), NLO (solid),
    and aMC@LO (crosses) are shown. Solid histograms in the right panel
    are relevant to a{\sc MC@NLO}, dashed ones to a pure NLO calculation.}
\label{amcatnlofig1}
\end{center}
\end{figure}
At moderate values of the Higgs transverse momentum, the scalar and
pseudoscalar cases are clearly distinguishable, while at larger values the
three distributions tend to coincide. Parton shower effects give in general
small corrections with respect to the a pure NLO calculation, except for
variables involving all produced particles, such as the transverse momentum of
the $ttH$ or $ttA$ system shown in the right panel of
Fig.~\ref{amcatnlofig1}.

\subsection{The $Vbb$ process at the LHC\label{amcatnlosec2}}
With $Vbb$ we understand $\ell \nu bb$ and
$\ell^+ \ell^- bb$ final states~\cite{Frederix:2011qg}, 
which are the main backgrounds to searches for 
SM Higgs production in association with
vector bosons ($WH/ZH$), with the subsequent Higgs decay into 
a $bb$ pair.
The a{\sc MC@NLO} framework allows a realistic study including
\begin{itemize}
\item NLO corrections;
\item bottom quark mass effects;
\item spin-correlation and off-shell effects;
\item showering and hadronization.
\end{itemize}
As an example we show, in Fig.~\ref{amcatnlofig2a}, the invariant mass of the pair of the two leading b-jets, compared with the signal distributions for a standard Higgs with $m_H = 120$ GeV. Fig.~\ref{amcatnlofig2a} is interesting 
because both signal and background are studied at the NLO accuracy.
\begin{figure}[h]
\begin{center}
\includegraphics[width=0.5\textwidth]{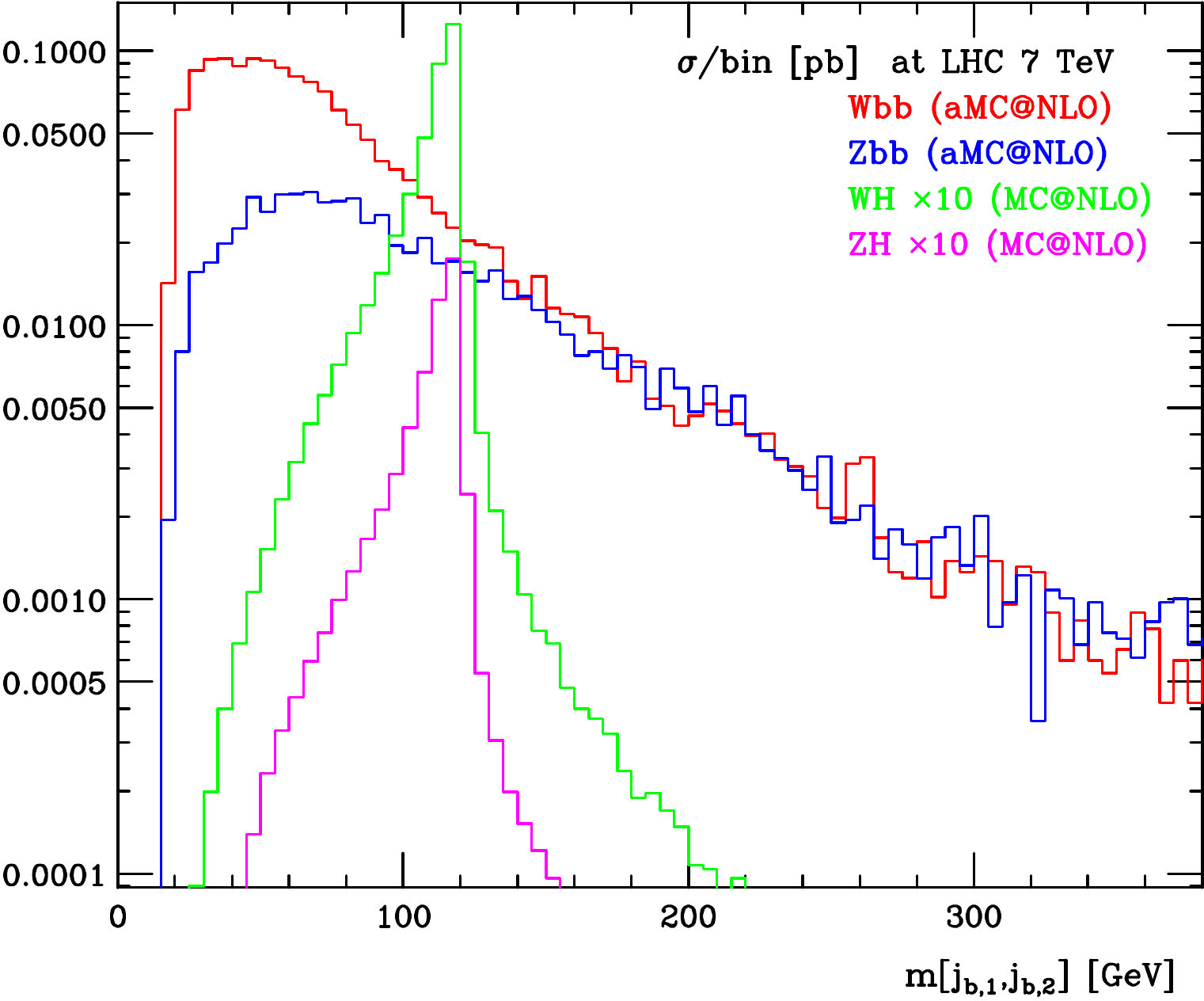}
 \caption{Invariant mass of the pair of the two leading $b$-jets.
$WH(\to \ell\nu bb)$, $ZH(\to\ell^+\ell^- bb)$, $\ell \nu bb$,
and $\ell^+ \ell^- bb$ results are shown, with the former two rescaled by a factor of ten.}
\label{amcatnlofig2a}
\end{center}
\end{figure}
It should be noted that, since completely hadronized events are simulated,
sophisticated studies of the jet sub-structure are possible within the a{\sc
MC@NLO} framework, as presented in Fig.~\ref{amcatnlofig2b}, where the
fractions of events containing zero $b$-jets, exactly one $b$-jet, and
exactly two $b$-jets are plotted.  The $b$-jet fractions are fairly similar
for $Wbb$ and $Zbb$ production, and the effects of the NLO
corrections are consistent with the fully-inclusive $K$ factors.  On the other
hand, the $bb$-jet contribution to the $b$-jet rate is seen to be more than
three times larger for $\ell^\pm \nu bb$ than for $\ell^+
\ell^- bb$ final states.  This fact is related to the different
mechanisms for the production of a $bb$ pair in the two processes.  At
variance with the case of $\ell^\pm \nu bb$ production, in a
$\ell^+ \ell^- bb$ final state the two $b$'s may come from the separate
branchings of two initial-state gluons, and thus the probability of them
ending in the same jet is much smaller than in the case of a $g\to bb$
final-state branching, which gives the only possible contribution to a
$\ell^\pm \nu  bb$ final state.
\begin{figure}[h]
\begin{center}
\includegraphics[width=0.5\textwidth]{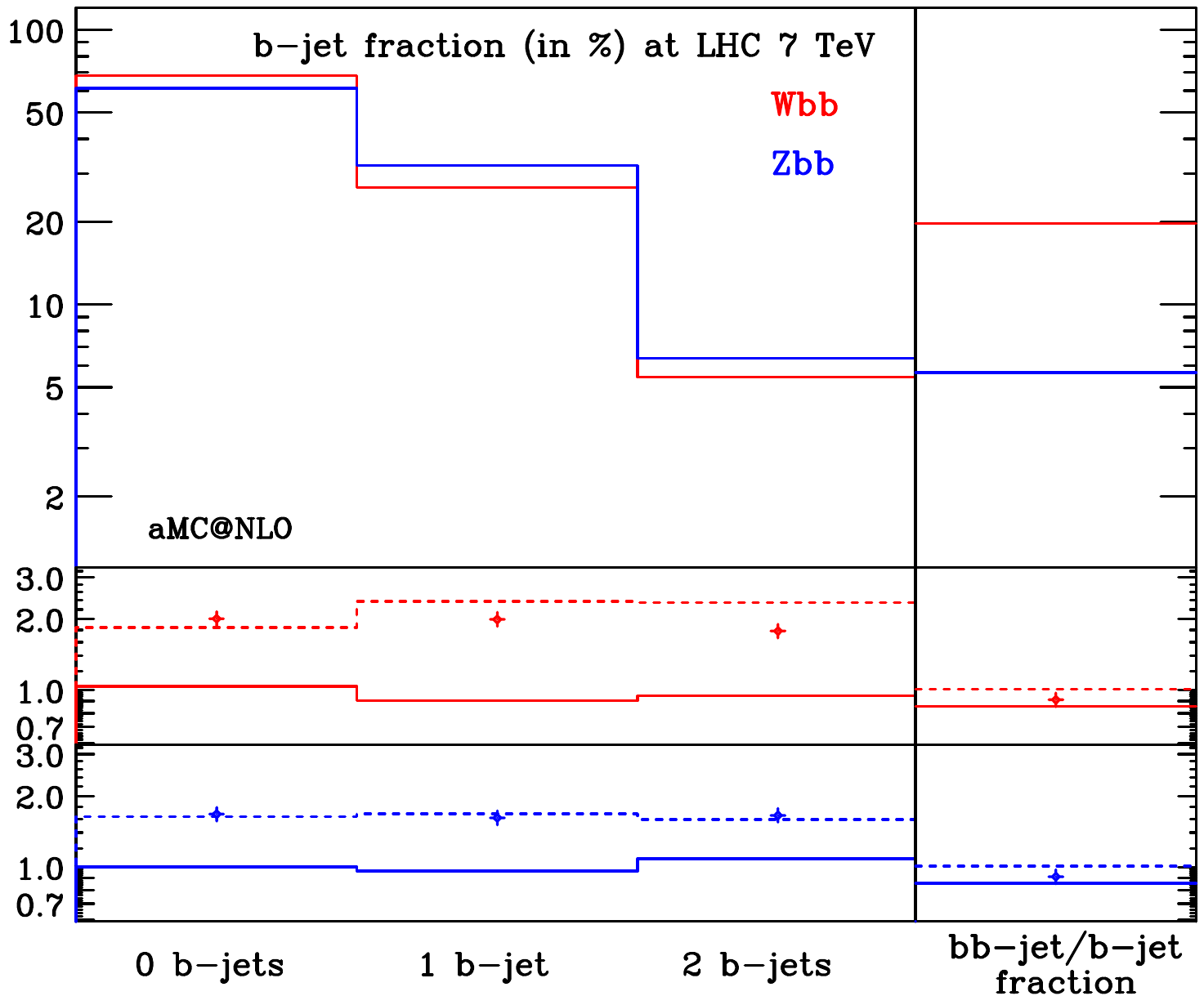}
 \caption{Fractions of events (in percent) that contain: zero $b$-jets,
  exactly one $b$-jet, and exactly two $b$-jets. The rightmost bin displays
  the fraction of $b$-jets which are $bb$-jets. The two insets show 
  the ratio of the a{\sc MC@NLO} results over the corresponding NLO (solid),
  a{\sc MC@LO} (dashed), and LO (symbols) ones, separately for $Wbb$ 
  (upper inset) and $Zbb$ (lower inset) production.}
\label{amcatnlofig2b}
\end{center}
\end{figure}
\subsection{Four-lepton production at the LHC \label{amcatnlosec3}}
Vector boson pair production is interesting in at least two respects. Firstly,
it is an irreducible background to Higgs signals, in particular through the
$W^+W^−$ and $ZZ$ channels which are relevant to searches for a
standard model Higgs of mass larger than about 140 GeV. Secondly, di-boson
cross sections are quite sensitive to violations of the gauge structure of the
Standard Model, and hence are good probes of scenarios where new physics is
heavy and not directly accessible at the LHC, yet the couplings in the vector
boson sector are affected.
We consider here the neutral process~\cite{Frederix:2011ss}
$$
pp \to (Z/\gamma^\ast) (Z/\gamma^\ast) \to  
\ell^+\ell^- \ell^{(\prime)+} \ell^{(\prime)-}\,,
$$
which, although smaller than the $W^+W^-$ channel, may provide a
cleaner signal due to the possibility of fully reconstructing the decay
products of the two vector bosons. a{\sc MC@NLO} predictions for the cross
sections are given in Tab.~\ref{amcatnlotab1}, which also includes a{\sc
MC@NLO} estimates for scale and PDF uncertainties.  The four-lepton invariant
mass and the transverse momentum distribution are presented in
Fig.~\ref{amcatnlofig3}, where comparisons between the results obtained with
a{\sc MC@NLO} matched to {\sc HERWIG} and to {\sc PYTHIA} are also given.
We stress that these results include the contributions due to 
$gg$-initiated processes, which have also been computed automatically.
These are formally of NNLO, but may play a non-negligible phenomenological
role owing to their parton-luminosity dominance at a large-energy
collider such as the LHC.
\begin{table}[t]
\begin{center}
\renewcommand*{\arraystretch}{1.5}
\begin{tabular}{|c|cc|c|}
\hline
 & \multicolumn{3}{c|}{Cross section (fb)}\\
\cline{2-4}  
 Process  & \multicolumn{2}{c|}{$q\bar q$/$qg$ channels} & {$gg$ channel} \\
\cline{2-4}
& ${\cal O}(\alpha_{\scriptscriptstyle s}^0)$ & ${\cal O}(\alpha_{\scriptscriptstyle s}^0)+{\cal O}(\alpha_{\scriptscriptstyle s})$ & ${\cal O}(\alpha_{\scriptscriptstyle s}^2)$ \\
\hline
$pp\to e^+e^-\mu^+\mu^-$&   
 9.19 & $12.90^{+0.27(2.1\%)+0.26(2.0\%)}_{-0.23(1.8\%)-0.22(1.7\%)}$  & 
 $0.566^{+0.162(28.6\%)+0.012(2.1\%)}_{-0.118(20.8\%)-0.014(2.5\%)}$     \\
$pp \to e^+ e^- e^+ e^-$      &   
 4.58 & $6.43^{+0.13(2.1\%)+0.11(1.7\%)}_{-0.13(2.0\%)-0.10(1.6\%)}$ & \\
\hline
\end{tabular}
\end{center}
\caption{Total cross sections for $e^+e^-\mu^+\mu^-$ and $e^+e^-e^+e^-$
production at the LHC ($\sqrt{S}= 7$~TeV) within the cuts 
$M(\ell^\pm\ell^{(\prime)\mp})\ge 30~{\rm GeV}$. The first and second errors 
affecting the results
are the scale and PDF uncertainties (also given as fractions of the
central values).}
\label{amcatnlotab1}
\end{table}
\begin{figure}[h]
\begin{center}
\includegraphics[width=0.49\textwidth]{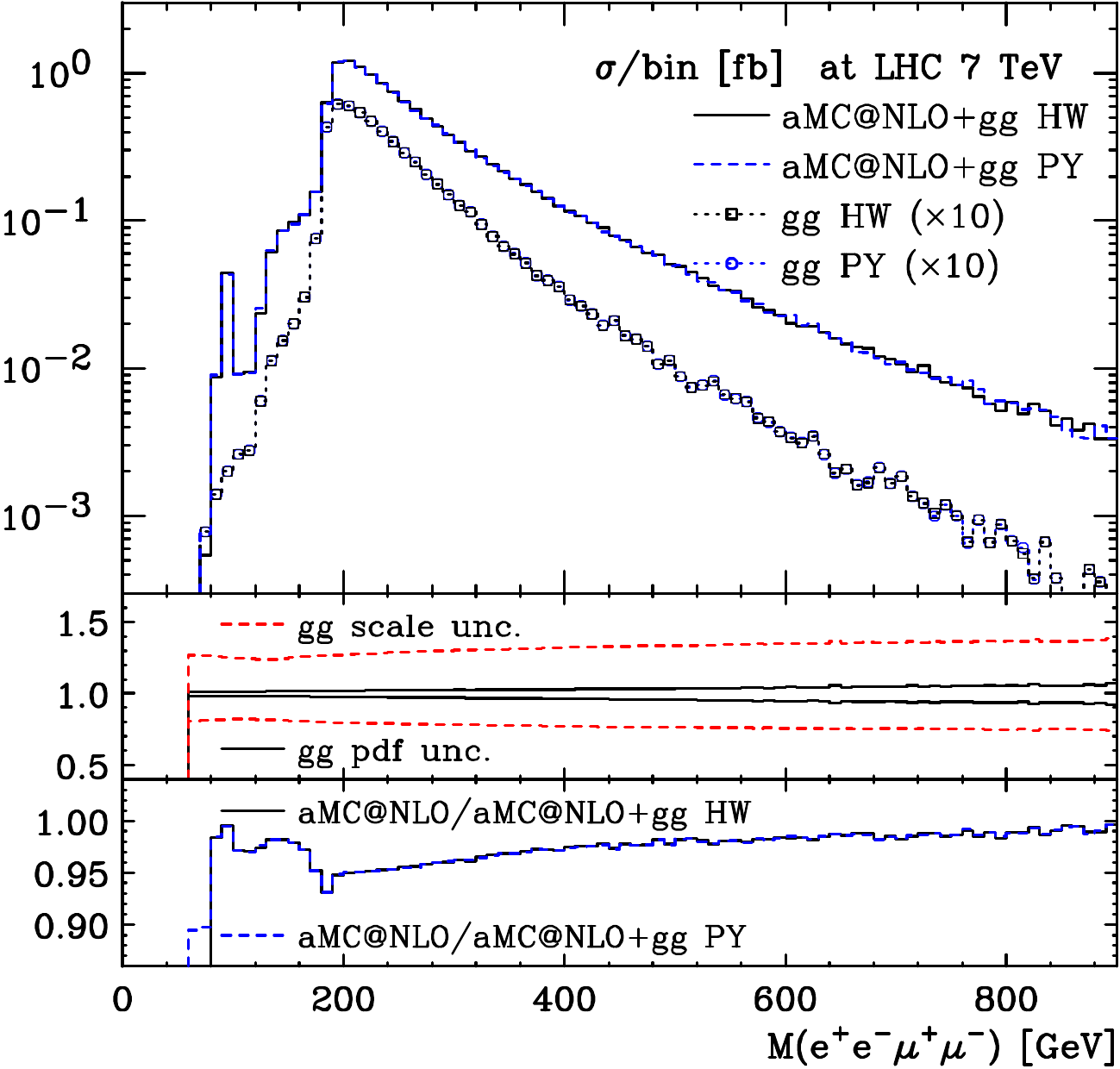}
\includegraphics[width=0.49\textwidth]{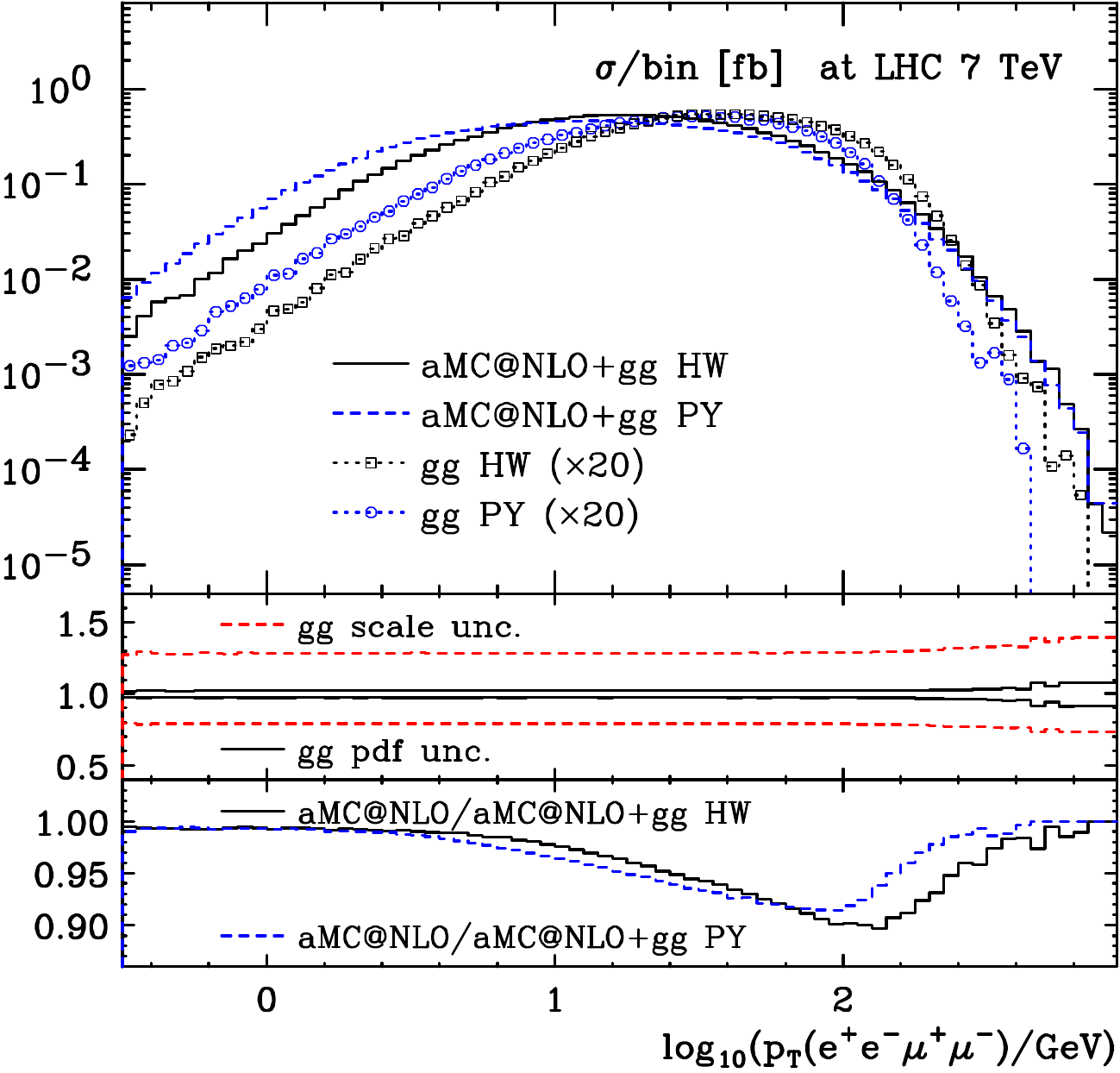}
 \caption{Four-lepton invariant mass and the transverse momentum distributions 
for a{\sc MC@NLO} +$gg$
{\sc HERWIG} (solid black) and {\sc PYTHIA} (dashed blue) results. The rescaled $gg$
contributions with {\sc HERWIG} (open black boxes) and {\sc PYTHIA} (open blue circles)
are shown separately. Middle insets: scale (dashed red) and PDF (solid black)
fractional uncertainties. Lower insets: a{\sc MC@NLO}/(a{\sc MC@NLO}+$gg$) with 
{\sc HERWIG} (solid black) and {\sc PYTHIA} (dashed blue).}
\label{amcatnlofig3}
\end{center}
\end{figure}

\subsection{$Wjj$ at Tevatron \label{amcatnlosec4}}
In~\cite{Aaltonen:2011mk} CDF reported an excess of events in two-jet
production in association with a $W$ boson, in the form of a broad peak
centered at $M_{jj}$ = 144 GeV in the dijet invariant 
mass~\footnote{Such an excess has so far failed to be confirmed by a very similar D0 analysis~\cite{Abazov:2011af}.}. Motivated by this
fact, we present in Fig.~\ref{amcatnlofig4} the a{\sc MC@NLO}
prediction~\cite{Frederix:2011ig} for the dijet invariant mass in $Wjj$
events, using the same cuts as CDF and D0 in the signal region, also comparing
with a pure NLO computation and with the {\sc Alpgen}~\cite{Mangano:2002ea}
findings (one-, two-, and three-parton multiplicities have been 
consistently matched to obtain the latter).  
Perturbative, parton-level results agree well with those obtained
after shower, and PDF and scale uncertainties (also reported in
Fig.~\ref{amcatnlofig4}) are well under control. In summary, we do not observe
any significant effects in the shape of distributions due to NLO corrections,
which therefore cannot be responsible for the excess of events observed by the
CDF collaboration.

\begin{figure}[h]
\begin{center}
\includegraphics[width=0.49\textwidth]{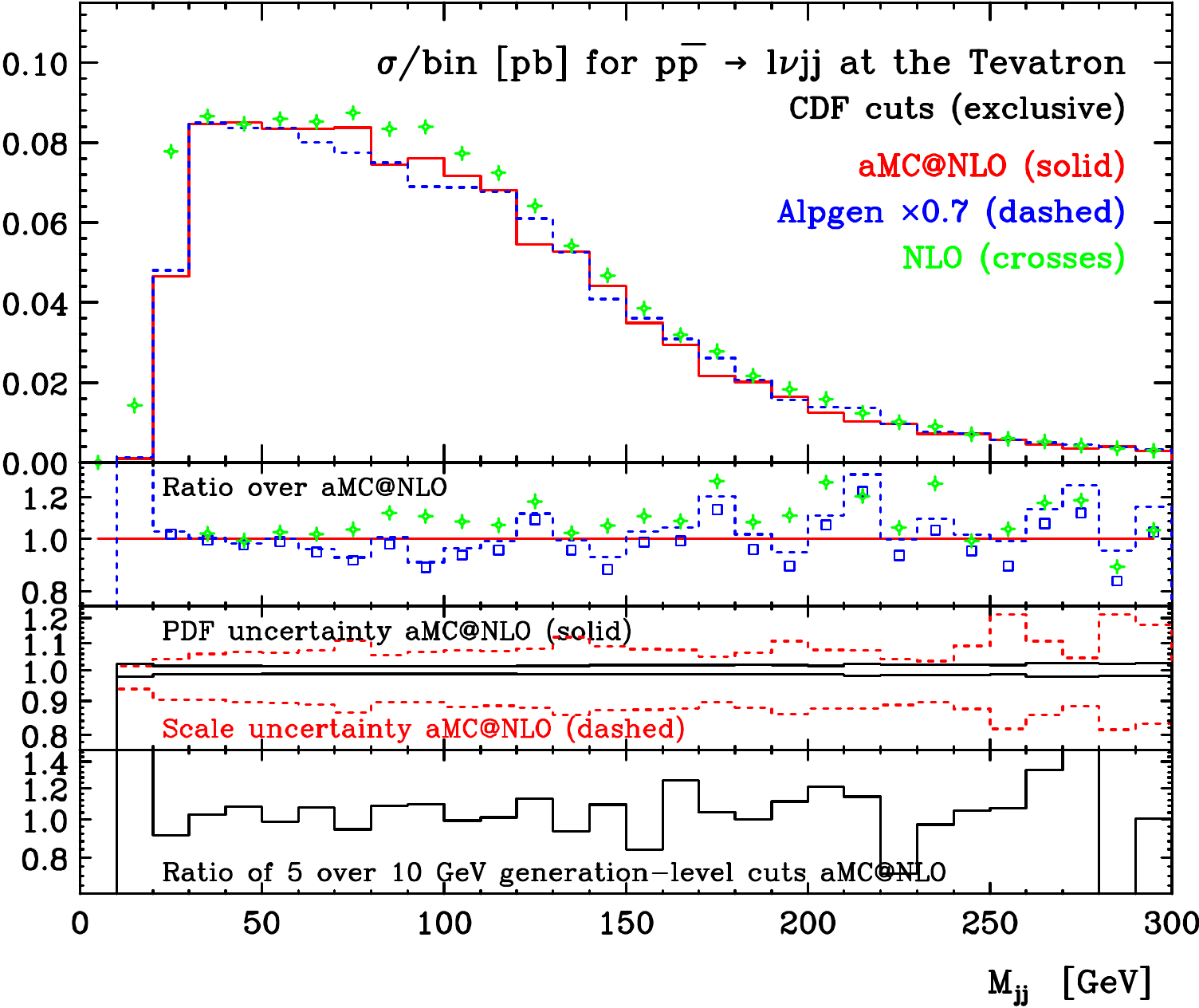}
\includegraphics[width=0.49\textwidth]{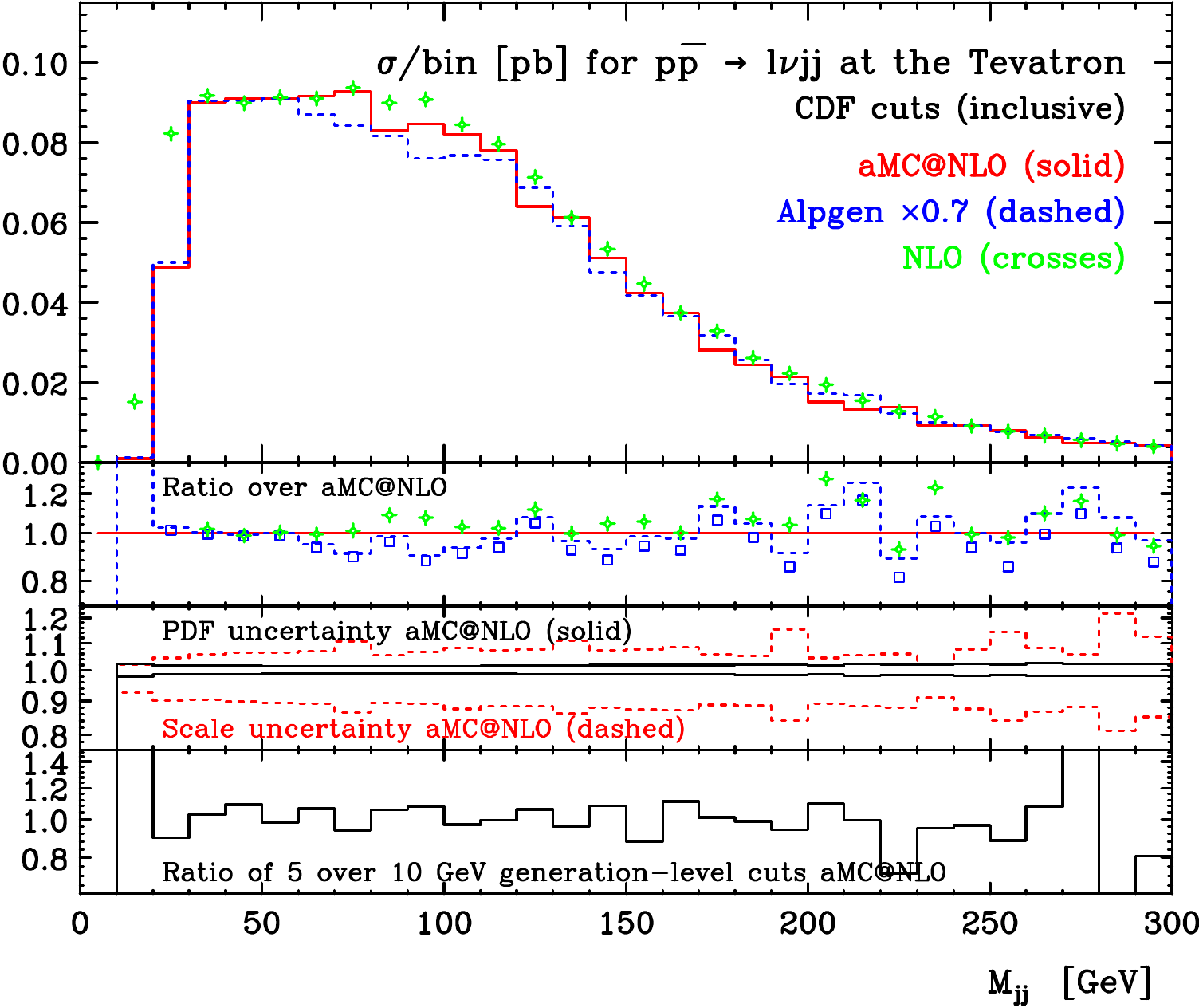}
 \caption{Invariant mass of the pair of the two hardest jets, with
  CDF/D0 cuts of~\cite{Aaltonen:2011mk} (left) and of~\cite{newwjjcuts} (right).}
\label{amcatnlofig4}
\end{center}
\end{figure}
\subsection{Conclusions \label{sec:concl}}
The results we have presented in this contribution are based on the strategic
assumption that, for the word {\it automation} to have its
proper meaning, the only operation required from a user is that of typing-in
the process to be computed, and other analysis-related information (such as
final-state cuts). In particular, the codes that achieve the {\it automation} 
may only differentiate between processes depending on their general
characteristics, but must never work on a case-by-case basis.  The a{\sc
MC@NLO} framework is based on such an assumption, providing a very powerful
tool to compare, at the NLO accuracy including showering and hadronization,
theory and experiment in high energy collisions.  As an example of the
flexibility of a{\sc MC@NLO} we have presented results for the processes $pp
\to ttH$, $pp \to Vbb$, $pp \to \ell^+\ell^- \ell^{(\prime)+}
\ell^{(\prime)-}$ at the LHC, and a study of $ p\bar{p} \to Wjj$ at Tevatron.

\subsection*{Acknowledgements}
This research has been supported by the Swiss National Science Foundation
(SNF) under contracts 200020-138206 and 200020-129513, by the Belgian IAP
Program, BELSPO P6/11-P and the IISN convention 4.4511.10.  S.F., F.M. and
R.P. acknowledge the financial support of the MICINN project FPA2011-22398
(LHC@NLO).


}

\section[Probing corrections to dijet production at the LHC]
{PROBING CORRECTIONS TO DIJET PRODUCTION AT THE LHC \protect\footnote{Contributed by: S.~Alioli, J.~R.~Andersen, C.~Oleari, E.~Re, J.~M.~Smillie}}
{\graphicspath{{Dijets/}}

\newcommand\sss{\mathchoice%
{\displaystyle}%
{\scriptstyle}%
{\scriptscriptstyle}%
{\scriptscriptstyle}%
}
\newcommand\pt{p_{\sss\rm T}}
\newcommand\pT{p_{\sss\rm T}}
\newcommand\HT{H_{\sss\rm T}}


\title{Probing corrections to dijet production at the LHC}

\author{S.~Alioli$^1$, J.~R.~Andersen$^2$, C.~Oleari$^3$, E.~Re$^4$, J.~M.~Smillie$^5$}
\institute{$^1$Ernest Orlando LBNL, University of California, Berkeley, CA 94720, USA\\
$^2$CP$^3$-Origins, University of Southern Denmark, Campusvej 55, DK-5230 Odense M, Denmark\\
$^3$Universit\`a di Milano-Bicocca and INFN, Sezione di Milano-Bicocca, 20126 Milan, Italy\\
  $^4$ IPPP, Department of Physics,
  University of Durham, Durham, DH1 3LE, UK\\
  $^5$School of Physics and Astronomy, University of Edinburgh, Mayfield Road,
  Edinburgh EH9 3JZ, UK}


\begin{abstract}
  We compare and discuss a few kinematic distributions for dijet production
  at the LHC, computed with a fixed next-to-leading order code, with the
  POWHEG BOX and with HEJ.  Previous experimental studies have dealt with
  kinematic distributions where the predictions of the three approaches were
  very similar. In this proceeding, we investigate kinematic distributions
  where the resummed effects in POWHEG and HEJ are clearly shown and enhanced
  with respect to the fixed NLO result, since different QCD-radiation regimes
  are probed.
\end{abstract}

\subsection{Introduction}
Dijet production is one of the cornerstone processes at the LHC.  The cross
section for jet production is very large, making it an important testing
ground for our understanding of QCD at high-energy scales.  In addition, jet
production is an important background for many searches for new physics. It
is therefore essential to probe and test our theoretical predictions.
Dijet-production studies can bring insights in jet production in
association with other particles too: for example, Higgs boson production
plus two jets in gluon fusion, a key process for assessing the CP properties
of the Higgs boson, can benefit from these studies.

There have been a number of very interesting experimental studies in dijet
production by both the ATLAS~\cite{Aad:2011jz, Collaboration:2011tq,Aad:2011fc}
and CMS~\cite{Chatrchyan:2011wn, Collaboration:2011j, Khachatryan:2011zj,
  CMS-PAS-FWD-10-014} Collaborations.  It is already clear that higher order QCD
contributions beyond a fixed order, low multiplicity calculation can be
important because the large available phase space for jet emission at the LHC
compensates for the suppression of extra powers of the strong coupling constant.

In this contribution, we compare two theoretical approaches to dijet
production that include higher order effects: POWHEG~\cite{Nason:2004rx,
  Frixione:2007vw, Alioli:2010xa, Alioli:2010xd} and
HEJ~\cite{Andersen:2009nu, Andersen:2009he, Andersen:2011hs}.  The POWHEG
method successfully merges a fixed next-to-leading order~(NLO) calculation
with a parton shower program, that resums leading logarithmic contributions
from collinear emissions. In this study, the POWHEG results obtained with the
POWHEG BOX~\cite{Alioli:2010xd} are interfaced with the
transverse-momentum-ordered shower provided by
PYTHIA~6.4.21~\cite{Sjostrand:2006za}.  In contrast, the starting point for
HEJ is an all-order approximation to the hard scattering matrix element in
the regime of wide-angle QCD emissions.  HEJ is accurate at leading
logarithmic precision in the invariant mass of any two jets.  This is then
supplemented with the missing contributions (through a merging and
reweighting-procedure) necessary to also ensure tree-level accuracy for final
states with up to four jets.  The tree-level matrix elements are taken from
Standalone Madgraph~\cite{Alwall:2007st}.

The POWHEG and HEJ approaches are clearly very different in their description
of QCD
radiation. Nevertheless, for several kinematic distributions (see for example
ref.~\cite{Aad:2011jz}) the predictions from POWHEG
and HEJ are very similar.
In this study, we investigate various observables which can expose the
differences in the two approaches and we compare them with the fixed NLO
results.

\subsection{A comparison between NLO, POWHEG and HEJ in dijet production}
In order to avoid biasing our event sample, we impose a minimal set of cuts,
avoiding symmetric cuts on the jet transverse momenta that would give an
unphysical cross section at fixed NLO level~\cite{Frixione:1997ks,Banfi:2003jj}, due to the
presence of unresummed logarithms.  Neither the {POWHEG} or {HEJ}
descriptions suffer from this instability. However, in order to have a
sensible fixed NLO cross section to compare with, we impose asymmetric cuts
\begin{equation}
  \label{eq:goodcuts}
  \pt^j > 35\ {\rm GeV},\quad \pt^{j_1} > 45\ {\rm
    GeV},\quad |y_j| < 4.7\,,
\end{equation}
i.e.~all jets are required to have a minimum transverse momentum of
35~GeV, and the hardest-jet transverse momentum, $\pt^{j_1}$, is
required to be greater than 45~GeV. In order to comply with the
experimental acceptance, all jets are further required to have an
absolute rapidity $|y_j|$ less than 4.7. Jets are defined according to
the anti-kt jet algorithm, with radius $R=0.5$. Only events with at
least two jets fulfilling Eq.~(\ref{eq:goodcuts}) are kept.

In the following, we compare the fixed NLO cross section with the POWHEG
first emission results, with the POWHEG results showered by PYTHIA and with
the HEJ predictions. The renormalization and factorization scales have been
chosen equal to the transverse momentum of the hardest jet in each event, for
the {HEJ} predictions.  For the NLO computation (and for computing the POWHEG
$\bar{B}$ function), scales are set to the transverse momentum of the so
called underlying Born configuration. Scale-uncertainty bands obtained by
varying these scales by a factor of two in each direction are shown for the
NLO and HEJ results.  The scales entering in the evaluation of parton
distribution functions and of the strong coupling in the POWHEG Sudakov form
factor are instead evaluated with a scale equal to the transverse momentum of
the POWHEG hardest emission~\cite{Frixione:2007vw, Alioli:2010xa}.

In Fig.~\ref{fig:Dijets_avgjets} we plot the average number of jets as a
function of the rapidity difference between the most forward and most
backward of the jets fulfilling Eq.~(\ref{eq:goodcuts}), $\Delta y_{fb}$, on the left-hand side, and as a
function of $\HT=\sum_j \pt^j$ on the right-hand side.
\begin{figure}
  \centering
  \includegraphics[width=0.48\textwidth, bb = 12 5 525 450, clip]{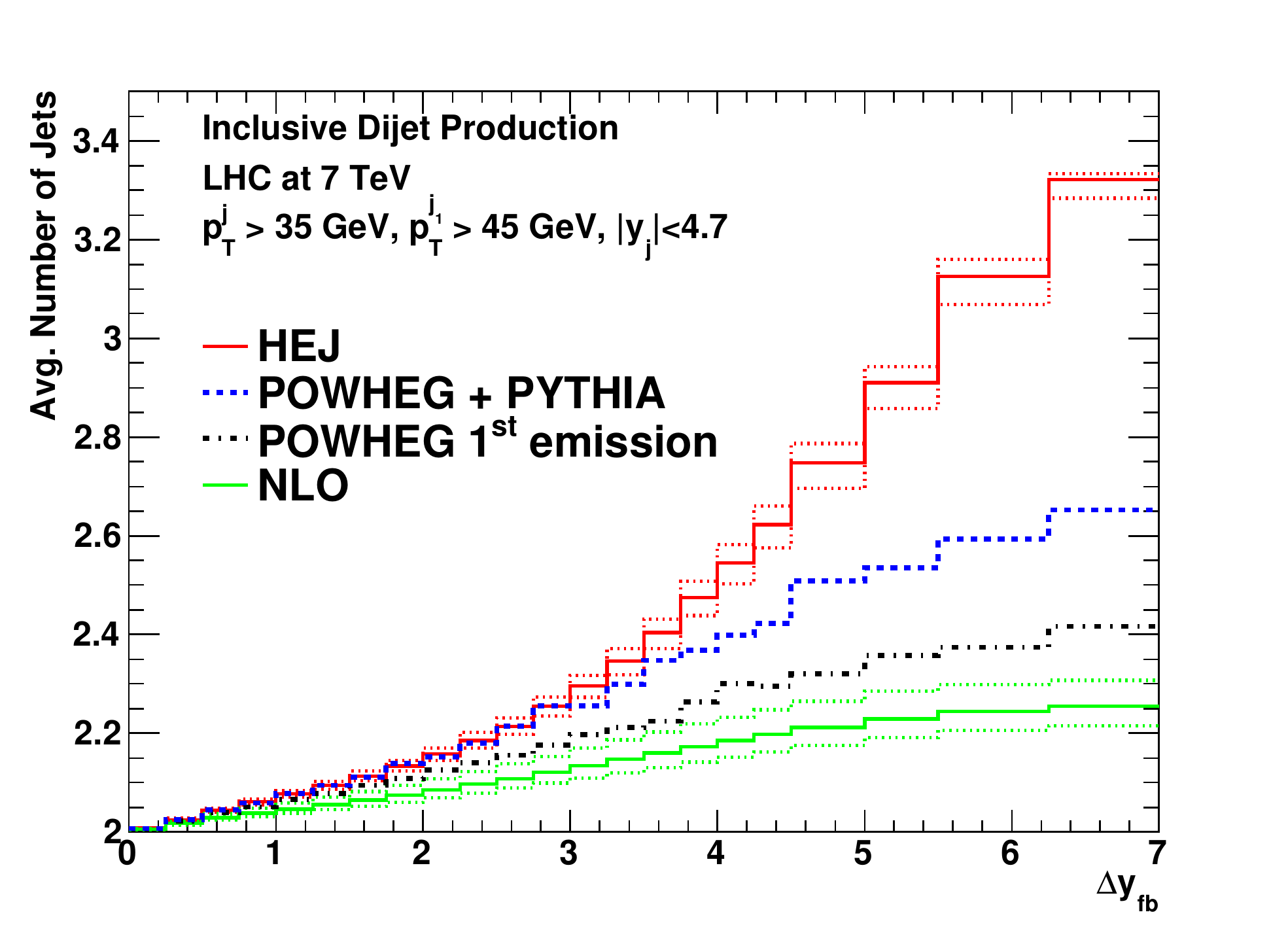}
  \includegraphics[width=0.48\textwidth, bb = 12 5 525 450, clip]{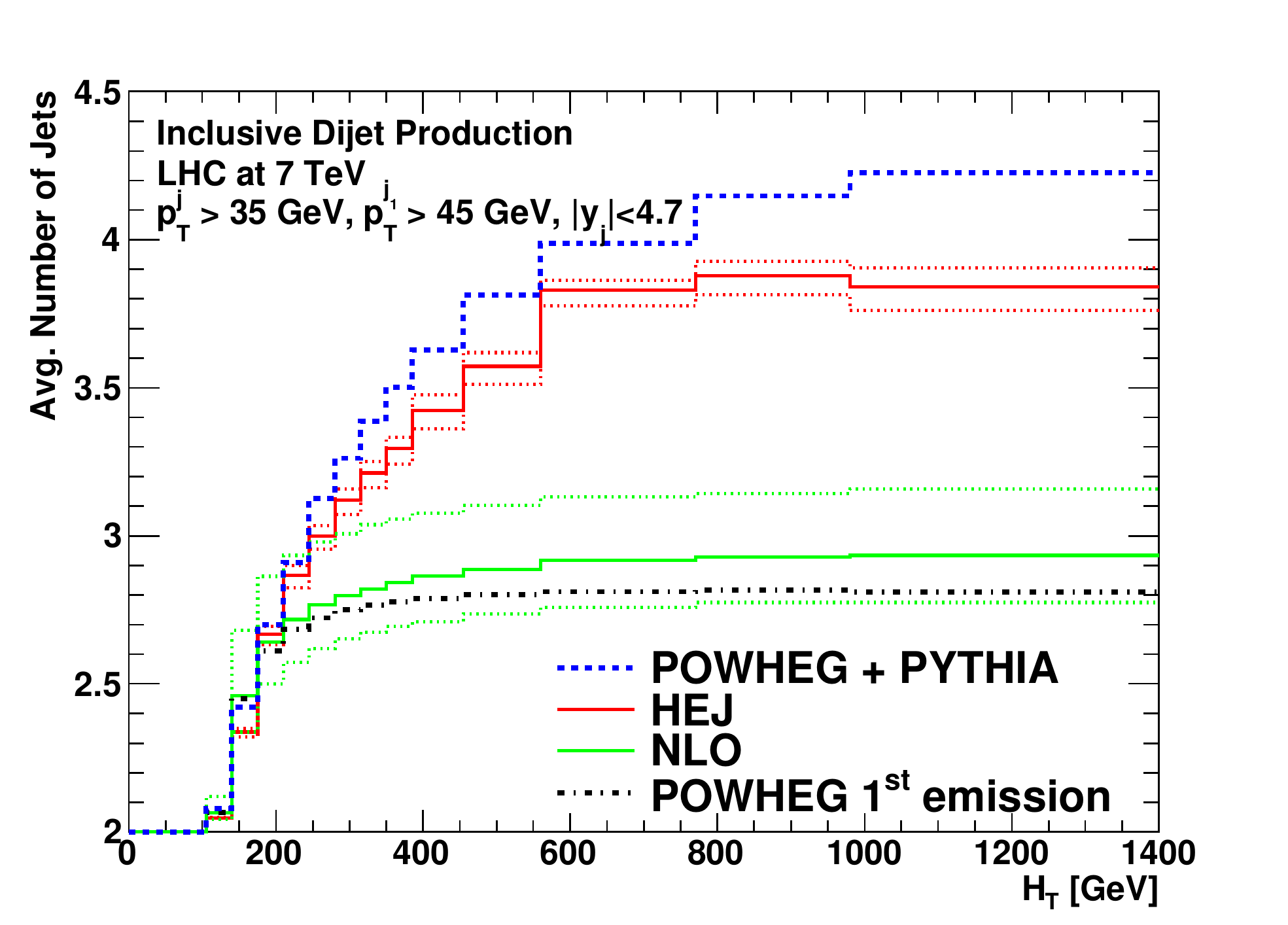}
  \caption{The average number of jets as a function of $\Delta y_{\rm fb}$
    (left plot) and of $\HT$ (right plot), as predicted by a fixed NLO
    calculation, by  POWHEG first emission, by POWHEG+PYTHIA and by HEJ.
    The dotted red lines around the {HEJ} prediction and the green ones
    around the NLO result are obtained by varying the renormalization and
    factorization scales by a factor of two around their central value.}
  \label{fig:Dijets_avgjets}
\end{figure}
The wide-angle resummation implemented in {HEJ} produces more hard jets than
POWHEG and the fixed NLO calculation, as the rapidity separation between the
most forward and the most backward jet in the event increases.  Both the NLO
and the first-emission POWHEG results have at most 3 jets, so that the
average number of jets cannot exceed 3, and give similar results. Additional
jets are instead produced by the PYTHIA shower, so that the average number of
jets is increased by roughly 20\% with respect to the NLO one, for $\Delta
y_{\rm fb}\approx 7$. For the same separation in rapidity, the HEJ prediction
is 45\% larger than the NLO result, with a chance to distinguish among
the three approaches.

The dependence of the average number of jets from $\HT$ (right plot) displays
a different behaviour: here the showered events have on average more jets than
HEJ and the NLO results, as the sum of the transverse momentum of all the
final-state jets increases.  It is interesting here to comment on the NLO
result obtained with the factorization and renormalization scales set to
$\pt^{\rm \scriptscriptstyle UB}/2$, i.e.~half of the transverse momentum of
the underlying Born configuration. In fact, from the plot, 
an unphysical behaviour of this quantity emerges: the average number of jets is
greater than 3 above $\HT\approx 270$~GeV.  This is due to the fact that the
high $\HT$ region is populated mostly by events with 3 jets, two of which
have approximately the same high transverse momentum, and the third one is
softer with respect to the other two (the cuts in Eq.~(\ref{eq:goodcuts})
are always in place).  In this configuration, the exclusive two-jet cross
section becomes negative, due to incomplete cancellation of the virtual
(negative) contribution, now enhanced by a higher value of the strong
coupling constant, evaluated at a lower renormalization scale. A more
detailed discussion can be found in ref.~\cite{AAORS}.

\begin{figure}
  \centering
  \includegraphics[width=0.70\textwidth]{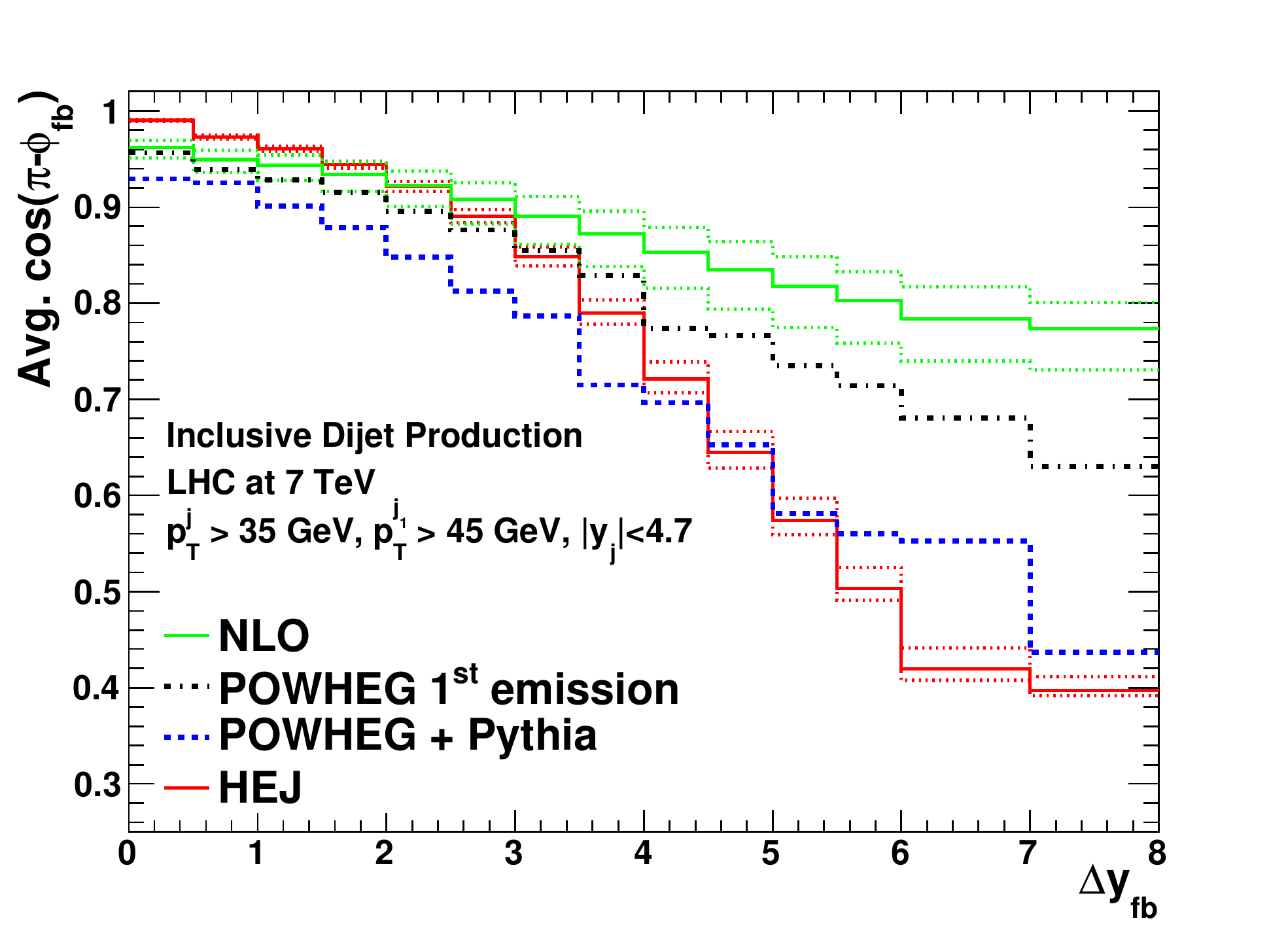}  
  \caption{The average value of $\cos(\pi-\phi_{\rm fb})$ as a function of
    $\Delta y_{\rm fb}$, where $\phi_{\rm fb}$ is the azimuthal angle
    separation between the most forward and most backward jet.  The dotted
    red and green lines are obtained by varying the renormalization and
    factorization scales by a factor of 2 in both directions.}
  \label{fig:Dijets_cosphi}
\end{figure}
As a last example of a kinematic distribution that displays different behaviour if
evaluated at NLO or using POWHEG or HEJ, we plot in
Fig.~\ref{fig:Dijets_cosphi} the average value of $\cos(\pi-\phi_{\rm
  fb})$, where $\phi_{\rm fb}$ is the azimuthal angle between the most
forward and backward jets, as a function of their rapidity separation $\Delta
y_{\rm fb}$. For dijet events at tree-level, $\phi_{\rm fb}=\pi$ since the two
jets must be back-to-back, and the average value of the cosine is 1.
Deviation from 1 then indicates the presence of additional emissions, so that
this kinematic distribution carries information on the decorrelation between
the two jets.  This quantity is more inclusive than the average number of
jets as it is sensitive also to emissions below the jet $\pt$ cut.  The
higher radiation activity in {POWHEG+PYTHIA} and in {HEJ}, with respect to
the fixed NLO and the POWHEG first-emission results, is clearly visible in
the figure: the stronger jet activity produced by HEJ at higher rapidity
separation (see the left plot of Fig.~\ref{fig:Dijets_avgjets}) lowers the
average value of the cosine below the POWHEG+PYTHIA result.  As expected, the
average value predicted by the POWHEG first-emission and the NLO calculation
is closer to 1, since they contain at most one radiated parton.

\subsection*{Conclusions}
In this proceeding, we have discussed the results obtained using a fixed NLO
calculation, HEJ and POWHEG+PYTHIA, in the description of three kinematic
distributions, selected in order to display more clearly the differences
among the three approaches: the average number of jets and azimuthal
decorrelation between the most forward and the most backward jet, plotted as
a function of the rapidity separation of the most forward and the most
backward jet, and the average number of jets plotted as a function of the sum
of the transverse momenta of all the jets in the event.

While the limitations of the NLO calculation are clearly visible when we
probe regions of the phase space where multi-jet emissions becomes important,
the predictions of POWHEG+PYTHIA and HEJ are distinguishable when dealing
with the average number of jets as a function rapidity span.  Less marked
differences are found as a function of Ht, and in the
study of the azimuthal decorrelation of the most forward and backward jet.

An experimental analysis of the dijet data, collected at the LHC, should then
follow to investigate to which extent our theoretical knowledge for these
kinematic distributions is under control.

\subsection*{Acknowledgments}
All of the authors would like to thank the organizers for an extremely
successful and inspiring workshop, and the staff of the Ecole de Physique des Houches for
their hospitality.  The work of ER and JMS is supported by the UK Science and
Technology Facilities Council (STFC).  ER and SA acknowledge financial
support from the LHCPhenoNet network under the Grant Agreement
PITN-GA-2010-264564 for travel expenses.


}

\section[W+jets production at the LHC: a comparison of perturbative tools]
{W+JETS PRODUCTION AT THE LHC: A COMPARISON OF PERTURBATIVE TOOLS \protect\footnote{Contributed by: J.~R.`Andersen, J.~Huston, D.~Ma\^itre, S.~Sapeta, G.~P.~Salam, J.~M.~Smillie, J.~Winter}}
{\graphicspath{{LH11Wjets/}}


\title{W+jets production at the LHC: a comparison of perturbative tools}

\author{J. R. Andersen$^1$, J. Huston$^2$, D. Ma\^itre$^{3,4}$, S. Sapeta$^{4,5}$, G.P. Salam$^{3,5,6}$, J. M. Smillie$^7$, J. Winter$^3$}
\institute{
 $^1$CP$^3$-Origins, Campusvej 55, DK-5230 Odense M, Denmark\\
  $^2$ Physics and Astronomy Department, Michigan State University, East Lansing, MI, 48824 USA\\
  $^3$PH-TH Department, Case C01600, CERN, CH-1211 Geneva 23, Switzerland\\
  $^4$IPPP, University of Durham, Science Laboratories, South Rd,
  Durham DH1 3LE, UK\\
  $^5$LPTHE, UPMC and CNRS UMR 7589, 75252 Paris cedex 05, France\\
  $^6$Department of Physics, Princeton University, Princeton, NJ 08544, USA\\
  $^7$School of Physics and Astronomy, University of Edinburgh, Mayfield Road,
  Edinburgh EH9 3JZ, UK
}


\begin{abstract}
  In this contribution, we discuss several theoretical predictions for
  $W$ plus jets production at the LHC, compare the predictions to recent
  data from the ATLAS collaboration, and examine possible improvements
  to the theoretical framework.
\end{abstract}

\subsection{Motivation}

Experimentalists are reliant on a number of tools, at LO and NLO, at
parton level and at hadron level, in order to understand both simple
and complex final states at the LHC. One of the benchmark processes,
for both signals to new physics and for their backgrounds, is the
production of $W$ plus jets. In this contribution, we discuss several
different predictions for the $W$ plus jets final state, concentrating on
the $H_T$ distribution.
%
We examine where
the predictions agree, and where they disagree and compare the
predictions to LHC data. We introduce the idea of NLO `Exclusive
Sums', 
and discuss the performance of this technique and consider also how
LoopSim may be able to improve the predictions.
We document the use of ROOT ntuples for $W$ plus jets
predictions produced by the BlackHat+Sherpa collaboration, indicating
how they can be used to examine the variation of the cross sections
with jet size/algorithm, PDFs, and scale choices.
We also study the possibility of using the LoopSim method together
with BlackHat+Sherpa type ntuples, since this may offer the
opportunity to improve on the results from NLO Exclusive Sums.


\subsection{Theory tools: strengths and weaknesses}

NLO is the first order at which the normalization (and sometimes the
shape) of LHC cross sections can be realistically calculated. The
state of the art is in parton-level programs such as BlackHat+Sherpa,
where $W+n$-jet cross sections are available, with $n$ up to 4 at NLO~\cite{Berger:2009zg,Berger:2009ep,Berger:2010zx}
(and soon up to 5~\cite{bern}). Of course, such parton-level final states do not
allow for the full comparisons to the data allowed by the full parton
shower Monte Carlo programs such as Sherpa. NLO matrix elements have
been included into parton shower Monte Carlos, but only for relatively
simple final states (although we note that the NLO matrix elements for
$W+2$ jets~\cite{Frederix:2011ig} and $W+3$
jets~\cite{Hoeche:2012ft} have recently been implemented in parton
shower Monte Carlo programs).

The Sherpa Monte Carlo
program~\cite{Gleisberg:2008ta,Gleisberg:2003xi} includes the exact LO
$W+n$-parton ($W+n$-jet) matrix elements, with $n$ up to 4 (in this
study), using the newer ME\&TS scheme as introduced in
Refs.~\cite{Hoeche:2009rj,Hoeche:2009xc,Carli:2010cg} for the addition
of states with different jet multiplicities with the correct
normalizations. The newer matrix-element plus parton-shower merging
scheme improves over the CKKW~\cite{Catani:2001cc,Krauss:2002up}
formalism by allowing for a better interplay between the matrix-element
and parton-shower descriptions. This in particular required the
implementation of truncated showers (`TS'). As before, additional jets
are, of course, then produced by the parton shower. Both BlackHat+Sherpa
and Sherpa rely on DGLAP-based evolution of gluon emission, on the
assumption that the gluon emissions are
strongly ordered in transverse momentum.  For an alternative
prediction, we use the program HEJ~\cite{Andersen:2009nu,Andersen:2009he,Andersen:2011zd}. The High Energy Jets
(HEJ) framework provides a leading-log resummation of the dominant
terms in the limit of large invariant mass between jets.  
%
%
%
In addition, HEJ contains a merging procedure to ensure tree-level
accuracy for final states with two, three or four jets.

A NLO $n$-jet prediction produces events with with either $n$ or $n+1$
partons. For observables for which higher multiplicities have a
significant impact, this limitation can be detrimental. If one has
predictions for different multiplicities, one can try to combine them
by avoiding double counting by requiring that the $n$-jet prediction
is used only to describe $n$-jet events (except for the highest
multiplicity where ($n+1$)-jets configurations are allowed). This
procedure is crude and does not increase the formal accuracy of the
prediction which is that of NLO of the smallest multiplicity. The idea
is that, in observables where higher multiplicities events dominate, a
better prediction might be obtained. This has been denoted as the `Exclusive
Sums' technique. The impact of the Exclusive Sums approach depends on
the kinematic variable under consideration. For this contribution, we
consider only the $H_T$ variable, defined as the sum of the transverse
momenta of all of the leptons (including neutrinos) and jets in the
event. 
The impact of the approach is expected to depend on the observable
under consideration and it may be more beneficial for variables sensitive to
multi-jet radiation, such as $H_T$, than for more inclusive variables
such as $p_{t,W}$. Comparisons for the latter are left to a
study now in progress.


\subsection{Use of BlackHat+Sherpa ntuples}

As has been partially detailed in these proceedings, there have been
many advances in the computation of the NLO corrections for multi-parton
final states. Often such calculations do not exist in a compact
user-friendly form, and other means must be taken to allow
experimentalists to have access to the results. The BlackHat+Sherpa
collaboration has chosen to make available ROOT tuples that contain
all of the parton-level information needed to form flexible
predictions. The ROOT ntuple framework is a very efficient way to store
such information and the use of ROOT tuples is very familiar to
experimentalists.

\begin{table}[b!]
\centering
\begin{tabular}{|c|c|p{10.5cm}|}
\hline&&\\[-2mm]
branch name & type & notes\\[2mm]
\hline&&\\[-2mm]
id & I & id of the event. Real events and their associated counter-terms share the same id. This allows for the correct treatment of statistical errors. \\
nparticle & I & number of particles  in the final state \\
px & F[nparticle] & array of the x components of the final state particles \\  
py & F[nparticle] & array of the y components of the final state particles \\  
pz & F[nparticle] & array of the z components of the final state particles \\  
E & F[nparticle] & array of the energy components of the final state particles \\  
alphas & D & $\alpha_s$ value used for this event \\
kf & I & PDG codes of the final state particles \\
weight & D & weight of the event \\
weight2 & D & weight of the event to be used to treat the statistical errors correctly in the real part\\
me\_wgt & D & matrix element weight, the same as weight but without pdf factors \\
me\_wgt2 & D & matrix element weight, the same as weight2 but without pdf factors \\
x1 & D & fraction of the hadron momentum carried by the first incoming parton \\
x2 & D & fraction of the hadron momentum carried by the second incoming parton \\
x1p & D & second momentum fraction used in the integrated real part \\
x2p & D & second momentum fraction used in the integrated real part \\
id1 & I & PDG code of the first incoming parton \\
id2 & I & PDG code of the second incoming parton \\
fac\_scale & D & factorization scale used \\
ren\_scale & D & renormalization scale used \\
nuwgt & I & number of additional weights \\
usr\_wgts & D[nuwgt] & additional weights needed to change the scale\\[2mm]
\hline
\end{tabular} 
\caption{Branches in a BlackHat+Sherpa ROOT file.}
\label{table:BranchList}
\end{table}


The ROOT ntuples store the four-vectors for the final state partons,
as well as their flavor information. The calculation is originally
performed using a specific choice of PDF, $\alpha_s(m_Z)$,
renormalization scale $\mu_R$ and factorization scale $\mu_F$, but
weight information is also stored in the ntuples that allows each
event to be easily re-weighted to any other (reasonable) values for
the above parameters. (PDFs are varied through calls to LHAPDF~\cite{LHAPDF}.) No
jet clustering has been performed on the final state partons; jet
reconstruction is left to the user, for any jet algorithm/size for
which the correct counter-events are present in the ntuple. For the
results presented here, the SISCone~\cite{Salam:2007xv}, $k_T$~\cite{Cacciari:2005hq} and
anti-$k_T$~\cite{Cacciari:2008gp} algorithms, with 
jet radii $R$ of $0.4$, $0.5$, $0.6$ and $0.7$ can be used. Each of the above jet
algorithms were run and the results stored in SpartyJet~
ntuples.\footnote{
SpartyJet~\cite{Delsart:2012jm} is a set of software tools for jet finding and analysis, built around the FastJet library of jet algorithms~\cite{Cacciari:2011ma}. SpartyJet provides four key extensions to FastJet: a simple Python interface to most FastJet features, a powerful framework for building up modular analyses, extensive input file handling capabilities, and a graphical browser for viewing analysis output and creating new on-the-fly analyses.} The
SpartyJet tuples were `friended' with the BlackHat+Sherpa ntuples,
allowing the analysis script access to all jet information. Such a
flexibility allows for an investigation of the dependence of the
physics on the details of the manner in which the partons are combined
into jets, in a manner difficult to achieve prior to this.

The four-vector information stored in the BlackHat+Sherpa ntuples is
shown in Table~\ref{table:BranchList}. Note the variety of entries needed for the
re-weighting of the cross section results, especially for the case of
the variation of the two scales $\mu_R$ and $\mu_F$.  Information is
stored in separate ntuples for the different categories of events,
which are typically Born, loop (leading color and sub-leading color),
real and subtraction terms. For large $n$, in $W+n$-parton final
states, there are many divergences present when two partons become
collinear or one parton becomes soft. These divergences are controlled
using the traditional Catani--Seymour approach~\cite{Catani:1996vz},
which involves the generation of many counter-events. Many of the events
have negative weights; only the sum is guaranteed to be
positive-definite. Predictions with reasonable statistical precision
may require the sum of billions of events. The resultant tuples may
amount to several Terabytes. However, the output can be subdivided
into ROOT files of order 5--10 GB, allowing for simultaneous parallel
processing of the events over multiple nodes, such as in the Tier3
facility at Michigan State University used for these comparisons.

\subsection{BlackHat+Sherpa predictions}

\begin{figure}[t!]
\begin{center}
\includegraphics[width=0.48\textwidth]{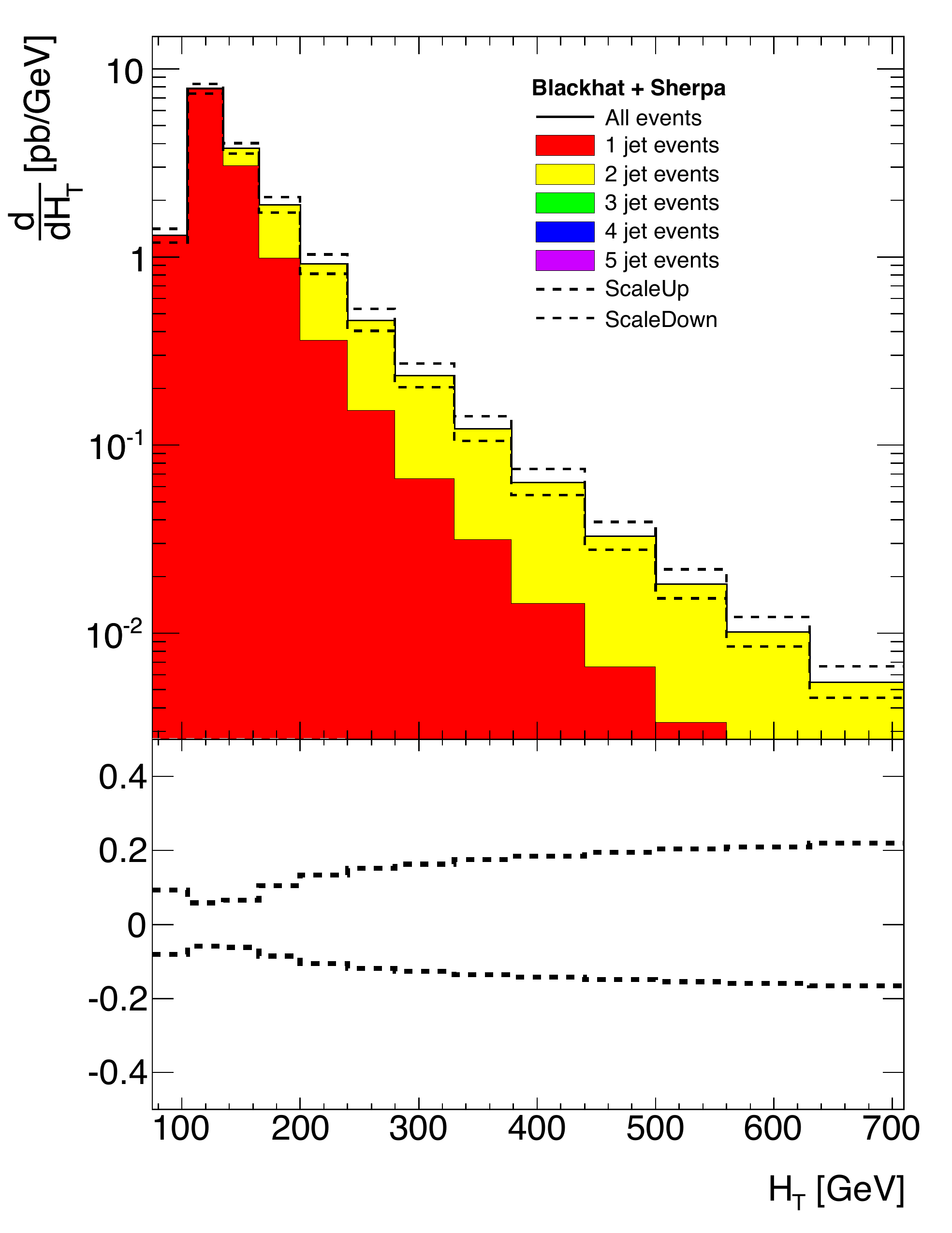}
\includegraphics[width=0.48\textwidth]{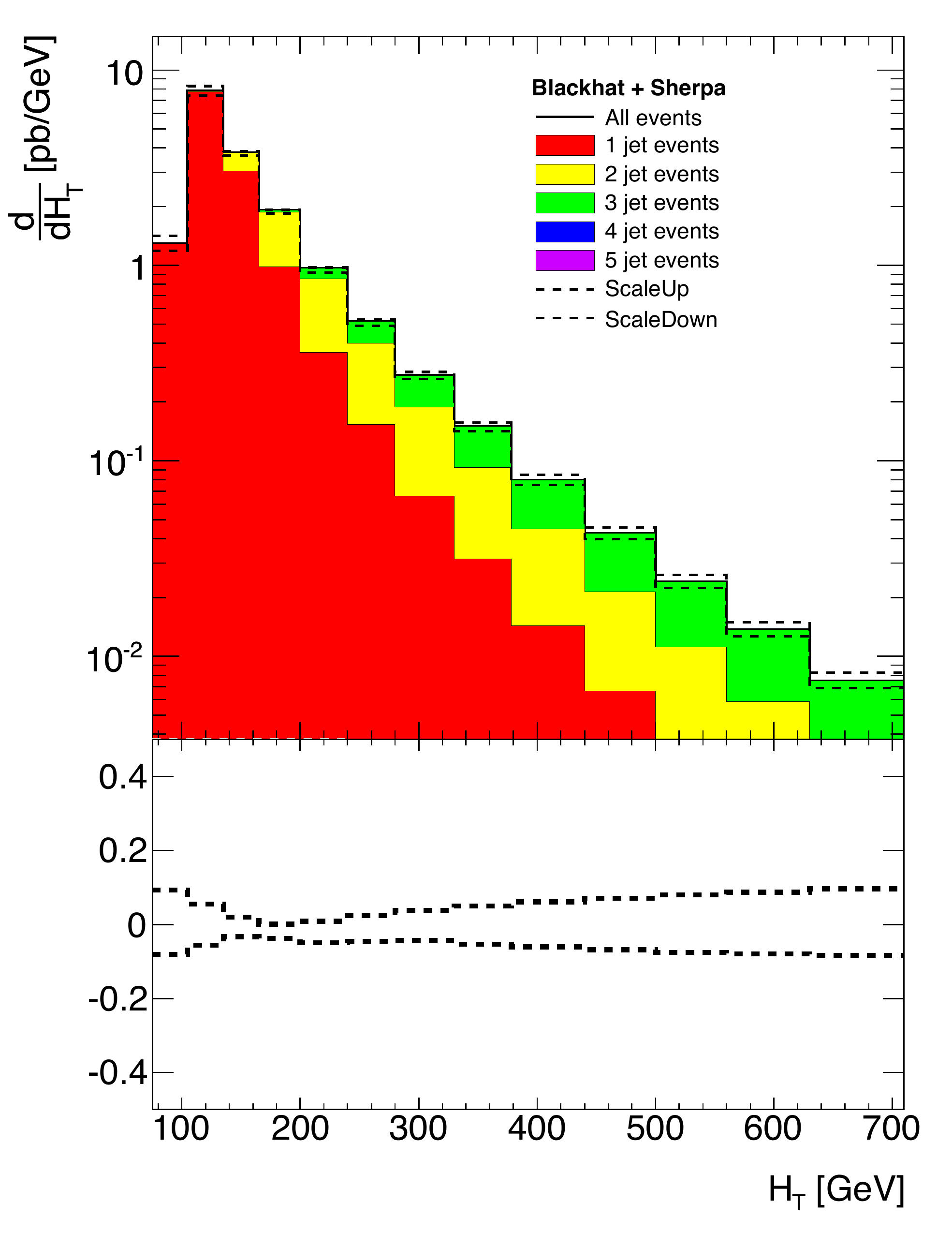}
\end{center}
\caption{The $W$ plus jets cross section, as a function of $H_T$, for
  the NLO inclusive $W+\ge1$-jet prediction (left) and for the
  Exclusive Sums approach, adding in $W+2$-jet production at NLO
  (right). The cross sections have been evaluated at a central scale
  of $H_T/2$ and the uncertainty is given by varying the
  renormalization and factorization scales independently up and down by
  a factor of 2, while ensuring that the ratio of the two scales is never
  larger than a factor of 2.}
\label{fig:HT1}
\end{figure}

\begin{figure}[t!]
\begin{center}
\includegraphics[width=0.48\textwidth]{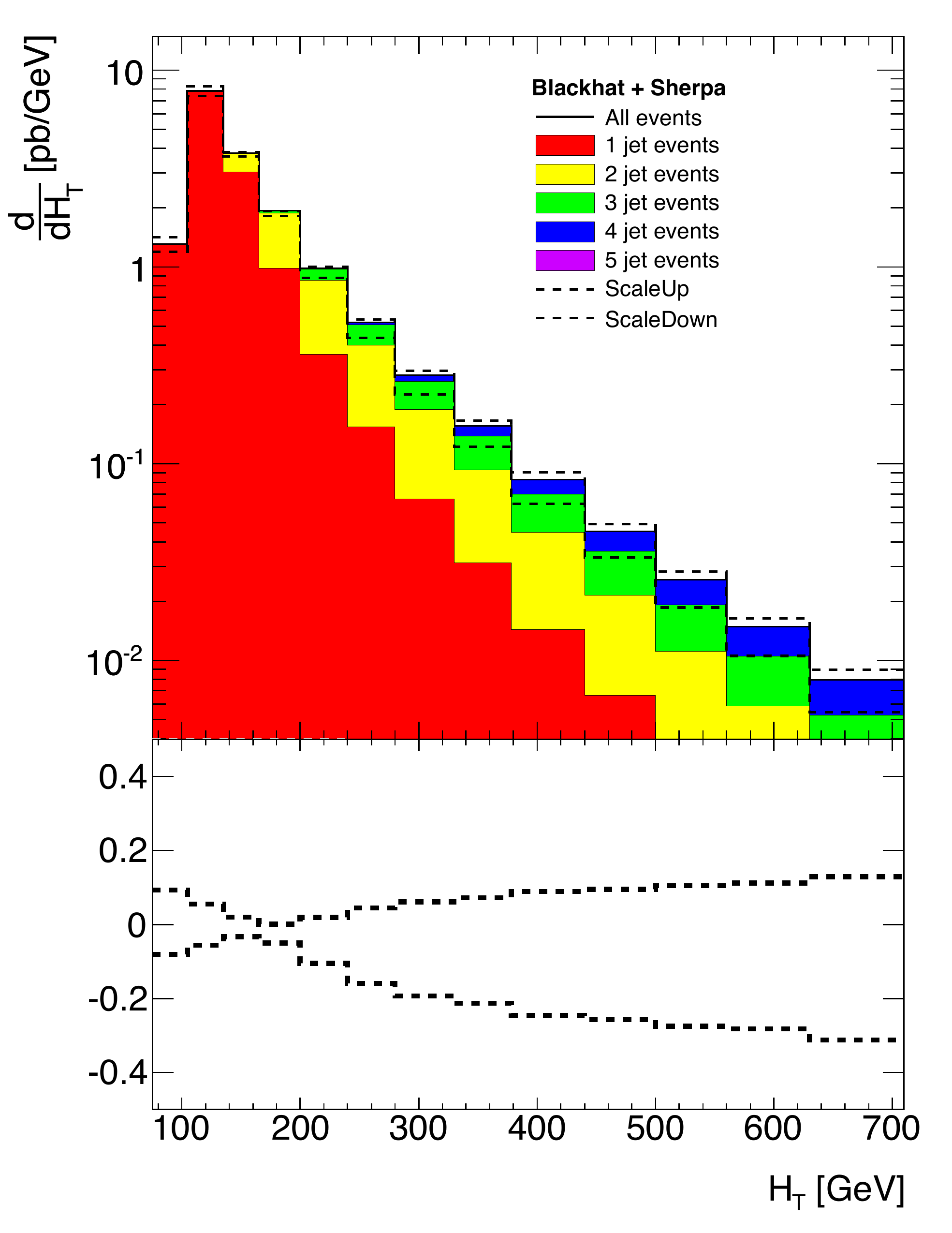}
\includegraphics[width=0.48\textwidth]{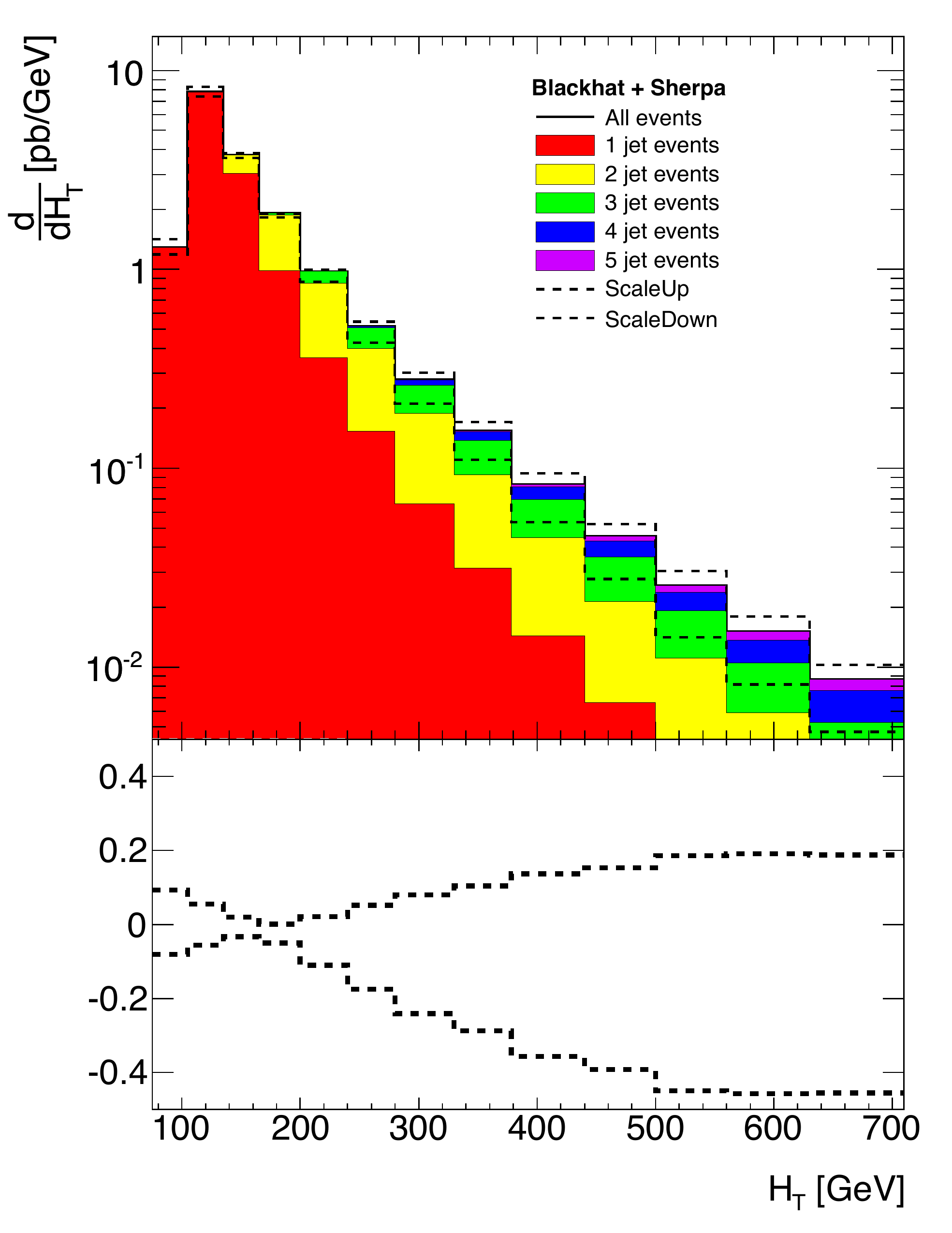}
\end{center}
\caption{ The $W$ plus jets cross section, as a function of $H_T$, for
  $W+\ge1$-jet production using the Exclusive Sums approach, and
  adding up to 3 jets at NLO (left) and 4 jets at NLO (right).  The
  cross sections have been evaluated at a central scale of $H_T/2$ and
  the uncertainty is given by varying the renormalization and
  factorization scales independently up and down by a factor of 2,
  while ensuring that the ratio of the two scales is never larger than a
  factor of 2.}
\label{fig:HT2}
\end{figure}

We have generated NLO predictions with the BlackHat+Sherpa predictions
implementing the cuts used in the 2010 ATLAS $W$ plus jets
paper~\cite{Aad:2012en}. For completeness, the cuts are reproduced
below:

\begin{itemize}
\item $p_T^\mathrm{lepton}\;>\;20$ GeV\,,
\item $|\eta^\mathrm{lepton}|\;<\;2.4$\,,
\item $E_T^\mathrm{miss}\;>\;25$ GeV\,,
\item $m_{T,W}\;>\;40$ GeV\,,
\item $p_T^\mathrm{jet}\;>\;30$ GeV\,,
\item $|y^\mathrm{jet}|\;<\;4.4$\,,
\item $\Delta R^\mathrm{lepton-jet}\;>\;0.5$\,.
\end{itemize}

In Figure~\ref{fig:HT1}, we show the NLO BlackHat+Sherpa prediction
for the $H_T$ distribution for $W+\ge1$ jets (left) using the
anti-$k_T$ jet algorithm with $R=0.4$. As the prediction is an
inclusive NLO calculation for $W+\ge1$ jets, there are contributions
from both the one-jet and the two-jet final states.  Note that as
$H_T$ increases, the contributions
from the $W+2$-jet subprocess also increases.  On the right, we
again show the $H_T$ distribution, but now compute the prediction
using the `Exclusive Sums' technique, adding in the NLO $W+2$-jet
information. Now there is a significant contribution at high $H_T$
from the $W+3$-jet final state as well. In Figure~\ref{fig:HT2}, the
$H_T$ prediction is shown using the Exclusive Sums approach, adding
1+2+3 jets at NLO (left) and 1+2+3+4 jets at NLO (right).
It is evident that as $H_T$ increases, contributions from higher jet
multiplicities that are only present implicitly in a traditional
inclusive NLO $W+\ge1$-jet calculation, become important. The
Exclusive Sums $H_T$ predictions agree with that for the inclusive NLO
$W+\ge1$-jet calculation at low $H_T$, but are larger at higher
$H_T$, and in better agreement with the ATLAS data (as discussed
below).

However, it can also be noticed that the scale dependences for the
Exclusive Sums predictions apparently get better when the 2-jet NLO
information is added, but significantly worse when the 3-jet and 4-jet
information is added. As discussed in the Appendix, the reduction in scale
dependence with the addition of the 2-jet NLO terms may be due to the stabilization of the predictions for the $qq \rightarrow Wq'q$ topologies. 
Adding the 3-jet and 4-jet NLO terms seems to destabilize the predictions. 
There are missing Sudakov terms needed to properly `stitch'
the different multiplicity samples together; it is hoped that the
LoopSim technique may offer one way in supplying those missing terms.

Below in Figure~\ref{fig:HT11}, we show the NLO BlackHat+Sherpa
predictions for the $H_T$ distribution for $W+\ge2$ jets: the
inclusive calculation to the left, the Exclusive Sums result adding
2+3-jet NLO information in the middle and the Exclusive Sums result adding
2+3+4-jet NLO information to the right. Over the kinematic range covered in
these plots, the Exclusive Sums technique adds less to the cross
section at high $H_T$, although there is still a degradation of the
scale dependence.

\begin{figure}[t!]
\begin{center}
\includegraphics[width=0.31\textwidth]{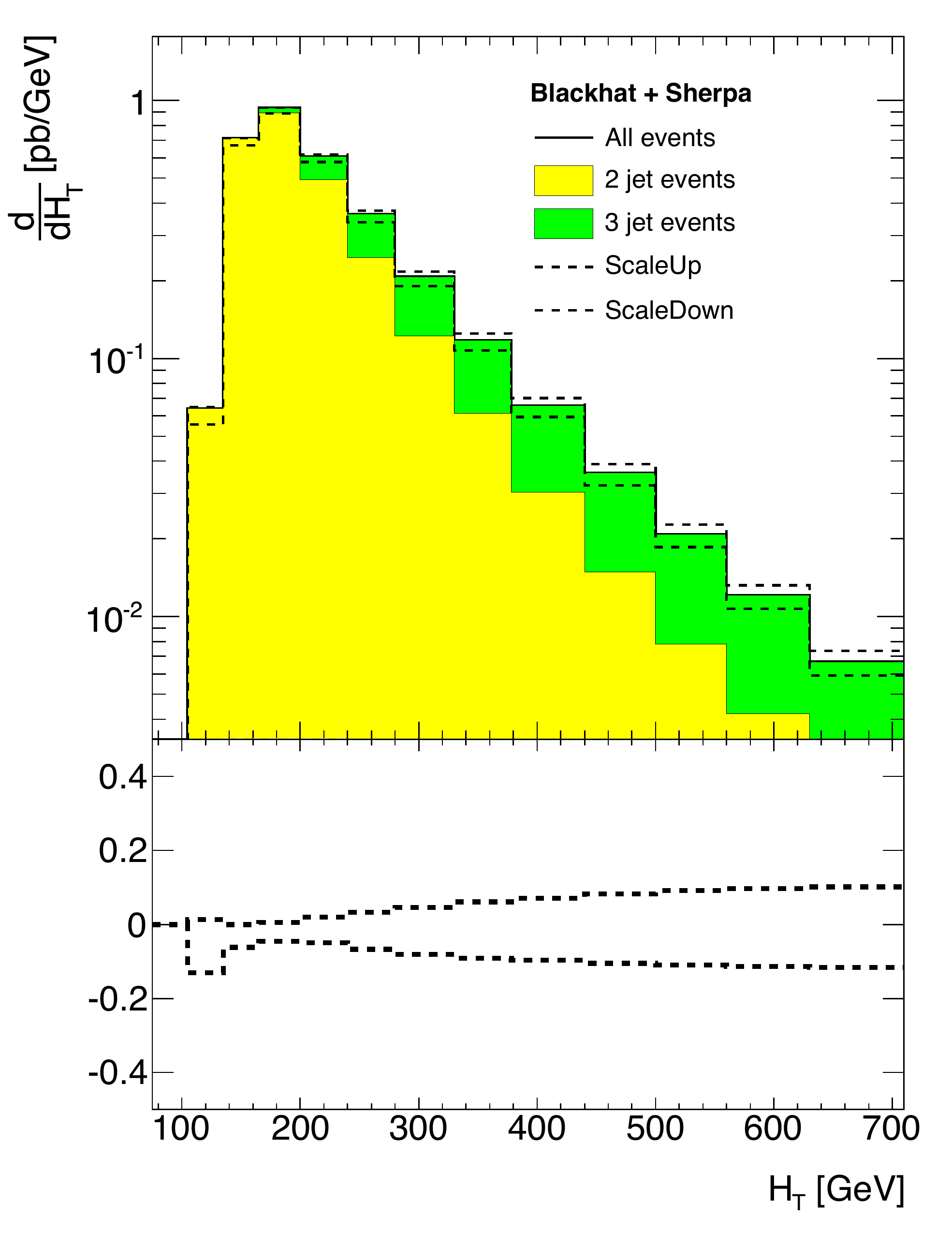}
\includegraphics[width=0.31\textwidth]{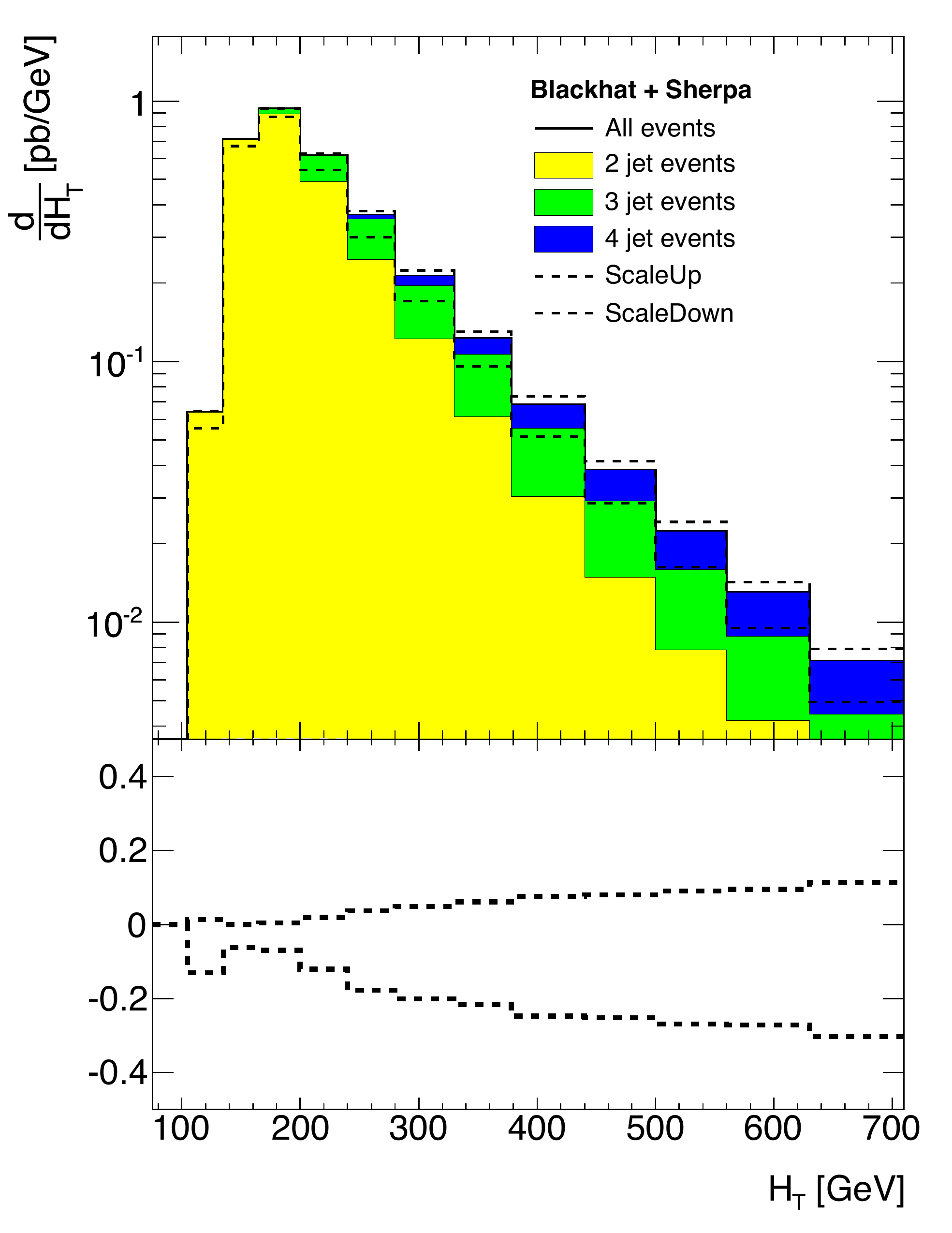}
\includegraphics[width=0.31\textwidth]{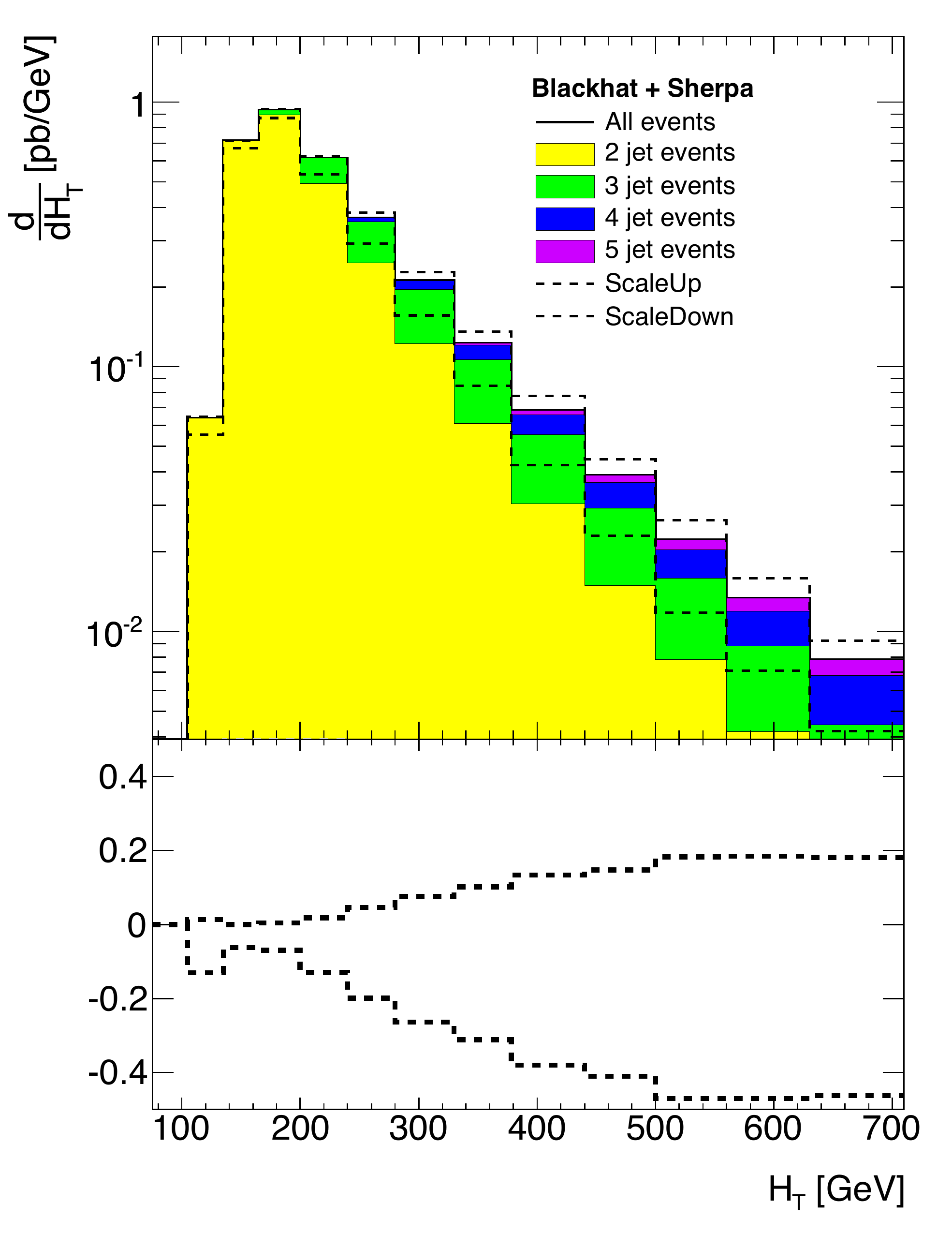}
\end{center}
\caption{The $W$ plus jets cross section, as a function of $H_T$, for
  $W+\ge2$-jet production using the inclusive NLO production (left)
  and the Exclusive Sums approach, adding up to 3 jets at NLO
  (center) and 4 jets at NLO (right).  The cross sections have been
  evaluated at a central scale of $H_T/2$ and the uncertainty is given
  by varying the renormalization and factorization scales independently
  up and down by a factor of 2, while ensuring that the ratio of the two
  scales is never larger than a factor of 2.}
\label{fig:HT11}
\end{figure}

\subsection{Towards interfacing BlackHat+Sherpa ntuples with LoopSim}

LoopSim is a method to simulate higher order QCD corrections, in particular
those beyond NLO.  It is expected to work best for processes with
large NLO-to-LO $K$-factor, however it was found to be
advantageous even in some cases where the $K$-factor is
moderate~\cite{Rubin:2010xp}. The method is based on unitarity and its
main ingredient is a procedure that takes events from a process with $n+m$
partons in the final state and produces counter-term events
with $n+m-1, n+m-2, \ldots, n$ particles, which approximate $1$-loop,
$2$-loop, etc. contributions.
In contrast to the Exclusive Sums method, it enables one to introduce
(approximate) virtual corrections beyond 1-loop, thus ensuring that
the $\alpha_s L^2$ type terms cancel for all the orders that are
included. While we will not show LoopSim results that are directly
comparable to the ATLAS data (the samples were generated before those
cuts were made public), we will examine below the dependence on the
$p_{t,\min}$ choice (which sets the size of $L=\ln O_t/p_{t,\min}$
where $O_t$ is a transverse observable) and see that it vanishes as
$p_{t,\min}\to0$.

To distinguish between the exact result at the order N$^p$LO and the result with
simulated loops we use a notation in which we replace N by $\bar n$ for the orders
simulated by LoopSim.  So for example, $W+1$ jet at $\bar n $LO has approximate
1-loop diagrams and is obtained by combining $W+1$ jet at LO with
$W+2$ jet at LO
where the latter is passed through LoopSim. Similarly $W+1$ jet at $\bar n $NLO has
exact 1-loop diagrams but simulated 2-loop contributions (by using
$W+2$ jet at NLO as an input to LoopSim).

As argued in the previous section, the BlackHat+Sherpa ntuples allow one to
efficiently perform a broad range of analyses. They have however a limitation.
In order to reduce the size of stored files, the only partonic events
that are recorded for the $W+n$-jet sample are those in which there are
at least $n$ jets above a $20$ GeV threshold.
%
%
Since this threshold is below the jet cuts used by ATLAS and CMS, it is
adequate for any NLO study of LHC jet cross sections.
%
%
The situation is slightly more complex if we want to use the
BlackHat+Sherpa ntuples to compute predictions beyond NLO using
LoopSim.  This is because the cut that is present in the $W+2$-jet
BlackHat+Sherpa NLO sample eliminates part of the real contribution to
the $W+1$-jet phase space at NNLO, for example $W+3$-parton events in
which the 3 partons all form part of a single jet, or in which 2
partons form part of one jet, while the third is well separated in
angle but below the $20$ GeV jet threshold.


Since we plan to use LoopSim interfaced to BlackHat+Sherpa ntuples in our future
study of multi-jet processes, it is important to directly check the effect of the
finite generation $p_t$ cut, $p_{t,\rm gen}^{\min}$, on the
predictions of the $p_t$ and $H_T$ distributions.
We have performed such a study for $W^-+1$ jet generated with MCFM,
where we varied a `parton'-$p_t$ generation cut from $1$ to
$20$~GeV.\footnote{Strictly speaking, the events destined for the $W+n$-jet
  sample were clustered with the $k_T$ algorithm with a very
  small separation, $R=0.03$, and accepted when at least $n$ small-$R$ jets were
  above the generation cut. These events were then passed through the
  standard analysis cuts (with or without LoopSim).}
This is not entirely equivalent to the cut in the BlackHat+Sherpa samples (which
is applied to the standard jets, not to the partons), but should be adequate
from the point of view of estimating the potential order of magnitude
of finite generation cuts.
The output from MCFM
was interfaced to LoopSim which produced the additional loop diagrams. Then, the
events were analyzed with the following set of cuts: $|y^\mathrm{lepton}|<2.5$,
$p_T^\mathrm{lepton}>20$~GeV , $|y^\mathrm{jet}|<4.5$, $p_T^\mathrm{jet}>25$~GeV ,
$m_{T,W}<20$~GeV, where the anti-$k_T$ algorithm with $R=0.4$ was used for
clustering.

\begin{figure}[t!]
  \includegraphics[width=0.495\textwidth,angle=0]{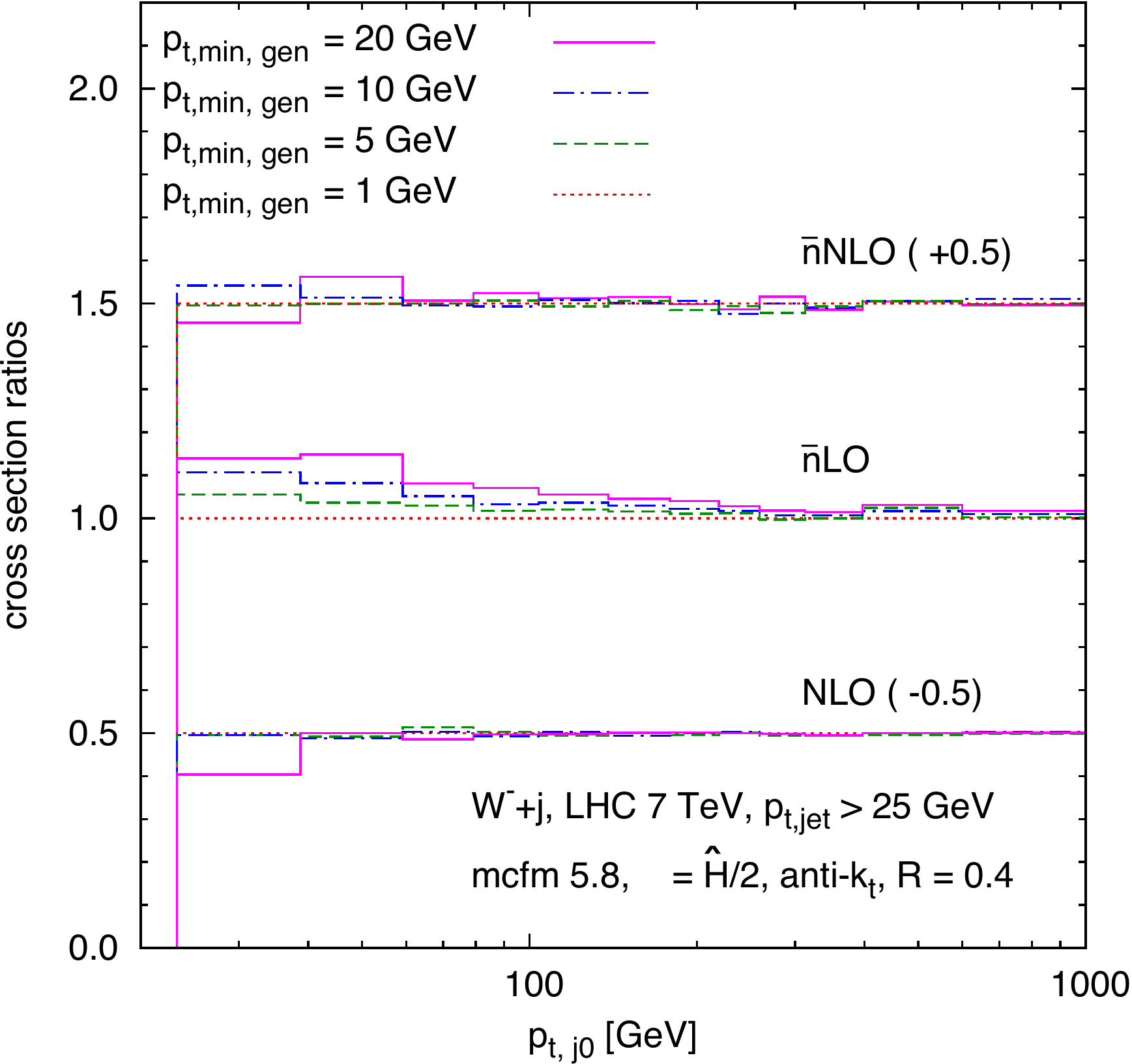}
  \includegraphics[width=0.495\textwidth,angle=0]{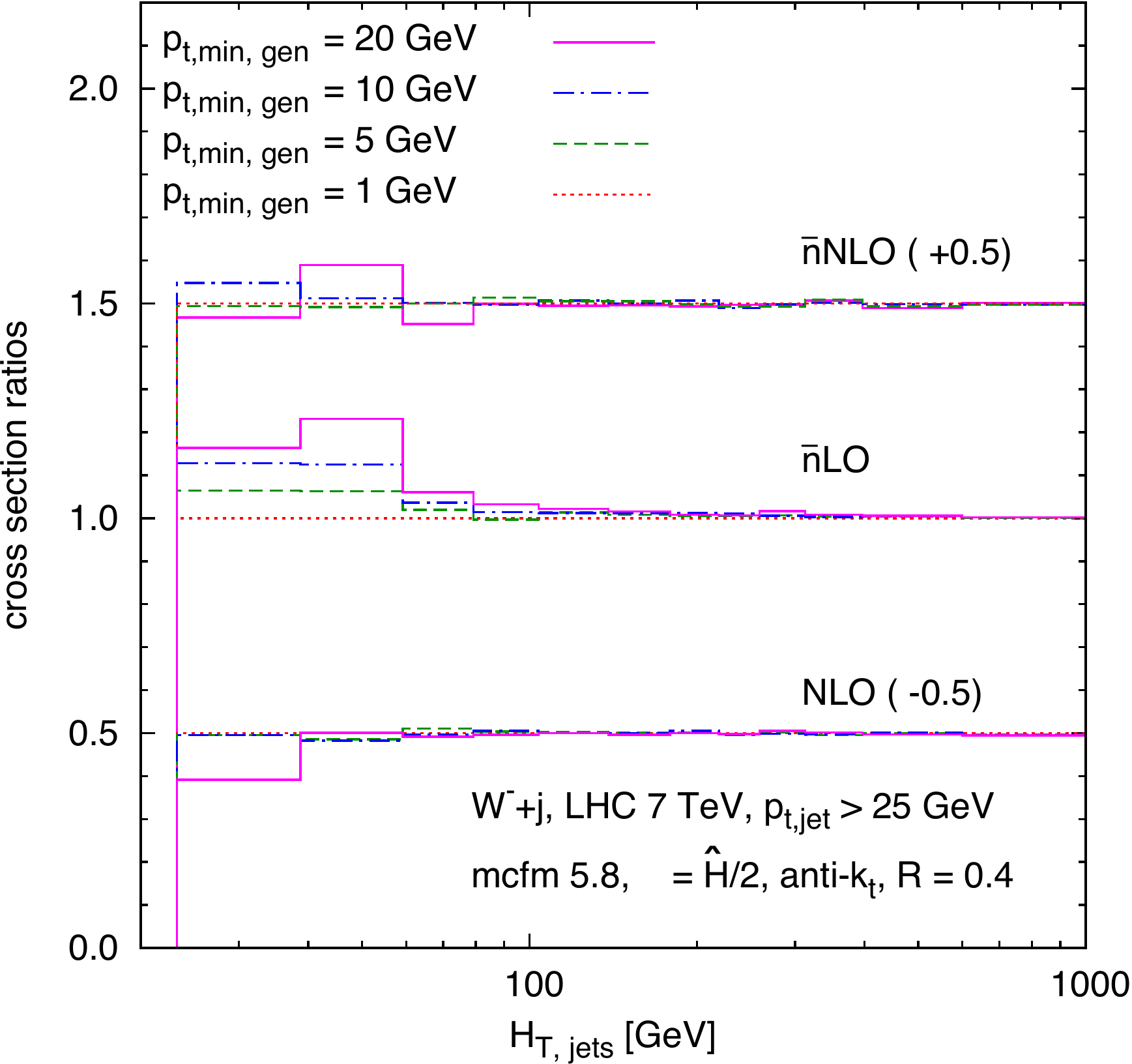}
  \caption{Ratios of cross sections from runs with a certain range of
    $p_{t,\rm gen}^{\min}$ values taken wrt.\ the cross section
    generated with $p_{t,\rm gen}^{\min}=1$~GeV for the distributions
    of $p_t$ of the leading jet (left) and the scalar sum of jets'
    transverse momenta, $H_{T,\rm{jets}}$ (right).
    \label{fig:finite-ptgen}}
\end{figure}

The results are presented in Figure~\ref{fig:finite-ptgen} where the ratios of
cross sections obtained with a range of generation cuts are shown as functions
of the $p_t$ of the leading jet and $H_{T,{\rm jets}}$. At NLO, the only
artefact we see is for the $p_{t,\rm gen}^{\min}$ of $20$ GeV in the bin
below $40$~GeV. This is as expected, since a $20$~GeV cut on each of
two partons can at most affect jets up to $40$~GeV (such an artefact
would not be present in the BlackHat+Sherpa samples).
At $\bar n$LO and $\bar n$NLO, however, the dependence on
$p_{t,\rm gen}^{\min}$ is extended to a larger range of
$p_{t,\rm lead.jet}/H_{T,\rm jets}$ and it is visible also for values
of $p_{t,\rm gen}^{\min}<20$~GeV.  However, even if the $p_{t,\rm gen}^{\min}$
dependence of the $\bar n$LO and $\bar n$NLO results is stronger than at NLO, it
dies out quickly with increasing $p_{t,\rm lead.jet}/H_{T,\rm jets}$ and becomes
irrelevant at  $\sim 100$ GeV, depending on the observable and the
order.
This appears to be consistent with the expectation that the effect of
the cut should vanish as a power of $p_{t,\rm gen}^{\min}/p_{t,\rm lead.jet}$ or
$p_{t,\rm gen}^{\min}/H_{T,\rm jets}$.

Therefore we conclude, that in spite of the finite generation cut one
should be able to trust the results obtained using BlackHat+Sherpa
ntuples, above a moderate $p_t$ limit, even for more complex analyses
such as those involving LoopSim.


\subsection{Comparisons to data, Sherpa and HEJ predictions}

\begin{figure}[t!]
\begin{center}
\includegraphics[width=0.495\textwidth]{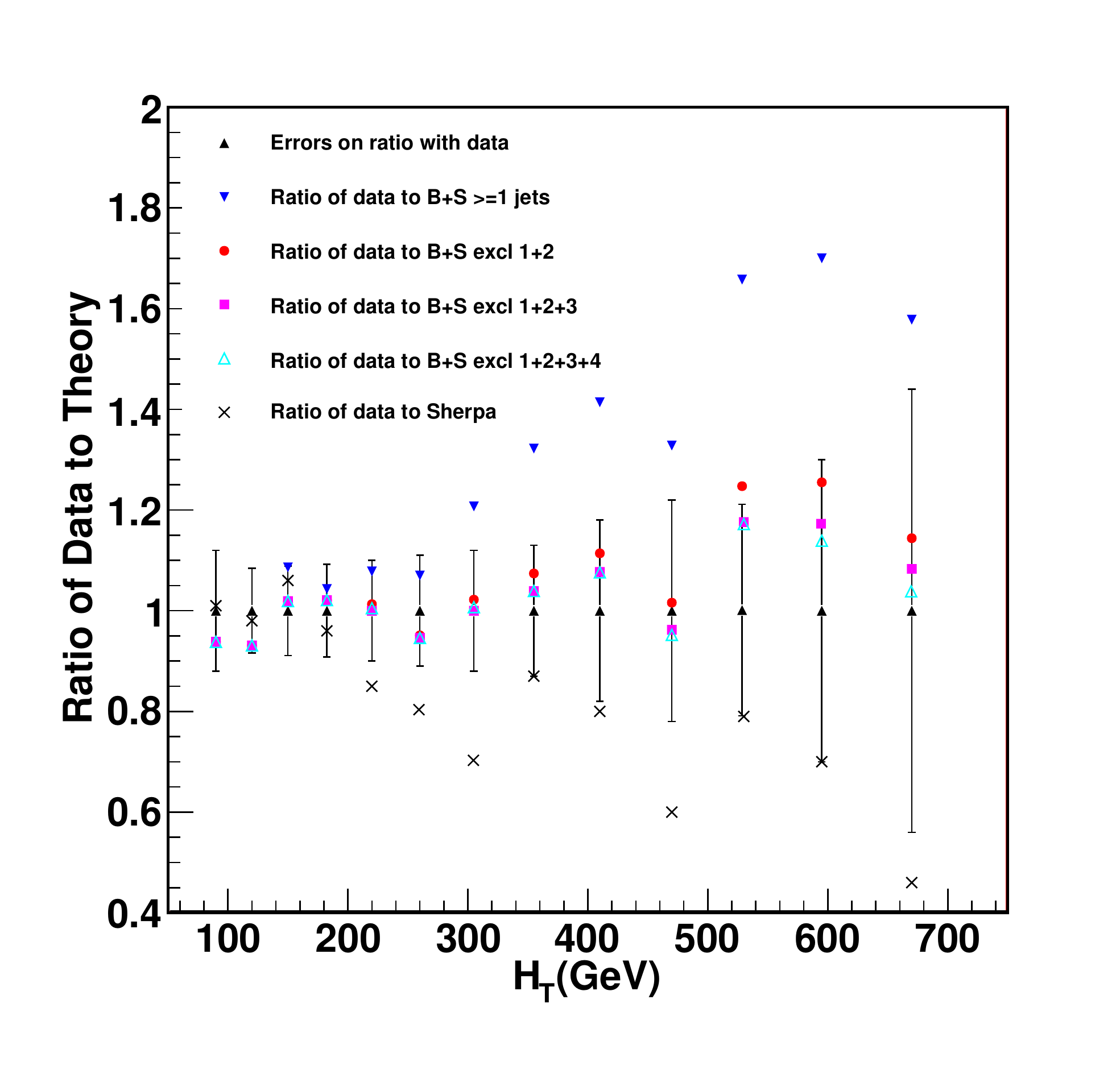}
\includegraphics[width=0.495\textwidth]{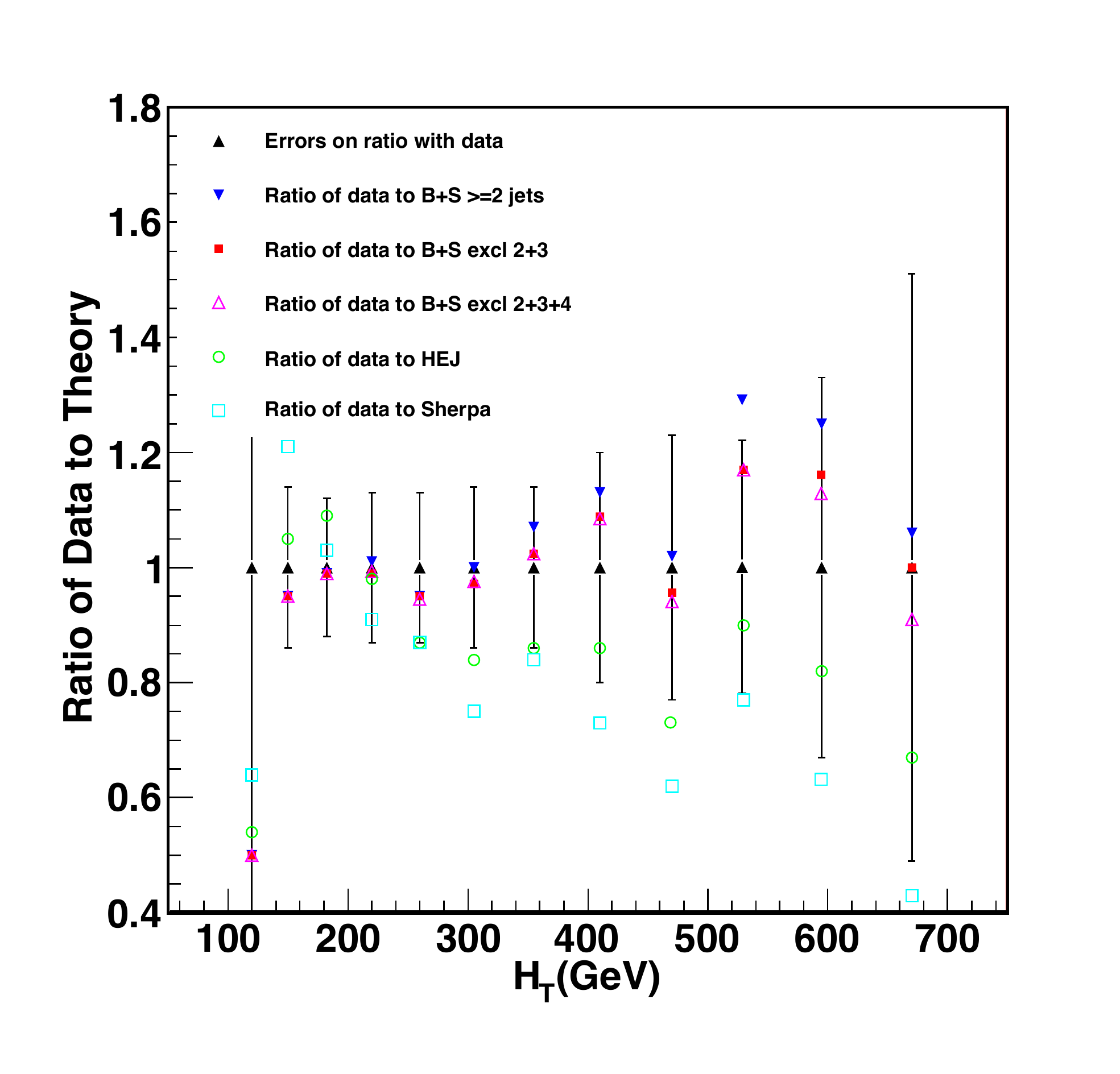}
\end{center}
\caption{The ratios of the ATLAS $W+\ge1$-jet (left) and $W+\ge2$-jet
  (right) cross sections, as a function of $H_T$, taken wrt.\ various
  theory predictions. The absolute normalization has been kept as
  given by the calculations. The error bars represent the total
  fractional error (statistical plus systematic added in quadrature)
  at each point.}
\label{fig:HT3}
\end{figure}

\begin{figure}[t!]
\begin{center}
\includegraphics[width=0.495\textwidth]{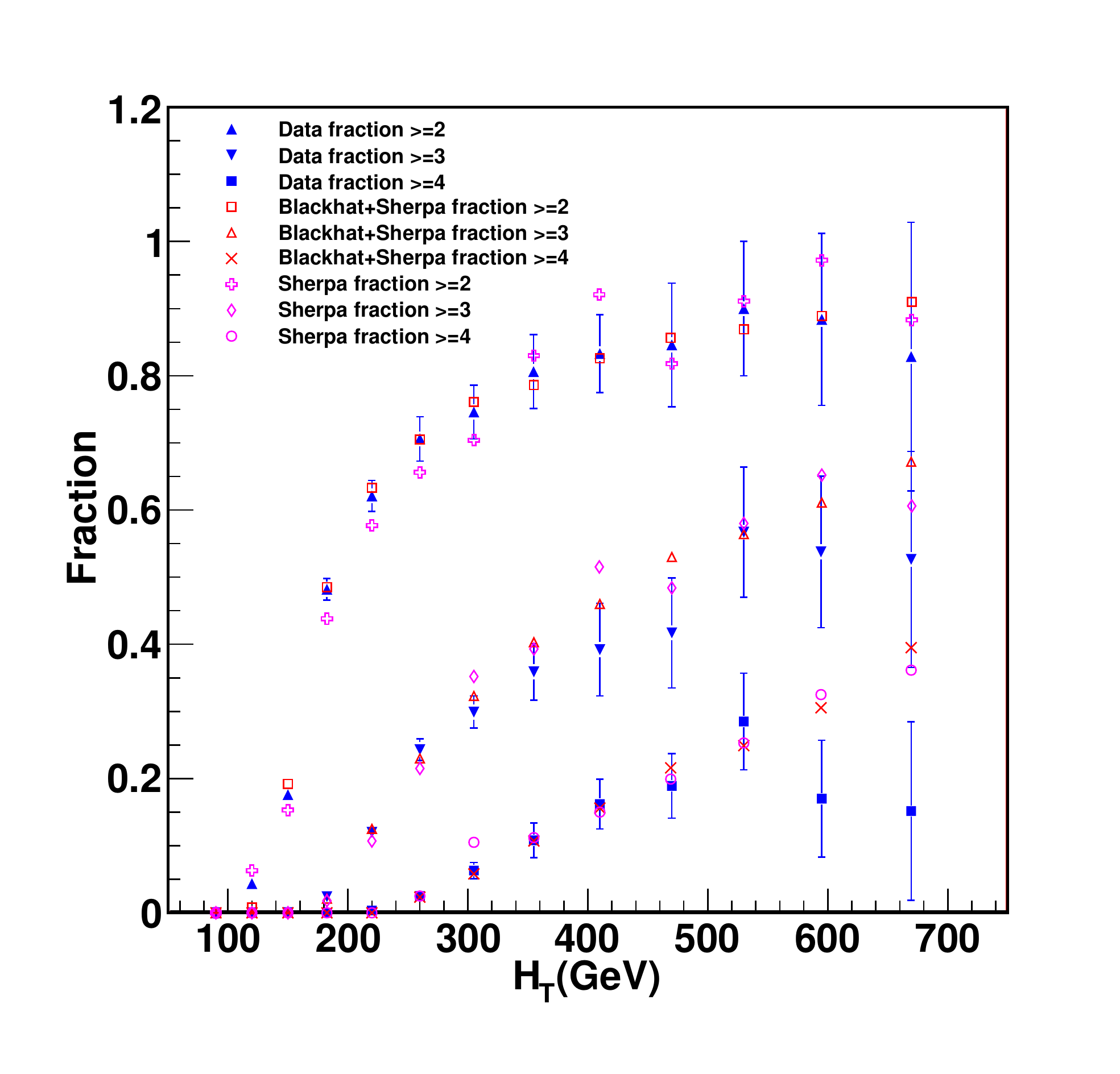}
\includegraphics[width=0.495\textwidth]{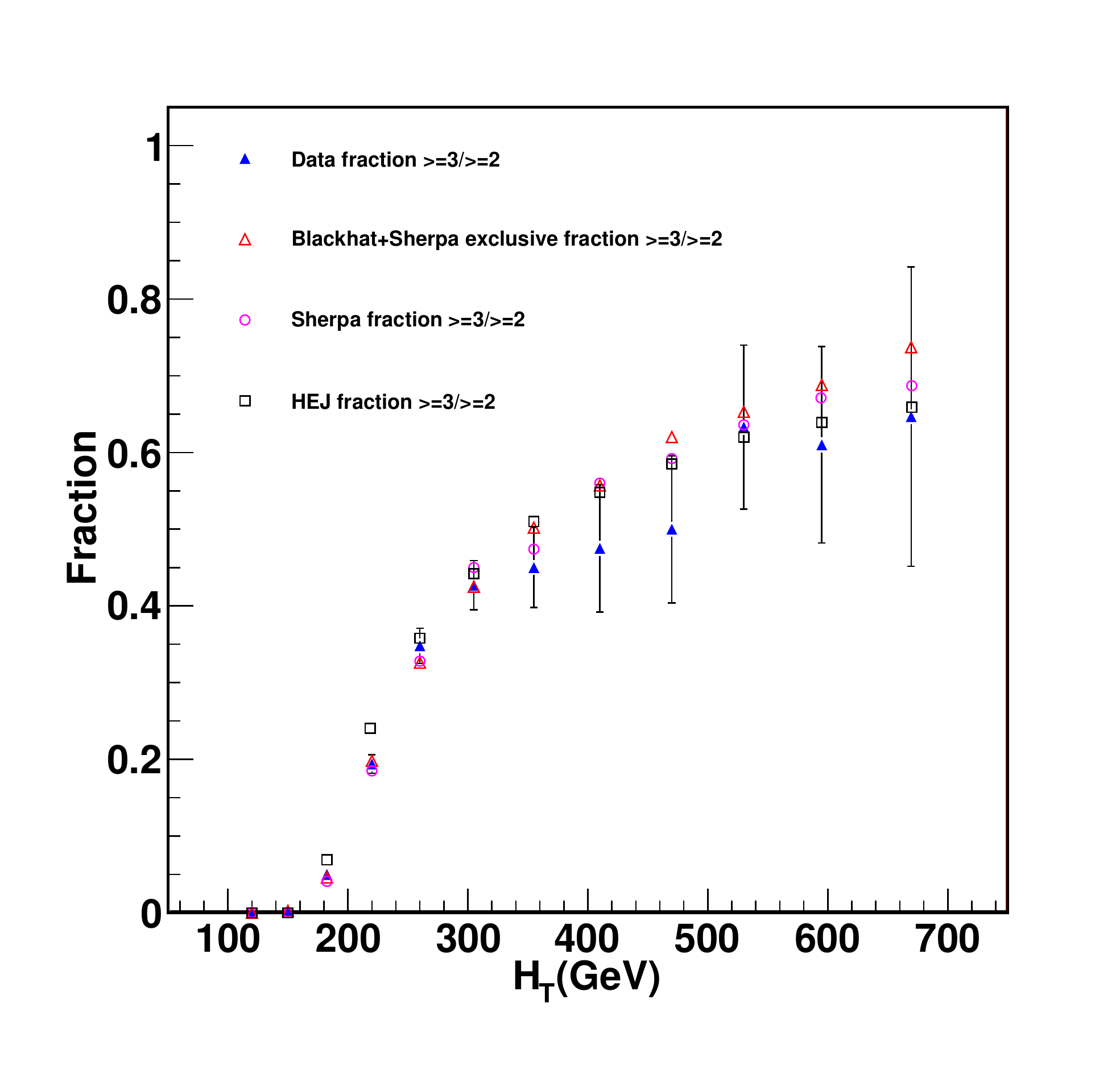}
\end{center}
\caption{ (left) The fraction of the $H_T$ cross section for
  $W+\ge1$-jet events arising from the $W+\ge2$-jet, $W+\ge3$-jet and
  $W+\ge4$-jet final states derived
  from the Exclusive Sums approach, from Sherpa and from HEJ, compared
  to the 2010 ATLAS data. (right) The ratio of the cross sections for
  $W+\ge3$ jets to $W+\ge2$ jets, as a function of $H_T$, using
  predictions from the Exclusive Sums approach, from Sherpa and from
  HEJ, compared to the ratio from the 2010 ATLAS data.}
\label{fig:HT4}
\end{figure}

In Figure~\ref{fig:HT3} (left), we compare the ratio of the 2010 ATLAS
$W$ plus jets data for the $H_T$ distribution for $W+\ge1$ jets to
predictions using the generic NLO calculation for $W+\ge1$ jet, the
Exclusive Sums approach adding up to 4 jets at NLO and the Monte
Carlo event generator Sherpa.
%
The agreement between the data and the pure NLO result is rather poor; 
it improves substantially with the inclusion of the Exclusive sums up
to two jets at NLO, with further small improvements coming from higher
multiplicities. As a reminder, we previously noted that the scale dependence 
improved when adding the 2-jet NLO information, but degraded when
adding higher jet multiplicities. The Sherpa prediction slightly
overshoots the data for $H_T$ in the inclusive $W+1$-jet bin. We
however note that the data versus Sherpa $H_T$ ratio has been formed
based on the absolute normalization as given by the Monte Carlo
simulation. Comparing the inclusive 1-jet cross sections, we find a
factor of $0.97$ between the data and the Sherpa result.

%
%
%
In Figure~\ref{fig:HT3} (right), we compare the ratio of the 2010 ATLAS
$W$ plus jets data for the $H_T$ distribution for $W+\ge2$ jets to
predictions using the generic NLO calculation for $W+\ge2$ jet, the
Exclusive Sums approach adding up to 4 jets at NLO, and to predictions
from HEJ and from Sherpa. As noted previously, there is some increase
in the predictions from the Exclusive Sums approach at the highest
$H_T$ values, but not nearly as much as in the $W+\ge1$-jet case.
These increases go in the direction of closer agreement with the data,
but the statistical error does not allow a clear judgement to be
made. The Sherpa and HEJ predictions for this ratio are in reasonable
agreement with the data but appear to fall off somewhat more rapidly
at large $H_T$ than either the data or the various BlackHat+Sherpa
predictions. Again this partly is the result of relying on the
absolutely normalized Monte Carlo predictions, which yield
$W+\ge2$-jet normalization factors of $0.95$ or $0.93$ between data
and Sherpa or HEJ, respectively.

In Figure~\ref{fig:HT4} (left), we show the predictions for the
fractions of the $H_T$ cross section in the inclusive $W+1$-jet bin
arising from the inclusive $W+2$-jet, $W+3$-jet and $W+4$-jet final
states as obtained from the Exclusive Sums approach and from Sherpa,
compared to the 2010 ATLAS data. In Figure~\ref{fig:HT4} (right), we
show the ratio of the cross sections for $W+\ge3$ jets to $W+\ge2$
jets, as a function of $H_T$, again using predictions from the
Exclusive Sums approach and from Sherpa but also from HEJ. We again
compare to the ratio given by the 2010 ATLAS data. All three
predictions agree with each other and with the data over the range
considered, despite the big differences in the approaches. There may
be an indication of some separation between the predictions at the
very highest $H_T$ values.

\subsection{Conclusions, outlook and future studies}

The advances achieved over the last few years in calculating NLO
corrections for multi-jet final states allow a more serious
consideration of the possibility to combine various $n$-jet NLO
predictions into an inclusive jet sample. The Exclusive Sums approach
discussed in this contribution is a first promising step into this
direction. More studies are required to understand the uncertainties
related to this procedure. One way of doing so would be to test the
stability of the predictions against variation of the jet algorithm
and/or parameters of the jet algorithm used to obtain and separate the
different NLO predictions for the fixed-multiplicity sets that
eventually make up the sum of exclusive $n$-jet contributions.%
\footnote{In this study, the same jet definition (anti-$k_T$ with
  $R=0.4$) was used for both establishing the separation of the fixed
  jet-multiplicity contributions and evaluating the cuts and
  observables.}

For the Exclusive Sums approach, outlined here for the
case of $W+\ge1$ jets, contributions are added proportional to
$\alpha_s^2$ ($W+1$ jet at NLO), $\alpha_s^3$ ($W+2$ jets at NLO),
$\alpha_s^4$ ($W+3$ jets at NLO) and $\alpha_s^5$ ($W+4$ jets at NLO),
i.e.\ this procedure mixes powers of $\alpha_s$ and thus is missing
essential Sudakov form factors that effectively bring each term to
the same power of $\alpha_s$.
%
%
One could imagine accomplishing this by embedding the NLO matrix
elements in a parton shower Monte Carlo framework, however the
technology for merging different multiplicities of NLO calculations
with a parton shower is still under development.
Note that at LO the tree-level matrix-element plus parton-shower
merging methods (e.g.\ as implemented in Sherpa) are designed to
satisfy this same-${\cal O}(\alpha_s)$ requirement by including the
(all-orders) leading-log effects to the `LO Exclusive Sums' exhibiting
the LO analog of the Exclusive Sums discussed here. Compared to the
matrix-element plus parton-shower merging, we see that the `NLO
Exclusive Sums' technique only accounts for Sudakov effects up to
${\cal O}(\alpha_s)$ while it describes each jet bin at full NLO
instead of LO accuracy.

Relying on the parton shower Monte Carlo framework is not the only way
to go in refining the Exclusive Sums strategy. Alternatively, the
LoopSim method can be used to provide approximations to the
higher-loop terms missing in the Exclusive Sums approach. As we have
seen here, prospects for using it together with BlackHat+Sherpa
ntuples seem promising.
A detailed comparison of the LoopSim results to LHC data is however
beyond the scope of this Les Houches contribution, though we look
forward to it being carried out in the near future.

\begin{figure}[t!]
\begin{center}
\includegraphics[width=0.77\textwidth]{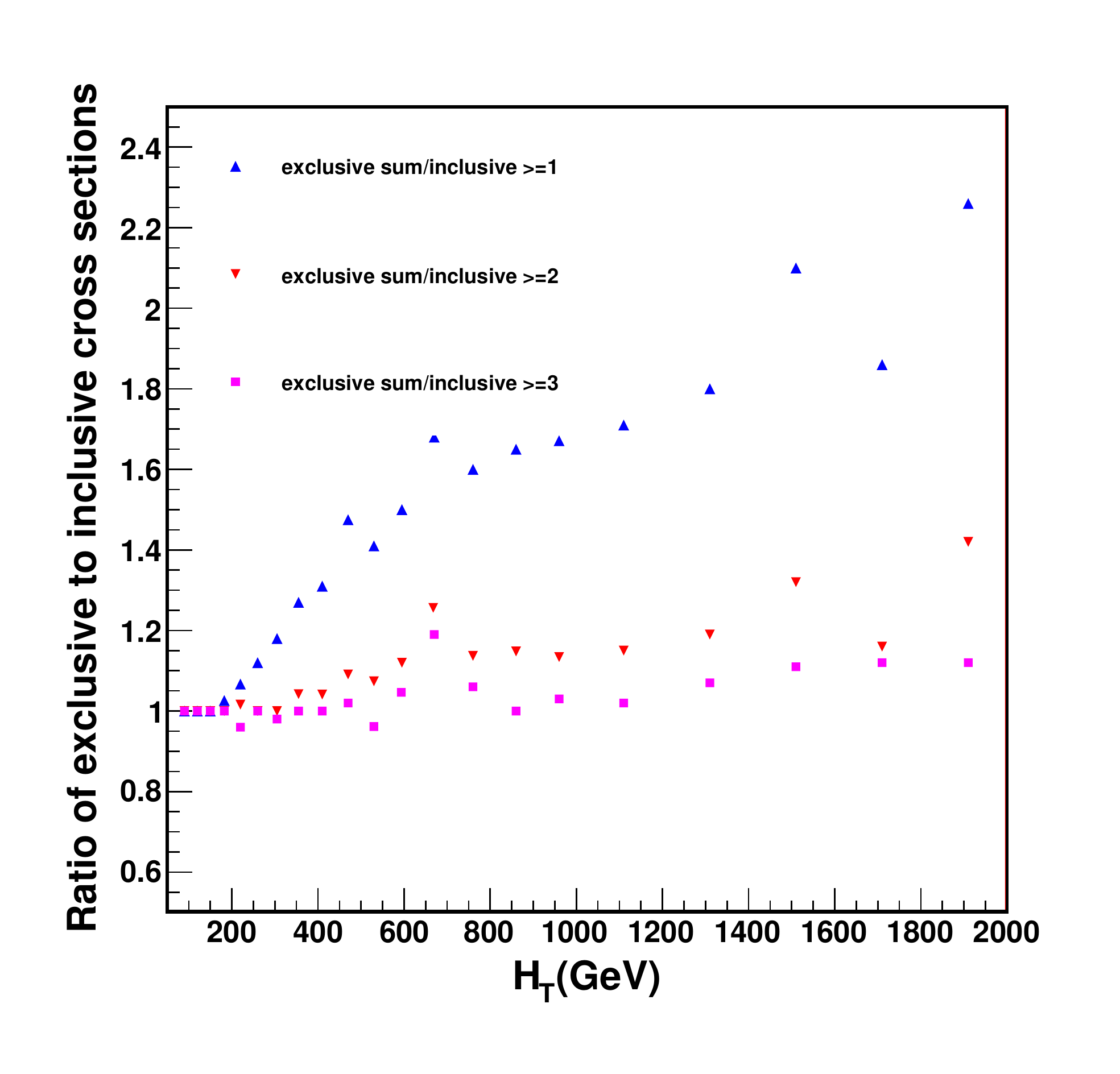}
\end{center}
\caption{The ratio of the predictions obtained from the NLO Exclusive
  Sums approach to the inclusive NLO predictions for $W\ge1$ jets,
  $W\ge2$ jets and $W\ge3$ jets. The `jitter' is due to the limited
  BlackHat+Sherpa statistics for these predictions.}
\label{fig:HT5}
\end{figure}

The ATLAS data taken in 2011 is about a factor of 130 times as large
as the data taken in 2010 (the only published data for $W$ plus jets so
far). This will allow a much further reach in all kinematic variables.
To get an idea, we show in Figure~\ref{fig:HT5} the ratio of the
predictions from the Exclusive Sums to the respective inclusive NLO
predictions for $W+\ge1,2,3$ jets. At an $H_T$ value of 2~TeV, the
ratio for $W\ge1$ jet is of the order of $2$; the ratio for $W\ge2$
jets rises to about $1.4$. The NLO-to-LO $K$-factor for $W+\ge1$ jet
rises rapidly with increasing $H_T$, while the $K$-factor for $W+\ge2$
jets increases only moderately (because no new subprocesses are being
introduced). It will be interesting to see if (a) the additional
factor of $2$ (for the $W+\ge1$-jet case) and (b) the additional
factor of approximately $1.4$ (for the $W+\ge2$ jet case) lead to
better agreement with the data. The LHC data from 2011 (and the higher
statistics expected in 2012) will reach these kinematic values and
should shed further light on the necessity and the efficacy of this
theoretical technique, not only for $W+\ge1$ jet, but for higher jet 
multiplicities as well.

\subsection*{Acknowledgments}
We would like to thank the Les Houches organizers for a very
stimulating workshop, and Zvi Bern, Lance Dixon and Kemal Ozeren for
useful discussions. GPS and SS gratefully acknowledge support from the
Agence Nationale de la Recherche under grant ANR-09-BLAN-0060. JMS is supported by the UK Science and Technology Council (STFC). JH acknowledges support from the National Science Foundation.

\subsection*{Appendix: a double logarithmic analysis of the Exclusive Sums method}

To help understand the structure of the Exclusive Sums method, it can
be useful to consider how it works in a simple double logarithmic
approximation.
We use $p_{t,\min}$ to represent the minimum $p_t$ for the jets in the
Exclusive Sums sample, and first study the cross section for $W$ production
as a function of $p_{t,W}$ at high $p_{t,W}$ ($\gg m_W$), considering
in particular the terms that go as $\alpha_s^n L^{2n}$ where $L = \ln
p_{t,W}/p_{t,\min}$.
The 0-jet sample does not contribute at all to non-zero $p_{t,W}$, so
the first term comes from the exclusive 1-jet contribution. If
calculated to all orders in the double logarithmic approximation
(DLA), it would have the form
\begin{equation}
  \sigma_{1,\mathrm{excl}}^{\text{DLA}}(p_{t,W})\;=\;
  \sigma_{1}^{\text{LO}}(p_{t,W})\,
  \exp\left(- \frac{2\,C \alpha_s}{\pi} L^2\right)\,,
\end{equation}
where $C = 2\,C_F + C_A$ for the (dominant) $qg\to W^{\pm} q'$
scattering process.
The $n$ exclusive jet rate would be given by
\begin{equation}
  \sigma_{n,\mathrm{excl}}^{\text{DLA}}(p_{t,W})\;=\;
  \sigma_{1}^{\text{LO}}(p_{t,W})\;
  \frac1{(n-1)!}\,\left(\frac{2\,C \alpha_s}{\pi} L^2\right)^{n-1} 
  \exp\left(- \frac{2\,C \alpha_s}{\pi} L^2\right),
\end{equation}
and one sees that the sum over all multiplicities is given by
\begin{equation}
  \label{eq:excl-sum-full}
  \sigma(p_{t,W})^{\text{DLA}}\;=\;
  \sum_{n=1}^\infty\sigma_{n,\mathrm{excl}}^{\text{DLA}}(p_{t,W}) 
  \,=\,\sigma_{1}^{\text{LO}}(p_{t,W})\,,
\end{equation}
i.e.\ in the double logarithmic approximation, there are no
corrections to the $p_{t,W}$ distribution at high $p_{t,W}$.
Now let us consider what happens if we expand each of the exclusive
sums to NLO. For the $n$-jet cross section, we have
\begin{equation}
  \sigma_{n,\mathrm{excl}}^{\text{NLO(DLA)}}(p_{t,W})\;\simeq\;
  \sigma_{1}^{\text{LO}}(p_{t,W})\;
  \frac1{(n-1)!}\,\left(\frac{2\,C \alpha_s}{\pi} L^2\right)^{n-1}
  \left(1 - \frac{2\,C \alpha_s}{\pi} L^2\right)\,.
\end{equation}
Performing the sum over $n$, which corresponds to summing an infinite
tower of NLO exclusive jet calculations, leads to
\begin{subequations}\label{eq:excl-sum-NLO}
\begin{align}
  \sigma(p_{t,W})^{\text{DLA}} 
                  &\;=\;\sum_{n=1}^\infty\sigma_{n,\mathrm{excl}}^{\text{NLO(DLA)}}(p_{t,W})\\
                  &\;=\;\sigma_{1}^{\text{LO}}(p_{t,W}) 
                    \exp\left(\frac{2\,C \alpha_s}{\pi} L^2\right) 
                    \left(1 - \frac{2\,C \alpha_s}{\pi} L^2\right)\\
                  &\;=\;\sigma_{1}^{\text{LO}}(p_{t,W})
                    \left(1 - \frac12\left(\frac{2\,C \alpha_s}{\pi} L^2\right)^2 + 
                          {\cal O}(\alpha_s^3 L^6)\right).
\end{align}
\end{subequations}  
As long as $L^2$ is not large, the difference between this and the
correct answer of Eq.~(\ref{eq:excl-sum-full}) is a straightforward
NNLO correction, i.e.\ small. However when $p_{t,W} \gg p_{t,\min}$
the logarithms become large, the $\alpha_s^2 L^4$ term can be of order
$1$ and the Exclusive Sums method may then no longer be a good
approximation.
A similar analysis can be performed for an exclusive sum truncated at
some finite order, as used in our study.

Given the above discussion, one may wonder then if there are any
circumstances in which the Exclusive Sums method will bring benefits.
For the observable studied in this contribution, $H_T$, the key
difference with respect to $p_{t,W}$ is that it is subject to a
`giant' $K$-factor at NLO.
This phenomenon is associated with `dijet' topologies in which a
soft or collinear $W$ is radiated off the dijet system, leading to a
double logarithmic (electroweak) enhancement.
In addition these topologies can be created by $qq$ type scattering
(whereas the LO process involves only $gq$ or $q\bar q'$ scattering),
leading to further enhancement in $pp$ collisions at large $H_T$.
Dijet type topologies contribute significantly to the $H_T$
distribution, even when the $W$ is soft, because the variable sums all
particles' transverse momenta (whereas the softness of the $W$ limits these
topologies' contribution to the $p_{t,W}$ distribution).

Because of the giant $K$-factor, for the $H_T$ variable the behaviour
of the Exclusive Sums method is more subtle than for $p_{t,W}$: while
the $\sigma_{W+2}^{\text{NLO}}$ contributions destabilize the prediction
for the $qg \to W q'$ type topologies, they instead stabilize the
prediction for the much larger $qq \to W q'q$ topologies (present only
at LO in a NLO $W+1$-jet calculation).
Going further in the exclusive sum, however, i.e.\ including
$\sigma_{W+3}^{\text{NLO}}$ and $\sigma_{W+4}^{\text{NLO}}$ contributions can
however destabilize the predictions for both kinds of topologies.
Traces of this behaviour were visible in the numerical studies
shown above.

%
%
%
%




}

\section[$\mathbold{W}$ Production in Association with Multiple Jets at the LHC]
{$\mathbold{W}$ PRODUCTION IN ASSOCIATION WITH MULTIPLE JETS AT THE LHC \protect\footnote{Contributed by: J.~R.~Andersen,
  D.~Ma\^itre,
  J.~M.~Smillie,
  J.~Winter}}
{\graphicspath{{WNJetsLHProc/}}

\title{$\mathbold{W}$ Production in Association with Multiple Jets at the LHC}
\author{Jeppe R.~Andersen$^1$,
  Daniel Ma\^itre$^{2,3}$,
  Jennifer M.~Smillie$^4$,
  Jan Winter$^2$}

\institute{%
  $^1$CP$^3$-Origins, Campusvej 55, DK-5230 Odense M, Denmark\\
  $^2$PH-TH Department, Case C01600, CERN, CH-1211 Geneva 23, Switzerland\\
  $^3$IPPP, University of Durham, Science Laboratories, South Rd,
  Durham DH1 3LE, UK\\
  $^4$School of Physics and Astronomy, University of Edinburgh, Mayfield Road,
  Edinburgh EH9 3JZ, UK
}


\begin{abstract}
  We compare the results from four different theoretical predictions for the
  production of a $W$ boson in association with at least two jets at the Large
  Hadron Collider.  We discuss a possible method for combining
  next-to-leading order samples with different jet multiplicity from
  \textsc{BlackHat+Sherpa}.  We then compare these results with the
  next-to-leading order $W$ plus two jet calculation, the leading
  order \textsc{ME\&TS} merged approach of \textsc{Sherpa} and the
  high-energy resummation approach of \textsc{High Energy Jets} in an
  attempt to determine if these approaches can be distinguished at the LHC.
\end{abstract}

\subsection{INTRODUCTION}

The production of a $W$ boson in association with jets at the Large Hadron
Collider (LHC) is an extremely important process.  It contributes to three
distinct areas of the rich physics program at the LHC.  Firstly, it is a key
Standard Model signal and therefore important to test our understanding of the
Standard Model in the TeV-scale energy range.  Secondly, it is an important
background in many searches for new physics where, for example, new heavy
coloured particles have cascade decay chains.  Thirdly, it provides an ideal
testing ground for experimental techniques such as a jet veto: what is learned
in the relatively well-understood treatment of $W$ plus jets can be directly
applied to Higgs searches for example.

It has been observed that the ratio of $W+(n+1)$-jet events to
$W+n$-jet events can be substantially larger than one might na\"ively
expect by considering the $\alpha_s$ suppression only.  This is
especially true in phase-space regions of large four-momenta, such as
the high-$H_T$ tail, because the available phase space for extra jet
emission at the LHC is extremely large.  It can therefore compensate
for the effect of an additional factor of the strong coupling.  This
effect is more visible in distributions where additional radiation
leads to a significant change in the value of the observable, as is
the case for the $H_T$ distribution, the scalar sum of the transverse
momenta of identified leptons, jets and missing energy. The change
will be more moderate in an observable like $H_{T,2}$, whose definition
differs from that of $H_T$ by truncating the jet sum to include only
the two hardest jets in the event.  To make an impact here requires
the radiation to lead to an additional jet with transverse momentum as
large as that of the second hardest jet, not only larger than the jet
$p_T$ threshold.  The effects will also be smaller in more inclusive
variables like the transverse momentum of the $W$ boson, $p_{T,W}$, or
the leading jet, $p_{T,j_1}$.

There are a number of different theoretical approaches to describing the
emission of large numbers of jets.  In order to probe to what extent the
differences in these will be accessible at the LHC, we will compare, in
this study, the predictions for the jet activity in inclusive
$W(\to e\nu)+2$-jet production from
(a) \textsc{BlackHat+Sherpa} (\emph{BHS})
\cite{Berger:2009zg,Berger:2009ep,Berger:2010zx},
(b) combined \emph{BHS} samples (to be described below),
(c) \textsc{Sherpa} \cite{Gleisberg:2008ta,Gleisberg:2003xi} run in
\textsc{ME\&TS} mode (\emph{S-MEPS})
\cite{Hoeche:2009rj,Hoeche:2009xc,Carli:2010cg} and
(d) \textsc{High Energy Jets} (\emph{HEJ}), an all-order
resummation of wide-angle radiation
\cite{Andersen:2009nu,Andersen:2009he,Andersen:2012tb}.

The current state-of-the-art next-to-leading order (NLO) predictions
for $W$ production in association with jets are those of \emph{BHS},
which have been calculated up to $W$ plus four jets with a leading-colour approximation for the virtual part \cite{Berger:2010zx}, and up to $W$ plus three
jets with a full color treatment \cite{Berger:2009zg,Berger:2009ep}.  In this study, we
consider the inclusive $W+2$-jet prediction at NLO accuracy, and
further, discuss and show predictions from an inclusive sample where
$W+2,3,4$-jet events generated by \emph{BHS} are combined in a simple
manner, nevertheless without introducing any double counting of
phase-space regions.

The \emph{S-MEPS} predictions are obtained from merging at leading
order (LO) tree-level Matrix Elements for $W+0,\ldots,n$-parton final
states with (Truncated) parton Showers (hence the name
\textsc{ME\&TS}) preserving the leading logarithmic accuracy to which
soft and collinear multiple emissions are described by the parton
shower. The newer \textsc{ME\&TS} merging scheme was introduced in
Ref.~\cite{Hoeche:2009rj} and optimised as documented in
Refs.~\cite{Hoeche:2009xc,Carli:2010cg} to improve
over the original \textsc{Sherpa} implementation based on the CKKW
approach \cite{Catani:2001cc,Krauss:2002up}. \textsc{ME\&TS}
guarantees a better matching regarding the usage of scales as
occurring in the evaluation of the matrix elements and those scales
driving parton showering. The \emph{S-MEPS} sample used in our study
was generated by including $W(\to e\nu)$ production matrix elements
with up to five extra partons (massless quarks, $u,d,s,c,b$, and
gluons).

The \emph{HEJ} framework is a resummation of the leading logarithmic terms
occurrung in pure, or $W$, $Z$ or $H$ plus, multi-jet production in the limit of
large invariant mass between each pair of jets, to all orders in $\alpha_s$.
This is then matched to tree-level accuracy for final states with two, three or
four jets.  In principle, the \emph{HEJ} framework can be merged with a parton
shower to add the collinear pieces which are not included in the \emph{HEJ}
description (\emph{HEJ} does include soft emissions down to around 2~GeV).
First steps in this direction for pure jet production were taken in
\cite{Andersen:2011zd}.  Here, the \emph{HEJ} predictions are calculated at the
parton level.

In the \ref{sec:nlo-exclusive-sums}~section, we will elaborate on the
method~(b) for combining NLO samples of different jet multiplicities.
Then, in the \ref{sec:comparison-results}~section, we will first show
explicit results of the sizable impact of large multiplicity events by
comparing predictions from the combined \emph{BHS} sample and the
\emph{S-MEPS} merged sample. Secondly, we will study variables chosen
to probe the differences in the treatment of the QCD radiation.  We
will show and compare the predictions for all four descriptions
mentioned above focusing on the following observables:
\begin{itemize}
\item the average number of jets as a function of
  $H_T=\sum_i p_{T,j_i}+p_{T,e}+p_{T,\nu}$ and $\Delta y$, the rapidity
  difference between the most forward and most backward jets, and also
\item the ratio of the inclusive $3$-jet rate to the inclusive $2$-jet
  rate as a function of $H_T$ and $\Delta y$.\footnote{Given the
    definition of $H_T$, we note
    $H_{T,2}=p_{T,j_1}+p_{T,j_2}+p_{T,e}+p_{T,\nu}$.}
\end{itemize}
We will then discuss the areas of agreement and difference that we
find, before we finally conclude in the \ref{sec:conclusions}~section.

\subsection{NLO EXCLUSIVE SUMS}
\label{sec:nlo-exclusive-sums}

An NLO $n$-jet prediction contains events with $n$ or $n+1$ partons. For
observables for which higher multiplicites have a significant impact, this
limitation can be detrimental. If one has predictions for different
multiplicities ($m,m+1,\ldots,M$), one can try to combine them by
avoiding double counting by requiring that the $n$-jet prediction is
used only to describe $n$-jet events (except for the highest
multiplicity where $(n+1)$-jets configurations are allowed). The total
cross section can be rewritten as a decomposition based on exclusive
(exc) and inclusive (inc) jet bins:
\begin{equation}\label{eq:totxsec}
  \sigma^\mathrm{tot}\;\equiv\;\sigma^\mathrm{inc}_m\;=\;
  \sum^{M-1}_{n=m}\sigma^\mathrm{exc}_n\,+\,\sigma^\mathrm{inc}_M\ .
\end{equation}
The exclusive-sums procedure describes each jet bin at NLO accuracy,
i.e.\ at $\mathcal{O}(\alpha^{n+1}_s)$, or, alternatively, only the
$(M+1)$-th (inclusive) jet bin is predicted with LO precision. We
hence note that the combination of the terms shown in
Eq.~(\ref{eq:totxsec}) occurs at different orders of the strong
coupling. Furthermore, the definition of an exclusive $n$-jet sample
requires a detailed treatment of jet vetoing. For these reasons, the
simple combination procedure is crude and does not increase the formal
accuracy of the prediction, which is that of NLO of the smallest
multiplicity. However, one can hope that the procedure will lead to a
better prediction in observables where higher multiplicity events
dominate.

More studies are required to understand the uncertainties related to this
procedure. One way of doing so would be to vary the jet algorithm and/or
parameters of the jet algorithm used to separate the different NLO predictions
into fixed multiplicities sets and test the stability of the
prediction.\footnote{In this study we used the same jet algorithm for
  both defining the partitioning in jet multiplicities as well as
  applying cuts and determining observables.}
This is left to a future study.

\subsection{RESULTS OF THE COMPARISON}
\label{sec:comparison-results}

\begin{table}[t!]
  \centering
  \begin{tabular}{clccrc}\hline&&&&\\[-1mm]
    &$|\eta_e|<2.5$&&&$p_{T,e}>20$~GeV&\\[1mm]
    &$M_{\perp,W}>20$~GeV&&&$p_{T,\nu}>20$~GeV&\\[1mm]
    &$|\eta_j|<4.5$&&&$p_{T,j}>25$~GeV&\\[3mm]\hline\hline
  \end{tabular}
  \caption{Summary of the cuts applied in the analysis.}
  \label{tab:cuts}
\end{table}

In this section, we compare the results of different theoretical
descriptions for $W+n$-jets production at the LHC.  The
number $n$ can take values from $2$ and above, as we will mostly
consider inclusive samples.  The four descriptions, which we will
compare here in more detail, are
\begin{itemize}
\item the \emph{BHS} calculation of $W+2$-jets at NLO,
\item the combined sample of $W+2,3,4$-jet events at NLO from
  \emph{BHS}, as described in the \ref{sec:nlo-exclusive-sums} section,
\item the \emph{S-MEPS} merged $W+n$-jets sample using LO tree-level
  matrix elements up to $n=5$, and
\item the approach of \emph{HEJ}.
\end{itemize}
Throughout this study, we will consider inclusive samples of $W^-$
boson production in association with at least two hard jets identified
by the anti-$k_T$ jet algorithm using $R=0.4$. The jets are required
to have $p_{T,j}>25$~GeV. We look only in the $W^-\to e^-\bar\nu_e$
decay channel and use the cuts given in Tab.~\ref{tab:cuts} where
$M_{\perp,W}$ is defined as
$M_{\perp,W}=\sqrt{(|\vec p_{T,e}|+|\vec p_{T,\nu}|)^2
  -(\vec p_{T,e}+\vec p_{T,\nu})^2}$.

The \emph{HEJ} predictions use the geometric mean of the jet
transverse momenta to determine the renormalisation and factorisation
scale, i.e.\ $\left(\prod p_{T,j}\right)^{1/n}$. This central choice
will be varied by a factor of two in either direction to provide an
envelope (marked by dotted lines in the corresponding figures) around the
\emph{HEJ} default prediction.  The \emph{BHS} predictions instead use
$\hat{H}'_T/2$ as the NLO calculation becomes unstable for a 
scale which is too low. In the \emph{S-MEPS} calculation, scales are chosen according
to the default prescription given by \textsc{ME\&TS} \cite{Hoeche:2009rj}.

\begin{figure}[t!]
  \centering
  \epsfig{width=.49\textwidth,angle=0,file=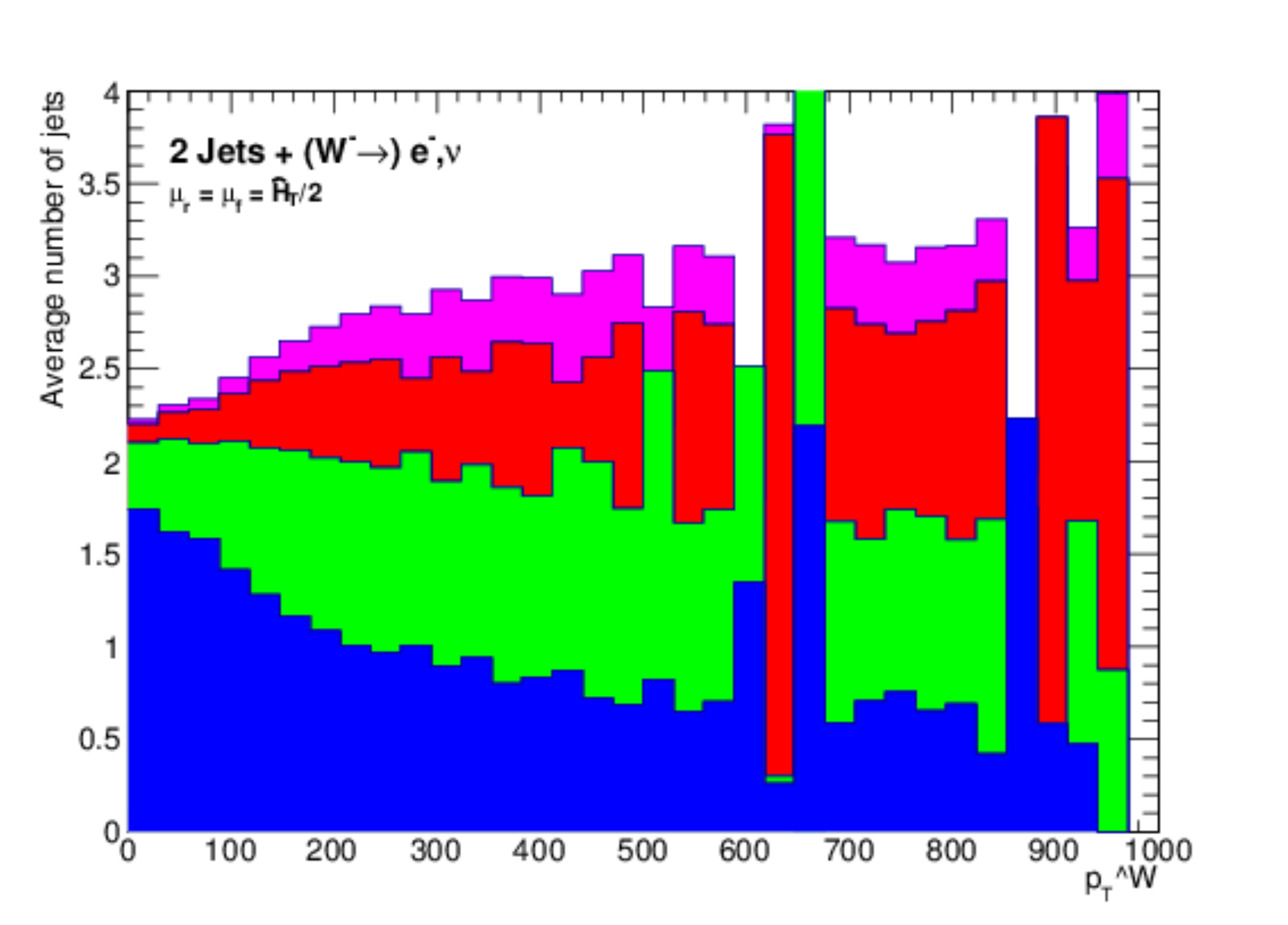}\hfill
  $\hphantom{x}$\\[-63mm]$\hphantom{x}$\hfill
  \epsfig{width=.384\textwidth,angle=-90,file=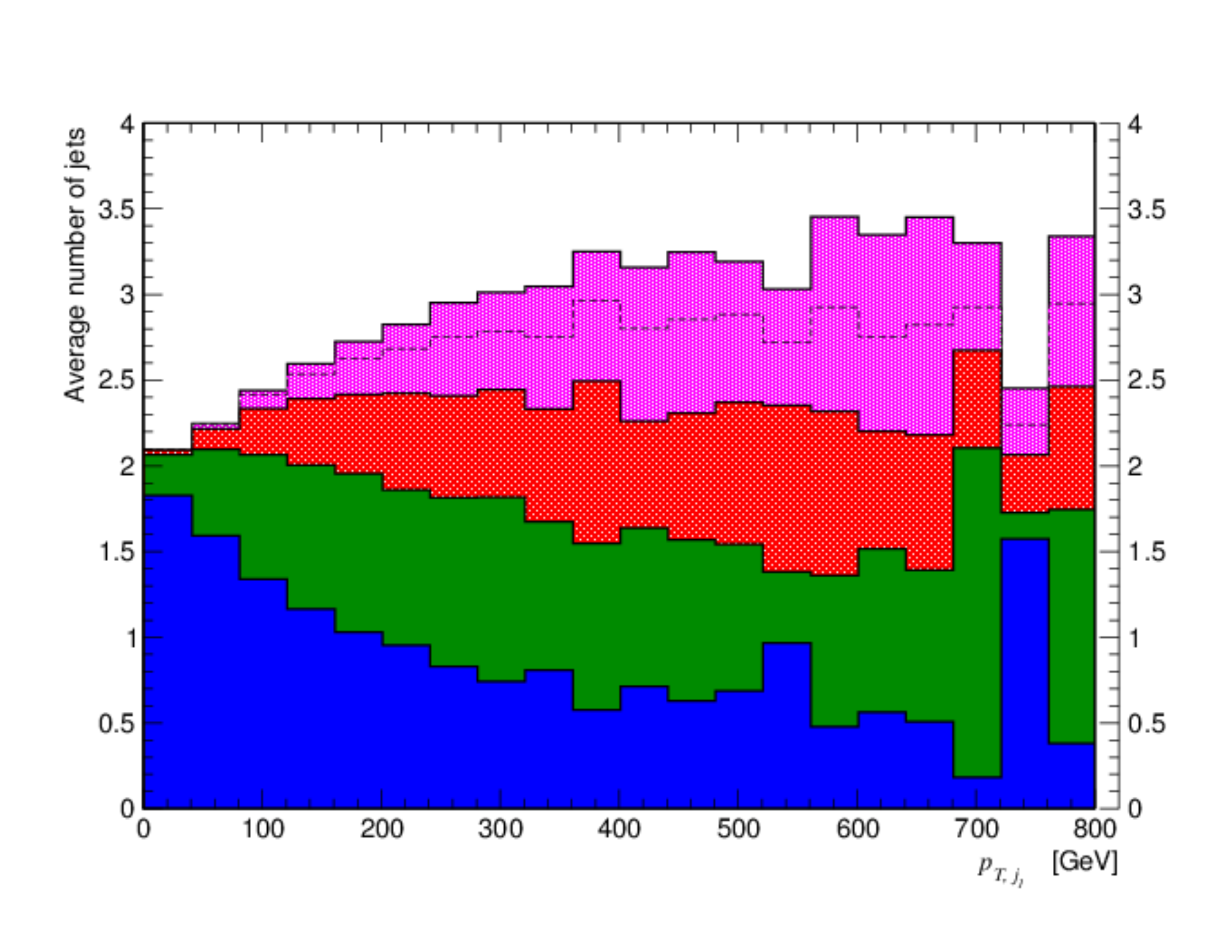}
  \caption{The average number of jets as a function of $p_{T,W}$
    (left) and $p_{T,j_1}$ (right). The $p_{T,W}$ plot shows the
    \emph{BHS} exclusive sums prediction, while the $p_{T,j_1}$ plot
    is obtained from \emph{S-MEPS}.}
  \label{fig:Wpt5j}
\end{figure}

\begin{figure}[t!]
  \centering
  \epsfig{width=.49\textwidth,file=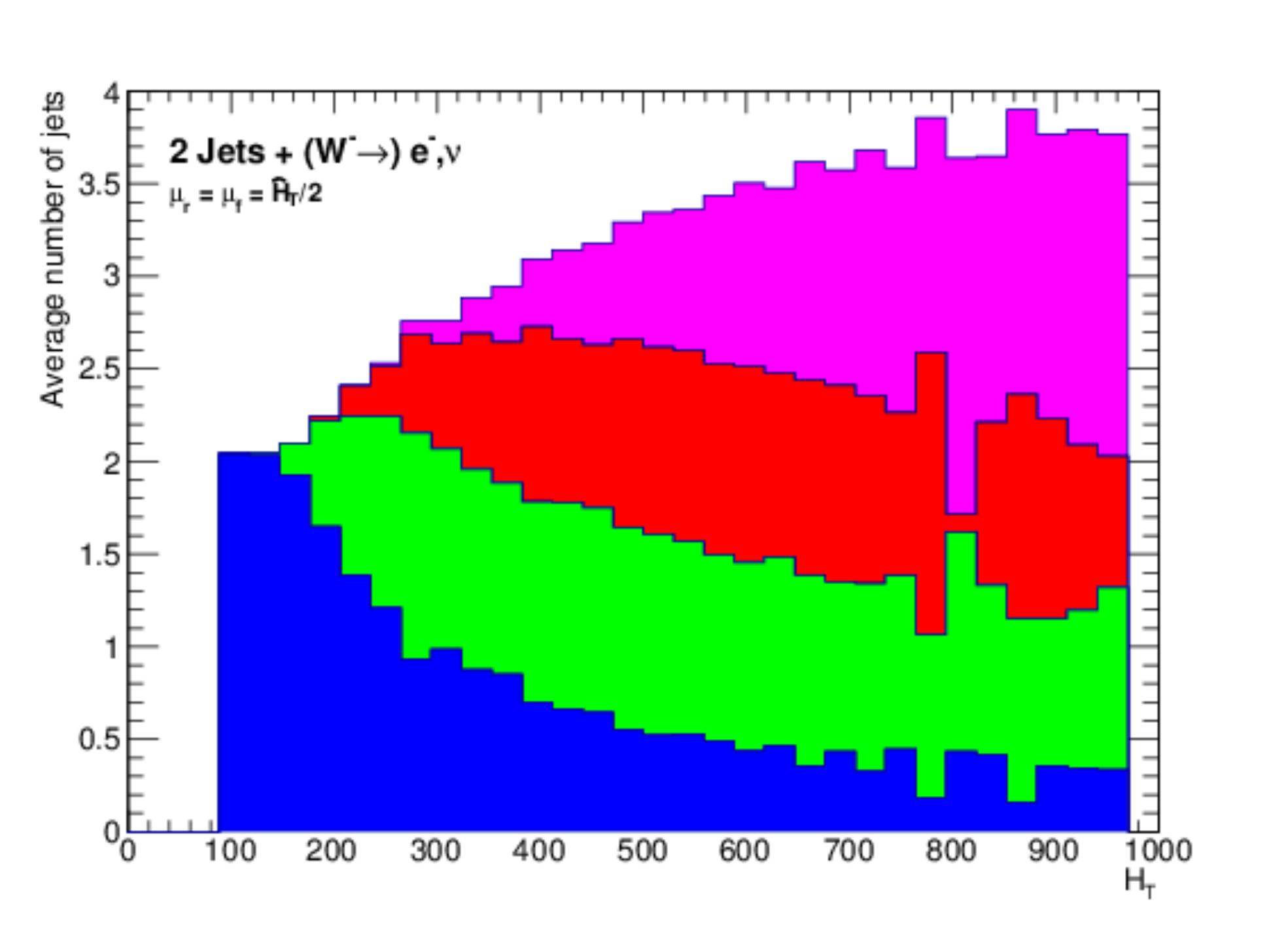}\hfill
  \epsfig{width=.49\textwidth,file=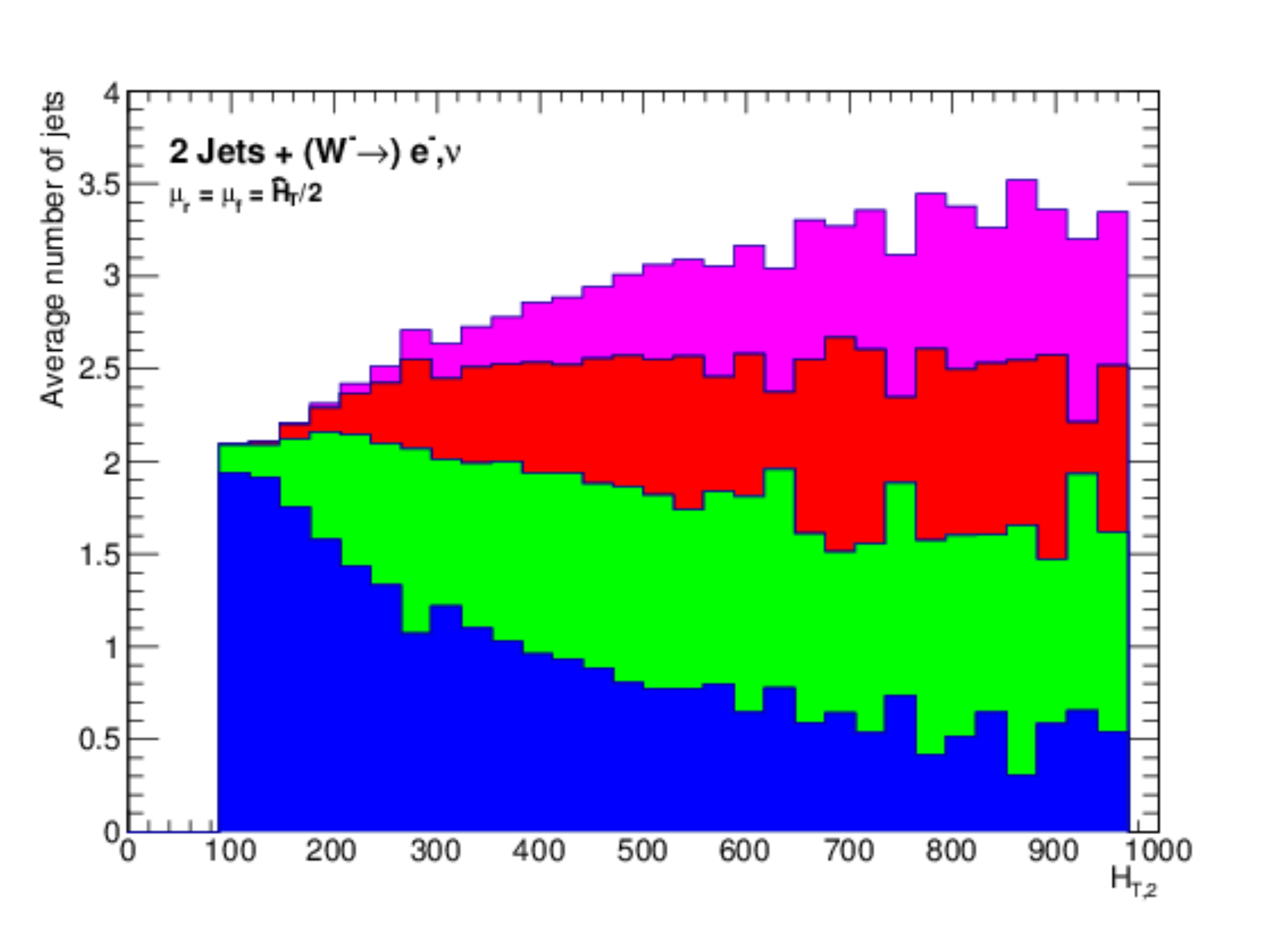}\\[-6mm]\hskip-1mm
  \psfig{width=.384\textwidth,angle=-90,file=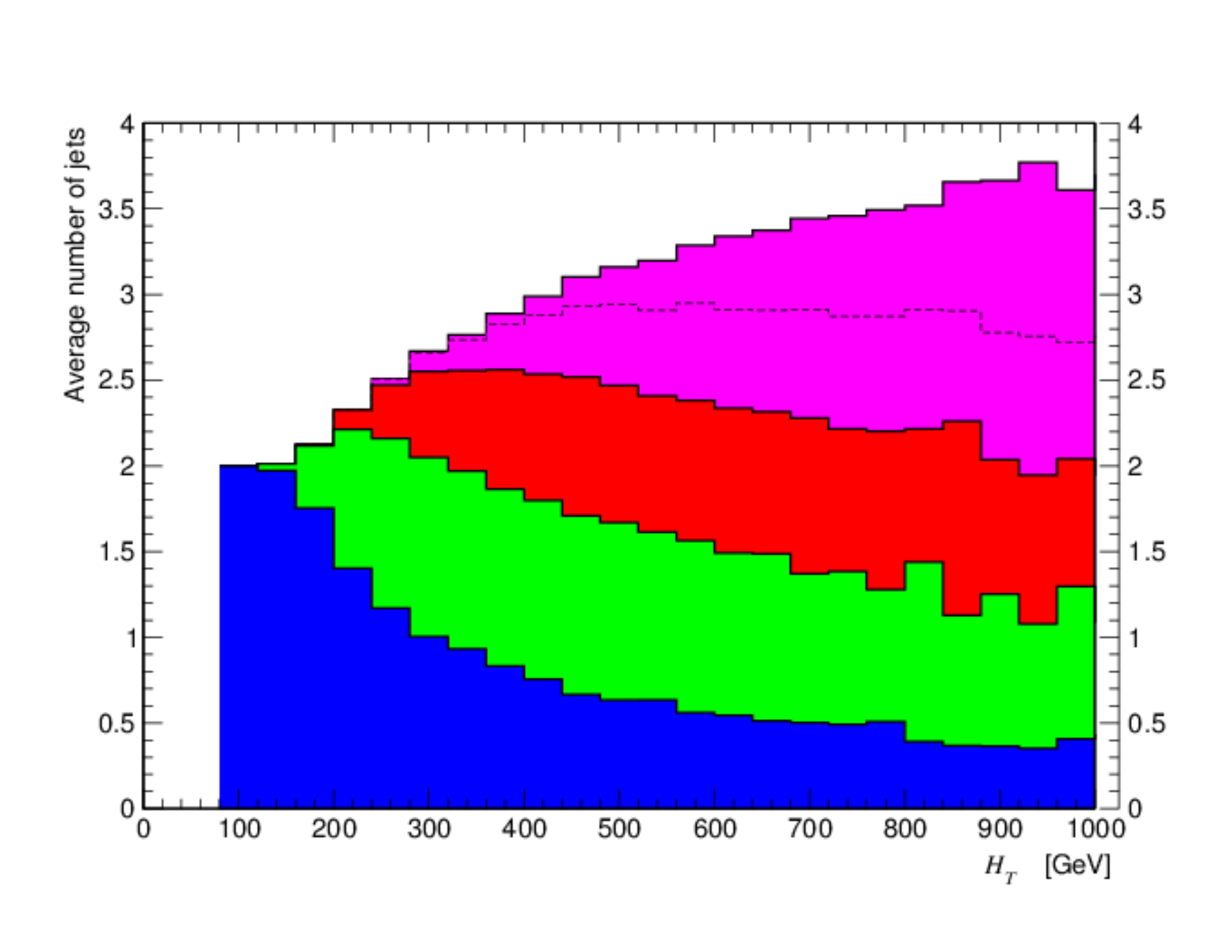}\hfill
  \psfig{width=.384\textwidth,angle=-90,file=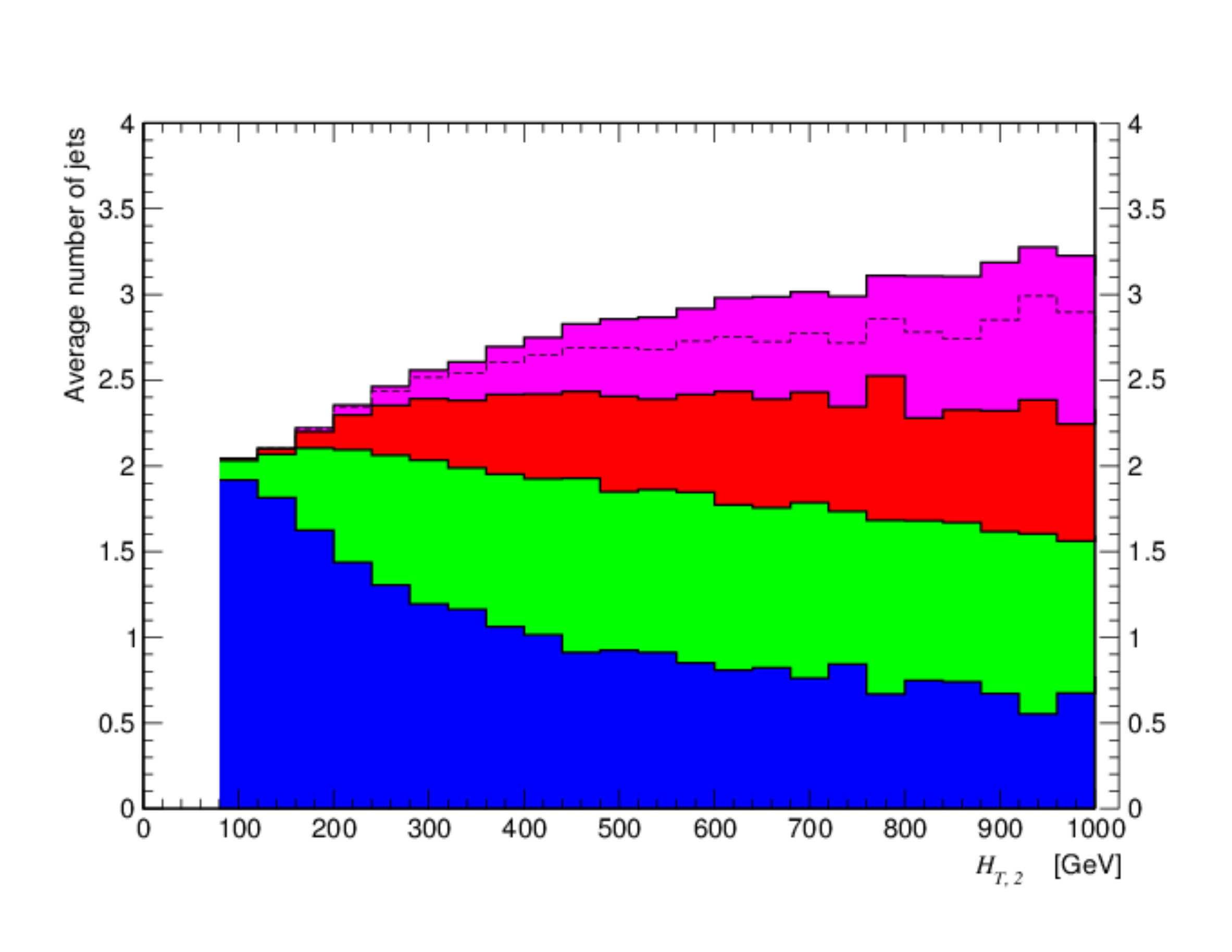}
  \caption{The contribution from different multiplicities to the
    average number of jets as a function of $H_{T}$ and $H_{T,2}$.
    The upper plots show the \emph{BHS} exclusive sums prediction,
    while the lower ones are extracted from \emph{S-MEPS}.}
  \label{fig:HT5j}
\end{figure}

The variables $H_{T,2}$, $p_{T,W}$ and $p_{T,j_1}$ are less sensitive
to the presence of additional radiation than $H_T$, as discussed in
the introduction. The plots, which we present in Figs.~\ref{fig:Wpt5j}
and \ref{fig:HT5j} address the alternative question: given a
particular value of $H_T$, $H_{T,2}$ etc.\ how many jets are typically
found in the event?

Figs.~\ref{fig:Wpt5j} and \ref{fig:HT5j} show the stacked results for
the average number of jets as a function of $p_{T,W}$, $p_{T,j_1}$,
$H_T$ and $H_{T,2}$ visualising the contributions from each exclusive
$2,3,4$-jet sample and the inclusive $5$-jet sample. The left (right)
plot in Fig.~\ref{fig:Wpt5j} and the upper (lower) rows of plots in
Fig.~\ref{fig:HT5j} depict the results as obtained from the combined
\emph{BHS} sample (the \emph{S-MEPS} sample). In all cases the
different colours correspond to the terms in the numerator of the
formula for the average number of jets,
\begin{equation}\label{eq:average}
  \left<N\right>_5\;=\;
  \frac{\sum\limits_{i=2,3,4} i\,n_i^\mathrm{exc}+5\,n_5^\mathrm{inc}}
       {\sum\limits_{i=2,3,4}    n_i^\mathrm{exc}+   n_5^\mathrm{inc}}\;=\;
  \frac{\sum\limits_{i=2,3,4} i\,n_i^\mathrm{exc}+5\,n_5^\mathrm{inc}}
       {n_2^\mathrm{inc}}\ ,
\end{equation}
where blue, green, red and magenta stand for $i=2,3,4$ and $i=5$,
respectively. The subscript to $\langle N\rangle$ clarifies that we
truncate the determination of the average after the fifth jet bin,
noting that $\langle N\rangle_k\to\langle N\rangle$ for a sufficiently
large number of jet bins. This makes no difference for the \emph{BHS}
predictions employed here since the jet multiplicity de facto is
limited to five, but it does for the \emph{S-MEPS} and \emph{HEJ}
computations where events with $i>5$ jets do occur. We have defined
$n^\mathrm{exc/inc}_k=d\sigma^\mathrm{exc/inc}_k/dO$ where $O$ denotes
an observable like $H_T$, or $\Delta y$ presented later on. Note
that in Fig.~\ref{fig:HT5j} the $5$-jet part contributes to the
average number of jets with a factor of $5$, while the $2$-jet part,
for example, contributes with a factor of $2$ only.

The layout of Fig.~\ref{fig:ratio5j} (including the colour coding) is
the same as before: here, we however display, wrt.\
$n_2^\mathrm{inc}$, the relative fractions of the different
multiplicities corresponding to the terms in the denominator of
Eq.~(\ref{eq:average}). In other words, in Fig.~\ref{fig:ratio5j} we
consider the partitioning of
\begin{equation}\label{eq:fraction}
  1\;=\;
  \frac{\sum\limits_{i=2,3,4} n_i^\mathrm{exc}+n_5^\mathrm{inc}}
       {n_2^\mathrm{inc}}\ .
\end{equation}
Although there is just a 30\% fraction of inclusive $5$-jet events to
the total cross section, we observe that their contribution to the
build-up of $\langle N\rangle(H_T)$ for very large $H_T$ gets close to
50\%. Also, for an $H_T\sim500$~GeV, the average number of jets is
composed evenly between the $2,3$-jet and $4,5$-jet contributions,
while the relative fraction of the $2,3$-jet events is nearly 70\%.
This emphasizes the dominance of multi-jet events in forming large
$H_T$ values. It also can be seen that for medium $H_T$ values,
$400<H_T<700$~GeV, all the multiplicities give roughly the same
contribution to the variable $\langle N\rangle(H_T)$, while for low
$H_T$, the average is primarily described by $2$-jet events.

\begin{figure}[t!]
  \centering
  \epsfig{width=.49\textwidth,file=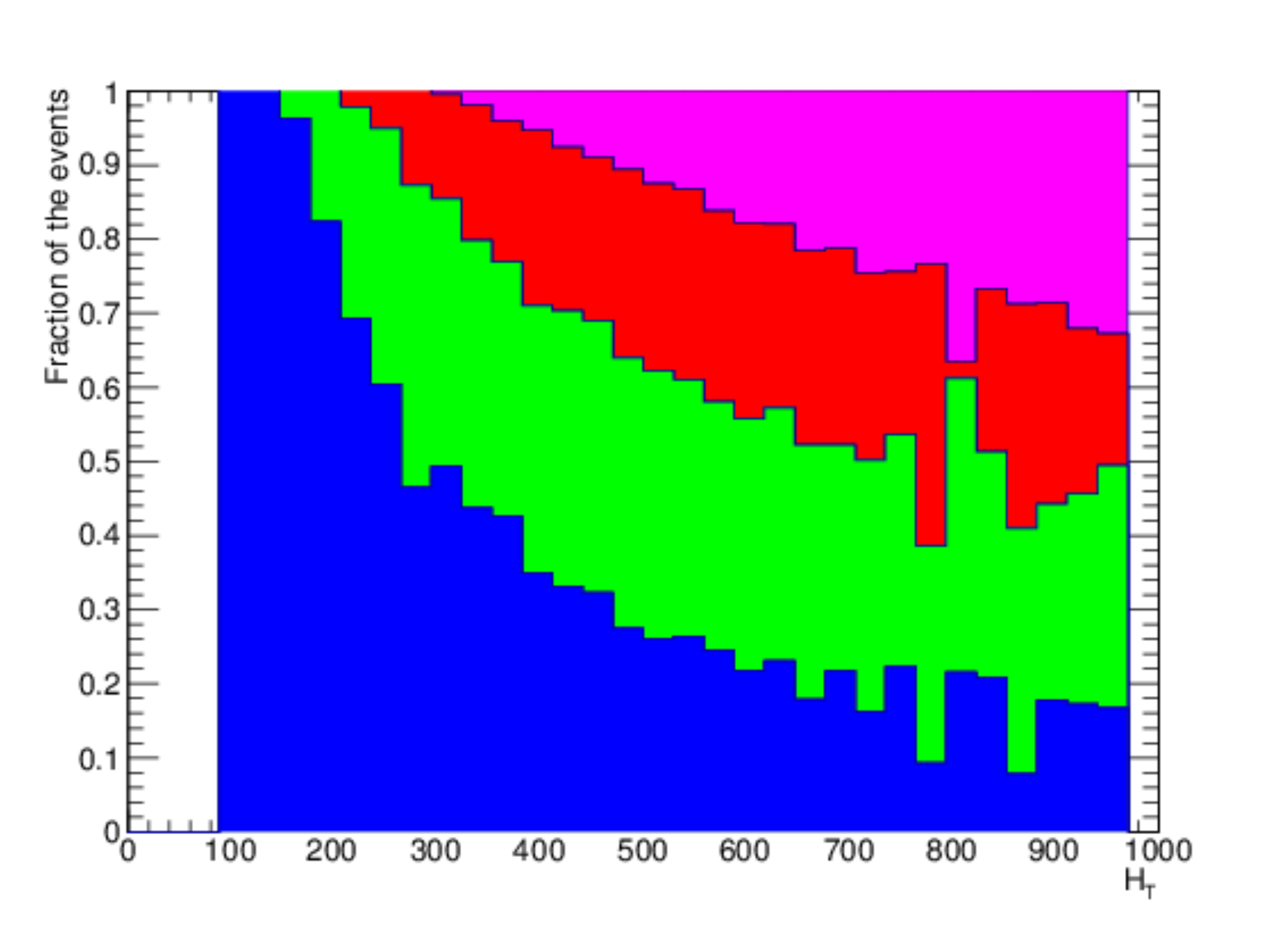}\hfill
  \epsfig{width=.49\textwidth,file=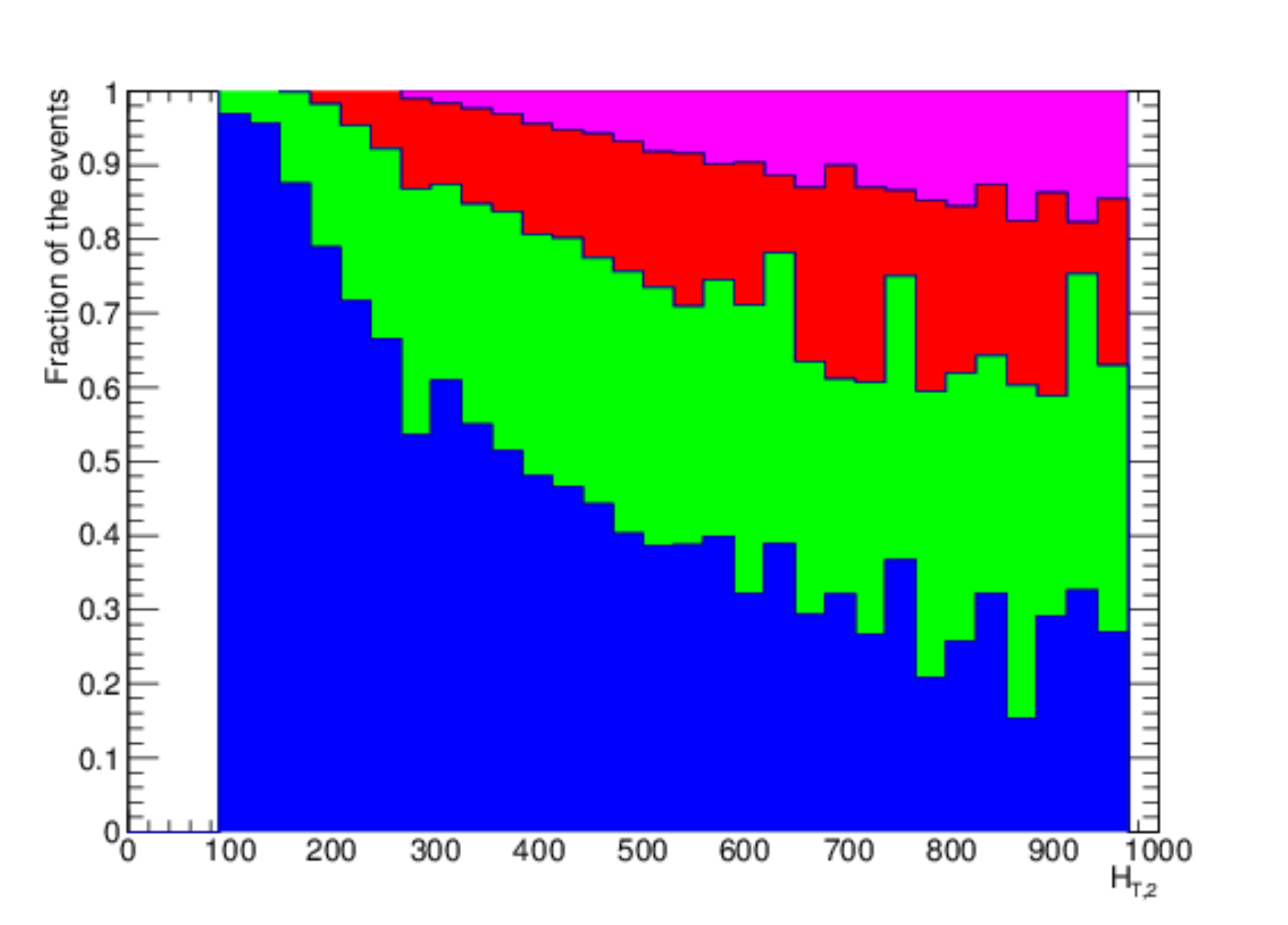}\\[-6mm]\hskip-1mm
  \epsfig{width=.384\textwidth,angle=-90,file=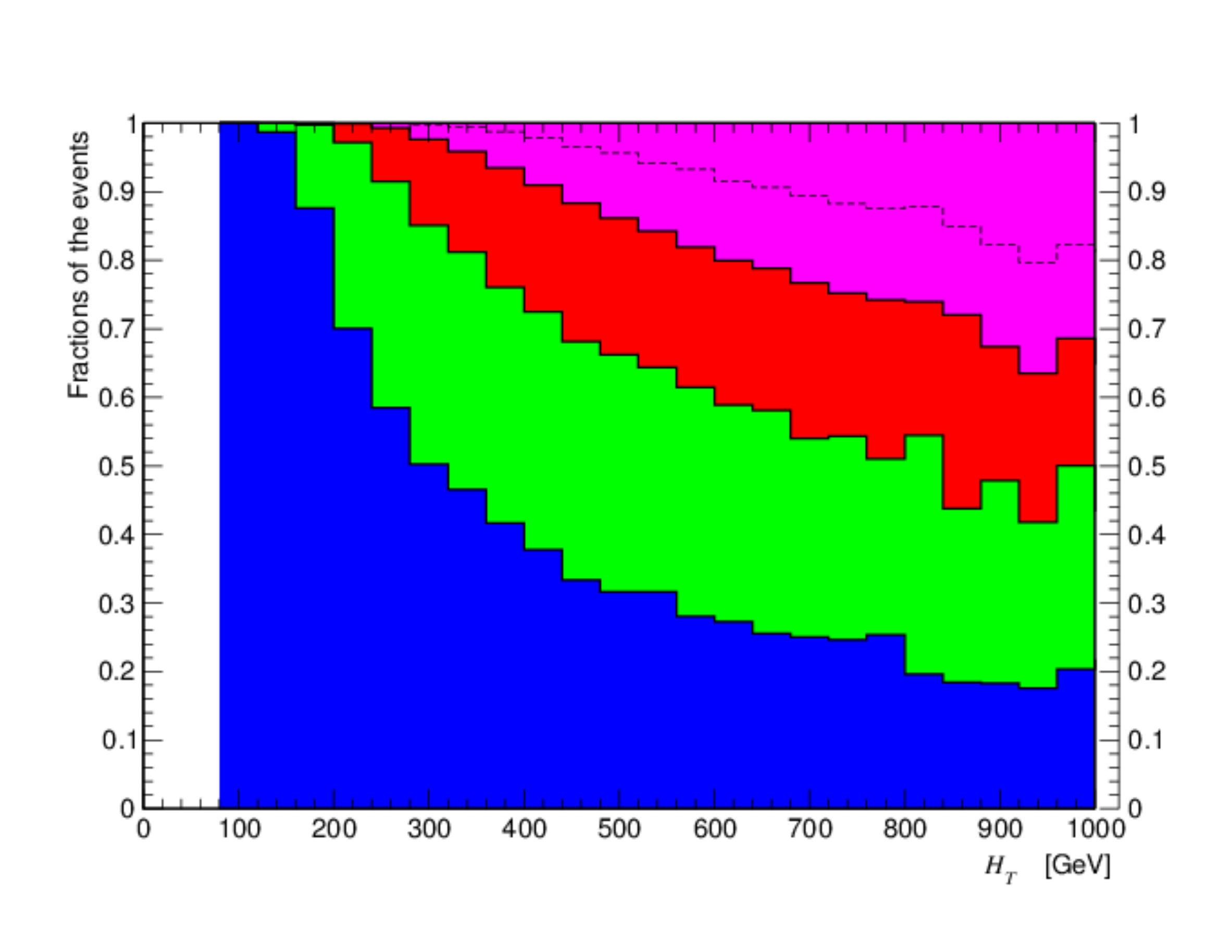}\hfill
  \epsfig{width=.384\textwidth,angle=-90,file=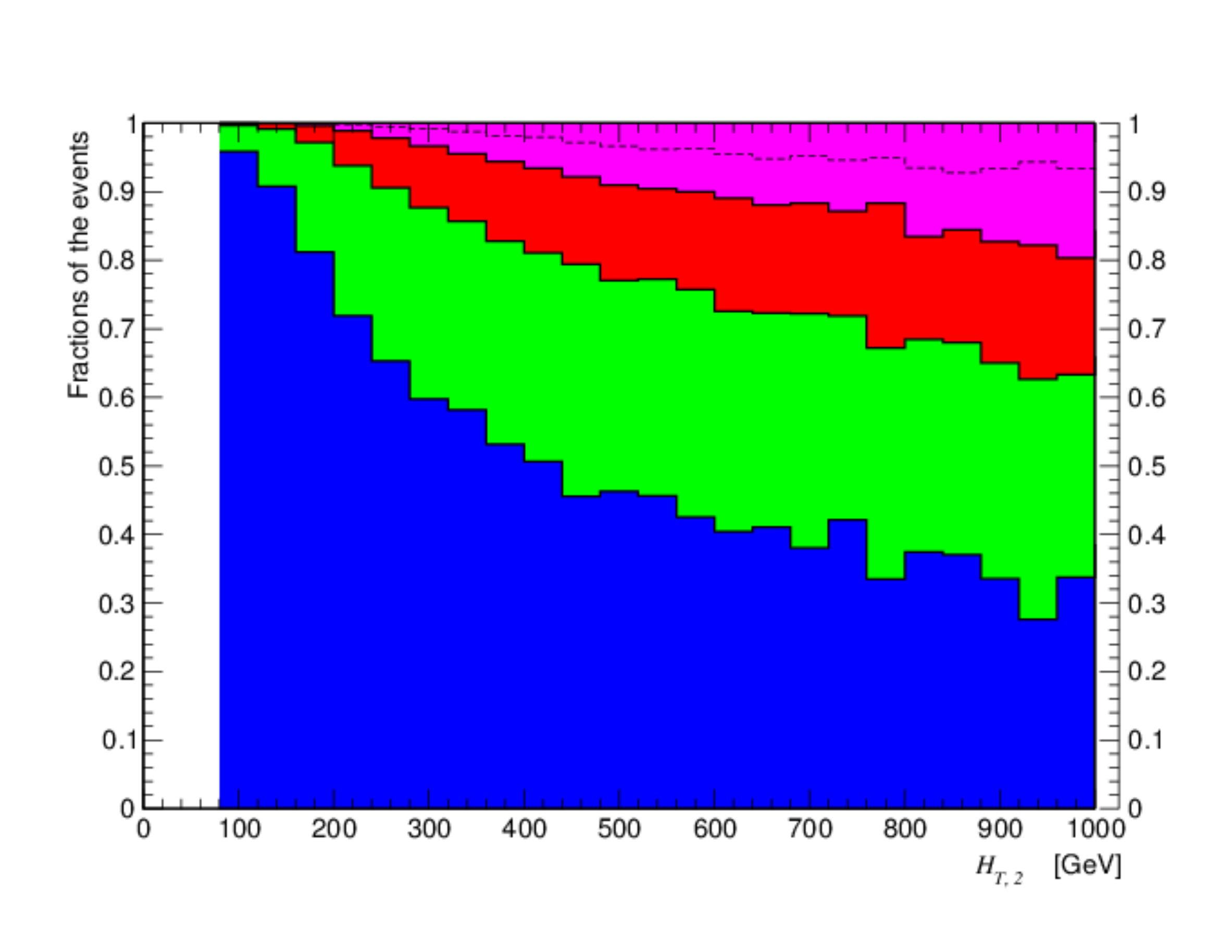}
  \caption{The fraction of the total rate from different multiplicities as a
    function of $H_T$ and $H_{T,2}$. The upper plots show the \emph{BHS}
    exclusive sums prediction, while the lower ones are extracted from
    \emph{S-MEPS}.}
  \label{fig:ratio5j}
\end{figure}

Going clockwise through Figs.~\ref{fig:Wpt5j} and \ref{fig:HT5j} we
see that the average number of jets is indeed sensitive to higher
multiplicities when considered as a function of $p_{T,W}$, $p_{T,j_1}$
and $H_{T,2}$, but in all these cases this happens to a lesser extent
as if considered as a function of $H_T$. As expected, the dependence is
mildest for $p_{T,W}$, the most inclusive observable studied here. We
also observe that the jet-bin decomposition of $p_{T,j_1}$ and
$H_{T,2}$ turns out very similar. Most strikingly we note the increase
in the contribution from the highest multiplicity events, the ones
containing more or at least five jets. For $H_{T,2}$, we furthermore
display to the right of Fig.~\ref{fig:ratio5j} the relative fractions
as done in the $H_T$ case. Even for largest $H_{T,2}$ values, the
fraction arising from $2,3$-jet events remains close to 65\% stressing
once more the lower sensitivity of $H_{T,2}$ versus $H_T$ regarding
multiple jet production.

Finally, we compare the plots from the combined \emph{BHS} samples in
all figures to the corresponding ones generated with the \emph{S-MEPS}
sample. Interestingly, the outcome looks very similar although
\textsc{ME\&TS} handles the single terms in Eq.~(\ref{eq:totxsec})
rather differently. They are calculated at least at leading
(soft/collinear) logarithmic accuracy improved by LO $n$-jet effects.
Presumably, for the exclusive jet bins, this description (which allows
a better treatment of jet vetoes) is not too far off the
exclusive sums approach, since the unresolved
${\cal O}(\alpha_s)$ corrections are also present in the Sudakov form
factors applied in the \textsc{ME\&TS} approach. Also, the combined
\emph{BHS} samples as well as the \emph{S-MEPS} sample use the same
tree-level matrix elements, namely up to $W+5$-parton matrix elements.
Clearly, it has to be studied further whether this similarity in the
results is a coincidence or not.

It is clear that the impact of the higher multiplicity samples is significant
throughout, especially in the high $H_T$ tail.  This is precisely the region,
which would be probed for signs of new physics, and therefore it is essential
that we fully understand our theoretical descriptions in this region.  This is
the subject of the remainder of this contribution, where we compare
all four different methods of modelling hard QCD radiation in
inclusive $W+2$-jet events.


\begin{figure}[t!]
  \centering
  \epsfig{width=.49\textwidth,file=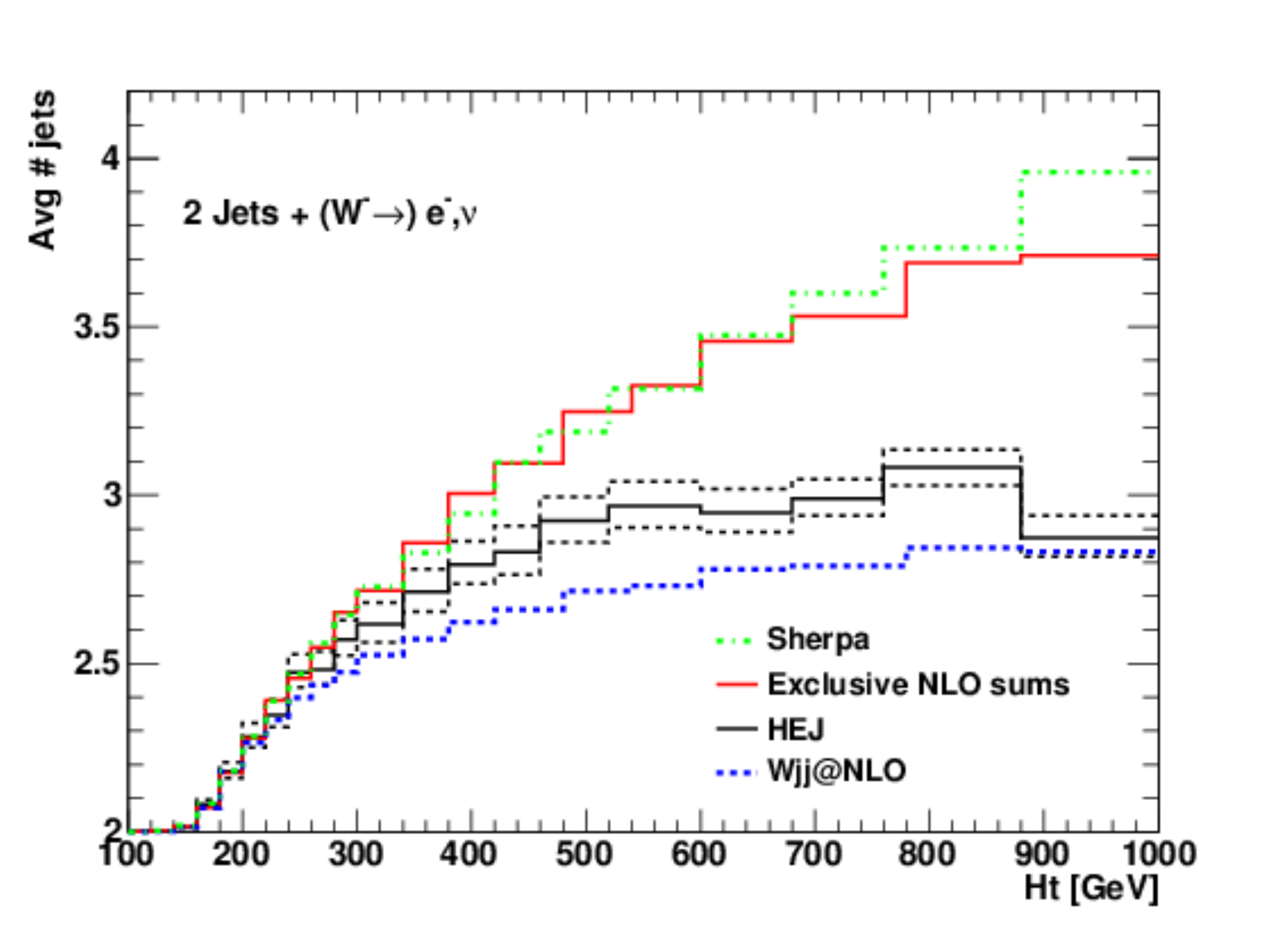}\hfill
  \epsfig{width=.49\textwidth,file=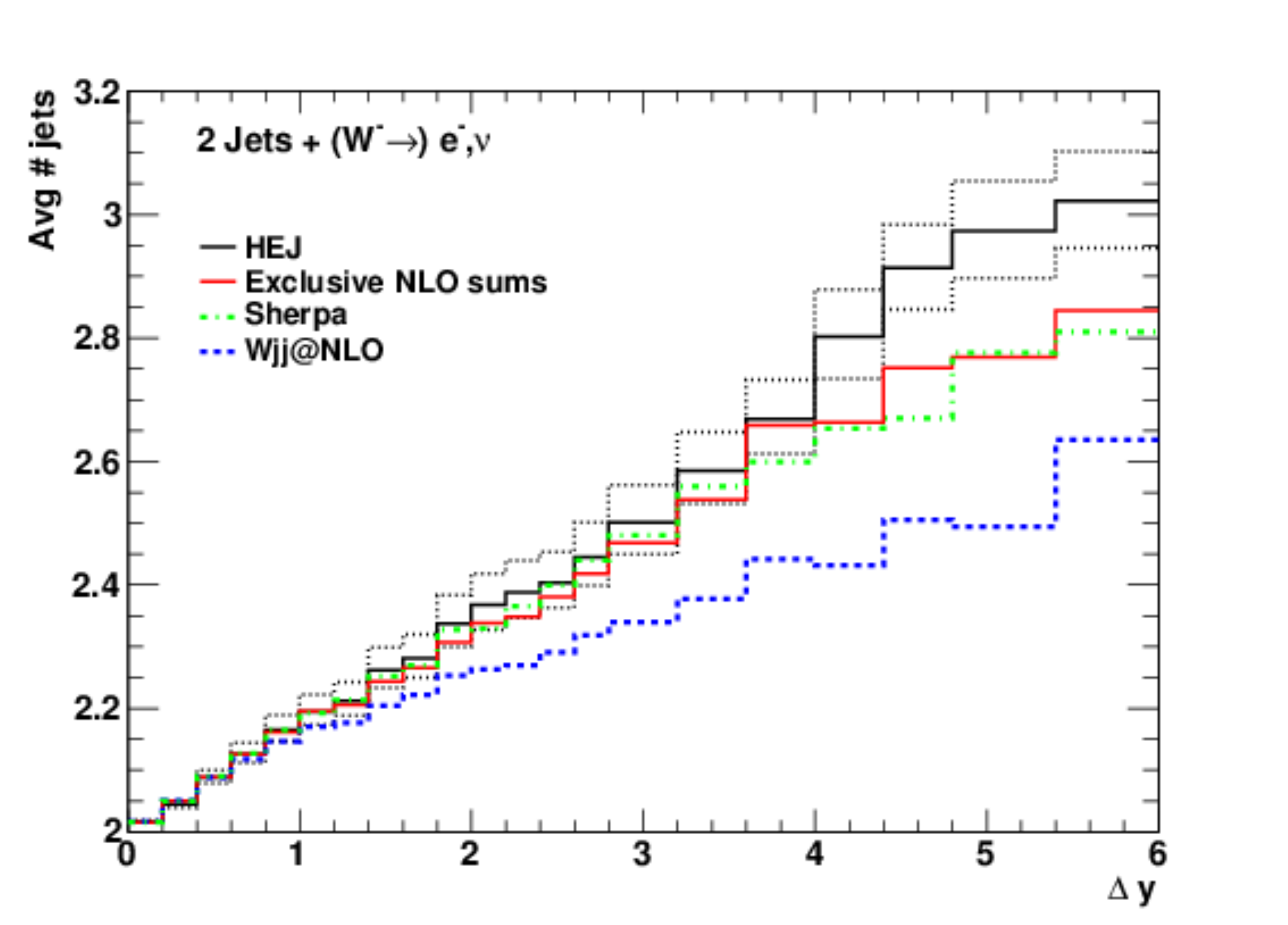}
  \caption{Average number of jets as a function of $H_T$ (left) and
    $\Delta y$ (right) in two \emph{BHS} descriptions, from \emph{HEJ}
    and from \emph{S-MEPS}, the latter using the $\langle N\rangle_7$
    definition. The bands shown with dotted lines for the \emph{HEJ}
    prediction are a result of varying the scale by a factor of $2$ in
    each direction.}
  \label{fig:njetsvsHT_both}
\end{figure}

The left plot of Fig.~\ref{fig:njetsvsHT_both} shows the final comparison plot
between the exclusive sums and inclusive $2$-jet \emph{BHS} results as well as
the \emph{HEJ} and \emph{S-MEPS} predictions for the average number of jets as a
function of $H_T$. The differences in the descriptions are significantly larger
than the scale uncertainty band on the \emph{HEJ} prediction. For the $W+2$-jet
NLO result, the number of jets rises to $2.6$ already at $H_T=500$~GeV but that
levels off significantly below the \emph{S-MEPS}, exclusive \emph{BHS} sum and
\emph{HEJ} results. The \emph{HEJ} results level off at a higher value of about
$3.0$, starting to clearly disagree with the exclusive sums and \emph{S-MEPS}
predictions above $500$~GeV, from where those two curves keep rising to a final
level of around $3.7$ to $4.0$. The \emph{S-MEPS} comes in highest at largest
$H_T$, where $\langle N\rangle_7$ is shown, cf.\ Eq.~(\ref{eq:average}), in
order to determine the average number of jets for this \emph{S-MEPS} result. The
reason for giving slightly higher $\langle N\rangle$ than the exclusive sums
lies in the contribution of additional parton-shower jets present in the
\emph{S-MEPS} calculation and more accurately accounted for by the use of the
$\langle N\rangle_7$ definition as compared to the earlier result based on
$\langle N\rangle_5$ presented in Fig.~\ref{fig:HT5j} to the lower left.

In the right panel of Fig.~\ref{fig:njetsvsHT_both}, we have plotted
the average number of jets as a function of the rapidity span,
$\Delta y$, instead of $H_T$ as before.  Again the differences are
larger than the scale variation shown on the \emph{HEJ} result, but
the ordering is different to that of the left plot of
Fig.~\ref{fig:njetsvsHT_both}. All four descriptions increase linearly
with $\Delta y$ but the gradient is steepest for the \emph{HEJ}
predictions where the average rises above $3.0$ for $\Delta y$ values
as large as $6.0$. The \emph{BHS} exclusive sum result is consistently
below this, reaching about $2.8$ at $\Delta y=6.0$, and agrees pretty
well with the \emph{S-MEPS} result based on $\langle N\rangle_7$. The
NLO $W+2$-jet prediction given by \emph{BHS} is lower still, between
$2.4$ and $2.5$ for $\Delta y\sim 5.0$.

It may seem surprising that on the plot on the left-hand side the
exclusive sums and \emph{S-MEPS} lie higher for most of the
distribution whereas on the right-hand side these approaches as well
as \emph{HEJ} give predicitions that are commensurate. The region of
high $H_T$ and that of high $\Delta y$ however are largely distinct as
it is very expensive to have both a large rapidity and large $p_T$ for
the jets. Also while radiating an additional jet automatically moves
an event towards the higher $H_T$ direction, radiating an additional
jet tends to not change the rapidity difference. So, we expect the
higher multiplicies to have a smaller effect on the average number of
jets as a function of $\Delta y$ compared to as a function of
$H_T$. This is indeed the case in Fig.~\ref{fig:njetsvsHT_both}.

Lastly, in Fig.~\ref{fig:3jo2jboth} we plot the ratio of the inclusive
$3$-jet to the inclusive $2$-jet rate as a function of $H_T$ (left)
and $\Delta y$ (right), again for all four descriptions used here. The
predicted $(d\sigma^\mathrm{inc}_3/dH_T)/(d\sigma^\mathrm{inc}_2/dH_T)$
all agree very well below $400$~GeV. The fixed order \emph{BHS} result
for $W+2$ jets is highest for large $H_T$,
however is known to become unreliable here, since the probability that
an inclusive $2$-jet event is at least a $3$-jet event turns too
large, being in conflict with the expected behaviour of an
${\cal O}(\alpha_s)$ correction. The \emph{BHS} exclusive sums, the
\emph{S-MEPS} and the \emph{HEJ} results, in this order, level off
considerably lower with the \emph{HEJ} fraction staying below 60\% to
70\%, which leaves the other predictions again above the \emph{HEJ}
uncertainty envelope. In contrast, when the same ratio of jet rates is
plotted against $\Delta y$, the \emph{HEJ} prediction is consistently
higher throughout. This again emphasises that differences in the
descriptions come to light in different kinematic regions. However, in
both cases here the magnitude of the differences is relatively small
and would be rather difficult to distinguish in present experimental
data.

\begin{figure}[t!]
  \centering
  \epsfig{width=.49\textwidth,file=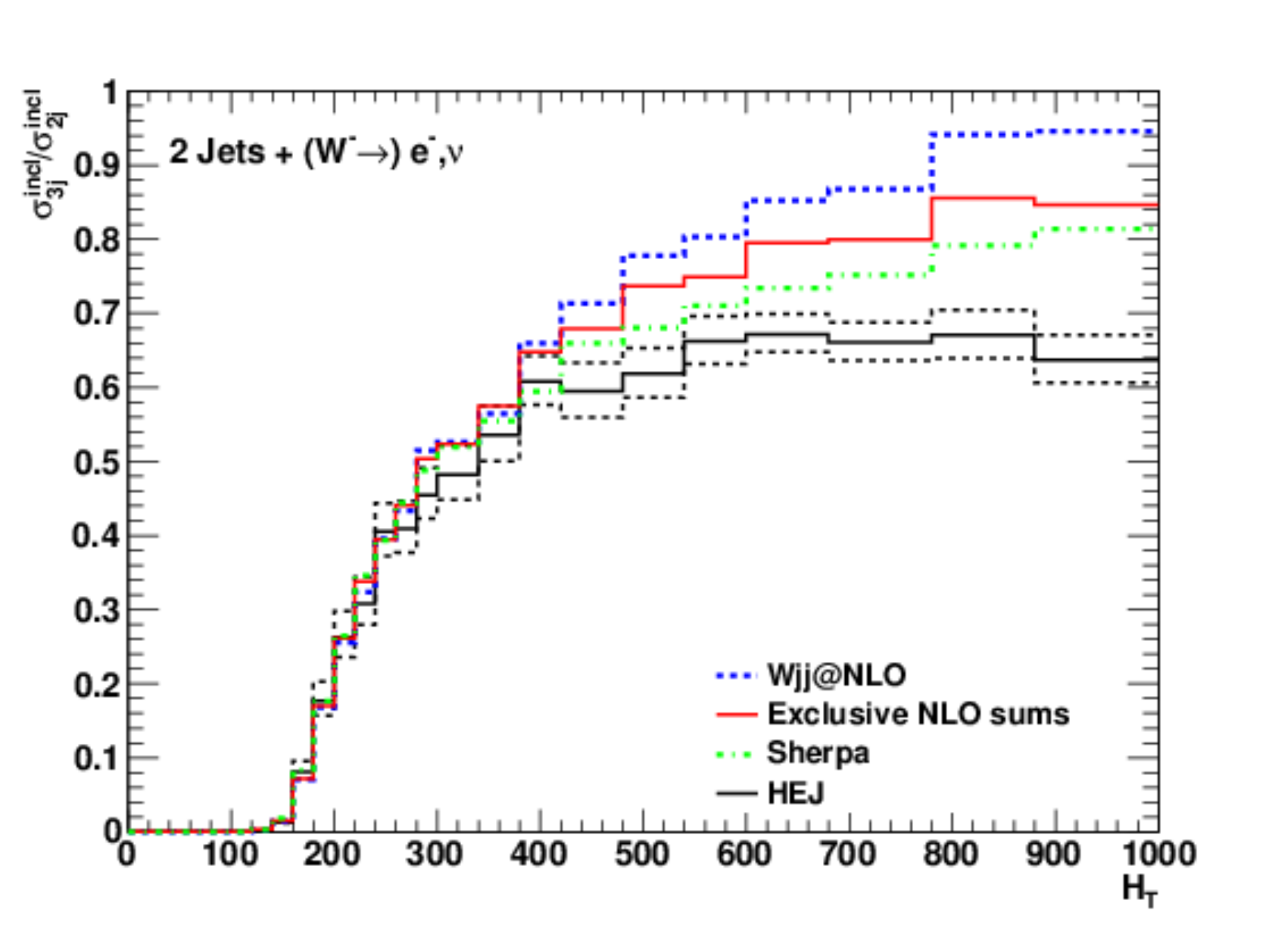}\hfill
  \epsfig{width=.49\textwidth,file=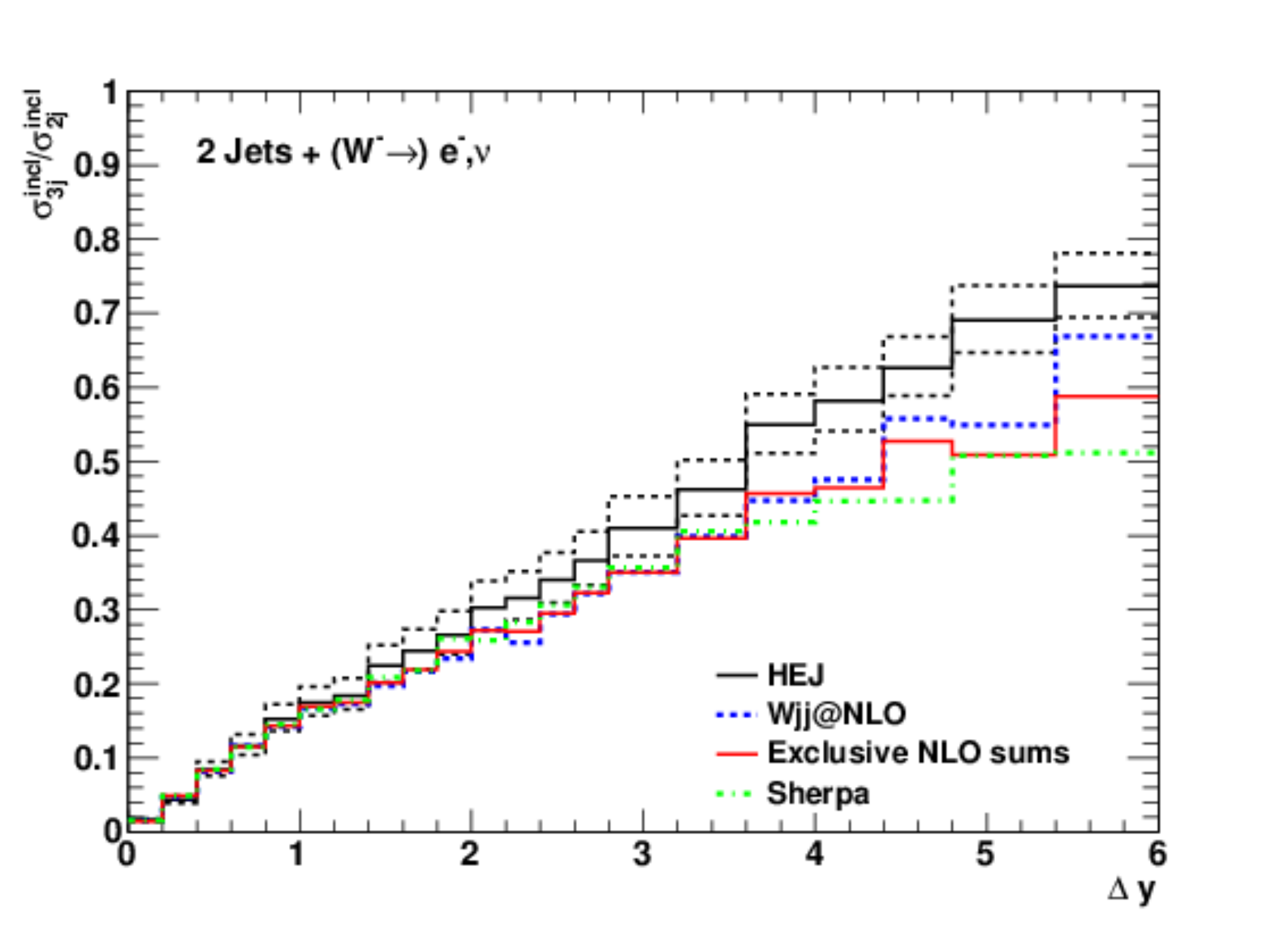}
  \caption{The ratio of the inclusive $3$-jet and $2$-jet rates in the
    inclusive $W+2$-jet NLO and exclusive sum description of
    \emph{BHS} as well as in the \emph{S-MEPS} and \emph{HEJ}
    approaches as a function of $H_T$ (left) and $\Delta y$ (right).
    Again, the dotted lines indicate the uncertainty band from varying
    the scale in \emph{HEJ} by a factor of $2$ in each direction.}
  \label{fig:3jo2jboth}
\end{figure}

\subsection{CONCLUSIONS}
\label{sec:conclusions}

We have compared a number of theoretical descriptions of $W^-$ production in
association with at least two jets.  After outlining one possible method of
combining NLO calculations of different multiplicities, we compared this with
a pure NLO calculation of $W+2$-jets production obtained by
\textsc{BlackHat+Sherpa}, a sample of leading-order events merged
using the \textsc{ME\&TS} method of \textsc{Sherpa}, and the
high-energy resummation of the \textsc{HEJ} framework.

We studied the average number of jets and the ratio of the $3$-jet and $2$-jet
inclusive cross sections as a function of $\Delta y$ and of $H_T$.  We find,
with these simple cuts, some clear differences in the predictions when we study
the average number of jets as a function of both $\Delta y$ and $H_T$.  Smaller
differences, which would be more difficult to disentangle
experimentally, are found when we study the ratio of inclusive rates.

It would be very valuable to have an experimental study, which probed
the average number of jets in $W$ production in association with at
least two jets, to test our different descriptions of these important
Standard Model processes.

\subsection*{ACKNOWLEDGEMENTS}
\label{sec:ack}

The authors would all like to thank the organisers and participants for the
extremely stimulating workshop.  DM’s work was supported by the Research
Executive Agency (REA) of the European Union under the Grant Agreement number
PITN-GA-2010-264564 (LHCPhenoNet).  JMS is supported by the UK Science and
Technology Facilities Council (STFC).



}


\section[Uncertainties in the simulation of $W+$ jets -- a case study]
{UNCERTAINTIES IN THE SIMULATION OF $W$+ JETS -- A CASE STUDY \protect\footnote{Contributed by: S.~Alioli, J.~R.~Andersen, V.~Ciulli, 
  F.~Cossutti, T.~Hapola, 
  H.~Hoeth, F.~Krauss, P.~Lenzi,
  L.~L{\"o}nnblad, G.~Luisoni, D.~Ma\^itre, 
  C.~Oleari, S.~Prestel, E.~Re, T.~Reiter, 
  M.~Sch{\"o}nherr, J.~Smillie, F.~Tramontano, J.~Winter, 
  K.~Zapp}}
\label{sec:case}
{\graphicspath{{MCuncert/}}
\numberwithin{equation}{subsection}
\newcommand{\LOPS}{LO$\otimes$PS\xspace}
\newcommand{\NLOPS}{NLO$\otimes$PS\xspace}
\newcommand{\MEPS}{M\protect\scalebox{0.8}{E}P\protect\scalebox{0.8}{S}\xspace}
\newcommand{\MENLOPS}{M\protect\scalebox{0.8}{E}N\protect\scalebox{0.8}{LO}P\protect\scalebox{0.8}{S}\xspace}
\newcommand{\MCatNLO}{M\protect\scalebox{0.8}{C}@N\protect\scalebox{0.8}{LO}\xspace}
\newcommand{\POWHEG}{P\protect\scalebox{0.8}{OWHEG}\xspace}
\newcommand{\aMCatNLO}{aM\protect\scalebox{0.8}{C}@N\protect\scalebox{0.8}{LO}\xspace}
\newcommand{\Ariadne}{A\protect\scalebox{0.8}{RIADNE}\xspace}
\newcommand{\Herwig}{H\protect\scalebox{0.8}{ERWIG}\xspace}
\newcommand{\Herwigpp}{H\protect\scalebox{0.8}{ERWIG++}\xspace}
\newcommand{\Jetset}{J\protect\scalebox{0.8}{ETSET}\xspace}
\newcommand{\PowhegBox}{P\protect\scalebox{0.8}{OWHEG} B\protect\scalebox{0.8}{OX}\xspace}
\newcommand{\Pythia}{P\protect\scalebox{0.8}{YTHIA}\xspace}
\newcommand{\PythiaEight}{P\protect\scalebox{0.8}{YTHIA}8\xspace}
\newcommand{\Sherpa}{S\protect\scalebox{0.8}{HERPA}\xspace}
\newcommand{\Alpgen}{A\protect\scalebox{0.8}{LPGEN}\xspace}
\newcommand{\Amegic}{A\protect\scalebox{0.8}{MEGIC++}\xspace}
\newcommand{\BlackHat}{B\protect\scalebox{0.8}{LACK}H\protect\scalebox{0.8}{AT}\xspace}
\newcommand{\Comix}{C\protect\scalebox{0.8}{OMIX}\xspace}
\newcommand{\GoSam}{G\protect\scalebox{0.8}{O}S\protect\scalebox{0.8}{AM}\xspace}
\newcommand{\HEJ}{H\protect\scalebox{0.8}{EJ}\xspace}
\newcommand{\Madevent}{M\protect\scalebox{0.8}{AD}E\protect\scalebox{0.8}{VENT}\xspace}
\newcommand{\Madgraph}{M\protect\scalebox{0.8}{AD}G\protect\scalebox{0.8}{RAPH}\xspace}
\newcommand{\MCFM}{M\protect\scalebox{0.8}{CFM}\xspace}
\newcommand{\Apacic}{A\protect\scalebox{0.8}{PACIC++}\xspace}
\newcommand{\CSS}{C\protect\scalebox{0.8}{SS}\xspace}
\newcommand{\Vincia}{V\protect\scalebox{0.8}{INCIA}\xspace}
\newcommand{\Diphox}{D\protect\scalebox{0.8}{I}P\protect\scalebox{0.8}{HOX}\xspace}
\newcommand{\FeynRules}{F\protect\scalebox{0.8}{EYN}R\protect\scalebox{0.8}{ULES}\xspace}
\newcommand{\LanHEP}{L\protect\scalebox{0.8}{an}H\protect\scalebox{0.8}{EP}\xspace}
\newcommand{\HqT}{H\protect\scalebox{0.8}{qT}\xspace}
\newcommand{\Jetphox}{J\protect\scalebox{0.8}{ET}P\protect\scalebox{0.8}{HOX}\xspace}
\newcommand{\Professor}{P\protect\scalebox{0.8}{ROFESSOR}\xspace}
\newcommand{\Resbos}{R\protect\scalebox{0.8}{ES}B\protect\scalebox{0.8}{OS}\xspace}
\newcommand{\Rivet}{R\protect\scalebox{0.8}{IVET}\xspace}
\newcommand{\SpringBases}{S\protect\scalebox{0.8}{PRING/}B\protect\scalebox{0.8}{ASES}\xspace}
\newcommand{\Vegas}{V\protect\scalebox{0.8}{EGAS}\xspace}
\newcommand{\Samurai}{S\protect\scalebox{0.8}{AMURAI}\xspace}
\newcommand{\Golem}{golem95\xspace}
\newcommand{\Ahadic}{A\protect\scalebox{0.8}{HADIC++}\xspace}
\newcommand{\Adicic}{A\protect\scalebox{0.8}{DICIC++}\xspace}
\newcommand{\Hadrons}{H\protect\scalebox{0.8}{ADRONS++}\xspace}
\newcommand{\Photons}{P\protect\scalebox{0.8}{HOTONS++}\xspace}
\newcommand{\Hera}{HERA\xspace}
\newcommand{\LEP}{LEP\xspace}
\newcommand{\LHC}{LHC\xspace}
\newcommand{\Tevatron}{Tevatron\xspace}
\newcommand{\Aleph}{ALEPH\xspace}
\newcommand{\Delphi}{DELPHI\xspace}
\newcommand{\Opal}{OPAL\xspace}
\newcommand{\DO}{D\O\ }
\newcommand{\CDF}{CDF\xspace}
\newcommand{\ATLAS}{ATLAS\xspace}
\newcommand{\CMS}{CMS\xspace}
\long\def\symbolfootnote[#1]#2{\begingroup%
\def\thefootnote{\fnsymbol{footnote}}\footnote[#1]{#2}\endgroup}
\newcommand{\EqRef}[1]{Eq.~\eqref{#1}}
\newcommand{\EqRefs}[2]{Eqs.~\eqref{#1}-\eqref{#2}}
\newcommand{\SecRef}[1]{Sec.~\ref{#1}}
\newcommand{\FigRef}[1]{Fig.~\ref{#1}}
\newcommand{\FigRefs}[2]{Figs.~\ref{#1}-\ref{#2}}
\newcommand{\TabRef}[1]{Tab.~\ref{#1}}
\newcommand{\TabRefs}[2]{Tabs.~\ref{#1}-\ref{#2}}
\newcommand{\AppRef}[1]{App.~\ref{#1}}

\newcommand{\abs}[1]{\left| #1\right|}
\newcommand{\rbr}[1]{\left( #1\right)}
\newcommand{\abr}[1]{\langle #1\rangle}
\newcommand{\cbr}[1]{\left\{ #1\right\}}
\newcommand{\sbr}[1]{\left[ #1\right]}
\newcommand{\im}{\imath}
\newcommand{\jm}{\jmath}
\newcommand{\ptmin}{p_{\perp,{\rm min}}}
\newcommand{\done}{{\rm d}}
\newcommand{\order}{\mathcal{O}}
\newcommand{\mc}[1]{\mathcal{#1}}
\newcommand{\mr}[1]{\mathrm{#1}}
\newcommand{\mb}[1]{\mathbb{#1}}
\newcommand{\dst}{\displaystyle}
\newcommand{\sst}{\scriptstyle}
\newcommand{\nnb}{\nonumber}
\newcommand{\bea}{\begin{eqnarray}}
\newcommand{\eea}{\end{eqnarray}}
\newcommand{\bi}{\begin{itemize}}
\newcommand{\ei}{\end{itemize}}
\newcommand{\hl}{\vphantom{$\int_A^B$}}
\newcommand{\alphas}{\ensuremath{\alpha_s}\xspace}
\newcommand{\MElevel}{matrix element level\xspace}
\newcommand{\PSlevel}{parton shower level\xspace}
\newcommand{\Hadlevel}{hadron level\xspace}
\newcommand{\UElevel}{hadron level including UE\xspace}
\newcommand{\QEDlevel}{hadron level including UE and QED final state radiation\xspace}

\newcommand{\todo}[1]{\textbf{\color{red}To do: #1}}
\newcommand{\note}[1]{\textbf{\color{red}Note: #1}}

\author{
  Simone Alioli$^a$, Jeppe R.~Andersen$^b$, Vitaliano Ciulli$^c$, 
  Fabio Cossutti$^d$, Tuomas Hapola$^b$, 
  Hendrik Hoeth$^e$, Frank Krauss$^e$, Piergiulio Lenzi$^f$,
  Leif L{\"o}nnblad$^g$, Gionata Luisoni$^e$, Daniel Maitre$^{e,h}$, 
  Carlo Oleari$^i$, Stefan Prestel$^g$, Emanuele Re$^e$, Thomas Reiter$^j$, 
  Marek Sch{\"o}nherr$^e$, Jennifer Smillie$^k$, Francesco Tramontano$^h$, Jan Winter$^h$, 
  Korinna Zapp$^e$}
\title{Uncertainties in the simulation of \texorpdfstring{$W+\,$}{W+}jets -- a case study}
\institute{
  $^a$ Ernest Orlando Lawrence Berkeley National Laboratory, University of California, Berkeley, CA 94720, USA\\
  $^b$ CP$^3$-Origins, University of Southern Denmark, Campusvej 55, DK-5230 Odense M, Denmark\\
  $^c$ Universit\`a di Firenze and INFN, Sezione di Firenze, via G. Sansone 1, 50019 Sesto F. (FI), Italy \\
  $^d$ INFN, Sezione di Trieste, Via Valerio 2, 34127 Trieste, Italia\\
  $^e$ Institute for Particle Physics Phenomenology,
  Durham University, Durham DH1 3LE, UK\\
  $^f$ PH-EP Department, CERN, CH-1211 Geneva 23, Switzerland \\
  $^g$ Department of Astronomy and Theoretical Physics, Lund University,
  S\"{o}lvegatan 14A, SE-22362 Lund, Sweden\\
  $^h$ PH-TH Department, CERN, CH-1211 Geneva 23, Switzerland\\
  $^i$ Universit\`a di Milano-Bicocca and INFN, Sezione di Milano-Bicocca, Piazza della Scienza 3, 20126 Milan, Italy\\
  $^j$ Max-Planck-Institut f\"ur Physik, F\"ohringer Ring 6, 80805 M\"unchen, Germany\\
  $^k$ School of Physics and Astronomy, University of Edinburgh, Mayfield Road,
  Edinburgh EH9 3JZ, UK\\
}
\begin{abstract}
  In this contribution, uncertainties in the simulation of a large variety of 
observables related to the production of $W$ in association with jets at the 
\LHC and the \Tevatron are discussed.  This work aims to
\begin{itemize}
\item serve as a compendium of currently publicly accessible tools in addition 
  to the ones presented in a previous publication \cite{Alwall:2007fs} with a 
  similar topic, and to compare their results;
\item discuss the origin and generic size of various uncertainties in the
  simulation of perturbative and non-perturbative aspects of this process;
\item trace the interplay of these uncertainties in various stages of the
  full event simulation;
\item hint at those uncertainties in each of the various tools considered here
  which the respective authors find relevant;
\item guide their users in how to assess the related uncertainties in a way
  the authors recommend. 
\end{itemize}

\end{abstract}
\subsection{Introduction}

The production of $W$-bosons in association with jets constitutes an important 
process at the Tevatron and the LHC, for a variety of reasons.  First of all, 
it represents a major background to Standard Model signatures such as top-pair
and single-top production,and it also plays a role in searches for the Higgs
boson in the Standard Model.  Furthermore, this reaction, together with the
fairly similar channel of $Z$-production in association with jets, provides 
one of the most important backgrounds in those searches for new physics where
large missing transverse energy and high jet multiplicities characterise the
respective signal.  Thirdly, this process has become a standard reaction for 
QCD studies at hadron colliders, ranging from the validation of simulation
tools for multijet signatures to measurements related to multiple parton
scattering.  Finally, this process also provides one of the main testbeds for
novel techniques in the automation of higher-order QCD corrections and their
matching or merging with subsequent parton showers in the framework of event 
generators.  

In the spirit of this last point, providing a testbed for the combination of 
fixed order calculations with the parton shower, this process has been analysed
in quite some depth in \cite{Alwall:2007fs} about five years ago.  A number of
reasons provide motivation to update and extend this previous study, namely
\begin{itemize}
\item the LHC being up and running and starting to provide highly precise 
  data such that a proper treatment of uncertainties becomes an important
  issue;
\item major improvements in the ability to calculate higher-order corrections 
  including up to four jets in the final state accompanying the $W$ bosons 
  \cite{Berger:2009ep,KeithEllis:2009bu,Berger:2010zx}; 
\item the advent of such next-to leading order calculations -- albeit for 
  lower final state multiplicities -- fully matched to the parton shower
  \cite{Alioli:2010qp,Frederix:2011ig,Hoeche:2012ft};
\item an improved understanding of the leading order merging prescription for
  towers of multijet multiplicities with the parton shower 
  \cite{Hoeche:2009rj,Lonnblad:2011xx}; 
\item the combination of matching and merging methods 
  \cite{Hamilton:2010wh,Hoche:2010kg};
\item and new methods to simulate multijet topologies based on the 
  high-energy limit \cite{Andersen:2011hs}.
\end{itemize}
Therefore this study aims at being a first step towards a more complete
update of \cite{Alwall:2007fs}, with a shift in focus towards a discussion of
theoretical uncertainties in different approximations, including perturbative
and non-perturbative effects.  Apart from tracing the origin and determining
the generic size of various uncertainties in the theoretical description of
various observables related to this process, also the interplay of them at
various stages of the simulation, from the matrix element to the hadron level
will be discussed.  Consequently, the most important causes for theory 
uncertainties in various tools are highlighted.  Therefore, one of the more
practically relevant goals is to also provide methods to reliably and robustly 
estimate such uncertainties for the various tools used in this study, as
recommended by authors or users.  

The outline is as follows: After briefly presenting the various tools included
in this study and discussing the way they have been used here in 
\SecRef{Sec:Codes}, example results for them will be presented individually,
tool by tool in \SecRef{Sec:Results}.  In \SecRef{Sec:Comparison} these results
are compared in order to see and quantify relative differences.  In this
endeavour, experimental results have not yet been included.  We reserve this
comparison with relevant data for a later, full-fledged analysis, which will
hopefully include even more tools.

\subsection{Codes}
\label{Sec:Codes}

In this work a variety of different codes has been employed, which allow
to study the process at various different stages:
\begin{enumerate}
\item Fixed order matrix elements:\\
  By now, the description of $W$ boson production in association with jets
  is possible for up to $4$ additional jets at NLO.
  Here, results from two NLO codes, \GoSam{}+\Sherpa~\cite{Cullen:2011ac,Gleisberg:2003xi,Gleisberg:2008ta} 
  and \BlackHat{}+\Sherpa~\cite{Berger:2009zg,Berger:2009ep,Berger:2010zx}, which are 
  either publicly available or provide publicly available event files, are 
  presented. The corresponding results therefore are on the \MElevel.
\item All-order resummed matrix elements:\\
  Approximations to the partonic matrix elements for the processes of $n$-jet
  production, and $W,Z,H+n$-jets, $n\ge 2$, was recently calculated to any
  multiplicity, and including all-order resummations for the leading virtual
  corrections. The all-order scheme~\cite{Andersen:2009nu,Andersen:2009he},
  implemented in the \HEJ~\cite{Andersen:2011hs} code, becomes exact in the
  limit of large invariant mass between each parton (the MRK limit of
  BFKL). The resummation scheme is merged with LO matrix elements (much like
  in \MEPS, see later). The resummation of \HEJ can also be interfaced to a
  parton shower~\cite{Andersen:2011zd}; the results presented here, however,
  are on the \MElevel. It should also be stressed that due to the nature of
  the approximation of \HEJ, the simulation here are relevant for the
  production of {\em at least} two jets in addition to the $W$ boson.
\item Parton showers:\\
  The pure parton shower code relies on the collinear approximation to
  produce additional jets.  By using a matrix element reweighting, however,
  in the process of $W$ production, typically one additional jet can correctly
  be described.  For this simulation, \PythiaEight~\cite{Sjostrand:2007gs} 
  has been used here, with results available on the \PSlevel, 
  \Hadlevel and \UElevel.
\item LO matrix elements merged with the parton shower (\MEPS):\\
  By now, the use of towers of multijet matrix elements with increasing 
  multiplicity merged to the parton shower following ideas presented 
  in~\cite{Mangano:2001xp,Catani:2001cc,Lonnblad:2001iq,Krauss:2002up}
  is common practise in the experimental collaborations.  In fact, a first
  comparison of different codes and implementation has been presented a while
  ago~\cite{Alwall:2007fs}.  Here, three implementations of these ideas are
  included, namely the ones in \Madgraph{}+\Pythia~\cite{Stelzer:1994ta,Maltoni:2002qb,Alwall:2007st,Alwall:2011uj,Sjostrand:2006za}, 
  \PythiaEight{}+ME~\cite{Lonnblad:2011xx} and 
  \Sherpa~\cite{Gleisberg:2008ta}.  
  Here results are available on all levels \MElevel, \PSlevel, \Hadlevel,
  \UElevel, and \QEDlevel in different combinations of codes. 
\item NLO matrix elements matched to the parton shower (\NLOPS and \MENLOPS):\\
  In principle two methods by now have been proposed and fully implemented 
  which consistently match full NLO calculations to the parton shower, namely 
  \MCatNLO~\cite{Frixione:2002ik} and 
  \POWHEG~\cite{Nason:2004rx,Frixione:2007vw}.  Here the latter is being used, 
  with its implementation in the \PowhegBox~\cite{Alioli:2010xd}, and 
  interfaced to the \Pythia~\cite{Sjostrand:2006za} parton shower in its 
  $k_T$-ordered version~\cite{Sjostrand:2004ef}.  In addition, a combination
  of such matching with the merging methods described in the previous point
  is available~\cite{Hamilton:2010wh,Hoche:2010kg}, ranging under the name
  \MENLOPS.  In this paper we use an implementation of such methods 
  provided in the \Sherpa framework.  In both cases, results are available on 
  all levels  \MElevel, \PSlevel, \Hadlevel, and \UElevel.  
\end{enumerate}

\subsubsection{\texorpdfstring{\protect\BlackHat}{BlackHat} + 
	    \texorpdfstring{\protect\Sherpa}{Sherpa}}

\label{Sec:Codes:BlackHat}

The NLO predictions are obtained by combining \BlackHat \cite{Berger:2008sj} for the virtual part and \Sherpa \cite{Krauss:2001iv,Gleisberg:2007md} for the real part. It is currently possible to obtain predictions at NLO for a $W$-boson in combination with up to four jets \cite{Berger:2009zg,Berger:2009ep,Berger:2010zx}.

The plots have been produced by re-analysing large event files produced by the combination of \BlackHat and \Sherpa. These files contain particle four-momenta as well as the coefficients of all scale dependent functions, including the PDFs so that it makes it possible to easily change factorisation and renormalisation scales as well as the PDF set. 

We used a common factorisation and renormalisation scale $\mu_F=\mu_R=\hat{H}'_T/2$ with $\hat{H}'_T=\sum_jp_T^j +E_T^W$ where the sum runs over all jets and $E_T^W=\sqrt{M^2_W+(p_T^W)^2}$.    
\paragraph{Estimation of uncertainties}
The estimation of the uncertainties for the NLO calculation obtained with \BlackHat{}+\Sherpa is obtained by combining in quadrature the pdf uncertainties obtained using the pdf error set and the uncertainties obtained by varying the factorisation and renormalisation scales simultaneously by factors of 1/2 and 2. To this error we also add in quadrature the integration error estimate. Another way of estimating the uncertainties due to the choice of scales is to compare predictions obtained using different choice of basis scales, but this has not been done for this study.

We used the CTEQ6.6 PDF set. The value of \alphas used for this calculation has also been taken as that provided with this PDF set. The PDF uncertainties are estimated using the hessian method and PDF 'error' set provided with the CTEQ6.6 PDF set. 
   

\subsubsection{\texorpdfstring{\protect\GoSam}{GoSam} + 
	    \texorpdfstring{\protect\Sherpa}{Sherpa}}
\label{Sec:Codes:GoSam}

\GoSam~\cite{Cullen:2011ac} is a new framework which allows the automated computation of one-loop scattering amplitudes for multi-particle processes. The one-loop scattering amplitudes are
generated in terms of algebraic $d$-dimensional unintegrated amplitudes, which are obtained via Feynman diagrams. This allows to perform symbolic manipulations of the expressions prior any numerical
step. For the reduction, the program offers the possibility to use either a $d$-dimensional extension of the OPP method \cite{Ossola:2006us,Ossola:2007bb,Ellis:2008ir}, as implemented in
\Samurai~\cite{Mastrolia:2010nb}, or tensor reduction as implemented in \Golem~\cite{Binoth:2008uq,Cullen:2011kv} interfaced through tensorial reconstruction at the integrand
level~\cite{Heinrich:2010ax}.

The \GoSam framework can be used to calculate one-loop corrections within both QCD and electroweak theory. Beyond the Standard Model theories can be interfaced using \FeynRules~\cite{Degrande:2011ua}
or \LanHEP~\cite{Semenov:2010qt}.

To produce results for a certain process specified by the user, the program must be fed with an ``input card'' with the details of the process. Alternatively, when interfacing the program with a
Monte Carlo (MC) event generator which supports the Binoth-Les-Houches-Accord (BLHA) interface~\cite{Binoth:2010xt}, the specific order file produced by the MC event generator can be passed to \GoSam.

The analysis presented here was performed using this latter generation mode and \Sherpa~\cite{Gleisberg:2003xi,Gleisberg:2008ta} was chosen as MC event generator. \Sherpa provides therefore the
matrix elements for the production of $W$ and exactly one jet at the Born-level and the NLO real corrections to it, together with the needed subtraction terms and their integrated counter-parts.
\GoSam provides the NLO virtual-part. The generation of the code follows the standards of the BLHA-interface~\cite{Binoth:2010xt}. During the first call of Sherpa an ``order file'' is written by the
MC program. This file is read-in by \GoSam to produce the code for the one-loop evaluation of needed process. If this happens successfully, a contract file with information on the different
possible subprocesses is produced by \GoSam and can be later read by the MC generator to recognize the numbering of the different partonic subprocesses. At running time all information between
\GoSam and \Sherpa is also passed using the BLHA-interface standards.

The steering of the event generation and the analysis interface with \Rivet~\cite{Buckley:2010ar} is done using \Sherpa cards. Each curve in the analysis consists of 100 combined runs of 50 million
events. The renormalisation and factorisation scales are set according to the choice made for this analysis in Les Houches to
$$\mu_{F}=\mu_{R}=\hat{H}_{T}'/2\,,$$
where $\hat{H}_{T}'$ is defined in the previous section.

\paragraph{Estimate of uncertainties}
The estimation of the uncertainties for the NLO calculation obtained with \GoSam{}+\Sherpa is done combining in quadrature the PDF uncertainties with the uncertainty coming from the separate variation
of factorisation and renormalisation scale by factors of 1/2 and 2. Ideally also the integration error should be added in quadrature to the previous estimate, however the MC integration error obtained
with \Rivet at NLO is not reliable because of the incapacity of \Rivet to take into account properly the correlation between real and subtraction events. For this reason and because of the very high
statistics of the MC sample, the MC integration error is neglected. To assess the PDF uncertainty we compute the envelope of the results obtained using the three different PDF sets
CT10~\cite{Lai:2010vv}, used as nominal set, MSTW08~\cite{Martin:2009iq} and NNPDF2.1~\cite{Ball:2011mu}. The total scale uncertainty is determined by adding in quadrature the factorisation and
renormalisation scale uncertainties. Each of them is found by computing the maximum between the nominal value and the up and down variations.


\subsubsection{\texorpdfstring{\protect\HEJ}{HEJ}}
\label{Sec:Codes:HEJ}

The \textit{High Energy Jets} (\HEJ) framework
\cite{Andersen:2009nu,Andersen:2009he} provides an alternative description of
collider events to the standard fixed order calculations (possibly interfaced
to a parton shower).
Instead, \HEJ uses approximations to the hard scattering matrix element to all
orders in \alphas which become exact in the High Energy limit.  The
approximation results in sufficiently simple matrix elements, that these can
be explicitly regulated, integrated and summed over any (relevant) multiplicity. This results
in an explicit \emph{all-order} resummation of the dominant contributions from wide-angle
QCD radiation.

The building blocks of the \HEJ framework ensure the correct leading
logarithmic behaviour in the Multi-Regge Kinematic limit (aka.~the High
Energy Limit) of large invariant mass between all partons, for both the real
and virtual corrections. The resummed $n$-jet rate is then further matched to
tree-level accuracy for events with up to and including four jets, using a
merging procedure for the soft radiation.

This procedure has so far been applied to the production of jets
\cite{Andersen:2011hs}, $W$ plus jets \cite{Andersen:2012tba}, $Z$ plus jets
and Higgs boson plus jets and has currently been implemented in a fully
flexible Monte Carlo for the first two of these processes.  The
implementation integrates explicitly over any number of QCD emissions from a
($W,Z,H$+) dijet system, and hence produces event samples for processes with two jets
or more.  Note 
that one has access to the momenta of all final state particles for every
event and it is therefore extremely simple to restrict to a subset of the
events if required, e.g.~3-jet exclusive events.

The \HEJ resummation includes emissions at large transverse momentum which are
increasingly important as the centre-of-mass energy of particle collisions
increases.  \HEJ is currently the only available flexible Monte Carlo generator
to obtain leading logarithmic accuracy in the limit of large invariant mass
between emissions.  However, the \HEJ framework does not include any
systematic resummation in the collinear limit.  This is included in a parton
shower, but a careful merging procedure is required to link one with \HEJ, as
there is significant overlap between the soft emissions included in each approach; the
first steps in this direction have been taken for jet production
\cite{Andersen:2011zd} and are ongoing.  In the current study though, only
parton level predictions are given.

\paragraph{Estimate of Uncertainty}

The \HEJ framework does not contain any tunable parameters other than the
choice of renormalisation and factorisation scale (just like any fixed order
calculation).  In this study, in common with other approaches, we choose both
of these to be given by the geometric mean of the transverse momentum of the
jets:
\begin{equation}
  \label{eq:HEJmuchoice}
  \mu_R\ =\ \mu_F\ =\ \left(\prod_{j=1}^n p_T^j\right)^{1/n},
\end{equation}
where the jets are defined according to the relevant cuts in each analysis.
This is however only an arbitrary choice, as the framework admits any choice for
the scale, including $H_T$, $p_T$ of the hardest jet and a fixed scale.  For a
given scale, \alphas is evaluated according to the relevant PDF.

In common with standard convention, we calculate the scale variation by
changing this scale by a factor of two in both directions.  In principle, one
could also include the PDF uncertainty, but this is not done in this study
(as the scale uncertainty dominates).  As described above, \HEJ contains
matching to tree-level accuracy for up to four jets.  However, unlike the
merging procedure in a showered sample, the merging scale here is not a free
parameter.  There is only one rational choice for the merging scale: the
minimum $p_T$ of a jet in the relevant analysis. However, in an inclusive
sample with at least two jets, one could use as a further estimator of
uncertainty the variation obtained when matching to three and four jet LO
matrix elements. This procedure will be studied in detail in
Ref.~\cite{Andersen:2012tba}, but in the present study, we quote the
uncertainty only from the scale variation.



\subsubsection{\texorpdfstring{\protect\Madgraph}{Madgraph} + \texorpdfstring{\protect\Pythia}{Pythia}}
\label{Sec:Codes:MadGraph}
\Madgraph~\cite{Stelzer:1994ta,Maltoni:2002qb,Alwall:2007st,Alwall:2011uj} is a
general purpose leading order matrix element (ME) generator, with a
broad variety of models available and easily extensible thanks to its
modular structure. The event generation is performed by the \Madevent
component, a tool implementing the Single Diagram Enhanced algorithm
for multi-channel phase space integration. When a user provided
process is specified, \Madgraph automatically generates the amplitudes
for all the relevant subprocesses and produces the mappings for the
integration over the phase space. This process-dependent information
is then used by \Madevent, where the process specific code generated
allows the user to calculate cross sections and to produce unweighted
events. Once the parton level events have been generated, a traditional
parton shower (PS) Monte Carlo library can be run on top of the \Madgraph
output to describe additional QCD radiation, and possibly
allow to produce hadron level generated events if a suitable
hadronisation model is then applied.

In order to avoid double counting of QCD radiation from the matrix
element and the parton shower, the MLM matching approach is used in
its ktMLM implementation provided by the \Madgraph
team~\cite{Alwall:2011uj}.

For the present study \Madgraph-5.1.1~\cite{Alwall:2011uj} has been used for the
matrix element generation, while the parton shower and hadronisation
has been provided by \Pythia 6.4.2.4~\cite{Sjostrand:2006za}. The 
$W + n$ jet process has been simulated up to 4 additional partons. The
PDF used in both calculations has been CTEQ6L1, and in the matrix
element calculation the strong coupling costant has
been setup to be equal to the one from the PDF used. The factorisation
scale and the hadronisation scale are set to the W transverse mass,
$m_{\perp,W}$. The parton level 
clusterisation scale {\it xqcut} has been set to 10 GeV, while the ME
- PS matching scale {\it qcut} has been set to the optimal value of 20
GeV, determined ensuring the smoothness of the differential jet rate.

The \Pythia settings have been defined according the so called Tune Z2,
an adjustment of Tune Z1 described in~\cite{Field:2010bc} for CTEQ6L1,
where the $p_\perp$ cutoff for the multiple parton interactions is set to
{\tt PARP(82)=1.832} obtained on top of LHC data as far as the
underlying event and multiple parton interactions are concerned,
while the fragmentation parameters are those optimized on LEP data by
the \Professor~\cite{Buckley:2009bj} team.

\paragraph{Estimate of uncertainties}

To estimate the uncertainties due to the factorisation scale and 
the renormalisation scale, which are set to $m_{\perp,W}$, we varied them
simultaneously by a factor two.
In addition we have independently varied by a factor two the ME - PS matching
scale.

While applying these modifications, the total cross-section is kept fixed
to the value obtained with the default parameters, 27.77 nb. That is because
we are only interested in shape variations of the distributions,
rather than in the total cross-section of the process calculated by
\Madgraph, which is accurate only at the leading order. 

\subsubsection{\texorpdfstring{\protect\PowhegBox}{PowhegBox} + 
	    \texorpdfstring{\protect\PythiaEight}{Pythia8}}
\label{Sec:Codes:PowhegBox}

The \PowhegBox~\cite{Alioli:2010xd} is a computer framework to ease
the \POWHEG{}~\cite{Nason:2004rx} implementation of new processes.  It
only requires as input the individual components of the NLO
calculation under consideration, {\it i.e.}, the Born process, its
virtual radiative corrections and the real emission contributions.
Then it automatically combines them, canceling the emerging soft and
collinear singularities in the Frixione-Kunszt-Signer (FKS)
subtraction scheme, and produces the required events. The \PowhegBox{}
is also a library, where previously implemented processes are
available in a common framework.  

For the present study we make use of the $W$+jet implementation
presented in~\cite{Alioli:2010qp}.

The produced events are passed to \PythiaEight~\cite{Sjostrand:2007gs}
through the Les Houches interface~\cite{Alwall:2006yp} and showered
with the default transverse-momentum ordered shower, vetoing further
emissions harder than the one already present in the input
events. This is achieved by setting the starting scale of the
shower as the transverse momentum of the hardest emission.\footnote{ 
An extra veto may be required at this stage, due to the different definitions of the transverse momentum used in the \PowhegBox{} -- either for initial or final state radiation -- and in \PythiaEight.}

When multiple partonic interactions (MPI) are turned on, these are
allowed to be harder than the first \POWHEG{}~emission. Indeed, since
the $W+1$ jet process is not accounted for in MPI, there is no
over-counting.

Eventually, the relevant distributions are evaluated by interfacing
the MonteCarlo output to the \Rivet~\cite{Buckley:2010ar} analysis,
for the two given sets of ATLAS and CMS cuts.

Since we have simulated events starting from a hard process where a
$W$ is produced in association with one jet, only observables
built from events where at least 1 jet is present will be shown.

\paragraph{Generation of predictions and estimate of uncertainties}

Predictions presented here are based on a merged sample of 4M $W^++
j$ and 4M $W^-+j$ weighted events, produced with the default
\PowhegBox choice of parameters. In particular, we have required a
minimum cut $p_{\rm \scriptscriptstyle T}= 5$~GeV on the associated jet at
the generation level and, in order to enhance the statistical sampling
of the high-$p_{\rm \scriptscriptstyle T}$ tail, we have further suppressed
the rapidly rising contribution at low jet $p_{\rm \scriptscriptstyle T}$ by
the factor $p^2_{\rm \scriptscriptstyle T} / \left( p^2_{\rm \scriptscriptstyle
    T,supp} + p^2_{\rm \scriptscriptstyle T} \right) $, with $p^2_{\rm
  \scriptscriptstyle T,supp} = 100$~GeV. The inverse of this factor enters
the event weight. 


We have adopted the \PowhegBox default values for EW parameters, namely
\begin{equation}
M_W=80.398~\rm GeV\,,\quad \Gamma_W=2.141~\rm GeV\,,\quad (\alpha_{\rm
  em})^{-1}=128.89\,,\quad \sin^2 \theta_W=0.222645
\end{equation}
and we have assumed a CKM matrix with a mixing between the first two generations only 
\begin{equation}
  |V_{ud}|=|V_{cs}|=0.975\,,\quad |V_{us}|=|V_{cd}|=0.222\,,\ {\rm
    and}\ |V_{tb}|=1\,.
\end{equation}
Finally, we have resctricted the integration region to the interval $0< M_W <2221$~GeV.      

For the computation of the \POWHEG{} $\bar{B}$ function, the
renormalisation and factorisation scale was chosen equal to
\begin{equation}
\mu_R=\mu_F= p_{\perp,j}\,,
\end{equation}
where $p_{\perp,j}$ corresponds to the transverse-momentum of the (single)
parton recoiling against the $W$ boson in the so-called underlying
Born kinematics~\cite{Frixione:2007vw}. We have also run the code
using
\begin{equation}
  \mu_R=\mu_F=1/2 \left( \sqrt{M_W^2+p_{\perp,W}^2} + p_{\perp,j}
  \right) \,,
\end{equation}
but no relevant differences were observed with respect to the
aforementioned choice, being the two scales similar for the $W+
1$ jet processes at hand.


The scales entering in the evaluation of parton distribution functions
and of the strong coupling in the \POWHEG{} Sudakov form factor are 
chosen to be equal to the transverse momentum of the \POWHEG{} hardest
emission~\cite{Frixione:2007vw,Alioli:2010xa}.

Scale-uncertainty bands obtained by varying the factorisation and
renormalisation scales entering the $\bar{B}$ function by a factor of
two in either directions are used as an estimate of the theoretical
error associated to higher order missing effects.

The uncertainty due to the PDF choice was estimated generating events
using three different sets (CT10 \cite{Lai:2010vv},
MSTW2008~\cite{Martin:2009iq}, and NNPDF2.1~\cite{Ball:2011mu}). The
value of the strong coupling constant at $M_Z$
is consistently read from the PDF table used.  The further showering
performed by \PythiaEight is instead performed with default PDF and
\alphas definitions, the difference being beyond the claimed
accuracy of the calculation.  In this study, we have used
\PythiaEight, version 8.153.



\subsubsection{\texorpdfstring{\protect\PythiaEight}{Pythia8}}
\label{Sec:Codes:Pythia}


\PythiaEight \cite{Sjostrand:2007gs} is the latest incarnation of event 
generators of the \Pythia family. At the heart of the generator are parton 
showers that evolve high-scale processes to the scale of hadronisation, by 
generating splittings with DGLAP splitting kernels. The splitting 
scales are ordered in relative transverse momentum 
\cite{Sjostrand:2007gs,Sjostrand:2004ef}, and the phase space is constructed in
a dipole-like manner in order to capture soft gluon coherence effects
\cite{Gustafson:1986db}. A key point of the evolution of partonic states in 
\PythiaEight is that all perturbative components are interleaved
\cite{Sjostrand:2007gs,Sjostrand:2004ef,Sjostrand:2004pf}, i.e.\ multiple 
partonic interactions, space-like and time-like showers are all generated in 
one transverse-momentum ordered evolution sequence. This means that due to the 
competition for phase space, all steps in the event generation are correlated. 
For a detailed discussion how parameters of the interleaved shower evolution 
are tuned to collider data, see \cite{Corke:2010yf}. \PythiaEight with additional 
matrix element corrections has so far not been tuned to data. Since in 
\cite{Lonnblad:2011xx}, only very small differences were seen for LEP between 
\PythiaEight with and without matrix element merging, we expect only small re-tuning 
effects in the parameters of the Lund string model \cite{Andersson:1983ia}. 
Similarly, since we keep the low-scale modelling of \PythiaEight largely 
intact, only small changes in the underlying event tuning are expected. We 
however expect that some re-tuning will be needed for jet shape data.

It should be noted that \PythiaEight includes a selection $2\rightarrow1$ and
$2\rightarrow2$ processes, as well as a limited variety of $2\rightarrow3$ 
processes, but does not contain a general ME generator. New processes, 
particularly for higher jet multiplicities, have to be made available in form
of Les Houches Event (LHE) \cite{Alwall:2006yp} files. By virtue of matrix 
element corrections, \PythiaEight describes the first emission in $\Wjets$ with
the full matrix element probability. When introducing matrix elements with one
additional jet within matrix element merging, this allows to fully cancel the 
merging scale dependence for the first emission, while small merging scale 
dependencies enter when including further jets. Current versions of 
\PythiaEight include a general implementation of the \textsc{Ckkw-l} matrix 
element merging prescription \cite{Lonnblad:2001iq}. Please consult 
\cite{Lonnblad:2011xx} for a detailed discussion of the implementation in 
\PythiaEight.


\paragraph{Generation of the predictions}

To generate predictions with stand-alone \PythiaEight (i.e.\ without inclusion
of matrix elements for $\W$ production in association with two or more jets),
the built-in $\mathrm{q}\bar{\mathrm{q}}\rightarrow \W$ matrix element in 
\PythiaEight was used to generate the initial configuration. This was then 
evolved with to the hadronisation scale and the ensemble of partons hadronised
using the Lund string model. For this study, we use the publicly available 
\textsc{Pythia} 8.157, with CTEQ6L1 parton distribution functions, and the 
associated Tune 4C. Since \cite{Lonnblad:2011xx} showed a large dependence of 
the quality of the matrix element merging on whether rapidity-ordered emissions
are explicitly forbidden in space-like showers, results are presented with and
without enforced rapidity ordering.



%

The inclusion of matrix elements for additional jets into \PythiaEight is 
achieved with \textsc{Ckkw-l} merging. All merging tasks are handled
internally in \textsc{Pythia} 8.157, allowing for a high degree of 
automation. This means that the user only needs to supply
\begin{itemize}
\item Matrix element configurations in form of LHE files.
\item An identifier giving the hard process of interest.
\item A value of the merging scale. Facilities to allow the user to implement
a her/his own merging scale definition are available.
\end{itemize}
For this report, matrix element configurations with additional jets were 
generated with \Madgraph/ \Madevent~\cite{Alwall:2007st}, and read into \PythiaEight in form of Les 
Houches Events. \PythiaEight then derives all possible parton shower histories 
for an event, probabilistically chooses a history, and uses the reconstructed 
states and splitting scales to perform a re-weighting with Sudakov factors 
and \alphas values. This means each event will have a weight
\begin{eqnarray}
w_{\textnormal{\tiny{CKKWL}}}
&=&  \frac{x_{n}^+f_{n}^+(x_{n}^+,\rho_n)}{x_{n}^+f_{n}^+(x_{n}^+,\mu_F^2)}
     \frac{x_{n}^-f_{n}^-(x_{n}^-,\rho_n)}{x_{n}^-f_{n}^-(x_{n}^-,\mu_F^2)}
     \nonumber\\
 &&\times
   \prod_{i=1}^{n} \Bigg[\frac{\alphas(\rho_i)}{\alpha_{\mathrm{sME}}}
     \frac{x_{i-1}^+f_{i-1}^+(x_{i-1}^+,\rho_{i-1})}
          {x_{i-1}^+f_{i-1}^+(x_{i-1}^+,\rho_i)}
     \frac{x_{i-1}^-f_{i-1}^-(x_{i-1}^-,\rho_{i-1})}
          {x_{i-1}^-f_{i-1}^-(x_{i-1}^-,\rho_i)}
   \Pi_{S_{+i-1}}(\rho_{i-1},\rho_i)\Bigg]\Pi_{S_{n}}(\rho_n,\tms)
   \nonumber
\end{eqnarray}
where $\rho_i$ and $x_{i}^\pm$ are the the reconstructed shower splitting 
scales and momentum fractions of the incoming partons in $\pm$z-direction, and 
$\Pi_{S_{+i}}(\rho_{i},\rho_{i+1})$ the parton shower no-emission probability
when evolving the state $S_{+i}$ from scale $\rho_{i}$ to $\rho_{i+1}$.
$\alpha_{\mathrm{sME}}$ gives the strong coupling used in the matrix element 
calculation. All reweighting factors are generated dynamically with 
help of the shower. The interleaved evolution of \PythiaEight is accommodated
by consistently including effects of multiple interactions into the no-emission
probabilities. A detailed description of the formalism is given in \cite{Lonnblad:2011xx}.

As input for the current analysis, we have produced LHE files for $\Wpjets$ 
with up to four (three) additional jets at Tevatron (LHC) energies. The 
renormalisation scale in \Madgraph was fixed to $\mu_R = \mz$. For hadronic
cross sections, CTEQ6L1 parton distributions (as implemented in LHAPDF 
\cite{Whalley:2005nh}) have been chosen at a factorisation scale $\mu_F = \mw$, and the strong
coupling in the ME was correspondingly fixed to $\alphas(\mz) = 0.129783$. To 
regularise QCD divergences and act as a merging scale, a cut in 
\begin{equation}
  k_\perp^2 =
  \min\left\{\min( p_{T,i}^2, p_{T,j}^2),\min( p_{T,i}^2, p_{T,j}^2) 
  \frac{(\Delta \eta_{ij})^2 + (\Delta\phi_{ij})^2}{D^2}\right\}
  \quad\mathrm{with}\quad D = 0.4
  \nonumber
\end{equation}
and a cut value of $k_{\perp,min} = \tms = 15$ GeV has been applied to the 
matrix element.

Merged \PythiaEight predictions are given for the default settings, i.e.\
using the parameters of Tune 4C, for Tune A2 \cite{ATLAS:2011dk},
and for Tune 4C without enforced rapidity ordering (dubbed Tune X).
Again, it should be noted that so far, no tuning including additional jets has
so far been conducted.

\paragraph{Estimate of uncertainties}

To estimate uncertainties of a merged prediction of $\Wjets$, it is interesting
to study the dependence on the merging scale value. For this, we have generated
LHE files with three different $k_{\perp,min} = \tms$ cuts ($\tms = 15, 30, 45$)
GeV, and performed \textsc{Ckkw-l} merging on these samples. Furthermore, to 
show the effect of tuning, the $\tms = 15$-GeV-sample was processed for two 
adequate tunes, Tune 4C and A2.

\paragraph{Uncertainties related to shower ordering}
\label{sec:Pythia8MEPS_ordering_choices}

In \cite{Lonnblad:2011xx}, it was shown that restricting shower emissions in
\PythiaEight to regions of phase space ordered both in transverse momentum and
rapidity leads to non-negligible effects in merged predictions. This can be
seen as an effect of limiting the shower accuracy by reducing the phase space
over which splitting kernels are integrated, meaning the accuracy of Sudakov
form factors is impaired. Loosely speaking, if above the merging scale, the 
matrix element, integrated over the full phase space\footnote{Particularly for
high jet multiplicities, it could be imagined that phase space integrators have
difficulties to fully sample the phase space, especially close to kinematic 
limits. By full phase space, we mean the region that the phase space 
generator actually filled.}, differs substantially from the splitting 
probabilities integrated over the allowed parton shower phase space, merged 
results will exhibit substantial merging scale dependencies. Such problems are 
obviously introduced if the parton shower phase space is heavily constrained. 

Changing the phase space regions in which the shower is allowed to radiate thus
allows us to estimate the uncertainties of the merging procedure in conjunction
with the underlying shower. Particularly, this procedure can test the quality
of the matrix element merging beyond the first few emissions, and give hints on
how the shower resummation may be improved.

To emphasise the impact of the shower transition probabilities, we choose a 
fairly small merging scale ($\tms = 15$ GeV) to regularise the tree-level
matrix elements for this investigation. Then, for each matrix element state, we 
generate \emph{all} possible parton shower histories for a matrix element 
state, by clustering emissions. This is achieved by inverting the shower 
momentum- and flavour-mappings.

When merging matrix elements with rapidity-ordered showers, we investigate two
ways of biasing the selection of a particular history, from which to generate
the necessary Sudakov form factors:
\begin{enumerate}
\item In a ``y-blind" sample, we do not include an additional
  discriminant based on rapidity. This means that -- just like in the standard
  case -- $\rho$-ordered will be preferred over $\rho$-unordered ones.
\item In a ``y-conscious" sample, we pick histories with 
  rapidity-unordered splittings only if no rapidity-ordered histories were
  found. Adopting this strict ordering criterion, histories ordered in $\rho$ 
  \emph{and} rapidity will be chosen predominantly, and only if no such history
  exists, histories un-ordered in either $\rho$ and/or rapidity are picked.
\end{enumerate}

It should be noted that to the accuracy of the parton shower, both these 
prescriptions are equivalent, and switching the choice of histories gives a 
real estimate of the quality of the merging in conjunction the underlying 
shower. We believe that including this uncertainty gives a pessimistic 
view on how wide the range of predictions of one merged calculation can be, 
indicating that although standard by now, matrix element merging in 
\PythiaEight should be applied with care. However, with reasonable 
settings, including additional jets can improve the description of multiple 
hard jets substantially. 

\subsubsection{\texorpdfstring{\protect\Sherpa}{Sherpa}}
\label{Sec:Codes:Sherpa}

\Sherpa~\cite{Gleisberg:2003xi,Gleisberg:2008ta} is a full-fledged event 
generator capable of simulating all aspects of particle collisions as they 
occur at particle accelerators such as the \Tevatron or the \LHC.  It includes
two independent matrix element generators, \Amegic~\cite{Krauss:2001iv} and 
\Comix~\cite{Gleisberg:2008fv}, to generate cross sections and distributions for
final state multiplicities of up to six to ten particles.  In the former one, 
methods to automatically generate dipole subtraction terms in the widely used
Catani--Seymour scheme~\cite{Catani:1996vz,Catani:2002hc} have been 
incorporated~\cite{Gleisberg:2007md}; the \Sherpa package also supports the
BLHA~\cite{Binoth:2010xt} for the interface to one-loop programs such as
\BlackHat or \GoSam.  For parton showering, \Sherpa employs an algorithm based 
on Catani-Seymour subtraction kernels, proposed in~\cite{Nagy:2006kb} and 
implemented in the \Sherpa framework in~\cite{Schumann:2007mg}.  
For the hadronisation, \Sherpa uses either its native hadronisation
scheme, based on the cluster fragmentation model~\cite{Webber:1983if}
and its implementation described in~\cite{Winter:2003tt} or an
interface to \Pythia~\cite{Sjostrand:2006za} providing access to the
routines of the Lund string model~\cite{Andersson:1983ia}.  Both
have been successfully tuned to \LEP data within the \Sherpa framework, with
a similar quality in describing the data.  The hadron decays are also
fully provided in the \Sherpa framework, as well as QED final state radiation 
to both the $W$-boson and the hadron decays, simulated using the YFS 
approach~\cite{Yennie:1961ad,Schonherr:2008av}.

In this work, the most recent, publically available \Sherpa version,
\Sherpa-1.3.1, has been used in two ways of running the simulation,
namely
\begin{enumerate}
\item in the \MEPS mode:\\
  In this method, towers of LO matrix elements with increasing jet 
  multiplicity, in the case at hand $W$, $W+1$, $W+2$, \dots, $W+n_J$ jets,
  are merged in the spirit of~\cite{Catani:2001cc,Krauss:2002up} to yield
  an inclusive sample.  In fact, codes relying on such algorithms have been
  compared in a previous publication~\cite{Alwall:2007fs}, which helped to
  establish and validate the methods and their various implementations.  In 
  contrast to the original implementation in \Sherpa~\cite{Schalicke:2005nv},
  which used analytical forms of Sudakov form factors etc., the current version
  of the method~\cite{Hoeche:2009rj} directly uses the parton shower for
  Sudakov rejections etc.\ and is thus closer in spirit to the variant
  presented in~\cite{Lonnblad:2001iq,Lavesson:2005xu} for multijet merging.  
\item in the \MENLOPS mode:\\  
  This method can be understood as the combination of a matching of the parton 
  shower to a NLO matrix element and a merging of additional towers of LO
  matrix elements with even higher jet multiplicities.  Thus, in the case at 
  hand, inclusive $W$ production calculated at NLO accuracy is merged, as above, 
  with LO matrix elements for $W+1$, $W+2$, \dots, $W+n_J$ jets. This method has
  been pioneered in~\cite{Hamilton:2010wh,Hoche:2010kg} where the
  implementation employed within \Sherpa has been detailed in the
  second reference.
\end{enumerate}
The respective settings and relevant details for both simulation modes
are described below.

\paragraph{\texorpdfstring{\protect\Sherpa}{Sherpa} in 
	       \texorpdfstring{\protect\MEPS}{MEPS} mode}
\label{Sec:Codes:Sherpa:MEPS}

In the \MEPS mode \Sherpa was run with up to $n_J=6$ jets in the
matrix element evaluation 
including all possible massless (anti-)quark and gluon initial and
final states. All matrix elements were 
generated using \Comix. The \MEPS-separation parameter was set to 
$Q_\text{cut}=20\text{ GeV}$, for its precise definition see 
\cite{Hoeche:2009rj}. The scales are chosen as 
\begin{equation}
  \alphas^{k+n}(\mu_\text{eff})\,=\,
  \alphas^k(\mu)\cdot
  \alphas(p_{\perp,1})\cdot\ldots\cdot\alphas(p_{\perp,n})\;,
\end{equation}
wherein the relative transverse momenta $p_{\perp,i}$ are the nodal
values of the final state partons of the $W+n$ parton matrix element
as obtained from recombining it using the inverted splitting
probablities given by the parton shower. The core scale $\mu$ is then chosen 
as the partonic centre-of-mass energy of the reconstructed core process, i.e.\ 
$\mu^2=\hat{s}_{2\to 2}$ where $k=0$ in the process at hand. In all stricly 
perturbative setups a parton shower cutoff of $t_0=(0.7\text{ GeV})^2$ has been 
used.

The parton shower cutoff and all fragmentation parameters of both the internal 
cluster hadronisation and the interfaced Lund string fragmentation models have 
been tuned to \LEP data and give a similarly good description.  
Similarly, the parameters of \Sherpa's MPI model have been tuned to \Tevatron 
and \LHC data using the CT10~\cite{Lai:2010vv} parton density parametrisation.  These parameters 
are given in \AppRef{App:Settings:Sherpa}.


\paragraph{\texorpdfstring{\protect\Sherpa}{Sherpa} in 
	       \texorpdfstring{\protect\MENLOPS}{MENLOPS} mode}
\label{Sec:Codes:Sherpa:MENLOPS}

In the \MENLOPS mode \Sherpa is run with essentially the same parameters as 
in the \MEPS mode, described in the previous subsection. Hence, $n_J=6$
and $Q_\text{cut}=20\text{ GeV}$. To be 
able to describe the inclusive $W$ production process at NLO accuracy, \Amegic 
was used for all parts of the NLO $W$ production matrix elements (supplemented 
with a hardcoded one-loop matrix element from the internal library) and the LO 
$W+1$ parton matrix element. Consecutively, the scales were chosen as above 
with $k=0$ for all tree-level parts and $k=1$ for the real and virtual 
corrections entering the next-to-leading order correction of the core process. 
All non-perturbative parameters remain unchanged wrt.\ the \MEPS mode.

\paragraph{Estimate of uncertainties}
\label{Sec:Codes:Sherpa:Uncertainties}
In order to estimate the uncertainites of the \Sherpa predictions, the
following procedures have been applied:
\begin{itemize}
\item[(A)] PDF uncertainties:\\
  Unlike in the PDF4LHC presciption~\cite{Botje:2011sn}, here only the
  central predictions of the three NLO PDFs, CT10~\cite{Lai:2010vv}, 
  MSTW2008~\cite{Martin:2009iq} and NNPDF2.1~\cite{Ball:2011mu}
  are compared to estimate the PDF uncertainties. The different
  parametrisations of PDFs as well as their corresponding value of
  \alphas, both its value at $M_Z$ and its running, enter in the
  calculation of the matrix elements, the parton shower and the
  underlying event.
\item[(B)] Scale uncertainties:\\
  In a global manner, all scales, renormalisation and factorisation
  scales are simultaneously modified by the canonical multiplication
  with 2 and 1/2.  This, however, is not only applied to the
  evaluation of the matrix elements but also to that of the parton
  shower, the hadronisation, the underlying event simulation and the
  hadron decays. Regarding the matrix-element evaluation, the \MEPS
  default scale choice forms the starting point for the scale
  variations to be executed.
\item[(C)] Hadronisation uncertainty:\\
  Here the intrinsic modeling uncertainties are evaluated by changing
  the hadronisation model operating on \Sherpa's parton shower final
  states, namely switching from \Sherpa's default cluster hadronisation
  to \Pythia's string fragmentation. For both schemes, an independently
  tuned set of parameters has been employed to perform the
  parton-to-hadron transition.
\item[(D)] Underlying event uncertainty:\\
  To this end the tune of the underlying event based on using the CT10
  PDF has been modified such that the plateau of the number of charged
  particles and sum of transverse momenta in the transverse region are
  increased or decreased by $10\%$. This change in the amount of MPI
  activity is accomplished by varying the $\sigma_\text{ND}$
  correction factor (SIGMA\-\_ND\-\_FACTOR) by $-0.04$ or $+0.05$,
  respectively.
\end{itemize}


\subsection{Results}
\label{Sec:Results}
In this section we compile results for the individual codes for a
number of representative observables at the different levels of the
simulation.  It should be noted, though, that in all results presented
in this section PDF uncertainties have been estimated by typically varying
only over a few different sets rather than employing the full procedure
as suggested by the PDF4LHC accord~\cite{Botje:2011sn}.

\subsubsection{\texorpdfstring{\protect\BlackHat}{BlackHat} + \texorpdfstring{\protect\Sherpa}{Sherpa}}
The following results have been obtained with
\BlackHat{}+\Sherpa. Uncertainties due to the factorisation/renormalisation
scale variation and the that due to the PDF uncertainties are shown.  The 
yellow band corresponds to the addition in quardature of these two 
uncertainties and the statistical estimation on the integration 
error. All observables are defined using the ATLAS cuts, cf.\ 
\AppRef{App:Observables}.

\begin{figure}[ht]
  \includegraphics[width=.48\textwidth]{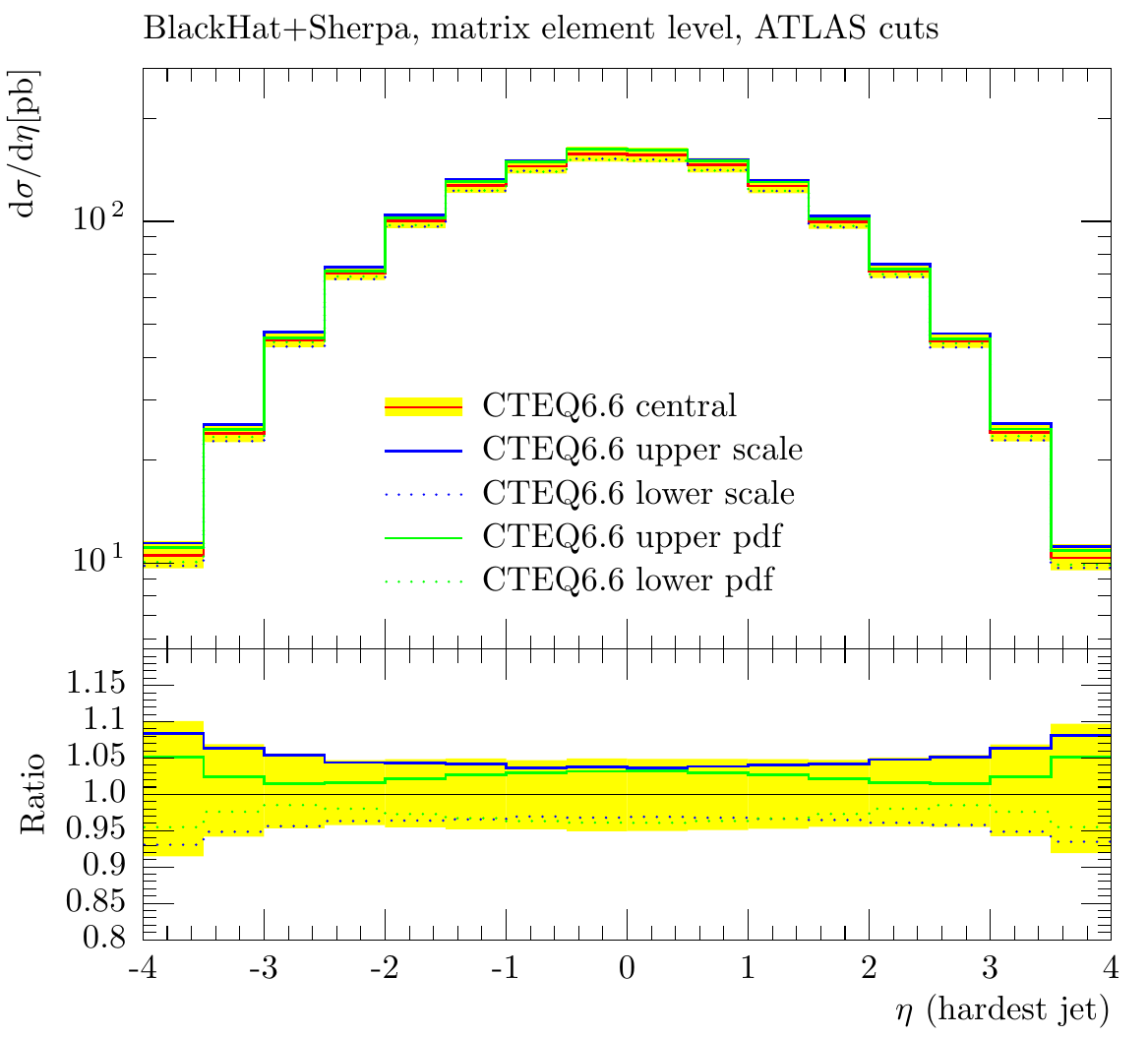}\hfill
  \includegraphics[width=.48\textwidth]{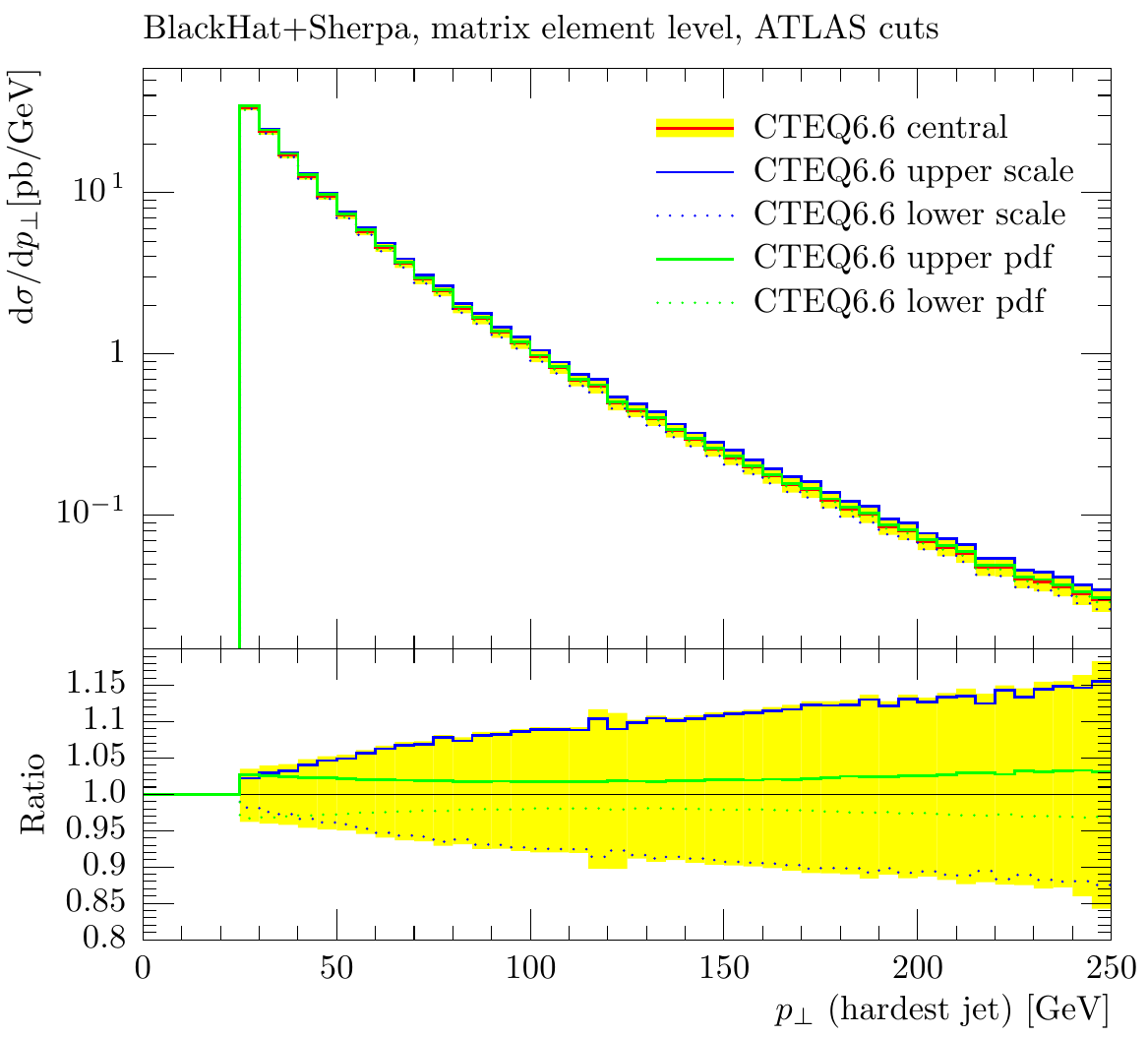}\\
  \includegraphics[width=.48\textwidth]{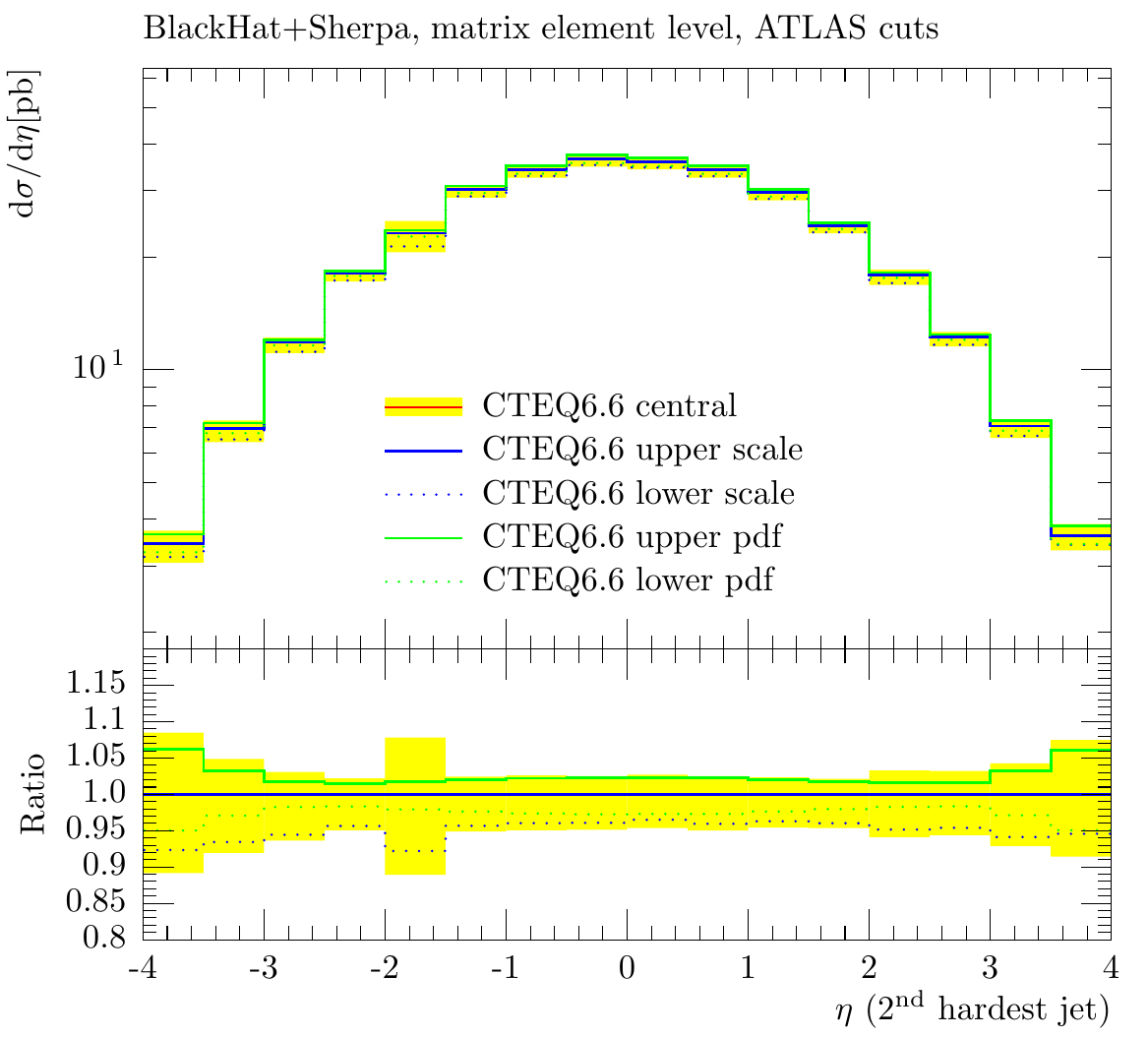}\hfill
  \includegraphics[width=.48\textwidth]{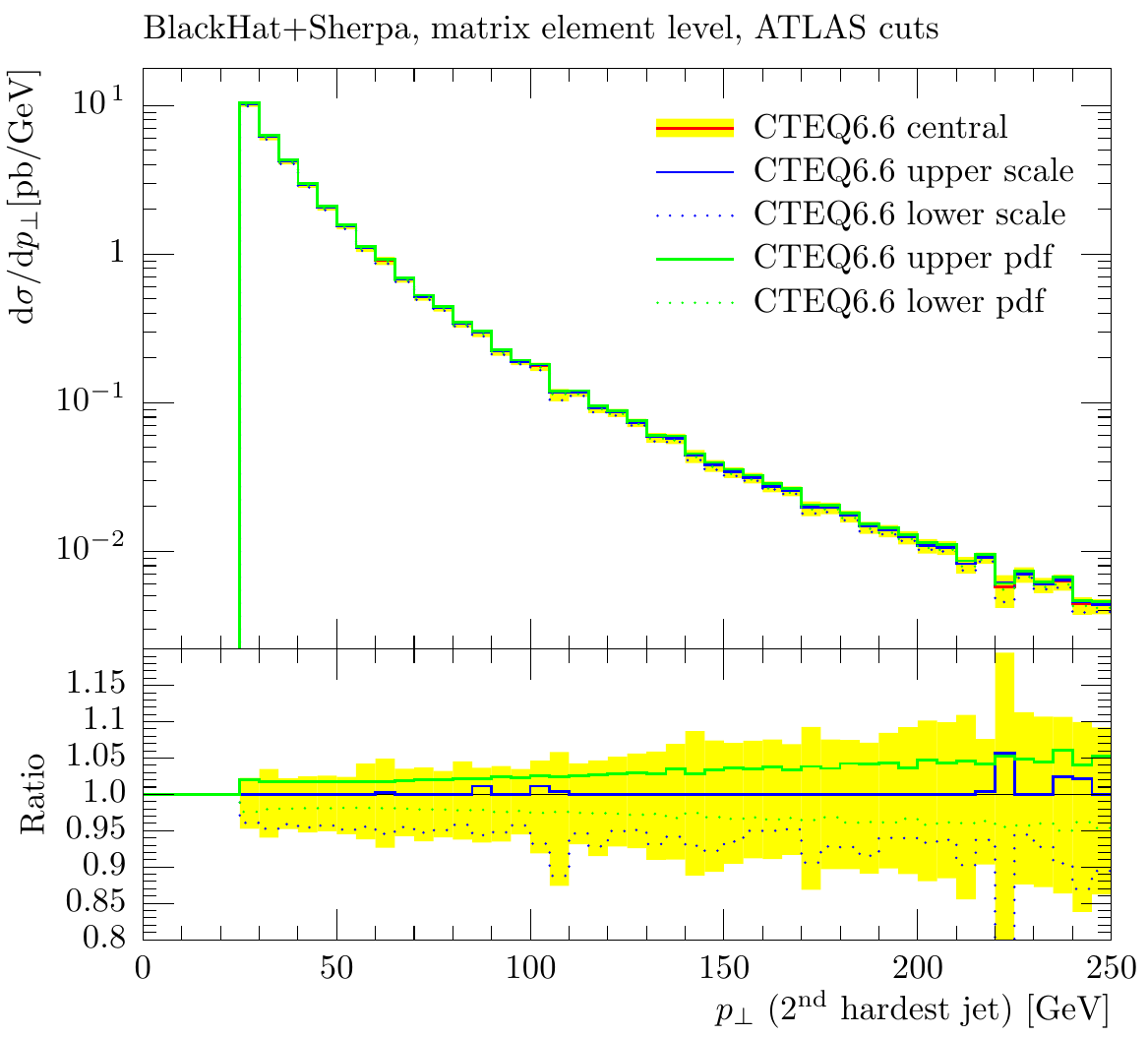}\\
  \includegraphics[width=.48\textwidth]{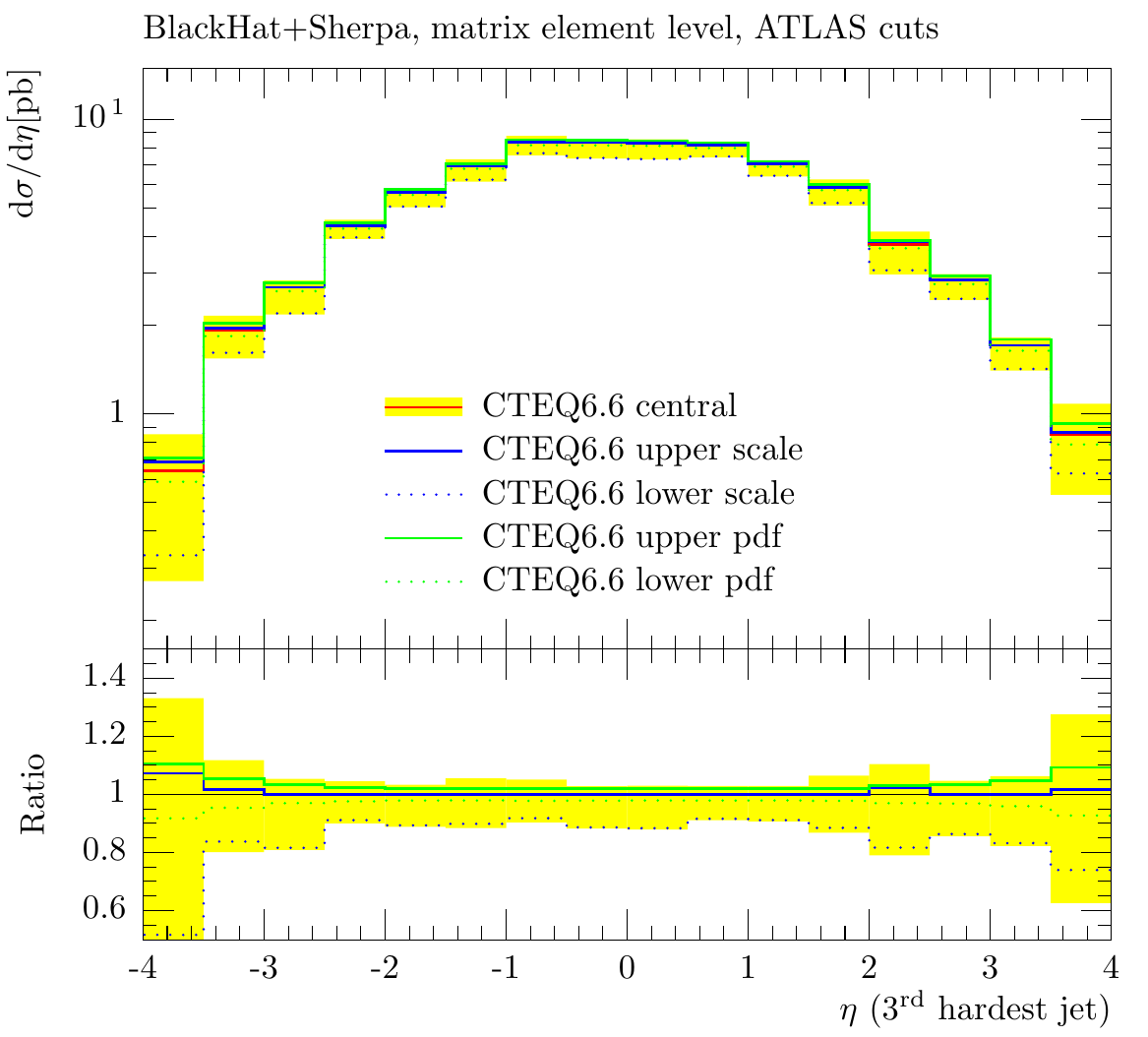}\hfill
  \includegraphics[width=.48\textwidth]{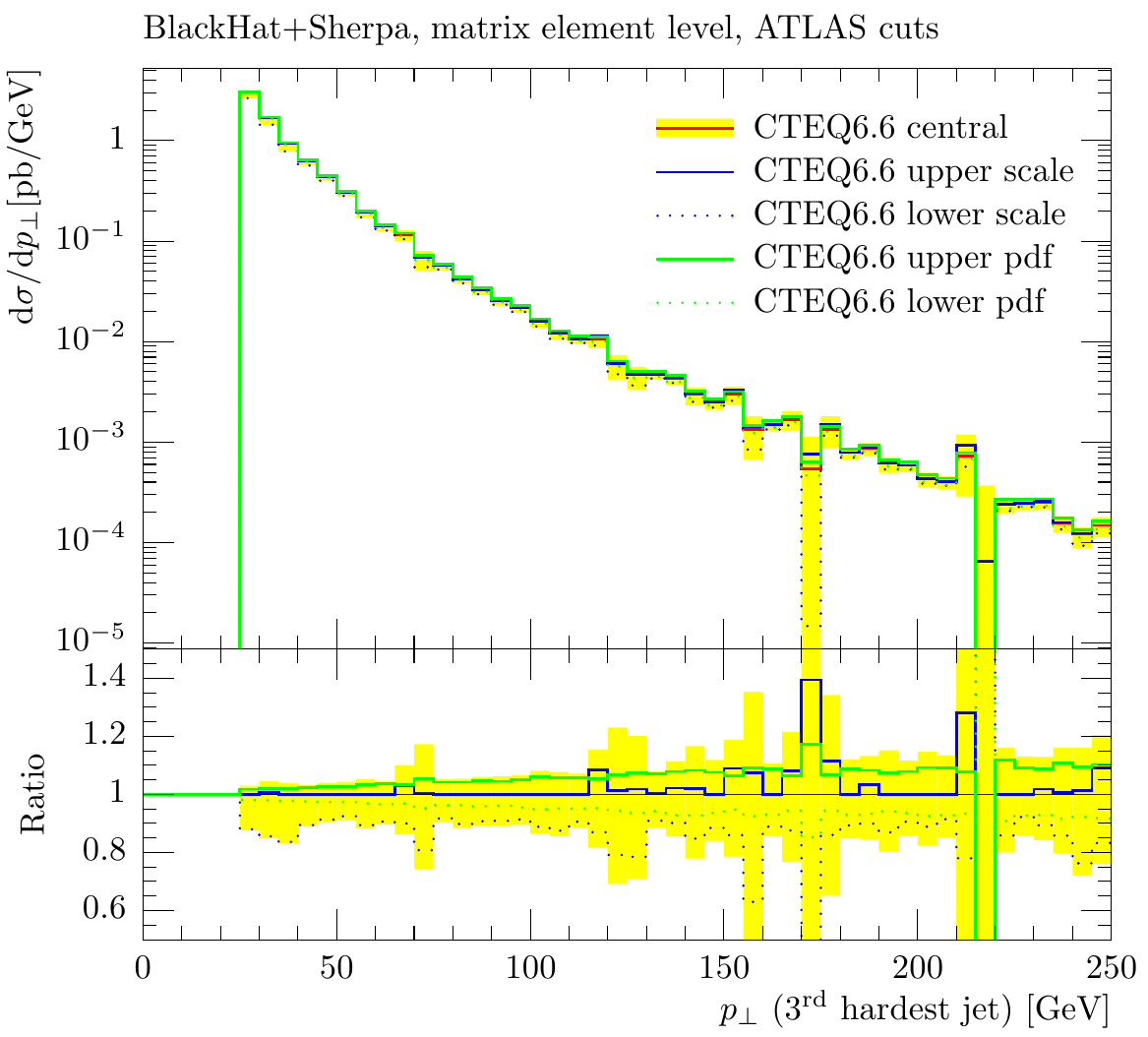}
  \caption{Pseudo-rapidity and transverse momentum distributions for the first 
	   jet in inclusive $W+1$ jet production (upper panel), for the second 
	   jet in inclusive $W+2$ jet production (central panel), anf for the 
	   third jet in inclusive $W+3$ jet production (lower panel).}
\end{figure}

\begin{figure}[t]
\includegraphics[width=.48\textwidth]{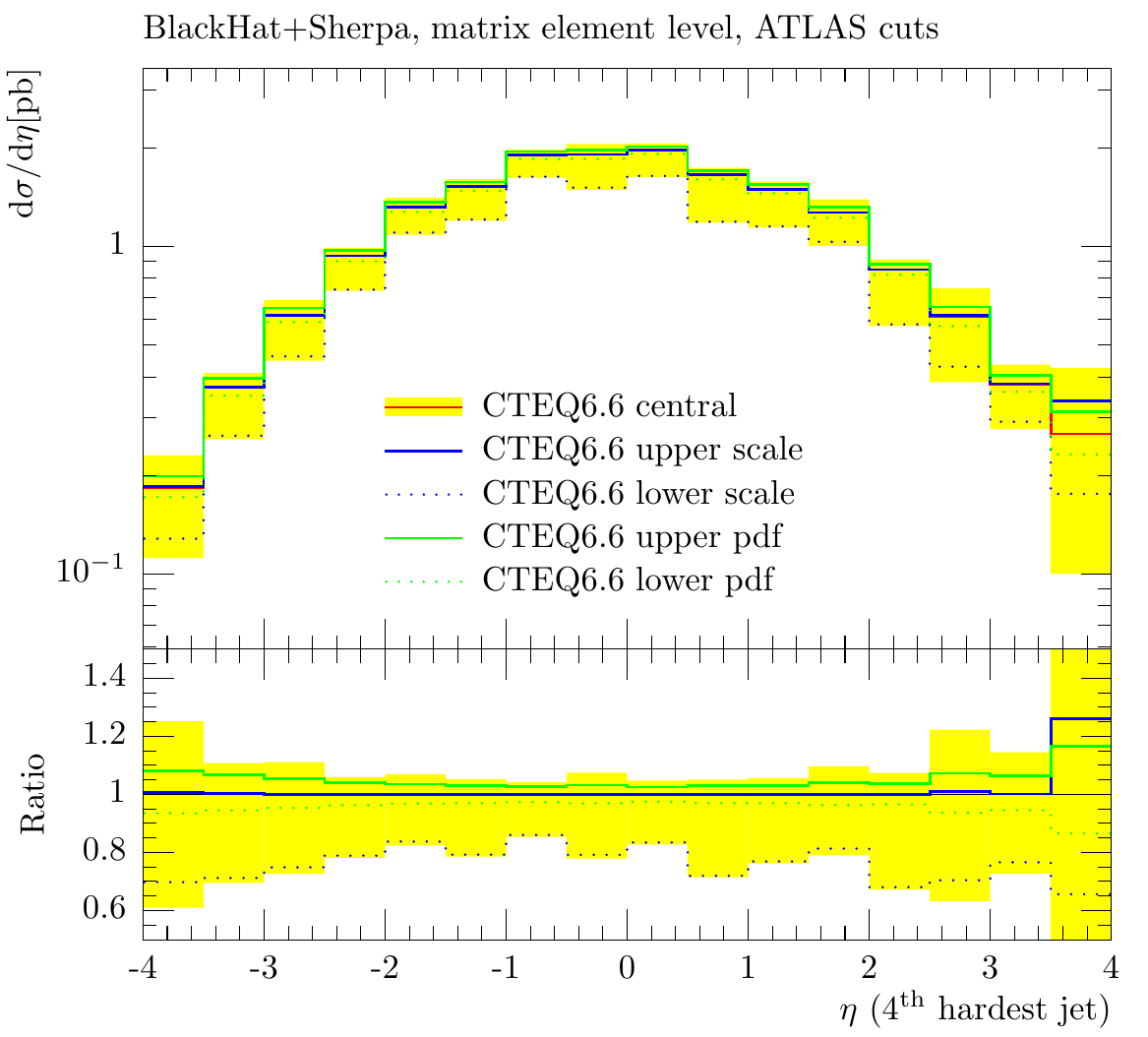}\hfill
\includegraphics[width=.48\textwidth]{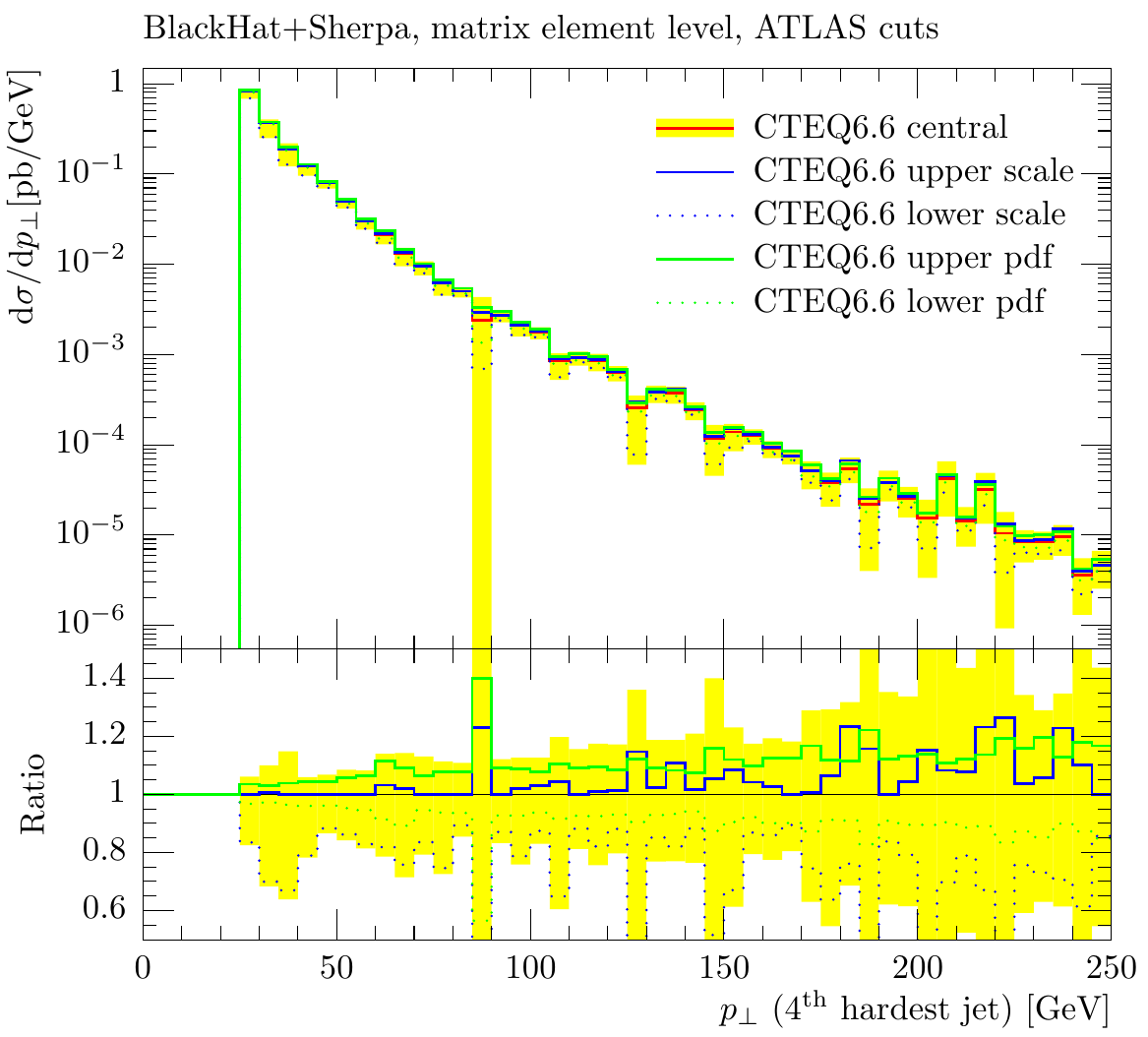}
\caption{Pseudo-rapidity and transverse momentum distributions for the fourth jet in $W+4$ jet production.}
\end{figure}

\begin{figure}[t!]
\includegraphics[width=.48\textwidth]{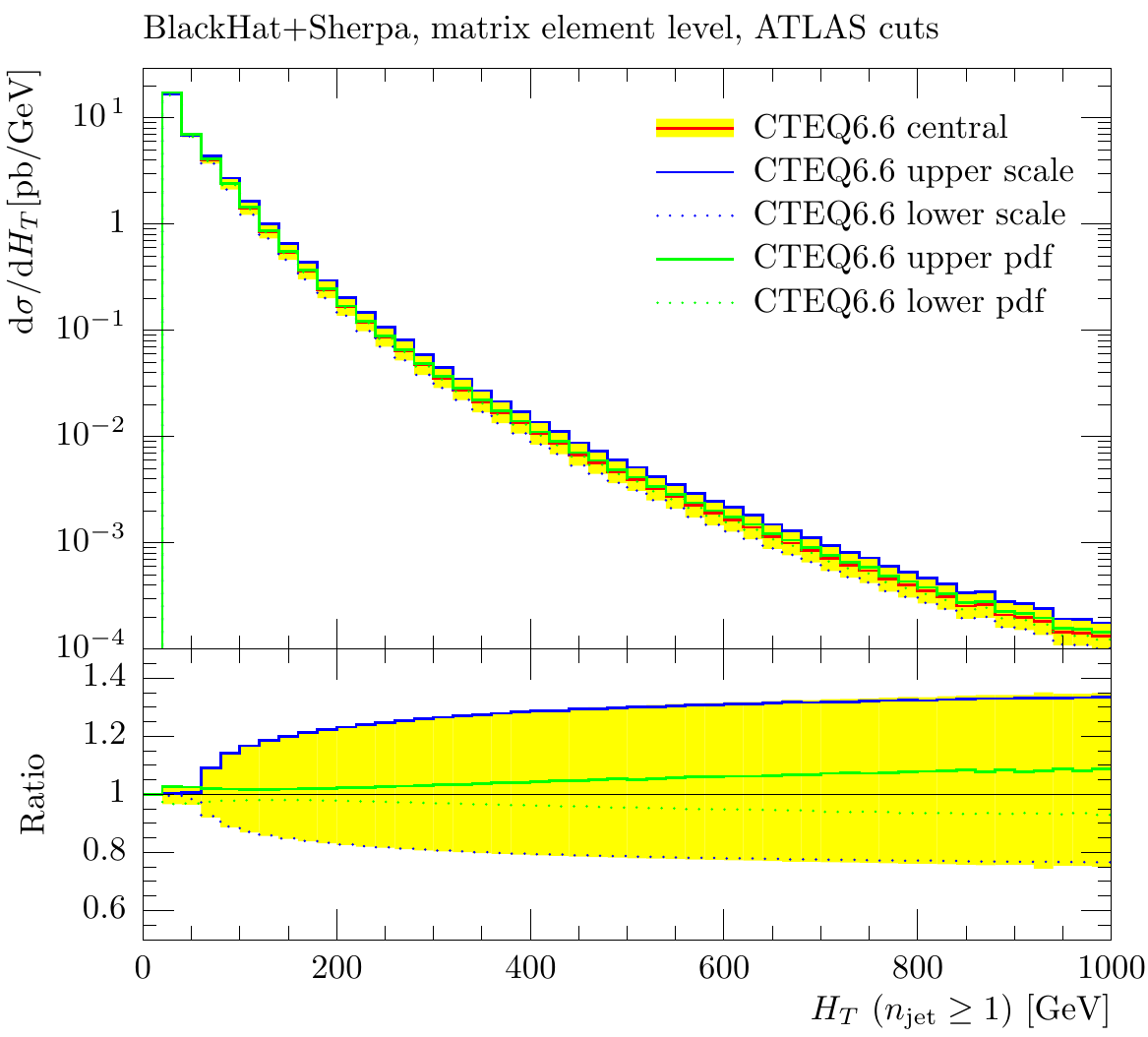}\hfill
\includegraphics[width=.48\textwidth]{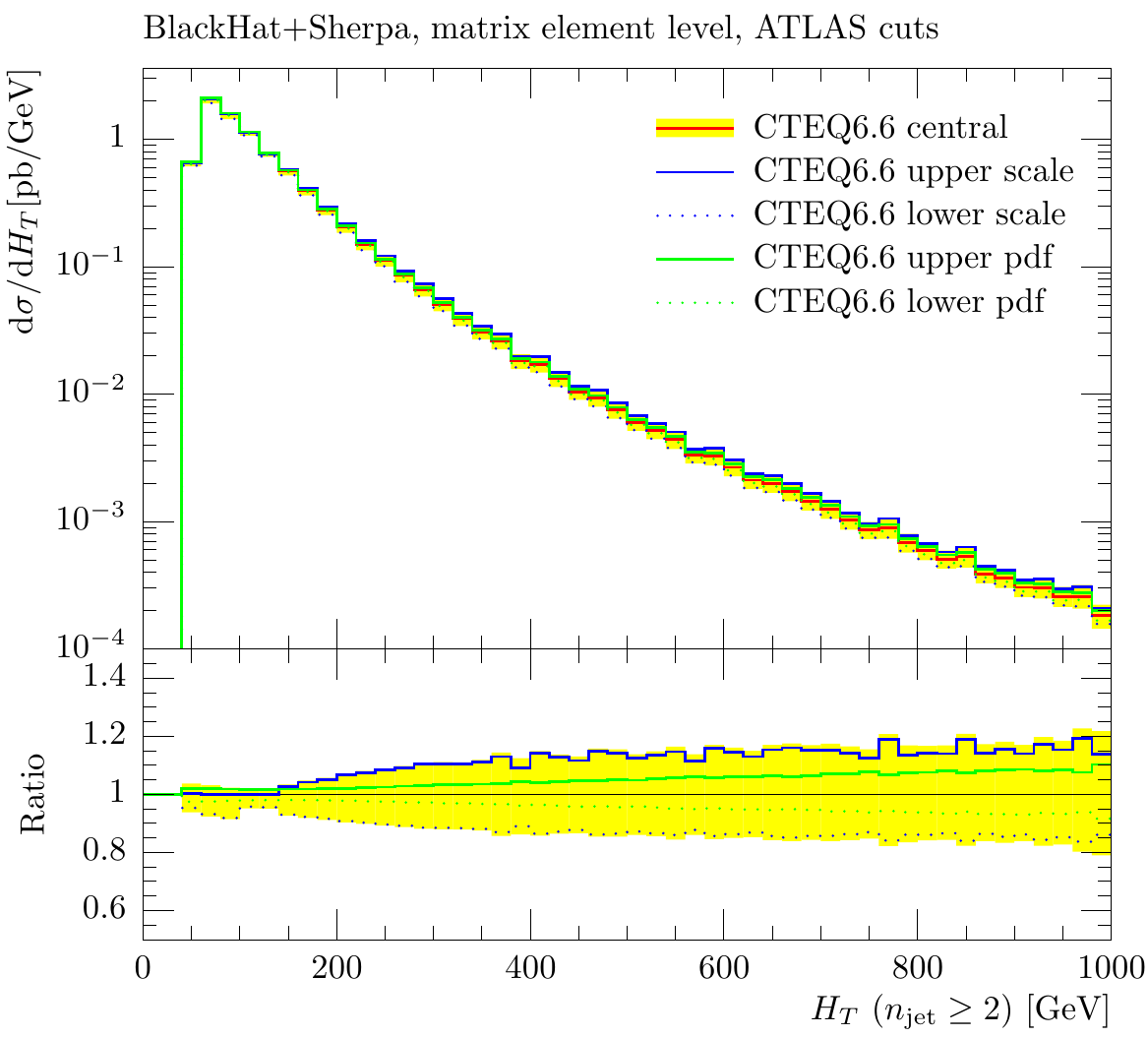}\\
\includegraphics[width=.48\textwidth]{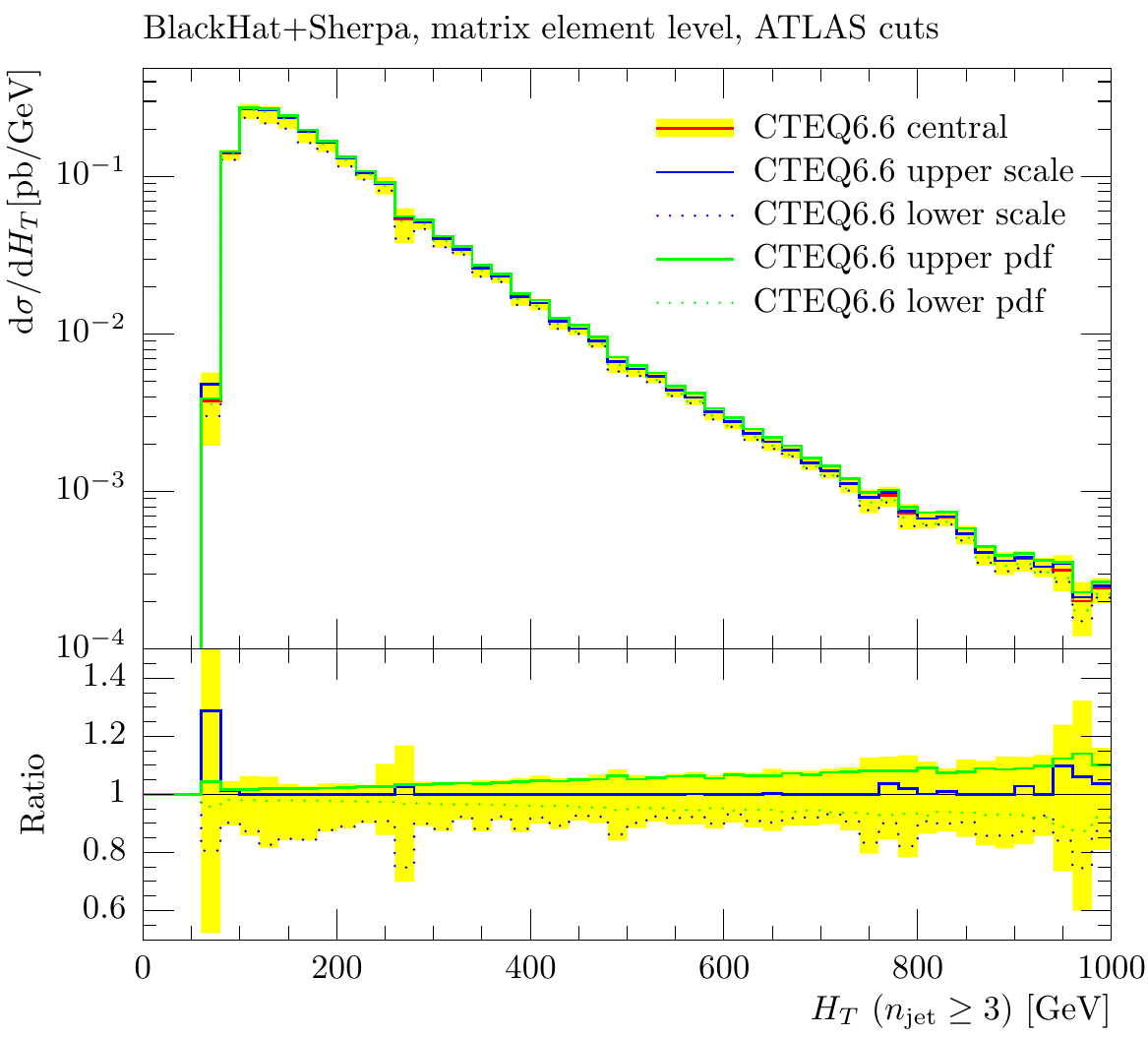}\hfill
\includegraphics[width=.48\textwidth]{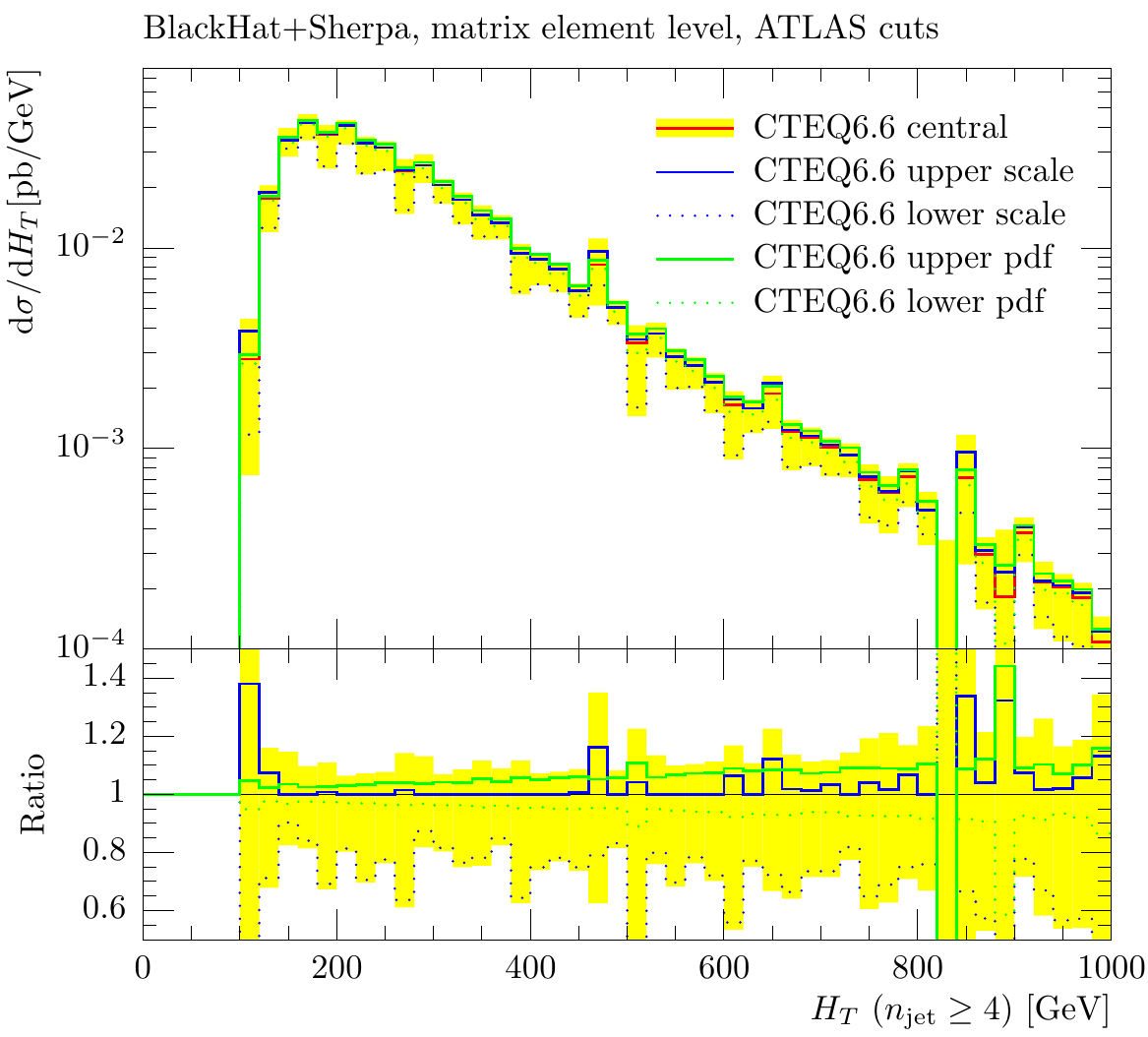}
\caption{HT distributions for event with at least one (top left), two (top right), three (bottom left) or four (bottom right) jets.}
\end{figure}

\FigRef{fig:BHS:njets} displays the inclusive cross section for a $W$ boson in 
association with $n$ jets, where $n$=1,2,3,4. A NLO computation of $W$+4 jets 
also provides a leading order calculation of the $W$+5 jets rate, but since it 
is not at NLO accuracy we refrain here from including it.  


In all the plots presented in this section the uncertainties are dominated 
by the uncertainty arising from the scale variation (it is not the case when the central scale of the process is chosen close to a local maximum, in which case the upper boundary of the scale variation is very close or identical with the central value, as can be seen from the plots corresponding to $W$+3,4 jets).  This is partially
due to the fact that for the assessment of the PDF uncertainty only error
sets have been employed that are closely related to the central set.  In 
addition, the functional form of the scale definition as given by the 
kineamtics of the final state has not been changed, but rather the emerging
scales $\mu_F$ and $\mu_R$ have been multiplied in parallel by factors of 2
and 1/2.

\FloatBarrier
\begin{figure}[t!]
\centering \includegraphics[width=.48\textwidth]{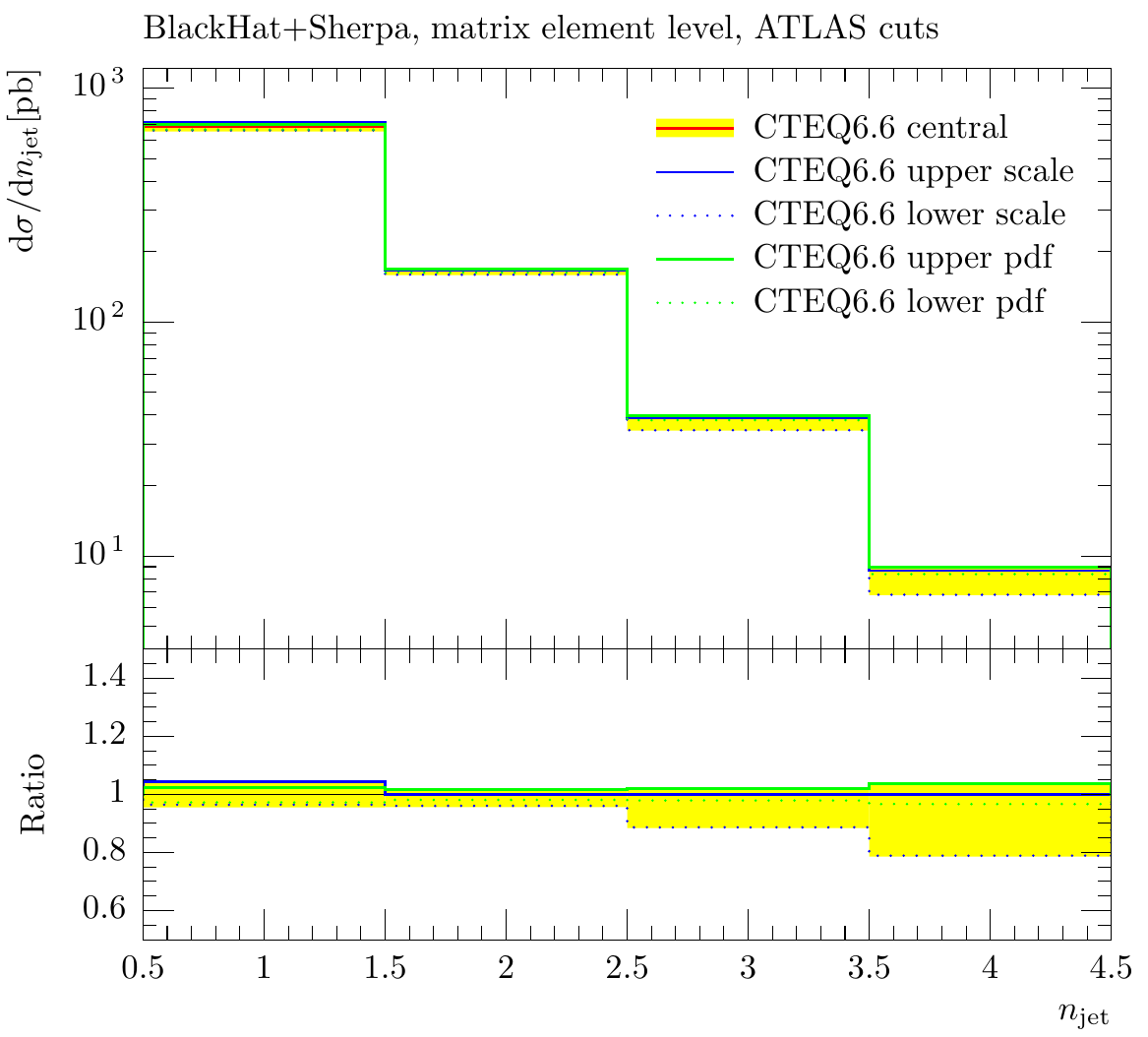}
\caption{Inclusive cross section for $W+n$ jet production.}\label{fig:BHS:njets}
\end{figure}

\vspace*{5mm}

\subsubsection{\texorpdfstring{\protect\GoSam}{GoSam} + \texorpdfstring{\protect\Sherpa}{Sherpa}}
The setup described in the previous section for the analysis using 
\GoSam{}+\Sherpa gives the following theoretical uncertainties. The 
plots show that in general the scale uncertainties are bigger then the 
PDF uncertainties and that the renormalisation scale dependence is usually 
bigger then the dependence on the factorisation scale. To illustrate the 
decrease in the scale uncertainty given by the NLO calculation we also 
include the distributions for the pseudo-rapidity and transverse momentum 
of the second hardest jet, which have only tree-level accuracy. All 
observables shown are defined using the ATLAS cuts, cf.\ 
\AppRef{App:Observables}.  Note that errors in the 2-jet configuration are
increased w.r.t.\ those provided by \BlackHat{}+\Sherpa, since here only
$W+1$ jet configurations are dealt with at NLO, and the 2-jet configurations 
therefore are descibed at LO only.

\begin{figure}[b!]
\includegraphics[width=.48\textwidth]{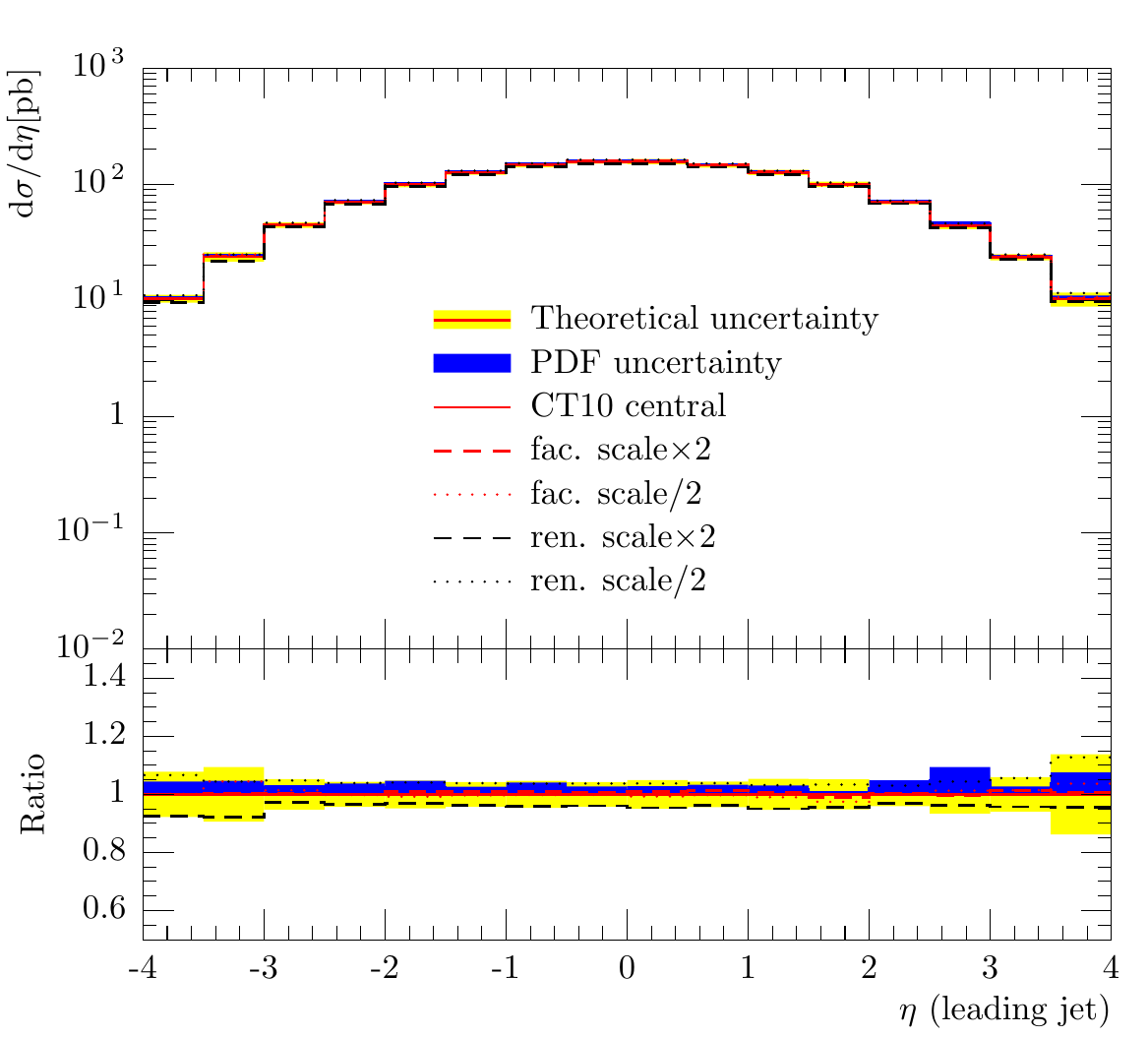}\hfill
\includegraphics[width=.48\textwidth]{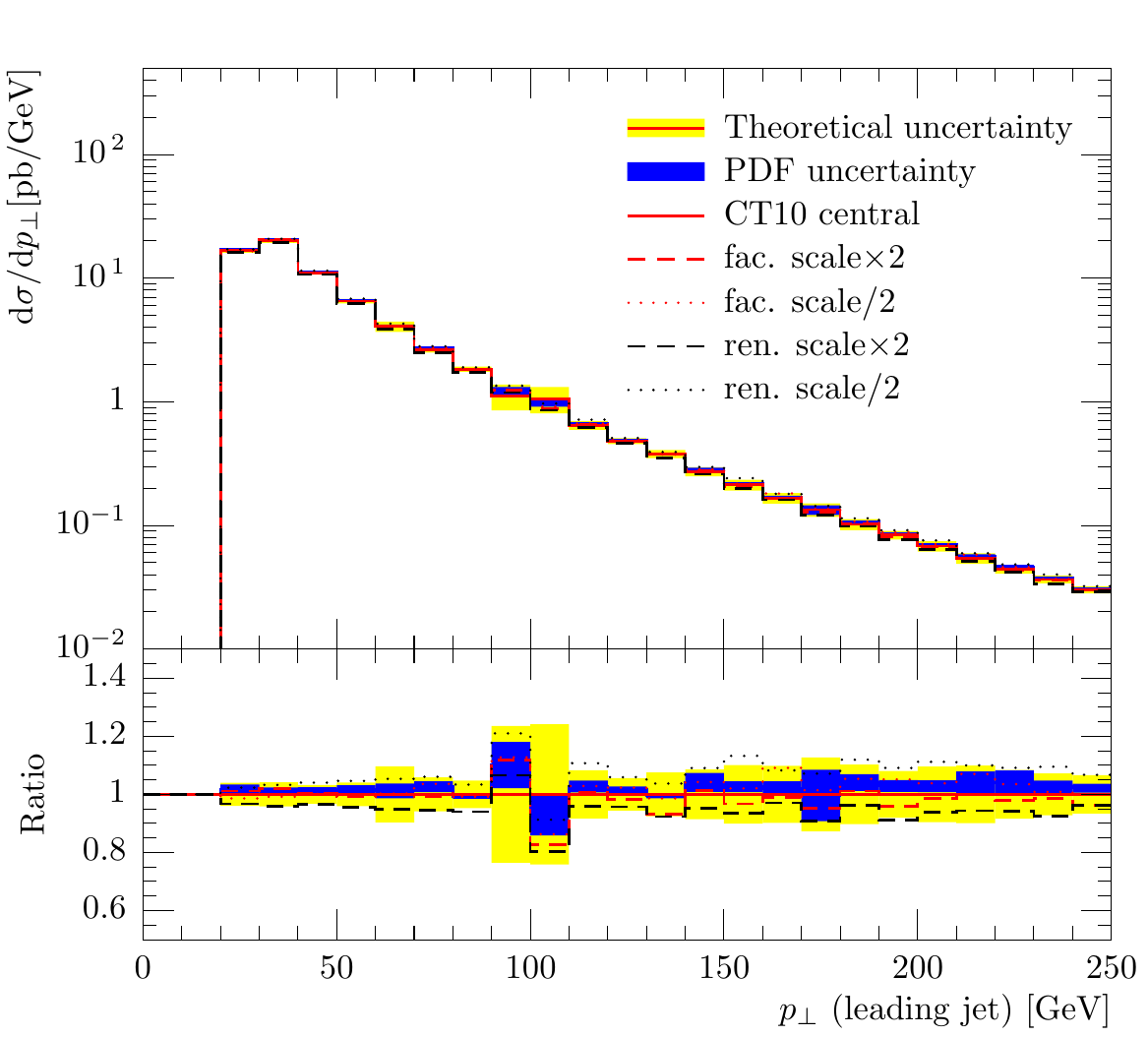}
\caption{Pseudo-rapidity and transverse momentum distributions for the hardest jet.}
\end{figure}

\begin{figure}[t!]
\includegraphics[width=.48\textwidth]{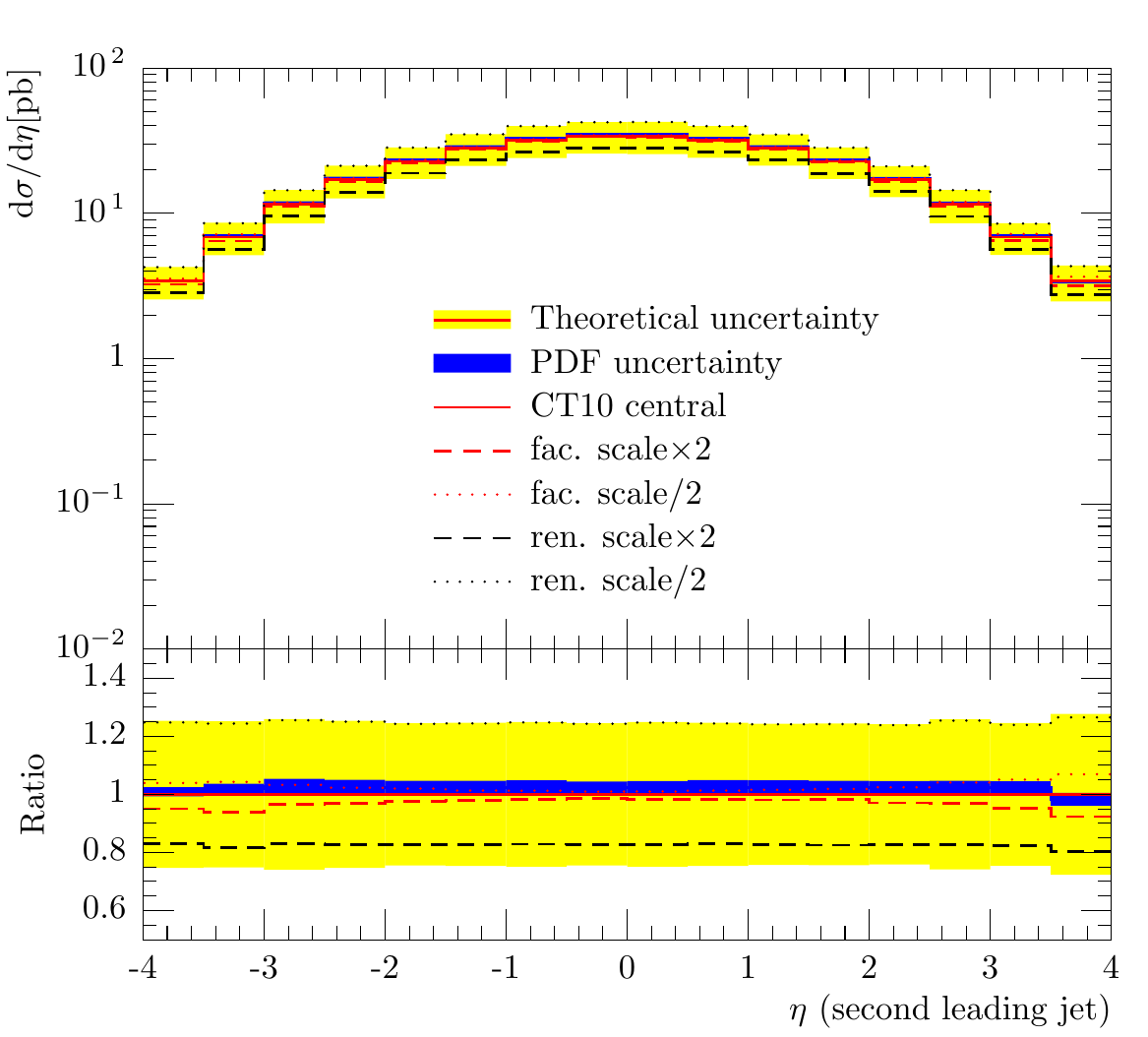}\hfill
\includegraphics[width=.48\textwidth]{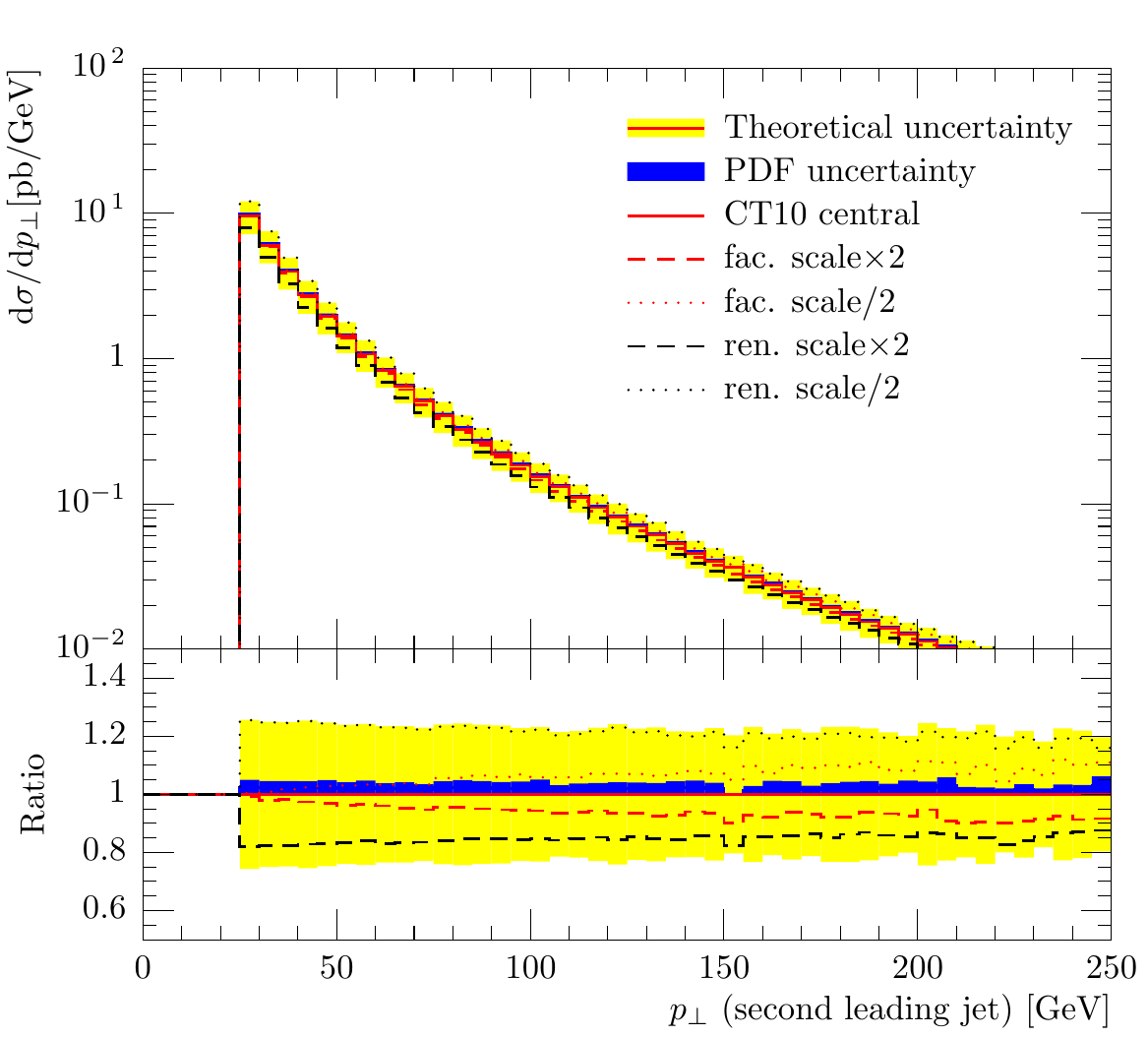}
\caption{Pseudo-rapidity and transverse momentum distributions for the second hardest jet. This distribution have formally leading order accuracy and have therefore a much larger scale dependence than
the same distribution for the hardest jet, for which a genuine NLO prediction is available.}
\end{figure}

%

\begin{figure}[t!]
\centering
\includegraphics[width=.48\textwidth]{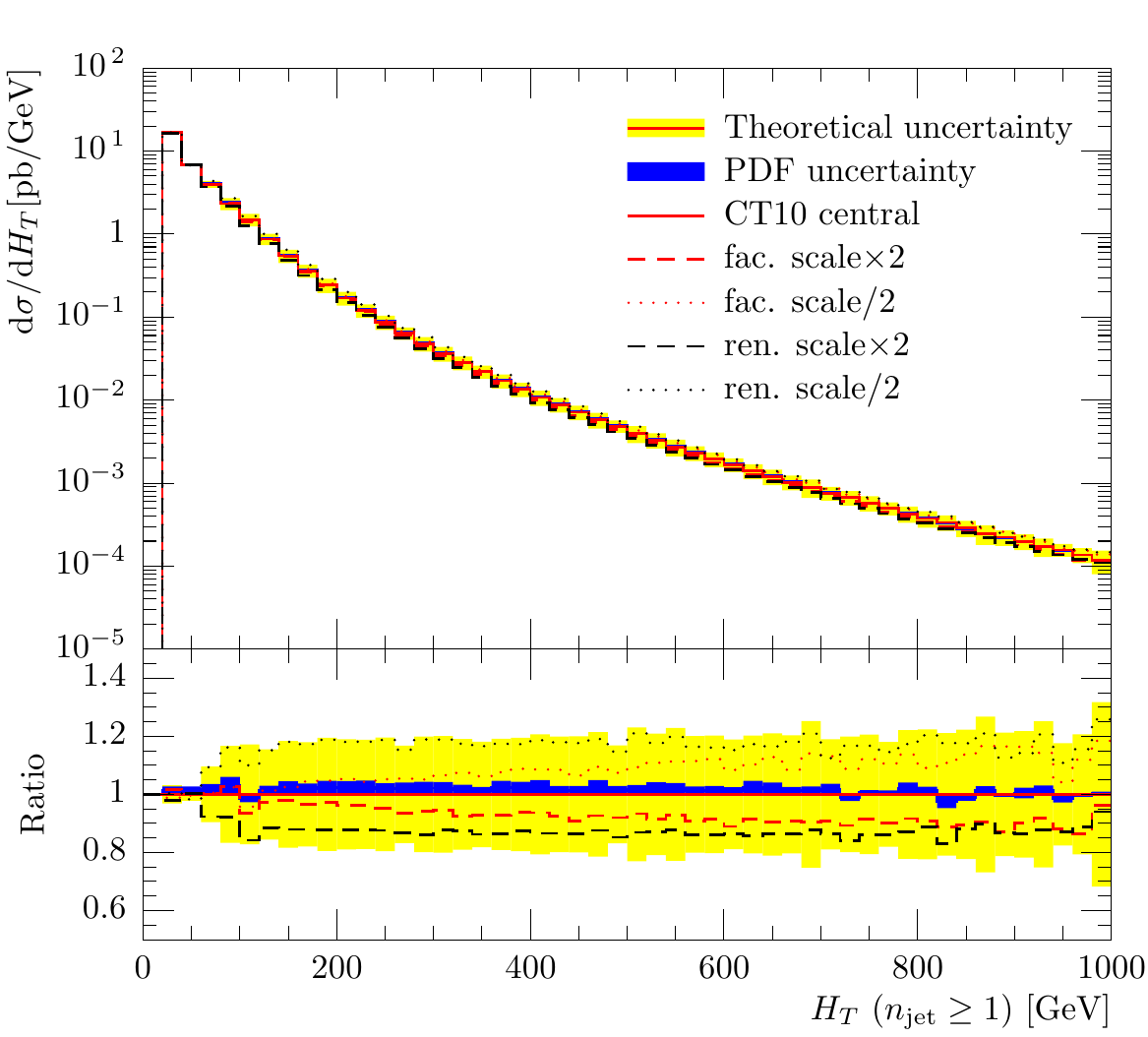}\hfill
\includegraphics[width=.48\textwidth]{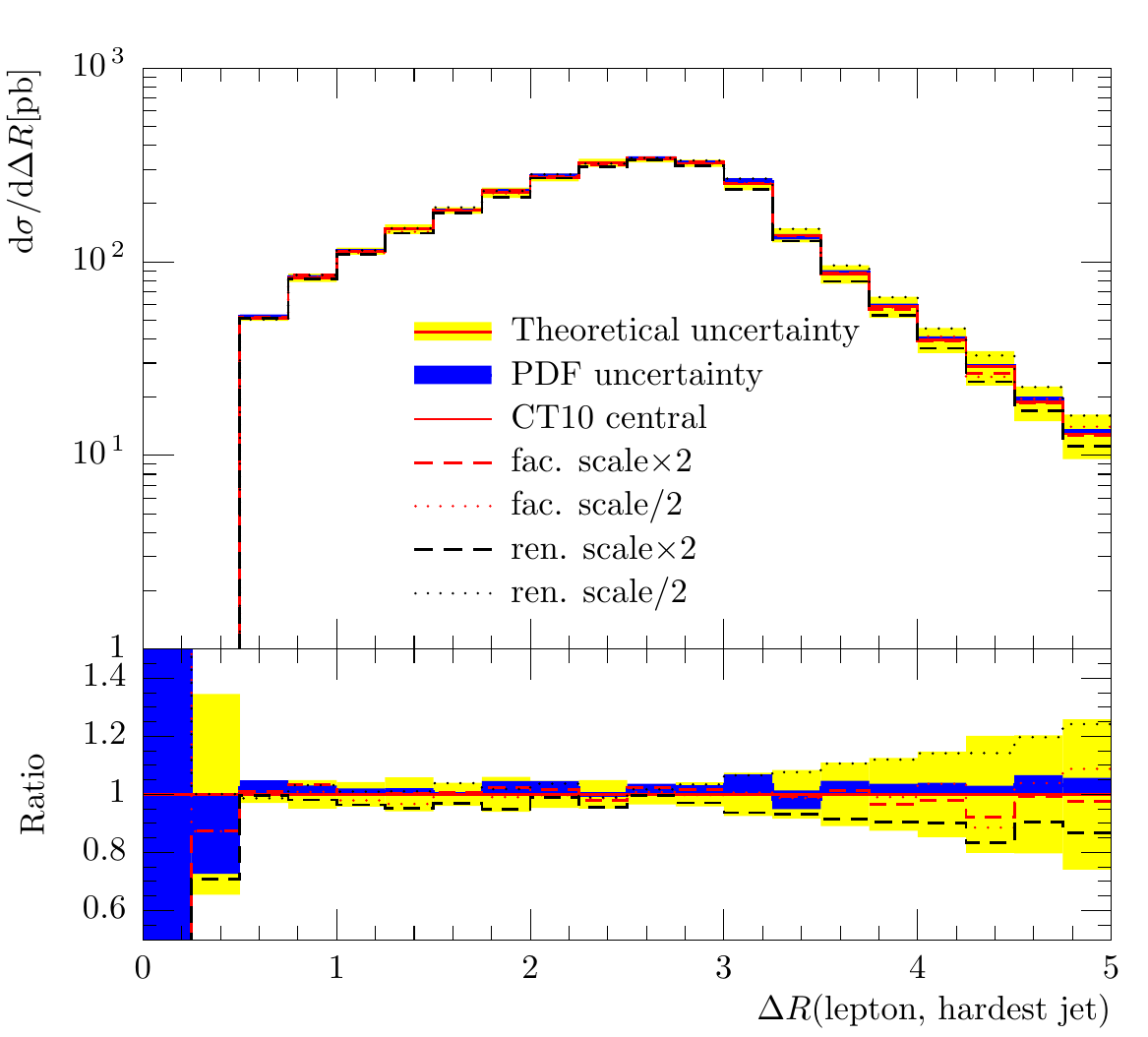}
\caption{HT distributions (left) and $\Delta R$ between lepton and hardest jet 
	 (right) for events with at least one jet.}
\end{figure}

\FloatBarrier

\subsubsection{\texorpdfstring{\protect\HEJ}{HEJ}}
\label{Sec:Results:HEJ}
\begin{figure}[p]
 \vspace*{-5mm}
 \includegraphics[width=.48\textwidth]{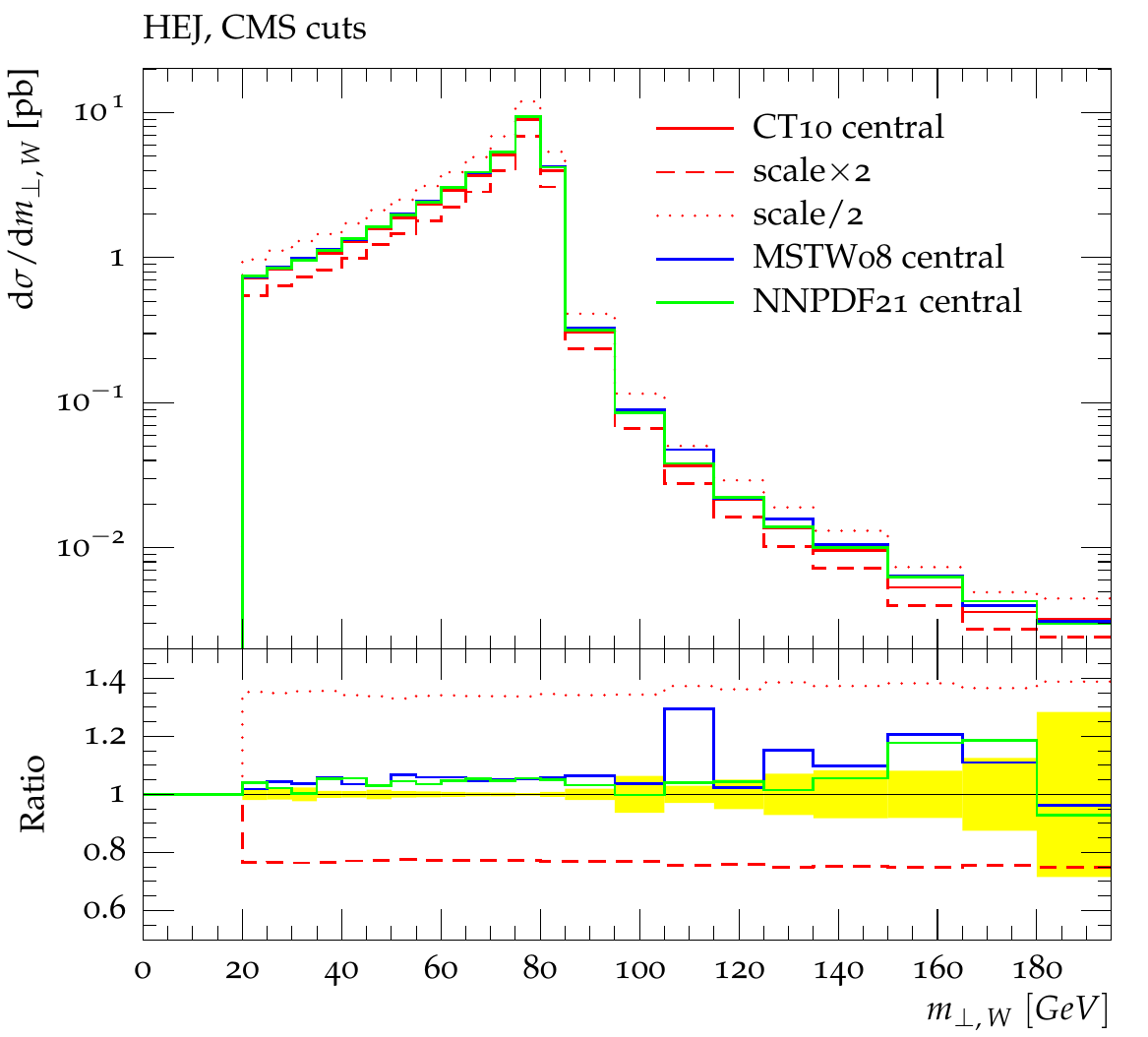}\hfill
 \includegraphics[width=.48\textwidth]{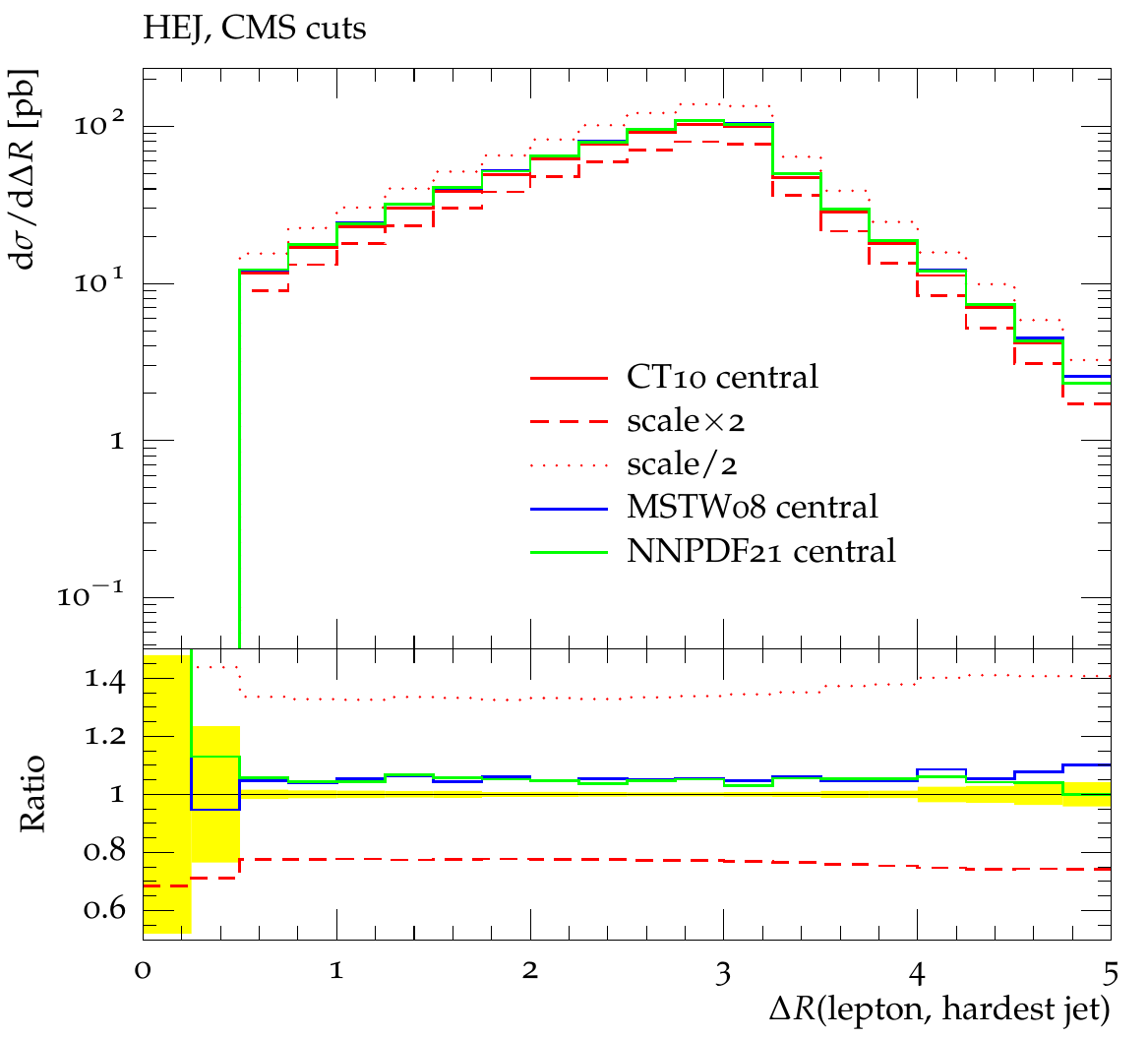}\\
 \includegraphics[width=.48\textwidth]{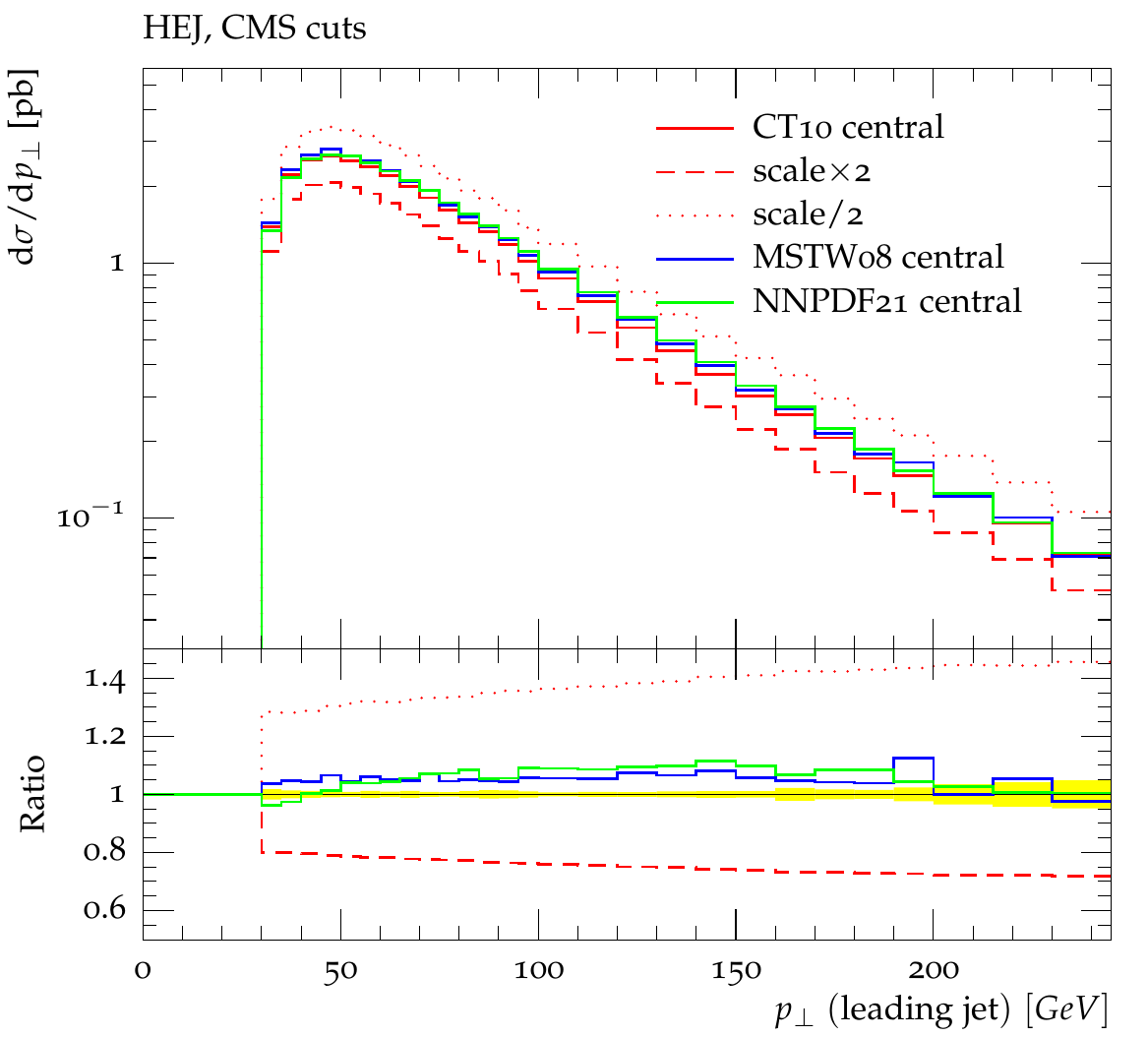}\hfill
 \includegraphics[width=.48\textwidth]{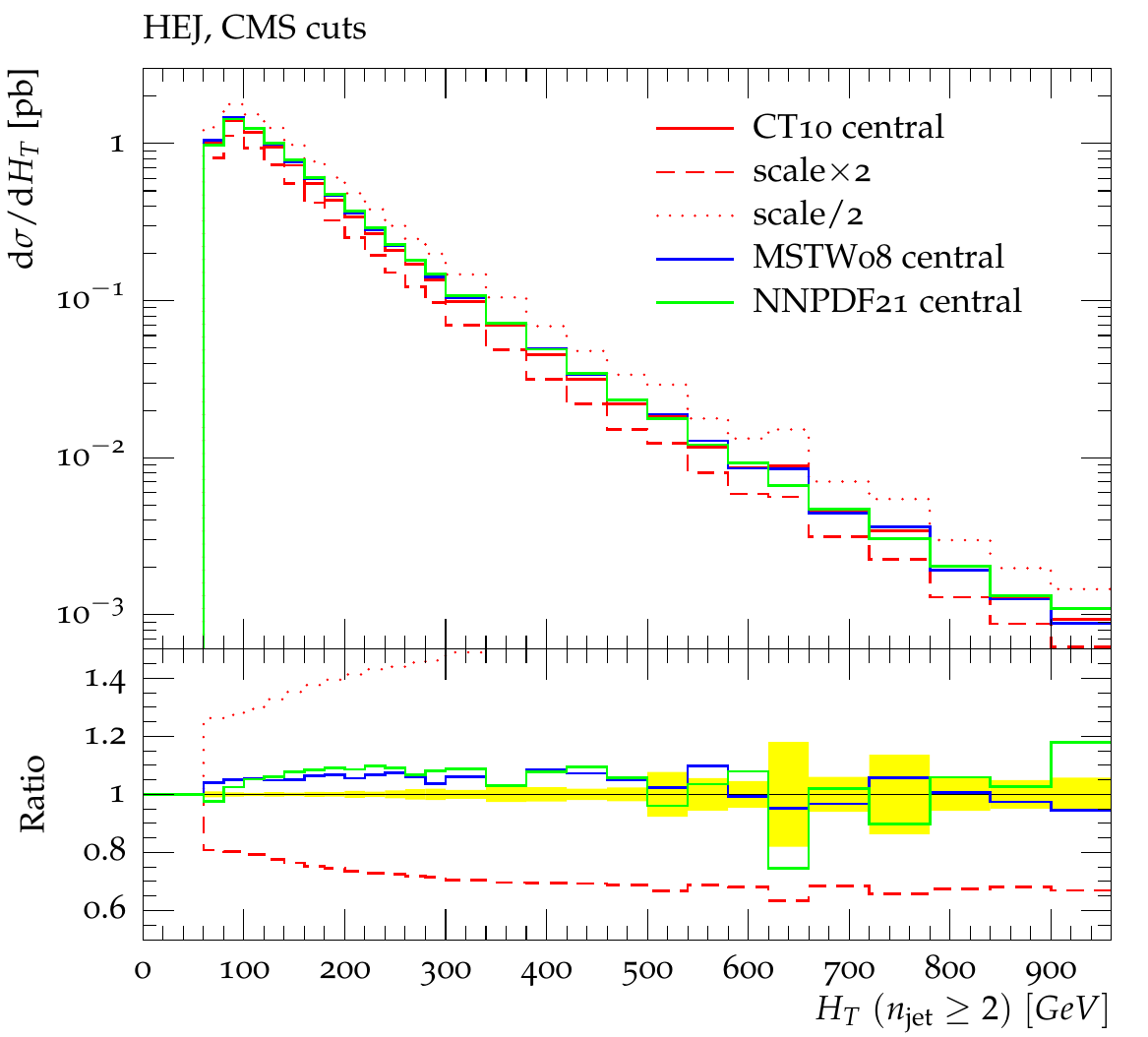}
 \caption{The \HEJ prediction for the distribution of the transverse mass of 
	  the $W$ boson (top left) and for the angle between the hardest jet 
	  and the charged lepton from the decay of the $W$ boson (top right), 
	  the transverse momentum of the hardest jet (bottom left) and for 
	  the $H_T$ distribution (bottom right) in events where a $W$ boson 
	  was produced in association with at least two jets.}
\end{figure}
\begin{figure}[p]
 \vspace*{-2mm}
 \centering
 \includegraphics[width=.48\textwidth]{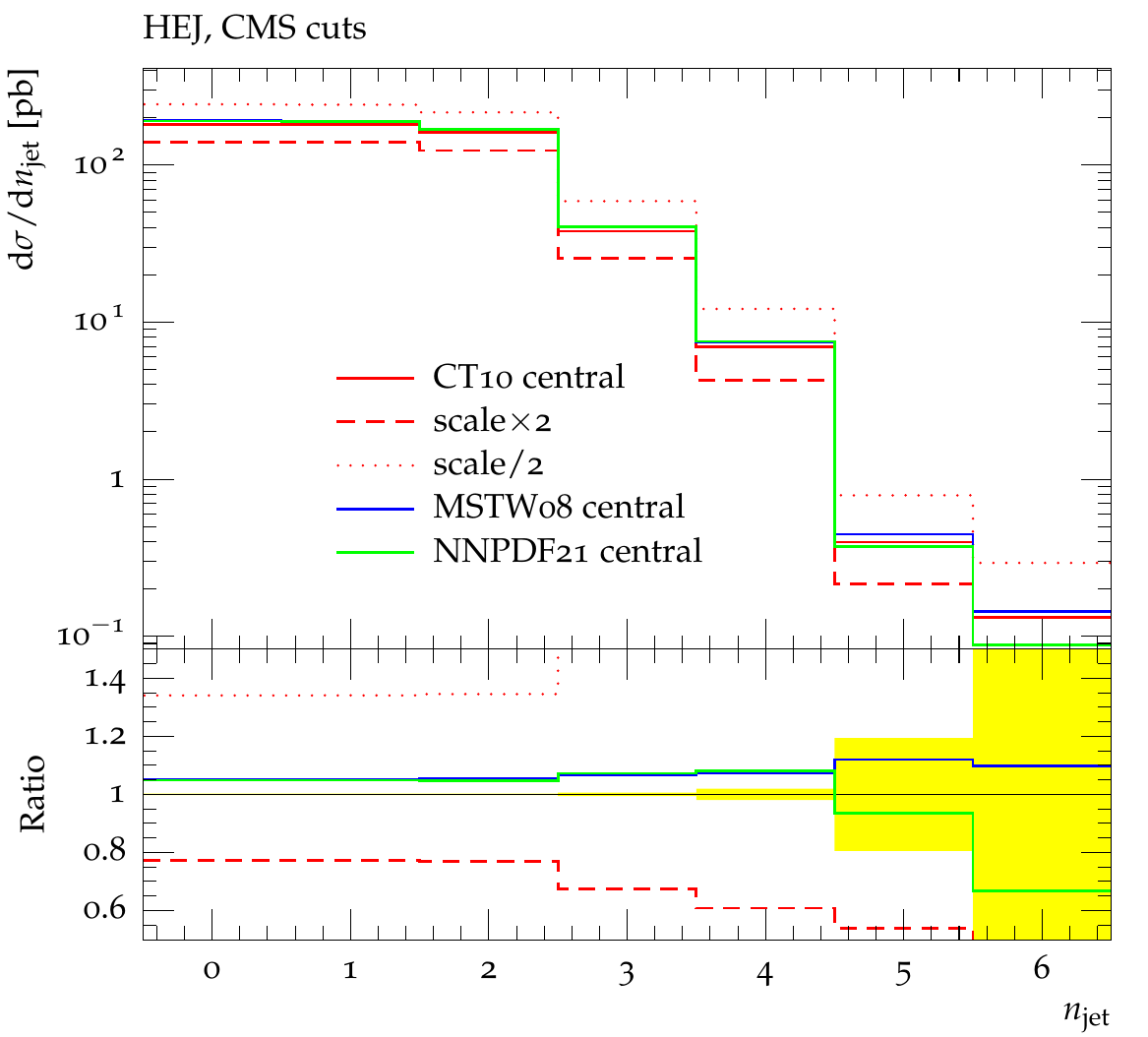}
 \caption{The \HEJ prediction for the cross sections of $W$ plus $n$ jets.
          \vspace*{-5mm}}
  \label{fig:hej_njsigmas}
\end{figure}

This section contains the predictions from the High Energy Jets (\HEJ) event
generator.  This gives predictions for the production of a $W$ boson in
association with \emph{at least two} jets.  Throughout, we show results for
CTEQ, MSTW and NNPDF parton distributions.  We show a scale uncertainty band
only for the first of these for clarity.  The results for the other two are very
similar.  The yellow band in the ratio panel shows the statistical uncertainty
in each case.  The scale variation is seen to be dominant over the statistical
uncertainty and the differences in choice of pdf. All observables are 
defined using the CMS cut definitions, cf.\ \AppRef{App:Observables}.

As discussed in \SecRef{Sec:Codes:HEJ}, the resummation contained in the \HEJ
framework is supplemented with a merging procedure to ensure tree-level accuracy
for events with up to and including four jets.  This leads to the larger drop
from the four jet to the five jet cross section, compared to the drop either
from three-jet to four-jet, or from five-jet to six-jet. This can be clearly
seen in \FigRef{fig:hej_njsigmas}.

%
%
%


\subsubsection{\texorpdfstring{\protect\Madgraph}{Madgraph} + \texorpdfstring{\protect\Pythia}{Pythia}}
\begin{figure}[b!]
 \includegraphics[width=.48\textwidth]{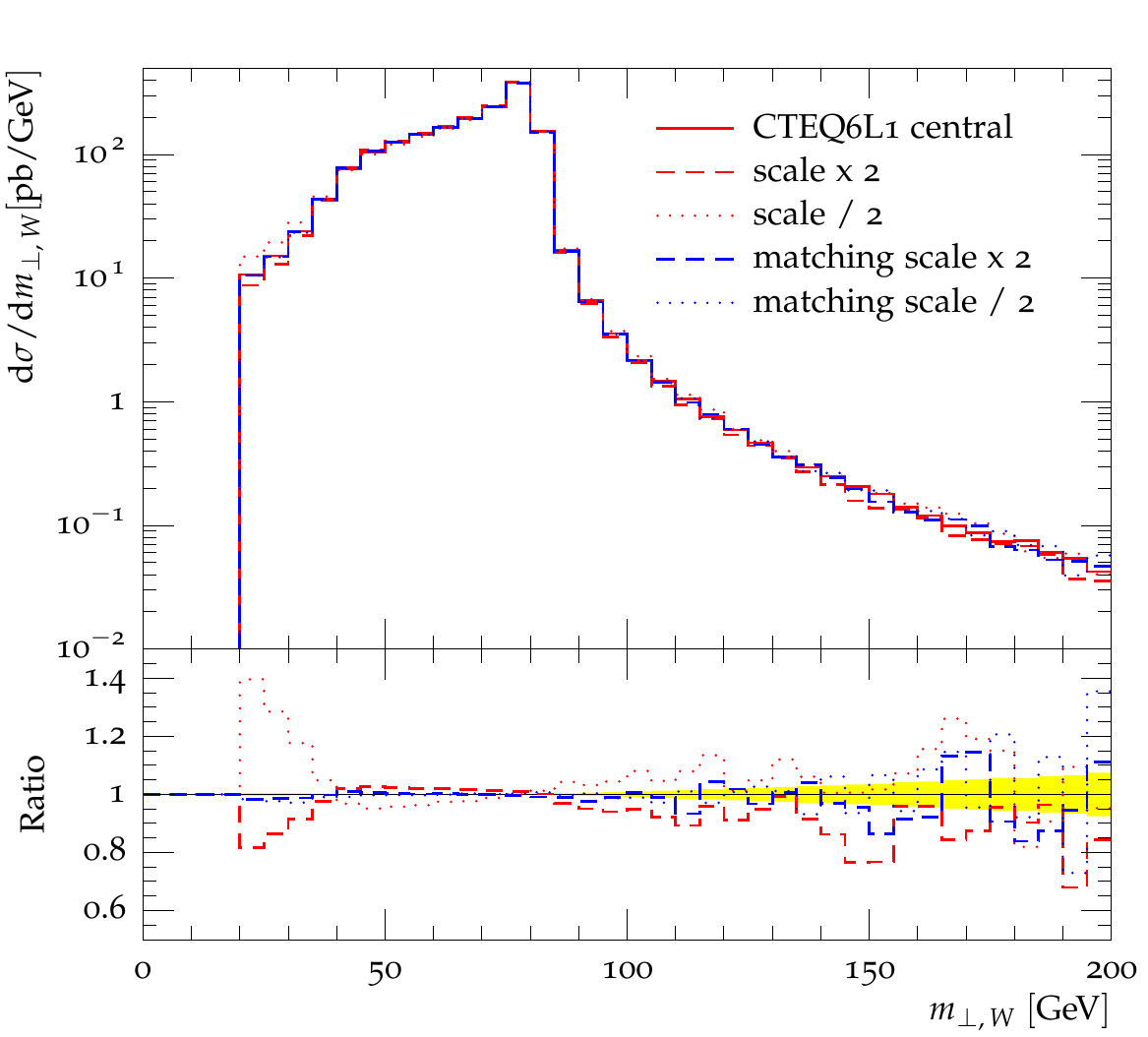}\hfill
 \includegraphics[width=.48\textwidth]{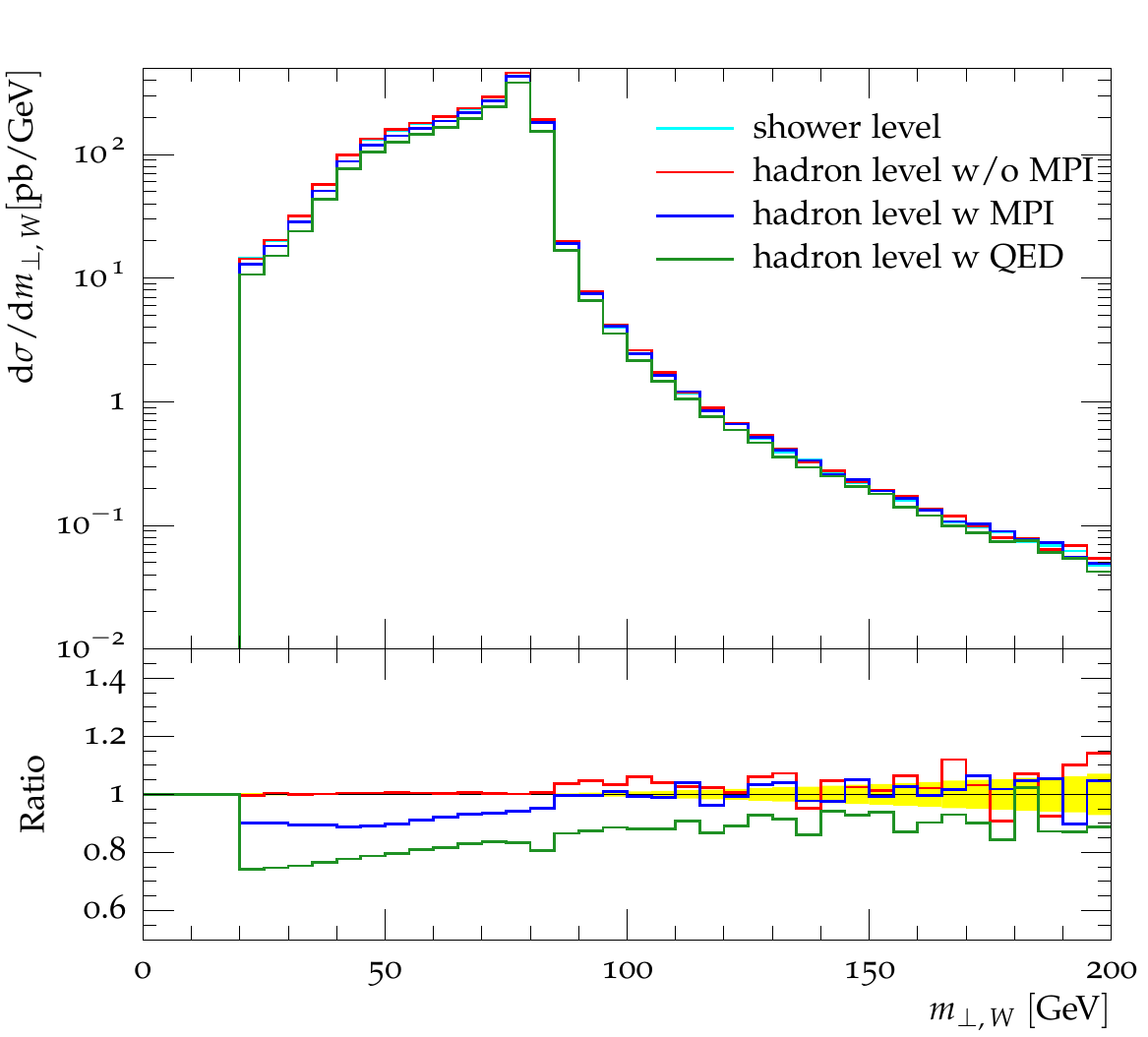}
 \caption{\Madgraph{}+\Pythia results for $W$ transverse mass.}
\end{figure}

\begin{figure}[b!]
\includegraphics[width=.48\textwidth]{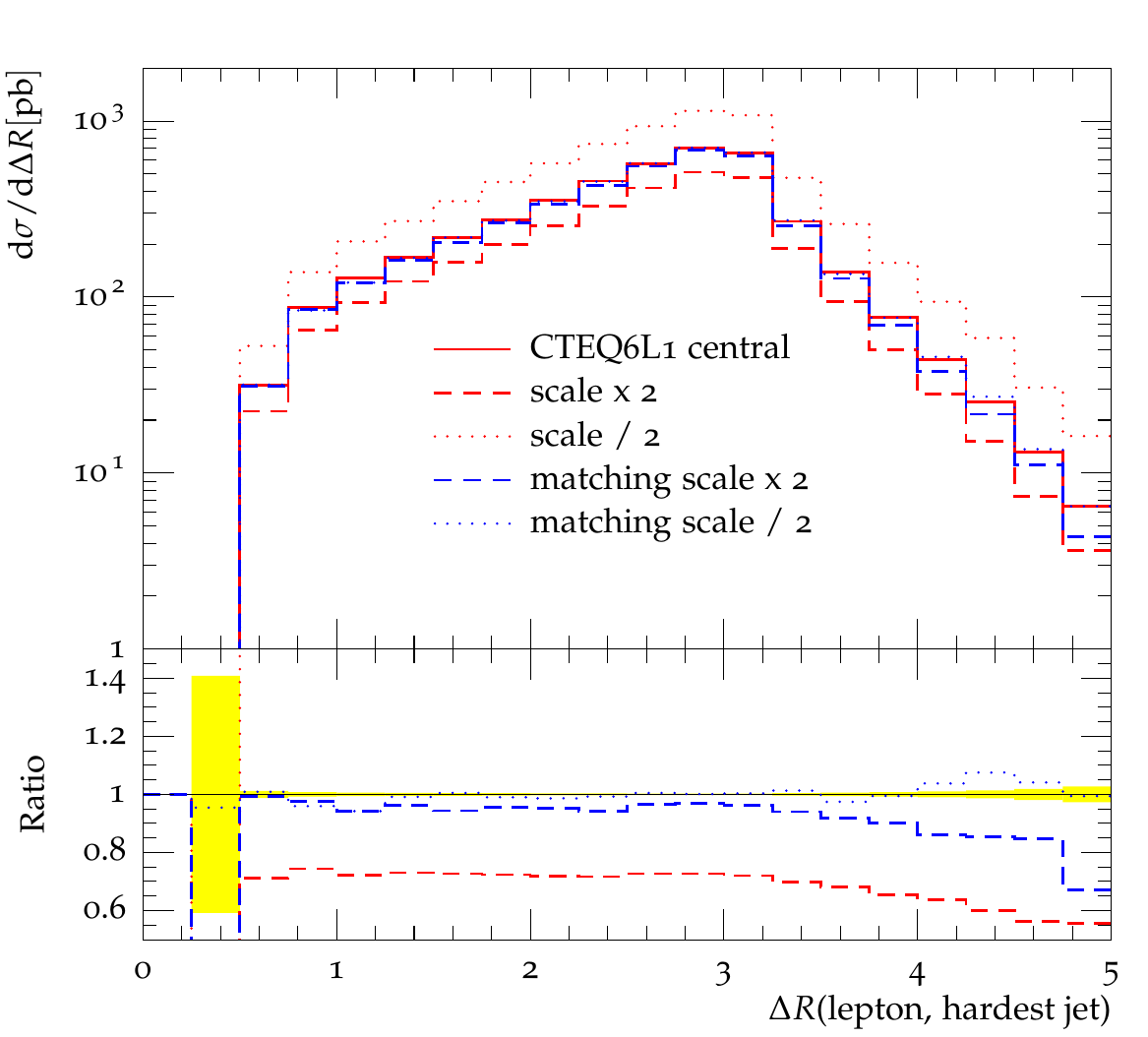}\hfill
 \includegraphics[width=.48\textwidth]{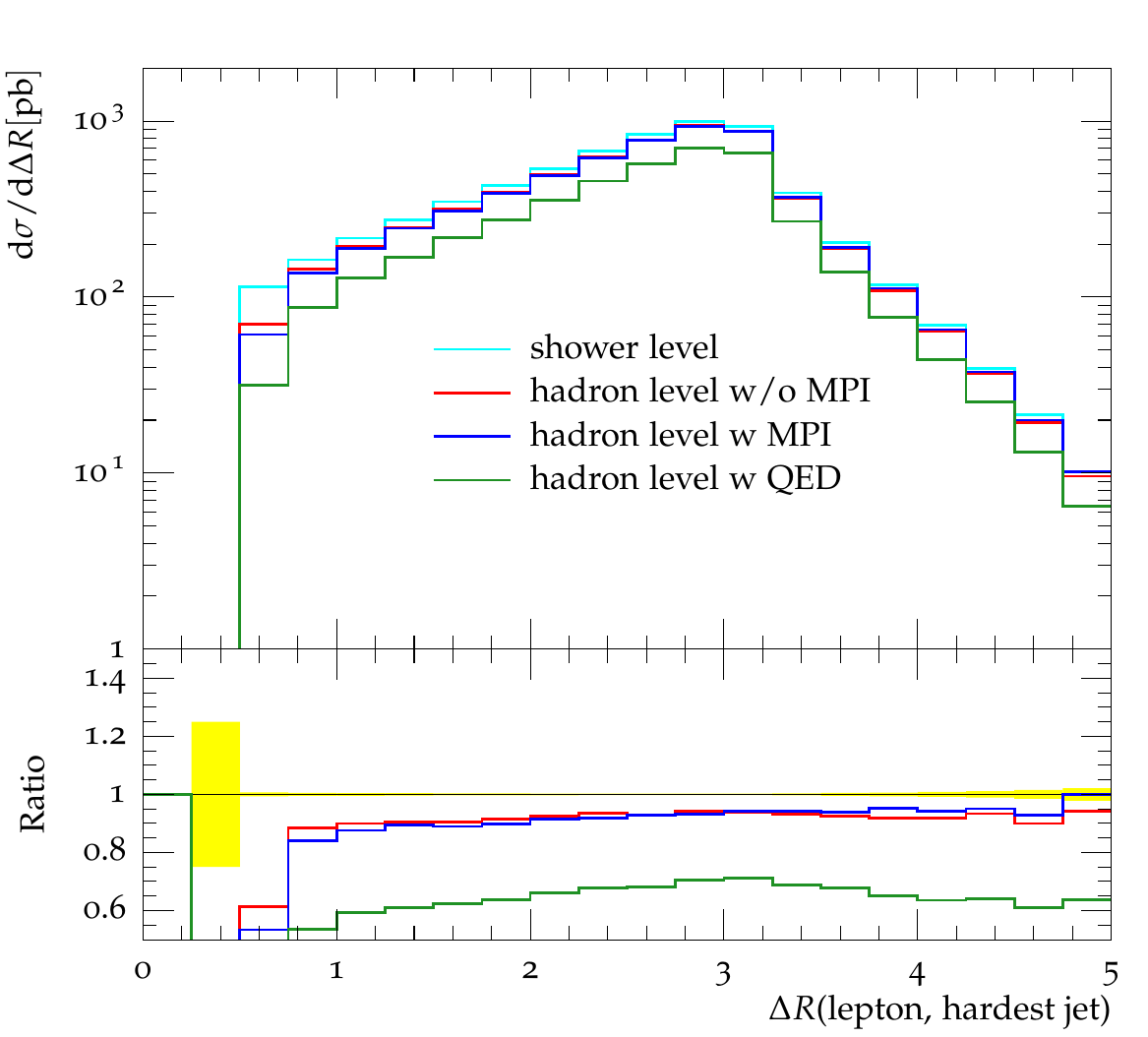}
 \caption{\Madgraph{}+\Pythia results for $\Delta R$
   between lepton and hardest jet.}
\end{figure}

The following results have been obtained with \Madgraph{}+\Pythia. 
Uncertainties due to the factorisation and renormalisation
scale and \MEPS matching scale are shown for results on \QEDlevel. 
A comparison of results on \PSlevel, \Hadlevel, \UElevel, and
\QEDlevel, is also presented.  All observables shown are defined 
using the CMS cuts, cf.\ \AppRef{App:Observables}.

From these results we can conclude that the largest uncertainty on all
observables is due to the factorisation and renormalisation
scale. In addition to that, a large effect is found by switching off the final state QED
radiation, while only a small difference is oberved between results
at the shower level and all other results prior to the QED radiation. 

\begin{figure}[ht]
 \includegraphics[width=.48\textwidth]{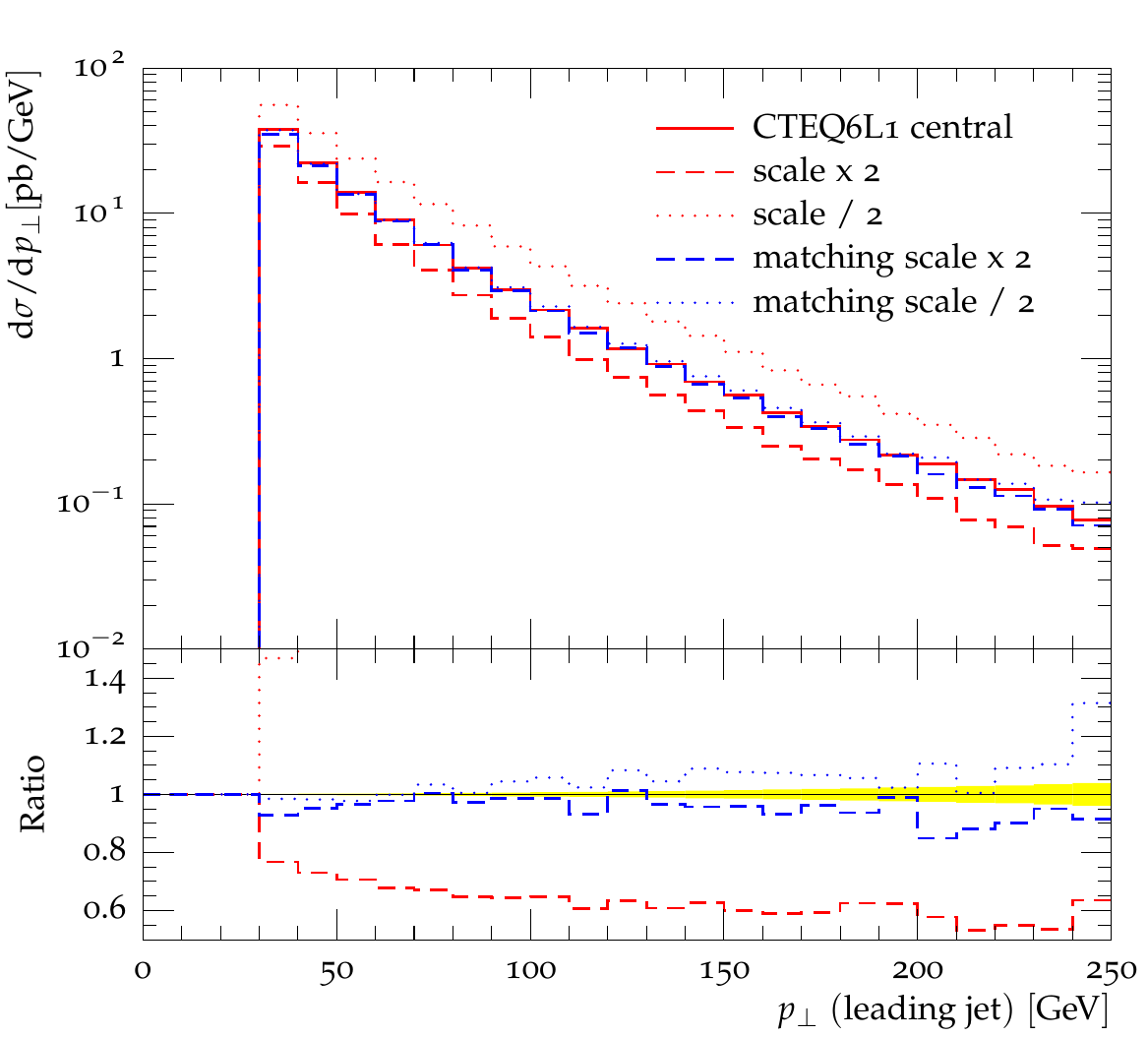}\hfill
 \includegraphics[width=.48\textwidth]{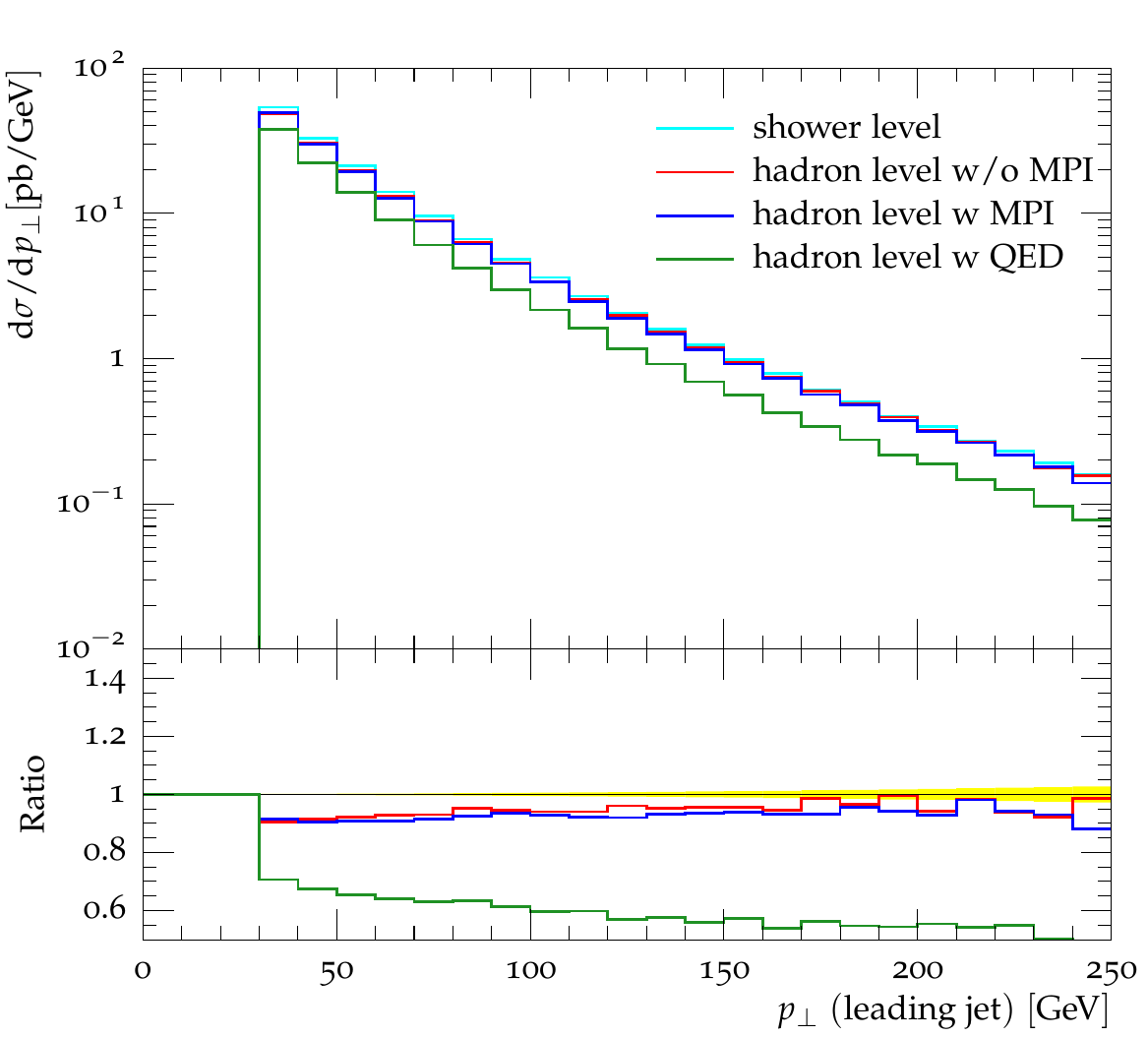} \\
 \includegraphics[width=.48\textwidth]{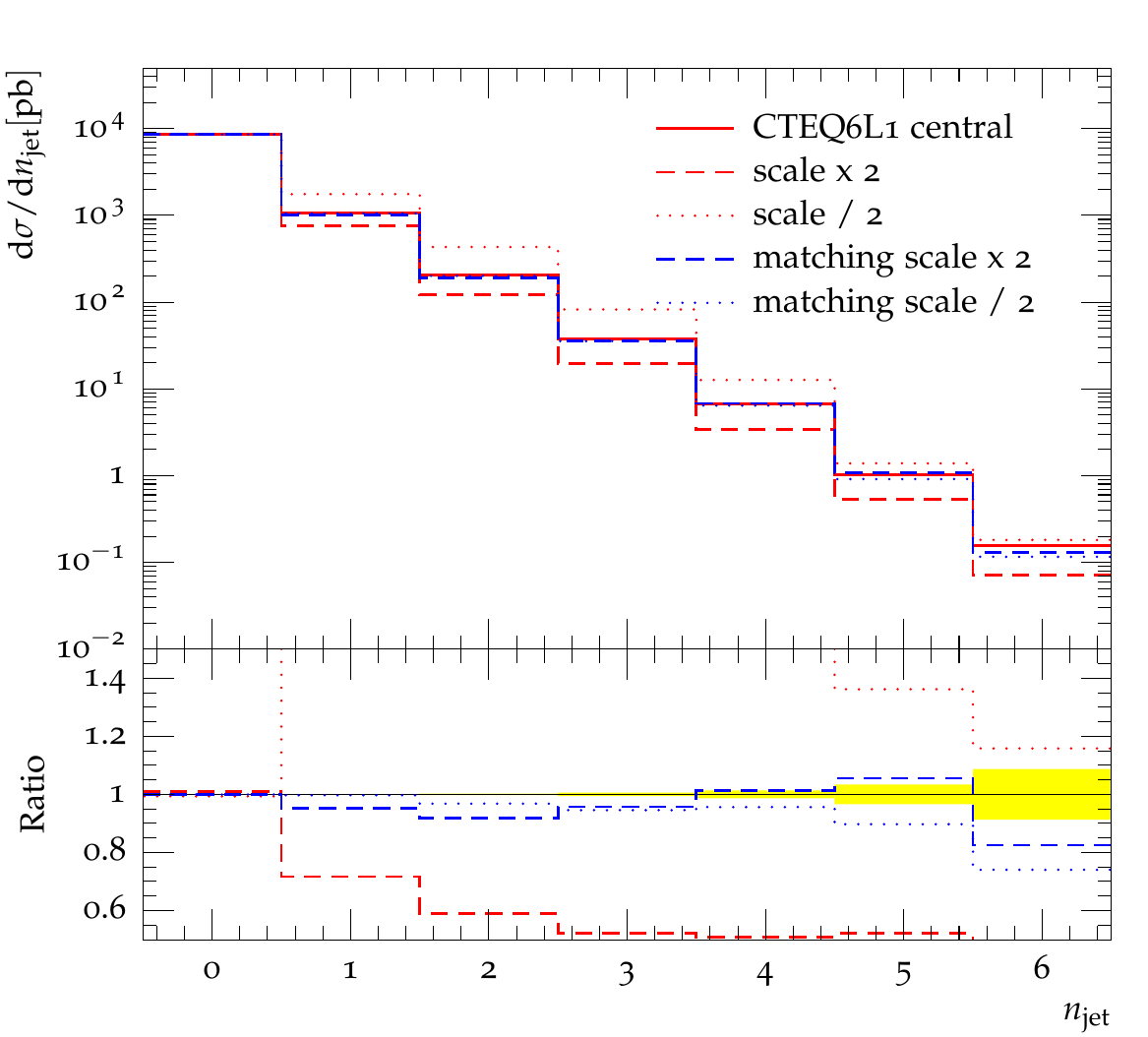}\hfill
 \includegraphics[width=.48\textwidth]{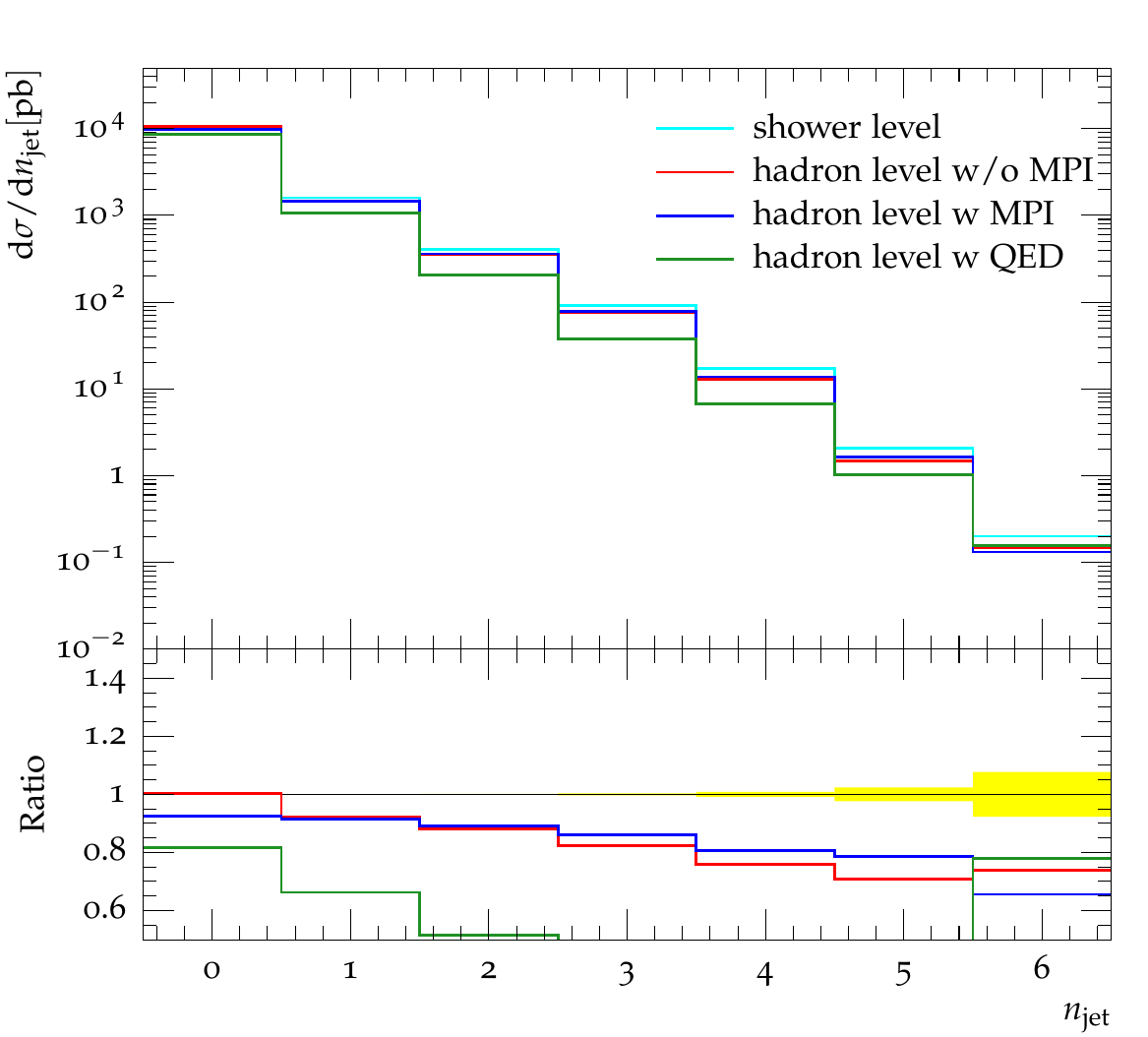} \\
 \includegraphics[width=.48\textwidth]{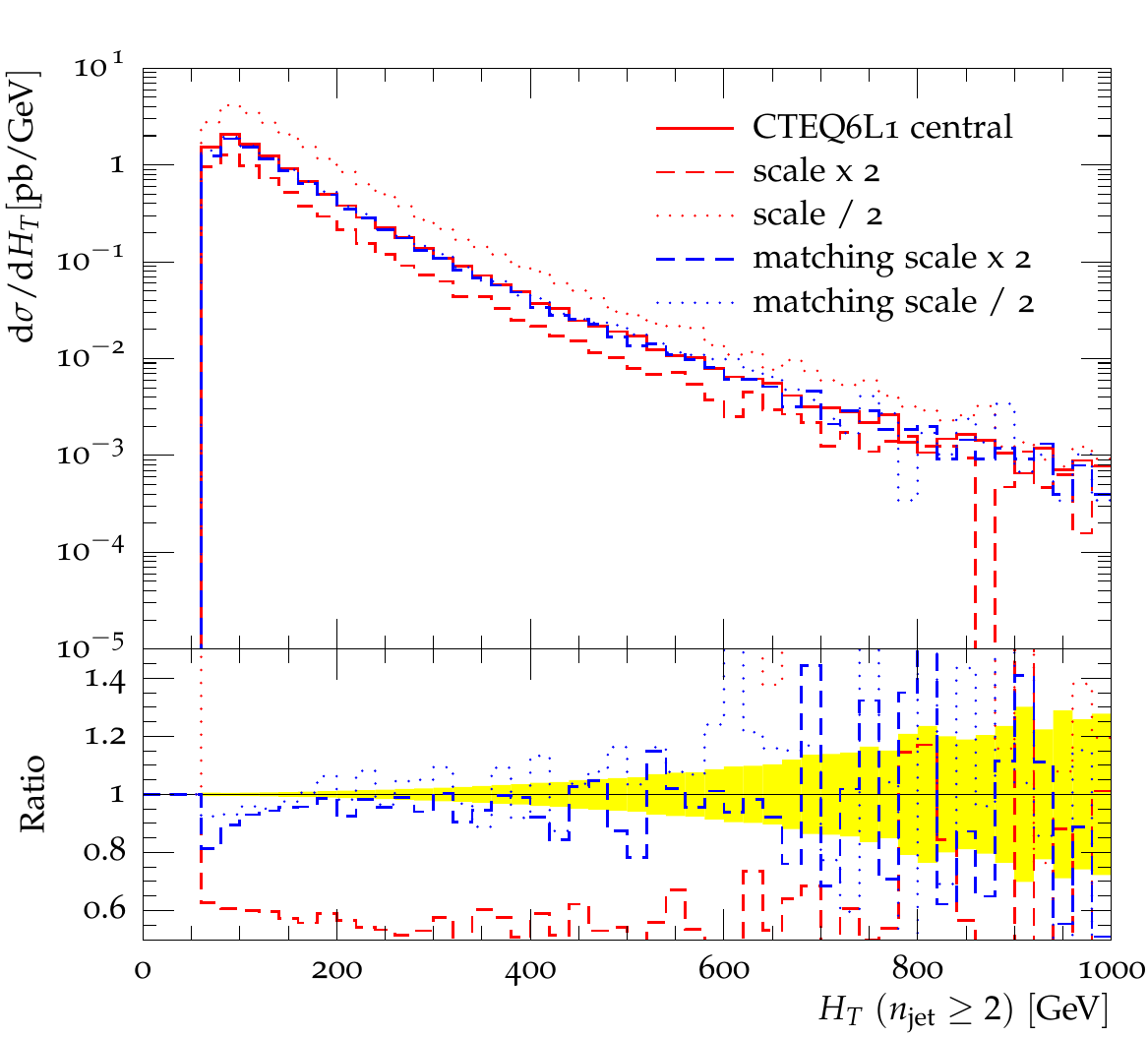}\hfill
 \includegraphics[width=.48\textwidth]{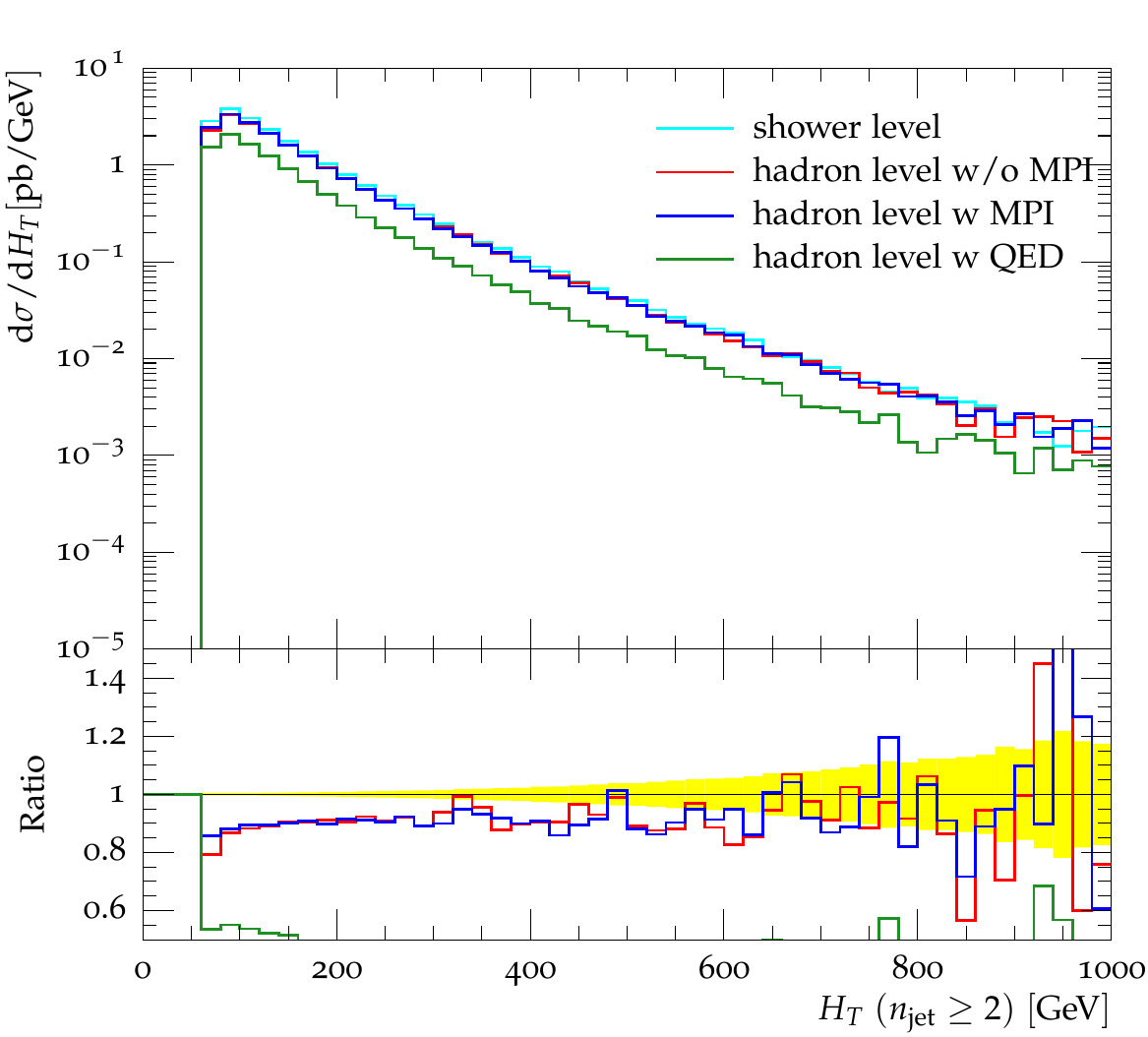}
 \caption{\Madgraph{}+\Pythia results for for $p_\perp$ of hardest
   jet (top), number of jets (middle) and $H_T$ of events with at least 2 jets (bottom).}
\end{figure}

\FloatBarrier

\subsubsection{\texorpdfstring{\protect\PowhegBox}{PowhegBox} + \texorpdfstring{\protect\PythiaEight}{Pythia8}}

In this section we show results obtained by running the \PowhegBox
implementation of $W+1$ jet together with \PythiaEight.  In all the
following plots of this section, CMS analysis cuts have been enforced, 
see \AppRef{App:Observables}.
For this study, in the left panels of
\FigRefs{fig:indiv_PowhegPythia8_njet}{fig:indiv_PowhegPythia8_jetpt1_HTjet2_SumET2}, 
we show uncertainties obtained
from variations of renormalisation and factorisation scales by a
factor of two in either directions and by choosing different PDF sets
in the computation of the hard scattering. Results are shown at the
final level, after the shower, the hadronisation and the inclusion of
MPI, all performed by \PythiaEight.  In general, we notice that the
uncertainty due to scale variations is greater than the changes in the
results due to different PDF choices.

In the right panels of \FigRefs{fig:indiv_PowhegPythia8_njet}
{fig:indiv_PowhegPythia8_jetpt1_HTjet2_SumET2} we show our
results at different stages of the simulation, for a fixed PDF set
(chosen to be CT10).  The stages considered include from the first
emission level up to the full showered events in \PythiaEight,
including MPI and also effects due to QED radiation off leptons and
quarks. Various stages of the simulation have been obtained setting the \PythiaEight
switches as reported in \SecRef{App:Settings:PowhegBox}.

We recall here that results should be considered to be physical only
after the the hadron level is reached (possibly including MPI and QED
effects).
In particular, we stress that the results at the parton level are obtained considering only the
\POWHEG first emission, and they are therefore only intermediate: 
indeed at this stage only the hardest radiation has been
generated and effects due to further showering are not yet taken into
account.

\begin{figure}[b!]
\centering
\includegraphics[width=.48\textwidth]{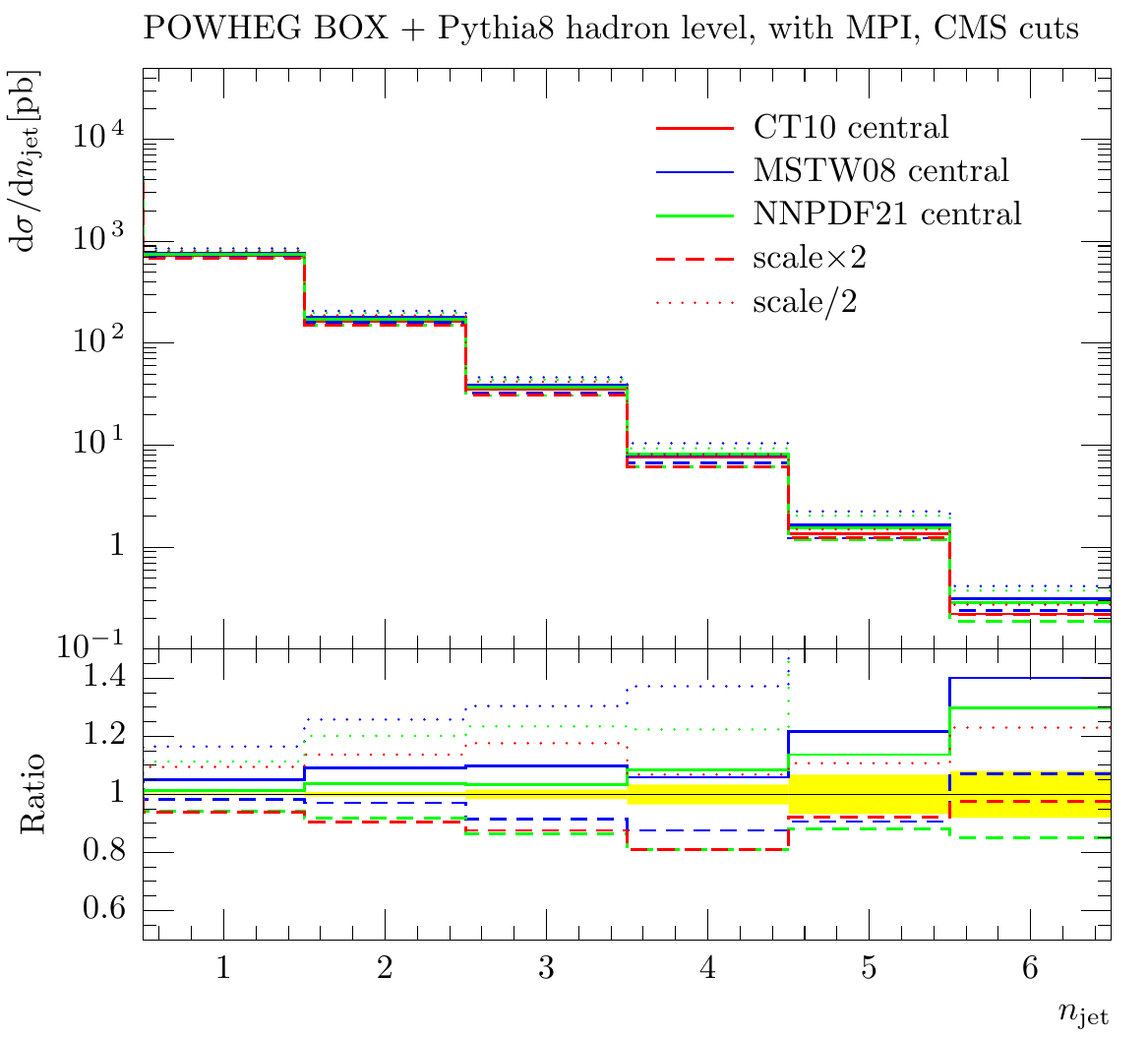}\hfill
\includegraphics[width=.48\textwidth]{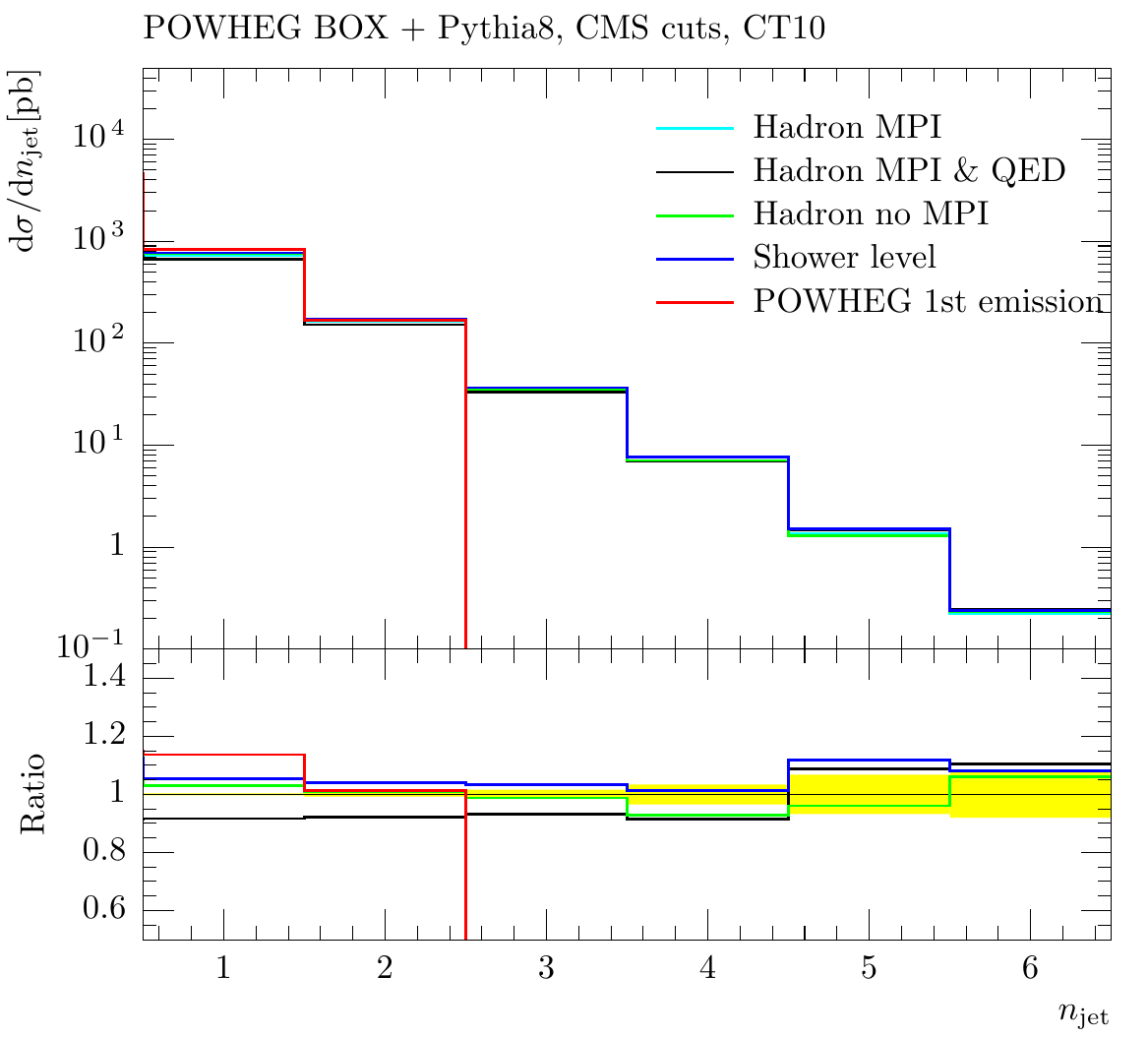} 
\caption{The number of jets, as predicted by \PowhegBox + \PythiaEight.}
\label{fig:indiv_PowhegPythia8_njet}
\end{figure}

\begin{figure}[p]
\centering
\includegraphics[width=.48\textwidth]{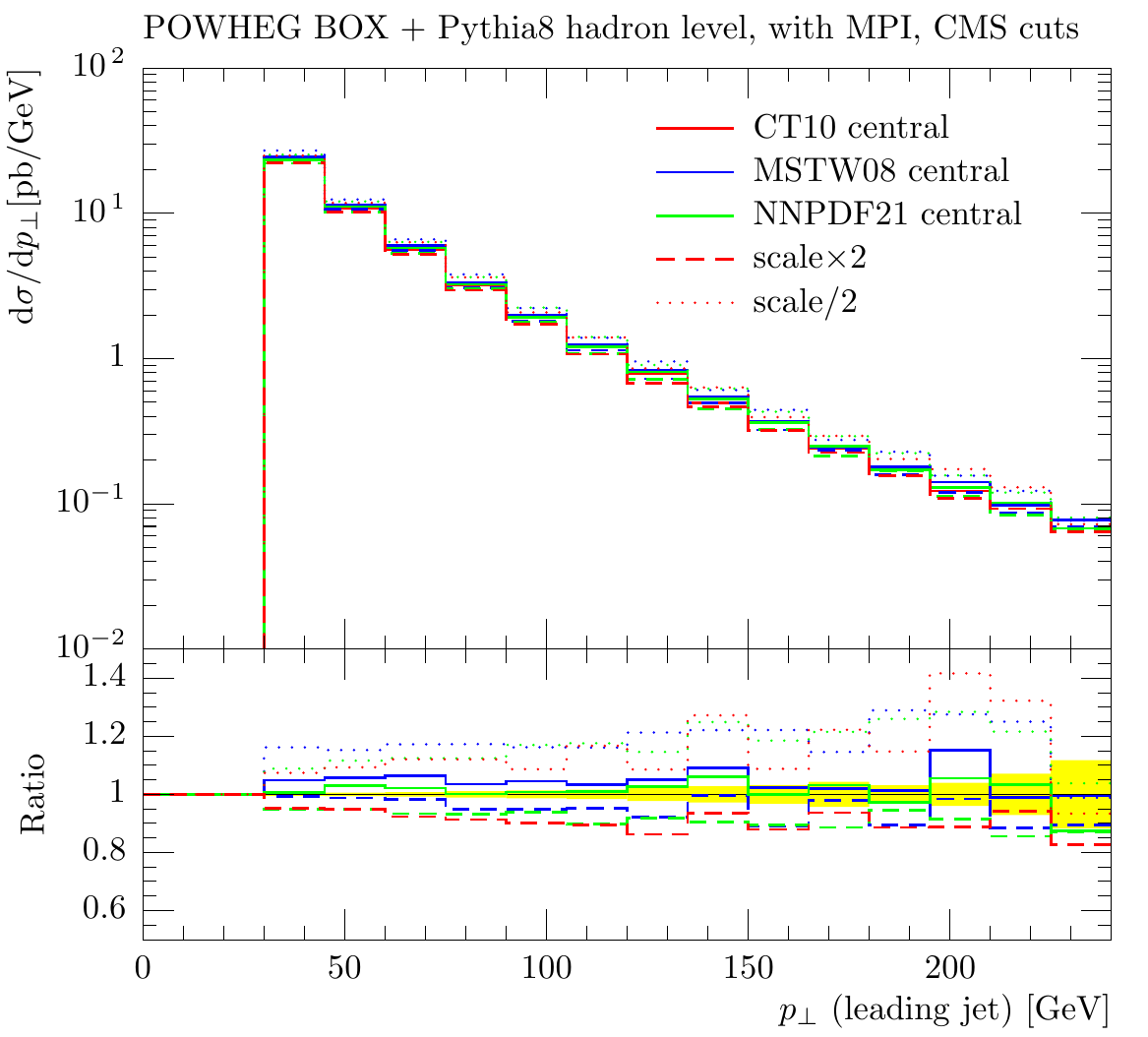}\hfill
\includegraphics[width=.48\textwidth]{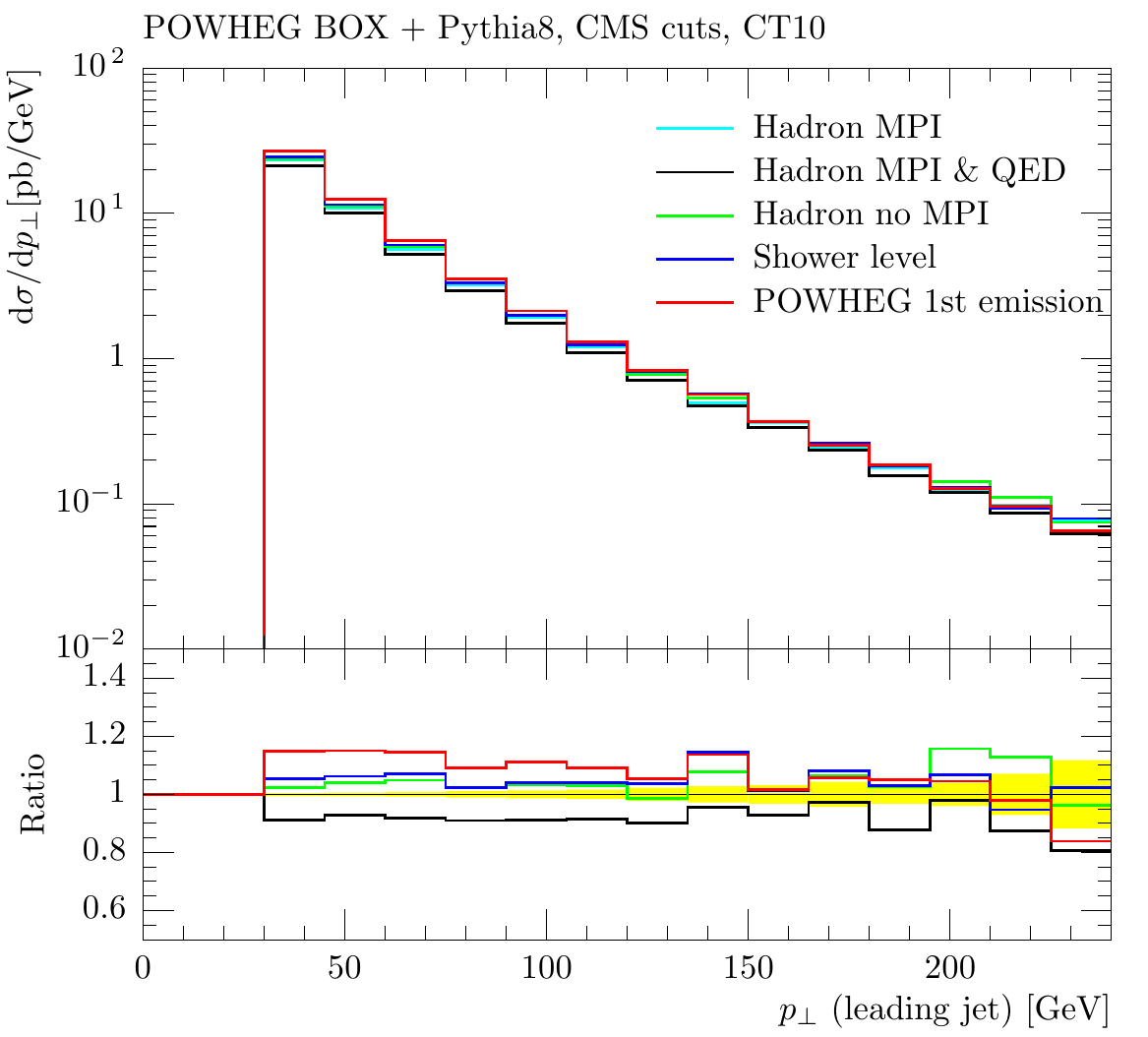} \\
\includegraphics[width=.48\textwidth]{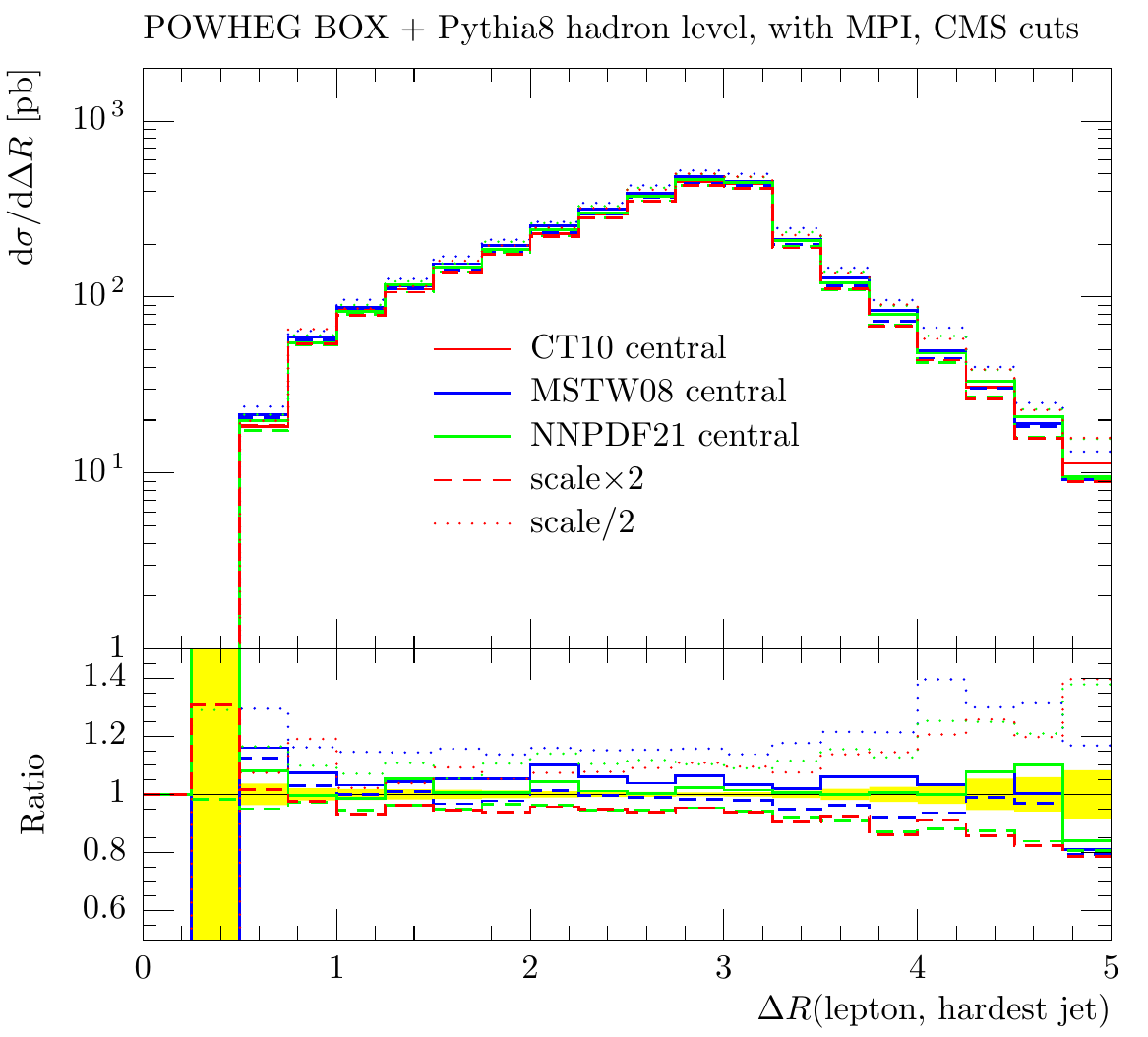}\hfill
\includegraphics[width=.48\textwidth]{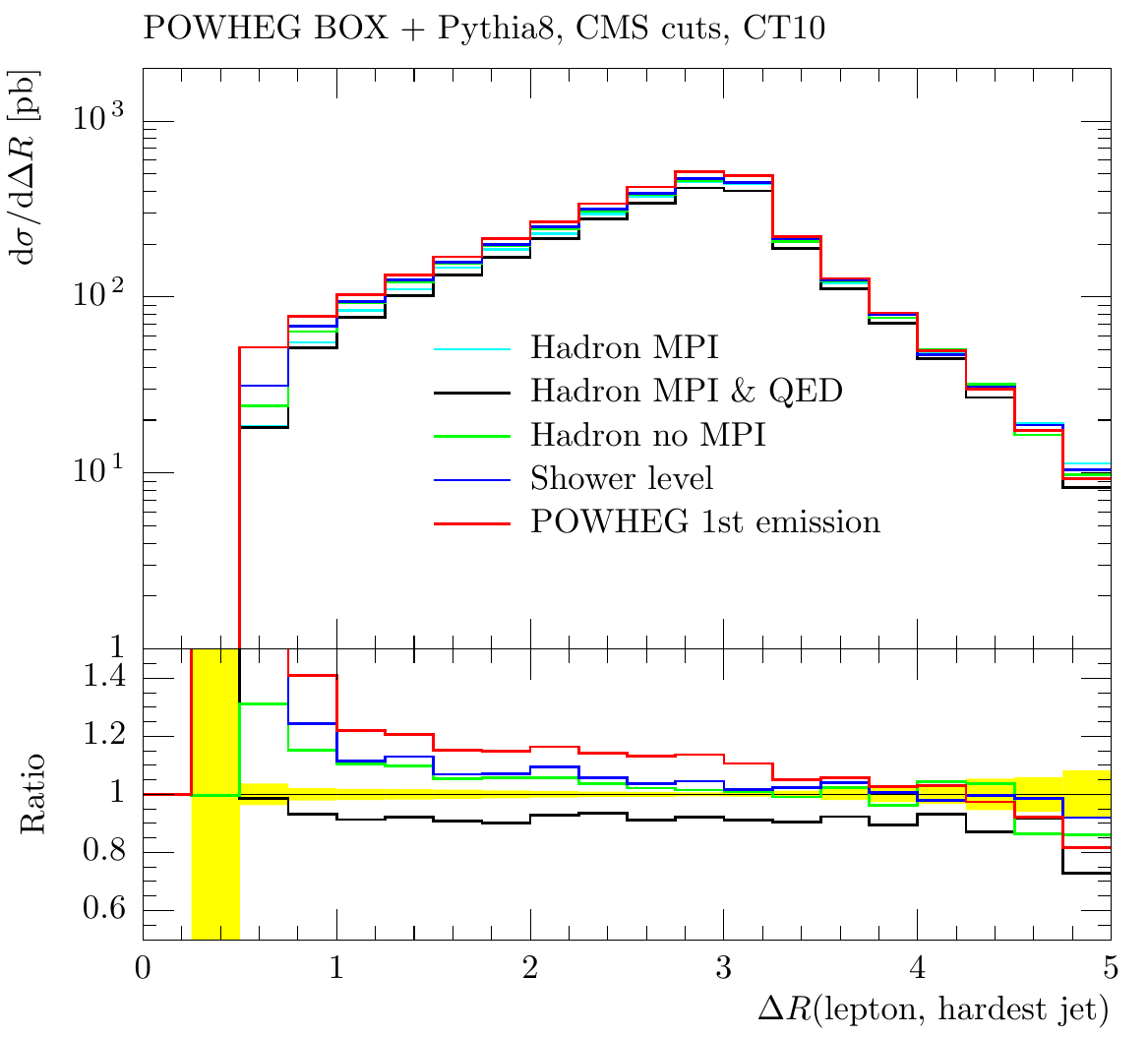} \\
\includegraphics[width=.48\textwidth]{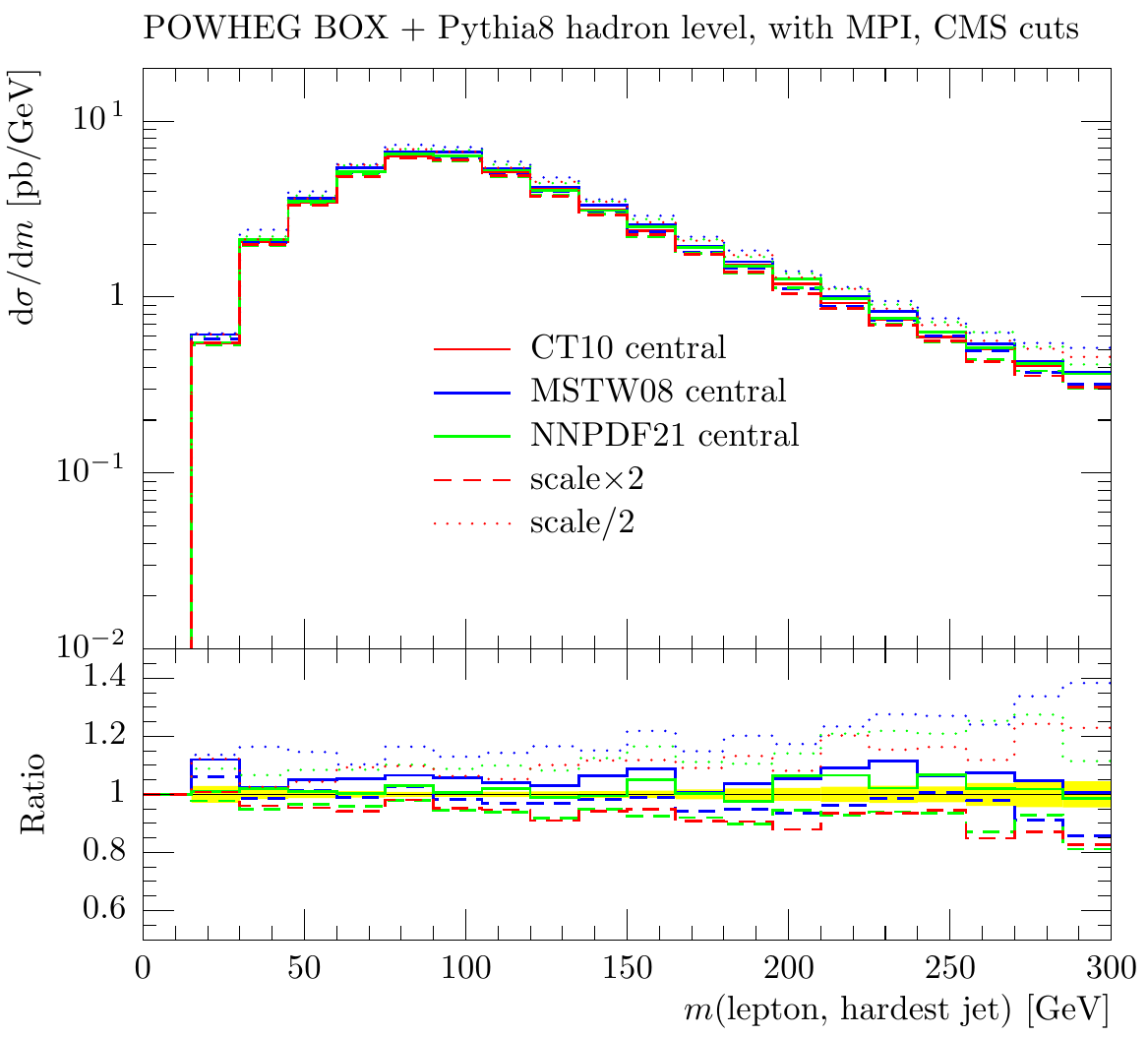}\hfill
\includegraphics[width=.48\textwidth]{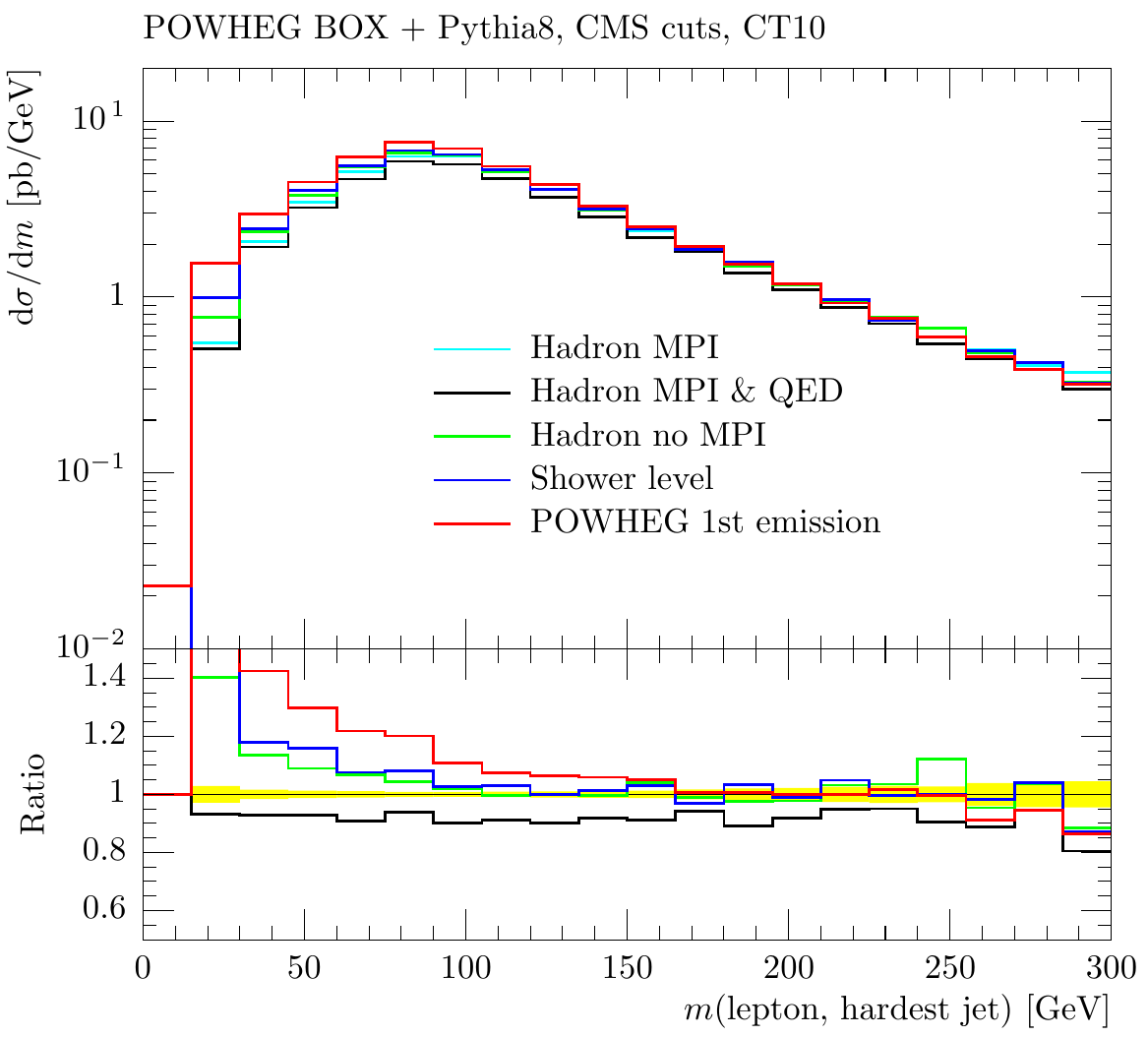} 
\caption{The hardest jet transverse momentum distribution (upper
  plots), the $\Delta R$ separation (middle plots) and the invariant
  mass $m$ (lower plots) of the hardest jets and the hardest lepton,
  as predicted by \PowhegBox + \PythiaEight.}
\label{fig:indiv_PowhegPythia8_ptj0_DR_mj0l}
\end{figure}

\begin{figure}[p]
\centering
\includegraphics[width=.48\textwidth]{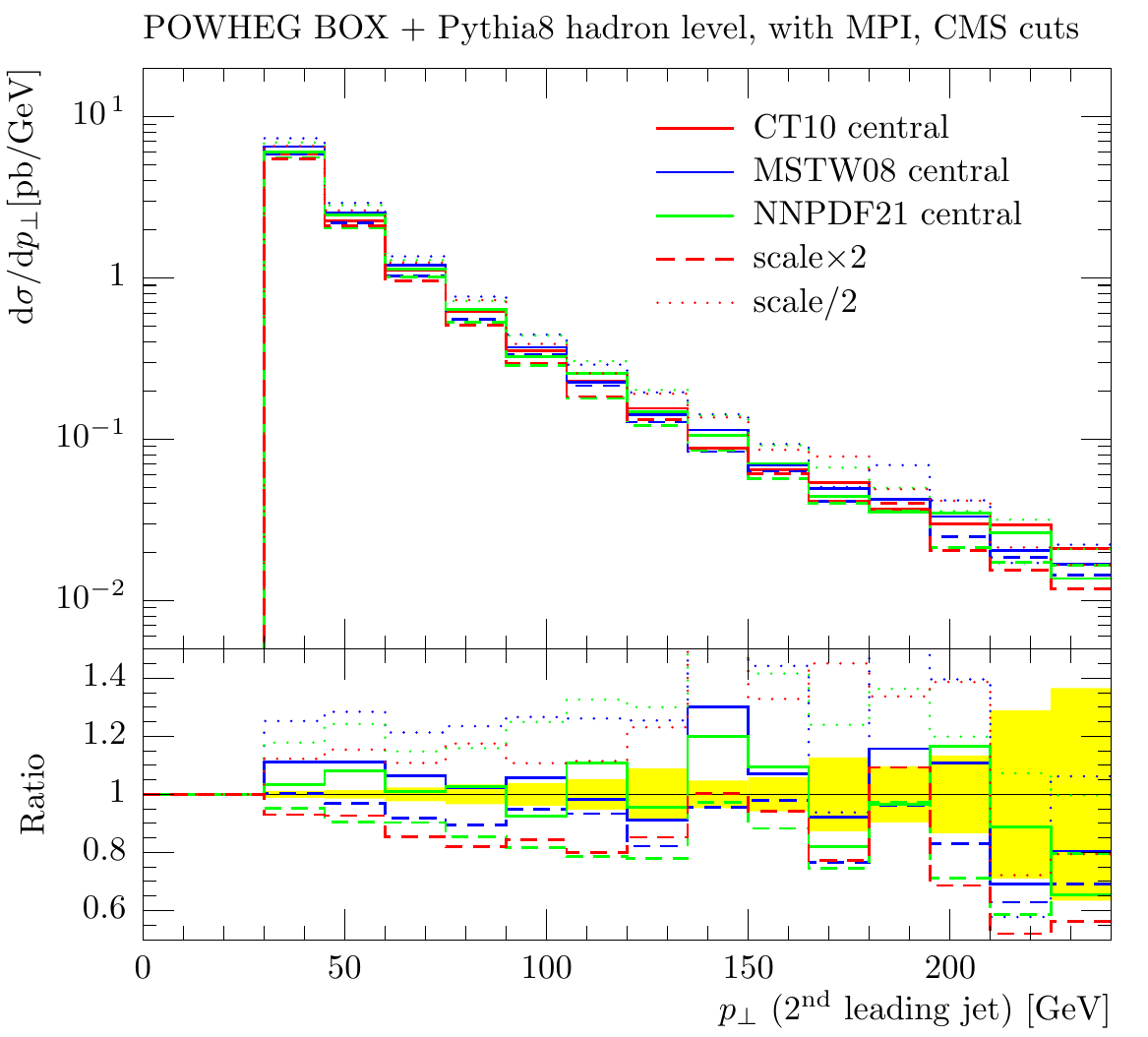}\hfill
\includegraphics[width=.48\textwidth]{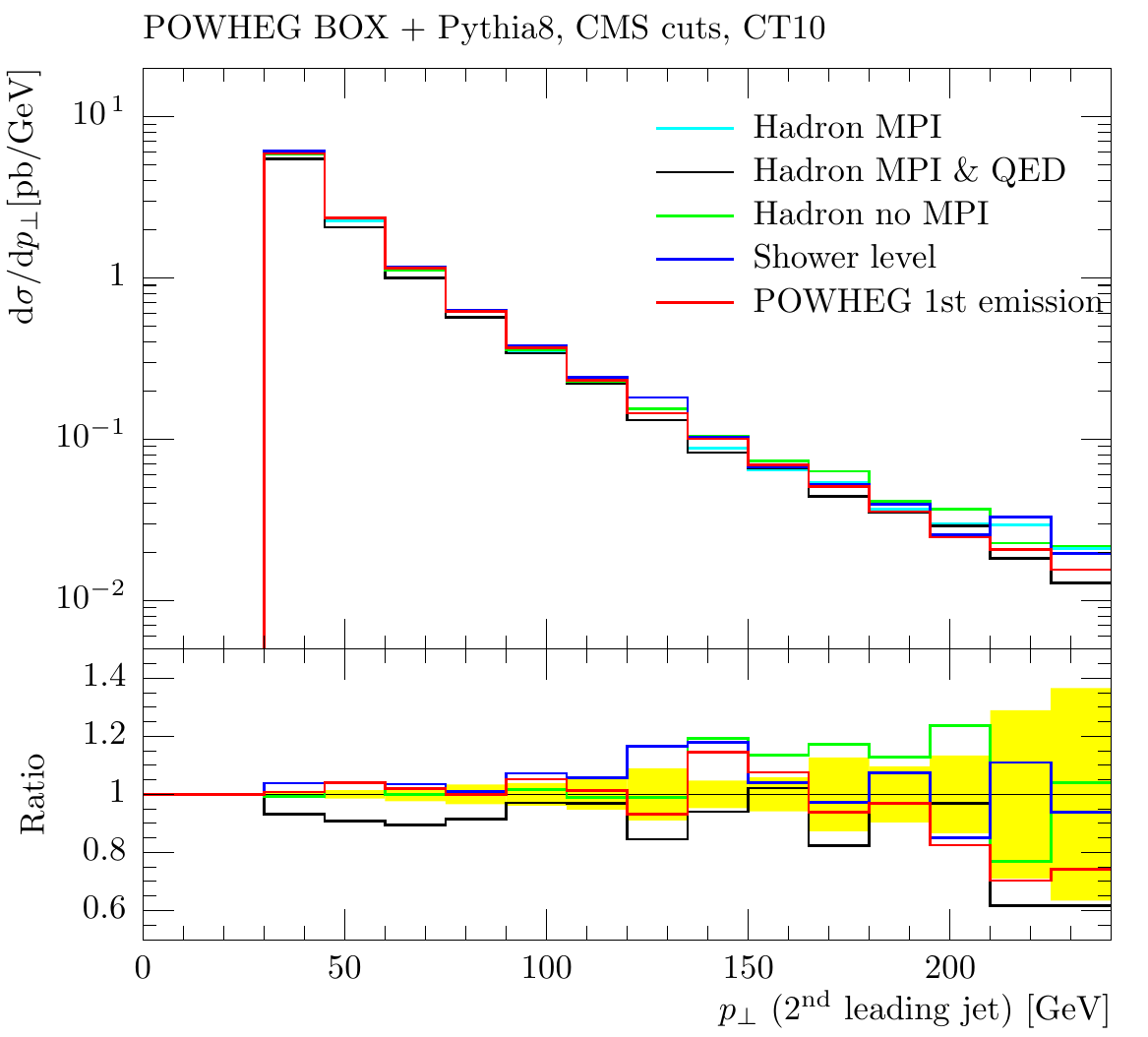} \\
\includegraphics[width=.48\textwidth]{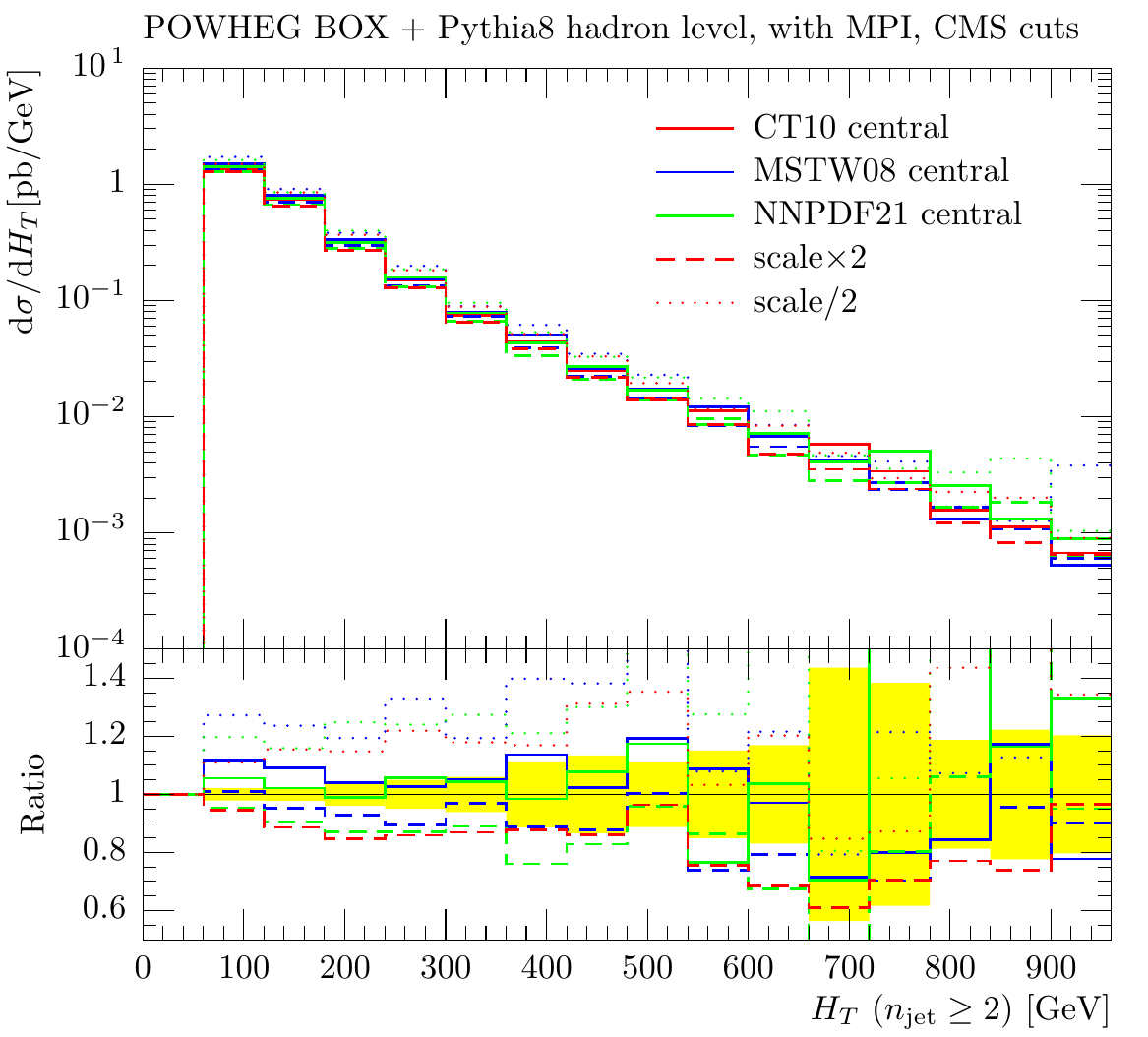}\hfill
\includegraphics[width=.48\textwidth]{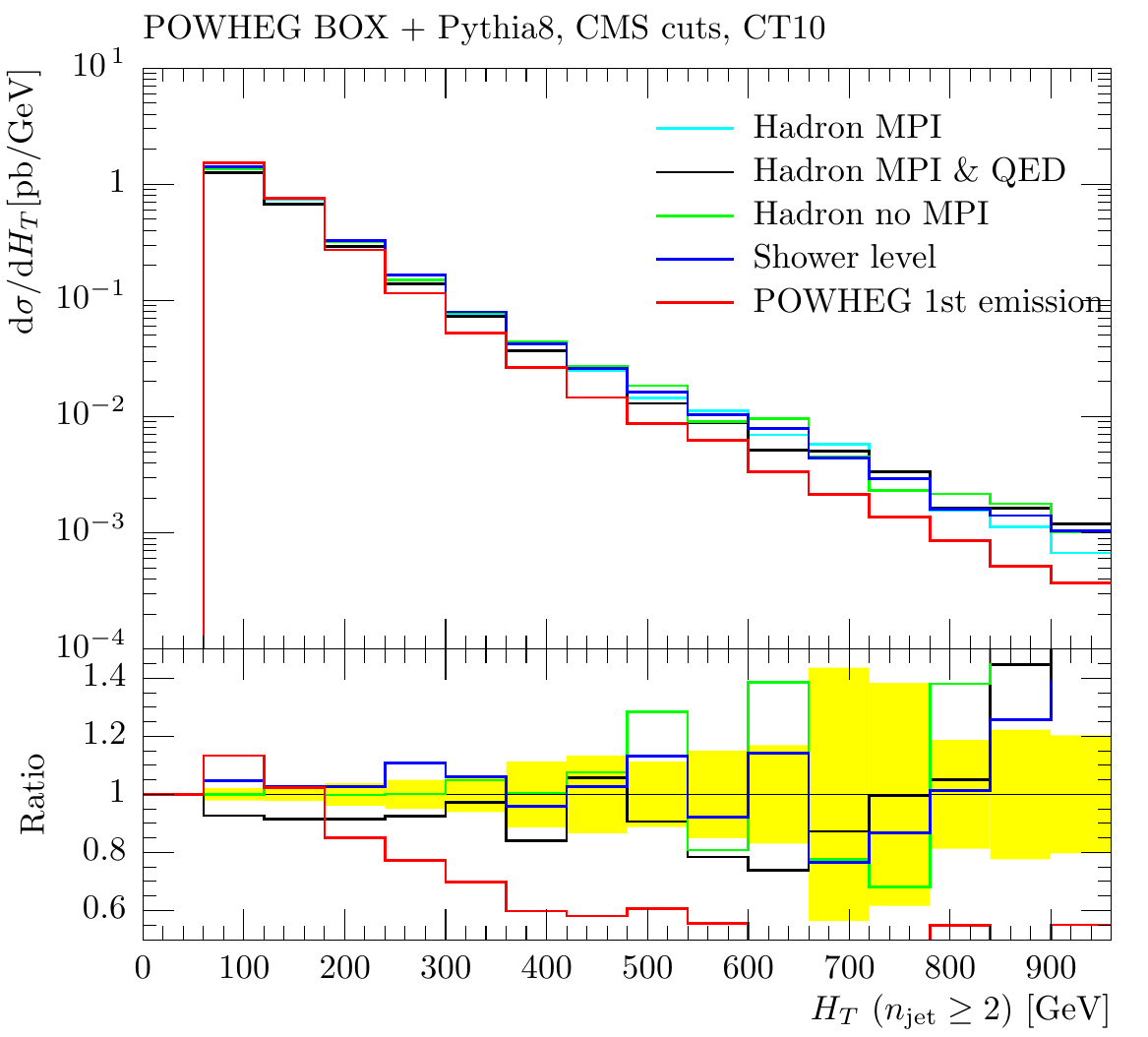} \\
\includegraphics[width=.48\textwidth]{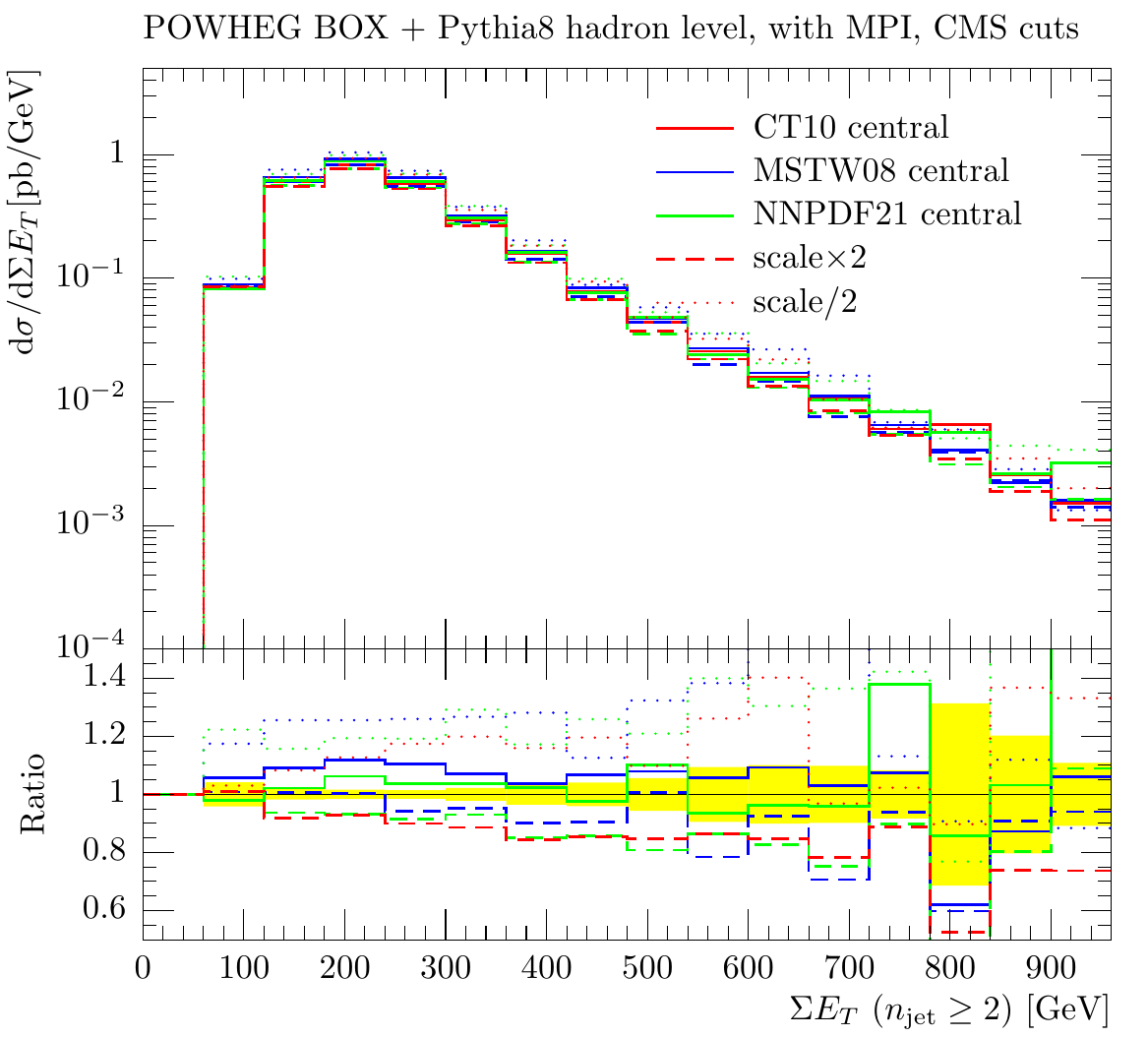}\hfill
\includegraphics[width=.48\textwidth]{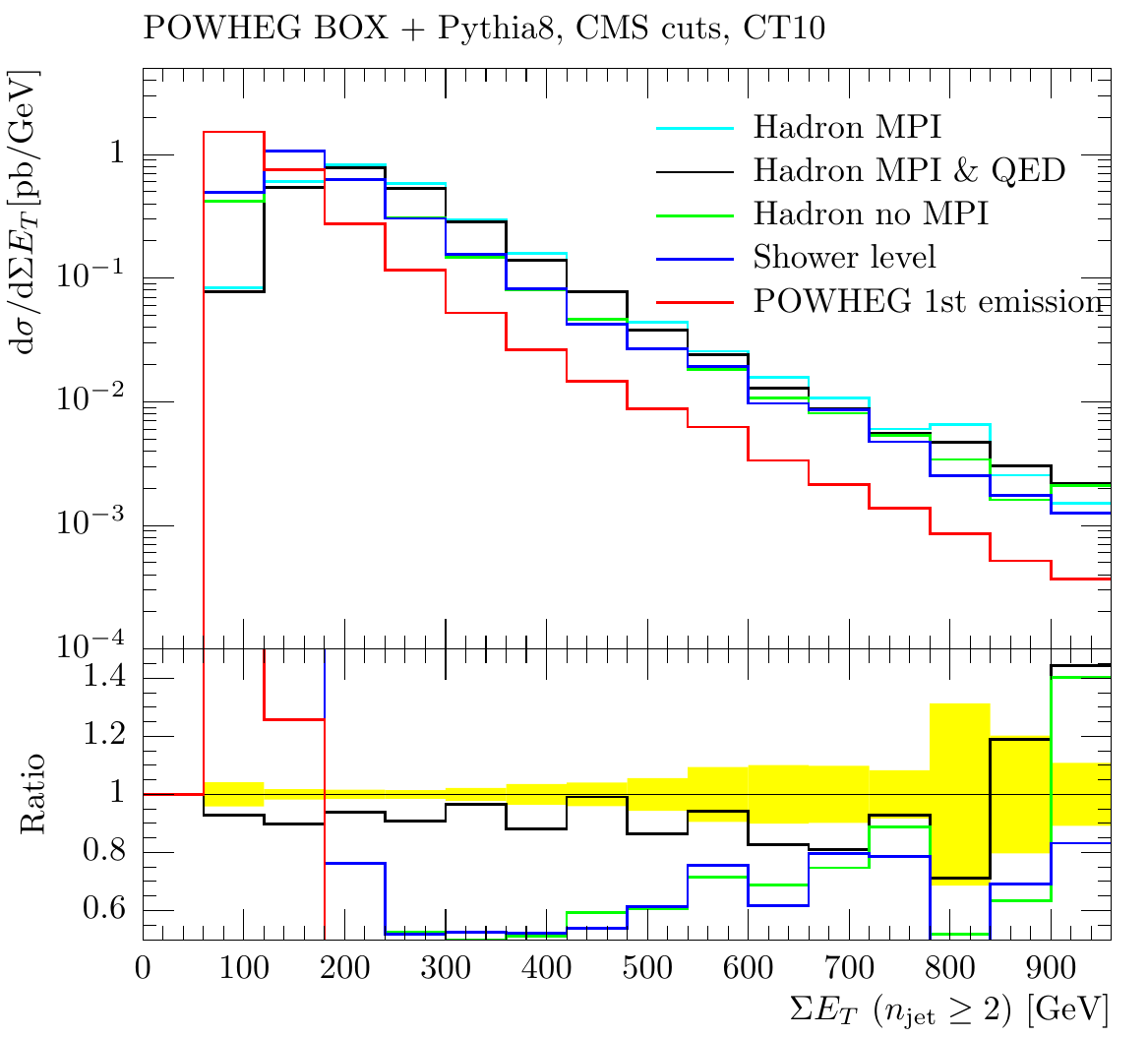} 
 \caption{The transverse momentum $p_\perp$ of the next-to-hardest jet, the scalar sum of the jet transverse energy $H_T$ of events with at least 2 jets and the sum of the transverse energies of all the particles in events with 2 or more jets, as predicted by \PowhegBox + \PythiaEight.}
\label{fig:indiv_PowhegPythia8_jetpt1_HTjet2_SumET2}
\end{figure}

For most of the observables results do not show large variations going
from a simulation level to another. In particular, for truly NLO
predictions such the plots in
\FigRef{fig:indiv_PowhegPythia8_ptj0_DR_mj0l} or the bin $n_{\rm
  jet}=1$ of \FigRef{fig:indiv_PowhegPythia8_njet}, the major
effects that arise at each successive stage of the simulation are a
change in the normalisation, due to a slightly different number of
events passing the analysis cuts when multiple emissions are allowed,
and a moderate shape distortion in the low end of the spectrum.  Both
these effects may be attributed to multiple QCD radiation due to
Sudakov effects introduced by the parton shower.  As expected,
these effects are of the same size, or smaller,
than the theoretical uncertainty due to scale and PDF's variations,
when propagated to the hadronic level.  Similar effects are also
observed when the QED radiation is turned on. In this case, results
are lowered as a consequence of the cuts on the lepton transverse
momentum and rapidity.

Due to the requirement of having at least two jets, the remaining
observables are predicted only at leading order or with leading log accuracy
by the \POWHEG simulation of $W+1$~jet. This is also reflected in the
larger band associated with the scale variations.

Observables such as $H_T$, the scalar sum of
the transverse energy of the jets for events with two or more jets,
show an enhancement in the high-$H_T$ tail. This effect mostly arise
as a consequence of the showering, since the successive stages do not
change the predictions any longer. The same behaviour, even more
enhanced, is also observed in the scalar sum of the transverse energy
of all particles, always in events with two or more jets.

In \FigRef{fig:indiv_PowhegPythia8_tunes} we instead compare the
effect of using different \PythiaEight tunes on our predictions,
obtained in this case at the hadron level, including MPI.  Essentially
all the observables turned out to be extremely stable under the
variations of the \PythiaEight tune, as shown in
\FigRef{fig:indiv_PowhegPythia8_tunes}.  Major differences only
appears for the beam thrust, when it is defined at the particle level 
(see \AppRef{App:Observables:Analysis}).  


\begin{figure}[t]
  \centering
\includegraphics[width=.48\textwidth]{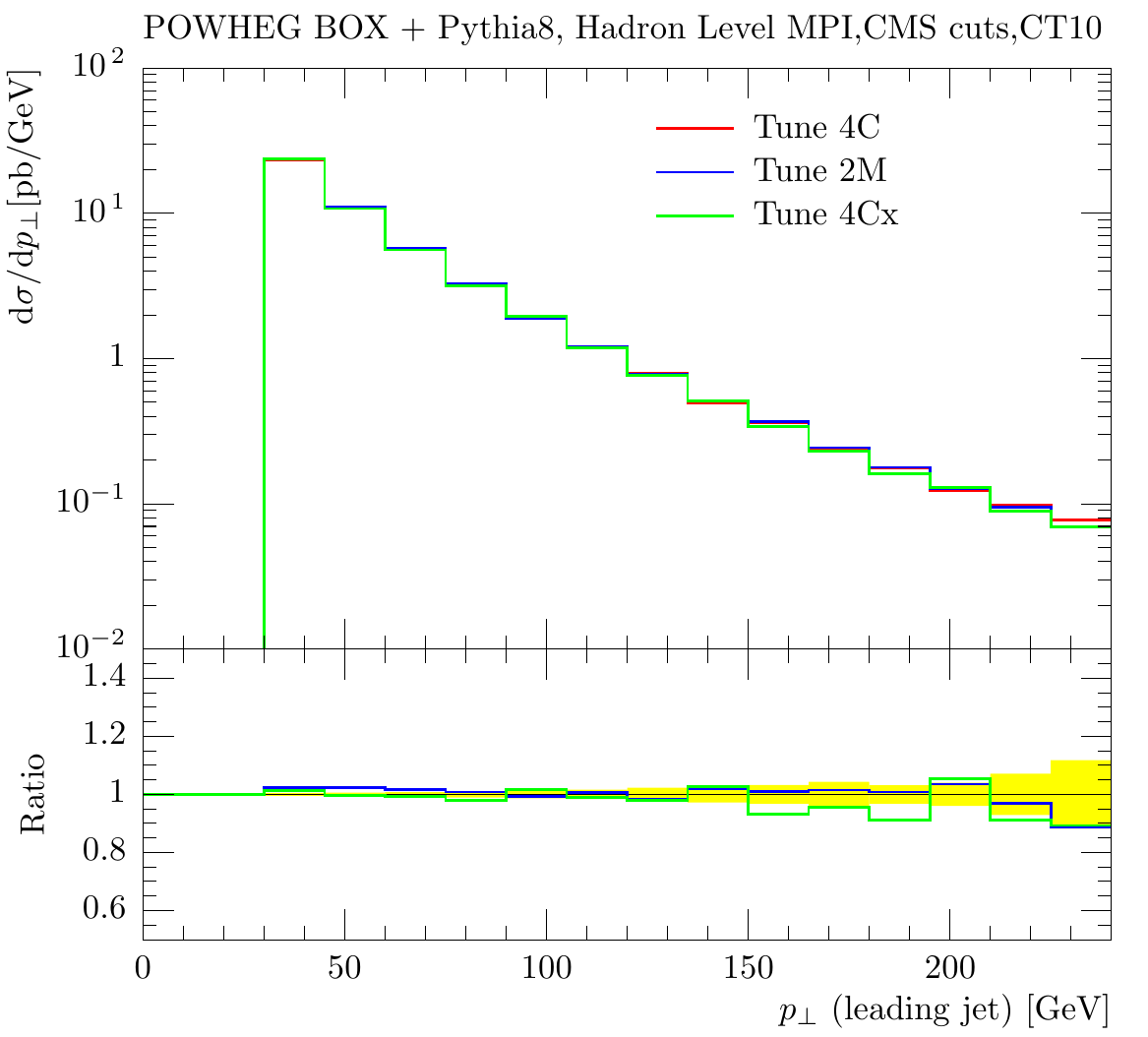}\hfill
\includegraphics[width=.48\textwidth]{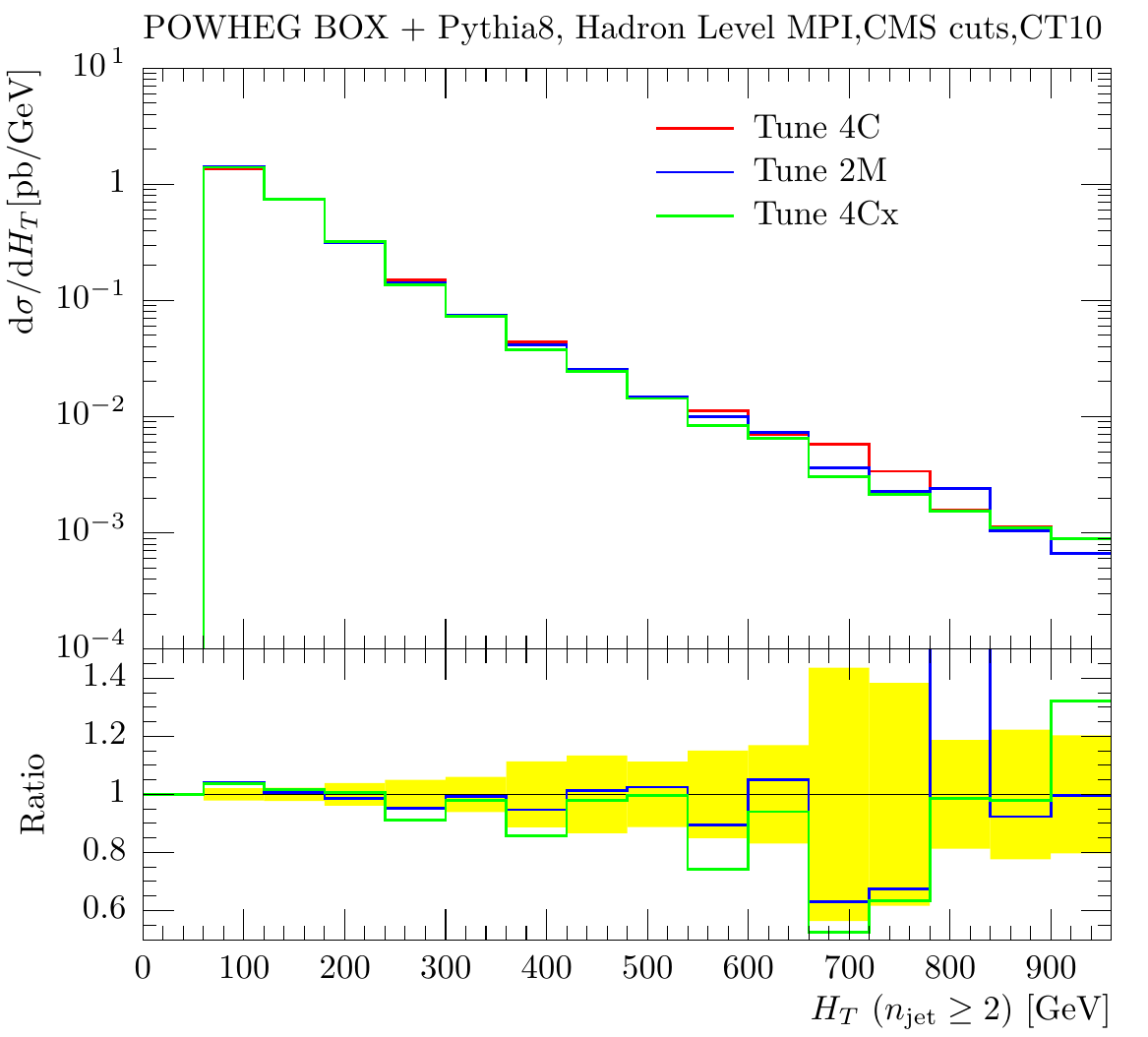}\\
\includegraphics[width=.48\textwidth]{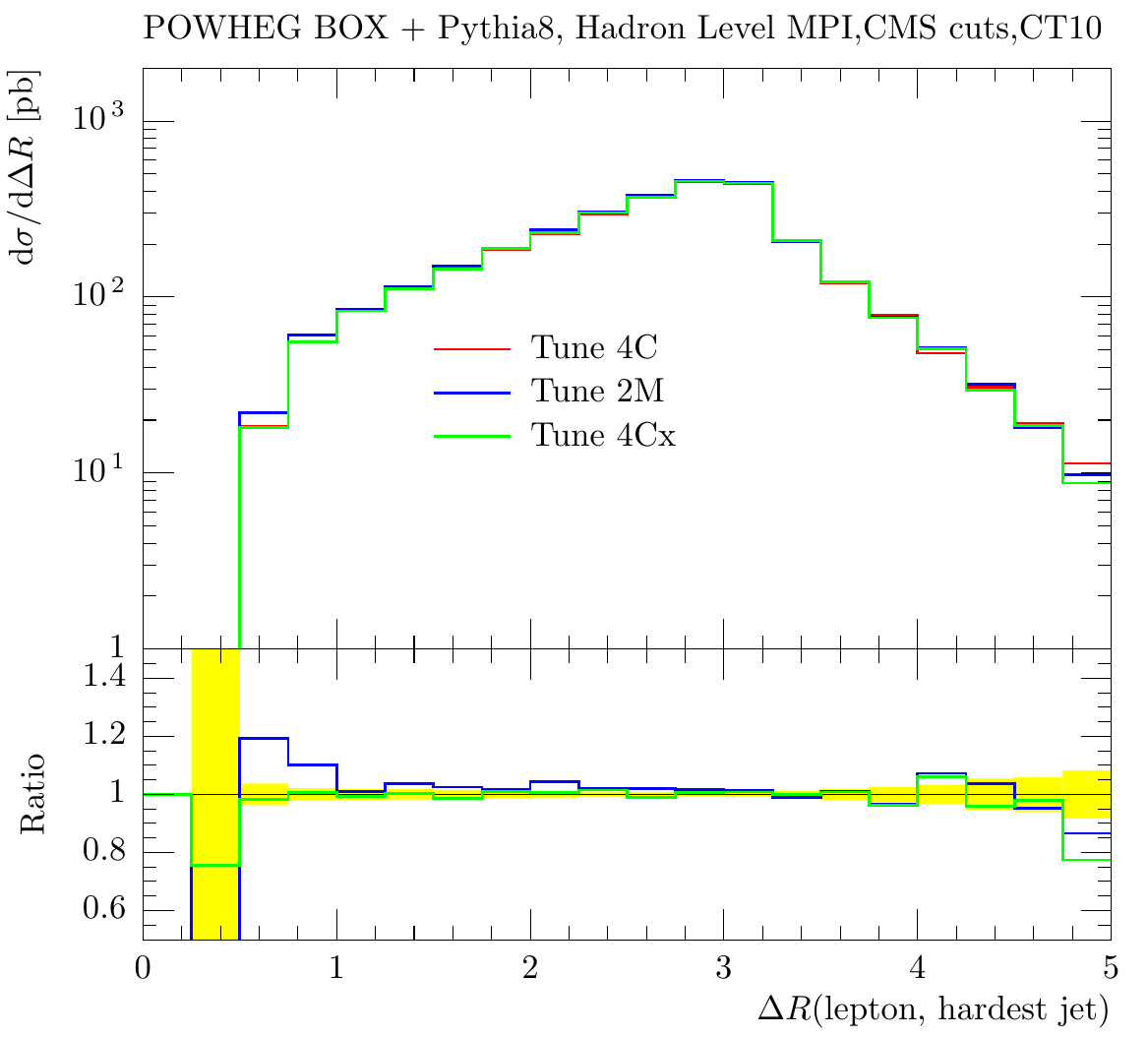}\hfill
\includegraphics[width=.48\textwidth]{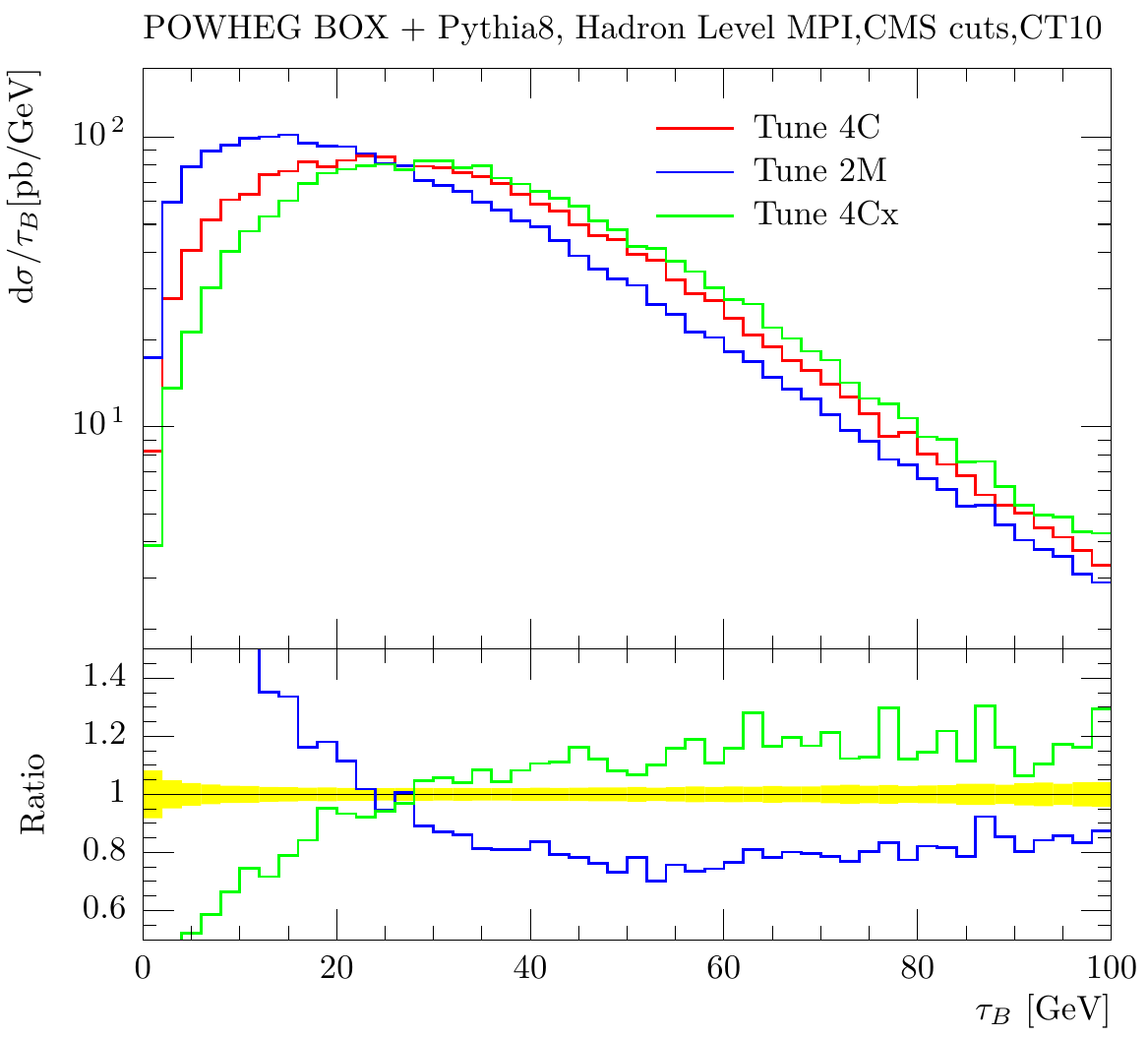}
\caption{Comparison between predictions using different \PythiaEight tunes,
  at hadron level with MPI, as predicted by \PowhegBox + \PythiaEight.}
\label{fig:indiv_PowhegPythia8_tunes}
\end{figure}

\FloatBarrier

\subsubsection{\texorpdfstring{\protect\PythiaEight}{Pythia8}}

For this study, \PythiaEight has been run stand-alone and including matrix 
elements with additional jets. Note that in \PythiaEight, multiple 
interactions are interleaved with space- and time-like showers, meaning that 
in general, MPI and parton showers cannot be disentangled by just 
switching off secondary scatterings. When referring to ``Hadron Level", we 
mean after the interleaved evolution (including QED splittings), and after 
hadronisation. For the sake of comparison, ``Shower Level"  indicates
results after (interleaved) final- and initial-state radiation, switching 
multiparton interactions off. All results presented in this section are 
generated with CTEQ6L1 parton distributions for protons colliding at 
$E_{\textnormal{\tiny{CM}}}=7000$ GeV. \textsc{Ckkw-l}-merged samples include
up to three additional jets, taken from \Madgraph/\Madevent.

\begin{figure}[t]
 \includegraphics[width=.48\textwidth]{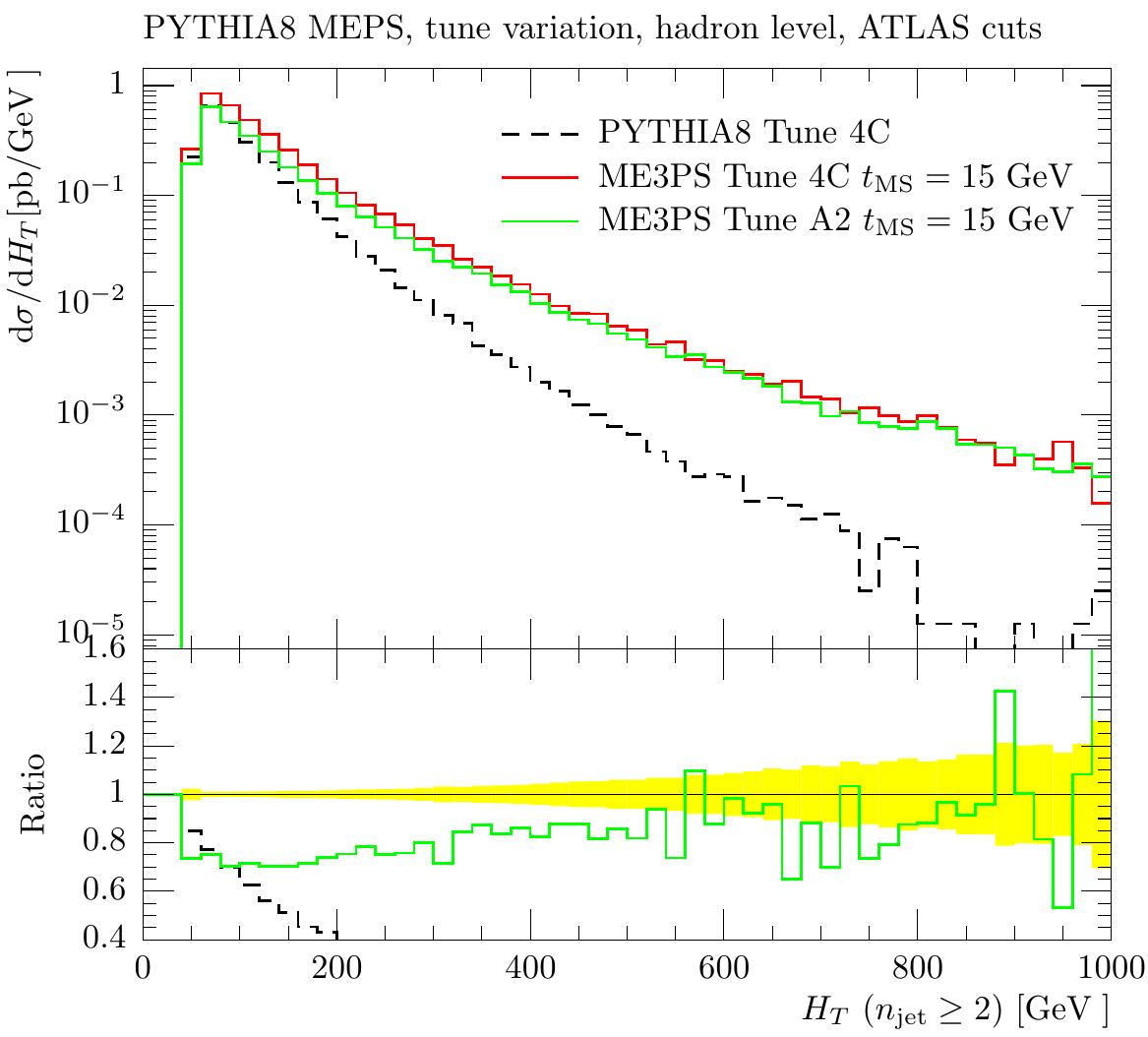}\hfill
 \includegraphics[width=.48\textwidth]{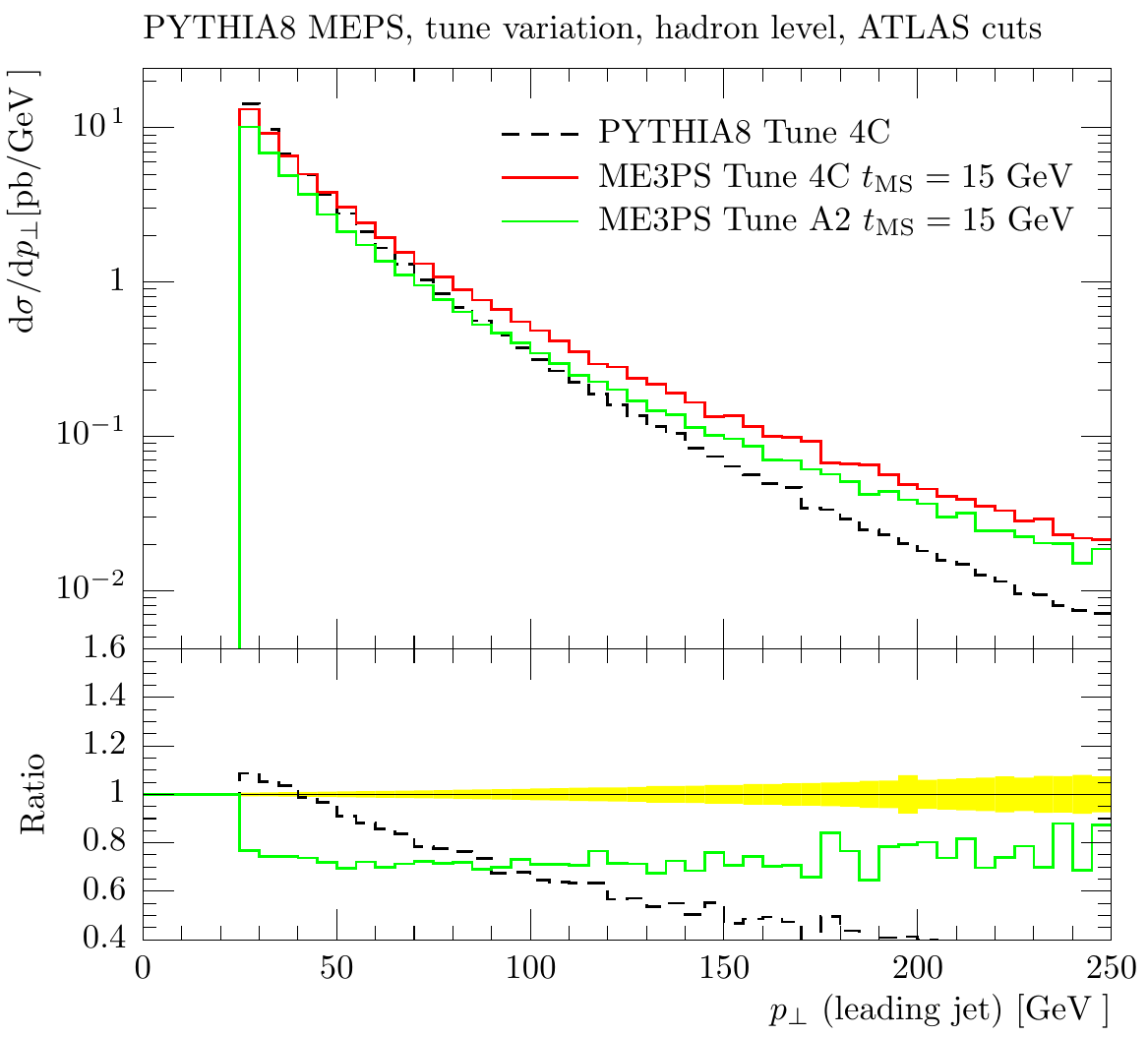}\\
 \includegraphics[width=.48\textwidth]{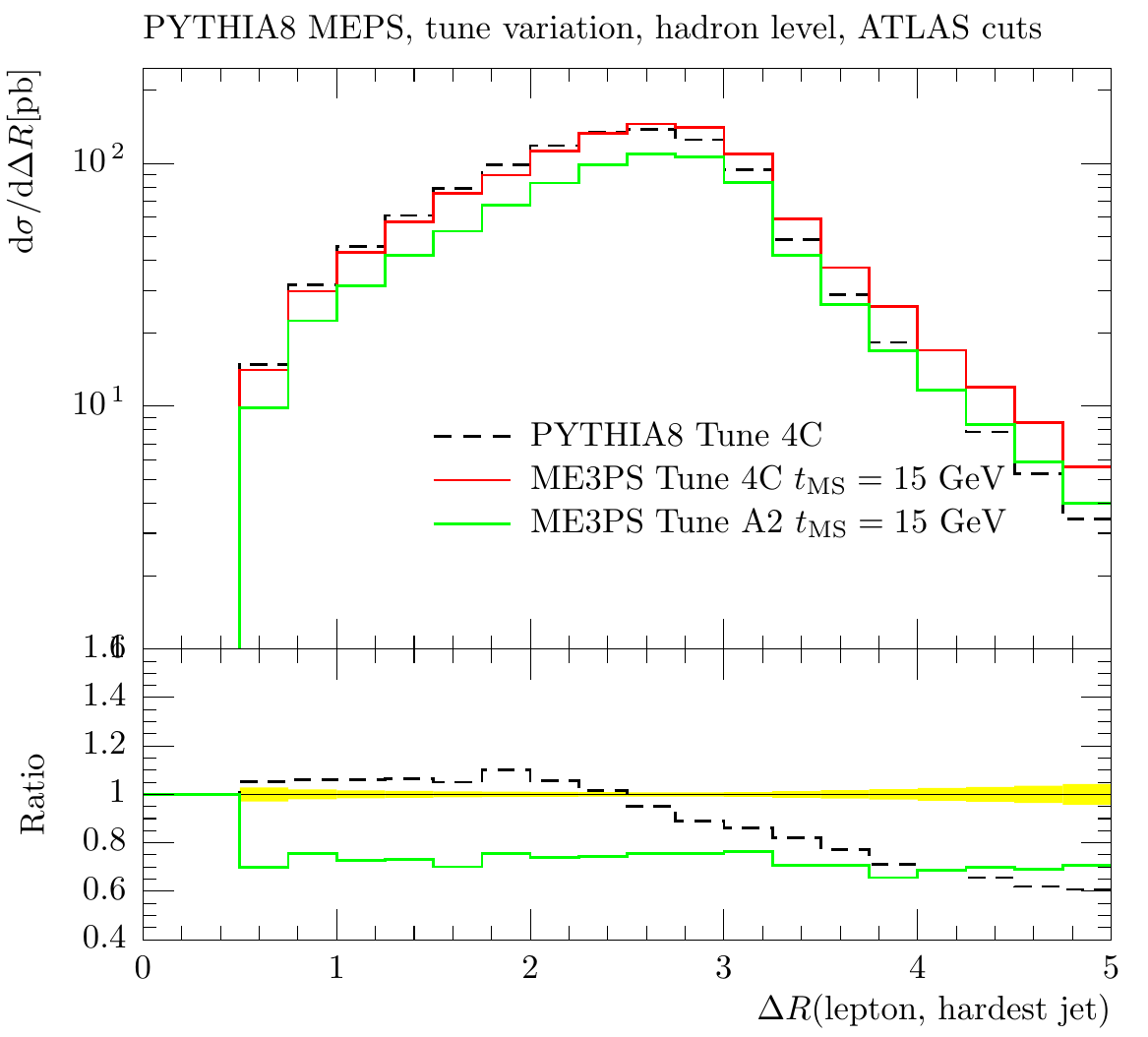}\hfill
 \includegraphics[width=.48\textwidth]{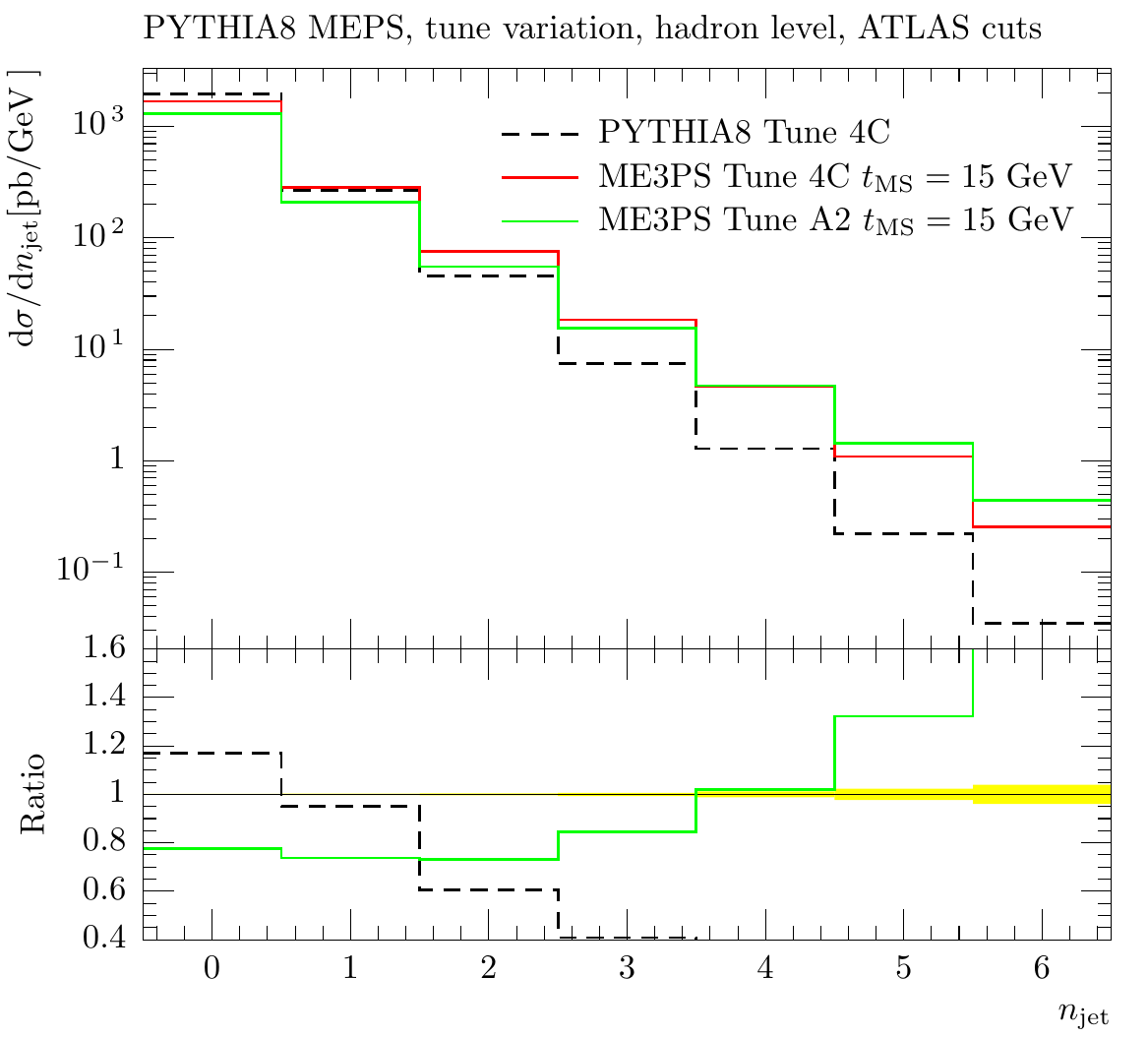}
 \caption{Tuning variations for \PythiaEight at hadron level.
 The plots show
 the $H_T$-distribution when requiring at least two jets (upper left), the 
 $p_\perp$ of the hardest jet (upper right), the $\Delta R$-separation of 
 lepton and the hardest jet (lower left), and the number of jets (lower right).
 The lower insets show the ratio of the samples in the upper half to 
 ME3PS (Tune 4C, y-blind treatment). All merged plots are produced with a 
 merging scale of $\tms=15$ GeV.}
\label{fig:indiv_Pythia8_tuneVariation_HTjet21_jetpt0_dRj0l_njet}
\end{figure}

\FigRef{fig:indiv_Pythia8_tuneVariation_HTjet21_jetpt0_dRj0l_njet} 
exemplifies how changes in the tuning of the event generator can affect the 
outcome of merged calculations in \PythiaEight. For this, we produce 
predictions for Tune 4C \cite{Corke:2010yf} and Tune A2 \cite{ATLAS:2011dk}. 
In general we observe only modest shape changes of up to about 20\% in 
observables, when comparing the two merged predictions, lending confidence to 
the statement that the tuning did not artificially produce hard scale physics.
Normalisation changes between 4C and A2 can be explained by a difference in 
Sudakov suppression: Since Tune 4C integrates the splitting kernels over a
smaller region of phase space, the suppression generated by trial showers is
less pronounced.
The increase in the number of jets in Tune A2 with respect to Tune 4C, after 
the third jet, is expected, because the generation of the fourth jet is handled
solely by the parton shower. Since 4C allows less phase space for these 
emissions by enforcing rapidity ordering, A2 will look harder. It is
debatable whether including rapidity ordering into the tuning makes the tune
mimic hard scale effects. The scales at which the fourth jet is produced
are certainly close to the scale of (hard) multiple interactions, which is in 
turn closely connect to soft physics. Although the enforced rapidity ordering
in Tune 4C might be considered questionable, we here take the pragmatic 
approach of considering the evolution both with and 
without enforced rapidity ordering. From the fact that up to three jets, the 
merged predictions of Tune 4C and Tune A2 only differ in normalisation, we 
anticipate that the effect of rapidity ordering will be reduced by merging 
more jets, since then, the number of jets above a cut-off will be dictated by 
the matrix element.

\begin{figure}[t]
 \includegraphics[width=.48\textwidth]{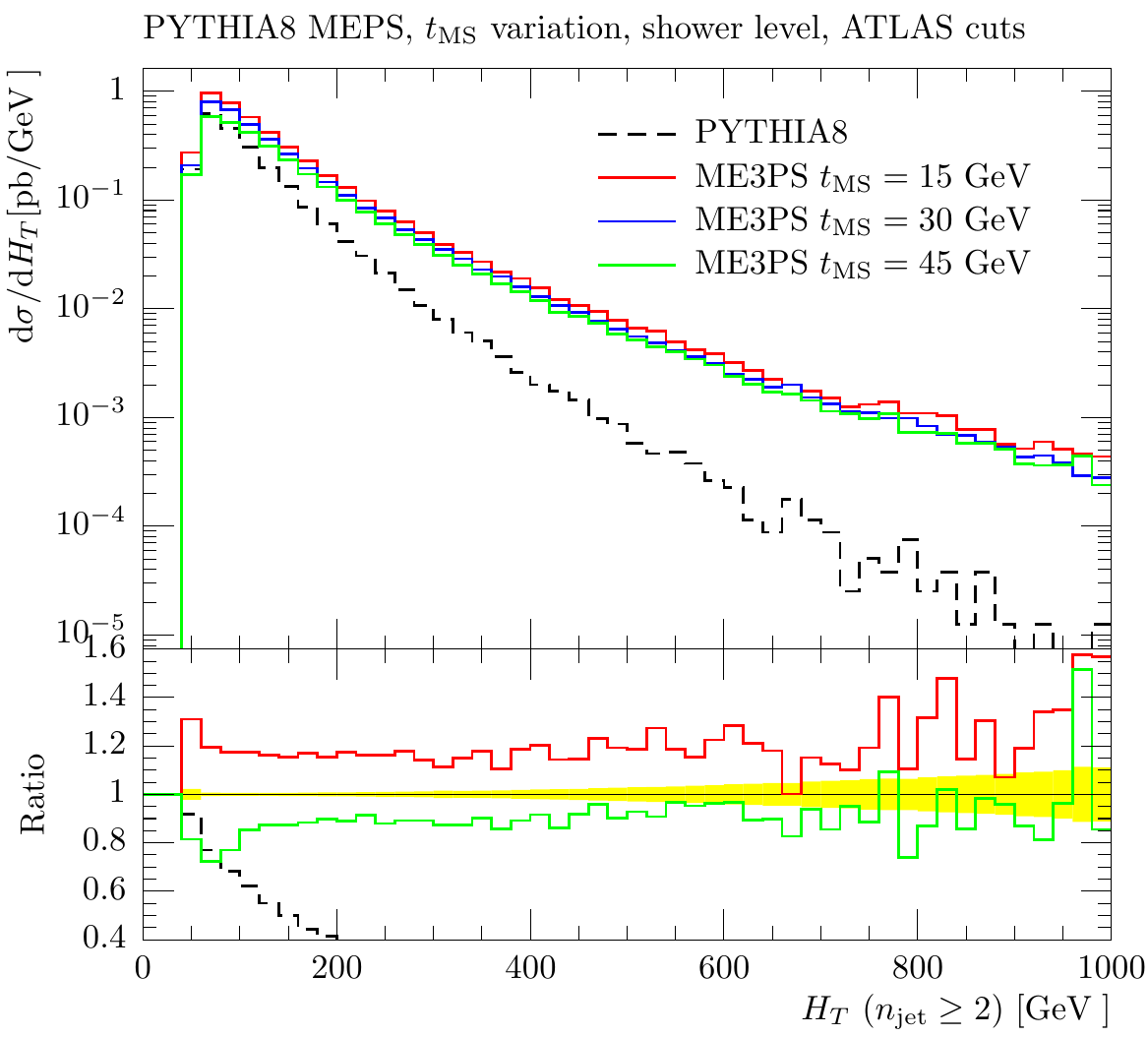}\hfill
 \includegraphics[width=.48\textwidth]{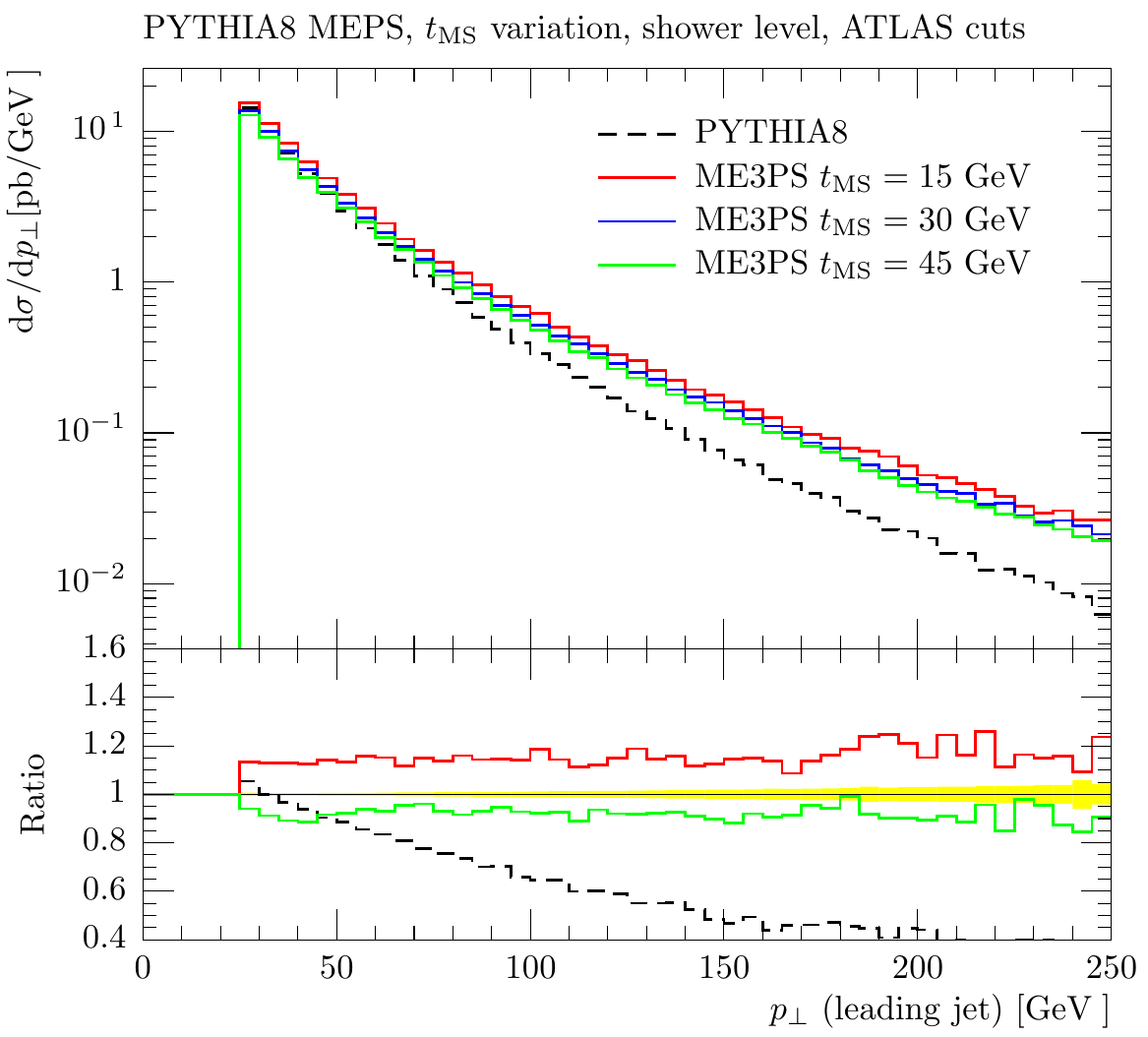}\\
 \includegraphics[width=.48\textwidth]{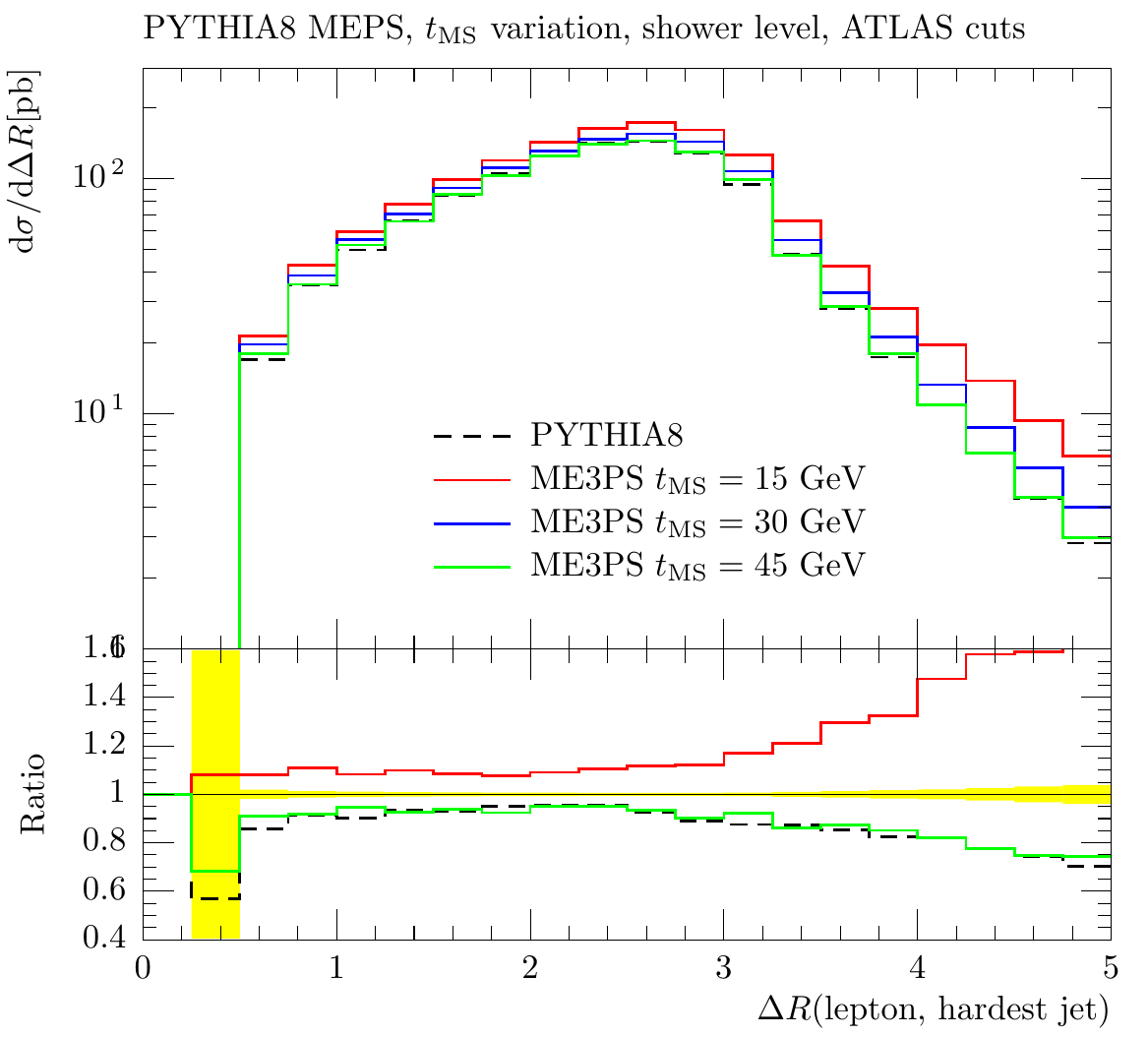}\hfill
 \includegraphics[width=.48\textwidth]{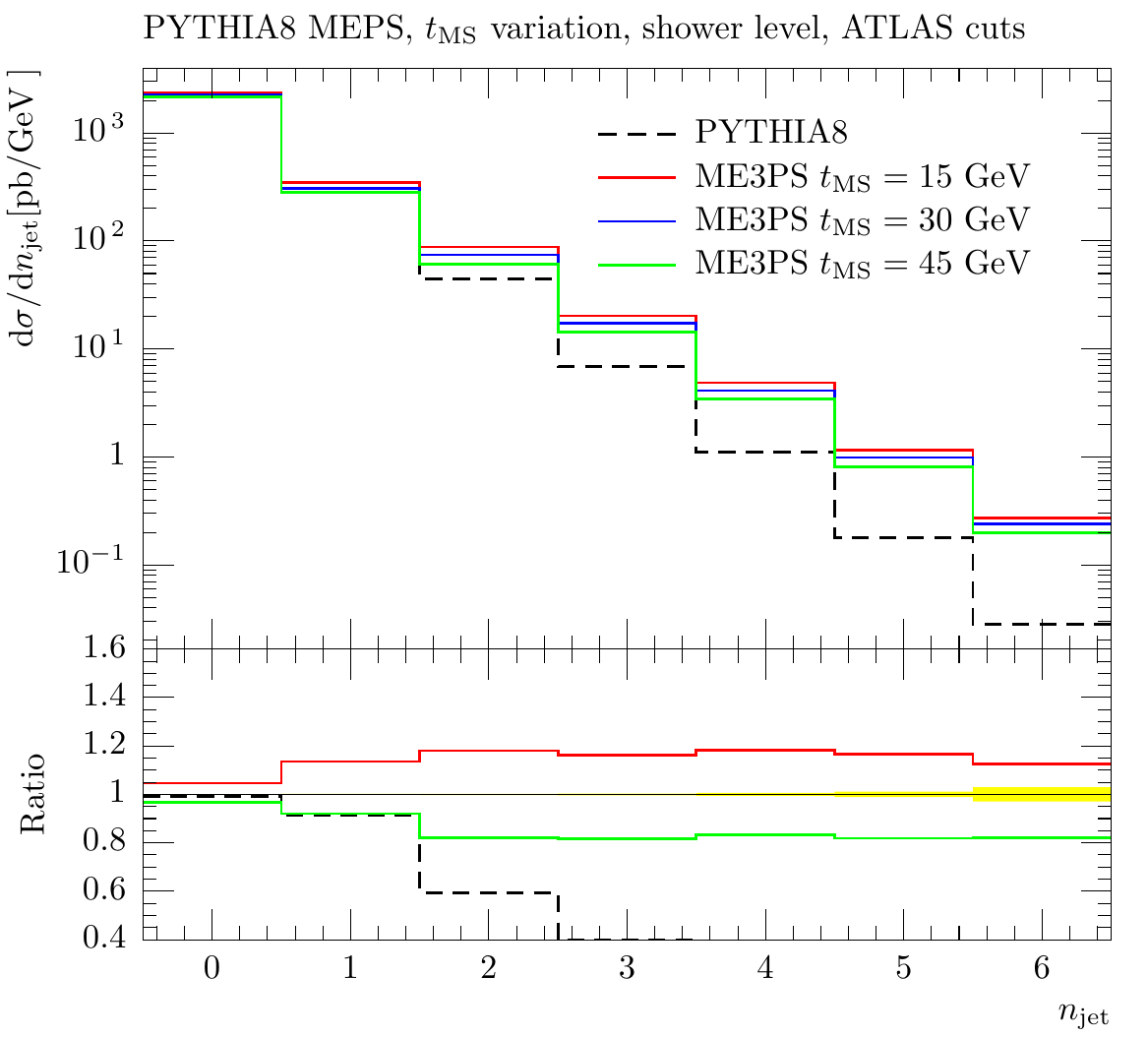}
 \caption{Variation of the merging scale value for \PythiaEight at shower level.
 The plots show
 the $H_T$-distribution when requiring at least two jets (upper left), the 
 $p_\perp$ of the hardest jet (upper right), the $\Delta R$-separation of 
 lepton and the hardest jet (lower left), and the number of jets (lower right).
 The lower insets show the ratio of the samples in the upper half to ME3PS for 
 $\tms=30$ GeV. All plots are generated using Tune 4C (y-blind treatment).}
\label{fig:indiv_Pythia8_scaleVariation_HTjet21_jetpt0_dRj0l_njet}
\end{figure}

\begin{figure}[t!]
 \includegraphics[width=.48\textwidth]{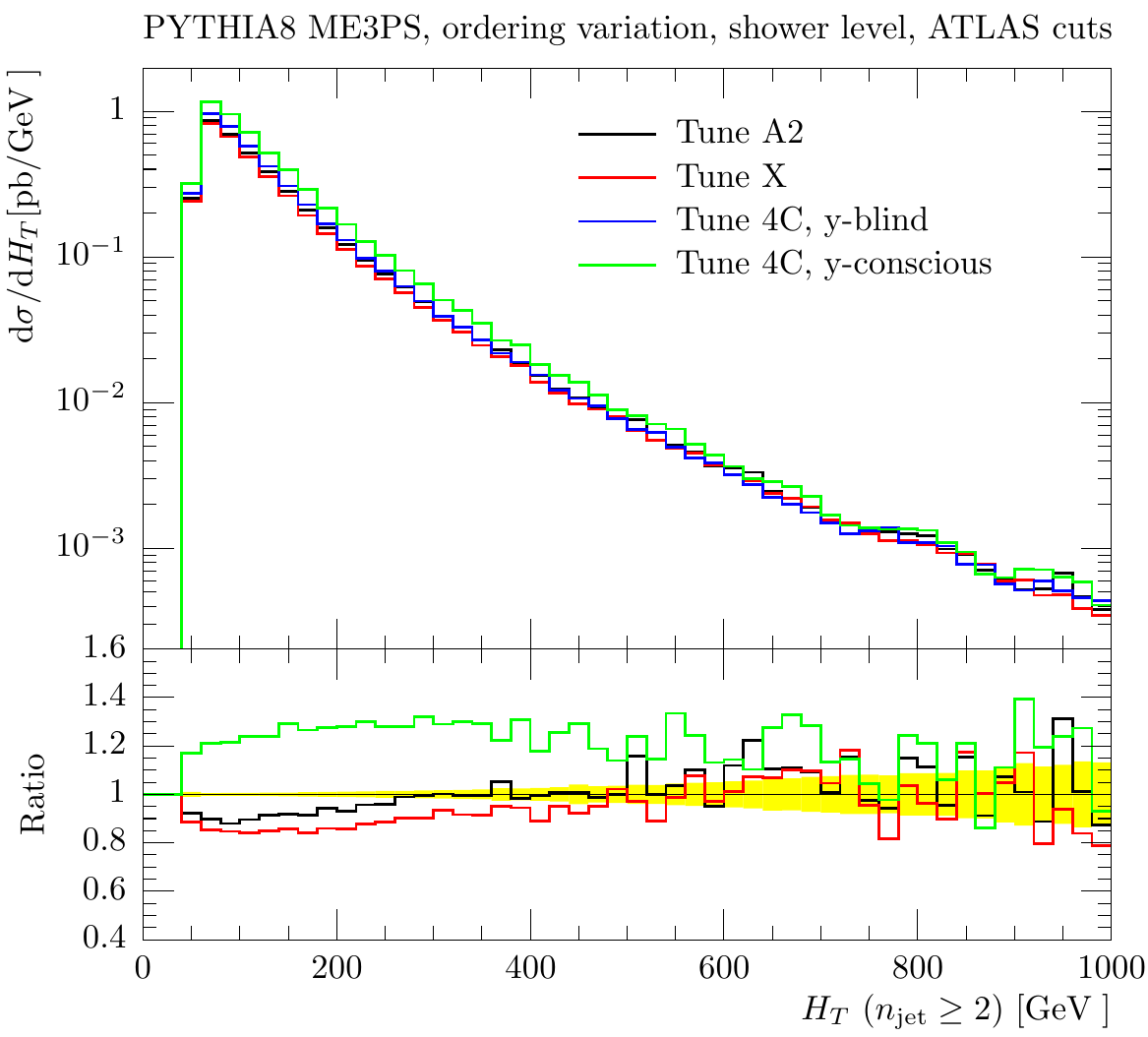}\hfill
 \includegraphics[width=.48\textwidth]{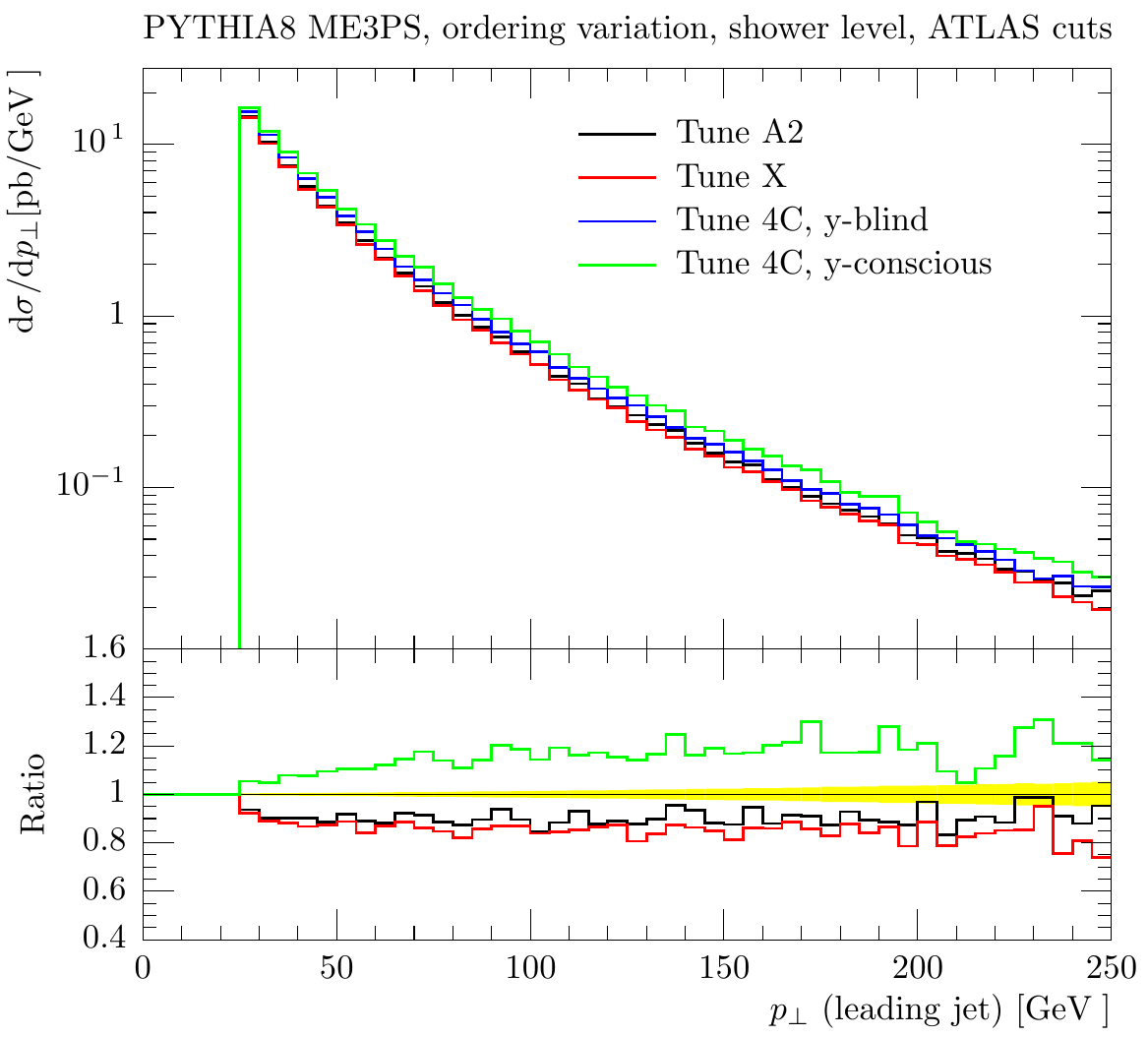}\\
 \includegraphics[width=.48\textwidth]{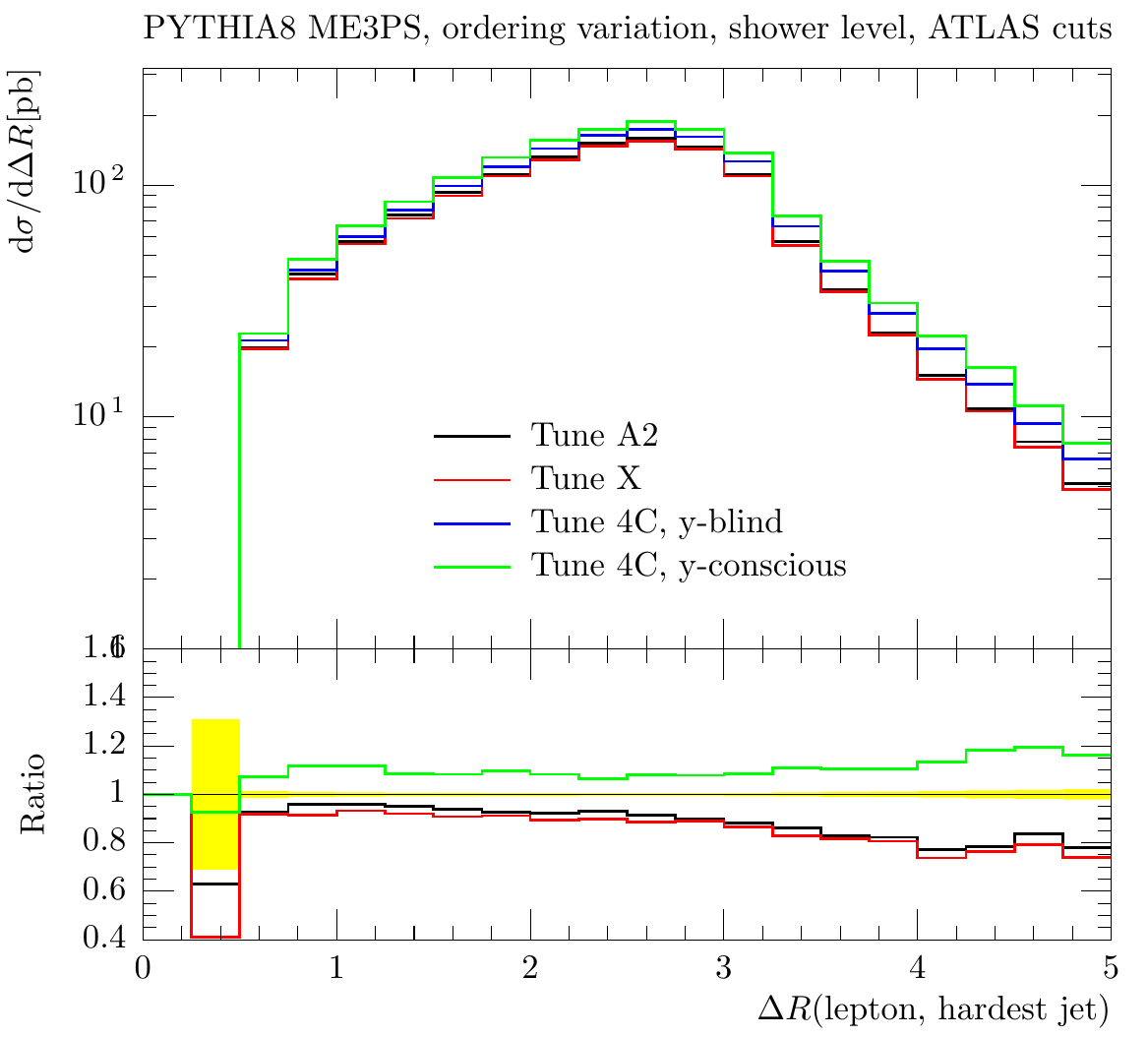}\hfill
 \includegraphics[width=.48\textwidth]{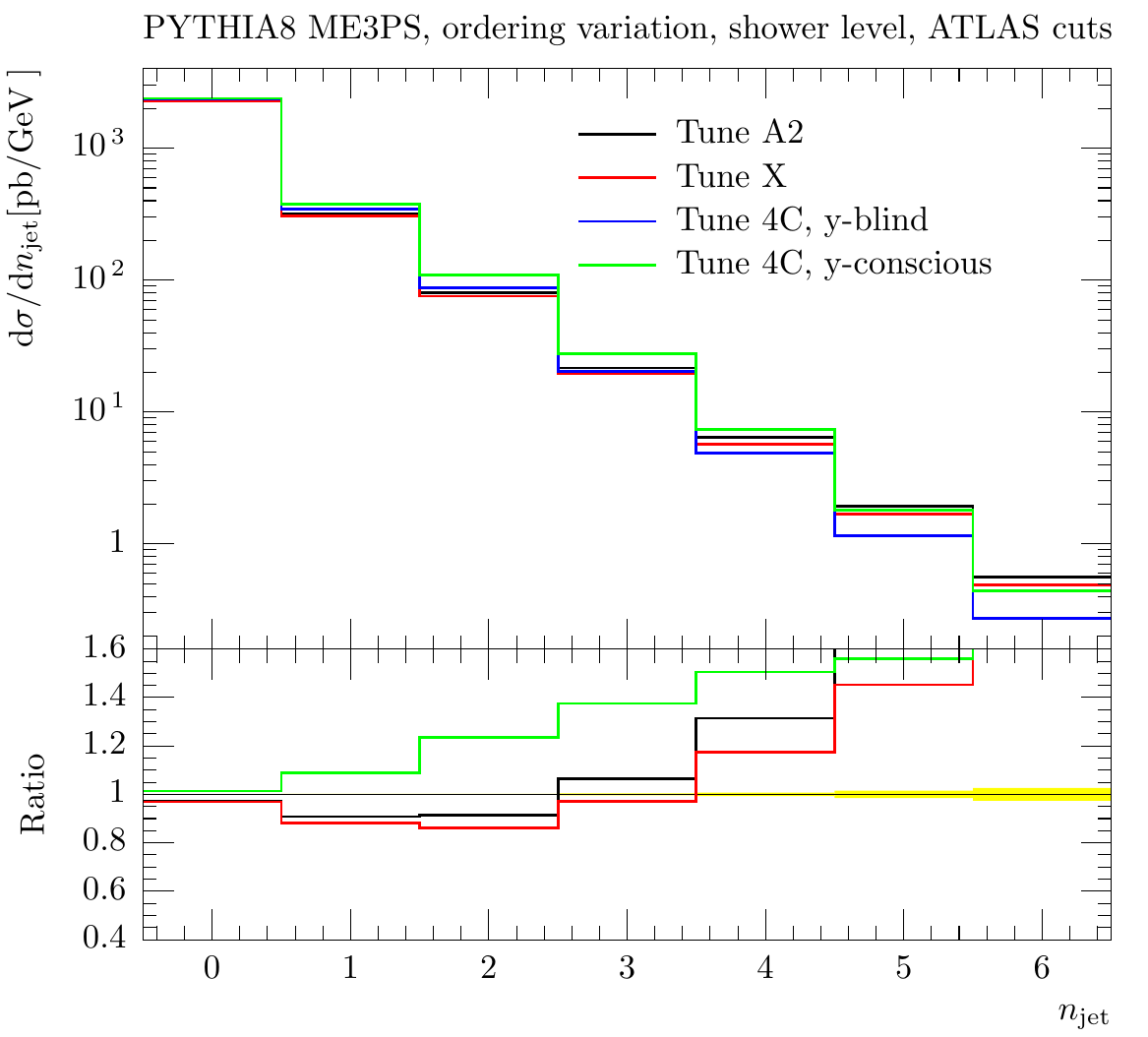}
 \caption{Variation of the criterion employed to favour ``ordered histories" in
 \PythiaEight at shower level.
 The plots show
 the $H_T$-distribution when requiring at least two jets (upper left), the 
 $p_\perp$ of the hardest jet (upper right), the $\Delta R$-separation of 
 lepton and the hardest jet (lower left), and the number of jets (lower right).
 The lower
 insets show the ratio of the samples in the upper half to ME3PS (Tune 4C, 
 y-blind 
 treatment). All merged plots are produced with a merging scale of $\tms=15$ 
 GeV.}
\label{fig:indiv_Pythia8_orderingVariation_HTjet21_jetpt0_dRj0l_njet}
\end{figure}

In \FigRef{fig:indiv_Pythia8_scaleVariation_HTjet21_jetpt0_dRj0l_njet}, we 
investigate the impact of changes in the merging scale value. Again, we mainly 
see normalisation changes and only small changes in shape, which in most cases 
are smaller than changes due to different tunes. The $R$-separation between 
lepton and hardest
jet $\Delta R(\textnormal{lepton,hardest jet})$ shows significant shape changes
above $\pi$. This again is an effect of Tune 4C, and is greatly reduced in 
Tune A2\footnote{For the sake of compactness, merging scale variation plots for
Tune A2 could not be included here. We hope the reader is nevertheless
willing to consider the following argument -- which is only supported by the
omitted results.}, as can be inferred from the tune variation. However, even in
Tune A2, small shape changes remain, with the change becoming less pronounced 
when comparing two large merging scales. We take this as an indication that the
shower splitting probability -- giving radiative contributions to 
$\Delta R(\textnormal{lepton,hardest jet}) > \pi$ for high $\tms$ -- and the 
the matrix element, which fills the same region in for the low merging 
scale case, are indeed different from the second jet on. This also explains the 
difference between Tune 4C and Tune A2, which differ by the phase space regions
over which the splitting kernels are integrated.

Finally, in 
\FigRef{fig:indiv_Pythia8_orderingVariation_HTjet21_jetpt0_dRj0l_njet}, we address
the interplay of matrix element merging and ordering in the underlying shower
more carefully. The effect of different choices manifests itself again mainly
in changes of the normalisation of the plots, and is comparable in magnitude 
to the impact of merging scale variations. At first, the changes may seem 
counter-intuitive, and need clarification. For this, it is important to
remember the definition of ``y-blind" and ``y-conscious" in section 
\ref{sec:Pythia8MEPS_ordering_choices}. The y-blind treatment will -- 
irrespectively of rapidity configurations -- mainly choose histories ordered in
the shower evolution variable $\rho$, and only pick $\rho$-unordered histories
if no other ones have been constructed. However, in the y-conscious approach, 
once no history ordered both in rapidity \emph{and} $\rho$ is found, one 
amongst all un-ordered histories is
chosen probabilistically, irrespectively of the history being y-/$\rho$-/or y- 
and $\rho$-unordered. Since the ordering criterion is
stricter, un-ordered histories will be chosen more frequently, meaning that
$\rho$-unordered ones will also contribute more, compared to the y-blind case.
Matrix element states with no ordered histories will have a 
number of jets at very similar scales, so that the Sudakov suppression 
generated by trial showers will be smaller.
Moreover, for matrix element states in which the last reconstructed splitting
is unordered, the parton shower will be started at the larger of the unordered 
scales\footnote{This is the default choice in \PythiaEight. Other choices are 
available to the user.}, which can result in a slightly harder spectrum of 
resolved parton shower jets.
Because $\rho$-unordered states are picked more often when requiring a tighter
ordering criterion, this leads in visible differences.
The y-conscious method might seem somewhat artificial,
considering that it introduces a larger dependence on states outside the range
of even the y-unordered shower variant. Nevertheless, the y-blind and 
y-conscious prescriptions are equivalent to the accuracy of the (y-ordered)
shower, so that both should be investigated when assessing the quality of the 
merging. From the visible changes, we can infer that different treatments of
formally sub-leading effects do matter. For the y-ordered evolution, these are more 
visible since the accuracy of the shower itself is worse, so that the effects of
including matrix element states cancel to a lesser degree. It is interesting to
note that the deviations between the different prescriptions are considerably
smaller if the merging scale is increased, again hinting at a reduced shower
accuracy if the evolution is ordered in multiple variables.

\FigRef{fig:indiv_Pythia8_orderingVariation_HTjet21_jetpt0_dRj0l_njet}
further shows distributions labelled Tune X, which have been generated
by using Tune 4C, removing the rapidity constraint on space-like emissions, and 
treating histories y-blind. Results of these runs, as expected, 
closely follow Tune A2. The outcome of both Tune A2 and Tune X differs only
slightly from the Tune 4C (y-blind) curves, consolidating the conclusion that 
shifting fractions of $\rho$-un-ordered histories are responsible for the 
deviations between the y-blind and y-conscious methods. As in the discussion
of tuning variation, the similarity in the results of the merged calculation
for Tune 4C and Tune A2 breaks down once we examine jets that are solely
produced by the shower, i.e.\ starting from the fourth jet.

\FloatBarrier

\subsubsection{\texorpdfstring{\protect\Sherpa}{Sherpa}}
\label{Sec:Results:Sherpa}
As described in \SecRef{Sec:Codes:Sherpa}, \Sherpa has been run in two 
modes for this comparison of \LHC predictions. The results for the
conventional merging of towers of tree-level matrix elements,
\Sherpa\MEPS, are presented in \SecRef{Sec:Results:Sherpa:MEPS} while
the results of its enhancement to NLO accuracy in the core $W$
production process, \Sherpa\MENLOPS, are displayed in
\SecRef{Sec:Results:Sherpa:MENLOPS}. As detailed earlier, all
parameters have been chosen identically otherwise. The precise
requirements regarding the event selection and the definitions of the
observables used in this comparsion follow the CMS cut specifications
and can be found in \AppRef{App:Observables}.

\paragraph{\texorpdfstring{\protect\Sherpa\MEPS}{Sherpa MEPS}}
\label{Sec:Results:Sherpa:MEPS}

\begin{figure}[p!]
  \includegraphics[width=.48\textwidth]{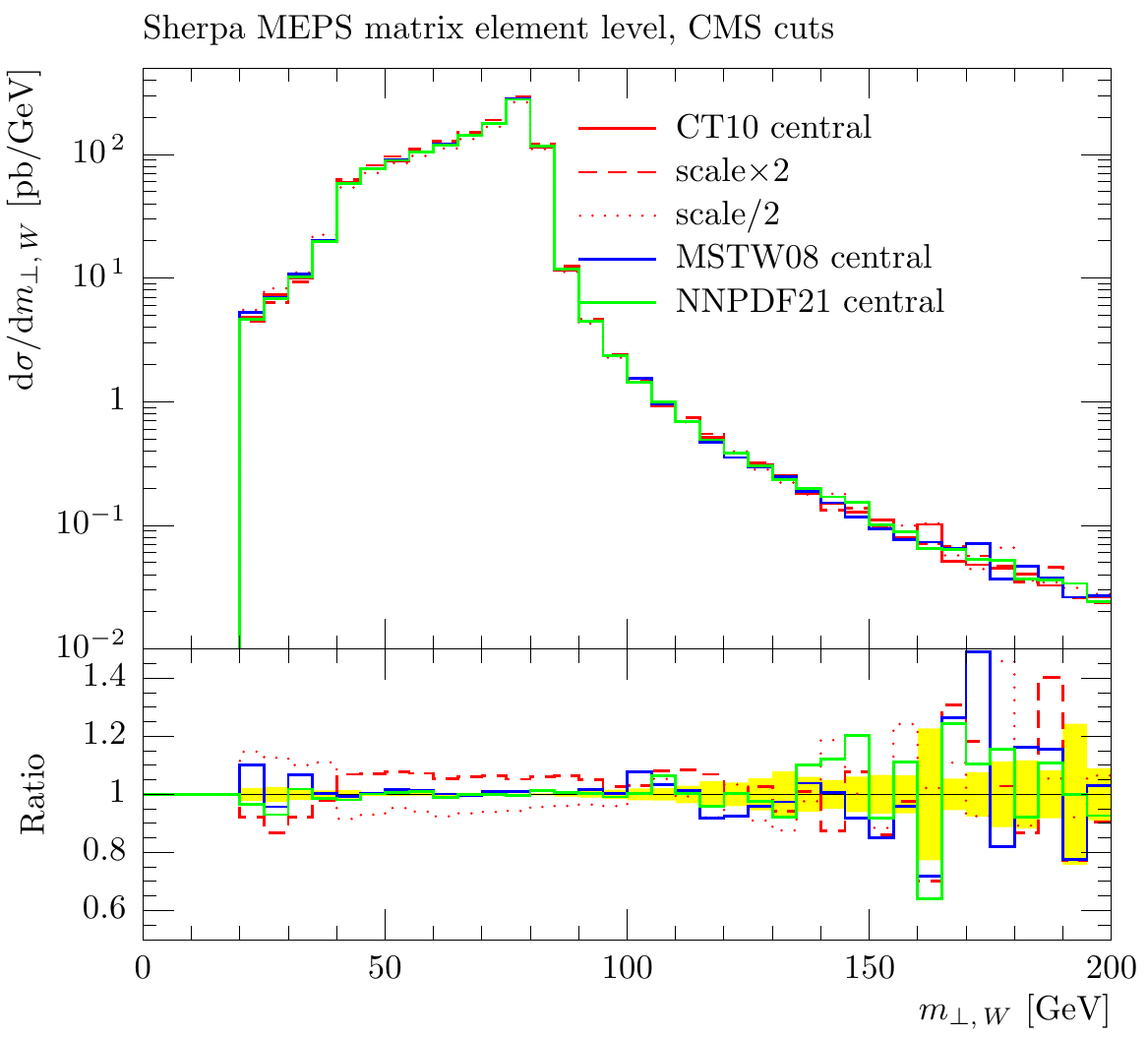}\hfill
  \includegraphics[width=.48\textwidth]{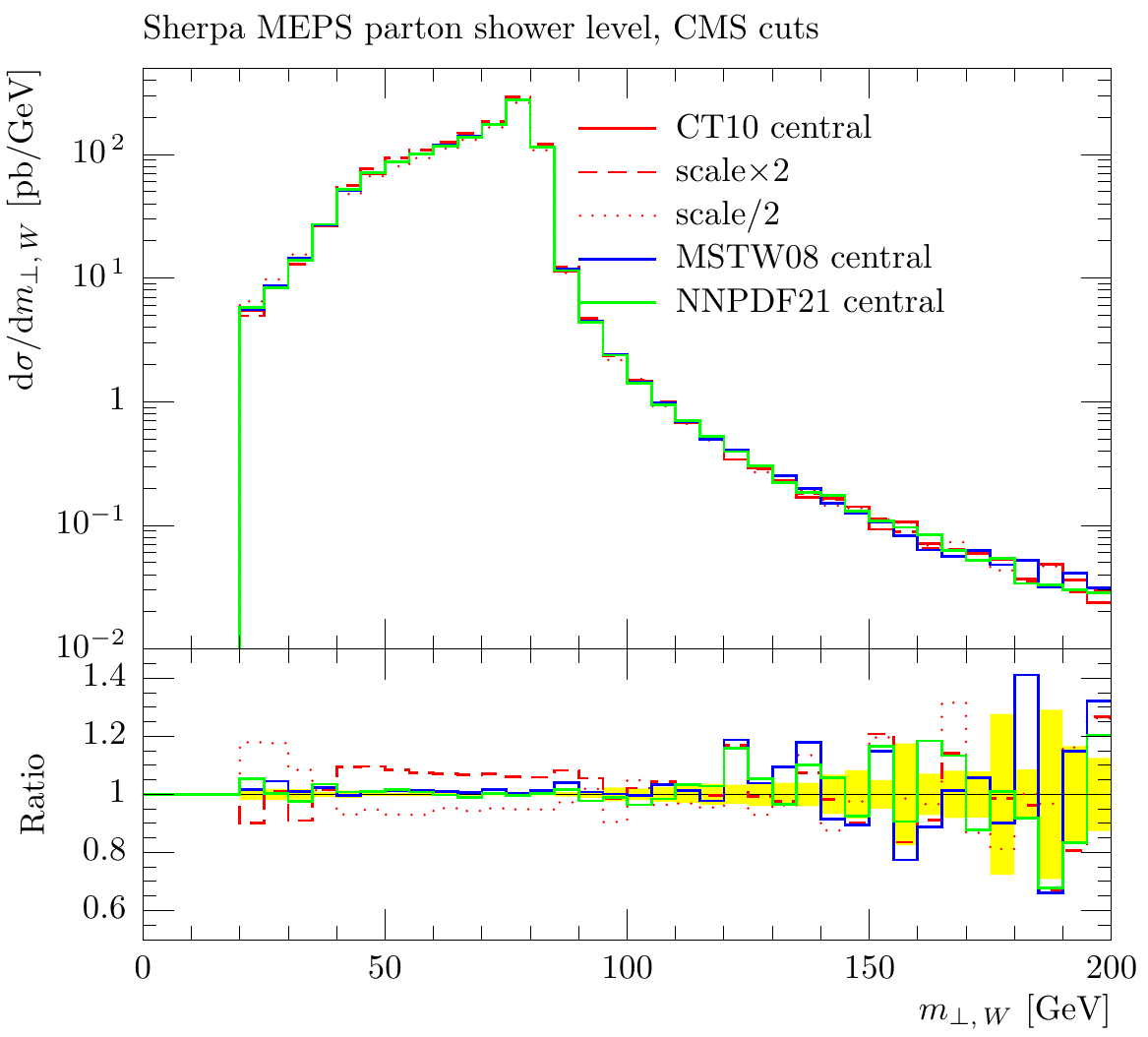}\\
  \includegraphics[width=.48\textwidth]{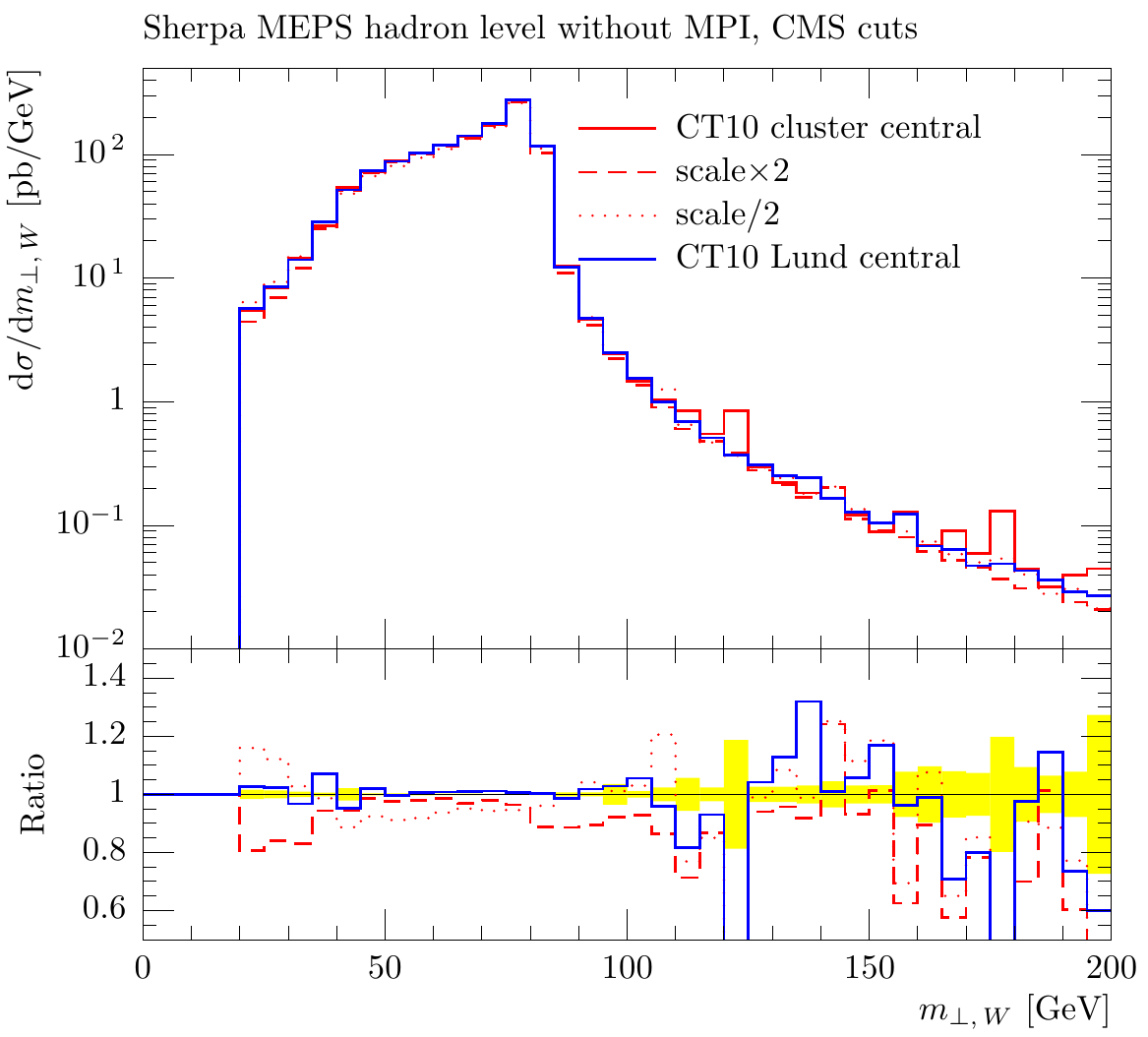}\hfill
  \includegraphics[width=.48\textwidth]{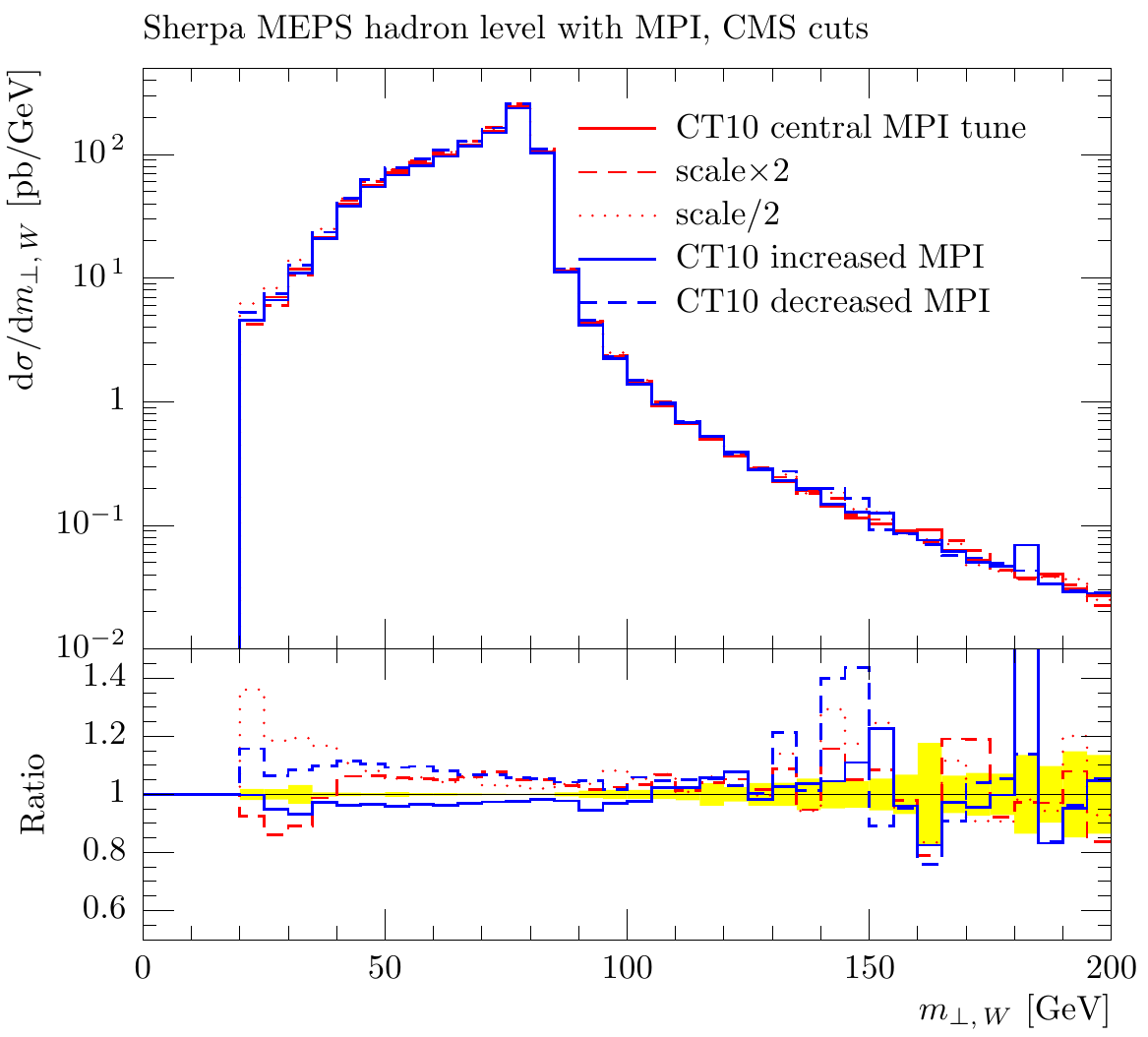}\\
  \includegraphics[width=.48\textwidth]{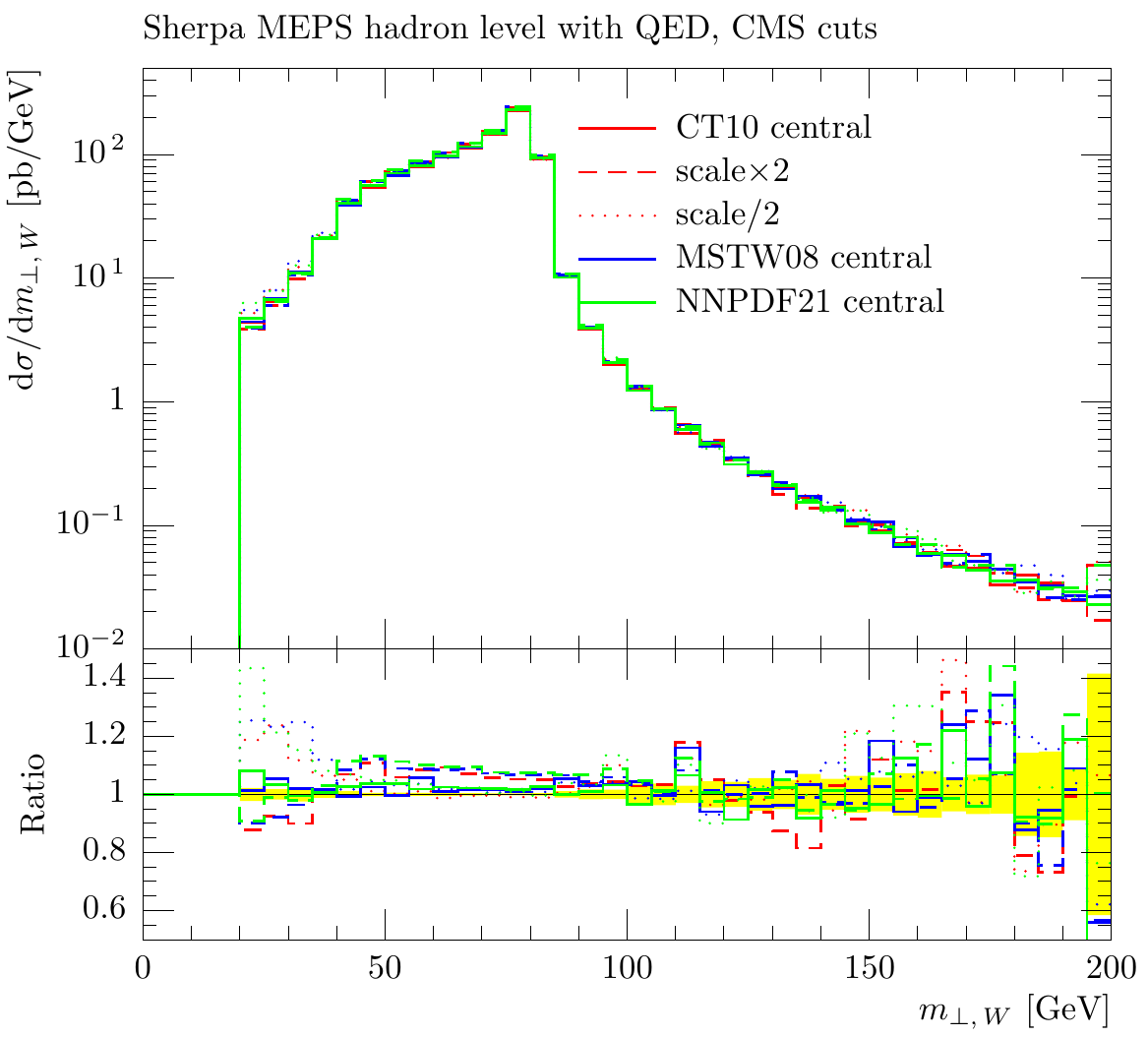}\hfill
  \includegraphics[width=.48\textwidth]{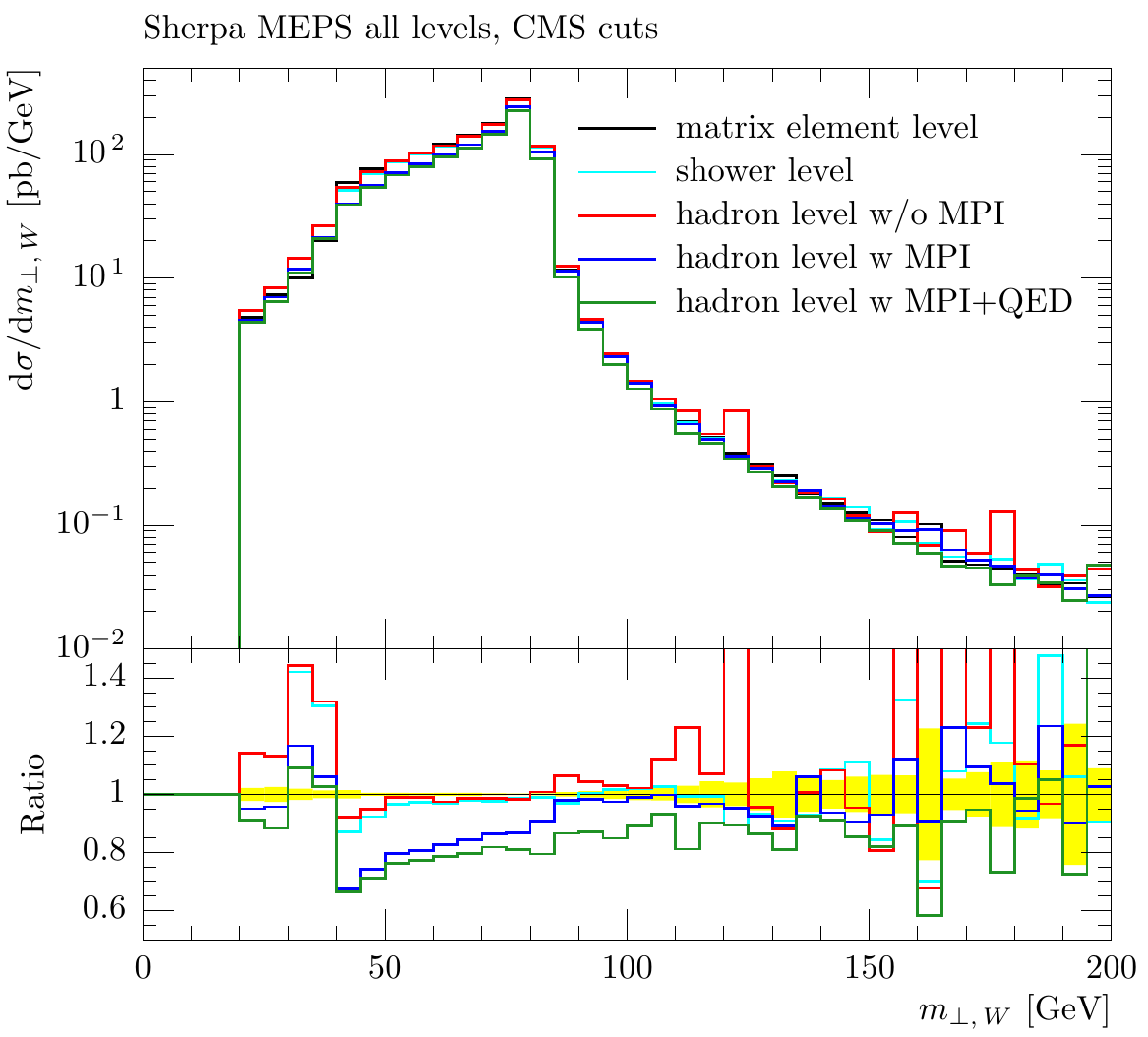}
  \caption{
	  \Sherpa\MEPS. Uncertainty of the transverse mass of the reconstructed $W$ on the 
	  matrix element level (upper left), after parton showering (upper 
	  right), including hadronisation correction (centre left), 
	  multiple parton interactions (centre right), and QED corrections 
	  (lower left). The lower right panel shows the evolution of the 
	  central value.
	  \label{Fig:Results:Sherpa:MEPS:Wmt}
  }
\end{figure}

\begin{figure}[p!]
  \includegraphics[width=.48\textwidth]{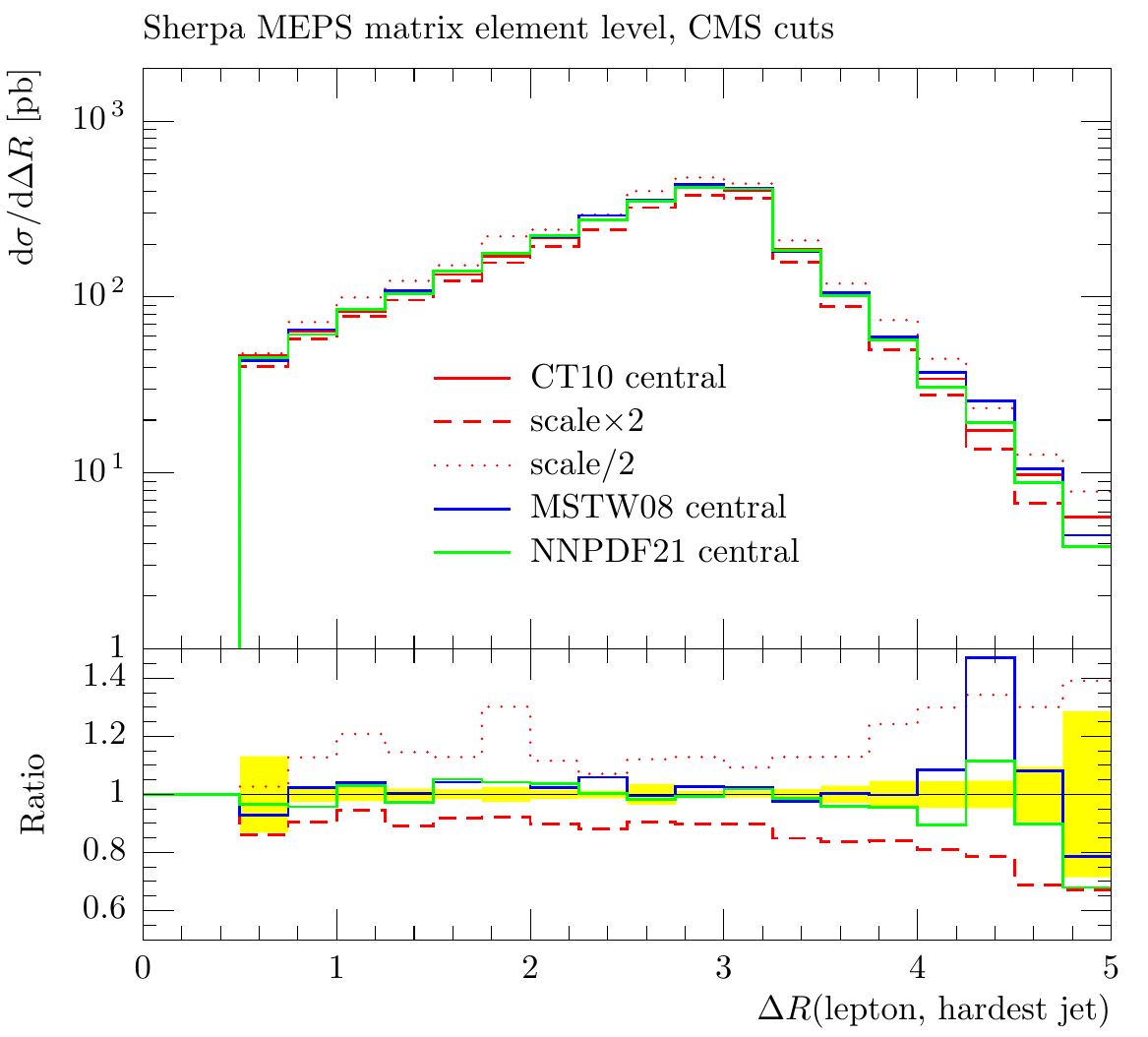}\hfill
  \includegraphics[width=.48\textwidth]{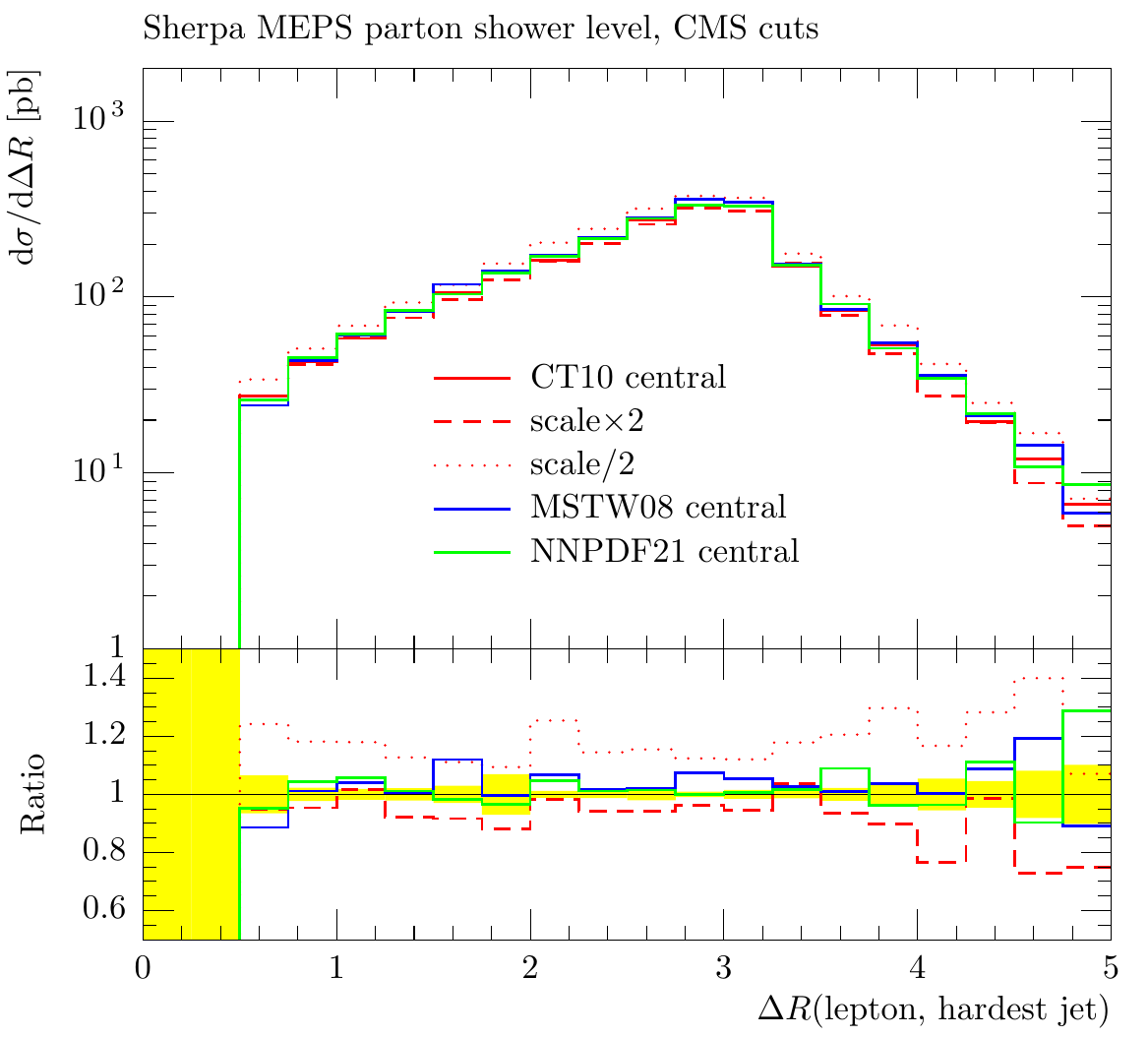}\\
  \includegraphics[width=.48\textwidth]{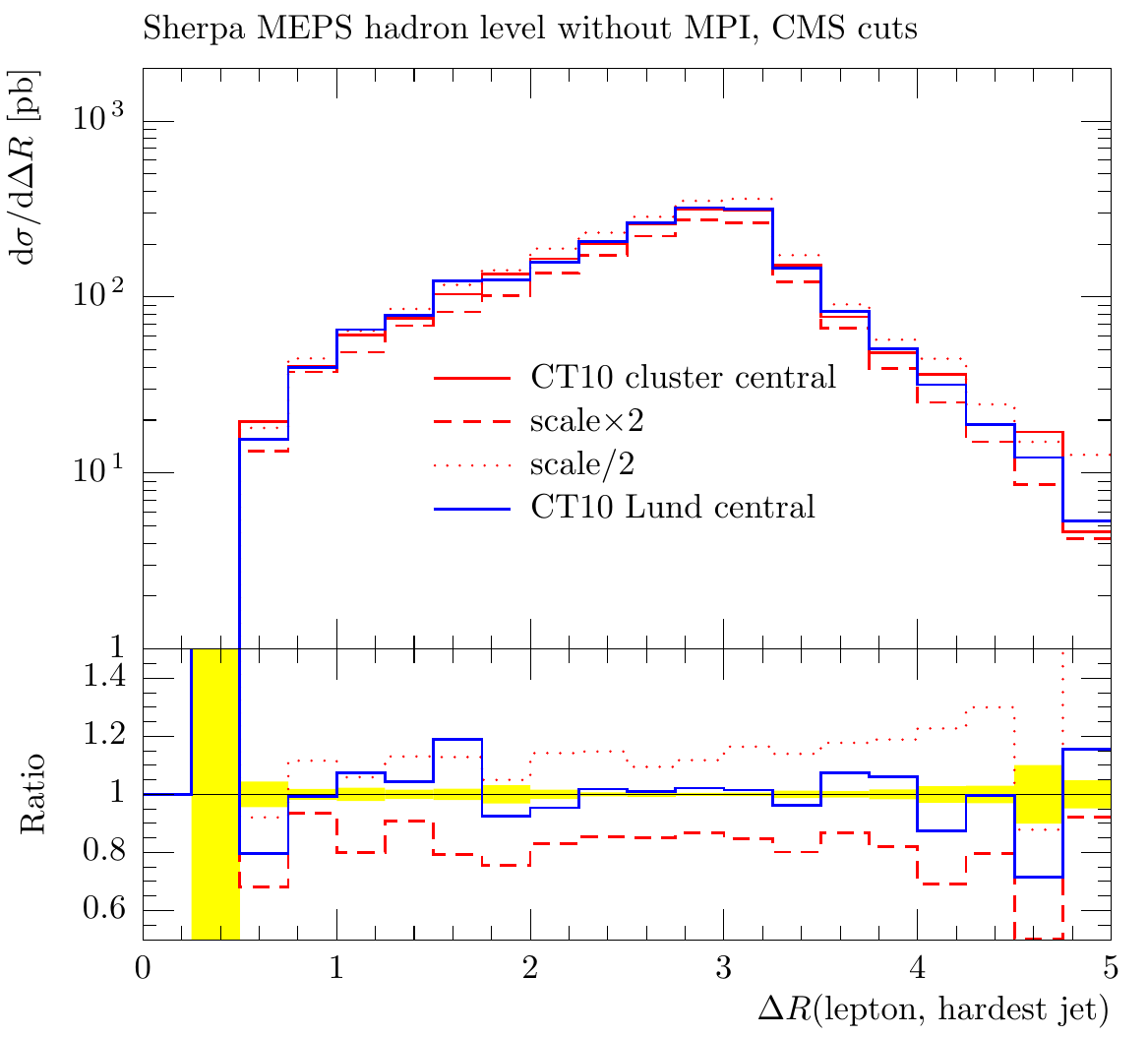}\hfill
  \includegraphics[width=.48\textwidth]{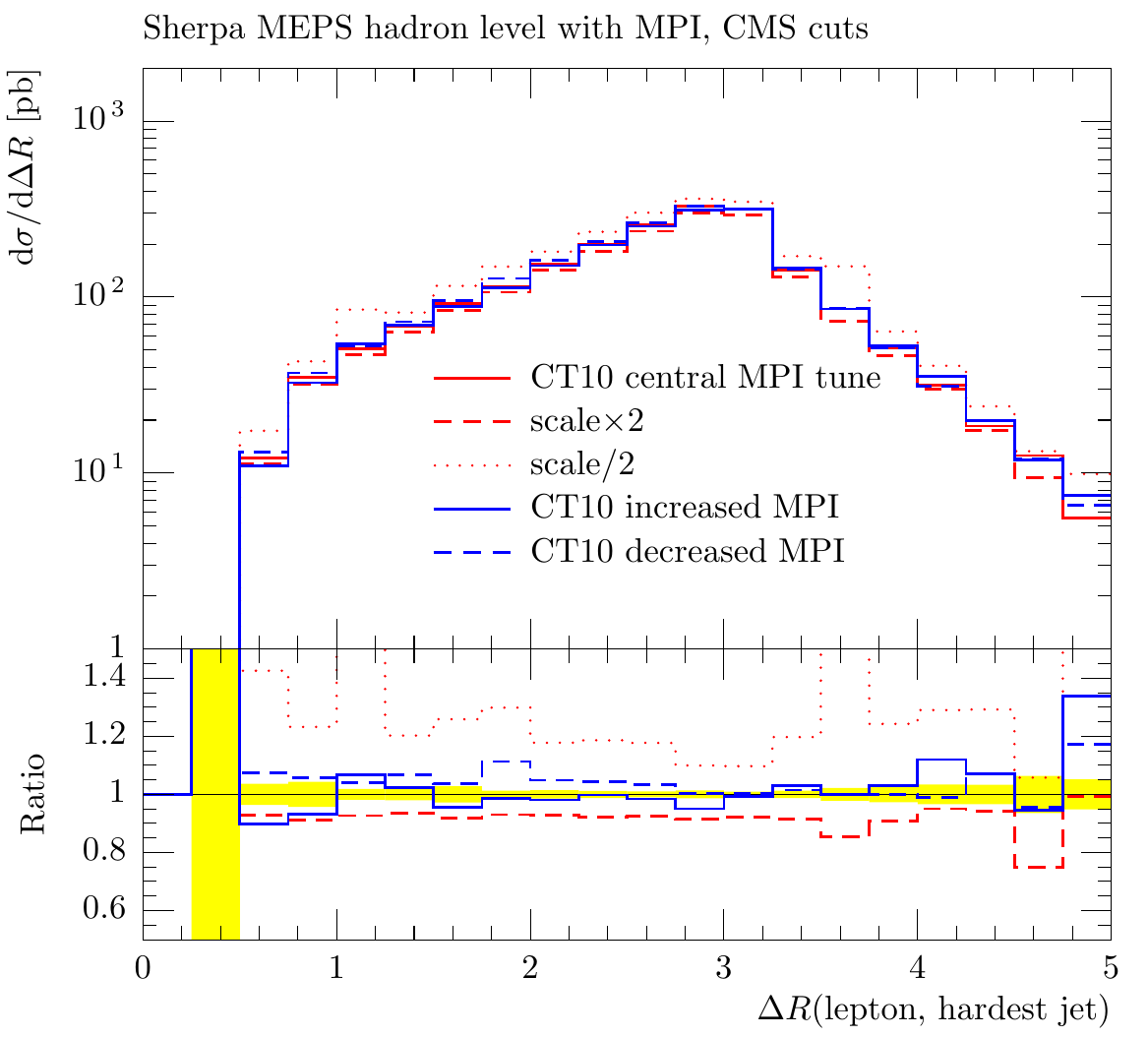}\\
  \includegraphics[width=.48\textwidth]{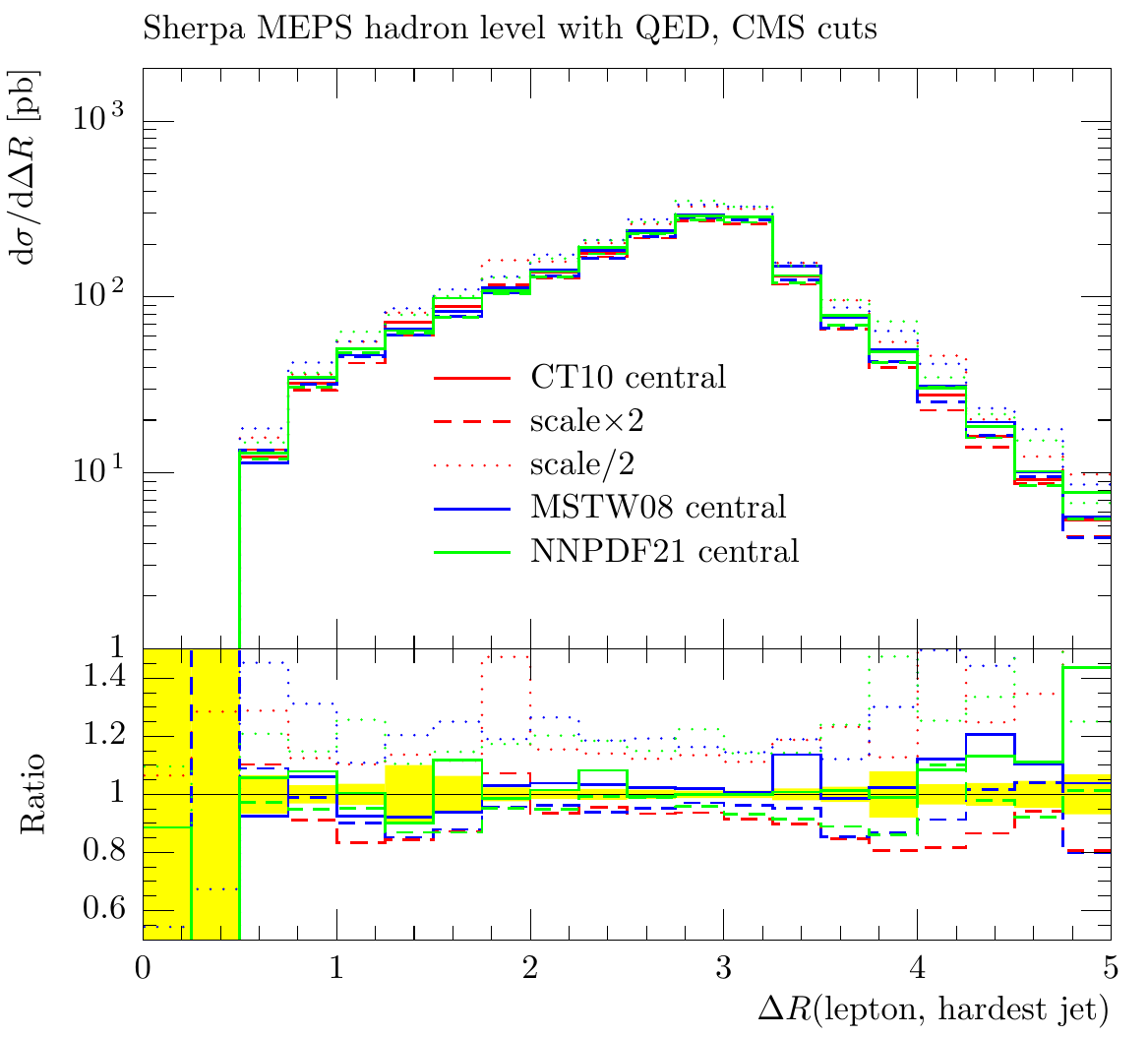}\hfill
  \includegraphics[width=.48\textwidth]{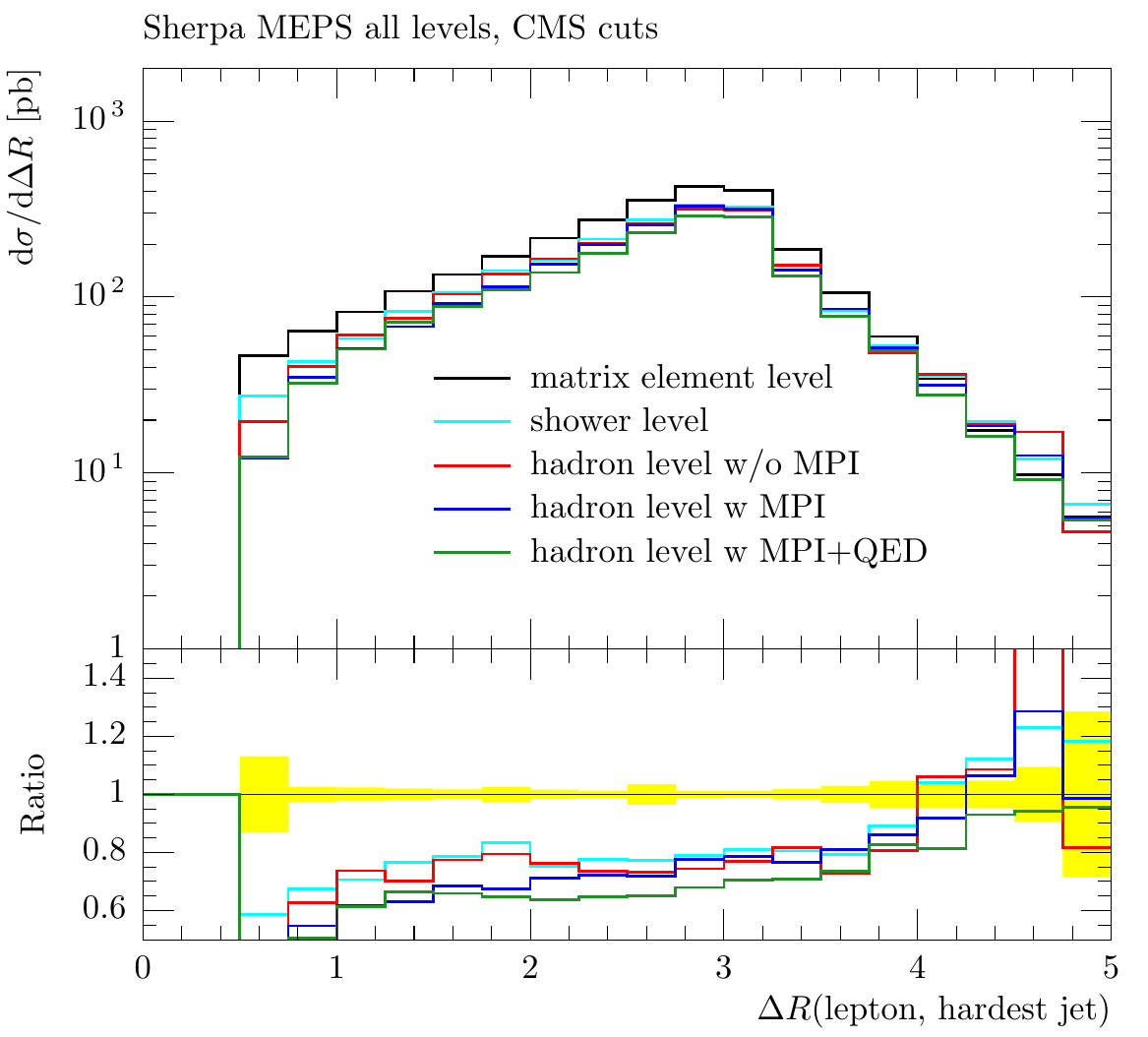}
  \caption{
	  \Sherpa\MEPS. Uncertainty of the angular separation of the charged lepton and the hardest jet on the 
	  matrix element level (upper left), after parton showering (upper 
	  right), including hadronisation correction (centre left), 
	  multiple parton interactions (centre right), and QED corrections 
	  (lower left). The lower right panel shows the evolution of the 
	  central value.
	  \label{Fig:Results:Sherpa:MEPS:dRj0l}
  }
\end{figure}

\begin{figure}[p!]
  \includegraphics[width=.48\textwidth]{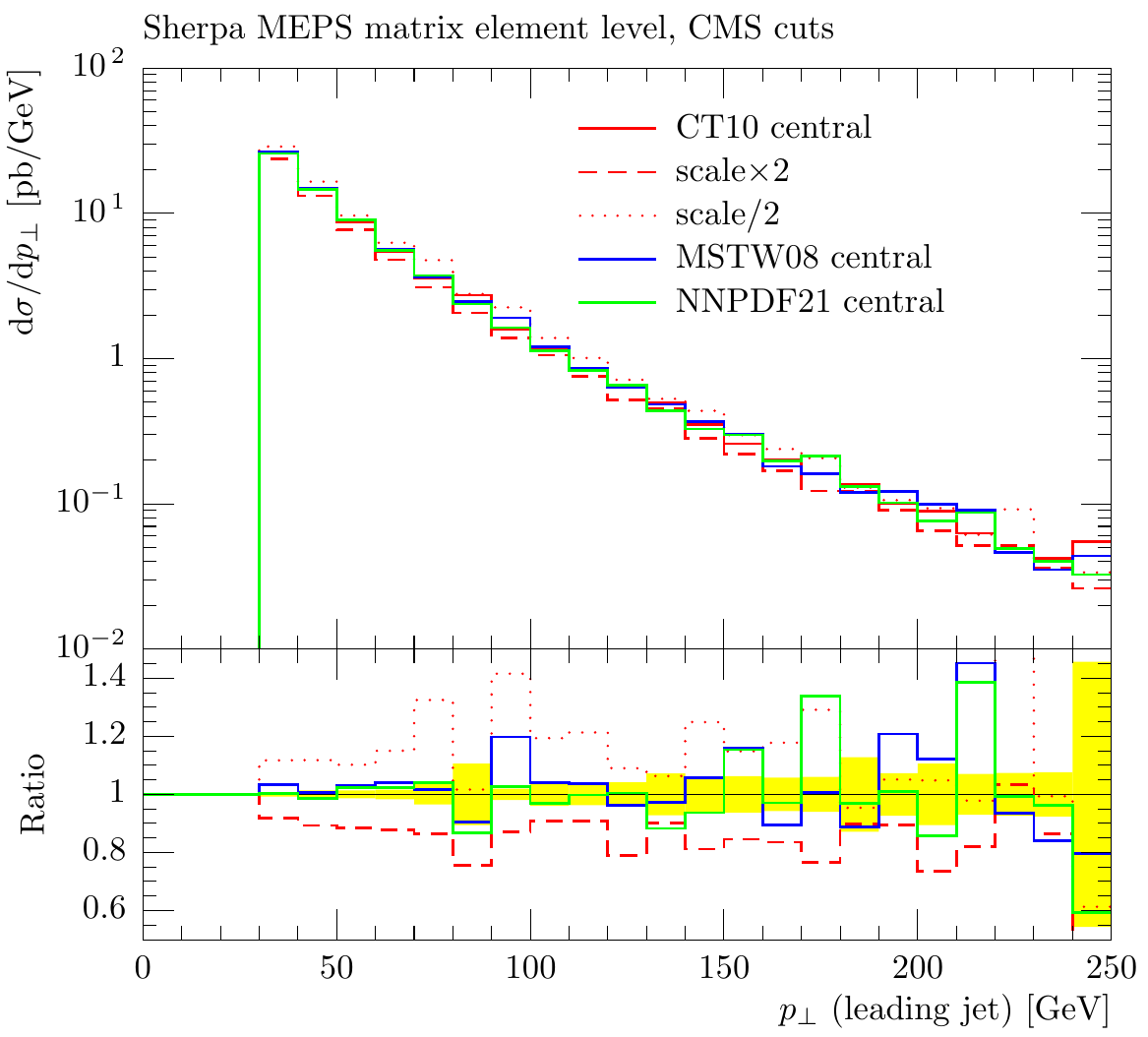}\hfill
  \includegraphics[width=.48\textwidth]{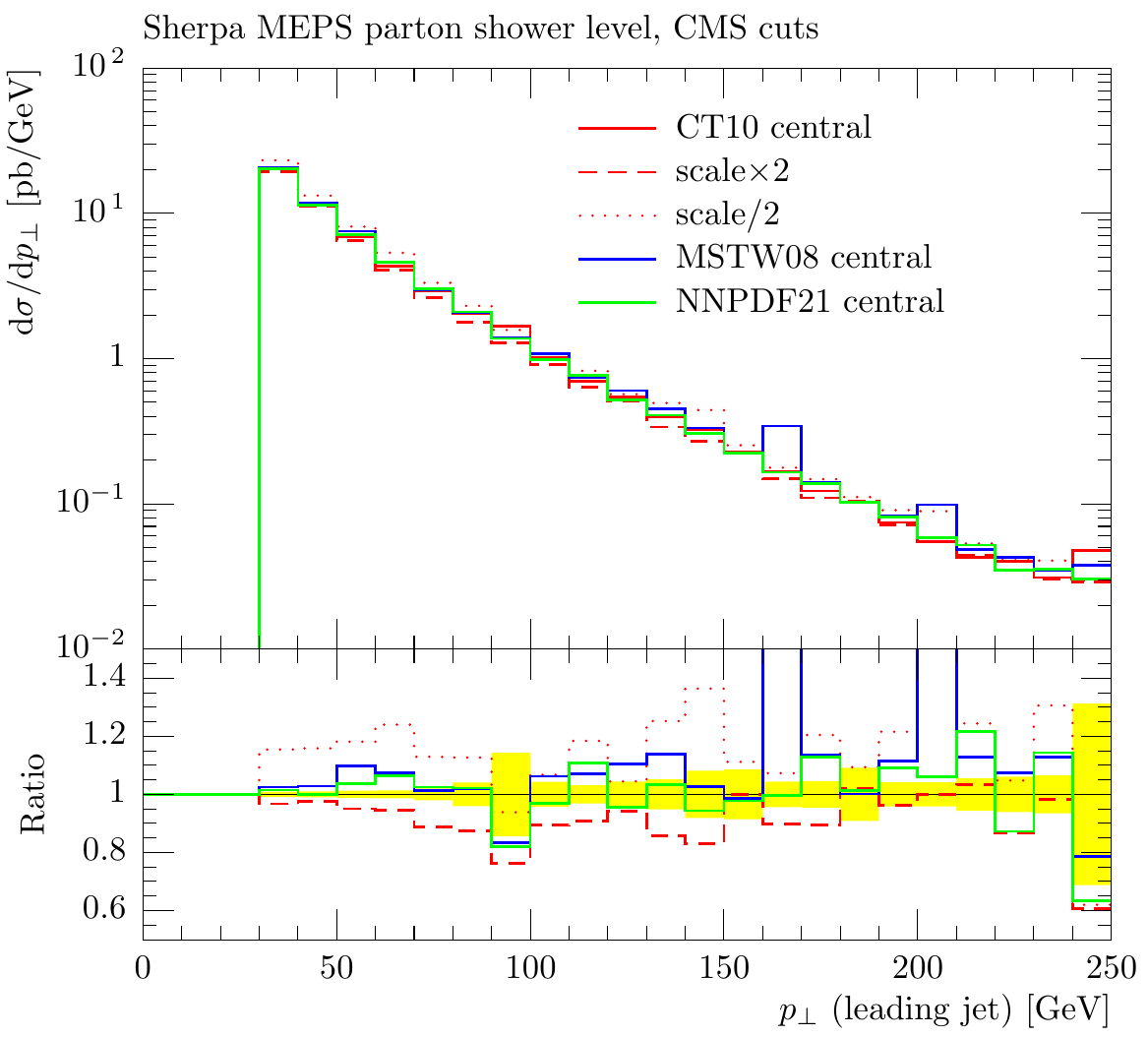}\\
  \includegraphics[width=.48\textwidth]{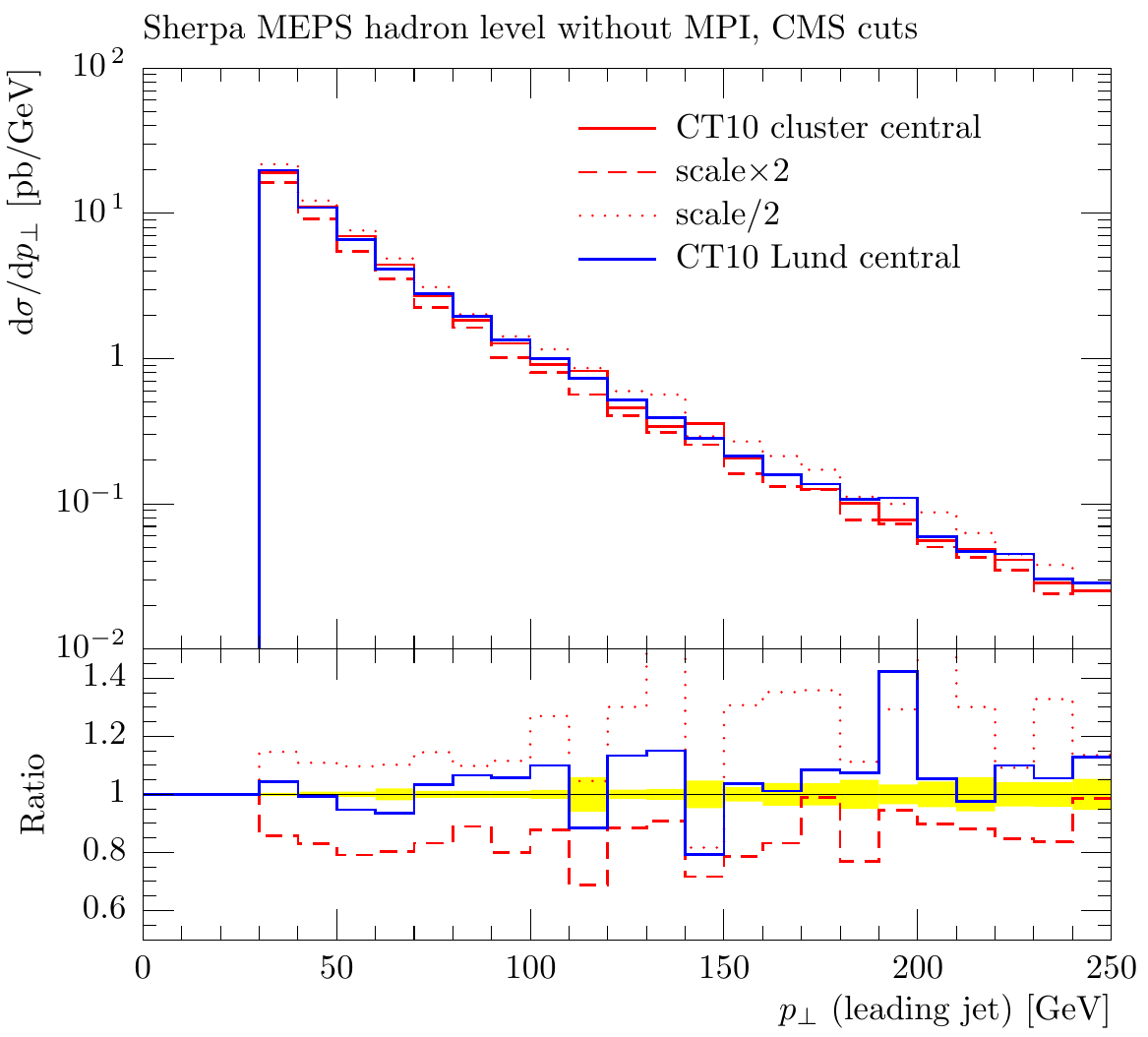}\hfill
  \includegraphics[width=.48\textwidth]{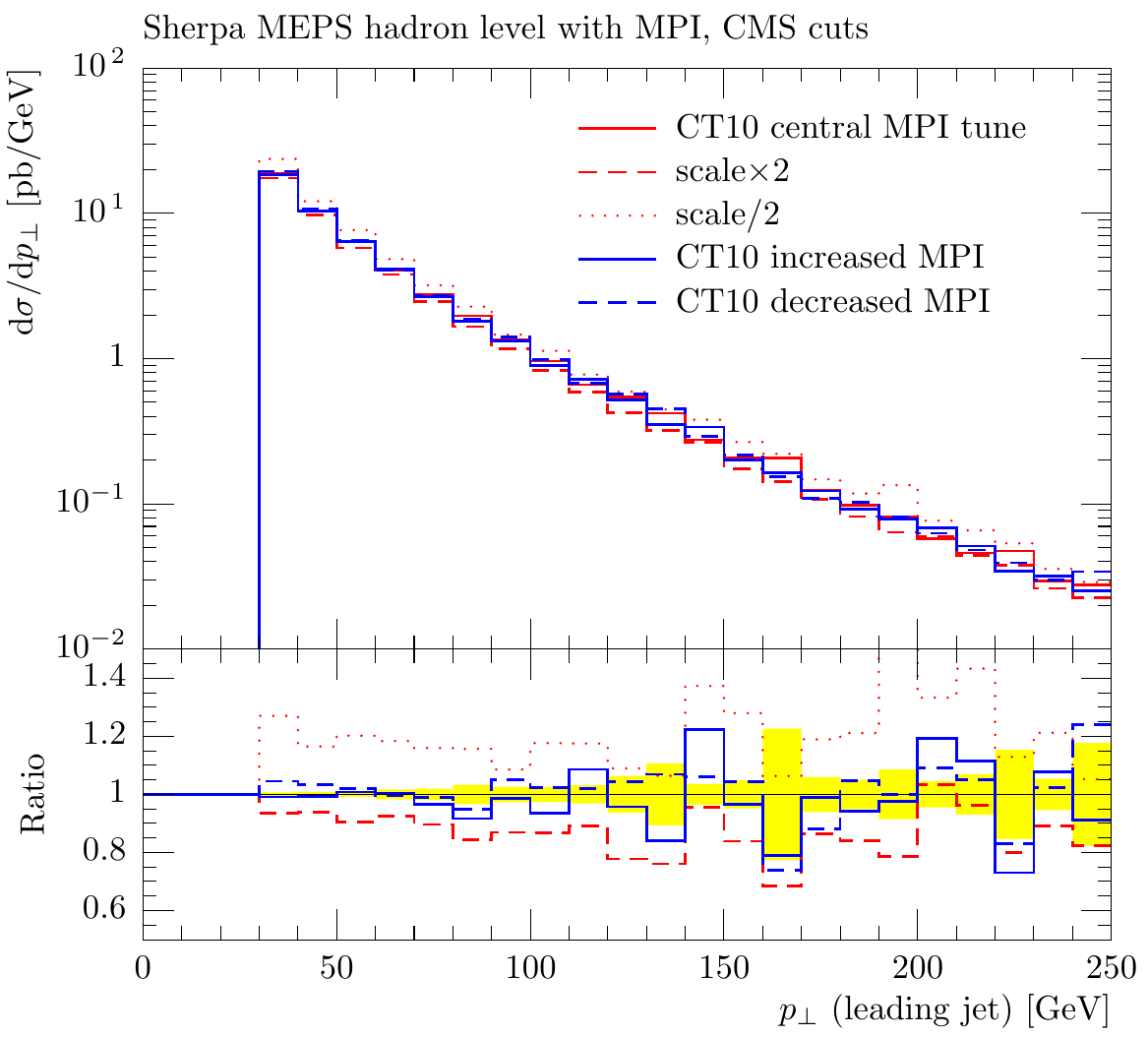}\\
  \includegraphics[width=.48\textwidth]{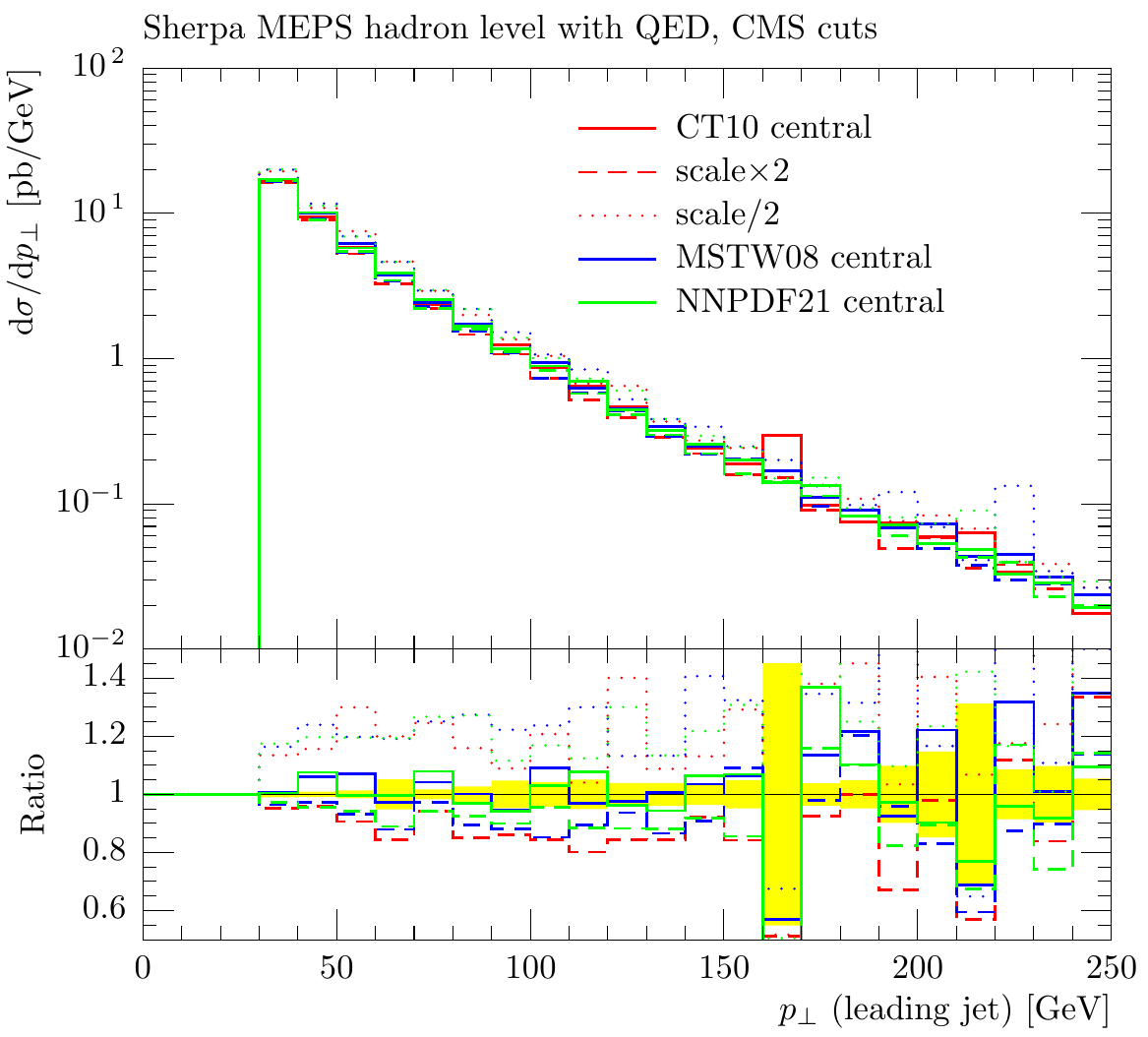}\hfill
  \includegraphics[width=.48\textwidth]{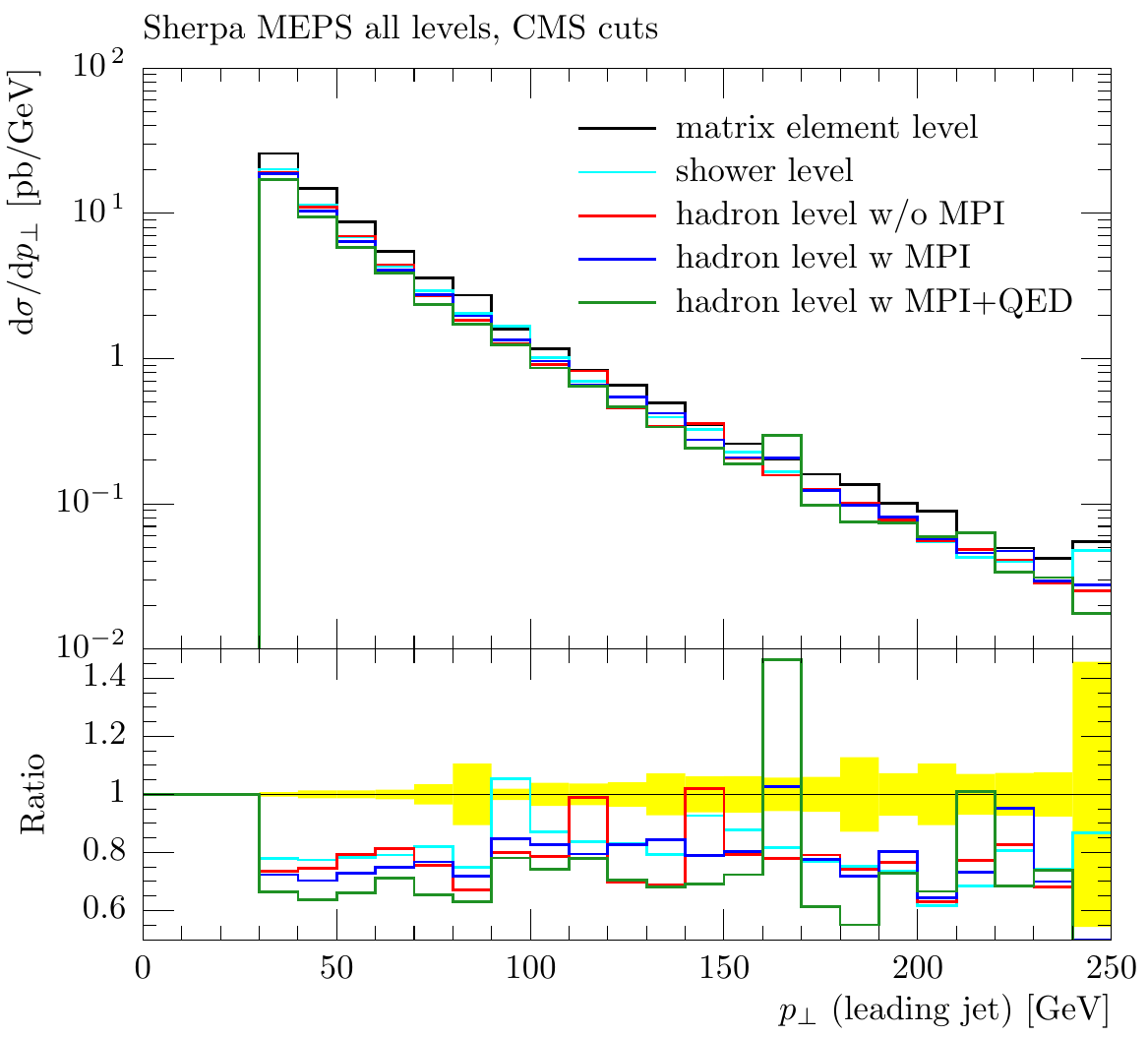}
  \caption{
	  \Sherpa\MEPS. Uncertainty of the transverse momentum of the hardest jet on the 
	  matrix element level (upper left), after parton showering (upper 
	  right), including hadronisation correction (centre left), 
	  multiple parton interactions (centre right), and QED corrections 
	  (lower left). The lower right panel shows the evolution of the 
	  central value.
	  \label{Fig:Results:Sherpa:MEPS:jetpt0}
  }
\end{figure}


\begin{figure}[p!]
  \includegraphics[width=.48\textwidth]{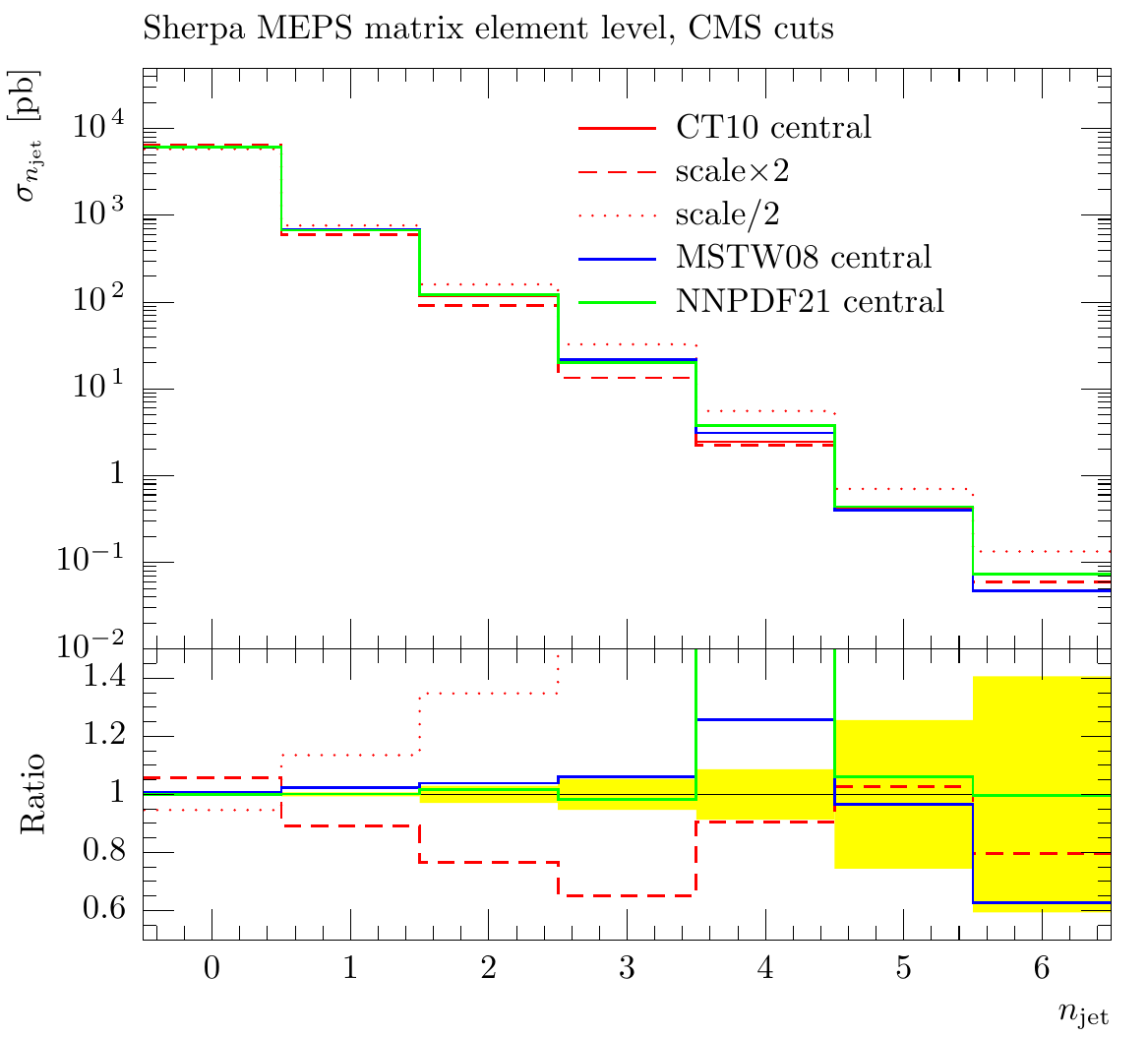}\hfill
  \includegraphics[width=.48\textwidth]{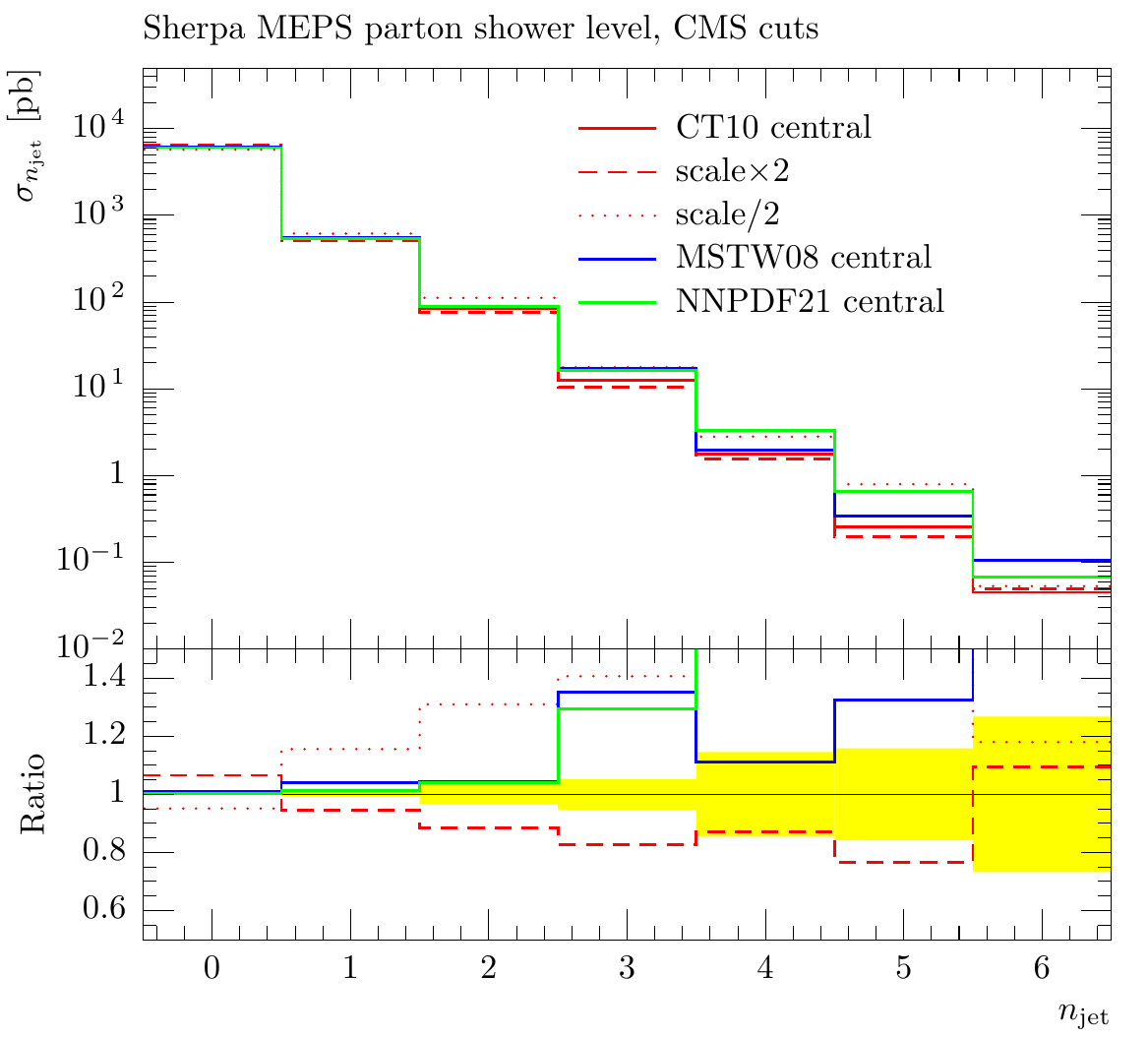}\\
  \includegraphics[width=.48\textwidth]{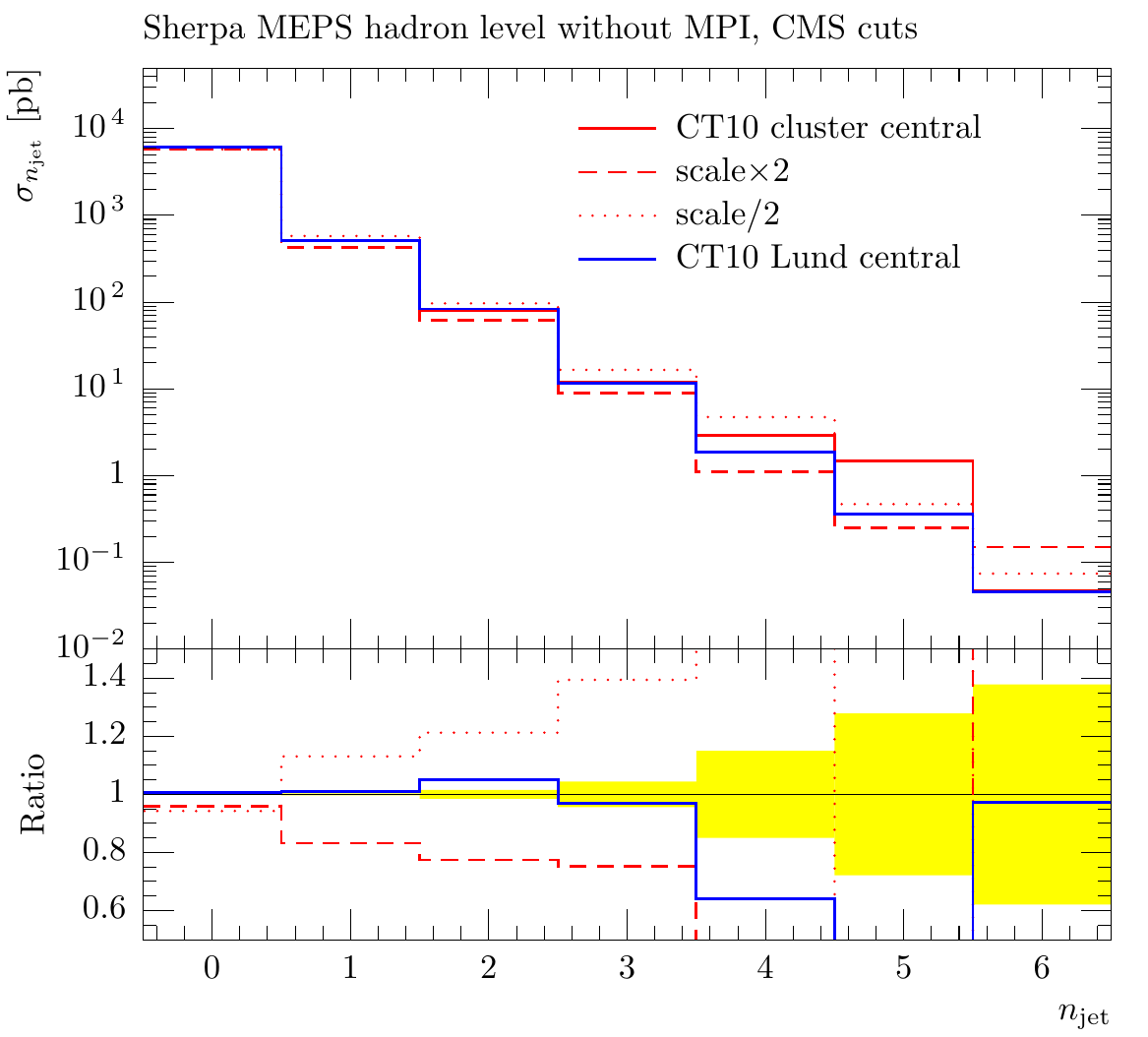}\hfill
  \includegraphics[width=.48\textwidth]{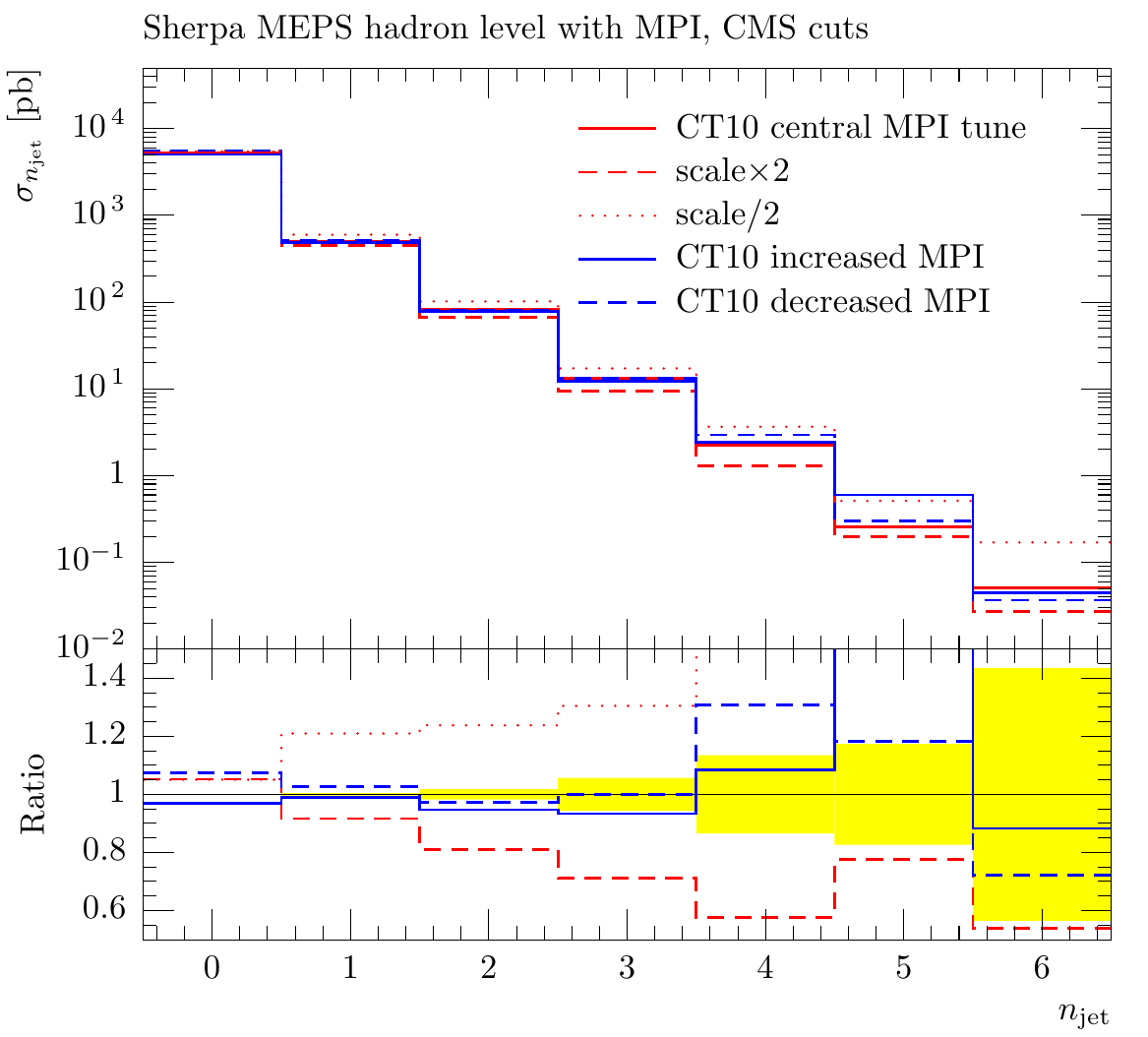}\\
  \includegraphics[width=.48\textwidth]{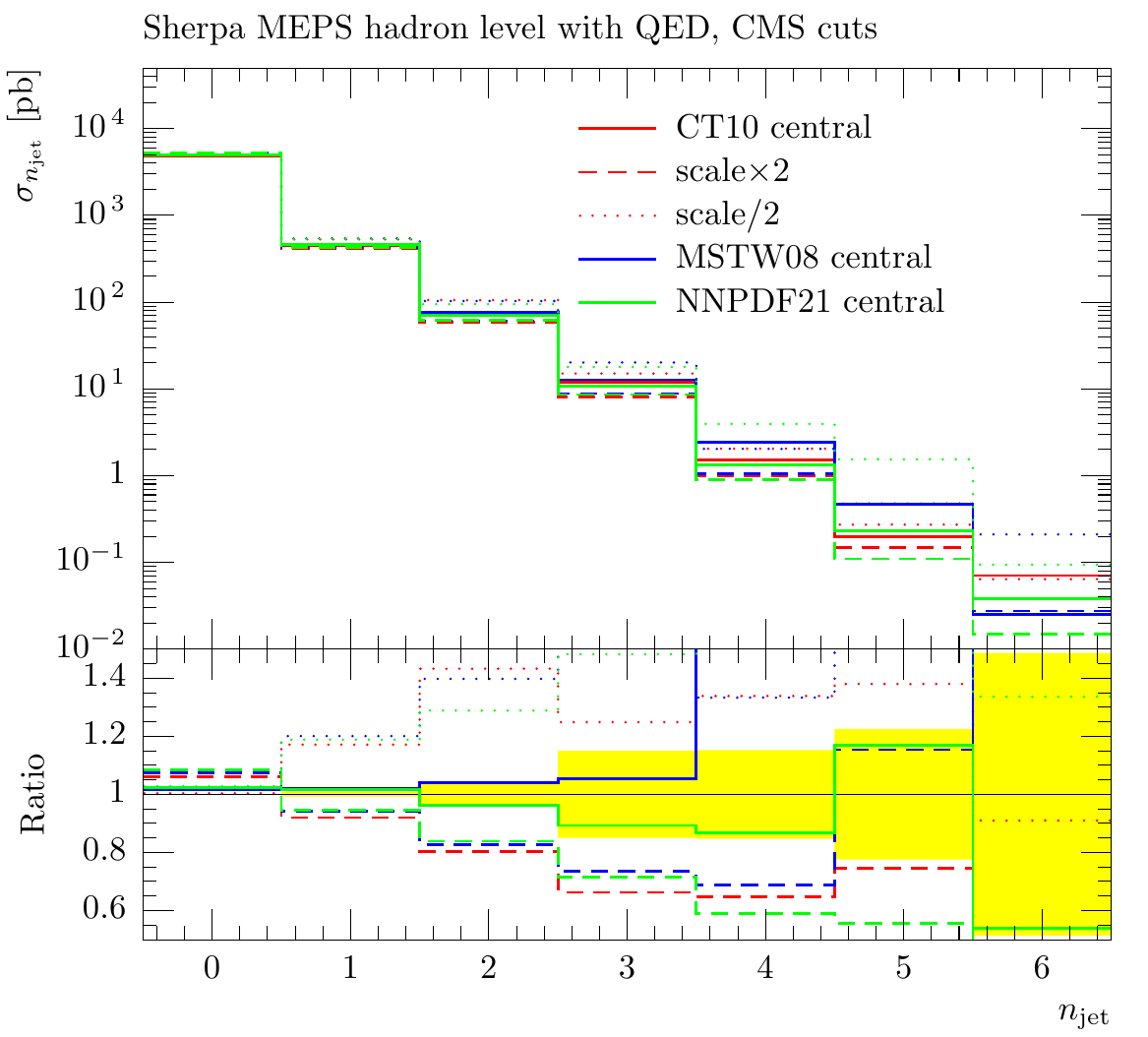}\hfill
  \includegraphics[width=.48\textwidth]{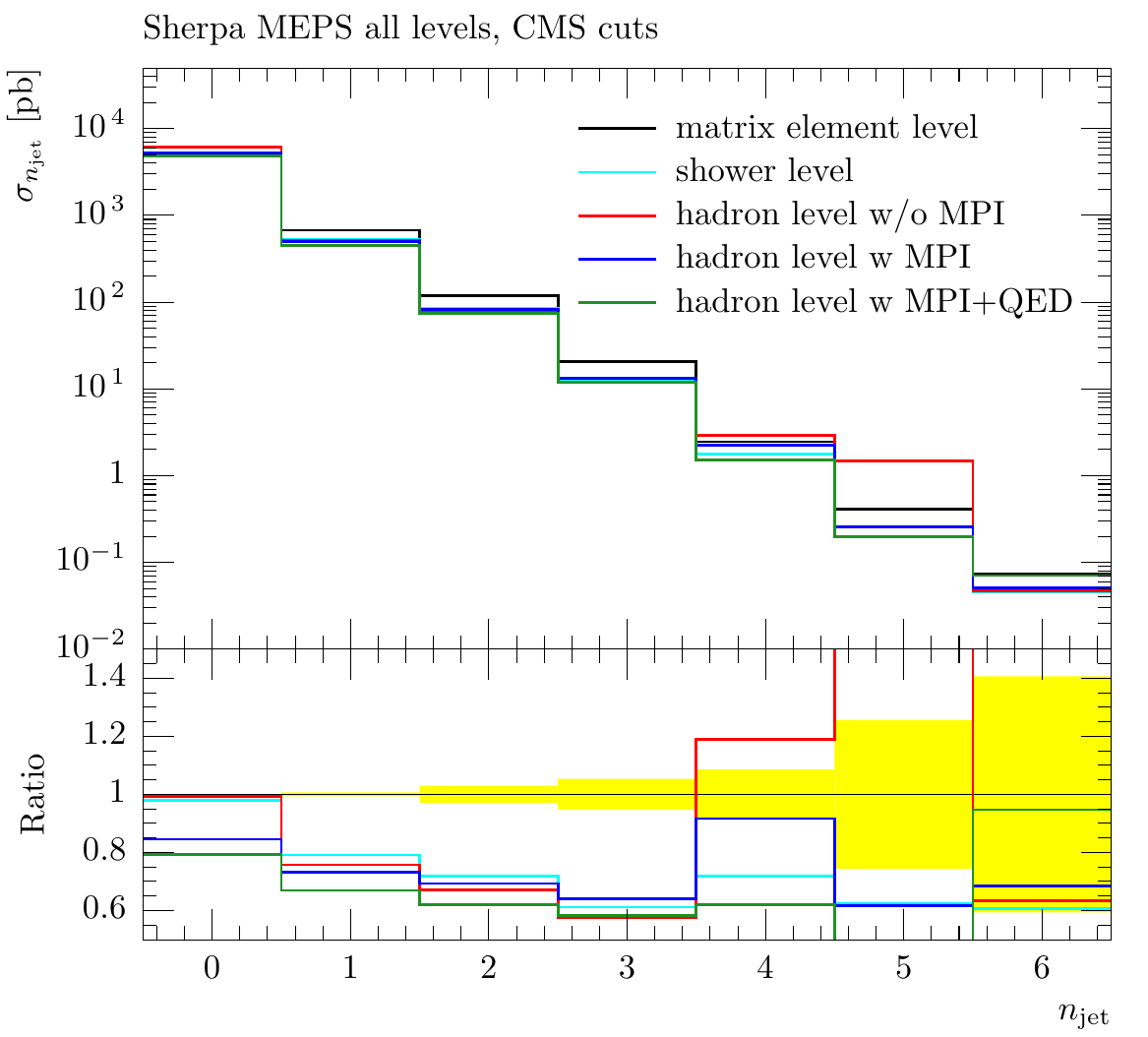}
  \caption{
	  \Sherpa\MEPS. Uncertainty of the inclusive jet multiplicity on the 
	  matrix element level (upper left), after parton showering (upper 
	  right), including hadronisation correction (centre left), 
	  multiple parton interactions (centre right), and QED corrections 
	  (lower left). The lower right panel shows the evolution of the 
	  central value.
	  \label{Fig:Results:Sherpa:MEPS:njet}
  }
\end{figure}

\FigRefs{Fig:Results:Sherpa:MEPS:Wmt}{Fig:Results:Sherpa:MEPS:njet} show
the results as obtained by running \Sherpa in the \MEPS mode for a
variety of inclusive and multi-jet observables at different levels of
the event generation. All central results are displayed together with
their respective uncertainties related to the different sources listed
in \SecRef{Sec:Codes:Sherpa:Uncertainties}. The layout in all figures
is the same: the upper left and right panels respectively show the
\MElevel and \PSlevel predictions for a given observable. The \MElevel
is defined as the event generation phase right before the parton
showering. For the \MEPS approach this means that modifications
necessary for the procedure to work like \alphas reweighting and
Sudakov rejection have been already included at this level. The
predictions presented in all centre panels were generated after
enhancing the event generation to include corrections induced by the
parton-to-hadron transition and decays of the therein produced
primordial hadrons. On top of these soft physics effects, one has to
also account for multiple parton interactions. The results in the
centre right panels of all figures incorporate these additional
corrections. Finally, all plots to the lower left show the most
complete \Hadlevel predictions, which were obtained by adding to the
event generation QED radiation effects as occurring in the decays of
the vector boson and the hadrons. To allow a direct comparison of the
impact of the consecutive event stages, the way the central results
change is summarized in all plots to the lower right of
\FigRefs{Fig:Results:Sherpa:MEPS:Wmt}{Fig:Results:Sherpa:MEPS:njet}.
In these, as in all other panels, the main plots are supplemented by
ratio plots stressing the magnitude of the differences and
uncertainties. Note that the yellow band throughout illustrates the
statistical uncertainty on the central event sample.

Apart from the summary plots at the lower right, all other cases
depict predictions documenting the uncertainty of the central
predictions at the different levels of event generation. These
uncertainty estimates are gained following the procedures outlined in
\SecRef{Sec:Codes:Sherpa:Uncertainties}. At all event simulation
phases, the scales are varied as described under this section's
point~(B). Note that the variation is applied to all phases used to
make up the respective central (or default) sample, which is taken as
the reference under all circumstances. For the \MElevel, \PSlevel and
full \Hadlevel results, PDF variations according to point~(A) are
shown in addition, whereas for the centre panel plots, the focus is on
the outcomes of the model and tune variations instead, as specified in
point~(C) and point~(D) of \SecRef{Sec:Codes:Sherpa:Uncertainties}.
Notice that the lower left panels also contain the outcomes of scale
variations utilizing the alternative PDFs mentioned under point~(A);
they are much alike the ones stemming from the default set.

As an example for an inclusive observable the transverse mass of the 
reconstructed $W$ boson is shown in \FigRef{Fig:Results:Sherpa:MEPS:Wmt}.
The scale uncertainties amount to $\sim$15\% at all generation levels,
whereas the uncertainties due to the choice of PDF are much smaller.
Similarly, the hadronisation uncertainty is negligible. The
$m_{\perp,W}$ observable however is more sensitive to the tuning of
the MPI model as can be seen from the $\pm$10\% envelope in the centre
right plot of \FigRef{Fig:Results:Sherpa:MEPS:Wmt}. The uncertainty is
of the same order as for the scale variations, which generally are
more pronounced in the soft region. When considering the impact of
each perturbative and non-perturbative event stage (see the plot to
the lower right), it is the MPI corrections that are largest in the
region of $m_{\perp,W}<m_W$, ranging up to $\sim$30\% wrt.\ the
\MElevel prediction. They are small above $m_W$. In this region the
dominant effect comes from the QED corrections, which themselves are
rather small, but they lead to a contamination of the electron
isolation. The application of the isolation cuts then yields a
reduction of the overall normalisation of the event sample. Finally
there is a small shift towards lower transverse masses, pronouncing
the deviation in the tail of the distribution somewhat further.

\FigRef{Fig:Results:Sherpa:MEPS:dRj0l} and
\FigRef{Fig:Results:Sherpa:MEPS:jetpt0} depict observables that
require the presence of at least one jet. In the former the geometric
separation, $\Delta R$, between the hardest jet and the electron is
shown, while in the latter, focus is on the transverse momentum,
$p_\perp$ of the hardest jet only. As before the dependence of the
predictions on PDF and hadronisation model changes remains negligible.
While the scale dependence of the $\Delta R$ and $p_\perp$ variables
increases to $\sim$30\%, the uncertainty due to the tuning of the
MPI model decreases to $\sim$5\% when compared to the findings
concerning the more inclusive observable considered above. In both
cases the reason for the sensitivity change obviously lies in
demanding at least one (hard) jet. The scale uncertainties primarily
result from changes in the overall cross section. Again, comparing the
results of the different event stages, one clearly observes the large
impact parton showering has on modifying the \MElevel predictions. The
non-perturbative effects go in the same direction amplifying the
parton shower effects, but as expected this amplification turns out to
be rather mild in the well separated and/or hard phase space regions.
QED corrections only play a minor role, and are far less important
than for the $m_{\perp,W}$ variable.

In \FigRef{Fig:Results:Sherpa:MEPS:njet} one of the simplest examples
of a multi-jet observable is presented, namely the distribution of the
inclusive $W+n$ jet cross sections as a function of $n_\text{jet}$.
Qualitatively, the parameter and model dependencies of the predictions
are found to behave as for the inclusive one-jet variables. As one
would expect, the scale uncertainties subsequently increase with the
order of the jet bin. The same can be noticed for the variation of
the PDFs used in the calculation -- even though here the effect is
considerably smaller.

\paragraph{\texorpdfstring{\protect\Sherpa\MENLOPS}{Sherpa MENLOPS}}
\label{Sec:Results:Sherpa:MENLOPS}

\begin{figure}[p!]
  \includegraphics[width=.48\textwidth]{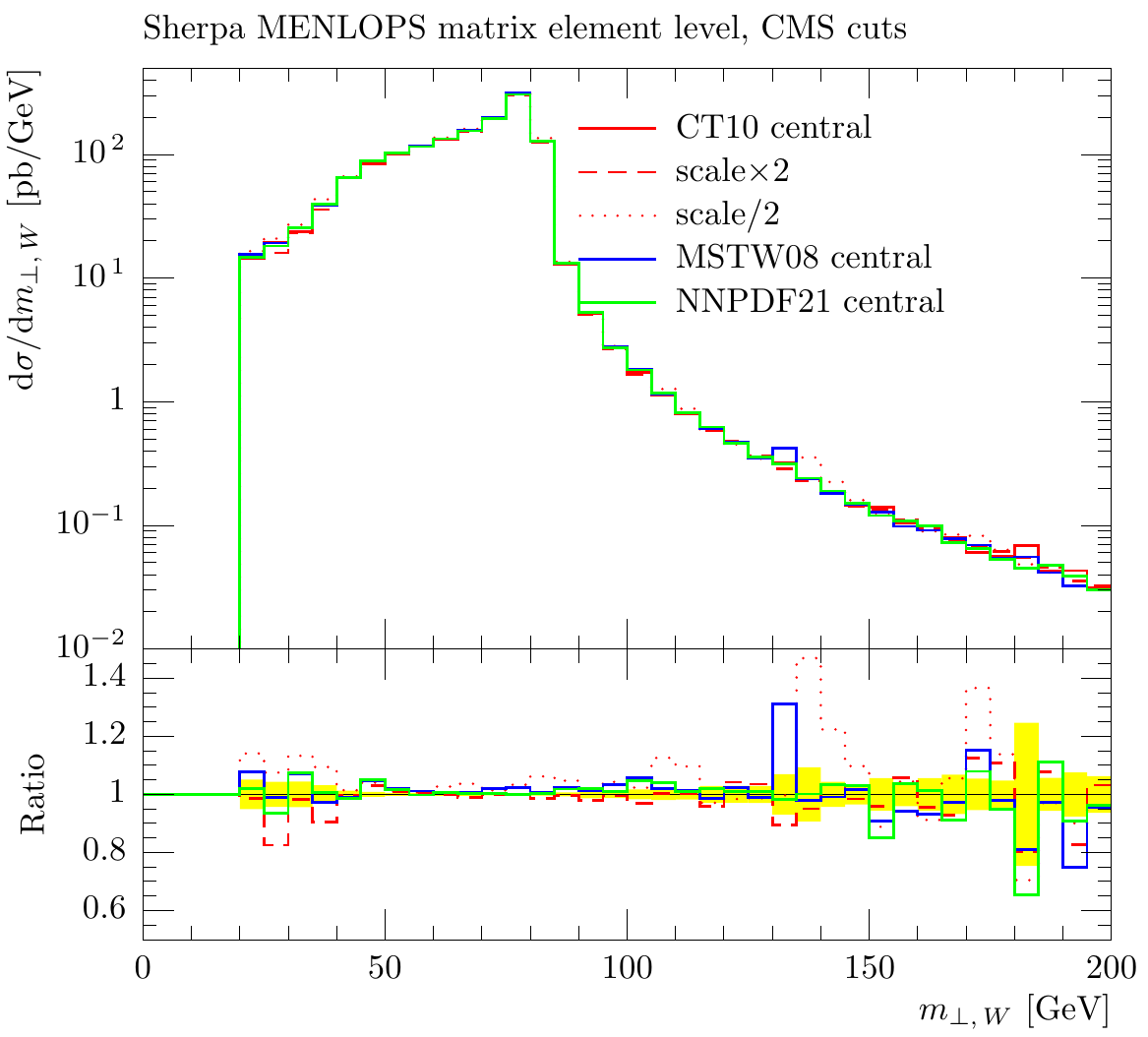}\hfill
  \includegraphics[width=.48\textwidth]{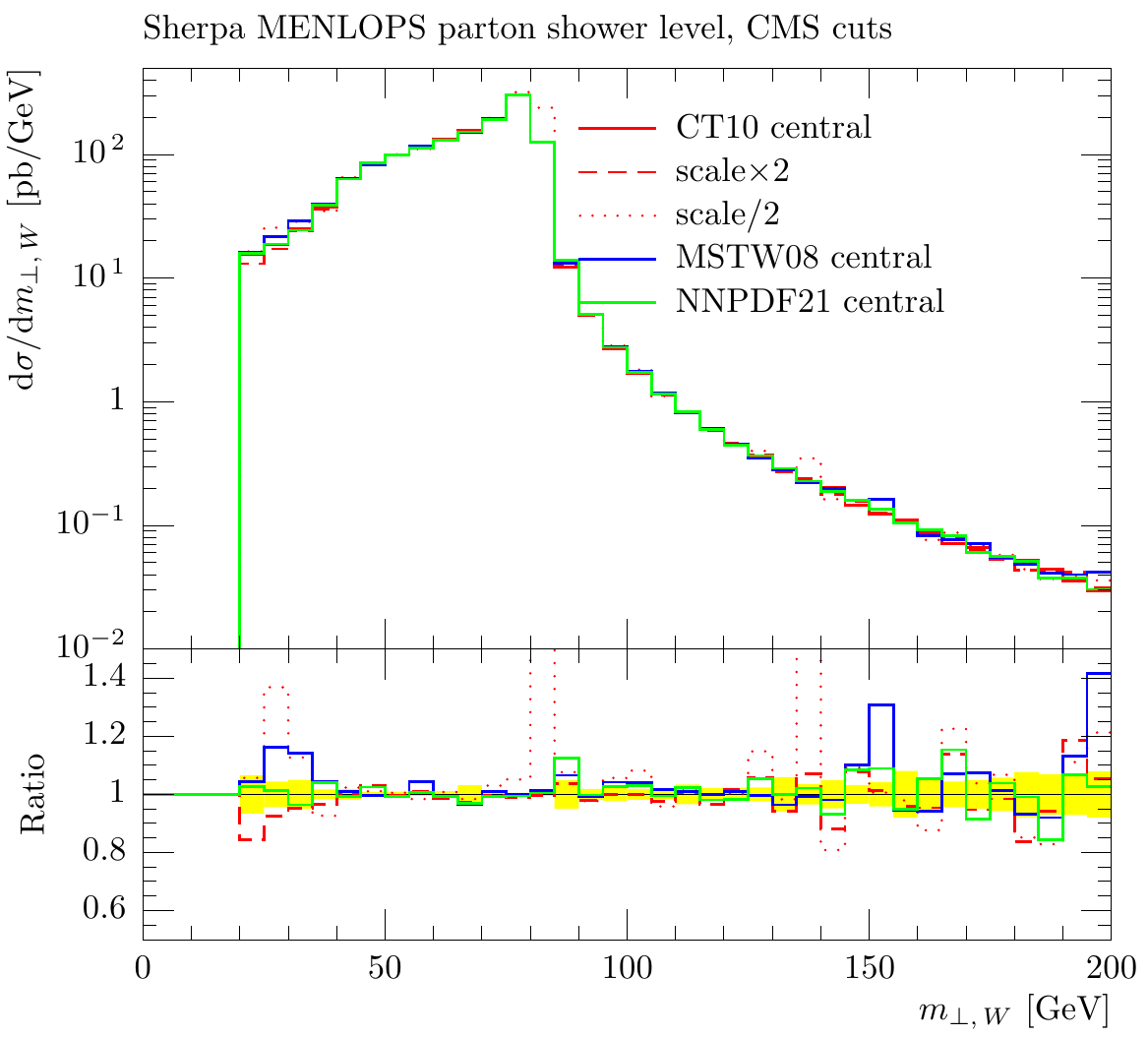}\\
  \includegraphics[width=.48\textwidth]{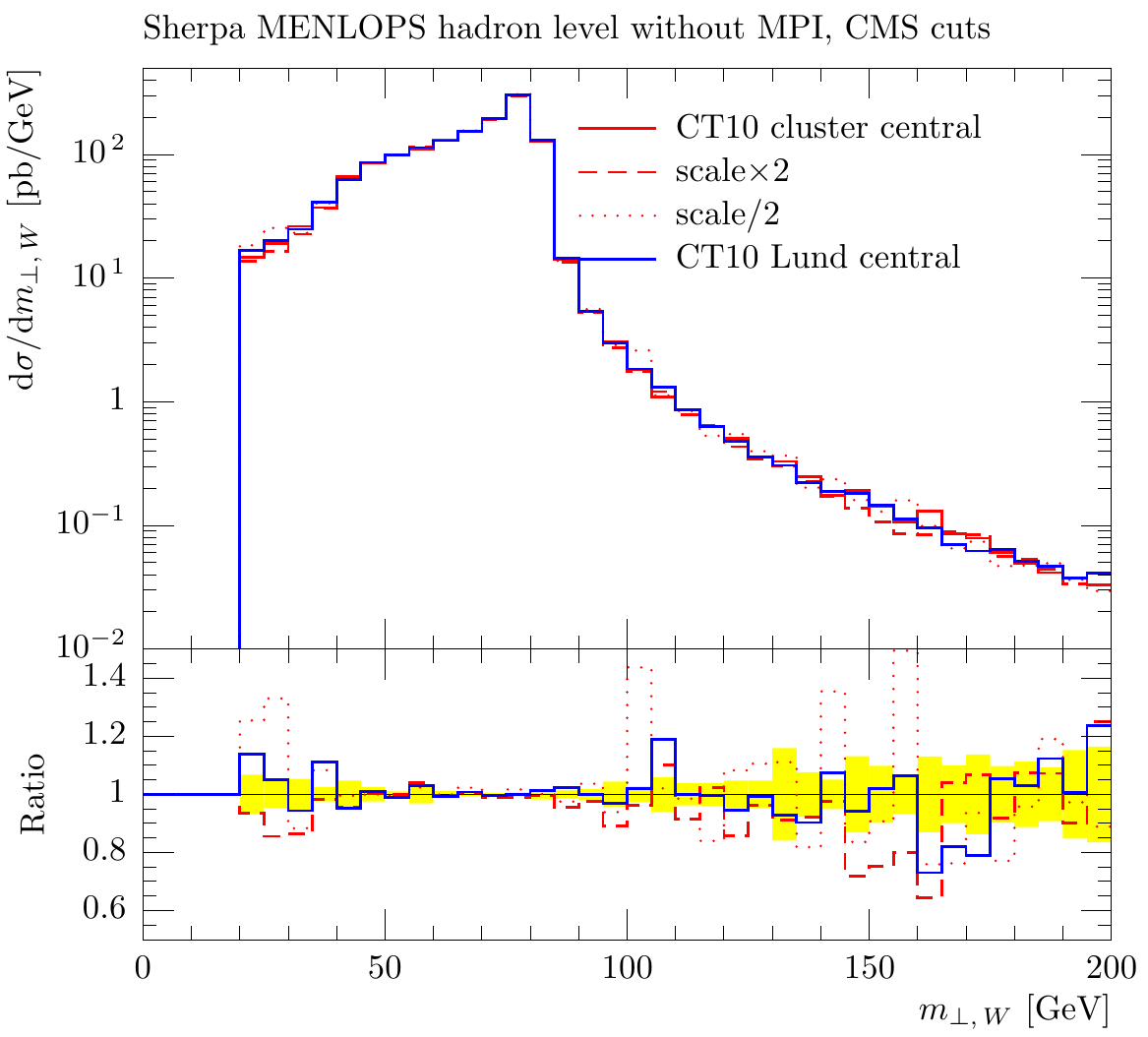}\hfill
  \includegraphics[width=.48\textwidth]{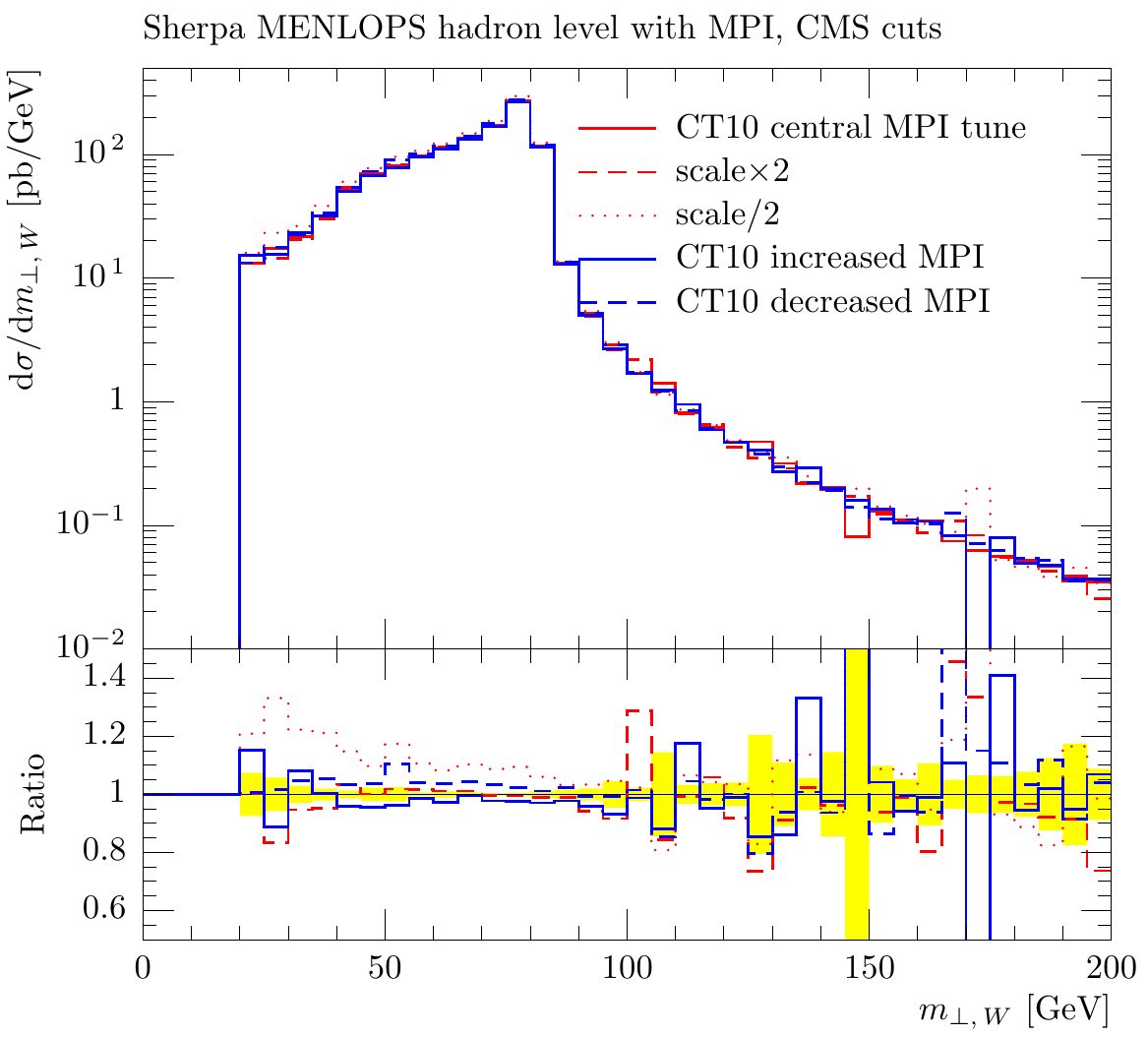}\\
  \includegraphics[width=.48\textwidth]{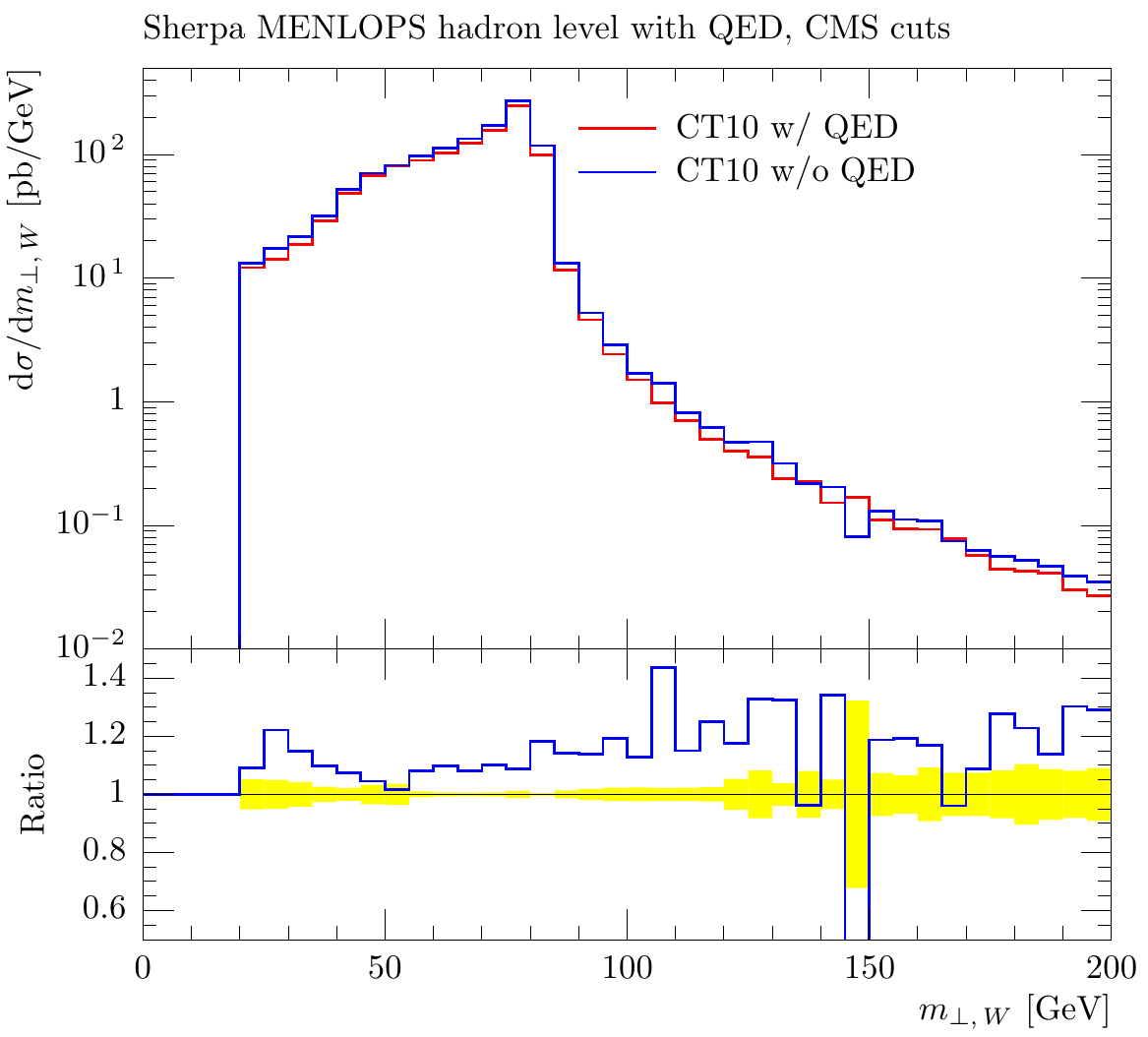}\hfill
  \includegraphics[width=.48\textwidth]{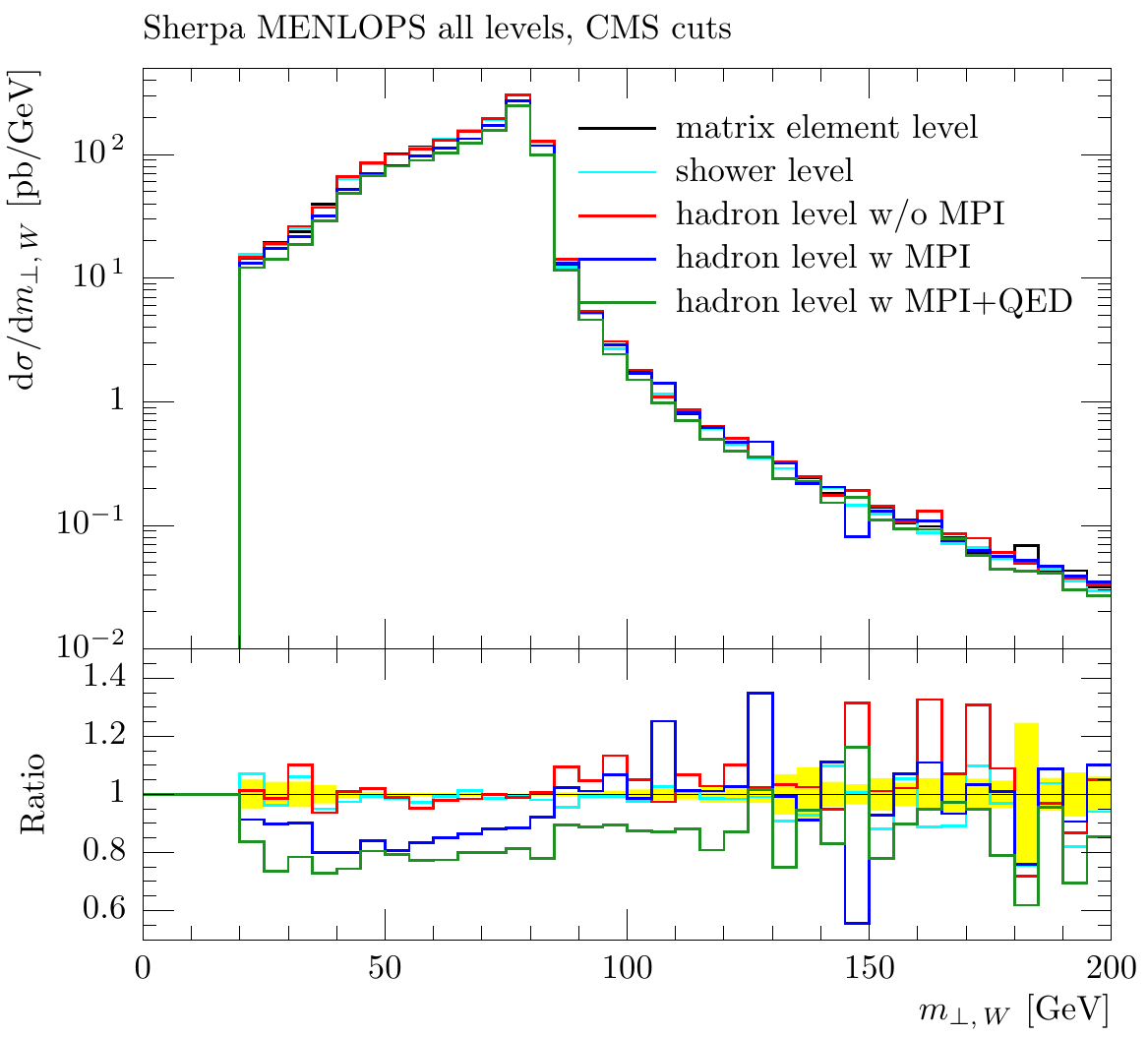}
  \caption{
	  \Sherpa\MENLOPS. Uncertainty of the transverse mass of the reconstructed $W$ on the 
	  matrix element level (upper left), after parton showering (upper 
	  right), including hadronisation correction (centre left), 
	  multiple parton interactions (centre right), and QED corrections 
	  (lower left). The lower right panel shows the evolution of the 
	  central value.
	  \label{Fig:Results:Sherpa:MENLOPS:Wmt}
  }
\end{figure}

\begin{figure}[p!]
  \includegraphics[width=.48\textwidth]{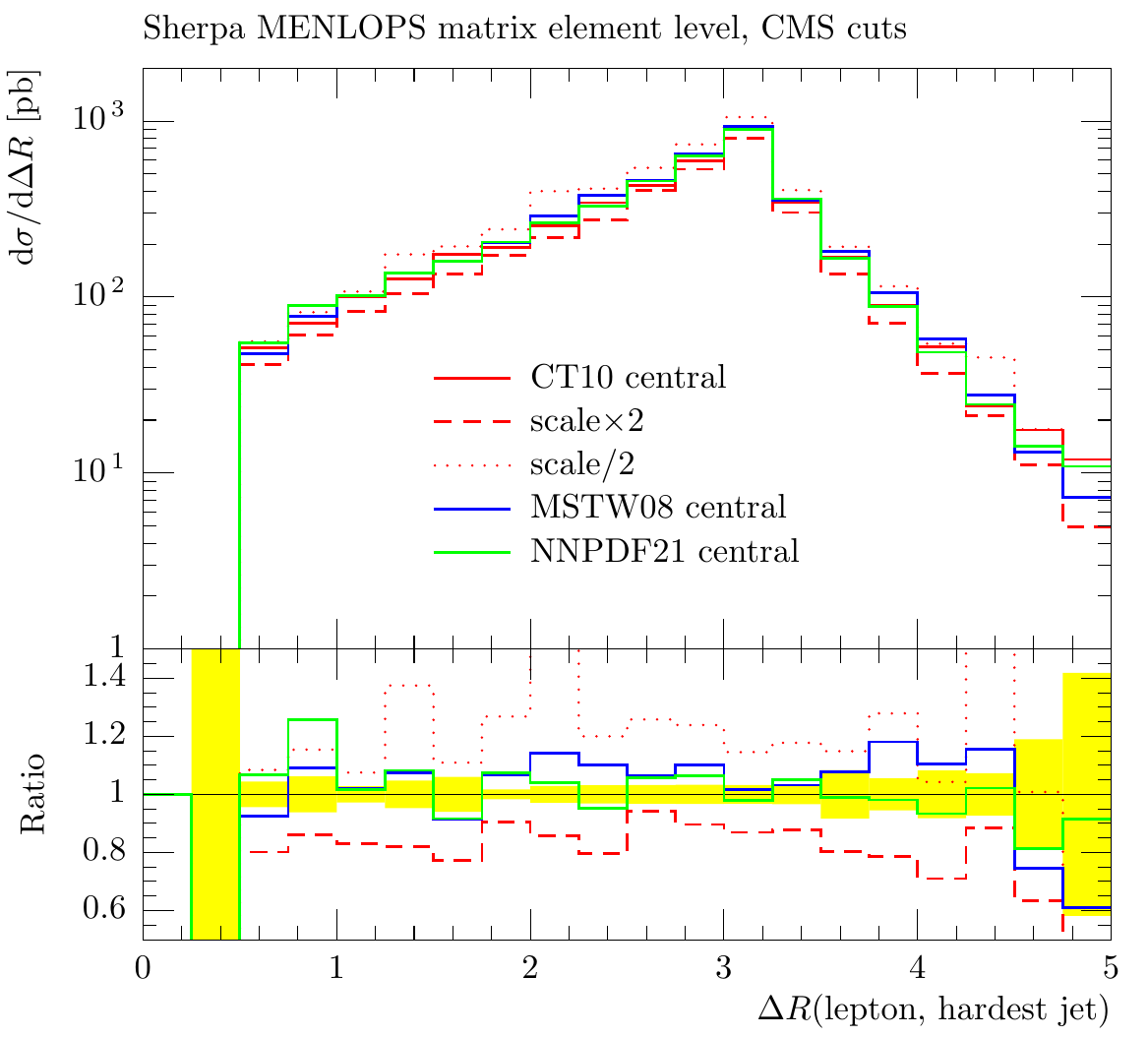}\hfill
  \includegraphics[width=.48\textwidth]{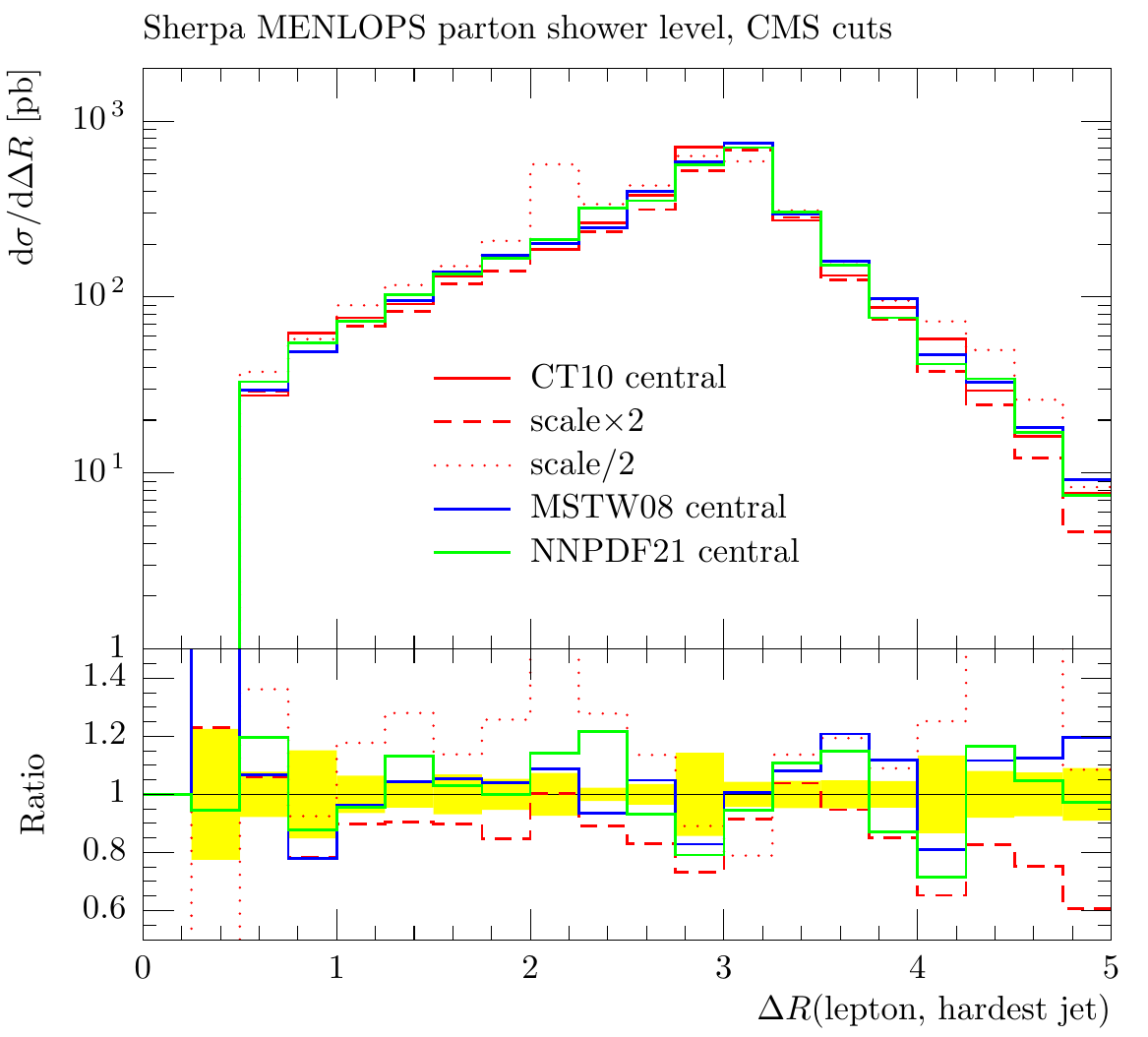}\\
  \includegraphics[width=.48\textwidth]{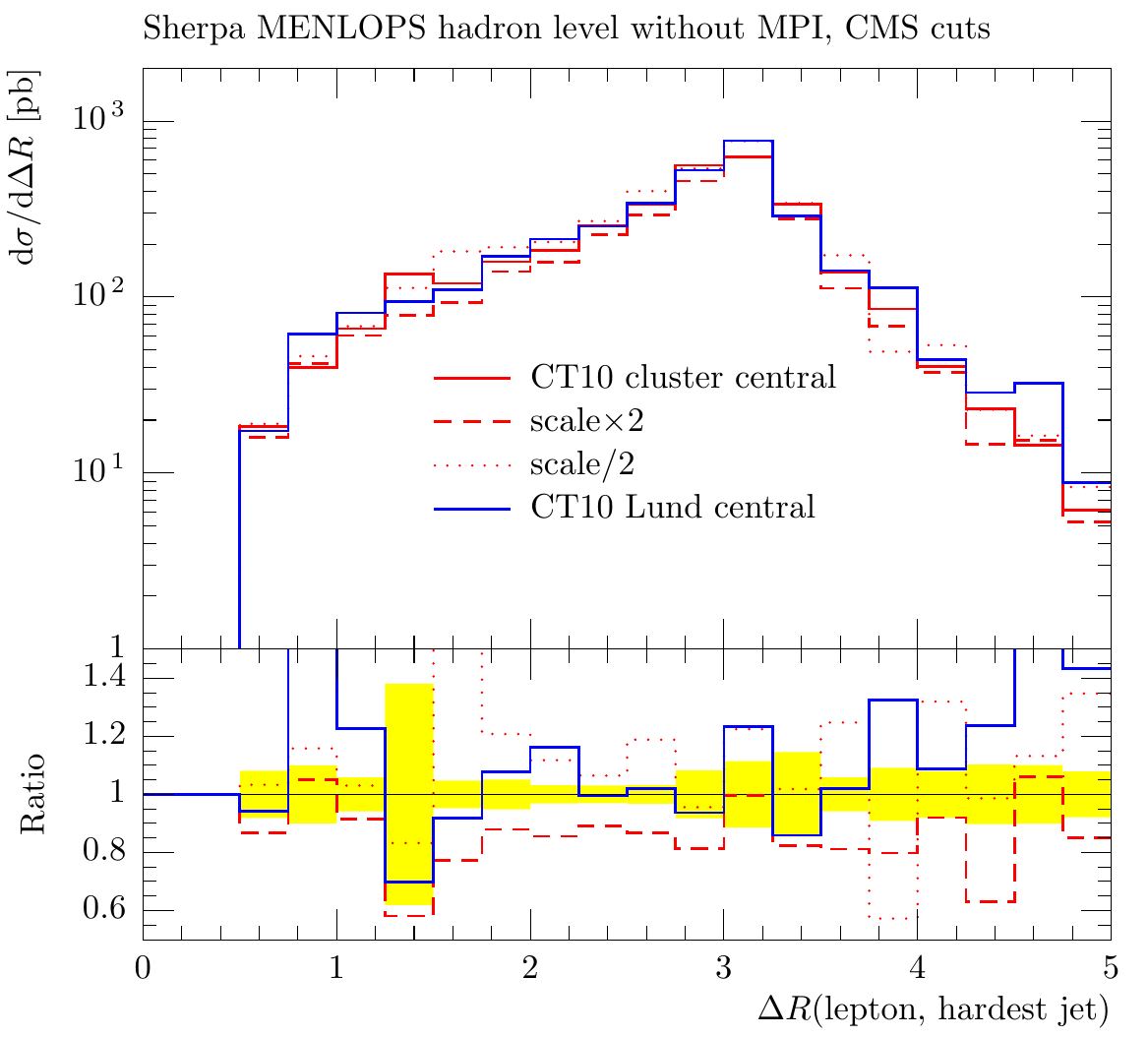}\hfill
  \includegraphics[width=.48\textwidth]{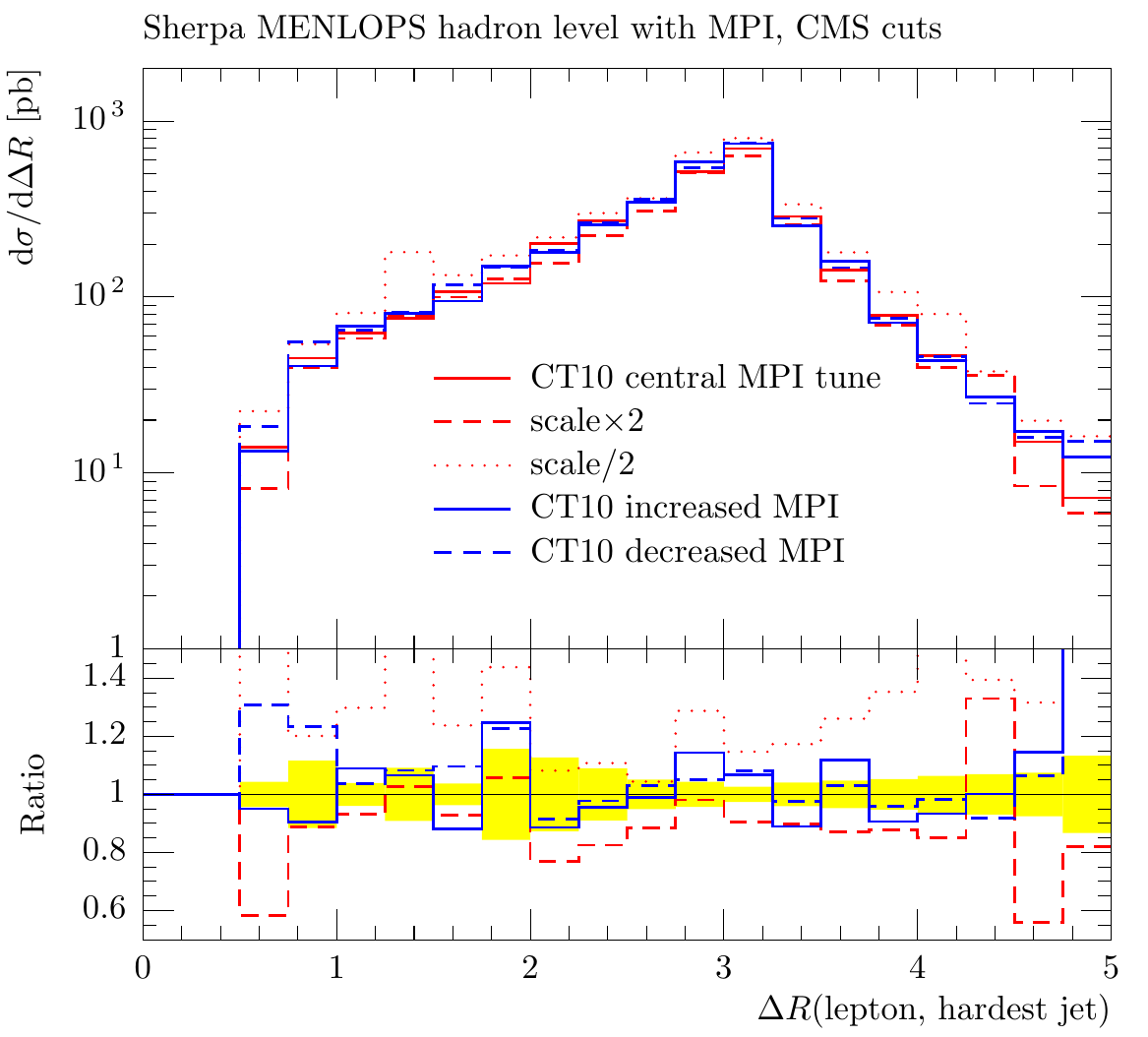}\\
  \includegraphics[width=.48\textwidth]{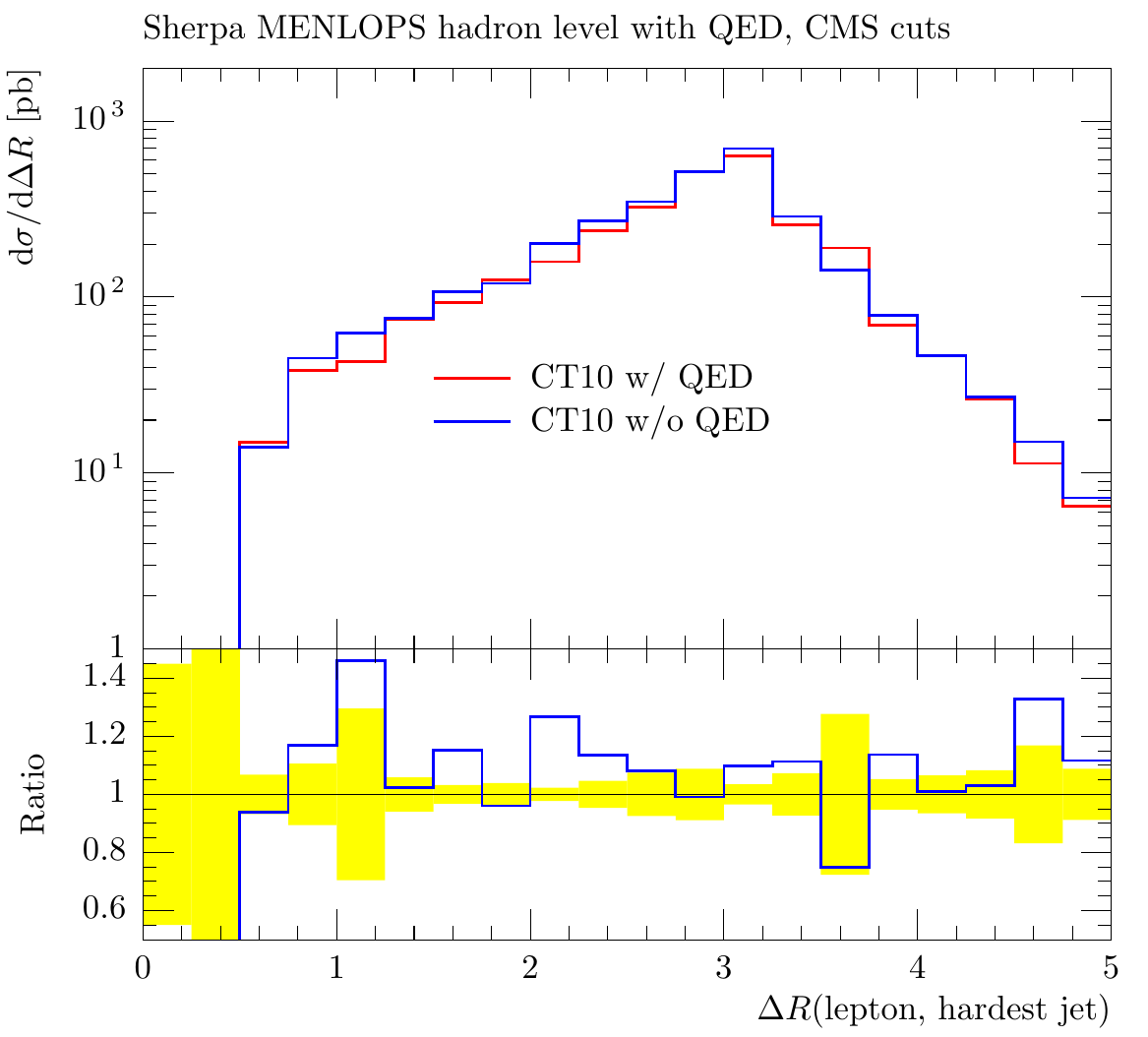}\hfill
  \includegraphics[width=.48\textwidth]{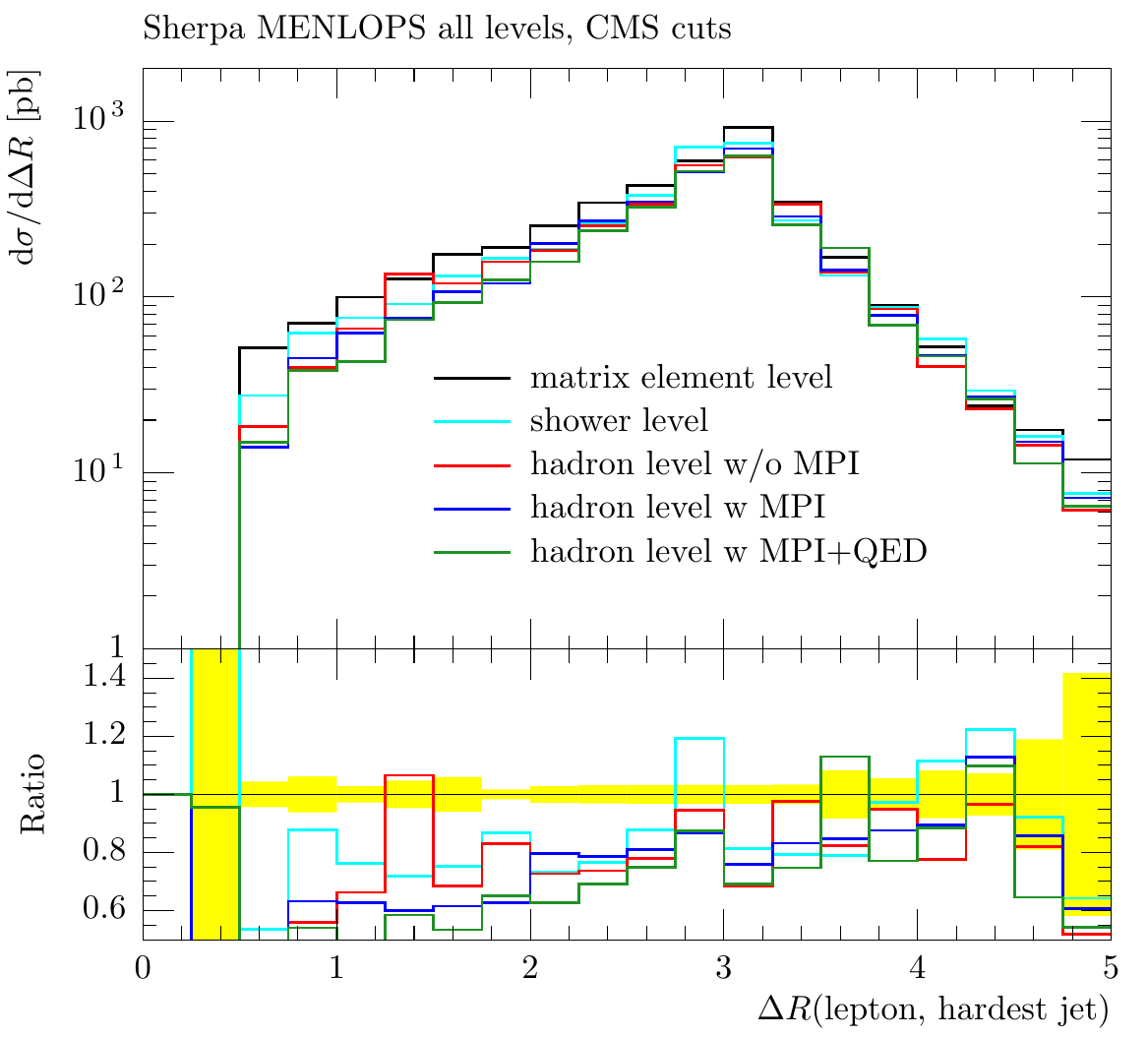}
  \caption{
	  \Sherpa\MENLOPS. Uncertainty of the angular separation of the charged lepton and the hardest jet on the 
	  matrix element level (upper left), after parton showering (upper 
	  right), including hadronisation correction (centre left), 
	  multiple parton interactions (centre right), and QED corrections 
	  (lower left). The lower right panel shows the evolution of the 
	  central value.
	  \label{Fig:Results:Sherpa:MENLOPS:dRj0l}
  }
\end{figure}

\begin{figure}[p!]
  \includegraphics[width=.48\textwidth]{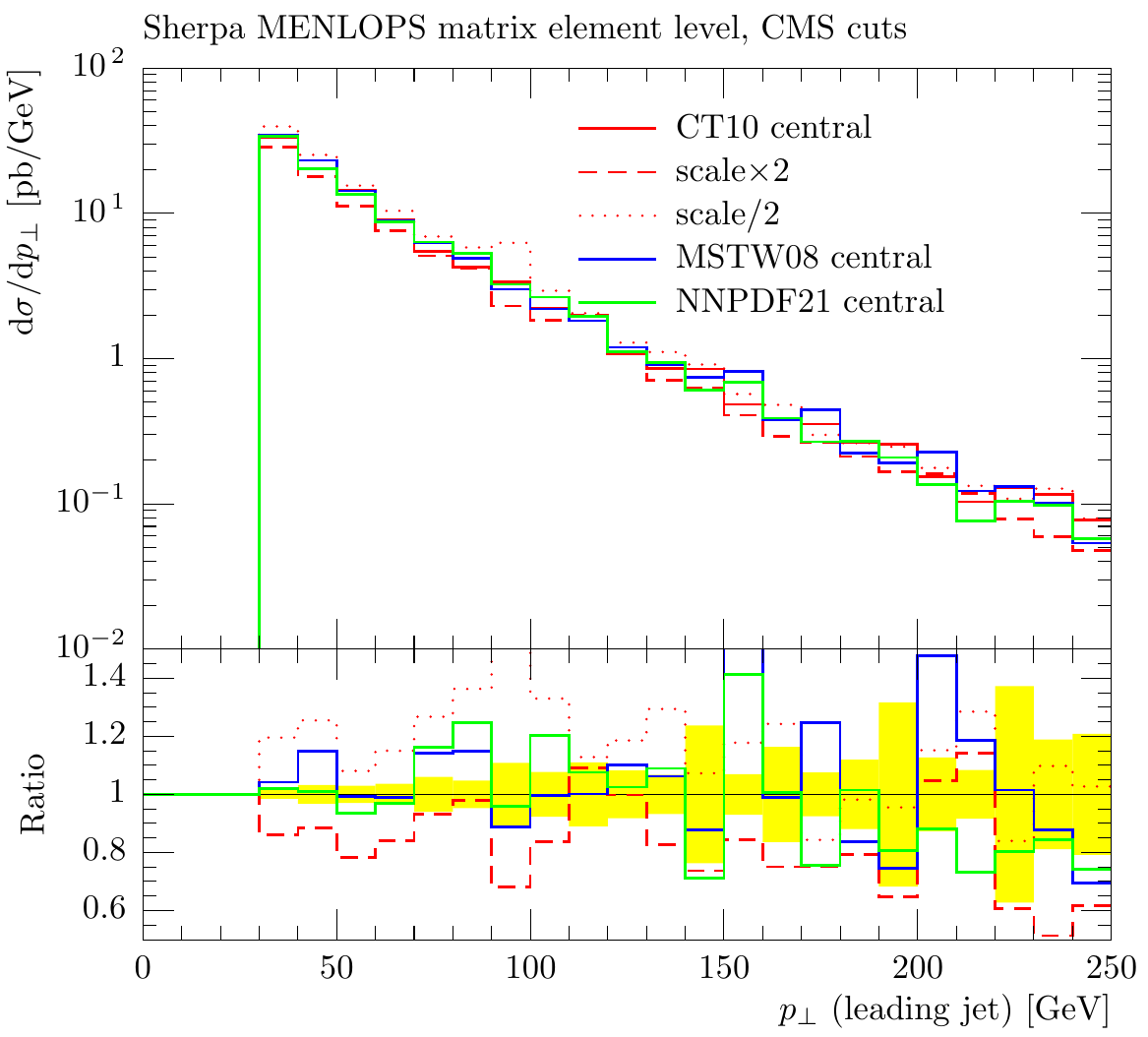}\hfill
  \includegraphics[width=.48\textwidth]{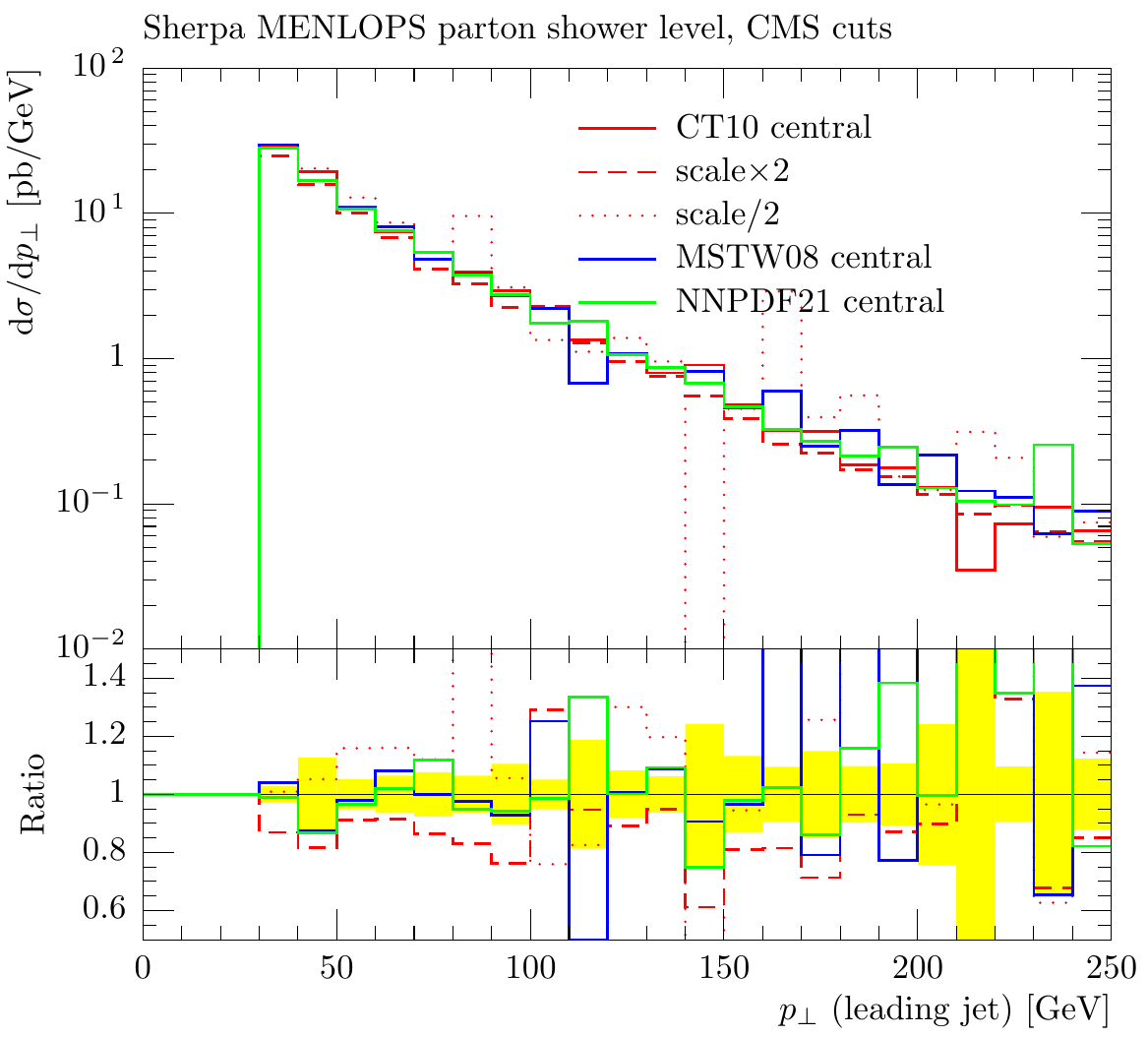}\\
  \includegraphics[width=.48\textwidth]{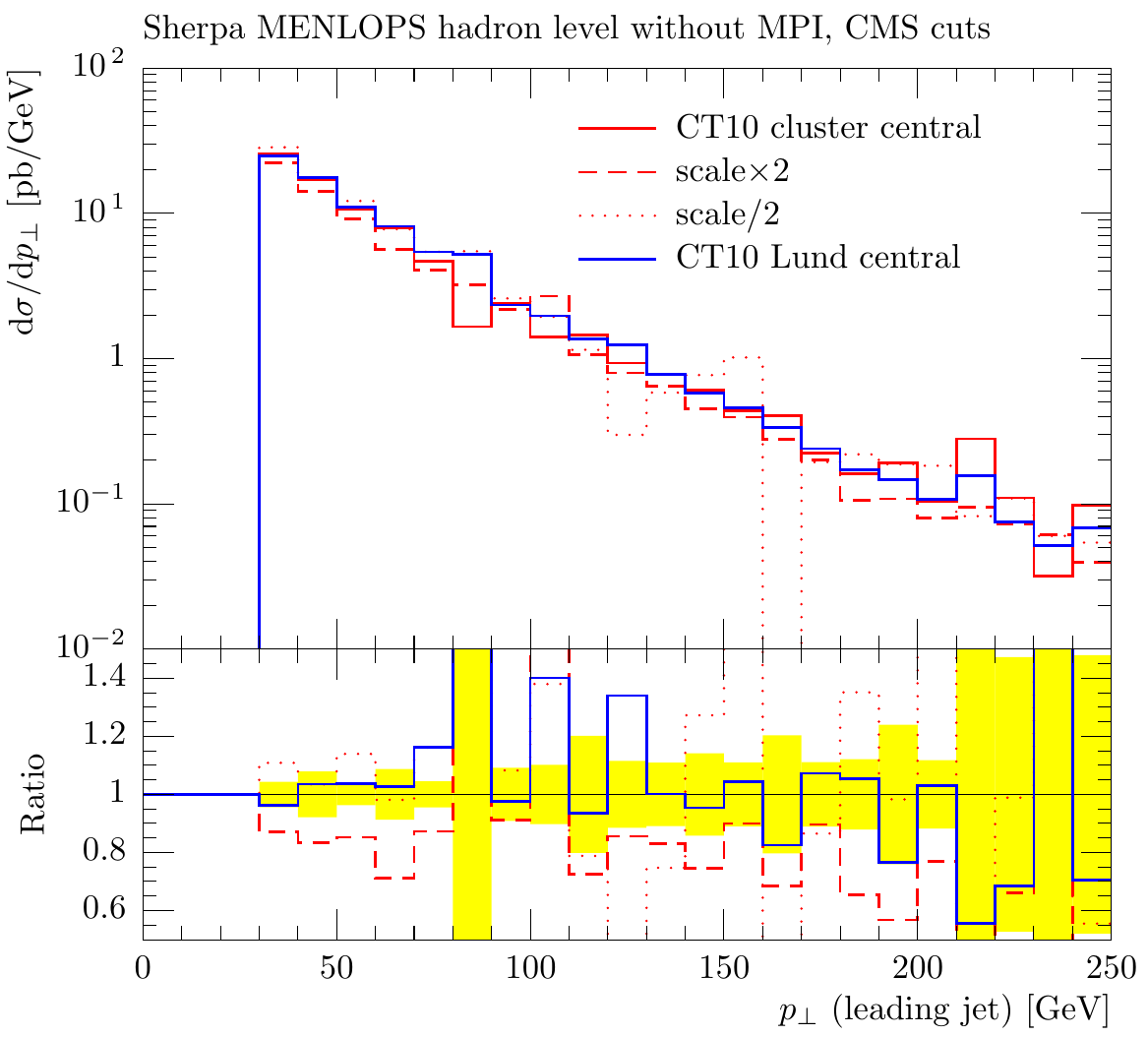}\hfill
  \includegraphics[width=.48\textwidth]{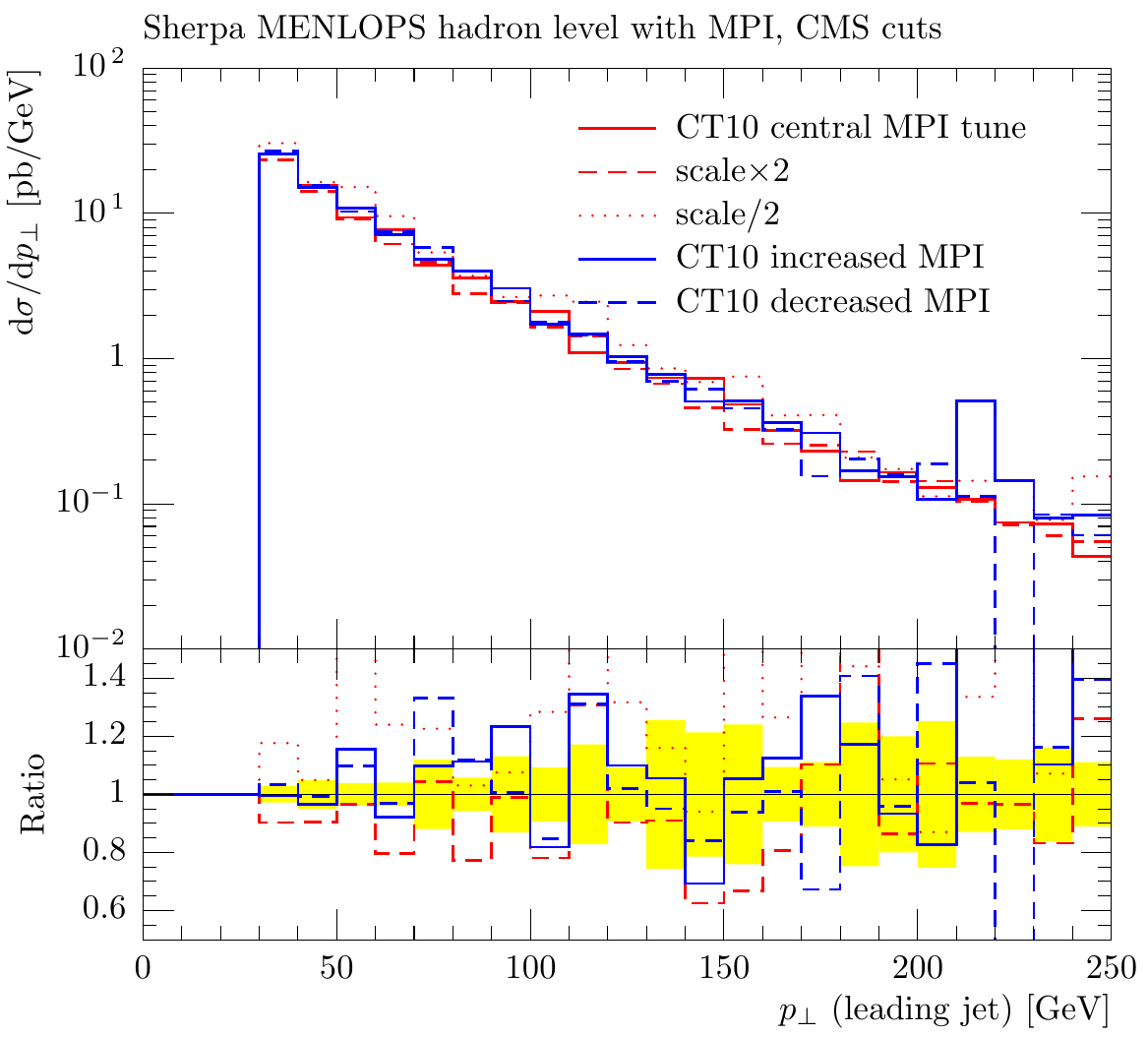}\\
  \includegraphics[width=.48\textwidth]{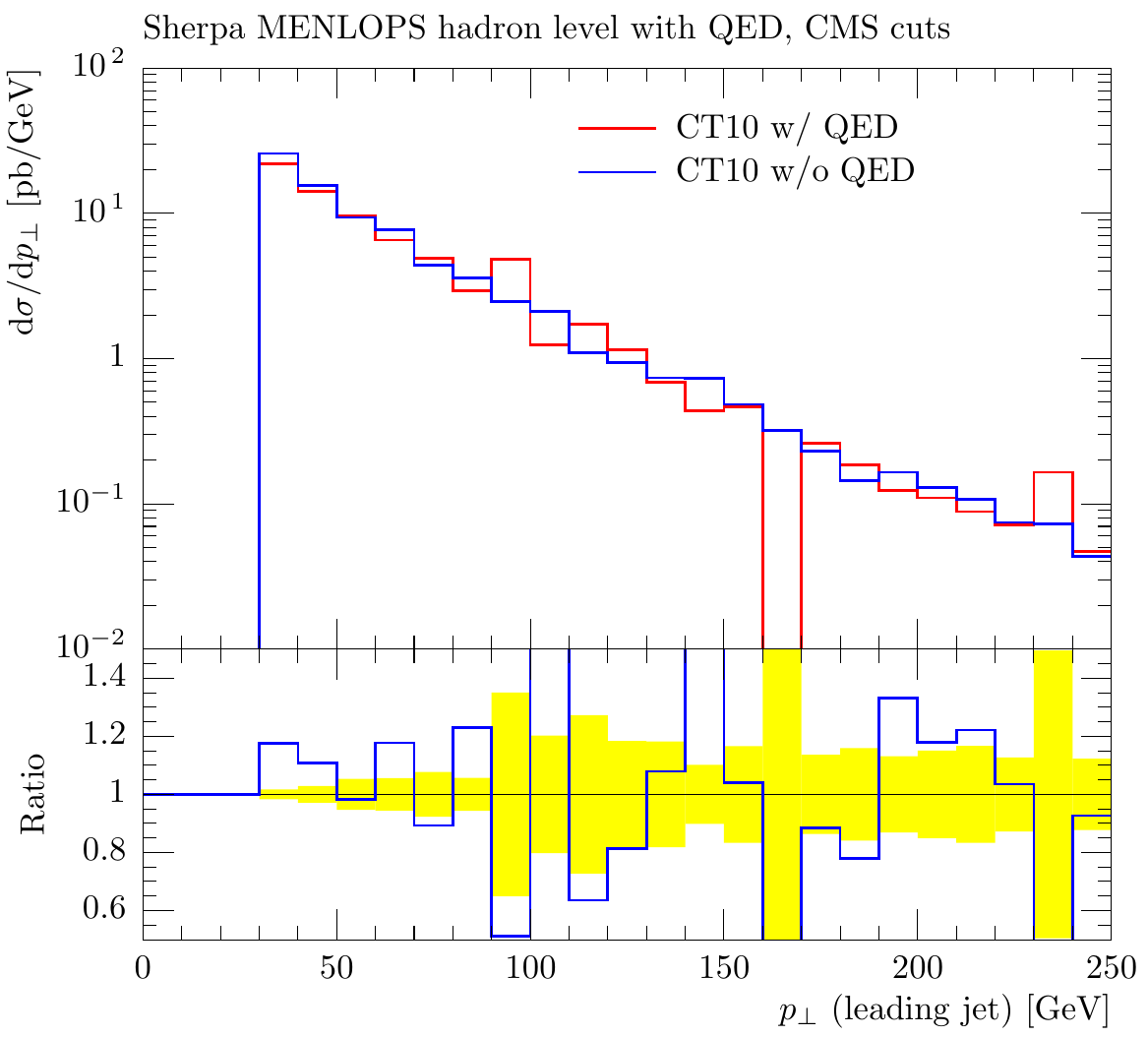}\hfill
  \includegraphics[width=.48\textwidth]{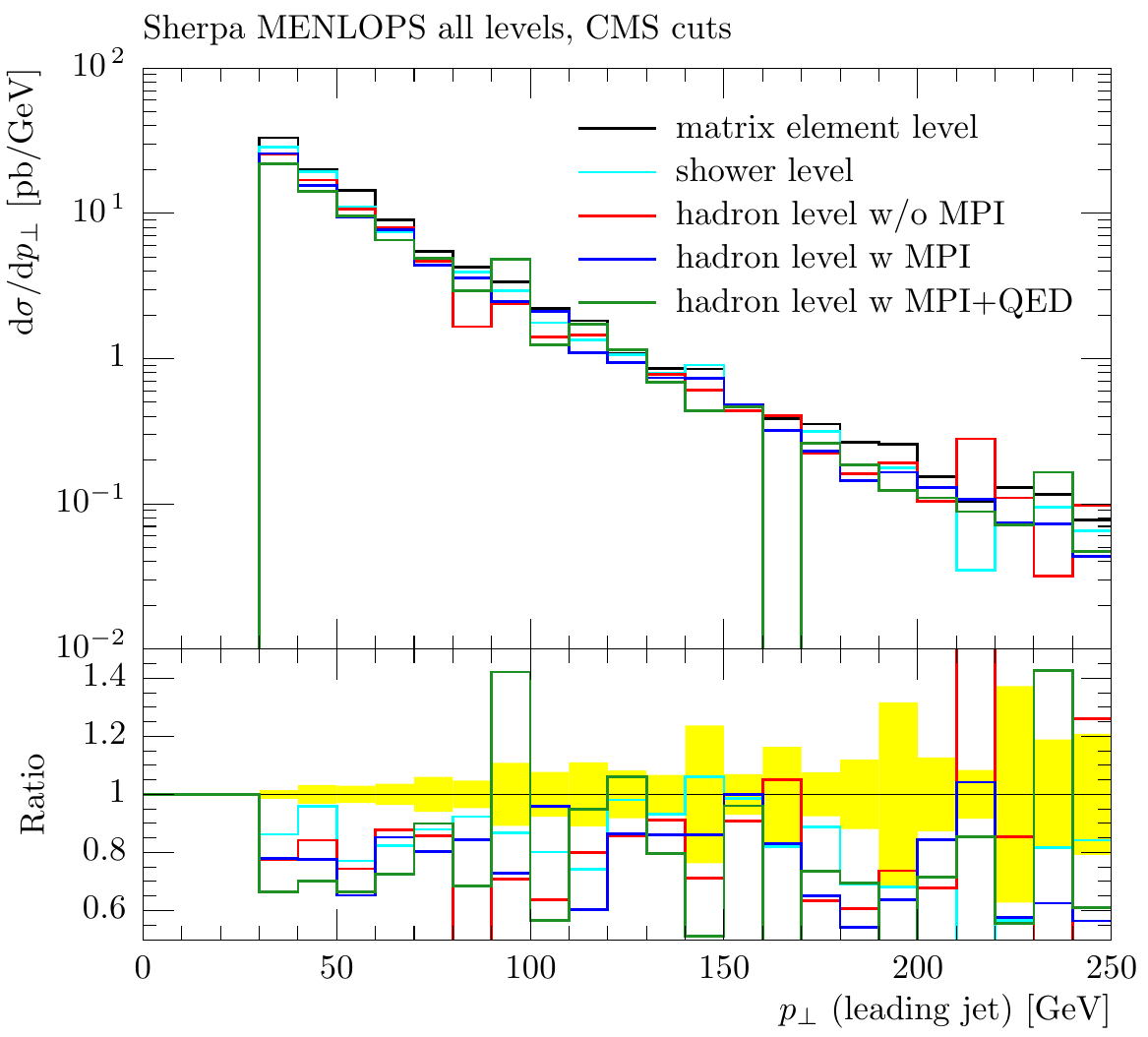}
  \caption{
	  \Sherpa\MENLOPS. Uncertainty of the transverse momentum of the hardest jet on the 
	  matrix element level (upper left), after parton showering (upper 
	  right), including hadronisation correction (centre left), 
	  multiple parton interactions (centre right), and QED corrections 
	  (lower left). The lower right panel shows the evolution of the 
	  central value.
	  \label{Fig:Results:Sherpa:MENLOPS:jetpt0}
  }
\end{figure}


\begin{figure}[p!]
  \includegraphics[width=.48\textwidth]{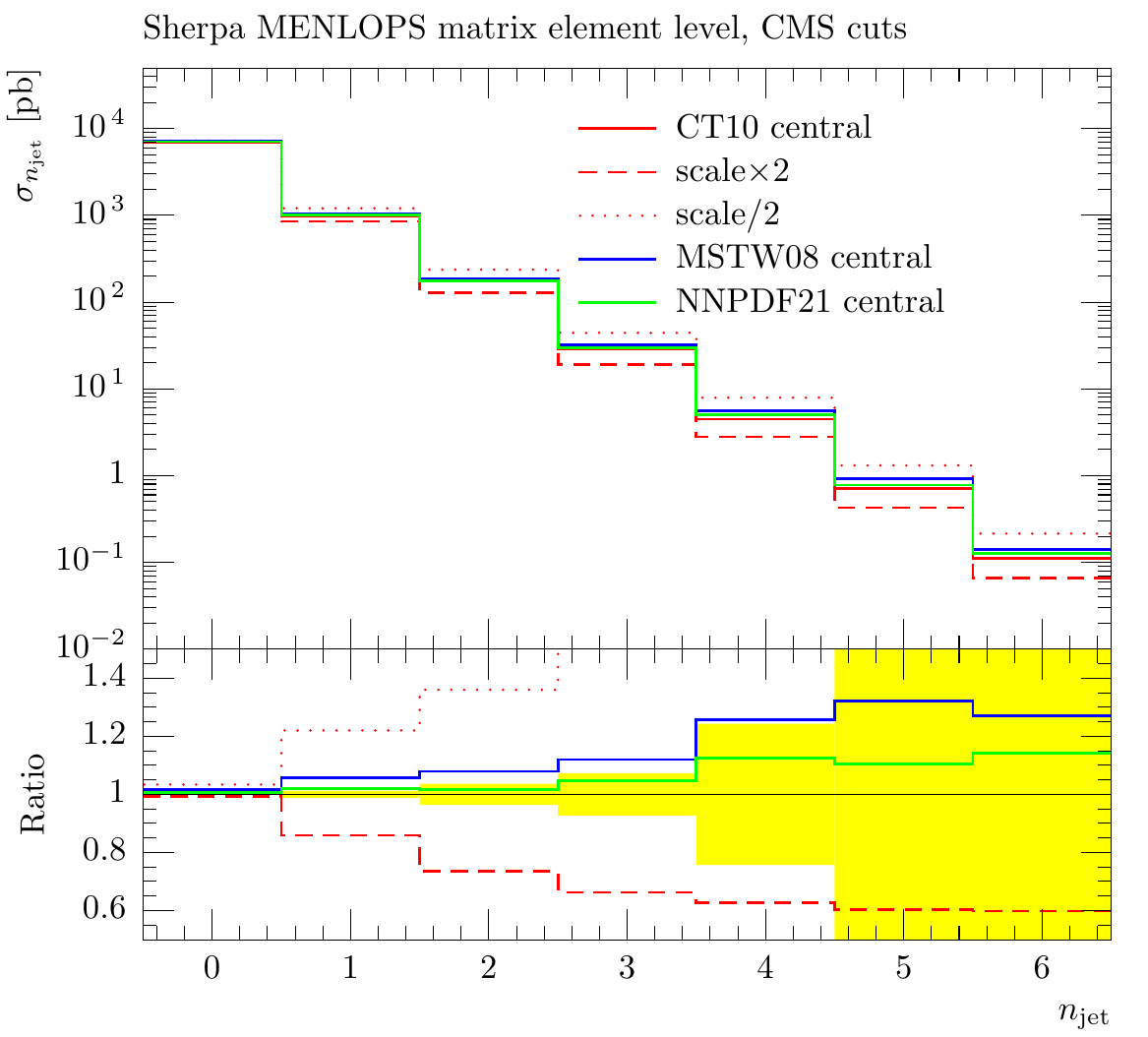}\hfill
  \includegraphics[width=.48\textwidth]{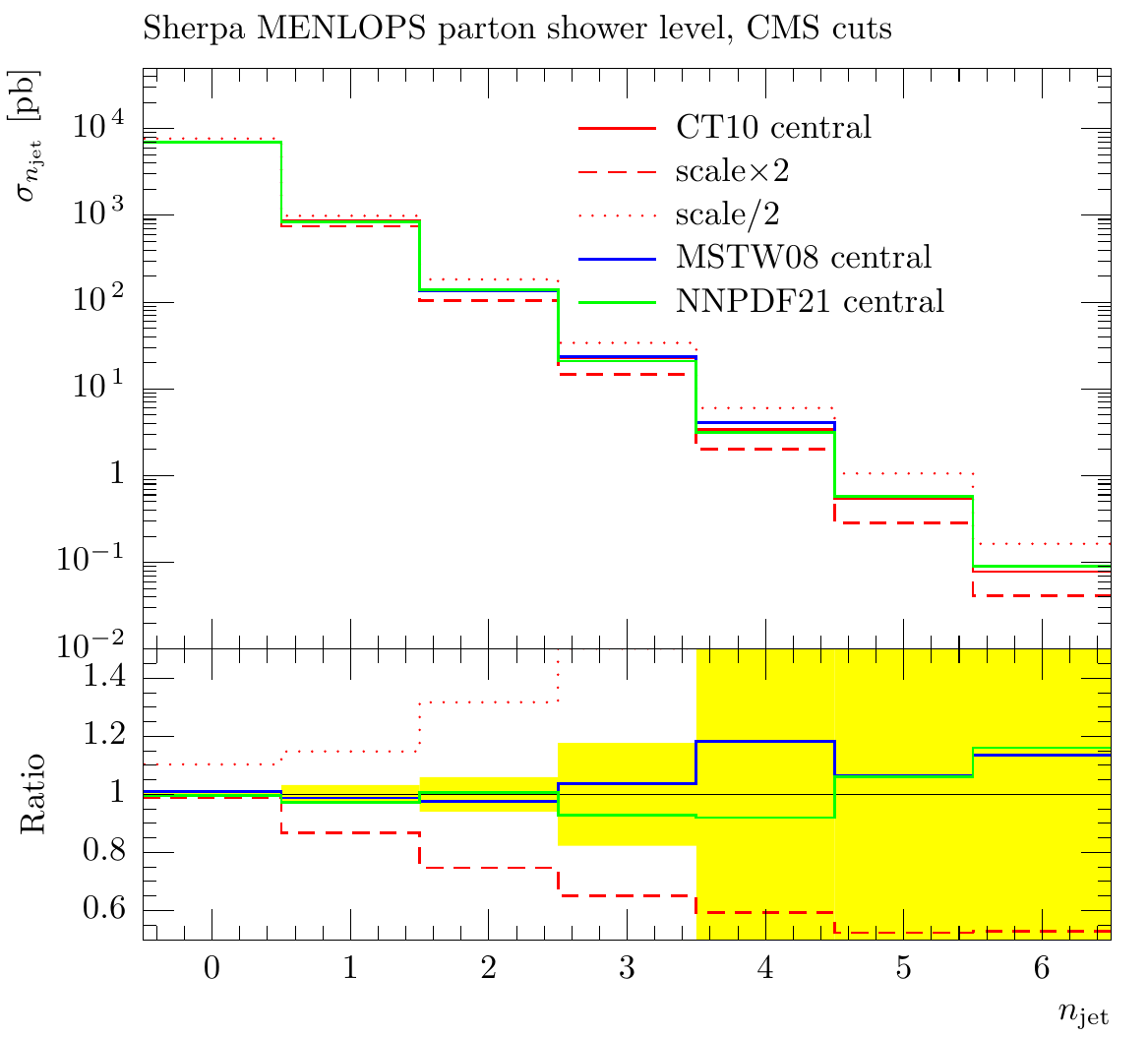}\\
  \includegraphics[width=.48\textwidth]{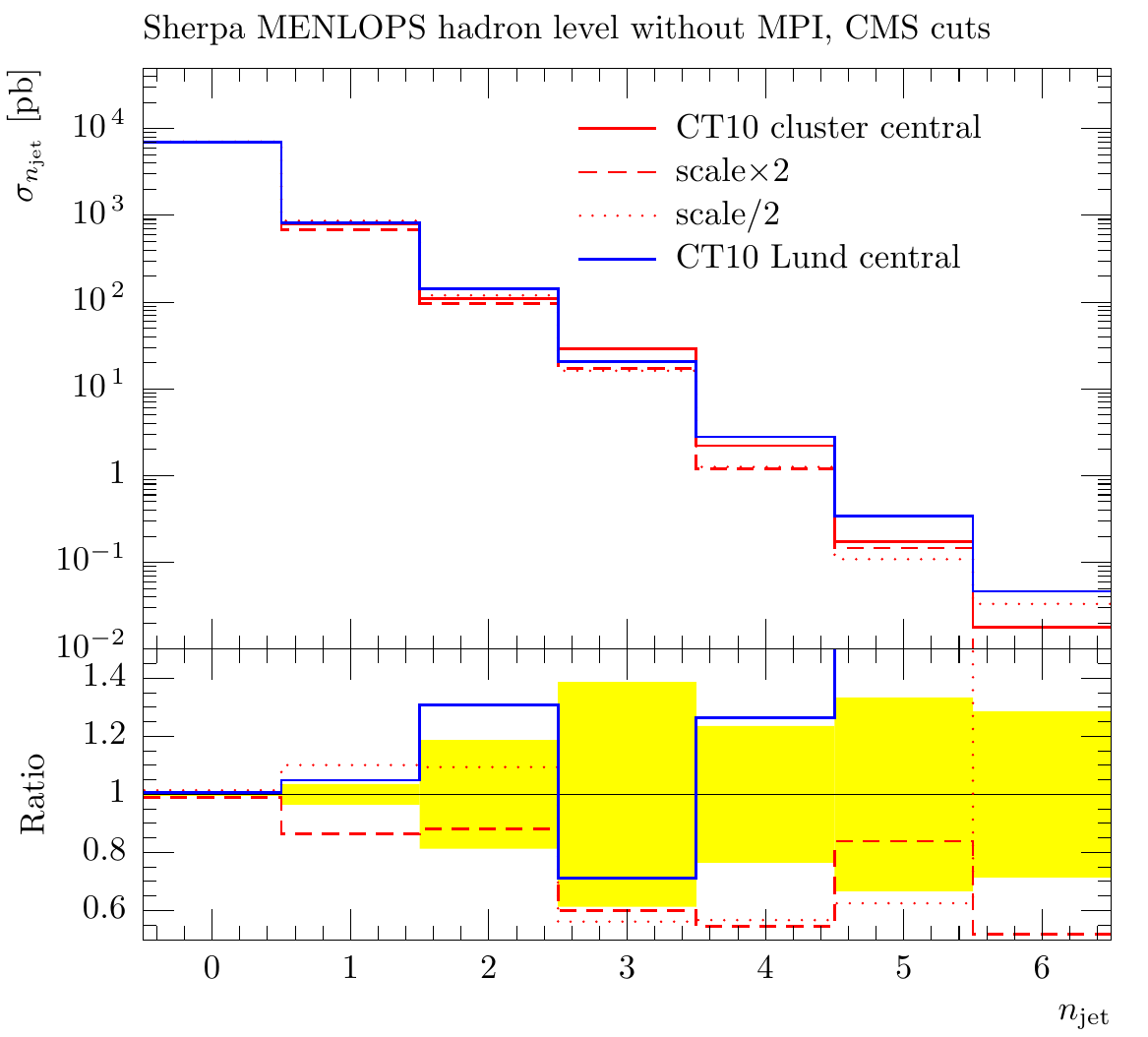}\hfill
  \includegraphics[width=.48\textwidth]{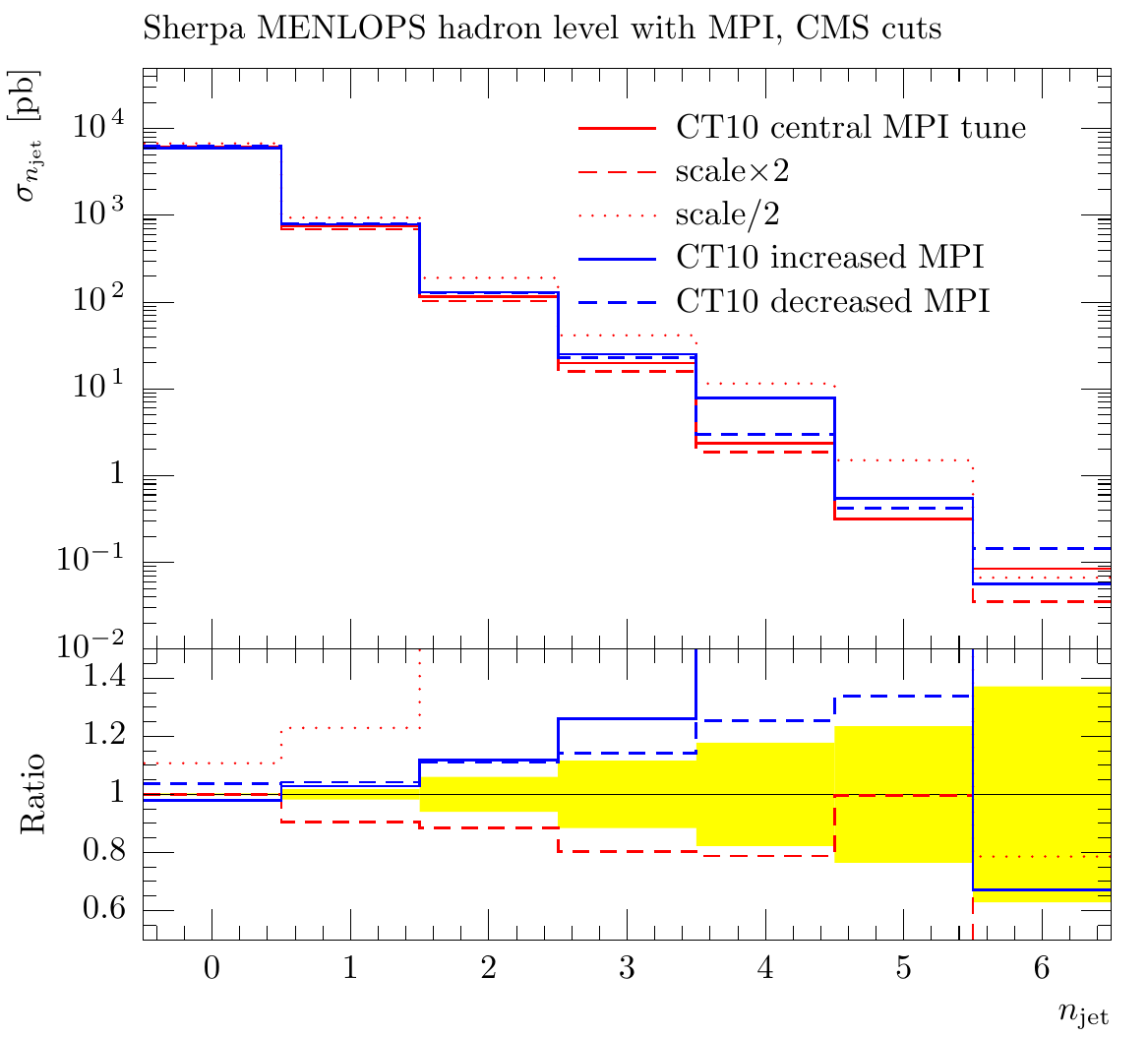}\\
  \includegraphics[width=.48\textwidth]{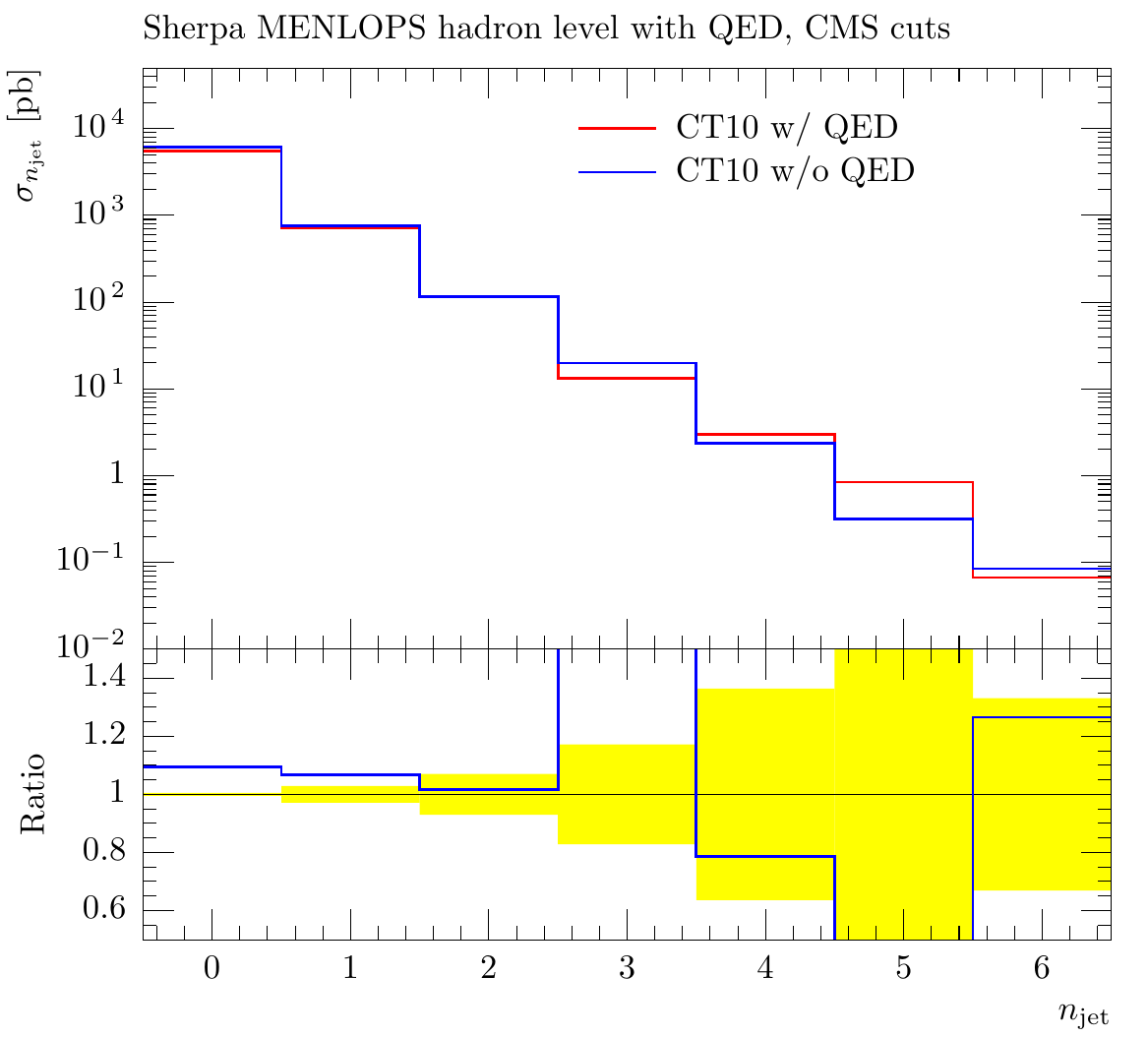}\hfill
  \includegraphics[width=.48\textwidth]{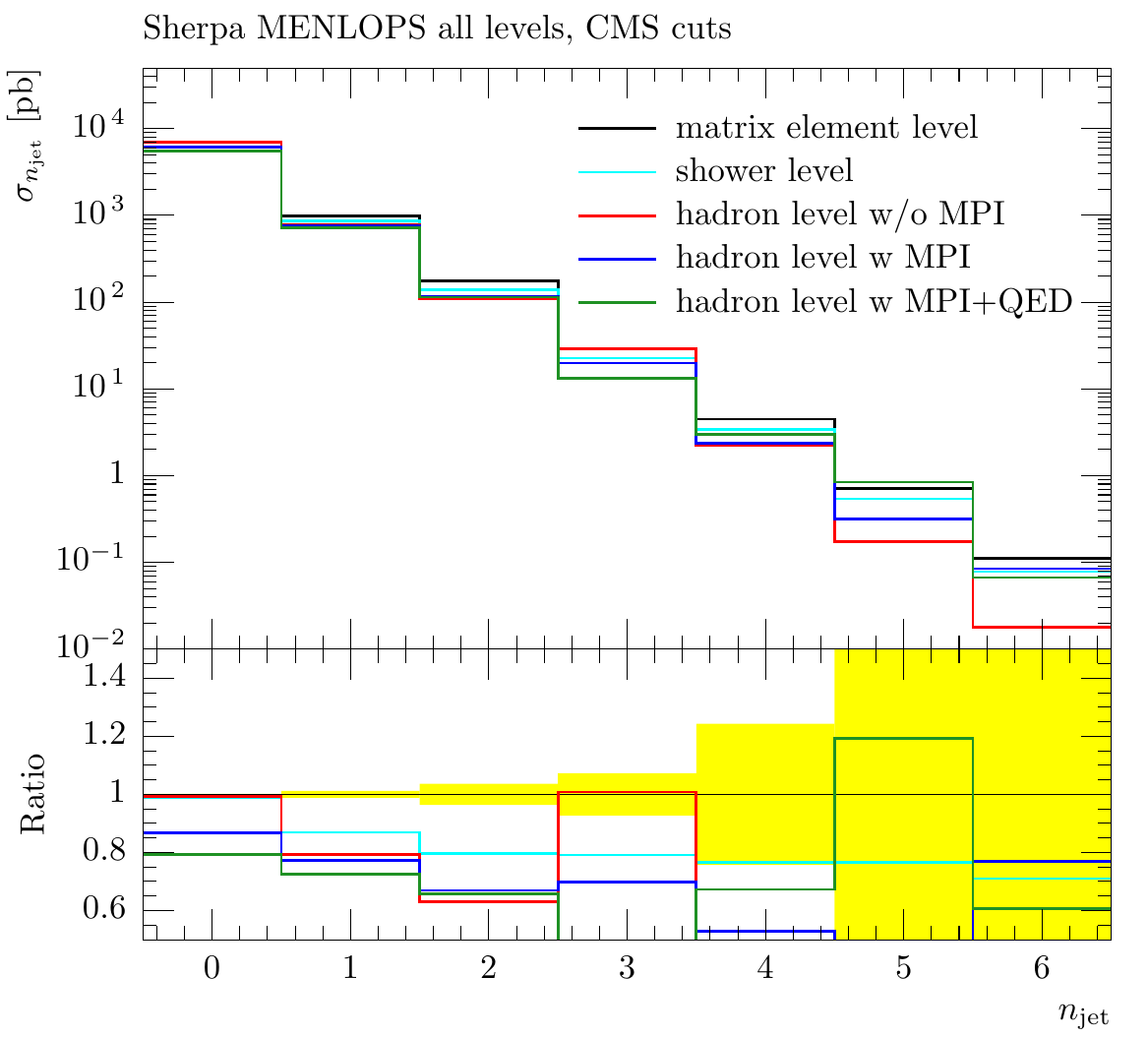}
  \caption{
	  \Sherpa\MENLOPS. Uncertainty of the inclusive jet multiplicity on the 
	  matrix element level (upper left), after parton showering (upper 
	  right), including hadronisation correction (centre left), 
	  multiple parton interactions (centre right), and QED corrections 
	  (lower left). The lower right panel shows the evolution of the 
	  central value.
	  \label{Fig:Results:Sherpa:MENLOPS:njet}
  }
\end{figure}

Following the outline of the previous subsection,
\FigRefs{Fig:Results:Sherpa:MENLOPS:Wmt}{Fig:Results:Sherpa:MENLOPS:njet}
compile the results, which were obtained by executing \Sherpa in the
\MENLOPS mode, cf.\ \SecRef{Sec:Codes:Sherpa:MENLOPS}. The
presentation is based on the same set of figures where the selection
of the observables has been taken as in the \MEPS case. Again, all
(central) predictions are examined towards their scale, PDF,
non-perturbative modeling and QED simulation dependence. One small
difference has to be pointed out: the plots to the lower left now
depict exclusively to what extent the additional QED corrections
modify the outcomes including multiple parton interactions and
hadronisation effects.

\FigRef{Fig:Results:Sherpa:MENLOPS:Wmt} shows the transverse mass of
the reconstructed $W$ boson. In the \MENLOPS approach, this observable
is described at NLO accuracy, which leads to a reduction of the
associated scale uncertainties. The scale variation results for
$m_{\perp,W}$ nicely confirm this expectation as can be seen in the
upper four plots of \FigRef{Fig:Results:Sherpa:MENLOPS:Wmt}. The
deviations from the central prediction are much smaller than those
found for the \MEPS scenario exhibited in
\FigRef{Fig:Results:Sherpa:MEPS:Wmt}; they now are of similar
magnitude as the PDF uncertainties. While the scale dependence is
reduced, PDF and MPI tune variations as well as QED corrections
manifest themselves as in the \MEPS case. In particular, the
discussion around \FigRef{Fig:Results:Sherpa:MEPS:Wmt} explaining the
effects of extra QED emissions (as being most relevant in the $W$
decay) can be used to understand the findings illustrated in the
bottom left panel of \FigRef{Fig:Results:Sherpa:MENLOPS:Wmt}.

The \MENLOPS method primarily improves the precision of the
description of the core process, here the description of the $W$
production process. One also benefits from improving the overall
normalisation. However, processes with additional partons in the final
state are described in the \MENLOPS approach at the same level of
accuracy as in the \MEPS approach -- in both cases by tree-level
matrix elements. Thus, the one-jet observables, $\Delta R$ between the
lepton and leading jet and the $p_\perp$ of the leading jet, and their
related uncertainties turn out to be predicted in a very similar
manner. This can be clearly observed by comparing 
\FigRefs{Fig:Results:Sherpa:MENLOPS:dRj0l}{Fig:Results:Sherpa:MENLOPS:jetpt0}
with \FigRefs{Fig:Results:Sherpa:MEPS:dRj0l}{Fig:Results:Sherpa:MEPS:jetpt0}.
Unlike the findings for $m_{\perp,W}$, it particularly can be noticed
that the scale dependence associated with the one-jet observables
shown here remains unchanged when compared to the respective \MEPS
results.

\FigRef{Fig:Results:Sherpa:MENLOPS:njet} depicts the distribution of
the inclusive $W+n$ jet cross sections as obtained for the \MENLOPS
case. Using the above reasoning, one can understand these results as
for the one-jet variables. Note that the scale dependence of the
zeroth jet bin shows the expected decrease owing to the NLO accuracy
underlying the description of the core process.

\FigRef{Fig:Results:Sherpa:MENLOPS:beamthrust} finally, highlights the 
evolution and uncertainties of two definition of the beamthrust, cf.\ 
\AppRef{App:Observables}: a physical observable summing over all final 
state particles excluding the $W$-constituent lepton and a pseudo-observable 
including the $W$ itself. For both observables small perturbative 
uncertainties are completely burried underneath much larger non-perturbative 
effects and modelling uncertainties, an effect also seen in results from the 
\PowhegBox{}+\Pythia simulation, cf.\ \FigRef{fig:indiv_PowhegPythia8_tunes}.  
This can only be interpreted as this observable being dominated by 
non-perturbative effects and in particular the underlying event, which 
somewhat invalidates statements about the merit of this observable in a clean 
determination of initial state radiation effects made 
in~\cite{Stewart:2010pd,Stewart:2011cf}.

\begin{figure}[t!]
  \includegraphics[width=.48\textwidth]{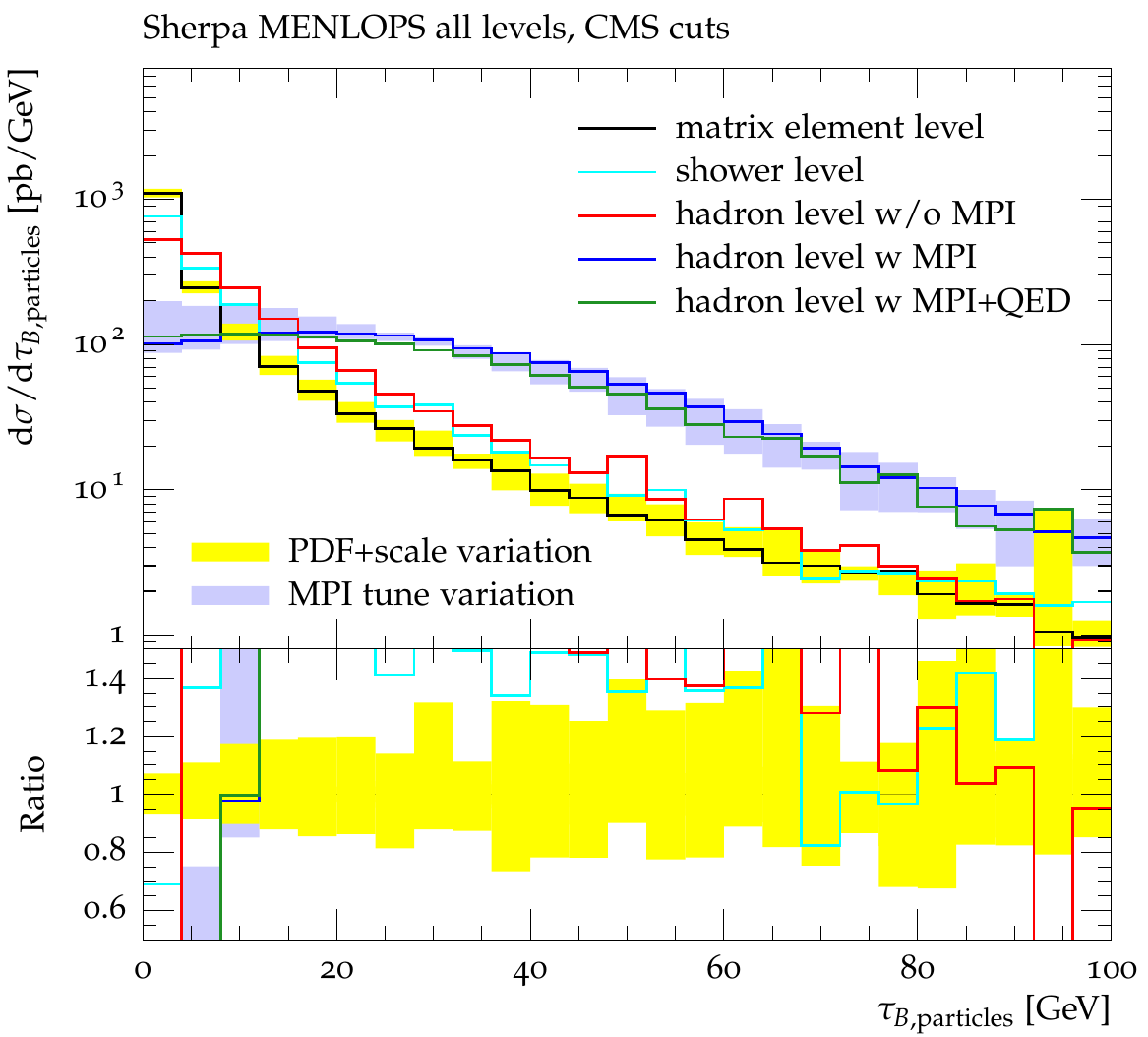}\hfill
  \includegraphics[width=.48\textwidth]{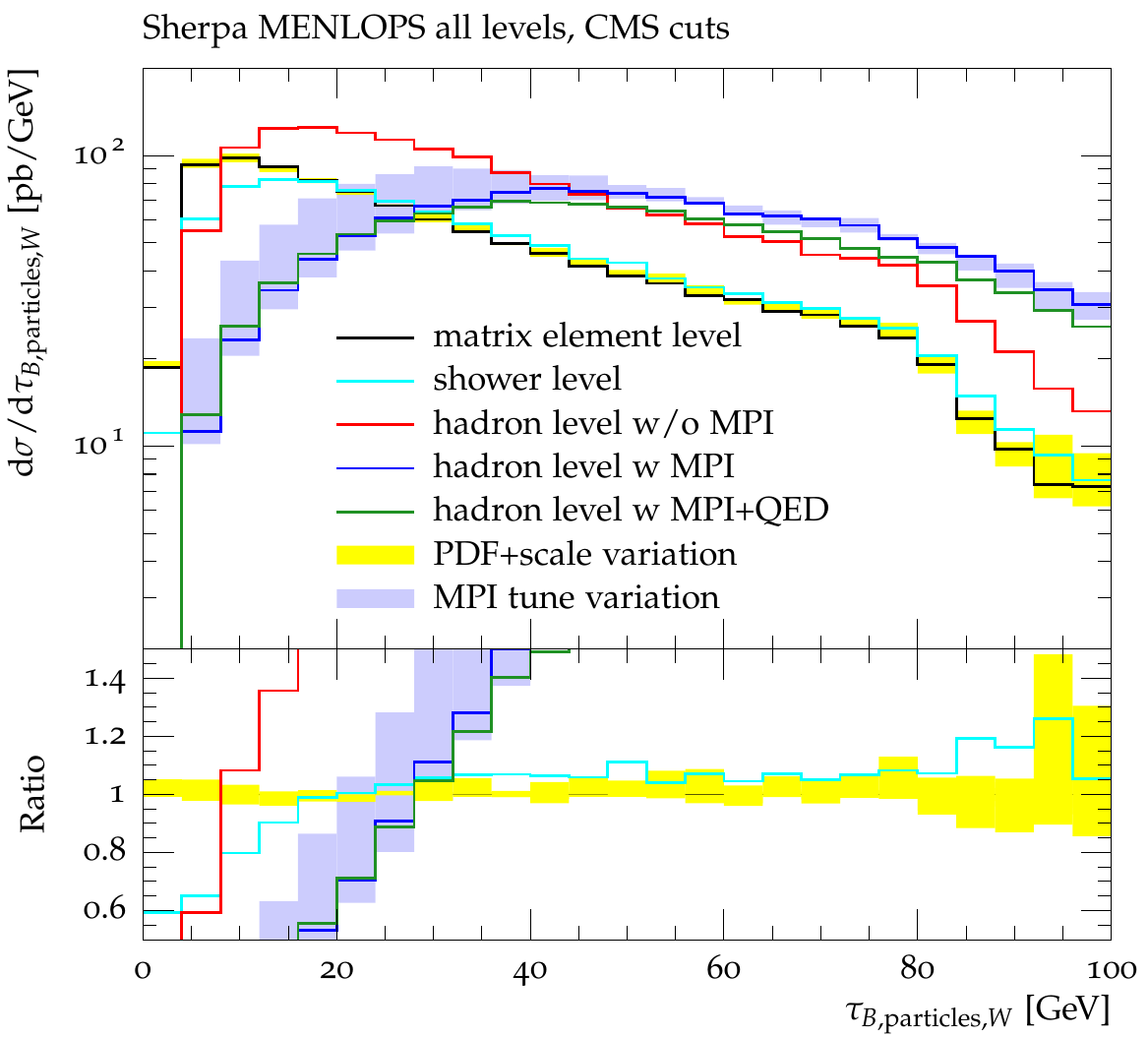}
  \caption{
	   \Sherpa\MENLOPS. Evolution and uncertainty of two definitions 
	   of the beamthrust, calculated using all particles not constituting 
	   the $W$ (left) and including the $W$ (right). Exemplary, the 
	   combined PDF and 
	   scale uncertainty on the \MElevel prediction (yellow) and the 
	   modeling uncertainty of the hadron level prediction (blue) are shown.
	   \label{Fig:Results:Sherpa:MENLOPS:beamthrust}
  }
\end{figure}

\FloatBarrier


\subsection{Comparisons}
\label{Sec:Comparison}

In this section we compare the results of different tools with each other.
While the aim of this study was to have a fairly tuned comparison with
as many aspects of the calculations as possible being centrally defined,
there are still important residual differences in the various results.
Obviously, the different codes produced results at different stages of
the simulation, which are not always directly comparable; in addition,
some of these stages are not very straightforward to obtain: for instance,
running \PythiaEight without multiple parton interactions included in the 
interleaved showering obviously changes the overall logic of the parton shower
model of this code.  In addition, other, more obvious differences occur, 
ranging from inconsistent choices of PDFs to different strategies in scale 
setting procedures.  For the case of the PDFs, by directly comparing results 
obtained with \BlackHat{}+\Sherpa using CTEQ6.6 and with \GoSam{}+\Sherpa
using CT10, it appears as if at NLO these differences are minor.  However, it
is not clear how much of the differences between \Madgraph{}+\Pythia and \PythiaEight, 
which both employ CTEQ6L1, and the other codes, which employ NLO PDFs, can be attributed 
to differences in PDFs.

In addition, results obtained with the NLO codes typically include at least one
jet - \PowhegBox{}+\PythiaEight and \GoSam{}+\Sherpa take $W+1$ jet at NLO as 
their core process - while \HEJ starts at $W+2$ jets, and \BlackHat{}+\Sherpa
presents results for up to 4 jets accompanying the $W$ boson in different
jet bins.  Obviously, on the other hand, the multijet merged samples
of \Madgraph{}+\Pythia, \PythiaEight\MEPS and \Sherpa include LO matrix 
elements for up to 3 to 6 jets.  

In the plots in this section each code is shown with a yellow error band, 
which is the envelope of the variations presented in \SecRef{Sec:Results}. The 
only exception is \BlackHat{}+\Sherpa, which is shown with a blue error band. In 
the ratio plots the codes are plotted relative to \BlackHat{}+\Sherpa, also 
at the parton shower level.

\subsubsection{Inclusive observables}
\begin{figure}[t]
  \includegraphics[width=.48\textwidth]{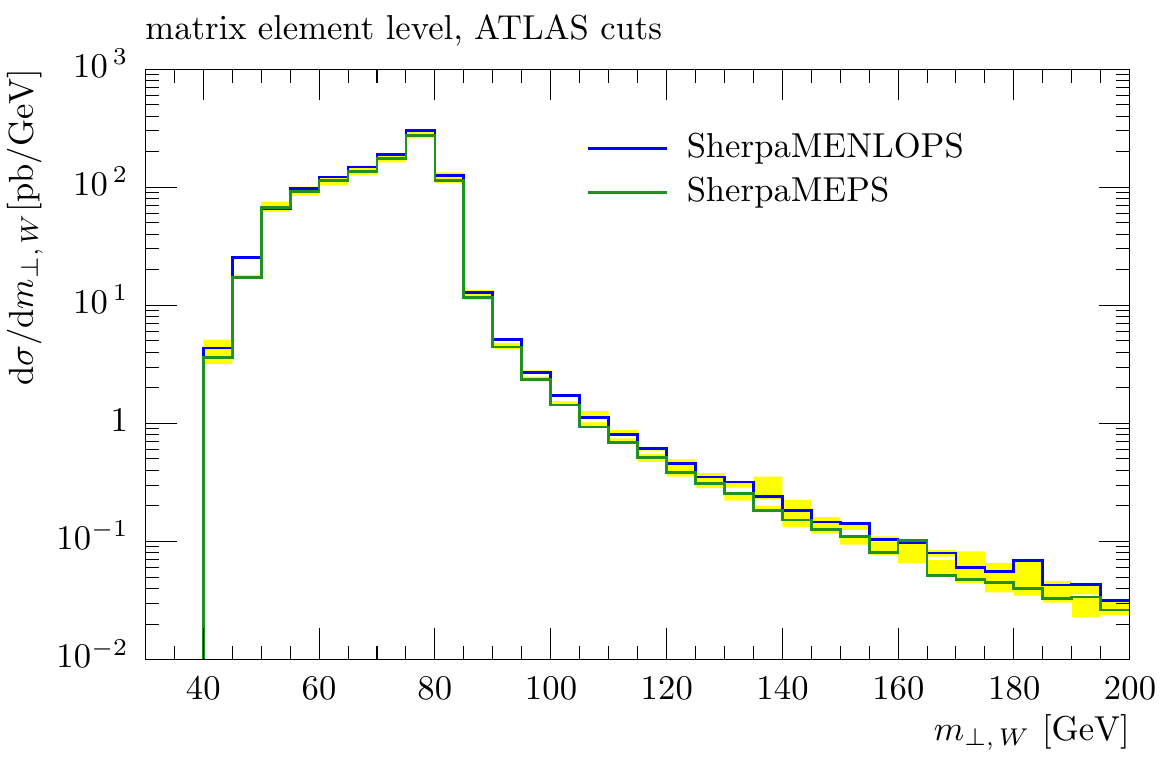}\hfill
  \includegraphics[width=.48\textwidth]{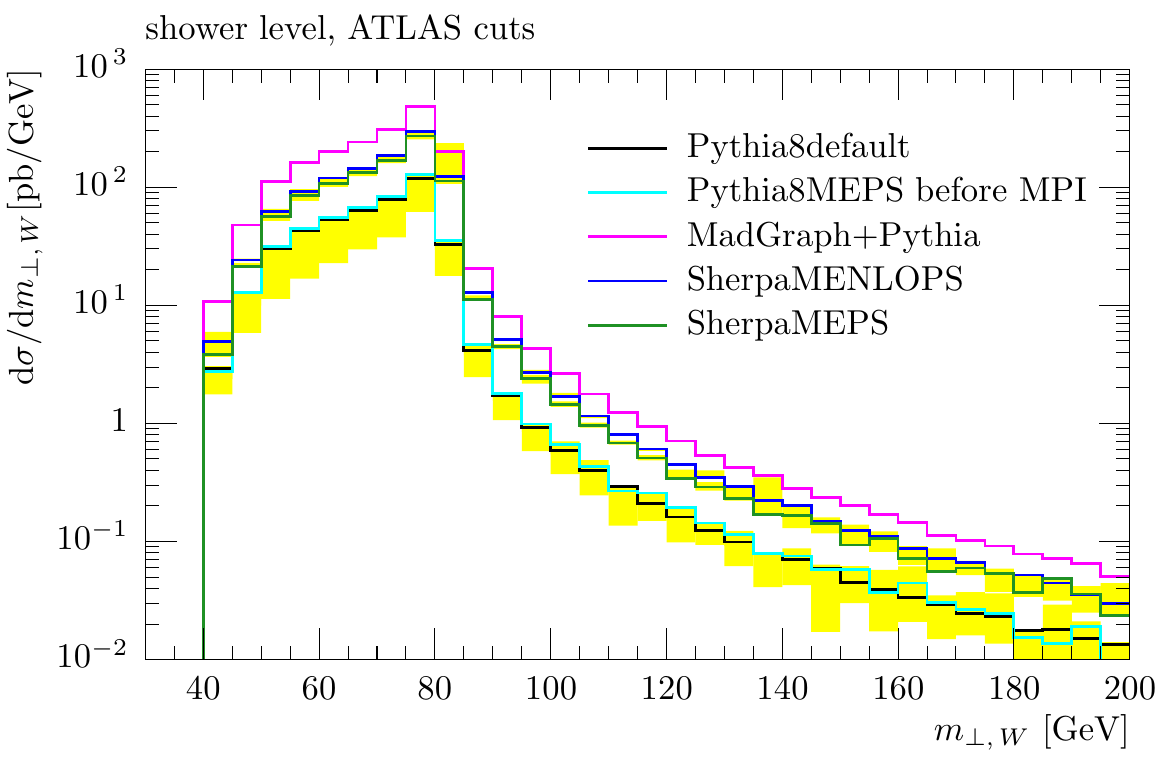}\\
  \includegraphics[width=.48\textwidth]{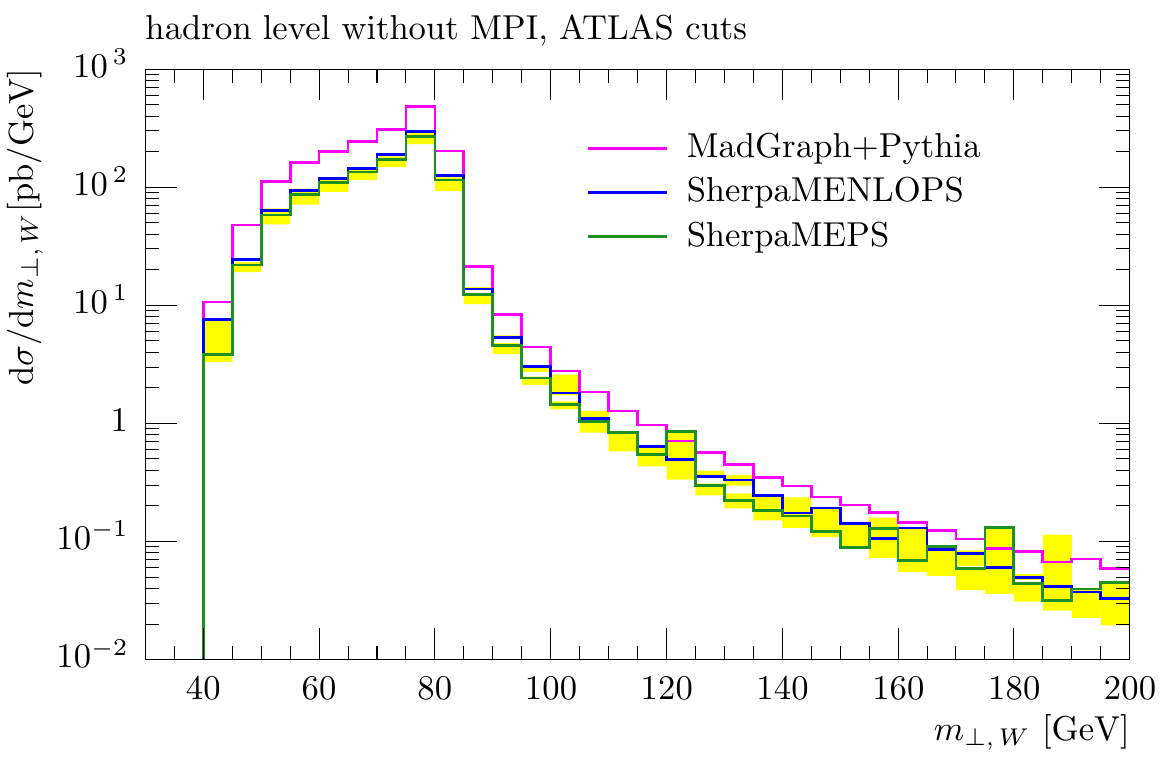}\hfill
  \includegraphics[width=.48\textwidth]{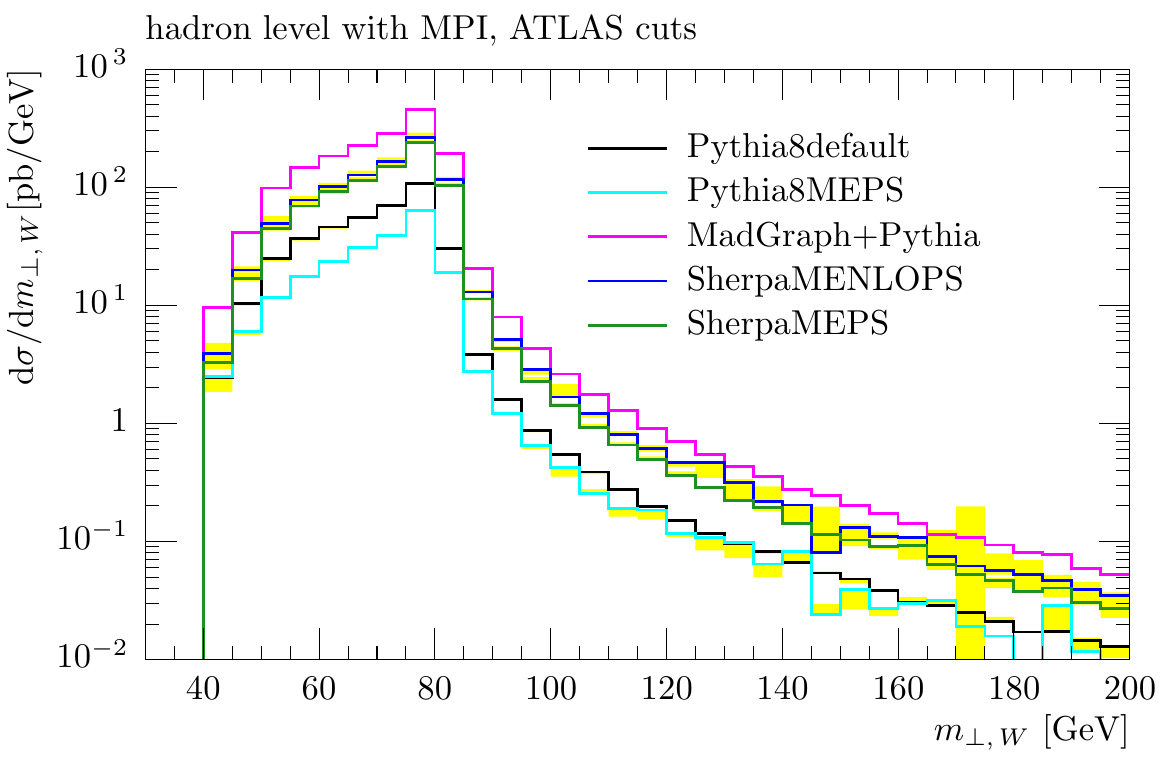} 
  \caption{Transverse mass of the reconstructed $W$ on all levels of the simulation, 
  for the exact definition see \AppRef{App:Observables:Analysis} and 
  for the cuts employed in the analysis \AppRef{App:Observables:Cuts}. 
  Note that \PythiaEight and \Madgraph{}+\Pythia use the CTEQ6L1 pdf, 
  while \Sherpa uses CT10.} 
\end{figure}

In this section we present some inclusive observables, which are typically
all obtained from codes employing multijet merging.  By and large, all
codes agree in the shapes of the $m_{\perp,W}$ distribution at different stages,
although there are sizable differences in the respective normalisation of the
samples.  

\subsubsection{Observables with at least one jet}
\begin{figure}[t]
 \includegraphics[width=.48\textwidth]{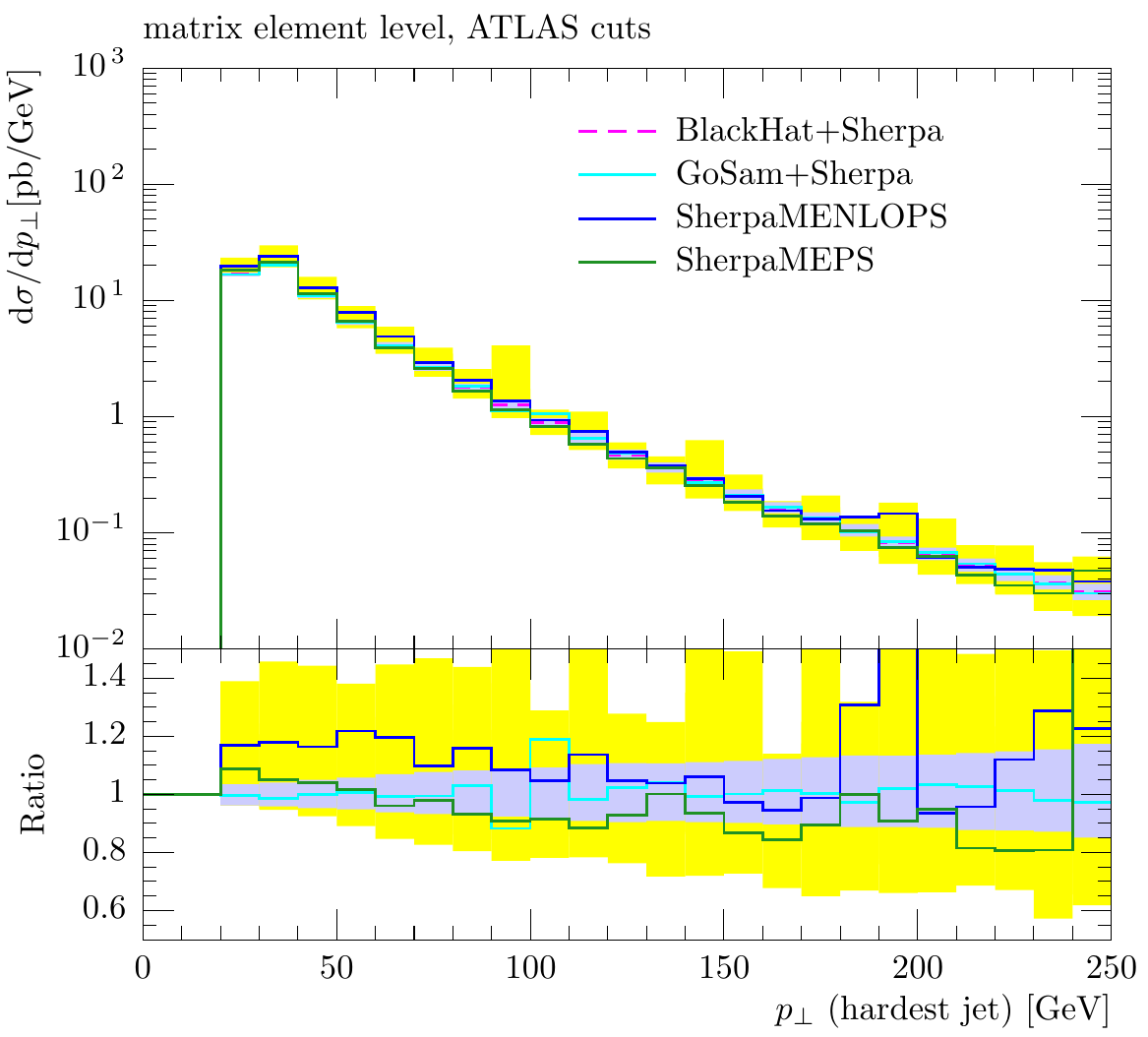}\hfill
 \includegraphics[width=.48\textwidth]{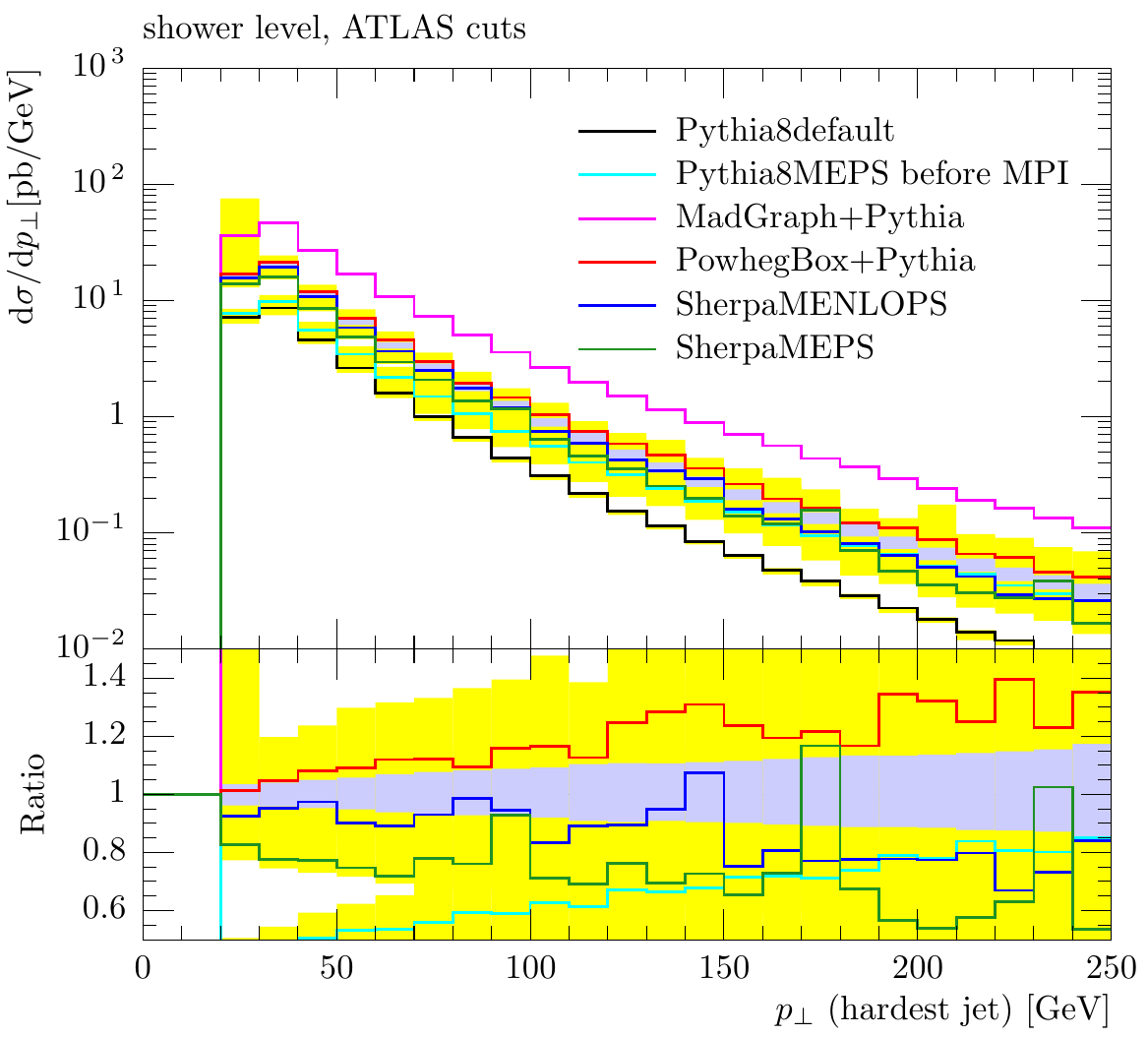}\\
 \includegraphics[width=.48\textwidth]{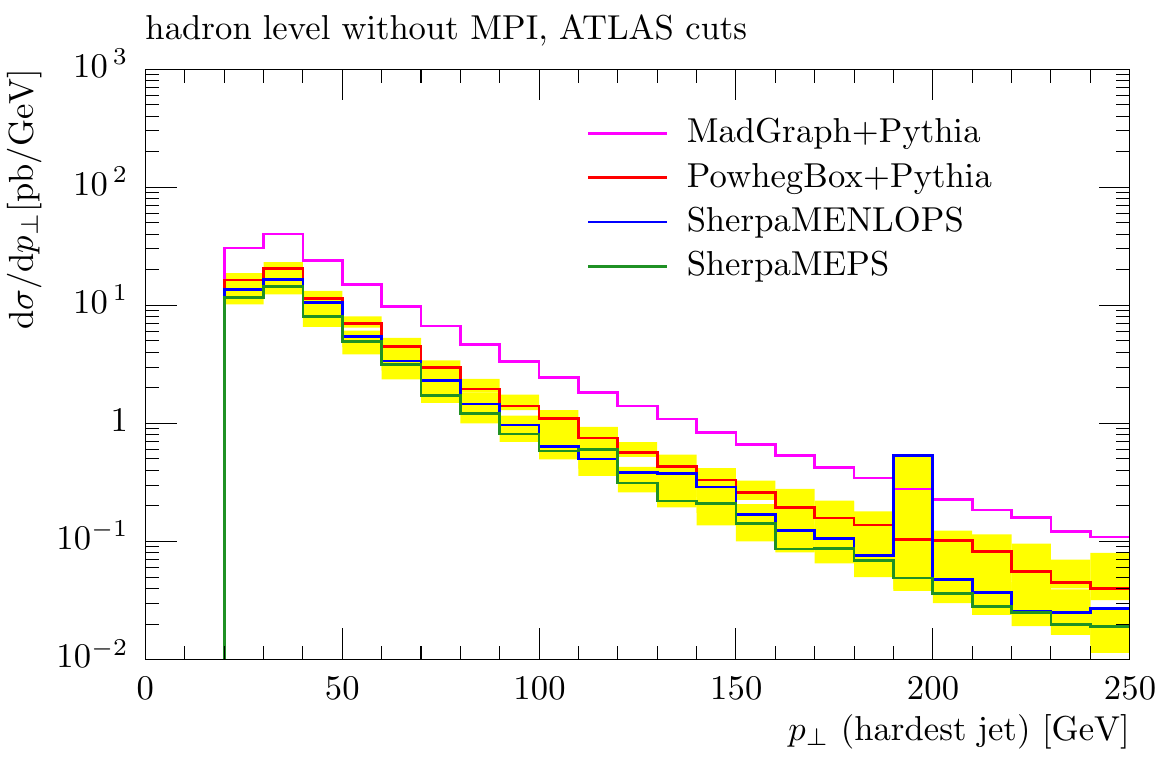}\hfill
 \includegraphics[width=.48\textwidth]{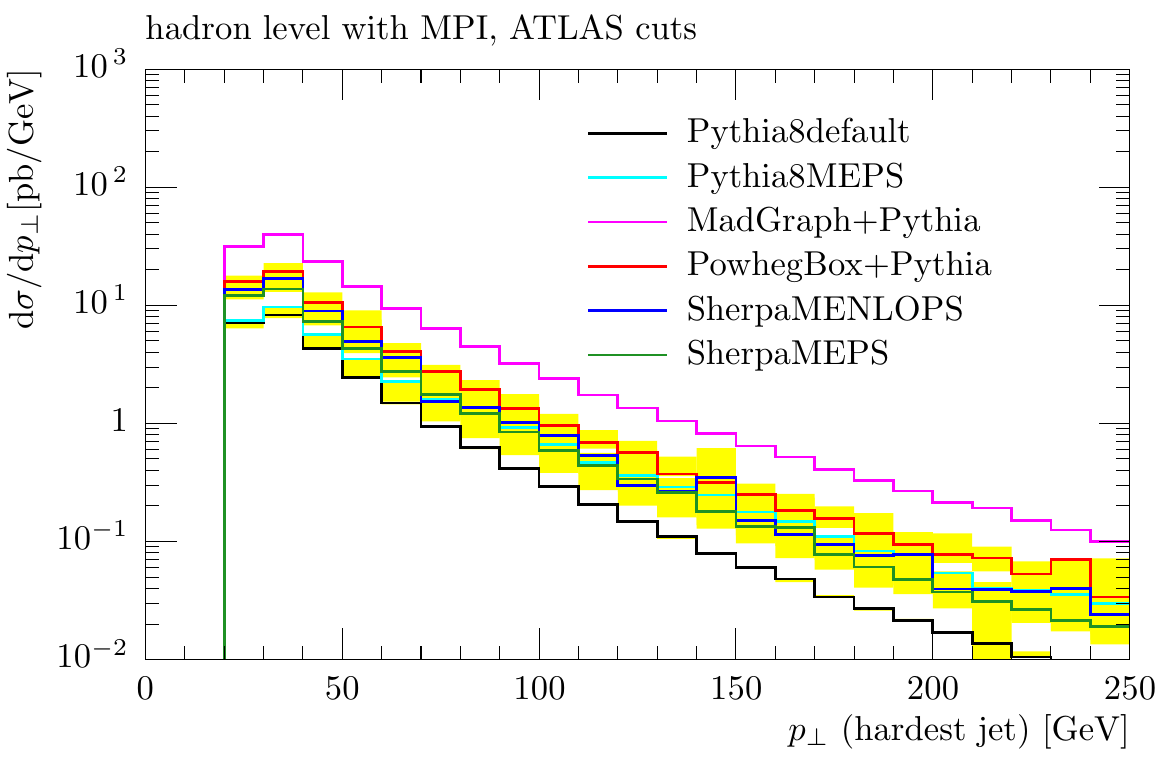}
 \caption{Transverse momentum of hardest jet on all levels of the simulation, where jets are reconstructed using the 
 anti-$k_\perp$ with $R=0.4$ within $|\eta|<4.4$ (for exact definitions and cuts see
 \AppRef{App:Observables:Cuts} and \AppRef{App:Observables:Analysis}). 
 Note that \BlackHat uses the CTEQ6.6 pdf, \PythiaEight and \Madgraph{}+\Pythia 
 CTEQ6L1 and all the others use CT10. In both ratio plots the ratio is taken
 with respect to \BlackHat{}+\Sherpa (on matrix element level).}
 \label{fig:comparison_ptjet1}
\end{figure}

\begin{figure}[t]
 \includegraphics[width=.48\textwidth]{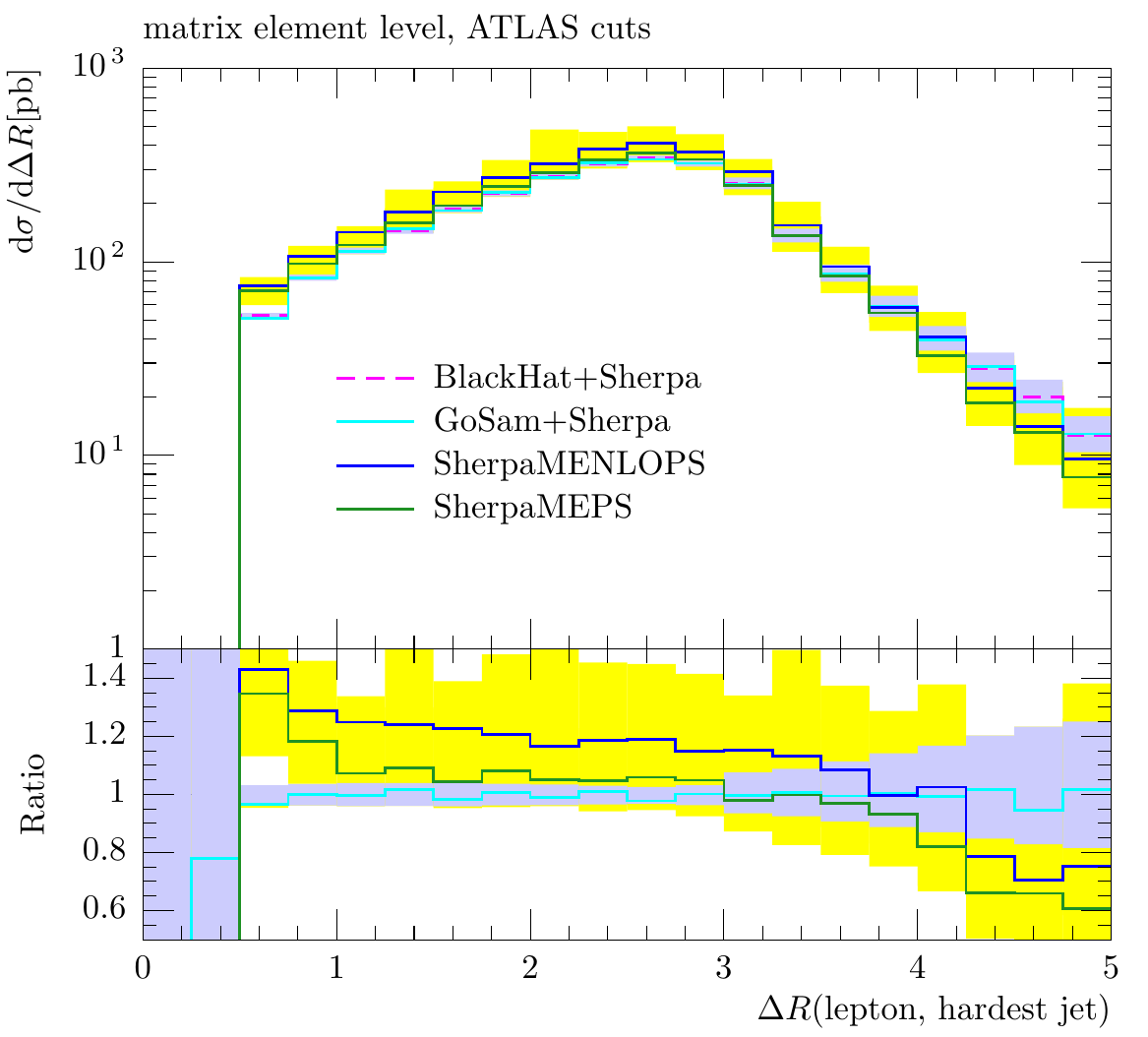}\hfill
 \includegraphics[width=.48\textwidth]{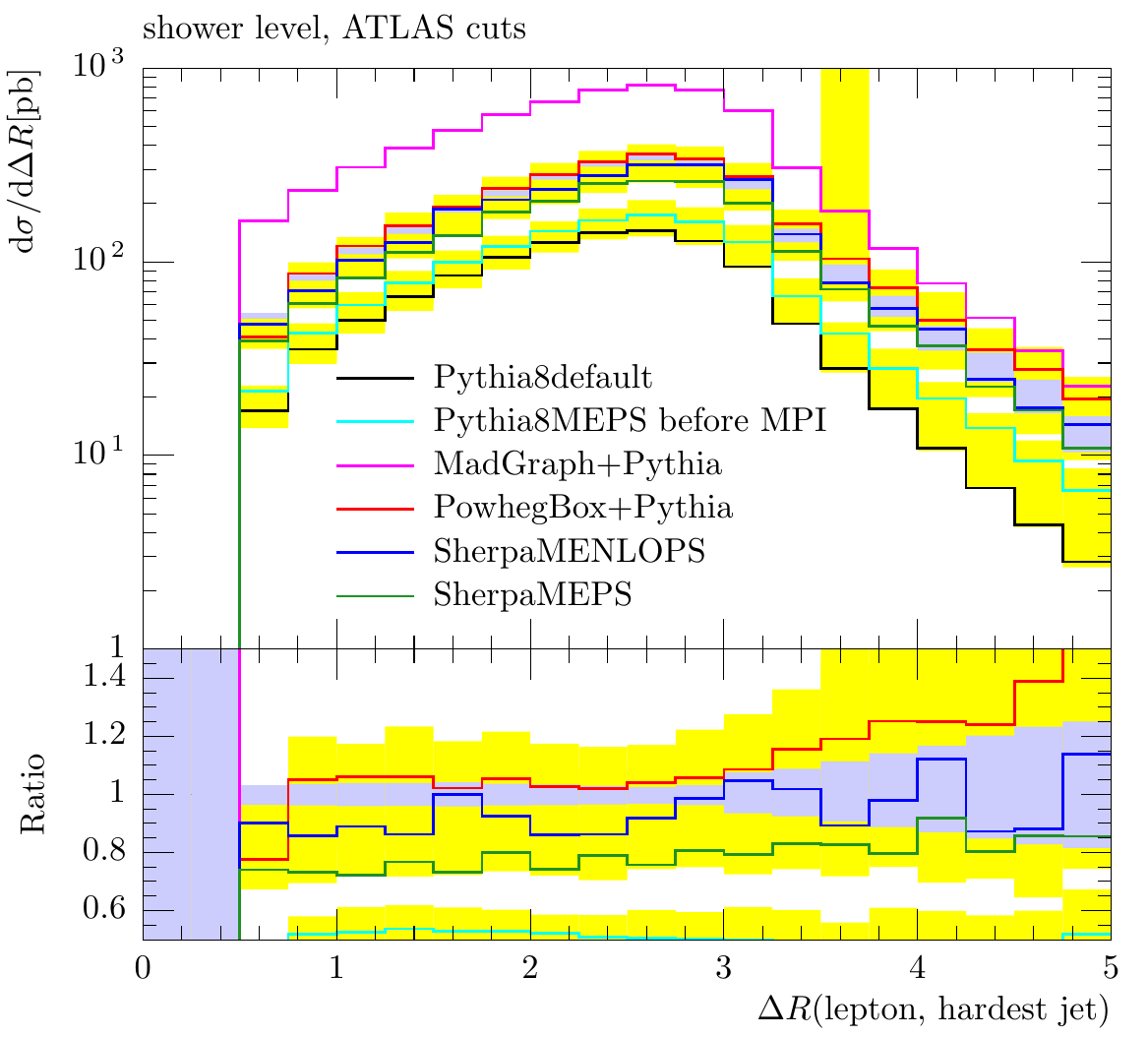}\\
 \includegraphics[width=.48\textwidth]{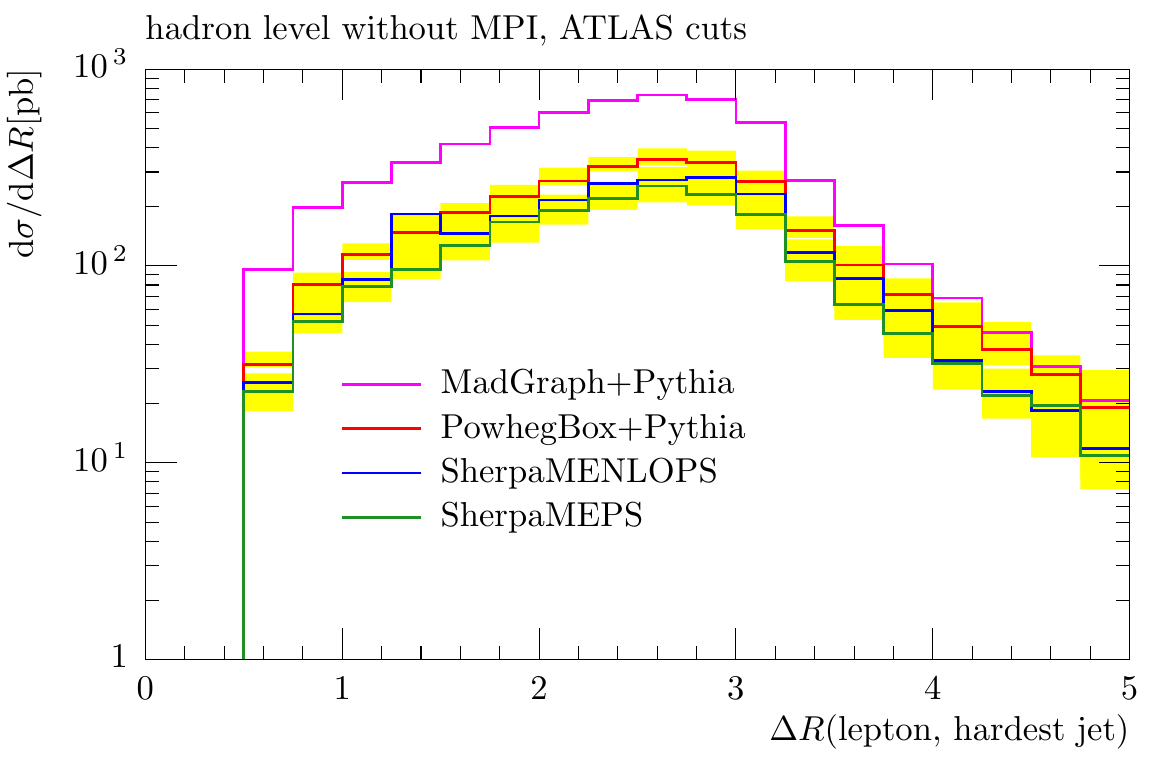}\hfill
 \includegraphics[width=.48\textwidth]{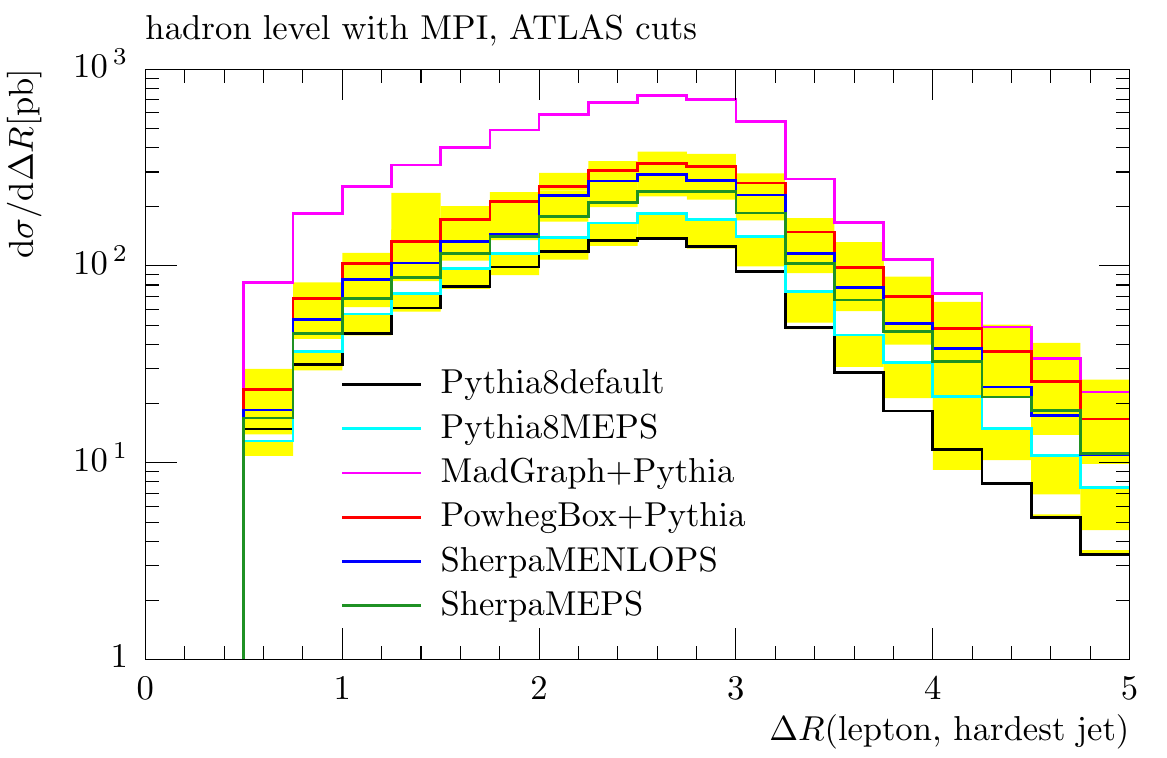}
 \caption{$\Delta R$ between hardest lepton and hardest jet on all levels of the simulation (for exact definitions and cuts see
 \AppRef{App:Observables:Cuts} and \AppRef{App:Observables:Analysis}). 
 Note that \BlackHat uses the CTEQ6.6 pdf, \PythiaEight and \Madgraph{}+\Pythia 
 CTEQ6L1 and all the others use CT10. In both ratio plots the ratio is taken
 with respect to \BlackHat{}+\Sherpa (on matrix element level).}
\label{fig:comparison_DeltaR}
\end{figure}

As a first and fairly telling observable the $p_\perp$-spectrum of the
hardest jet is compared, cf.\ \FigRef{fig:comparison_ptjet1}.  At the
parton level, the results of the NLO calculations --
\BlackHat{}+\Sherpa and \GoSam{}+\Sherpa -- agree nearly perfectly
with each other and within about 20\% with the multijet merged samples
of \Sherpa, both at LO (\Sherpa\MEPS) and in the \MENLOPS
(\Sherpa\MENLOPS) sample.  The increase of the latter with respect to
the former at relatively low transverse momenta of about 50 GeV or
below can probably be related to the different scale definition in the
argument of the strong coupling, where the NLO calculations choose
$\mu_R^2 = \left(H_T'/2\right)^2\approx M_W^2/4+p_{\perp, j}^2$ while
in the \Sherpa simulation the transverse momentum of the jet has been
chosen.  Clearly, for small transverse momenta this will lead to
visible differences.  Going from the matrix element to the parton
shower level typically leads to the jets becoming softer and to
losing some of them, due to partons emitted outside the jet and a
corresponding energy loss.  This explains why the \Sherpa distribution
at the shower level is softer than the NLO result, and thus the
\Sherpa result at the matrix element level, although the size of the
difference seems to be larger than one would na\"ively expect.  This
finding is, however, somewhat at odds with the results obtained from
\Madgraph{}+\Pythia, which seem to be slightly harder in shape and
significantly larger in normalisation.  The \PythiaEight\MEPS sample,
on the other hand, has a smaller one-jet inclusive cross section than
\Sherpa, but the jet spectrum exhibits a somewhat harder tail,
corresponding to a shape difference of about 30-40\% with respect to
both the \Sherpa results.  The same finding, a somewhat harder tail,
is also true for the \PowhegBox{}+\PythiaEight results.  The same
trends can be also found at the hadron and hadron + MPI level. 
For the \PowhegBox result the difference can be attributed to the
usage of a scale defined at the ``underlying-Born'' level (cf.\
\SecRef{Sec:Codes:PowhegBox} for more details). Indeed it has been
checked explicitly that a NLO computation performed with the same
scale choice used in \PowhegBox{} gives a result in complete agreement
with the \PowhegBox{} result shown here.
Clearly, the differences between different calculations and codes
exhibited here deserve a more in-depth study, which, unfortunately, is
beyond the scope of this comparison.

Similar findings are also true for the next observable, the $\Delta R$ 
distribution between the lepton stemming from the $W$ decay and the hardest jet
displayed in \FigRef{fig:comparison_DeltaR}.  Again, the two \Sherpa samples 
are compared with the two NLO samples, this time exhibiting a sizable shape 
difference towards an increase at smaller and a decrease at larger distances 
of about 40\% relative cross section.  While higher jet configurations 
typically tend to be a bit more central, it seems far-fetched to attribute
this difference only to them.  At the same time, large differences in $R$
are most likely due to jets which are pretty much forward\footnote{
  Assuming the lepton and the jet to be back-to-back, $\Delta\phi=\pi$,
  one still needs $\Delta\eta \approx 4$ to obtain 
  $(\Delta R)^2 = (\Delta\phi)^2+(\Delta\eta)^2\approx 5^2$.}.  
This region of phase space for jet production, however, is known to be quite 
susceptible to mismatches in scale and/or PDF definitions.  However, it is 
worth noting that this difference vanishes almost completely at the parton 
shower level.  The \PythiaEight\MEPS sample, despite a sizable difference in cross 
section, appears to follow the shape of the NLO and \Sherpa results.  Further 
comparing these results to those of the other codes at the shower level 
suggests that the \Madgraph{}+\Pythia merged sample, apart from a drastically 
enhanced cross section, also shows an enhancement in shape at smaller 
$\Delta R\le 2$ w.r.t.\ the NLO result.  Interestingly enough, the 
\PowhegBox{}+\PythiaEight sample exhibits the opposite behaviour: while the 
cross section seems fairly consistent with the \Sherpa and the NLO ones, the 
shape shows some enhancement of up to 40\% at large distance $\Delta R$, which
following the reasoning for the jet-$p_\perp$ spectrum may also hint at being
due to a difference in the definition of scales.  As before, the same trends 
visible at the parton shower level can also be found at the hadron and 
hadron + MPI level.

\FloatBarrier

\subsubsection{Multi-jet observables}
\begin{figure}[t]
 \includegraphics[width=.48\textwidth]{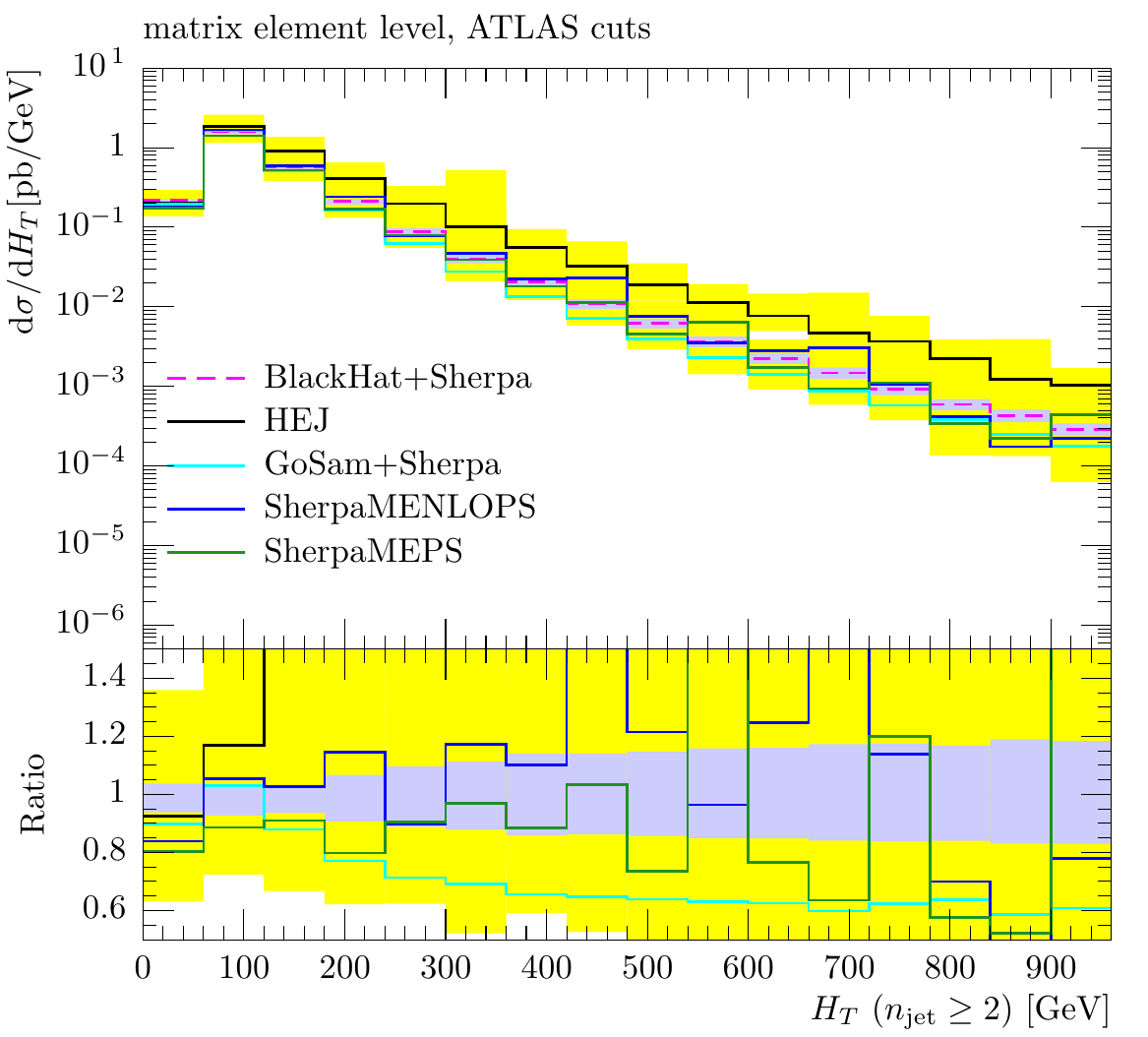}\hfill
 \includegraphics[width=.48\textwidth]{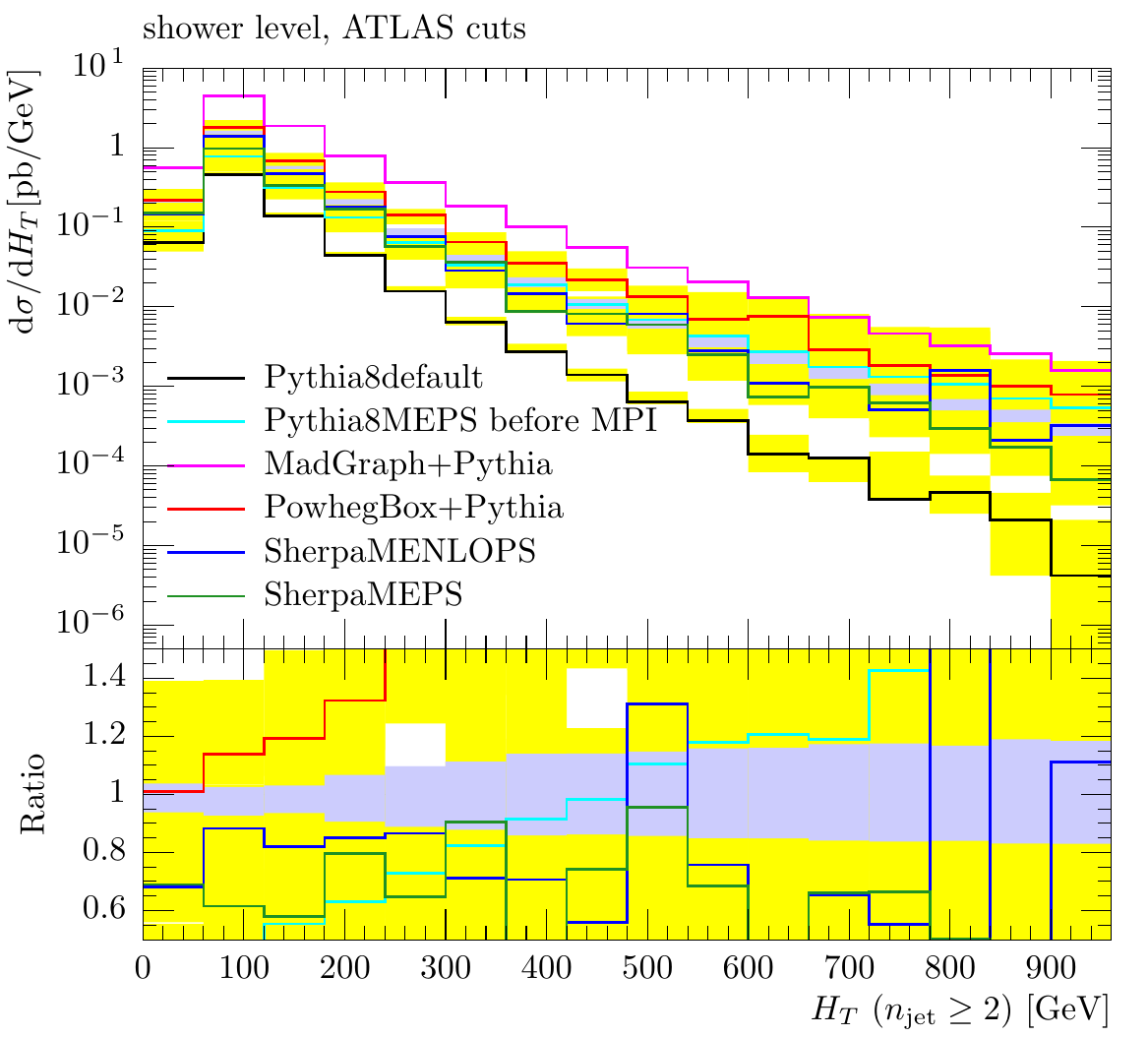}\\
 \includegraphics[width=.48\textwidth]{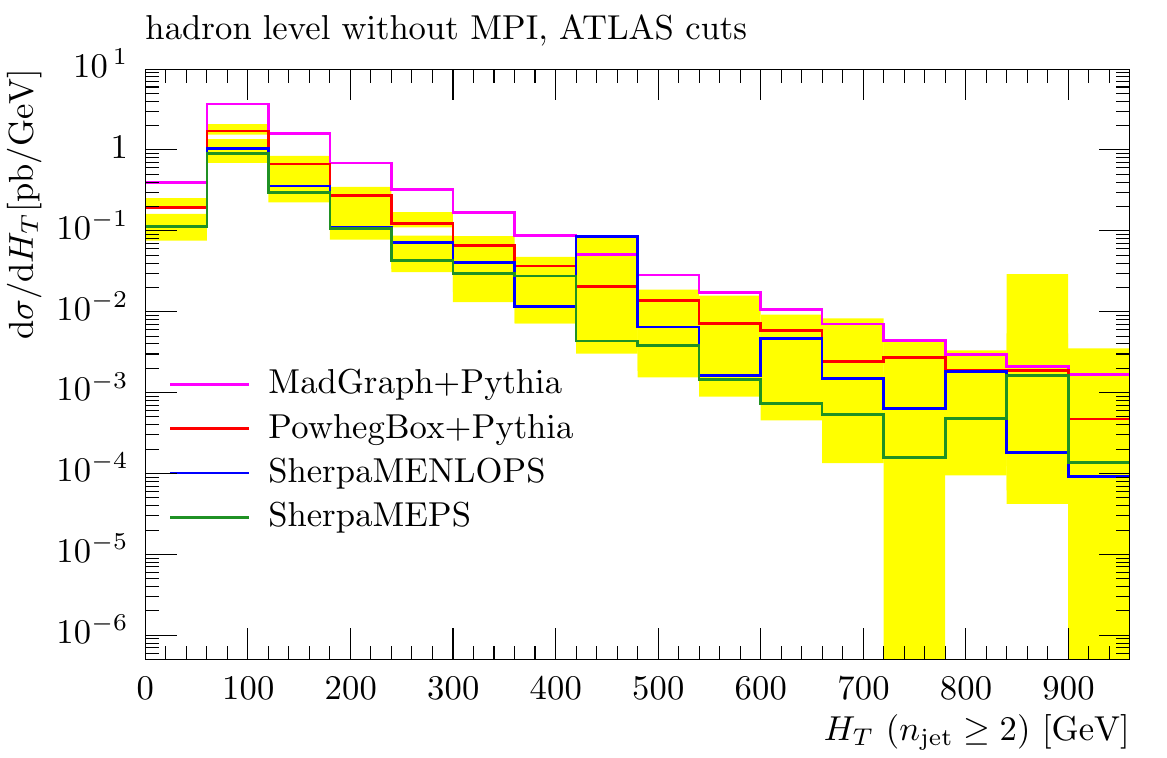}\hfill
 \includegraphics[width=.48\textwidth]{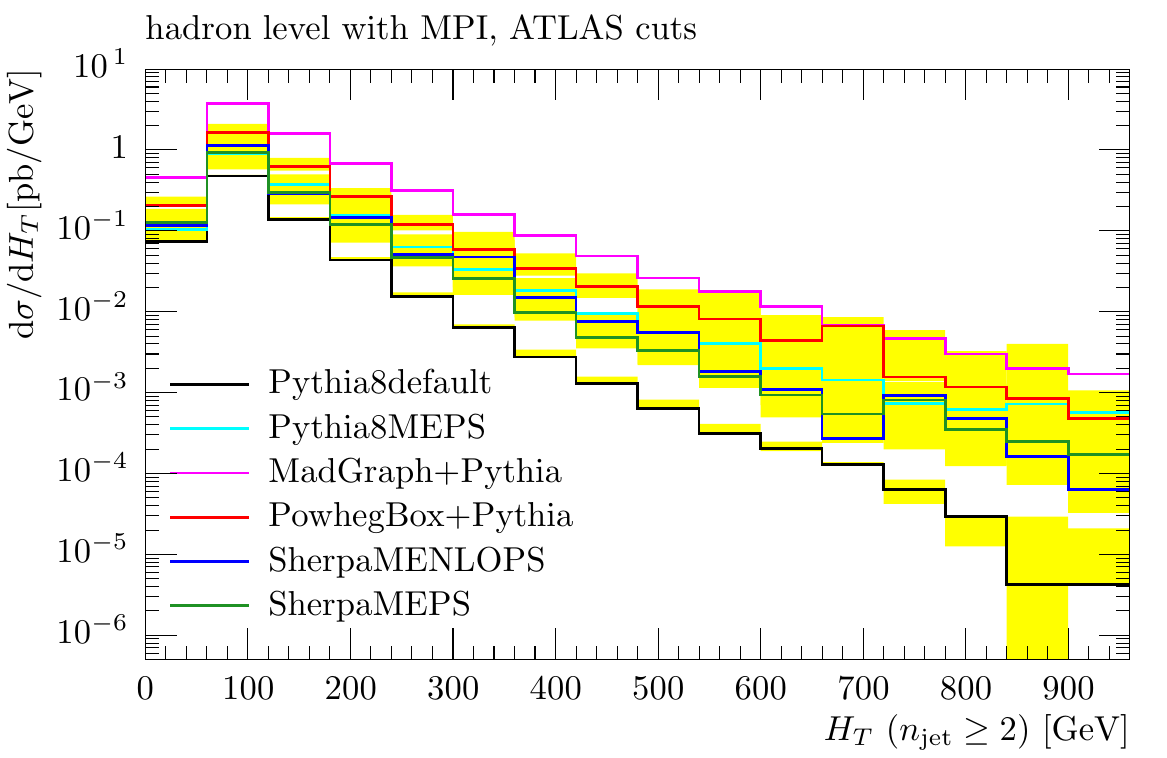}
 \caption{$H_T = \sum_{i\in \{\text{jets}\}} E_{\perp\, i}$ of events with at 
 least 2 jets on all levels of the simulation (for exact definitions and cuts see
 \AppRef{App:Observables:Cuts} and \AppRef{App:Observables:Analysis}).
 Note that \BlackHat uses the CTEQ6.6 pdf, \PythiaEight and \Madgraph{}+\Pythia 
 CTEQ6L1 and all the others use CT10. In both ratio plots the ratio is taken
 with respect to \BlackHat{}+\Sherpa (on matrix element level).}
  \label{fig:multijetht}
\end{figure}

In observables including at least two jets, consider first the case of the
$H_T$ distribution depicted in \FigRef{fig:multijetht}.  Over the full
range and obscured by large statistical fluctuations both \Sherpa samples
seem to follow the NLO prediction from \BlackHat{}+\Sherpa.  The LO
result from \GoSam{}+\Sherpa, on the other hand, appears to fall off
at the hard end of the distribution.  The prediction from \HEJ is a bit
more subtle to judge: at low $H_T$ (around 100~GeV), we see that it is in 
good agreement with the predictions from the other approaches.  However, 
as higher values of $H_T$ are probed, the \HEJ prediction becomes noticeably 
larger than the fixed-order descriptions, including those from \Sherpa where 
different multiplicities are merged.  This is the region in $H_T$ where we 
would expect high multiplicities to have a noticeable effect, and therefore 
where we would expect to see the impact of the resummation in \HEJ.  This is,
however, slighlty at odds with the fact that the \Sherpa prediction included
up to 6 jets and that the multijet rates and the $p_\perp$ distributions of
the fifth and sixth jet from \HEJ undershoot those from \Sherpa, cf.\
\FigRef{fig:multijetnjet} and \FigRef{fig:multijetptj}.  However, a similar
trend concerning the hard tail of this distribution appears also on the
shower level in the \Madgraph{}+\Pythia sample, which includes up to 4 extra 
jets, and in the \PowhegBox{}+\PythiaEight sample, which includes 2 jets at LO 
and 1 jet at NLO.  The trend is even more pronounced with an even harder tail
for the \PythiaEight\MEPS sample, which includes 3 extra 
jets.  At this level, \Sherpa more or less follows the NLO result.  It should 
be noted, though, that all approaches remain within the scale variation band 
indicated on the BlackHat prediction.  This findings are consistently carried
over to the hadron and hadron+MPI level.

\begin{figure}[t]
 \includegraphics[width=.48\textwidth]{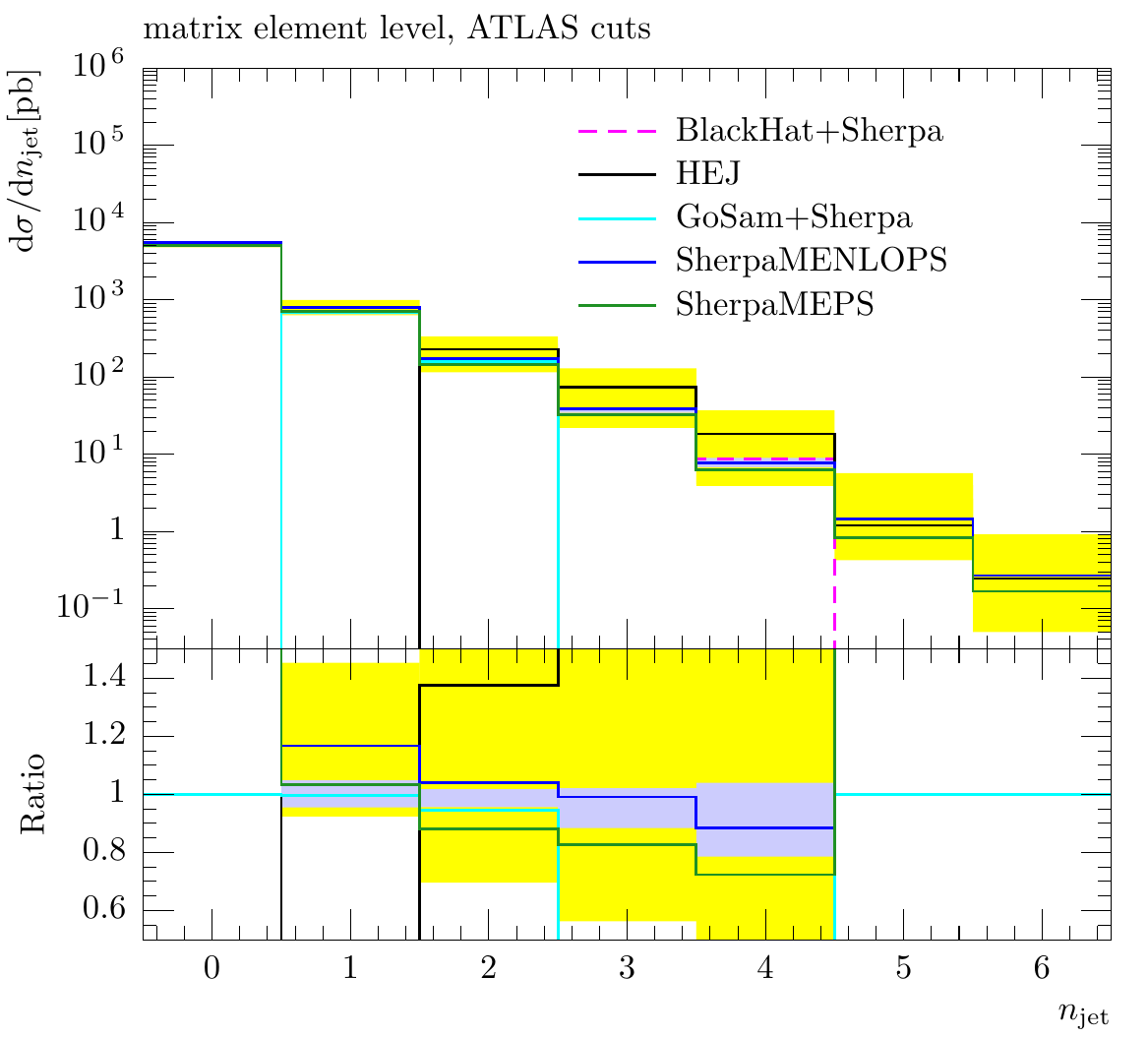}\hfill
 \includegraphics[width=.48\textwidth]{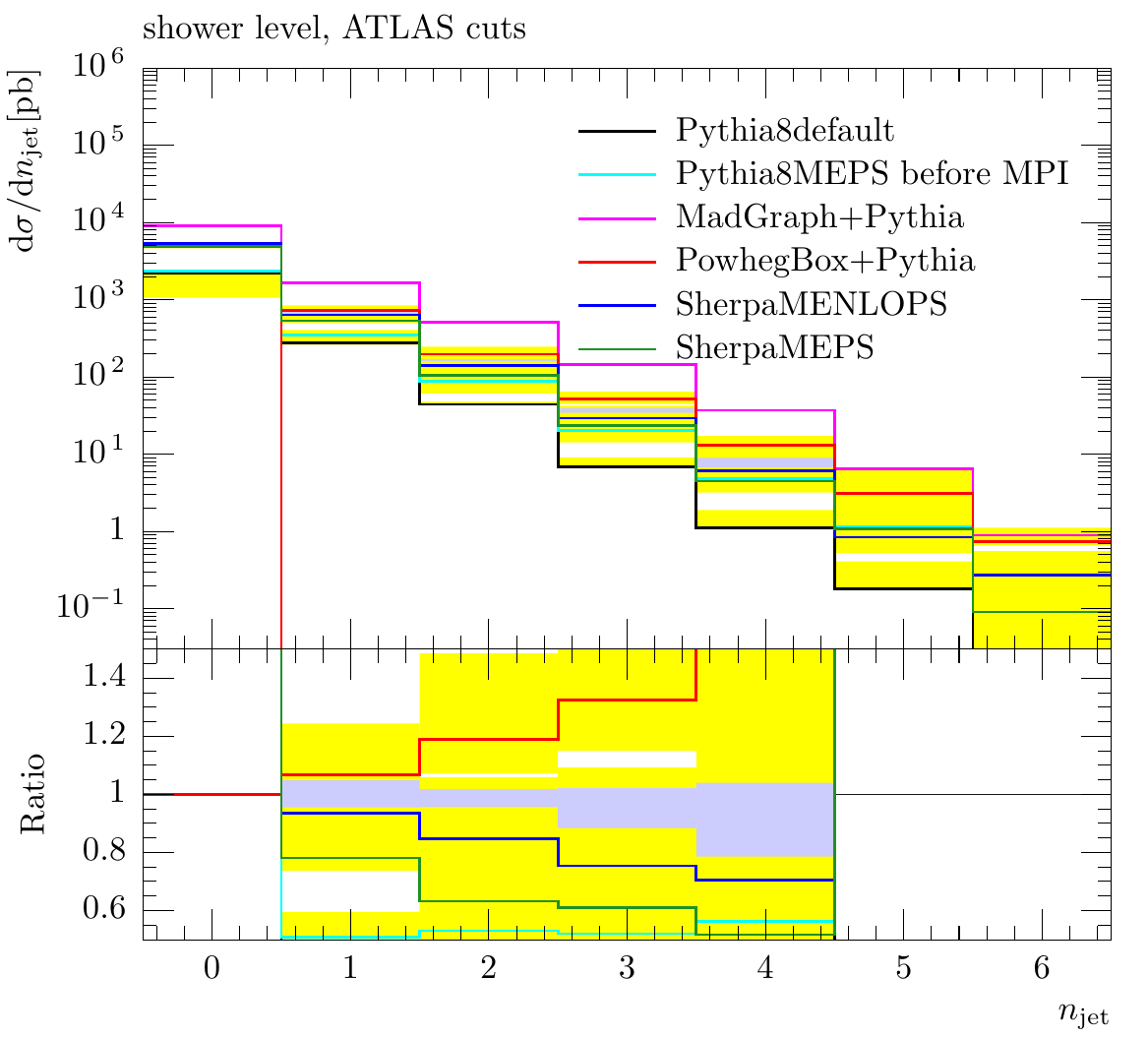}\\
 \includegraphics[width=.48\textwidth]{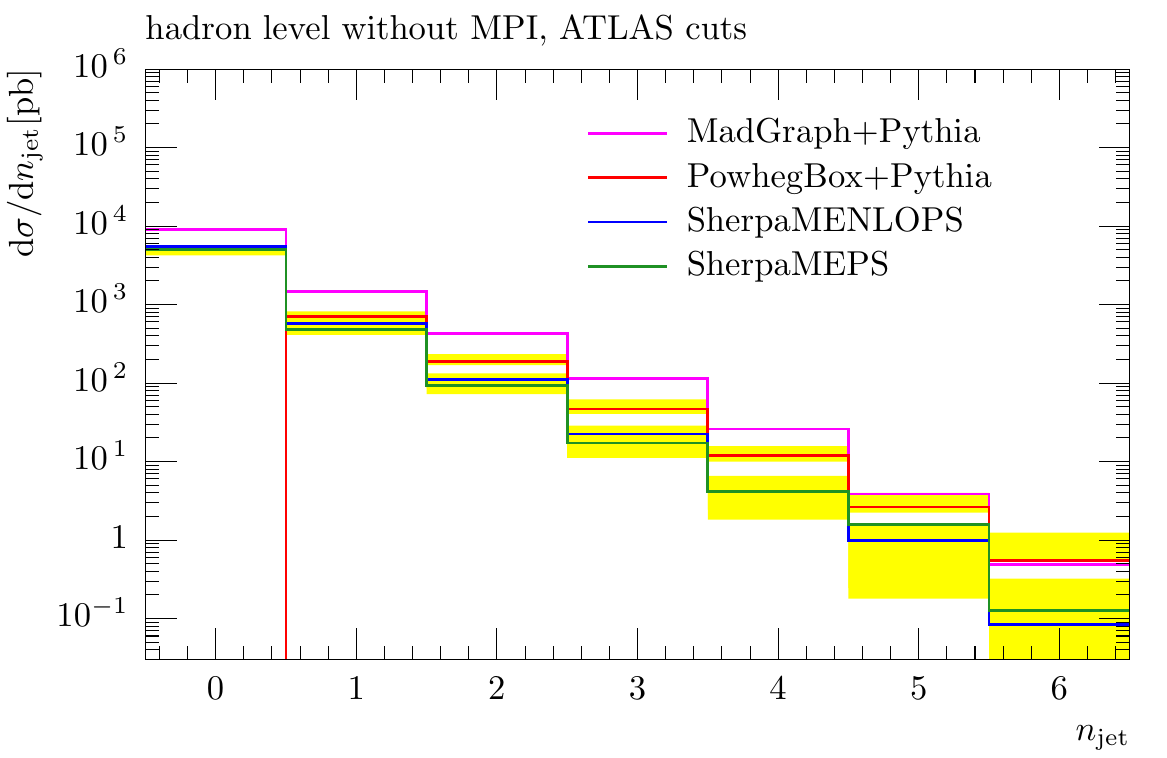}\hfill
 \includegraphics[width=.48\textwidth]{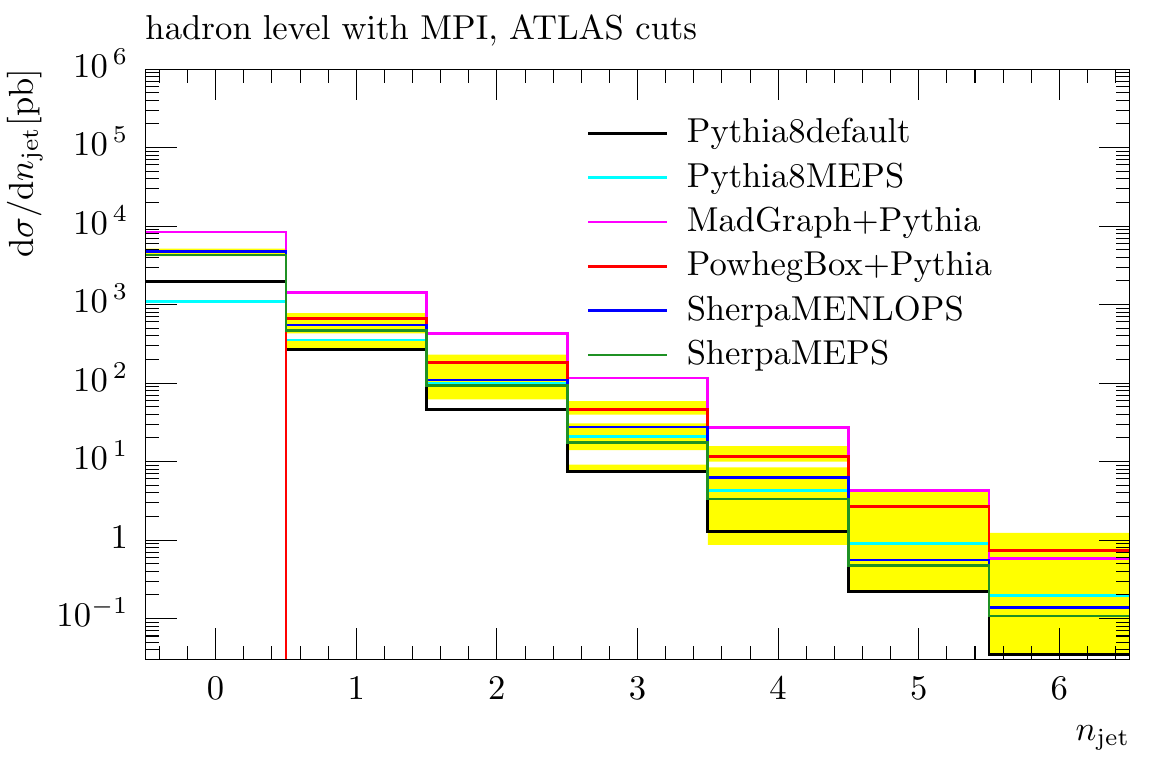}
 \caption{Number of jets on all levels of the simulation (for exact definitions and cuts see
 \AppRef{App:Observables:Cuts} and \AppRef{App:Observables:Analysis}).
 Note that \PowhegBox{}+\PythiaEight and \GoSam calculate $W+\text{1\ jet}$ on matrix element level, 
 while \HEJ starts with $W+\text{2\ jets}$ and that \BlackHat uses the CTEQ6.6 pdf, \PythiaEight and \Madgraph{}+\Pythia 
 CTEQ6L1 and all the others use CT10.  In both ratio plots the ratio is taken
 with respect to \BlackHat{}+\Sherpa (on matrix element level).}
 \label{fig:multijetnjet}
\end{figure}

Turning to the $n$-jet rates, at the matrix element level, \Sherpa follows
fairly closely the NLO results in different jet multiplicity bins, while
\HEJ seem to overshoot the central value in the 3- and 4-jet bin, but staying
inside the NLO scale uncertainty band.   going back to the tree-level
result of \Sherpa in the 5- and 6-jet bins.  As discussed in \SecRef{Sec:Results:HEJ}, the 
\HEJ framework includes tree-level matching for final states with up to and 
including four jets in the final state.  Therefore it is fair to assume that
the absence of matching for five jets and above leads to the larger drop in 
cross section observed in \FigRef{fig:multijetnjet} from four-jet to five-jet 
as compared to that from either three-jet to four-jet or from five-jet to 
six-jet and lends support to the suspicion that in \HEJ a matched sample would
also provide larger 5- and 6-jet multiplicities.  At the shower level, the
trend already visible at the $H_T$ distribution repeats itself.  The smaller
cross section in the \PythiaEight\MEPS sample is mainly due to
the low multiplicity bins, such that the shape of the $n$-jet distribution
also has a relatively harder tail than the \Sherpa sample.  In contrast,
the \PowhegBox+\PythiaEight result, starting consistently at 1 jet, appear to
be at the upper end of the NLO uncertanities throughout.  

\begin{figure}[p]
 \vspace*{-3mm}
 \includegraphics[width=.48\textwidth]{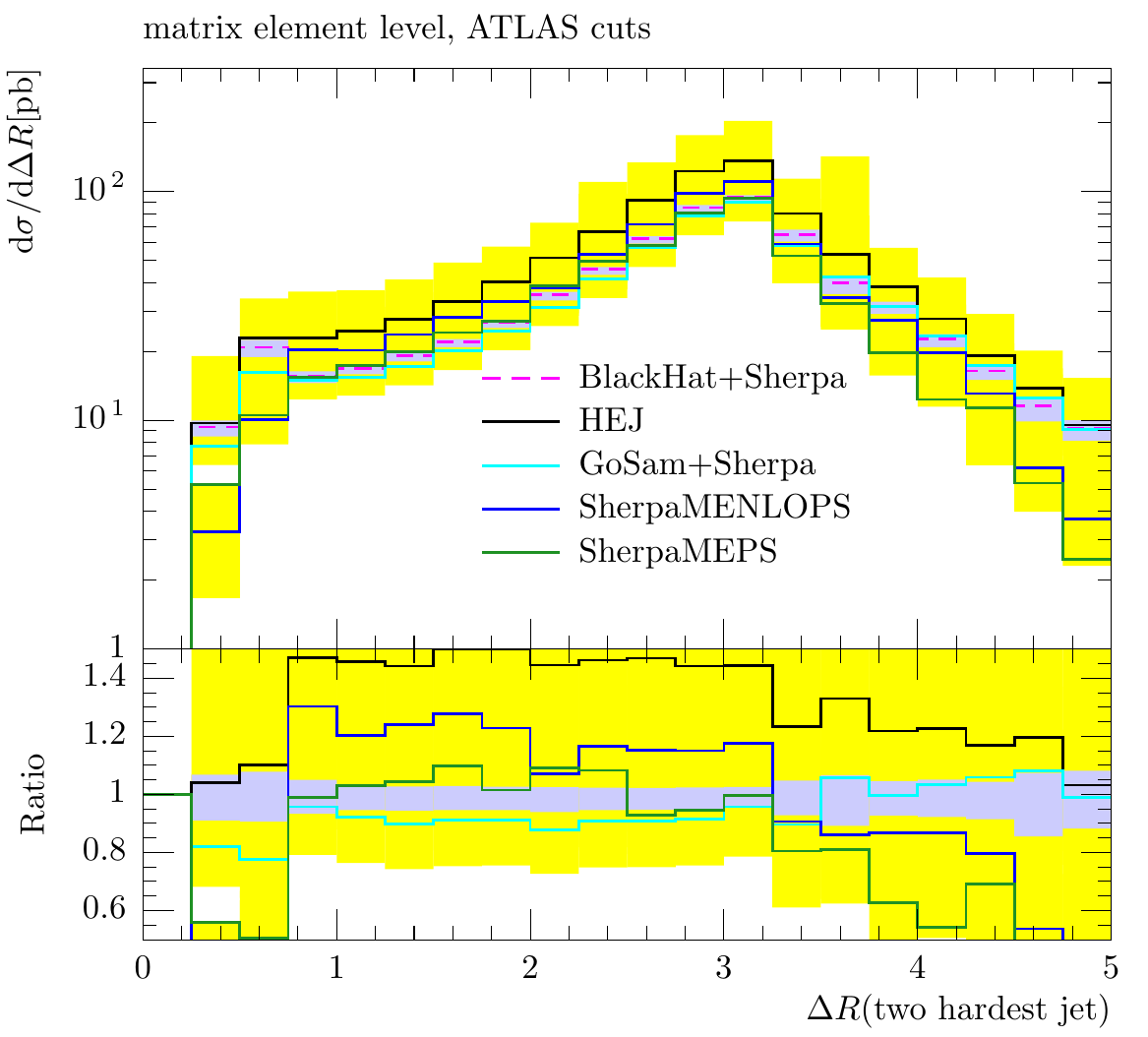}\hfill
 \includegraphics[width=.48\textwidth]{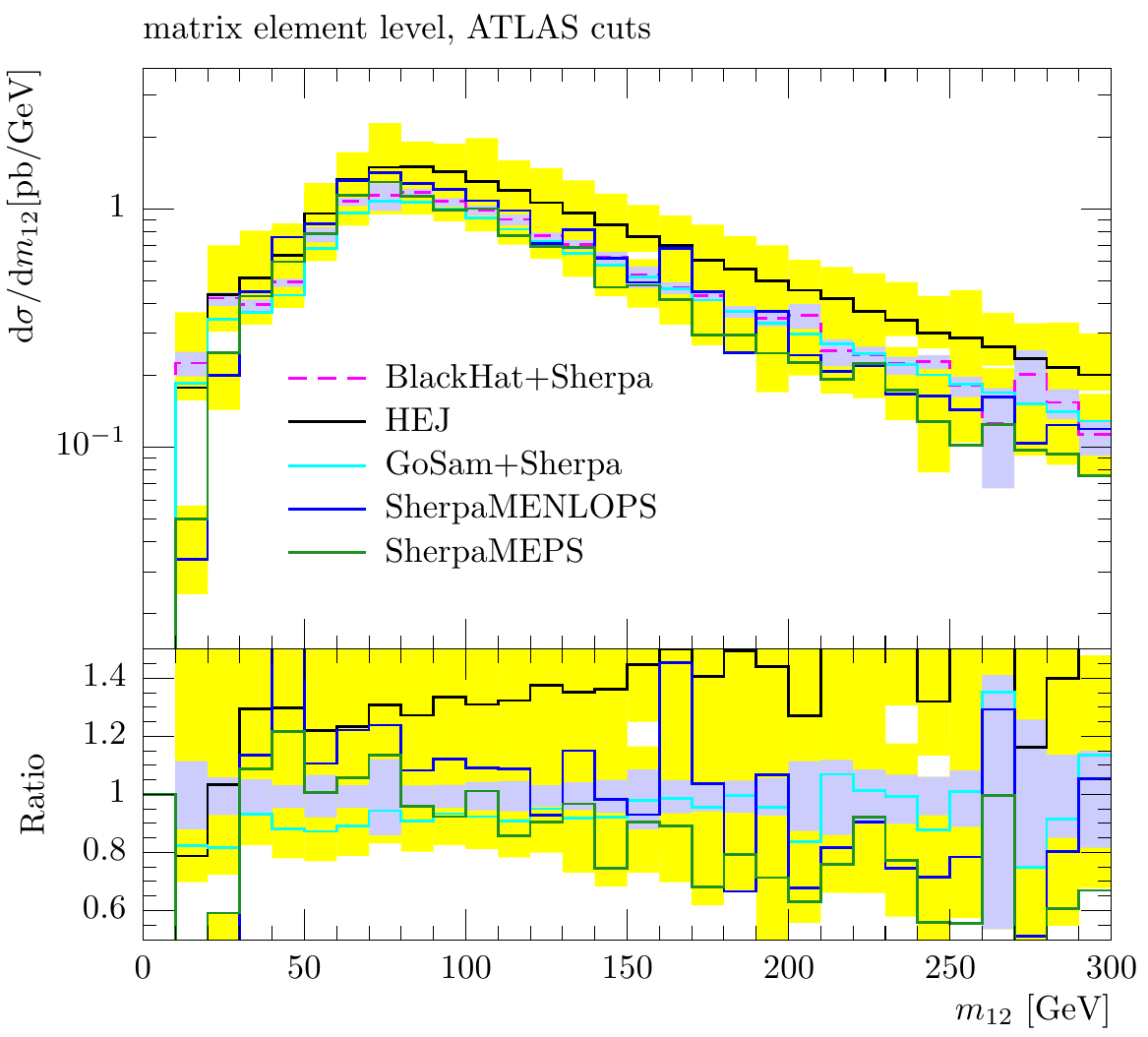}\\
 \caption{$\Delta R$ of two leading jets (left) and invariant mass of 
 two hardest jets (right) \MElevel (for exact definitions and cuts see
 \AppRef{App:Observables:Cuts} and \AppRef{App:Observables:Analysis}). 
 Note that \BlackHat uses the CTEQ6.6 pdf, \PythiaEight and \Madgraph{}+\Pythia 
 CTEQ6L1 and all the others use CT10. In both ratio plots the ratio is taken
 with respect to \BlackHat{}+\Sherpa (on matrix element level).\vspace*{-1mm}}
 \label{fig:2jetcorrels}
\end{figure}

\begin{figure}[p]
 \includegraphics[width=.48\textwidth]{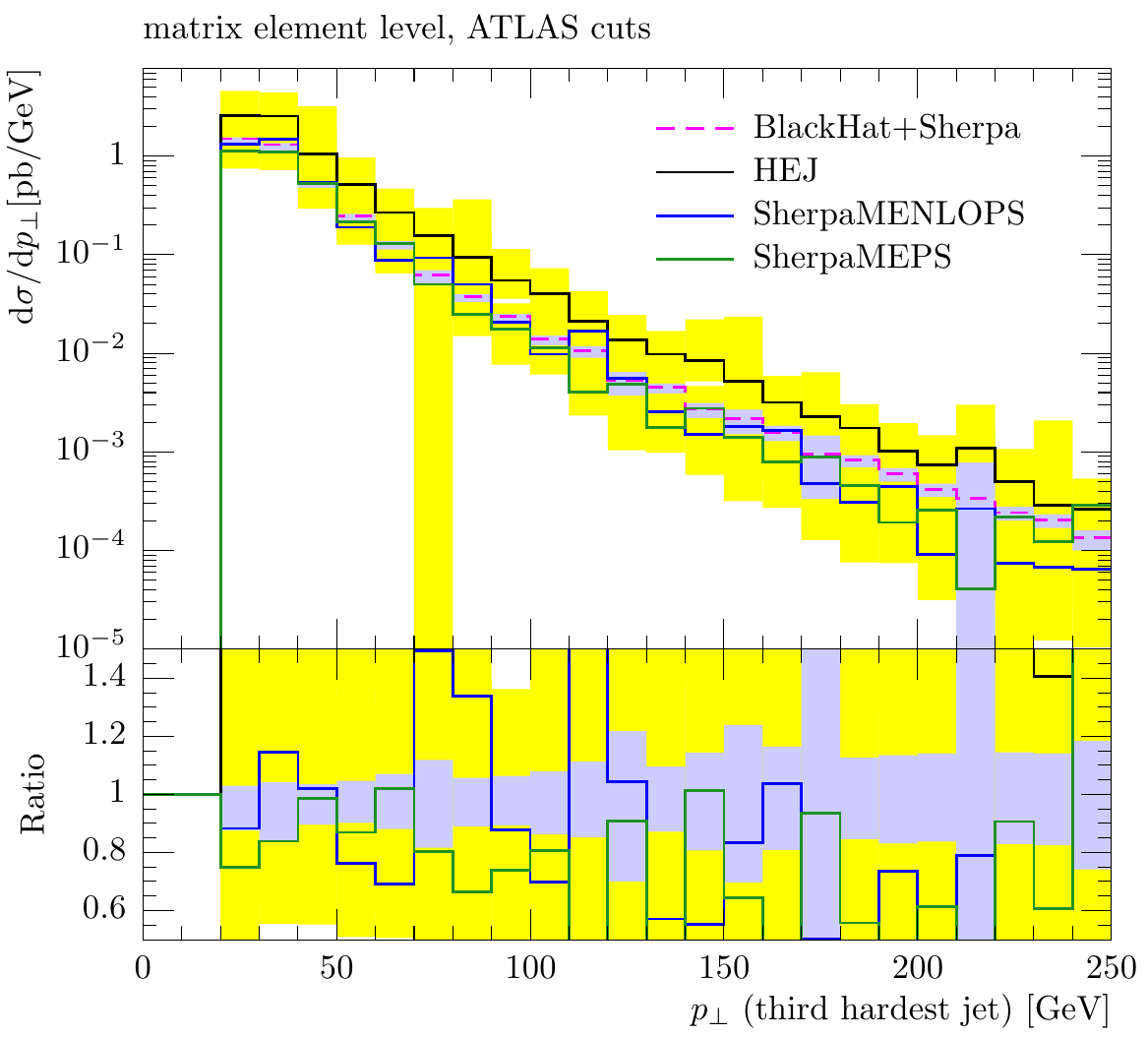}\hfill
 \includegraphics[width=.48\textwidth]{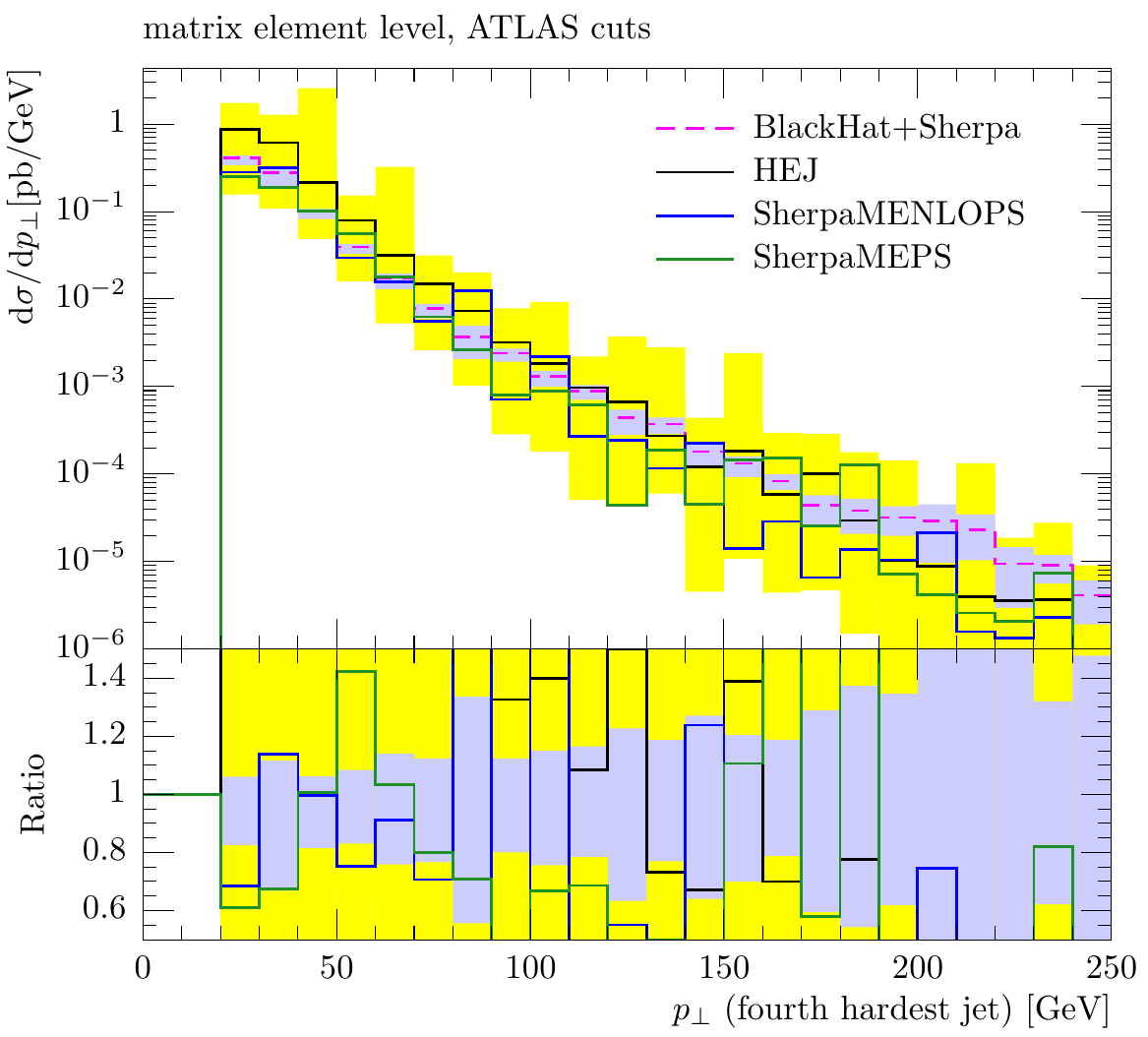}\\
 \includegraphics[width=.48\textwidth]{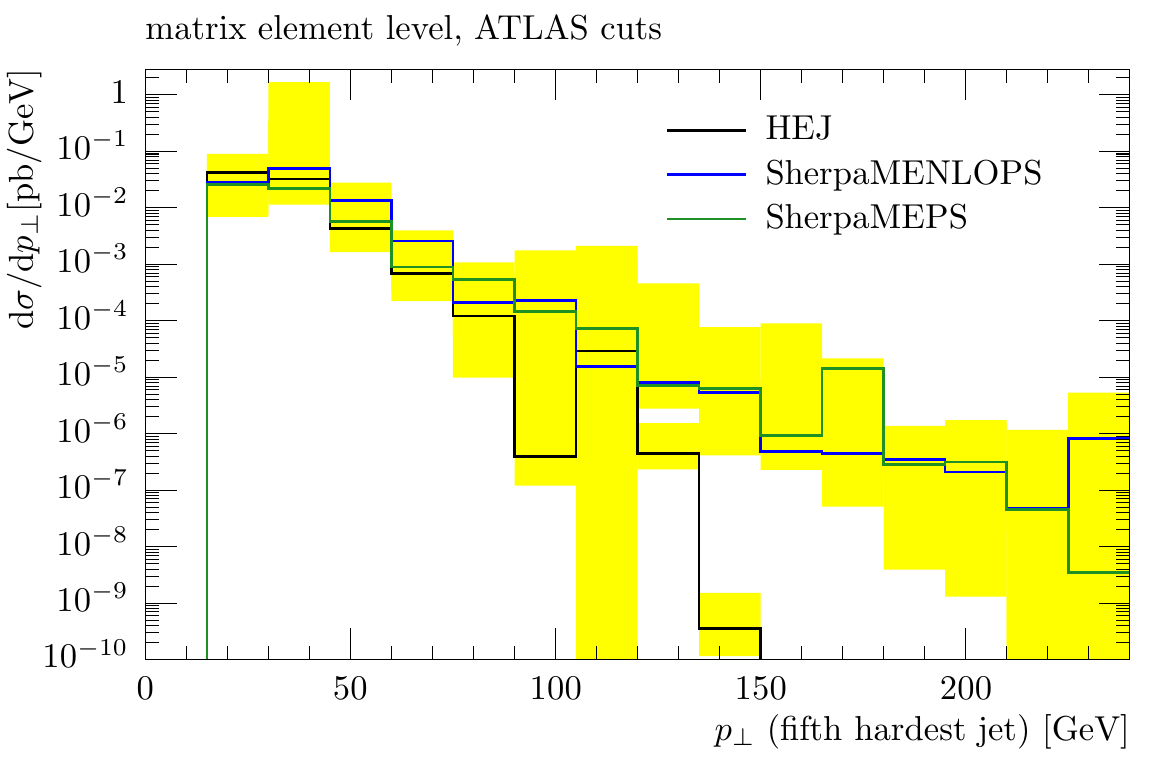}\hfill
 \includegraphics[width=.48\textwidth]{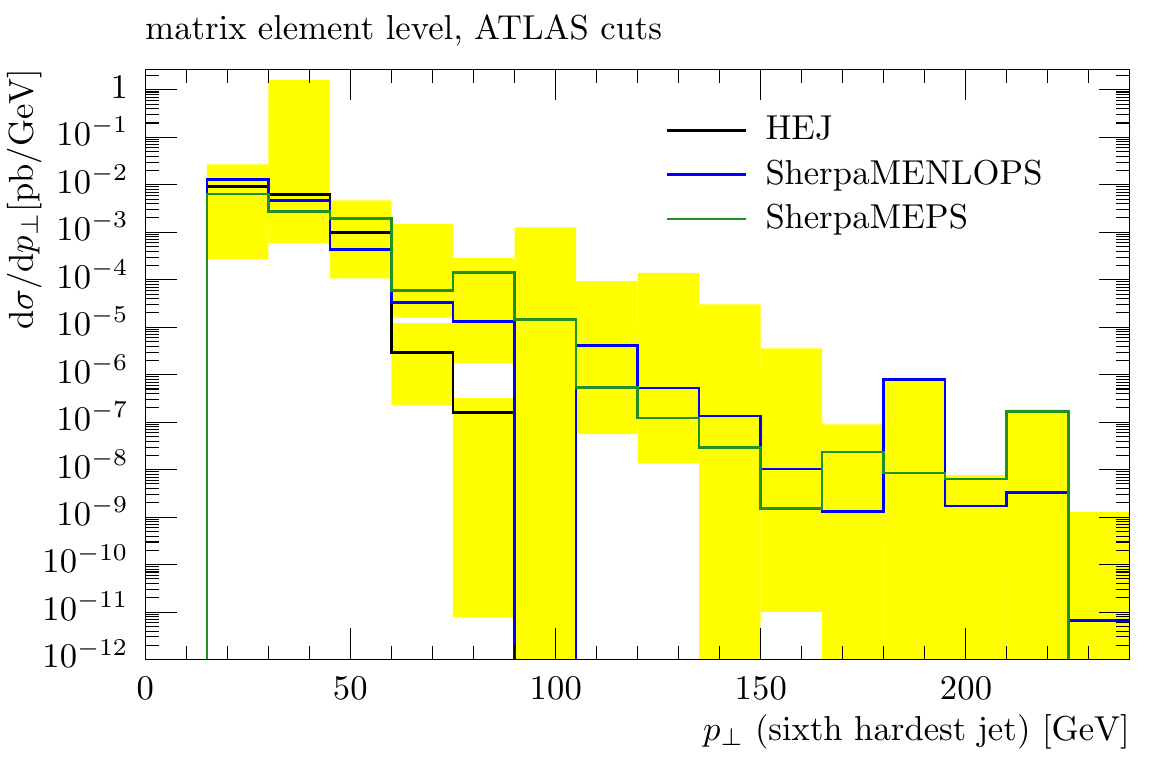}
 \caption{Transverse momentum of third to sixth hardest jet on \MElevel (for exact definitions and cuts see
 \AppRef{App:Observables:Cuts} and \AppRef{App:Observables:Analysis}).
 Note that \BlackHat uses the CTEQ6.6 pdf, \PythiaEight and \Madgraph{}+\Pythia 
 CTEQ6L1 and all the others use CT10. In both ratio plots the ratio is taken
 with respect to \BlackHat{}+\Sherpa (on matrix element level).\vspace*{-3mm}}
\label{fig:multijetptj}
\end{figure}

Looking at the correlation of the two leading jets in \FigRef{fig:2jetcorrels}
at the matrix element only, both the $\Delta R$ and the $m_{12}$ distribution 
provided by \Sherpa have a slight tilt against the NLO prediction from 
\BlackHat{}+ \Sherpa, undershooting the latter result by up to about 40\% for 
large $\Delta R$ and by up to about 20\% for large $m_{12}$.  While \HEJ
seems to roughly follow the shape of \Sherpa for $\Delta R$, it is
significantly harder than \Sherpa and the NLO result for large values of
$m_{12}$.  In addition, in both cases, \HEJ also predicts a larger cross
section that the other tools. 

\FigRef{fig:multijetptj} shows the transverse momentum distributions for the
third to sixth jets ordered in $p_\perp$, and at the matrix element level.  
For the third hardest jet, the prediction from \HEJ is similar in shape but 
higher in cross section than the results obtained at NLO from 
\BlackHat{}+\Sherpa or the two \Sherpa samples.  For the fourth jet, the 
\HEJ cross section still seems higher than the other ones, but this discrepancy
seems to be mainly around comparably low jet $p_\perp$.  For larger values of
$p_\perp$ all tree-level type or resummed predictions are below the NLO
result.  Surprisingly, for the fifth and sixth hardest jets, the \HEJ 
predictions follow the \Sherpa ones for low values of $p_\perp$ below about 
60 GeV, before they fall off nearly instantly.  This again may be an artefact 
of tree-level matching not being included in \HEJ for the production of 
five- and six-jets or of missing statistical support in this region of phase 
space.

\FloatBarrier

\subsection{Conclusions}

In this study first steps towards an update and extension of the comparison
in~\cite{Alwall:2007fs} have been made.  In contrast to the older study, a
larger variety of tools including fixed-order and resummation tools as well
as NLO matched and tree-level merged simulations have been included.  Not
surprisingly, some observables appear to be described fairly consistently
between different tools, while others exhibit large deviations, sometimes
clearly beyond the formal accuracy claimed by the different methods, and
also beyond the best estimates of intrinsic modelling or calculational
uncertainties provided by the authors.  In some instances the relative
differences are way beyond na\"ive expectations by most of the authors of this 
study.  This clearly hints at the need to carefully cross-validate different 
tools before deploying them for large scale simulations, and it also 
necessitates an increased collaboration of the authors of such tools in order 
to arrive at a more consistent picture.

We hope that this study triggered some future work towards the latter goal.

\subsection*{Acknowledgements}
We would like to thank the organisers of Les Houches 2012 for repeatedly
hosting such a fantastic and fruitful workshop.  We would also like to
thank all participants for a lively and stimulating atmosphere and lots
of inspiring discussions. 

MS's and DM's work was supported by the Research Executive Agency (REA) of the 
European Union under the Grant Agreement number PITN-GA-2010-264564 
(LHCPhenoNet). SA and ER also acknowledge funding from LHCPhenoNet for
travel expenses.
LL's and SP's work was supported by the Swedish research
council (contract 621-2009-4076).  JMS and GL are supported by the UK Science and
Technology Facilities Council (STFC). T.R. is supported by the Alexander von
Humboldt Foundation, in the framework of the Sofja Kovaleskaja Award Project
``Advanced Mathematical Methods for Particle Physics'', endowed by the German
Federal Ministry of Education and Research. F.T. is supported by Marie-Curie-IEF, project:
``SAMURAI-Apps''.

\subsection{Cuts and observables}
\label{App:Observables}

\subsubsection{Cuts}
\label{App:Observables:Cuts}

\begin{table}[h!]
\begin{center}
\begin{tabular}{|l|l|l|}
\hline
                    & ATLAS                & CMS \\ \hline
lepton $p_\perp$    & $> \unit[20]{GeV}$   & $> \unit[20]{GeV}$         \\
lepton $|\eta|$     & $< 2.5\ (e,\,\mu)$   & $< 2.5\ (e), 2.1\ (\mu)$   \\
$\slashed{E}_\perp$ & $> \unit[25]{GeV}$   & no cut                     \\
$m_{\perp,\,W}$     & $> \unit[40]{GeV}$   & $> \unit[20]{GeV}$         \\
jet $p_\perp$       & $> \unit[25]{GeV}$   & $> \unit[30]{GeV}$         \\
jet $|\eta|$        & $< 4.4$              & $< 2.4$                    \\
jet radius          & 0.4 (anti-$k_\perp$) & 0.5 (anti-$k_\perp$)       \\
lepton isolation    & $< \unit[10]{\%}$ of lepton energy &
                      $< \unit[10]{\%}$ of lepton energy \\
                    &  in cone with R=0.5  & in cone with R=0.5 \\
\hline
\end{tabular}
\end{center}
\caption{Cuts used in this study inspired by common ATLAS and CMS cuts.
	 \label{Tab:Cuts}}
\end{table}

\TabRef{Tab:Cuts} presents the cuts applied to define the event 
selection in both the ATLAS and CMS specifications. 

\subsubsection{Analysis procedure and definition of observables}
\label{App:Observables:Analysis}

A common analysis was implemented within the \Rivet framework and 
used by all codes providing individual events.  This analyses is 
carried out as defined in the following:

\begin{enumerate}
 \item remove all neutrinos from all final states (i.e. 'all particles' from 
	now on means 'all particles without neutrinos')
 \item find hardest isolated lepton (electron or muon) ('lepton' from now on 
        means 'hardest isolated lepton')
 \item cut on lepton $p_\perp$ and $|\eta|$      
 \item compute missing transverse energy $\slashed{E}_\perp$:
    \begin{enumerate}
     \item sum the three-momenta of all particles within $|\eta|<10$, 
	this yields $-\slashed{\mathbf{p}}$
     \item compute missing energy as $\slashed{E} = |\slashed{\mathbf{p}}|$
     \item assume resulting four-vector $\slashed{p}$ corresponds to neutrino
    \end{enumerate}
 \item for ATLAS cut on $\slashed{E}_\perp$
 \item resonstruct $W$ four-momentum as $p^W = p^\text{lepton} + \slashed{p}$
 \item compute $W$ transverse mass squared as  $m_{\perp,\,W}^2 = 
       (p_\perp^\text{lepton}+\slashed{p}_\perp)^2 - (p_\perp^{W})^2$
 \item cut on $W$ transverse mass
 \item remove lepton from final state
 \item cluster into jets keeping only those passing the $p_\perp$ and $|\eta|$
      cuts
 \item compute $H_T = \sum_{i\in \{\text{jets}\}} E_{\perp\, i}$ 
 \item compute beam thrust $\tau_B = \sum_{i\in \{\text{particles}\}} 
      \left(E_i-\left|p_i^z\right|\right)$ using all visible particles
\end{enumerate}

It should be noted that this defintion of the $W$ is infra-red safe only for
transverse observables.
\subsection{Detailed settings}
\label{App:Settings}

\subsubsection{\texorpdfstring{\protect\Sherpa}{Sherpa}}
\label{App:Settings:Sherpa}

For this study \Sherpa-1.3.1 was used. Except for the underlying event, which 
was tuned for the CT10~\cite{Lai:2010vv} parton distribution functions and 
whose parameters are given below, all other 
non-perturbative parameters were kept at their default values. The underlying 
model was tuned for the cluster hadronisation.

\begin{center}
 \begin{tabular}{|l|l|}
 \hline
  K\_PERP\_MEAN\_1 	& 1.17 			\\
  K\_PERP\_MEAN\_2 	& 1.17			\\
  K\_PERP\_SIGMA\_1 	& 0.760			\\
  K\_PERP\_SIGMA\_2 	& 0.760			\\
  PROFILE\_PARAMETERS 	& 0.576, 0.353		\\
  RESCALE\_EXPONENT 	& 0.238			\\
  SCALE\_MIN 	        & 2.52			\\
  SIGMA\_ND\_FACTOR 	& 0.465  		\\
  \hline
 \end{tabular}
\end{center}

\subsubsection{\texorpdfstring{\protect\PythiaEight}{Pythia8}}
\label{App:Settings:Pythia8}

To produce the results, we have used two tunes of \PythiaEight, Tune 4C and 
Tune A2, both of which use CTEQ6L1 parton distributions. Tune 4C is the 
default tune in \PythiaEight -- no additional input settings are necessary. For
completeness, below we list all parameters that are implicitly set by choosing
the default Tune 4C.
\begin{verbatim}
PDF:pSet = 8
SigmaProcess:alphaSvalue = 0.135
SigmaDiffractive:dampen = on 
SigmaDiffractive:maxXB = 65.0
SigmaDiffractive:maxAX = 65.0 
SigmaDiffractive:maxXX = 65.0
TimeShower:dampenBeamRecoil = on
TimeShower:phiPolAsym = on
SpaceShower:alphaSvalue = 0.137
SpaceShower:samePTasMPI = false
SpaceShower:pT0Ref = 2.0
SpaceShower:ecmRef = 1800.0
SpaceShower:ecmPow = 0.0
SpaceShower:rapidityOrder = on
SpaceShower:phiPolAsym = on
SpaceShower:phiIntAsym = on
MultipartonInteractions:alphaSvalue = 0.135
MultipartonInteractions:pT0Ref = 2.085
MultipartonInteractions:ecmRef = 1800.
MultipartonInteractions:ecmPow =  0.19
MultipartonInteractions:bProfile = 3
MultipartonInteractions:expPow = 2.0
BeamRemnants:primordialKTsoft = 0.5
BeamRemnants:primordialKThard = 2.0
BeamRemnants:halfScaleForKT = 1.0
BeamRemnants:halfMassForKT =  1.0
BeamRemnants:reconnectRange = 1.5
\end{verbatim}
A detailed discussion of these choices can be found in \cite{Corke:2010yf}.
All other parameters remain with their default values. For our purposes, it 
might be interesting to remark that the starting value for \alphas-evolution
in time-like splittings is given by
\begin{verbatim}
SpaceShower:alphaSvalue = 0.1383
\end{verbatim}
To investigate the impact of rapidity ordering in space-like showers, we chose
to remove enforced rapidity ordering by setting
\begin{verbatim}
SpaceShower:rapidityOrder = off
\end{verbatim}
If rapidity ordering is enforced in ISR, the question arises how it should be 
treated when picking histories. For this purpose, \PythiaEight supplies the
switch
\begin{verbatim}
Merging:enforceStrongOrdering
\end{verbatim}
When switched ``on", this parameter will result in picking non-rapidity-ordered 
histories only if no rapidity-ordered paths where found, thus disfavouring 
non-rapidity-ordered parton shower histories for matrix element states. To have
a more complete understanding of the impact of tuning, we also changed to the 
recently proposed Tune A2 \cite{ATLAS:2011dk}. For this, we have to set
\begin{verbatim}
Tune:pp = 7
\end{verbatim}
\PythiaEight will then reset the following parameters:
\begin{verbatim}
PDF:pSet = 8  
SigmaProcess:alphaSvalue = 0.135  
SigmaDiffractive:dampen = on
SigmaDiffractive:maxXB = 65.0
SigmaDiffractive:maxAX = 65.0
SigmaDiffractive:maxXX = 65.0  
TimeShower:dampenBeamRecoil = on  
TimeShower:phiPolAsym = on  
SpaceShower:alphaSvalue = 0.137
SpaceShower:samePTasMPI = false
SpaceShower:pT0Ref = 2.0   
SpaceShower:ecmRef = 1800.0
SpaceShower:ecmPow = 0.0   
SpaceShower:rapidityOrder = false
SpaceShower:phiPolAsym = on  
SpaceShower:phiIntAsym = on  
MultipartonInteractions:alphaSvalue = 0.135
MultipartonInteractions:pT0Ref = 2.18  
MultipartonInteractions:ecmRef = 1800.
MultipartonInteractions:ecmPow = 0.22   
MultipartonInteractions:bProfile = 4   
MultipartonInteractions:a1 = 0.06 
BeamRemnants:primordialKTsoft = 0.5   
BeamRemnants:primordialKThard = 2.0   
BeamRemnants:halfScaleForKT = 1.0   
BeamRemnants:halfMassForKT = 1.0   
BeamRemnants:reconnectRange = 1.55  
\end{verbatim}
Apart from not enforcing rapidity ordering in space-like splittings, this tune
differs from Tune 4C in that the proton size is considered $x-$dependent. This
is in the spirit of Tune 4CX, which was introduced in \cite{Corke:2011yy}. In
general, since we include matrix element states for two and three jets, we do 
not apply additional matrix element corrections in \PythiaEight after the first
emission, by setting 
\begin{verbatim}
SpaceShower:MEafterFirst = off
TimeShower:MEafterFirst = off
\end{verbatim}

\subsubsection{\texorpdfstring{\protect\PowhegBox + \PythiaEight}{PowhegBox + Pythia8}}
\label{App:Settings:PowhegBox}
For this study we used  \PowhegBox rev1282 and \Pythia 8.153. Except for the specific
subprocess requested, the parton distribution functions set and the
renormalisation/factorisation scale factors chosen, all the other
parameters were kept fixed below during all the runs. Here is a sample
\PowhegBox input file: 
\begin{verbatim}
! W^+ + jet production parameter
idvecbos 24         ! PDG id of vector boson (24: W+, -24: W-)
vdecaymode 1        ! decay channel (1: electron, 2: muon, 3: tau) 
numevts 4000000     ! number of events to be generated
ih1 1               ! hadron 1 (1 for protons, -1 for antiprotons)
ih2 1               ! hadron 2 (1 for protons, -1 for antiprotons)
ebeam1 3500d0       ! energy of beam 1 in GeV
ebeam2 3500d0       ! energy of beam 2 in GeV
lhans1  192800      ! pdf set for hadron 1 (LHA numbering)
lhans2  192800      ! pdf set for hadron 2 (LHA numbering)
ncall1 100000       ! number of calls for initializing the ...
itmx1    5          ! number of iterations for initializing the ...
ncall2 250000       ! number of calls for computing the integral ...
itmx2    4          ! number of iterations for computing the ...
foldcsi   1         ! number of folds on csi integration
foldy     1         ! number of folds on  y  integration
foldphi   1         ! number of folds on phi integration
nubound 100000      ! number of bbarra calls to setup norm of ...
icsimax  1          ! <= 100, number of csi subdivision when ...
iymax    1          ! <= 100, number of y subdivision when ...
xupbound 2d0        ! increase upper bound for radiation generation
renscfact  1d0      ! (default 1d0) ren scale factor: muren  = ...
facscfact  1d0      ! (default 1d0) fac scale factor: mufact = ...
withdamp    1       ! (default 0, do not use) use Born-zero ...
iseed    12345679   ! initialize random number sequence 
bornktmin  5d0      ! (default 0d0) kt min at Born level for ...
bornsuppfact 100d0  !  (default 0d0) mass param for Born ...
withnegweights 1    ! (default 0) allows negative weighted ...
runningscale 1      ! (default 0) ren. and fact. scales set to ...
masswindow_low 1000 ! restricts phase space to ...
masswindow_high 1000! restricts phase space to ...
\end{verbatim}

When interfacing to \PythiaEight we have changed the following settings with respect to \PythiaEight defaults,
 for the various stages under investigations:   

\begin{verbatim}
///Hadron Level w MPI and QED
BeamRemnants:reconnectRange = 1.50000
MultipleInteractions:alphaSvalue = 0.13500
MultipleInteractions:bProfile = 3
MultipleInteractions:ecmPow = 0.1900
MultipleInteractions:expPow = 2.0000
MultipleInteractions:pT0Ref = 2.0850
PDF:pSet = 8
SigmaDiffractive:dampen = on
SigmaDiffractive:maxAX = 65.0000
SigmaDiffractive:maxXB = 65.0000
SigmaDiffractive:maxXX = 65.0000
SigmaProcess:alphaSvalue = 0.13500
SpaceShower:MEafterFirst = off
SpaceShower:MEcorrections = off
SpaceShower:pTmaxMatch = 0
SpaceShower:rapidityOrder = on
TimeShower:MEcorrections = off
TimeShower:MEafterFirst = off
TimeShower:pTmaxMatch  = 0

//Hadron Level w MPI (added)
SpaceShower:QEDshowerByQ = off
SpaceShower:QEDshowerByL = off
TimeShower:QEDshowerByQ = off
TimeShower:QEDshowerByL = off

//Hadron Level w/o MPI (added)
PartonLevel:MI = off

//Shower Level (added)
HadronLevel:All = off

//Parton Level (added)
PartonLevel:ISR = off
PartonLevel:FSR = off
PartonLevel:Remnants = on

\end{verbatim}
and, most important, we have vetoed shower emissions with a transverse
momentum greater than the value of {\tt SCALUP} read from the Les
Houches event file for the corresponding event.

\subsubsection{\texorpdfstring{\protect\Madgraph}{Madgraph} + \texorpdfstring{\protect\Pythia}{Pythia}}
\label{App:Settings:MadGraph}

For this study \Madgraph/\Madevent 5.1.1.0 and \Pythia 6.4.2.4 is
used. The LHE files are generated for events with a $W$ and 
up to four additional partons, {\it i.e.} for the process:
\begin{verbatim}
pp>w- -> l-vl~ ; l-vl~~j ; l-vl~~jj ;l-vl~~jjj ; l-vl~~jjjj ; 
l-vl~ ; l-vl~j ; l-vl~jj ; 
l-vl~jjj ; l-vl~ jjjj () 
\end{verbatim}
The mass of the b quark is set to zero. The strong constant
$\alphas(M_Z^2)$ is set to 0.1300 both in the matrix element
calculation and in the proton PDF, that is the CTEQ6L1.
 
\Pythia is used for the parton shower and the hadronisation with
the following parameters modified according to tune Z2. 

\begin{verbatim}
MSTU(21)=1      ! Check on possible errors during program execution 
MSTJ(22)=2      ! Decay those unstable particles 
PARJ(71)=10 .   ! for which ctau  10 mm 
MSTP(33)=0      ! no K factors in hard cross sections 
MSTP(2)=1       ! which order running alphaS 
MSTP(51)=10042  ! structure function chosen (external PDF CTEQ6L1) 
MSTP(52)=2      ! work with LHAPDF 
PARP(82)=1.832  ! pt cutoff for multiparton interactions 
PARP(89)=1800.  ! sqrts for which PARP82 is set 
PARP(90)=0.275  ! Multiple interactions: rescaling power 
MSTP(95)=6      ! CR (color reconnection parameters) 
PARP(77)=1.016  ! CR 
PARP(78)=0.538  ! CR 
PARP(80)=0.1    ! Prob. colored parton from BBR 
PARP(83)=0.356  ! Multiple interactions: matter distribution para...
PARP(84)=0.651  ! Multiple interactions: matter distribution para... 
PARP(62)=1.025  ! ISR cutoff 
MSTP(91)=1      ! Gaussian primordial kT 
PARP(93)=10.0   ! primordial kT-max 
MSTP(81)=21     ! multiple parton interactions 1 is Pythia default 
MSTP(82)=4      ! Defines the multi-parton model
PMAS(5,1)=4.8   ! b quark mass 
PMAS(6,1)=172.5 ! t quark mass 
MSTJ(1)=1       ! Fragmentation/hadronization on  
MSTP(61)=1      ! Parton showering on   
\end{verbatim}

For additional studies, we set  
\begin{verbatim}
MSTJ(41)=3      ! switch off lepton FSR
MSTP(81)=20     ! switch off MPI
MSTJ(1)=0       ! Fragmentation/hadronization off
\end{verbatim}
to switch off, respectively, final state QED radiation, multi-particle interactions, and hadronisation.
}

\part[EXPERIMENTAL DEFINITIONS AND CORRECTIONS]{EXPERIMENTAL DEFINITIONS AND CORRECTIONS}
\label{part:corr}
\section[Photon isolation and fragmentation contribution]
{PHOTON ISOLATION AND FRAGMENTATION CONTRIBUTION \protect\footnote{Contributed by:N.~Chanon, S.~Gascon-Shotkin, J.~P.~Guillet, J.~Huston, E.~Pilon, M.~Schwoerer, M.~Stockton, M.~Tripiana }}
{\graphicspath{{Photons/}}

\title{Photon isolation and fragmentation contribution}

\author{N. Chanon$^1$, S. Gascon-Shotkin$^2$, J.P. Guillet$^3$, J. Huston$^4$, E. Pilon$^3$, M. Schwoerer$^5$, M.Stockton$^6$, M.Tripiana$^7$}

\institute{$^1$Institute of particle physics, ETH Z\"urich, Switzerland\\
        $^2$Universit\'e de Lyon, Universit\'e Claude Bernard Lyon 1/Institut de Physique Nucl\'eaire de Lyon IN2P3-CNRS\\
           $^3$LAPTH, Annecy-le-Vieux, France\\
           $^4$Department of Physics and Astronomy, Michigan State University, United States of America\\
           $^5$LAPP,  Annecy-le-Vieux, France\\
	       $^6$Department of Physics, McGill University, Montreal QC, Canada\\
	       $^7$Instituto de F\'isica de La Plata, UNLP-CONICET, La Plata, Argentina}


\begin{abstract}

Photon isolation and its link with the fragmentation contribution is explored via NLO matrix-element generator and parton-shower Monte-Carlo.

Firstly the dependence of the inclusive photon and di-photon NLO cross sections to the choice of isolation criteria are investigated. The isolation criteria used is the
discretized version of the Frixione isolation, with parameters chosen for those
most practical at an experimental level. As an extention, a more generalized version of the standard Frixione isolation is also studied. The selection of scale is also investigated in search of the `saddle point', which would give the optimal scale choice. In addition the choice of jet algorithm is investigated for the photon with associated jet cross section.

Secondly, properties of the fragmentation contribution in parton-shower Monte-Carlos are investigated. The distance profile of the photon to the other generator level particles in the event is explored in the case of neutral mesons, fragmentation photons and direct photons. Next the impact of a ``hollow'' or ``crown'' isolation criterion, expected to enhance the fragmentation contribution, is explored. Then, to complement the NLO inclusive studies, the impact of typical Frixione isolation criteria on the fragmentation component are investigated in the parton-shower Monte-Carlos.

Finally conclusions are made comparing the properties of the fragmentation contribution in NLO generators and parton-shower generators.

\end{abstract}


\subsection{INTRODUCTION}

Experimental measurements of single photons and di-photons require the application of isolation cuts to reduce the copious backgrounds arising from jet fragmentation. Such cuts also have the impact of reducing the fragmentation contributions of photon production. On the theoretical side, including fragmentation contributions of photon production can greatly increase the complexity of the calculations, while the application of appropriate isolation cuts can effectively remove those fragmentation contributions.

\subsubsection{Frixione isolation}

In the following we will study the Frixione isolation criterion~\cite{Frixione:1998hn}, which was designed to suppress the fragmentation contribution. It has been shown to reduce the fragmentation contribution in NLO generators~\cite{Binoth:2010ra}. The question is to know whether or not the behavior is still applicable using parton shower Monte-Carlo and if it can be used experimentally.

We consider the following function for the isolation criterion :
\begin{equation}\label{eq_frixione}
E_{T}^{iso}(R) < f(R) = \epsilon \cdot p_{T,\gamma} \cdot \Bigg( \frac{1-cos(R)}{1-cos(R_0)} \Bigg)^n
\end{equation}
where $E_{T}^{iso}(R)$ is the isolation sum of all particles inside a cone of $R = \sqrt{\Delta \phi^2 + \Delta \eta^2}$ around the photon, $\epsilon$ is the strength or the scale of the isolation criterion, $p_{T,\gamma}$ is the transverse energy of the photon, $R_0$ is the first considered cone and $n$ the power of the isolation criterion. This formula can be altered by replacing $p_{T,\gamma}$ with a fixed threshold. Functional forms $f(R)$ for different $\epsilon$ and $n$ are shown Fig.~\ref{fig:FrixioneFunction}.

\begin{figure}[hbtp]
  \begin{center}
    \resizebox{6cm}{!}{\includegraphics{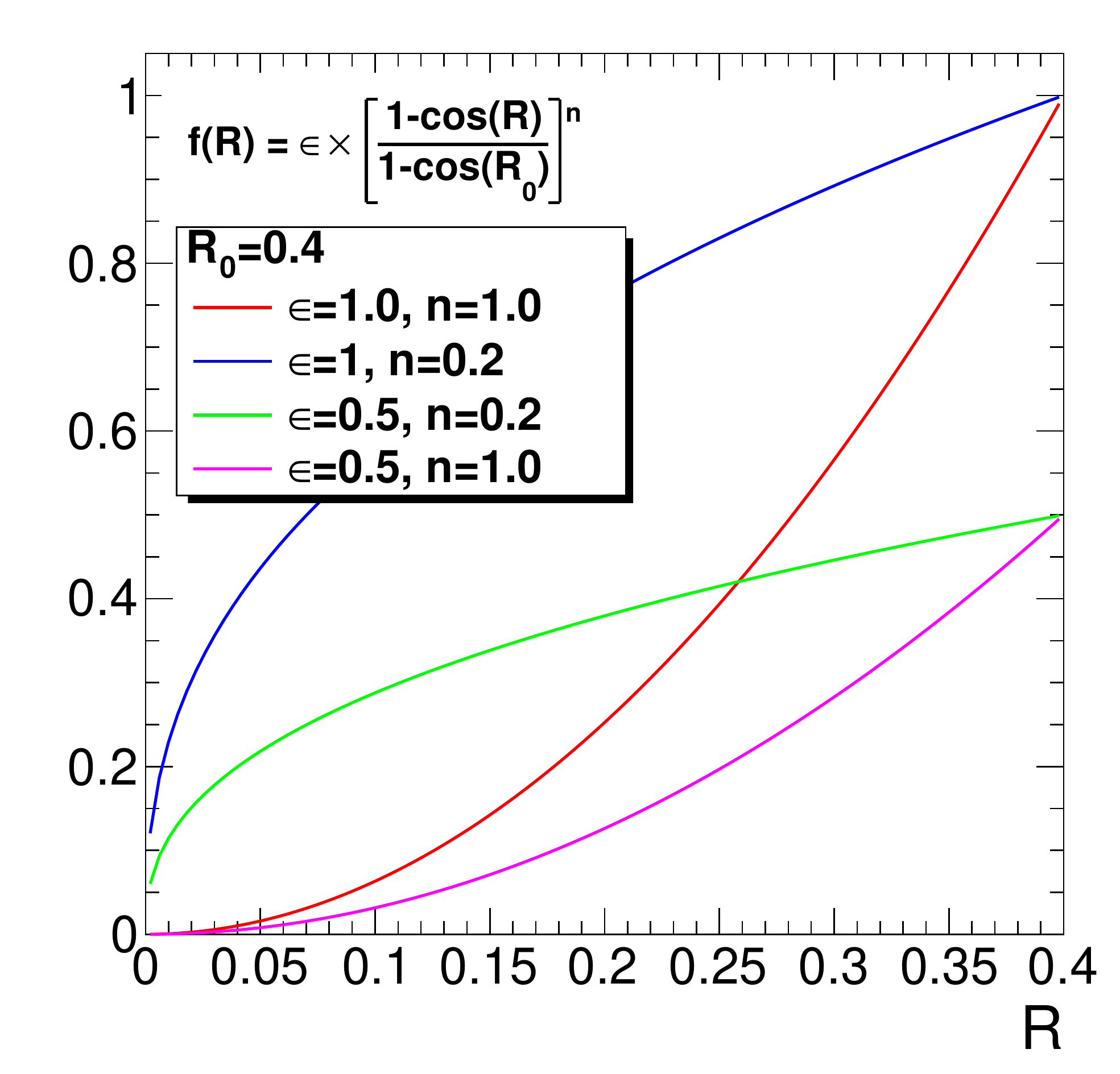}}
    \caption{Examples of Frixione functional form for different parameters.}
    \label{fig:FrixioneFunction}
  \end{center}
\end{figure}

The Frixione isolation is tighter and tighter when decreasing the $R$ cone size. With matrix element NLO generators (as in Jetphox~\cite{jetphox1,jetphox2} and Diphox~\cite{Binoth:1999qq}), the only contribution possible at a given $R\neq 0$ is the one coming from the suplementary hard jets in the event, while the fragmentation debris are emitted colinearly to the photon at $R=0$ (angle information is lost due to the fragmentation function which is integrated over the angle). With parton-shower generator, fragmentation photons are emitted off quarks at a non-zero angle during the showering process. In the following sections we will study the link between isolation and fragmentation in NLO generators, then with parton-shower Monte-Carlo.

\subsubsection{Experimental complications}

There are various mismatches between isolation cuts applied to theoretical calculations and isolation cuts applied to data (or to Monte Carlo). First of all, we wish to apply the isolation cut only to energy related to the hard scatter. Experimentally, most of the energy inside an isolation cone is due to the underlying event associated with the hard scatter, or the remnants of additional interactions in the same crossing.  Techniques such as  jet area subtraction~\cite{Cacciari:area,Cacciari:UE} can be used to remove an amount of energy from the isolation cone roughly equal to the expected contamination from underlying event/pileup, leaving only energy related to the hard scatter and specifically to fragmentation processes. Since there is no underlying event/pileup in partonic level theory calculations, only the isolation cut needs to be applied to the theory. 

Full details of the use of this correction within the ATLAS collaboration can be
found in the inclusive cross section measurement~\cite{InclusivePaper2010}. The
ATLAS isolation definition uses a cone around the cluster of cells that are
identified as a photon. These photon cluster cells are not included in the sum,
so there is first a correction for any leakage of the photon shower into the
surrounding cells (typically a few percent of the photon $p_T$). The
pile-up/underlying event correction is then applied by calculating per event the
ambient energy from the jet activity in that specific event. This follows the
jet area corrections method mentioned above, where all jets are reconstructed
without any minimum momentum threshold. The energy density of each jet is
calculated and the median density is used for the correction. In 2010 this
typically resulted in a correction of around $900 \rm{MeV}$. 


The original Frixione isolation scheme assumed that an isolation cut could be applied continuously as a function of R (distance from the photon). Actual detectors have a finite granularity. A solution to this was the adoption of a discretized version of Frixione isolation, allowing this granularity to be taken into account~\cite{Binoth:2010ra}. However, it is not possible to place an isolation cut on the inner-most cone (typically $R\sim0.1$), because of the presence of the photon itself. While the separation between the fragmentation photon and the jet remnants is finite in data (and in Monte Carlo) , fragmentation is treated as a collinear process in partonic cross sections. The inability to apply the isolation cut down to $R=0$, results in a greatly reduced ability to discriminate against fragmentation processes in the partonic level theory. 

To rectify this the Frixione calculation could be modified into a `crown'
isolation, whereby the last cone is missed from the calculation. Unfortunately
as most of the radiation is collinear in the fragmentation events, it is likely
to reduce its effect of removing these events. Other studies~\cite{MartinPrivate} have shown that the photon quality cuts applied by
the experiments will reduce the fraction of fragmentation photons accepted,
where substantial fragmentation energy is collinear with the photon. However,
the rejection is not 100\%, so we are still left with a smaller reduction of
fragmentation contributions in the partonic level theory than are actually
(presumably) present in the data. In these proceedings, we will discuss how to
more properly incorporate the correct level of rejection in the theory.


\subsubsection{Choice of fragmentation scale}

In addition to the experimental difficulties with applying the isolation criteria there are
also difficulties in choosing appropriate scales for the theoretical
calculation. This is discussed in the following text, along with other
considerations for applying Frixione isolation at a theoretical level.

Fragmentation is treated as a collinear process in partonic calculations. In this framework, the original ``continuous" Frixione criterion \cite{Frixione:1998hn} was designed to inhibit the appearance of final state photon-parton collinear singularities which otherwise require absorption in a fragmentation function $D(z,M_F)$.  Thus, cross sections for the production of prompt photons isolated with this criterion involve no fragmentation contribution. Discretized variants of this criterion have been proposed which aim at matching better what can be actually implemented experimentally \cite{Binoth:2010ra}. They consist in a limited number of nested cones ${\cal C}_{j=1,\cdots, n}$ with respective radii $R_{1} = R_{min} < R_{2} < \cdots < R_{n} = R_{max}$ defined in the 
azimuthal and rapidity differences with respect to the photon direction, and requiring recursively that the accompanying hadronic transverse energies inside every successive cone ${\cal C}_{j}$ be less than an ordered sequence of maximum values\footnote{Whether the $E_{T \; j}^{iso}$ take fixed values or are functions of the photon's transverse momentum does not matter here.}  $E_{T \; j}^{iso}$  such that $0 < E_{T \; 1}^{iso} < \cdots <  E_{T \; n}^{iso}$. 
However, in contrast with the continuous criterion, such discretized variants still involve a fragmentation contribution, though the latter is expected to be small, since the situation in the innermost cone shares some similarity with the standard cone criterion. When quantifying the magnitude of the fragmentation contribution with such discretized criteria, a potentially tricky issue concerns the fragmentation scale dependence and the ``best choice" of scale. 

This issue matters for isolation with the standard cone criterion when the radius $R$ of the cone is $\ll 1$ while $E_{T}^{iso}$ is kept fixed. Whereas the natural fragmentation scale $M_F$ in the non-isolated case is $\sim p_{T}^{\gamma}$, this choice can lead to very poor theoretical estimates at Next to Leading Order (NLO) in perturbative QCD when $R \ll 1$ \cite{jetphox1}. The scale dependence near the choice $M_{F} \sim p_{T}^{\gamma}$ is then large and, 
worse, the theoretical prediction may eventually exhibit an unphysical violation of unitarity whereby the predicted NLO cross section for photons becomes {\em larger} than the inclusive one, so that even for only moderately small $R$ the reliability of the prediction is questionable. On the other hand, as\footnote{The logarithmic behaviour holds in when $M_F \gg \Lambda_{QCD}$. In the non-perturbative regime instead no logarithm is expected to develop and one expects rather a power-suppressed behaviour; see ref. \cite{Bourhis:1997yu} for more details.} $D(z,M_F) \sim \log (M_F/\Lambda_{QCD})$, with 
the choice $M_F \sim R p_{T}^{\gamma}$ the fragmentation contribution is suppressed compared with $M_F \sim p_{T}^{\gamma}$. The situation is improved regarding both scale dependence and unitarity, although it does not solve the problem completely. One actually faces a multiscale problem: 
$\Lambda_{QCD} \ll R p_{T}^{\gamma} \ll p_{T}^{\gamma}$, and a one-scale compromise is possibly insufficient depending on the kinematical regime explored. The atypical choice $M_F \sim R p_{T}^{\gamma}$ has in principle to be supplemented by a resummation of the logarithmic $R$ dependence coming form outside the cone, if at all possible. At leading-$\log R$ 
(LLR) accuracy at least, such a resummation is actually feasible, which furthermore allows to solve the apparent puzzle why scale choices should be very different in the cases with isolation in a narrow cone vs. broad cone or without isolation.

The concern about the discretized Frixione criteria is that the innermost cone size is quite small. The choice for the fragmentation scale $M_F$ shall then arguably be $M_F \sim {\cal O}(R_{min} p_{T}^{\gamma})$. On the other hand, as the allowed transverse energy deposit $E_{T}^{iso}(R_{min})$ inside this cone is correspondingly small, the width of the interval in the fragmentation variable on which the fragmentation function is convoluted with the partonic cross section is restricted to a rather narrow range $0 <1-z < E_{T}^{iso}(R_{min})/p_{T}^{\gamma} \sim \epsilon (R_{min}/R_{max})^{n}$. This leads to a quite suppressed fragmentation contribution. The combination of the two effects: a low fragmentation scale and a narrow $z$-range, is the discrete counterpart of the inhibition of fragmentation by the continuous criterion. We may thus expect that the issue of the narrow cone is less worrying for the reliability of the NLO calculation in this case than if only $R$ were taken small while keeping $E_{T}^{iso}$ fixed. 
In order to assess the uncertainty on the fragmentation contribution we may perform the calculation for the ``arguably better" scale $M_F \sim R_{min} \, p_{T}^{\gamma}$ and compare it to the expectedly larger result for the standard choice $M_F \sim p_{T}^{\gamma}/2$.


\subsection{ISOLATION FOR INCLUSIVE PHOTONS AND DIPHOTONS AT NLO}

The study at NLO uses the Jetphox generator to calculate the inclusive photon cross section and Diphox for the di-photon cross section. Details of how to
use the software and to obtain predictions with errors can be found
in~\cite{Blair:1379880} and the selection criteria used are listed in the appendix. Previous results from Les Houches~\cite{Binoth:2010ra} showed that the discretized Frixione
isolation criteria did manage to reduce the fragmentation contribution, here we extend that study in several ways. Firstly the cross section returned has
been compared to that calculated from using the standard cone isolation, as used
in current measurements. A generalized form of the Frixione isolation is discussed, aimed
to satisfy both the experimental and the theoretical requirements on the isolation cut for different $p_T$
regimes. In addition the effects of changing the number of cones
used in the calculation and of choosing an $E_T$ cut, rather than relating it to
the photon $p_T$, are investigated. In addition, further complications to
comparing theoretical and experimental isolation calculations are discussed. Finally there are further brief studies using Jetphox to look at scale and jet algorithm choices.

\subsubsection{Discretized prescription}

The parameters used to define different selections, according to Eq.~\ref{eq_frixione}, were:\\

\begin{tabular}{llll}
$a$: $\epsilon$=0.05 $n$=0.2 & 
$b$: $\epsilon$=1 $n$=0.2 & 
$c$: $\epsilon$=1 $n$=1 & 
$d$: $\epsilon$=0.5 $n$=1 \\
$e$: $\epsilon$=0.05 $n$=1 &
$f$: $\epsilon$=1 $n$=0.1 &
$g$: $\epsilon$=1 $n$=0.5 \\
\end{tabular}\\

where all but the last two were based on the previous
study. In all cases $R_0$ was chosen to be 0.4, with $R$ being set to either:
0.4, 0.3, 0.2, 0.1 or 0.4, 0.35, 0.3, 0.25, 0.2, 0.15, 0.1.

The comparison to the cone isolation in Fig.~\ref{isolation} shows that out of
the chosen parameters only 1 set removes the fragmentation contribution more
than what is removed by the cone algorithm, although two are lower until high
$p_T$. It also shows that criteria $b$ and $f$ are not much better than applying no isolation criteria at all. When altering Eq.~\ref{eq_frixione} to use a fixed $E_{T}=4\rm{GeV}$ instead of $p_{T,\gamma}$ the results are more
promising but this is  because it applies a cut in the 0.1 cone that is below the experimental accuracy (of the order 100 MeV due to detector resolution/noise). Unfortunately Fig.~\ref{isolation}
also shows that this is also the case for the $p_T$ requirements, as case $e$ (the
only criteria to perform better than the standard cone) also applies a cut that
is not viable experimentally in the 0.1 cone.


\begin{figure} \begin{center}
\includegraphics[width=\textwidth]{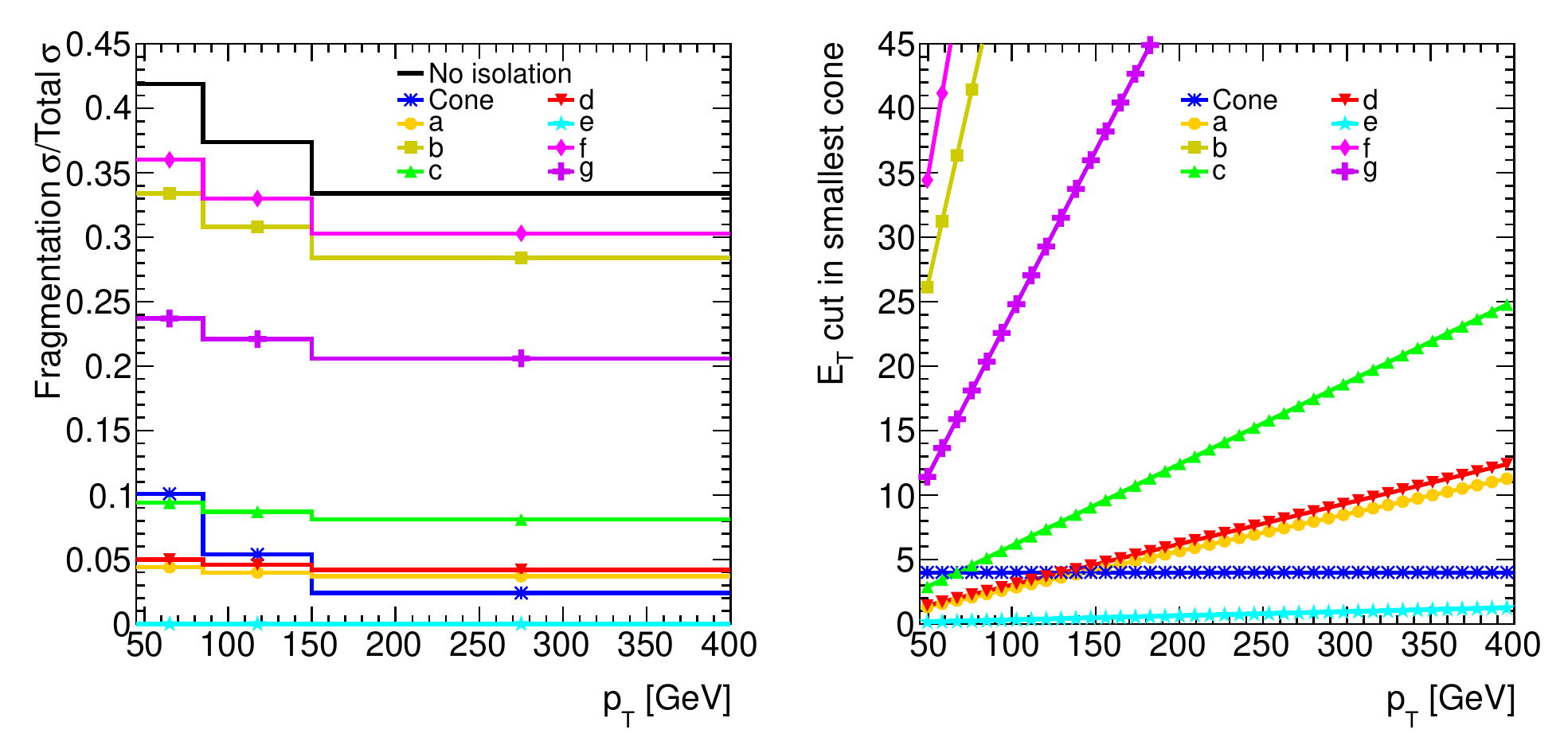} 
\caption{Left: Fragmentation fraction for the cone and Frixione isolation
criteria. Right: The applied $E_{T}$ cut on the isolation sum as a function of $p_{T}$ for the 0.4
cone or 0.1 cone in the Frixione criteria.} 
\label{isolation}
\end{center} \end{figure}

There are some positive outcomes from these studies, firstly the Frixione
criteria $b$ and $d$ maybe useful criteria to use experimentally as they keep the
fragmentation contribution similar and low in all bins, which could help with
understanding of the systematic errors/correlation between bins. Secondly the 
comparison of the number of cones used in the Frixione criteria resulted in a
difference of around $1\%$ on the total cross section and almost no effect on
the fragmentation fraction. This means that it is fine to use the lower number
of cones case, and that the discrete Frixione criteria is most likely very
similar to that of the continuous version. 
\subsubsection{Generalized prescription}

As seen previously, to remove the fragmentation contribution in the theory, a small value of $\varepsilon$ is needed. However, given the effects of finite resolution
and granularity on the experimental description of the isolation energy, a minimum threshold has to be allowed in the isolation cone, especially at low
$p_T$. A typical value of 2-4 GeV is used as experimental cut, to optimize the rejection of hadronic background coming from the decay of light 
mesons. Now, at high $p_T$ this cut might result too tight, particularly on the theoretical side given that an isolation cut much smaller than the 
photon $p_T$ can cause large logs in the calculations, this effect was not observed in the previous Les Houches study. 

As a good compromise of these two requirements, it has been proposed~\cite{EricPrivate} to extend the original Frixione prescription (Eq. \ref{eq_frixione})
to a more general form:
\begin{equation} 
E_{T}^{iso} < \left( (E_{0})^{k} + (\varepsilon.p_{T})^k\right)^{1/k} \left(\frac{1-\cos{R}}{1-\cos{R_{0}}}\right)^n 
\end{equation}
where: 
\begin{itemize}
\item[$R_0$] is the maximum cone size
\item[$E_0$] is the minimum energy pedestal allowed in a cone of size $R_0$
\item[$\varepsilon$] is the fraction of the photon $p_T$ allowed in the cone of size $R_0$
\item[$k$] determines the shape of the isolation profile in $p_T$ 
\item[$n$] determines the shape of the isolation profile in R (see Fig. \ref{fig:profiles}[right])
\end{itemize}


The extra parameters give enough flexibility to ensure a (finite) tight cut at low $p_T$ ($\sim E_0$) and, at the same time, a loose cut at high end of the spectrum driven 
by the photon $p_T$. The $k$ parameter controls how quickly/smoothly is the transition from one regime to the other (see Fig. \ref{fig:profiles}[left]). 

\begin{figure} \begin{center}
\includegraphics[width=0.49\textwidth]{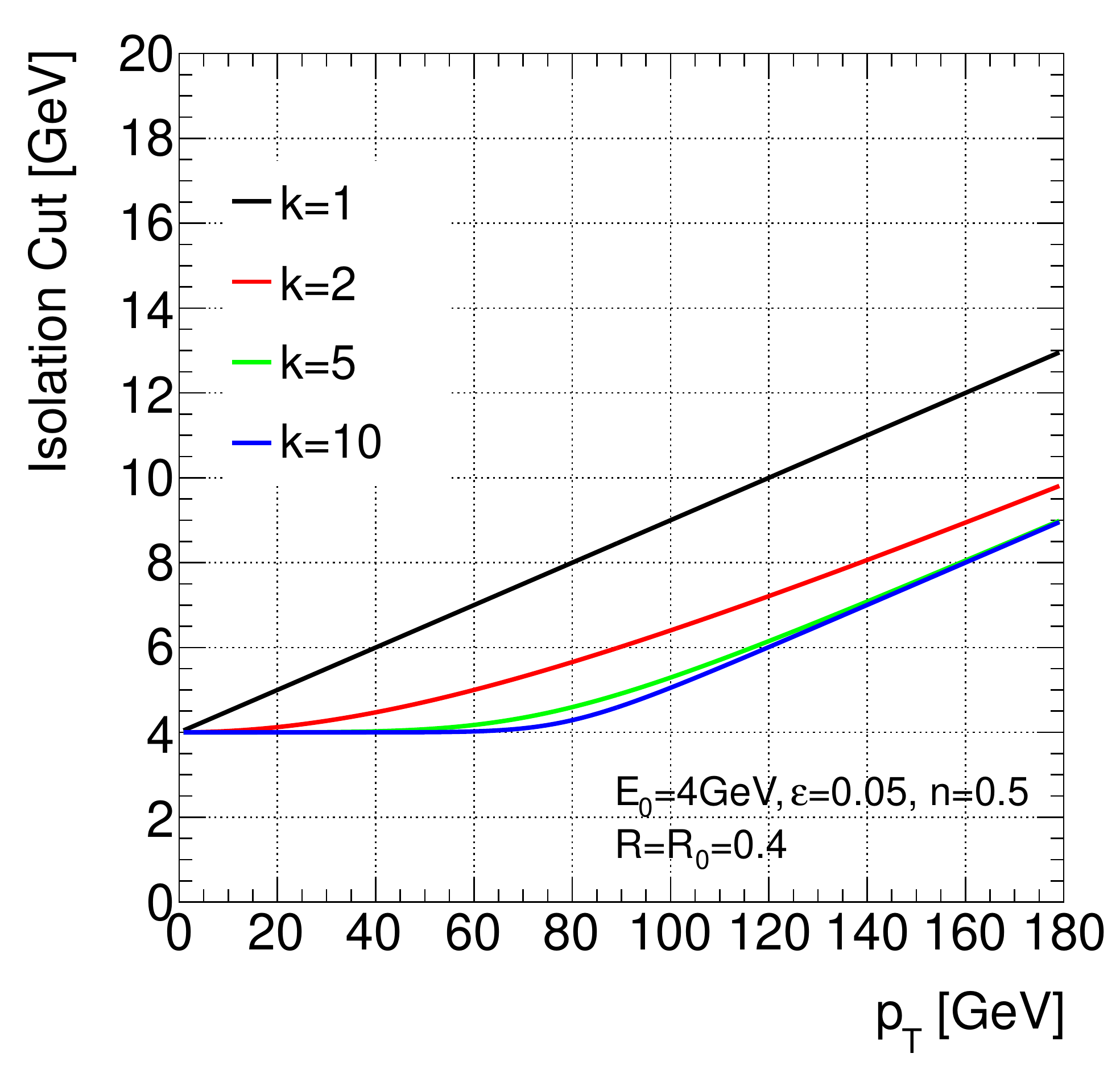} 
\includegraphics[width=0.49\textwidth]{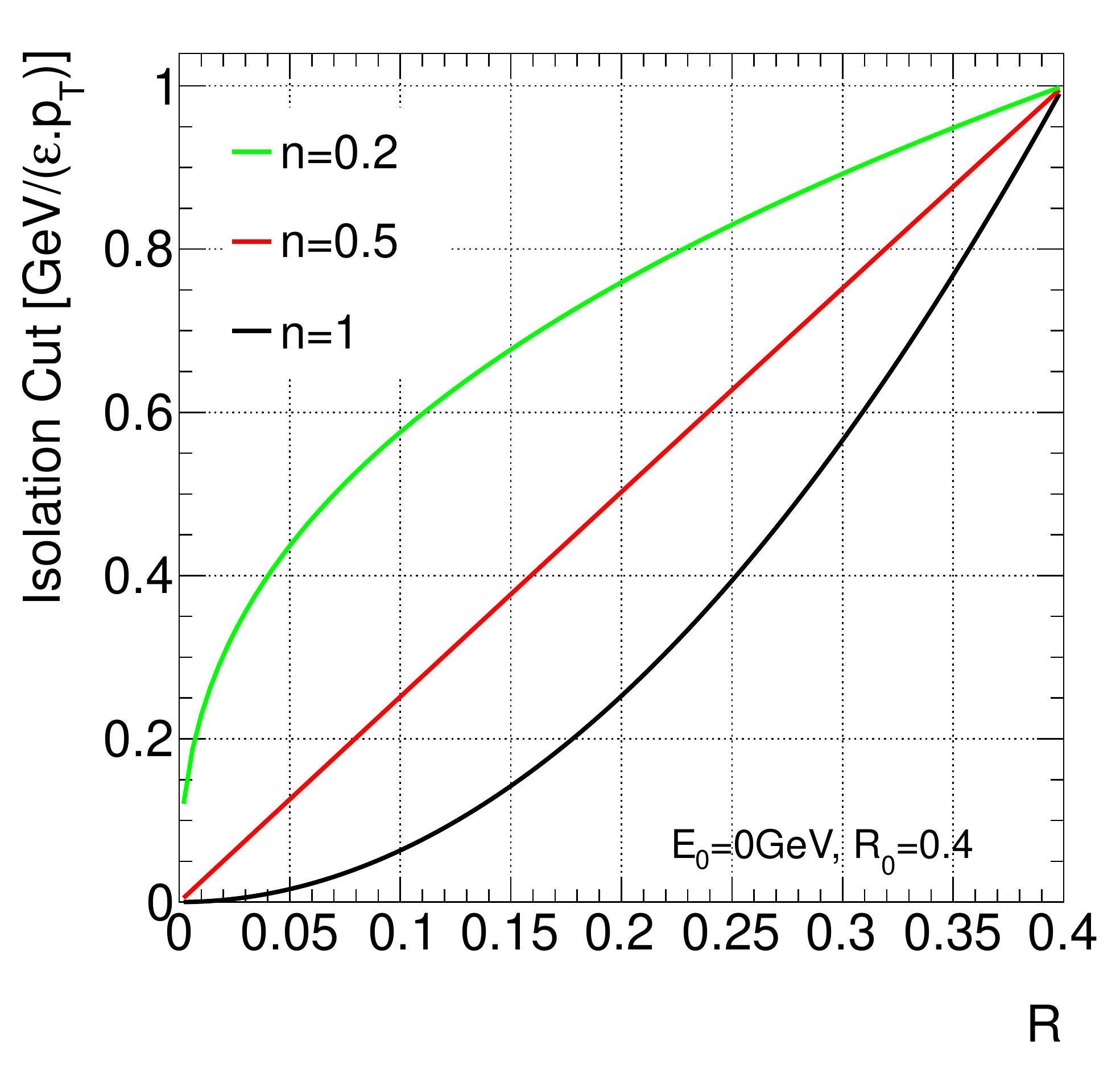} 
\caption{Effect of the two terms in the modified Frixione isolation prescription  effects of the different pieces of the generalized Frixione isolation prescription on 
the isolation cut as a function of $p_T$ (left) and the cone radius R (right).}
\label{fig:profiles}
\end{center} \end{figure}

This generalized prescription\footnote{Indeed for $E_0=0$ the original Frixione prescription is restored.} has been implemented in Jetphox recently 
and some possible configurations are explored here. The studied configurations vary $\varepsilon$ (=0.05,1) and $k$ (=2,5,10), and have a fixed 
value for $E_0=4\rm{GeV}$ (the typical cut applied in ATLAS) and $n=0.5$ (given the linear behaviour of isolation distribution width observed for direct photons in ATLAS~\cite{MikePrivate}).

The high-$\varepsilon$ configurations ($\varepsilon$=1), show a worse performance at removing the fragmentation contribution with respect to the fixed cone approach 
and are practically insensitive to the value of $k$ in the formula. The remaining fragmentation fraction is $\sim25\%$ at 45 GeV decreasing to $20\%$ in the highest
$p_T$ bin. On the other hand, as seen in Fig. \ref{fig:modified}, all the configurations for a low value of $\varepsilon$ (=0.05) show an improvement in fragmentation rejection compared to both the no isolation and fixed cone cases, in the whole $p_T$ region ($45\rm{GeV}<p_T<600\rm{GeV}$). The $p_T$ 
profile for the smaller cone ($R=0.1$) in this case (Fig. \ref{fig:modified}[right]) looks also more promising in terms of its applicability at the experimental level.

\begin{figure} \begin{center}
\includegraphics[width=0.49\textwidth]{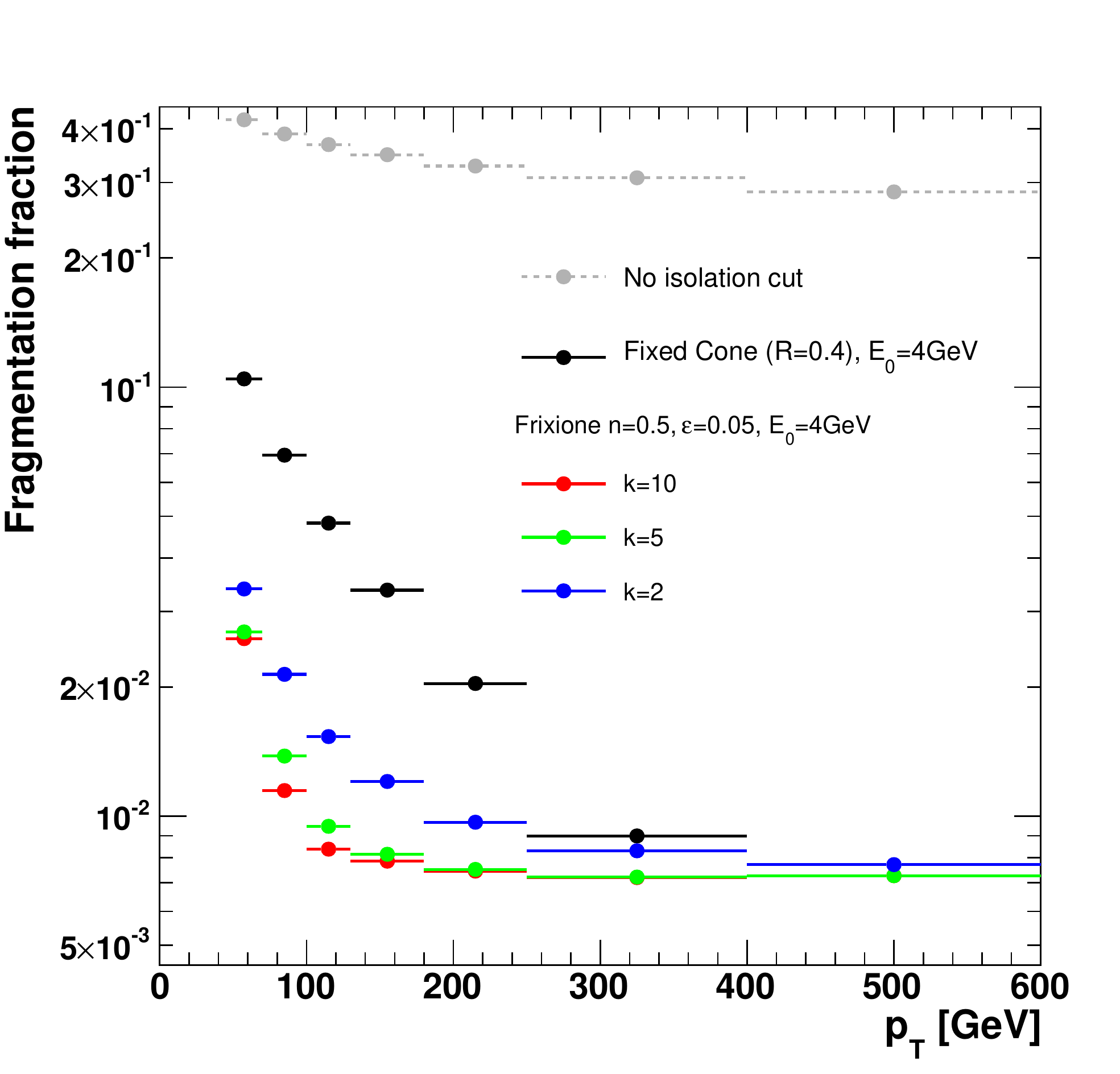} 
\includegraphics[width=0.49\textwidth]{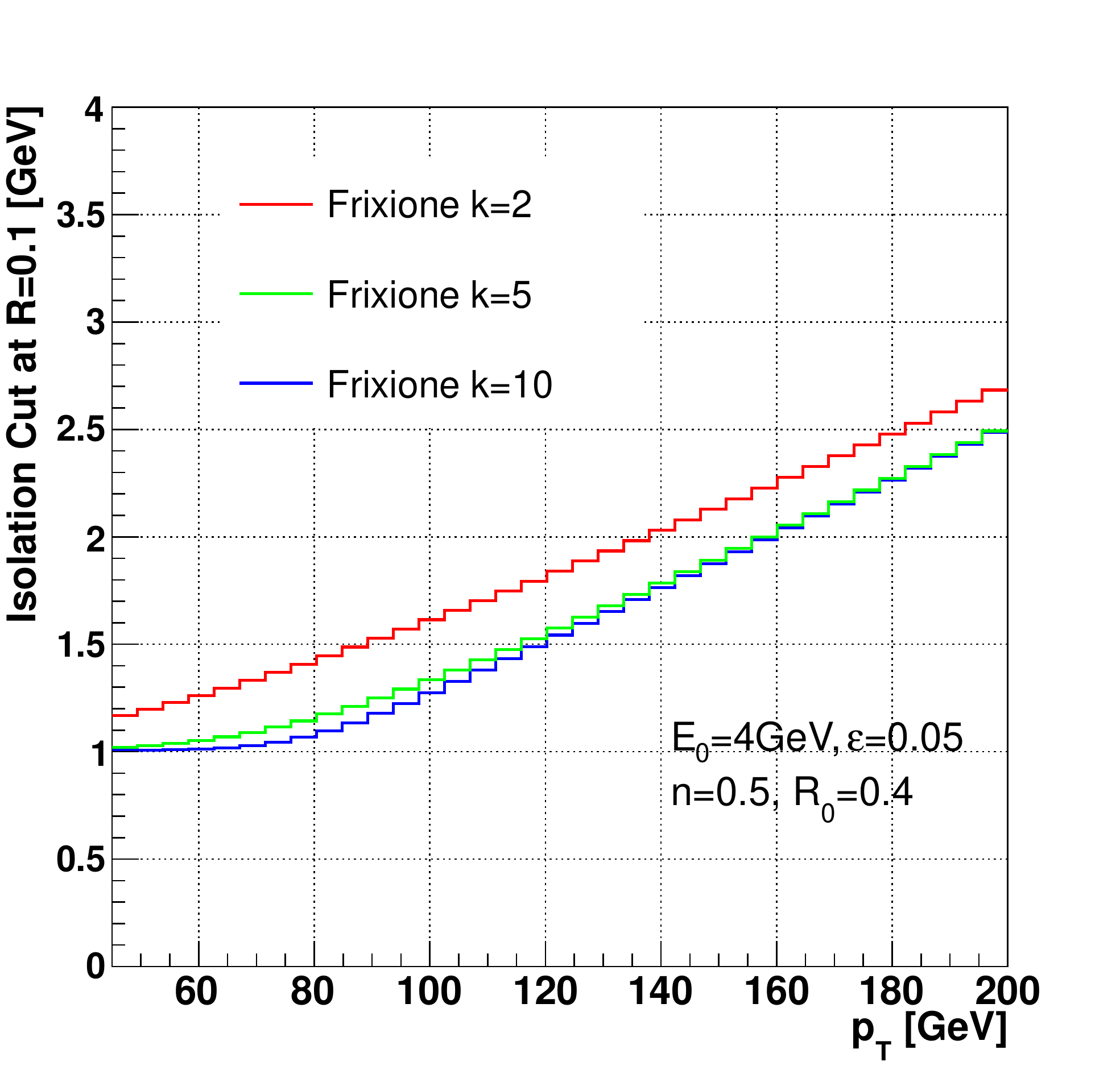} 
\caption{Left: Fragmentation fraction for the cone and the (generalized) Frixione isolation
criteria. Right: The applied isolation cut as a function of $p_{T}$ for 0.1 cone in the (generalized) Frixione criteria.} 
\label{fig:modified}
\end{center} \end{figure}

\subsubsection{Continuous and discretized Frixione criteria in di-photon events}



Following the previous studies with inclusive photons, we now consider the production of photon pairs.
The aims of this study are i) to assess the effect on the magnitude of the fragmentation 
contribution by comparing results from using the continuous Frixione criterion with those using several variants of the 
discretized version, implemented in the NLO programme Diphox, thereby 
providing a NLO assessment of how much fragmentation may be missing in the NNLO 
calculation of Catani et al. \cite{Catani:2011qz} which includes no fragmentation and 
therefore uses the continuous criterion; ii) to probe the dependence 
of the prediction with respect to the fragmentation scale choice. It supplements a 
similar comparison which had been performed for inclusive photon production in 
\cite{Binoth:2010ra}.

Fig. \ref{f1a} provides a comparison of the original continuous criterion to the discretized version of the criterion based 
on four nested cones with respective radii $R_{min} =0.1, R_{2} = 0.2, R_{3} = 0.3$ and 
$R_{max} = 0.4$. 
Four variants of the energy profile $E_{T}^{iso}(R)$  
as defined in Eq.~\ref{eq_frixione} have been considered: $(\epsilon, n) = (0.05,0.2), (0.05,1), (0.5,1)$ and $(1,1)$.
\begin{figure}
\begin{minipage}[t]{.5\linewidth}
\begin{center} 
\includegraphics[scale=0.4,angle=90]{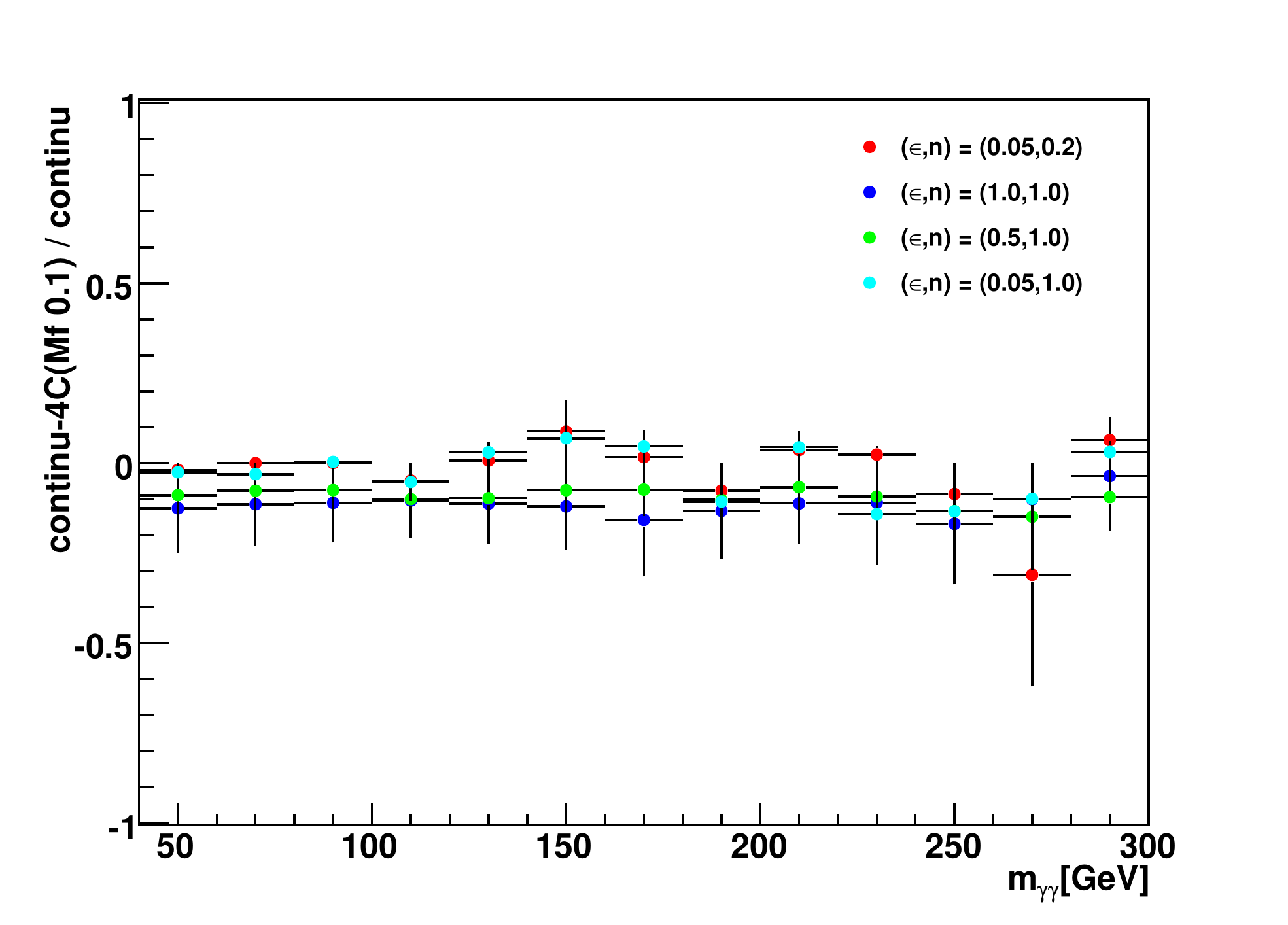}
\end{center} 
\end{minipage}
\begin{minipage}[t]{.5\linewidth}
\begin{center} 
\includegraphics[scale=0.4,angle=90]{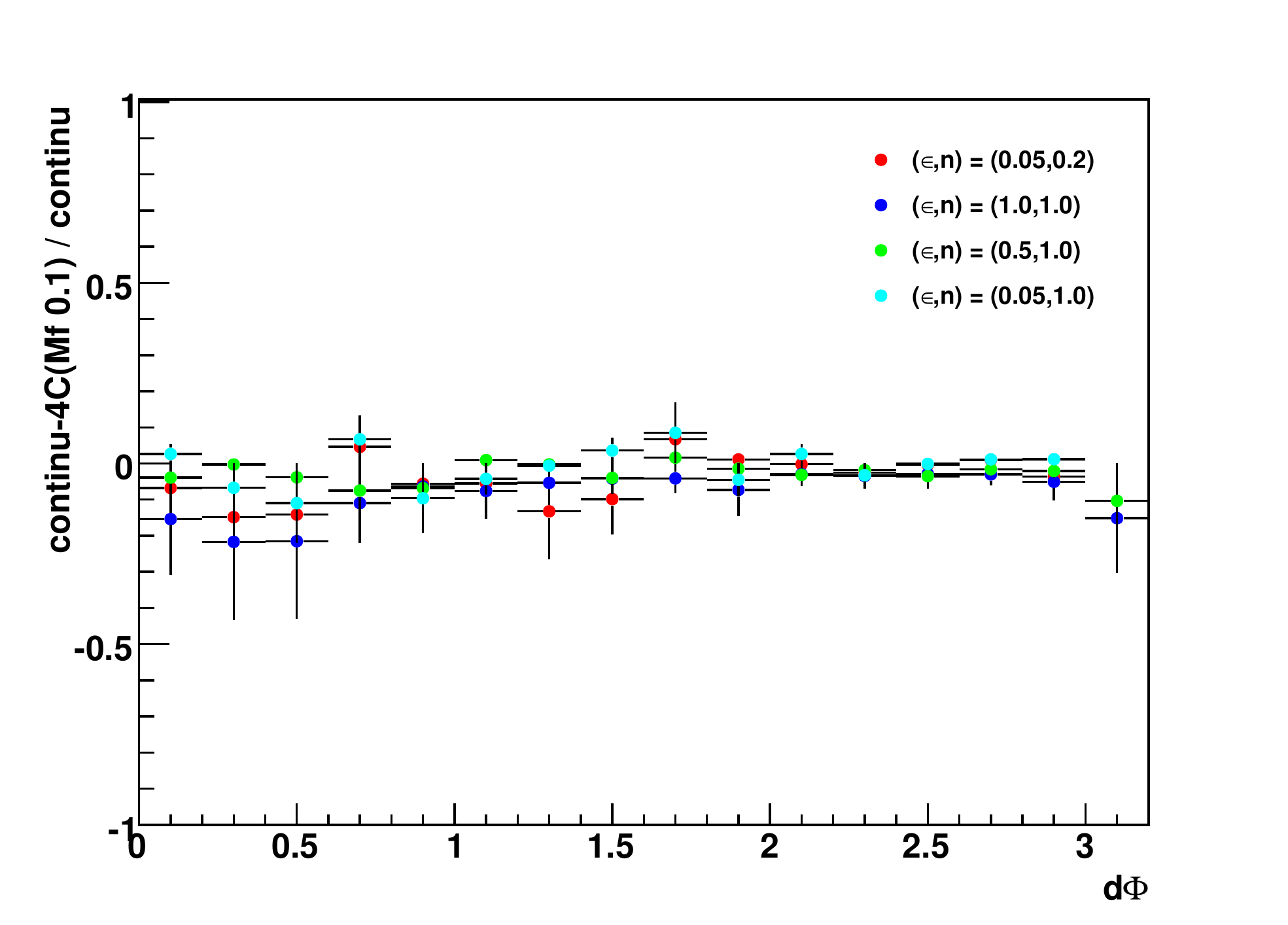}
\end{center} 
\end{minipage}
\caption{\label{f1a} Comparing continuous and discretized Frixione criterions for the
distributions in $m_{\gamma \gamma}$ (left) and $\Delta \phi$ (right) of photon pairs, for four variants of the criterion.}
\end{figure}
Fig. \ref{f1a} (left) presents the distribution in invariant mass of photon pairs in the
range $40$ GeV $ \leq m_{\gamma \gamma} \leq 300$ GeV. 
The discretized criterion $(0.05,1)$ suppresses fragmentation so much that there is practically 
no difference between the discretized and continuous versions. 
With the criterion $(1,1)$, the discretized version leads to
a distribution $O(10-12\%)$ larger than the continuous one. The choice $(0.5,1)$
displays a similar feature, though quantitatively less important. The energy profile
of the fourth choice is not suited for an efficient isolation unless $\epsilon$ is 
chosen very small. A similar comparison is shown on Fig. \ref{f1a} (right) for the distribution 
in the difference in azimuthal angle $\Delta \phi$ between the two photons. 
Whereas the distribution in
invariant mass is dominated by the direct contribution, the tail of the distribution 
in $\Delta \phi$ tail at low $\Delta \phi$ is more sensitive to the fragmentation 
contribution. Therefore, the conclusions are qualitatively similar to the one drawn 
for the distribution in invariant mass, yet the effects are quantitatively larger.  

Fig. \ref{f3a} assesses the dependence on the fragmentation scale $M_{F}$, for the 
distributions in invariant mass (left) and in $\Delta \phi$ (right) respectively. 
Two choices were considered: 
$M_{F} = R_{min} \,  
\mbox{min} \, \{p_{T}^{\gamma \, 1}, \, p_{T}^{\gamma \, 2}\}
= 0.1 \, \mbox{min} \, \{p_{T}^{\gamma \, 1}, \, p_{T}^{\gamma \, 2}\}$ vs.
$M_{F} = R_{max} \, 
\mbox{min} \, \{p_{T}^{\gamma \, 1}, \, p_{T}^{\gamma \, 2}\} 
= 0.4 \, \mbox{min} \, \{p_{T}^{\gamma \, 1}, \, p_{T}^{\gamma \, 2}\}$ closer to a standard
choice\footnote{Fragmentation scale options depending only on the $p_{T}$ of the photon from 
fragmentation are not available because Diphox encodes the two 
photons in a symmetrized way.}. 
\begin{figure}
\begin{minipage}[t]{.5\linewidth}
\begin{center} 
\includegraphics[scale=0.4,angle=90]{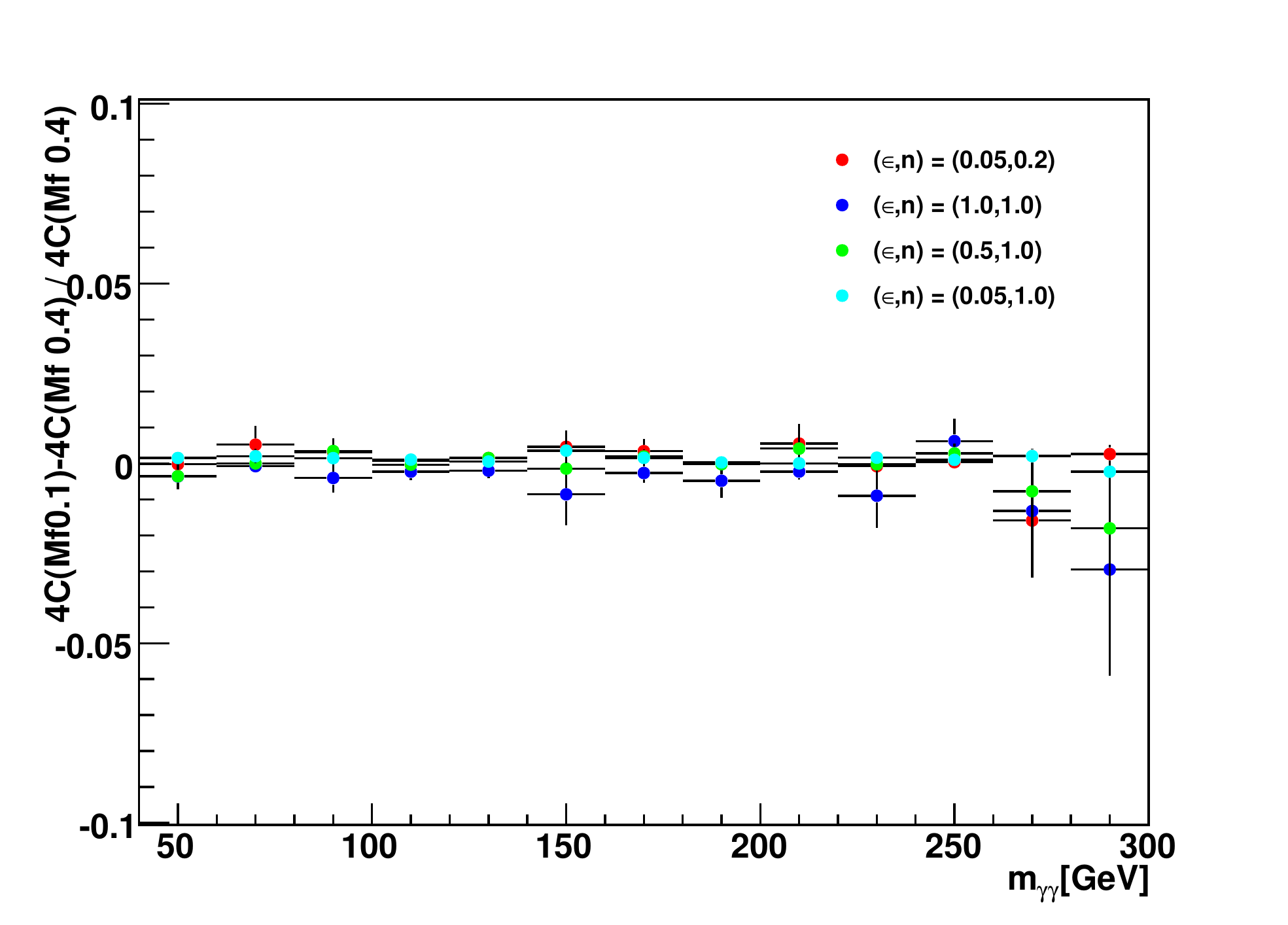}
\end{center}
\end{minipage}
\begin{minipage}[t]{.5\linewidth}
\begin{center} 
\includegraphics[scale=0.4,angle=90]{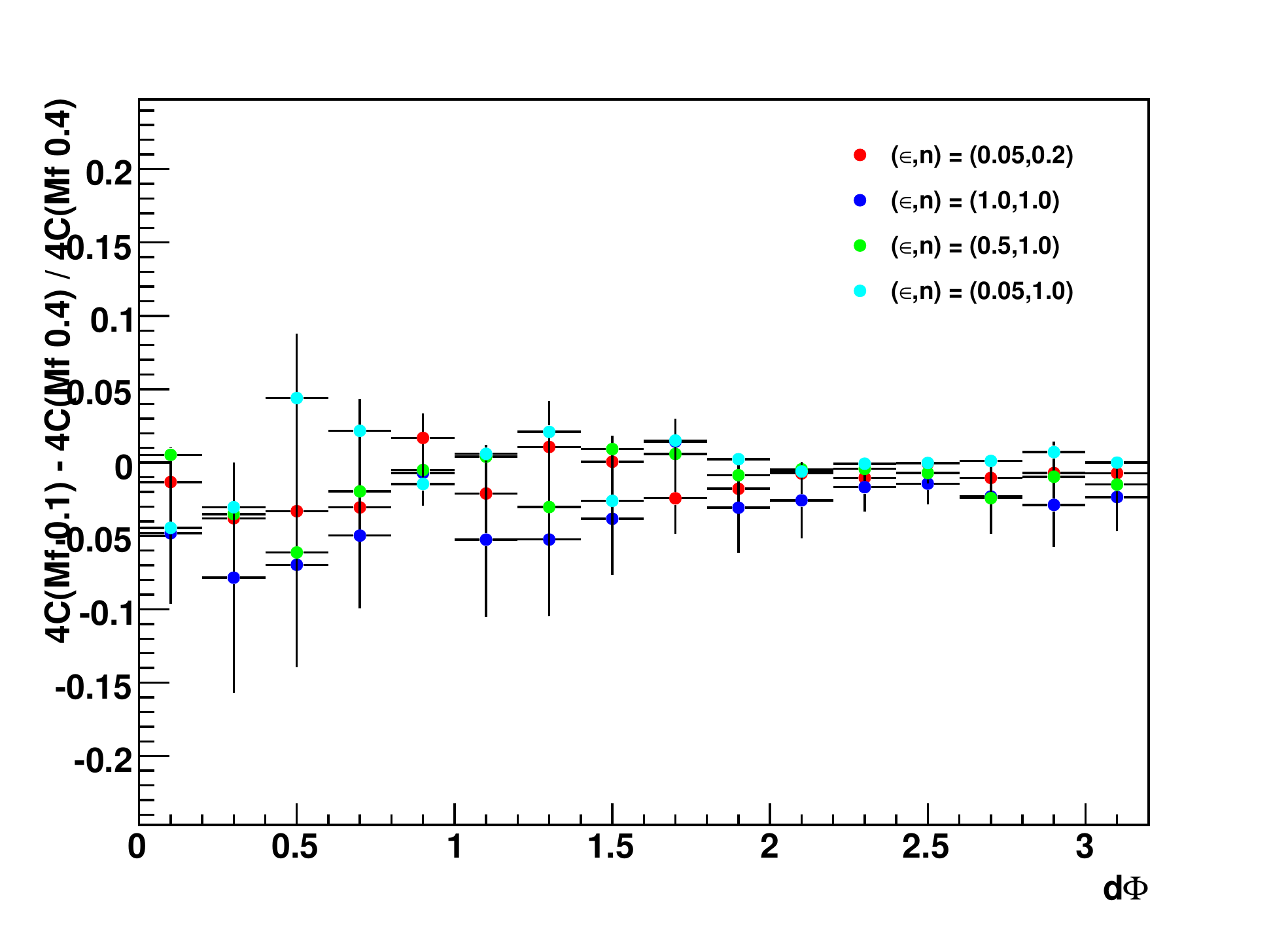}
\end{center}
\end{minipage}
\caption{\label{f3a} Fragmentation scale dependence of the distribution in $m_{\gamma \gamma}$ 
(left) and $\Delta \phi$ (right) of photon pairs, for four variants of the discretized Frixione criterion.}
\end{figure}
As expected, the distribution in invariant mass, which is not very sensitive to the 
fragmentation contribution, is practically not impacted by the choice. The 
distribution in $\Delta \phi$ is more sensitive to the fragmentation component and
the sensitivity to the fragmentation scale choice is larger than for the distribution
in invariant mass. The sensitivity to the fragmentation scale choice is the largest 
in the case of the criterion $(\epsilon,n) = (1,1)$, for which the predictions are 
5-7\% smaller, rather uniformly, with the lower scale choice than with the more 
standard one.

In conclusion this preliminary study shows that the impact of the fragmentation 
contribution passing the discretized criterion seems to be almost negligible on the 
distribution in invariant mass, and remains small even on the tail of the distribution in 
azimuthal angle. Notwithstanding, the conclusions shall have limited use depending on
how isolation is actually implemented experimentally in the innermost cone. We here stick to
a discretized version of the Frixione criterion which respects the original idea of a transverse 
energy deposit decreasing towards zero with the cone radius. If instead any experimental
constraint would allow a more permissive condition in the innermost cone, a dedicated study would
be mandatory.


\subsubsection{Additional studies at NLO}

In addition to the isolation studies with Jetphox, we present here two brief studies as an attempt to reduce the theoretical errors from the NLO calculation. These study the choice of renormalizaion and factorization scale parameter and secondly the jet algorithm parameters.


As studied in~\cite{Blair:1379880}, the scale choice is set to a fraction of the
photon $p_T$. By altering this fraction around the central value of 1.0, it is
hoped to gain an uncertainty on the terms missed in the NLO calculation. The
best selection for this central value would be to be at a `saddle point', where
moving in any direction from this point gives similar changes in the cross
section. However, it is found that as the scale is reduced (in steps: 2.0, 1.0,
0.5, 0.25, 0.1, 0.05 and 0.01) the cross section increases, when moving the two scales coherently or independantly. One difference in
this result to the previous study was that it was carried out in three $p_T$
bins, but the result remained the same for all (only the highest bin was able to be
calculated with a scale of 0.01). Similarly the addition of using Frixione
isolation instead of the standard cone isolation also resulted in the same cross
section behaviour. The summary of this is that there must be large contributions
needed from NNLO. However, on the positive side, in all 3 $p_T$ bins, the
variation between 0.5-1.0-2.0 resulted in differences of similar magnitude
around 1.0, so this is likely a safe estimate of the uncertainty.


After the inclusive photon measurements, the next step experimentally is to
require the addition of at least one jet. Using a jet of $10\rm{GeV}$ the cross
section was calculated for two algorithms each for multiple sizes:
\begin{itemize} 
\item Kt algorithm with $\Delta R=$ 0.3, 0.4, 0.5 or 0.6 
\item Cone with $\Delta R=$ 0.4, 0.5 or 0.6 
\end{itemize}
These choices had an affect of $<1\%$
on the cross section computed in 3 photon $p_T$ bins, suggesting that this will
not increase the error for the NLO calculation when moving from the inclusive cross section to that with an additional jet.


\subsection{FRAGMENTATION PHOTONS IN PARTON-SHOWER MONTE-CARLO}

The second part of this study continues to investigate photon isolation, but now in di-photon events using parton-shower Monte-Carlo 
generators; again the selection used is listed in the appendix. The study begins by investigating the distance between 
the photon and other particles. It then moves into studying several different styles of isolation criteria, including 
Frixione criteria as done in the inclusive NLO studies.

\subsubsection{Topology of fragmentation photons}

We consider three sets of parton-shower Monte-Carlo samples for the $\gamma\gamma$+X process:
\begin{itemize}
\item Pythia~\cite{Sjostrand:2006za} $\gamma\gamma$ Born and Box direct processes, plus the Pythia $\gamma$+jet process 
with the jet fragmenting into a photon (20 million events were generated for the $\gamma$+jet sample and 1000 times 
more would have been needed for the dijet fragmenting to two photons due to the low $q\rightarrow\gamma$ branching 
ratio for isolated photons).
\item Pythia $\gamma\gamma$ Born and Box direct processes, plus the Pythia $\gamma$+jet process with the jet 
fragmenting into a photon and the Pythia dijet process with the two jets fragmenting into photons. Both Pythia 
$\gamma$+jet and dijet samples were generated with a filter which enhances the presence of events with 
isolated electromagnetic particles.
\item Madgraph~\cite{Stelzer:1994ta} $\gamma\gamma$ + up to two supplementary hard jets, with 
fragmentation/hadronization done with Pythia.
\end{itemize}

The fragmentation contribution is included as a bremsstrahlung contribution in Madgraph at matrix element level, 
while it is included as a showering contribution in Pythia $\gamma$+jet and dijet (in the PYTHIA samples we identify 
fragmentation photons as those having a quark or gluon\footnote{Photon radiation directly by a gluon is of course not 
physically possible; the representation as such in PYTHIA is a technical shortcut for the actual physical process} 
as parent). The fragmentation fraction found is compatible with ref~\cite{PhysRevLett.107.102003}.  
We consider additionally the case 
where the two jets fragment into boosted neutral mesons ($\pi^{0}$, $\eta$, $\rho$ and $\omega$) that can experimentally 
mimic direct or fragmentation photons at reconstructed level because of the finite granularity of the detector. These 
samples include an underlying event but were generated without pile-up.

\begin{figure}[hbt]
  \begin{center}
    \resizebox{8cm}{!}{\includegraphics{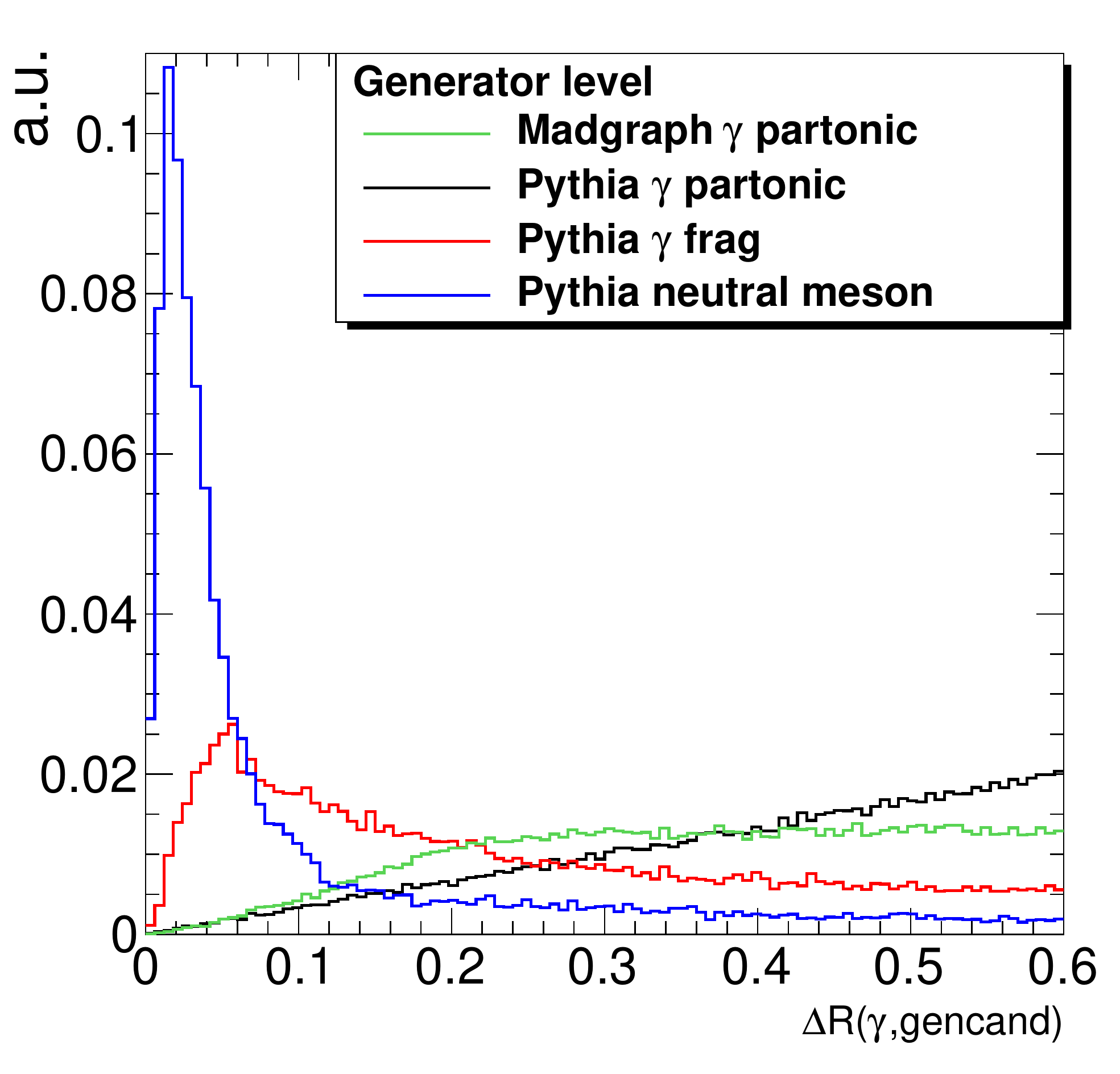}}
    \caption{$\Delta R$ distance distribution between the photon and the other generator-level particle candidates in 
the event, for neutral mesons, fragmentation photons and partonic photons.}
    \label{fig:DeltaRprofile}
  \end{center}
\end{figure}

Fig.~\ref{fig:DeltaRprofile} shows the $\Delta R$ distance between the photon or neutral mesons and the other particle 
candidates in the event. Partonic photons, fragmentation photons and neutral mesons have different properties as a 
function of $\Delta R$. Partonic photons in Pythia have a linear behavior, which is expected because the only 
contribution that can enter in the isolation sum is the underlying event and pile-up (with also a small contribution 
from QCD radiation at the shower level) which is expected to be uniform 
in space. As each bin consists of an annulus with radius growing linearly as a function of $R$, the quantity of particles grows 
linearly with $R$ in the area of the annulus. Neutral mesons have a radically different profile, with a peak of the 
$\Delta R$ distribution close to 0. The peak is caused by the decay of particles resulting from jet fragmentation close 
to the neutral meson direction. Pythia fragmentation photons have a behavior somehow in between that of neutral mesons 
and partonic photons. The peak at low $\Delta R$ is still present but much reduced with respect to that of neutral 
mesons. Madgraph partonic photons exhibit a modulation of the Pythia partonic photon $\Delta R$ distribution, probably 
because Madgraph includes fragmentation as a bremsstrahlung contribution.


From this we can expect that the smaller the $\Delta R$ cone used in Frixione isolation (until $\Delta R \simeq 0.1$), 
the higher the discrimination against the neutral mesons and fragmentation photons. The discrimination against neutral 
mesons is higher than that against fragmentation photons (as is well-known experimentally). This can be seen in
Fig.~\ref{fig:IsolationProfiles}, which shows the isolation sum profile divided by the transverse energy of the photon 
for different cone sizes.

\begin{figure}[hbt]
  \begin{center}
    \resizebox{7cm}{!}{\includegraphics{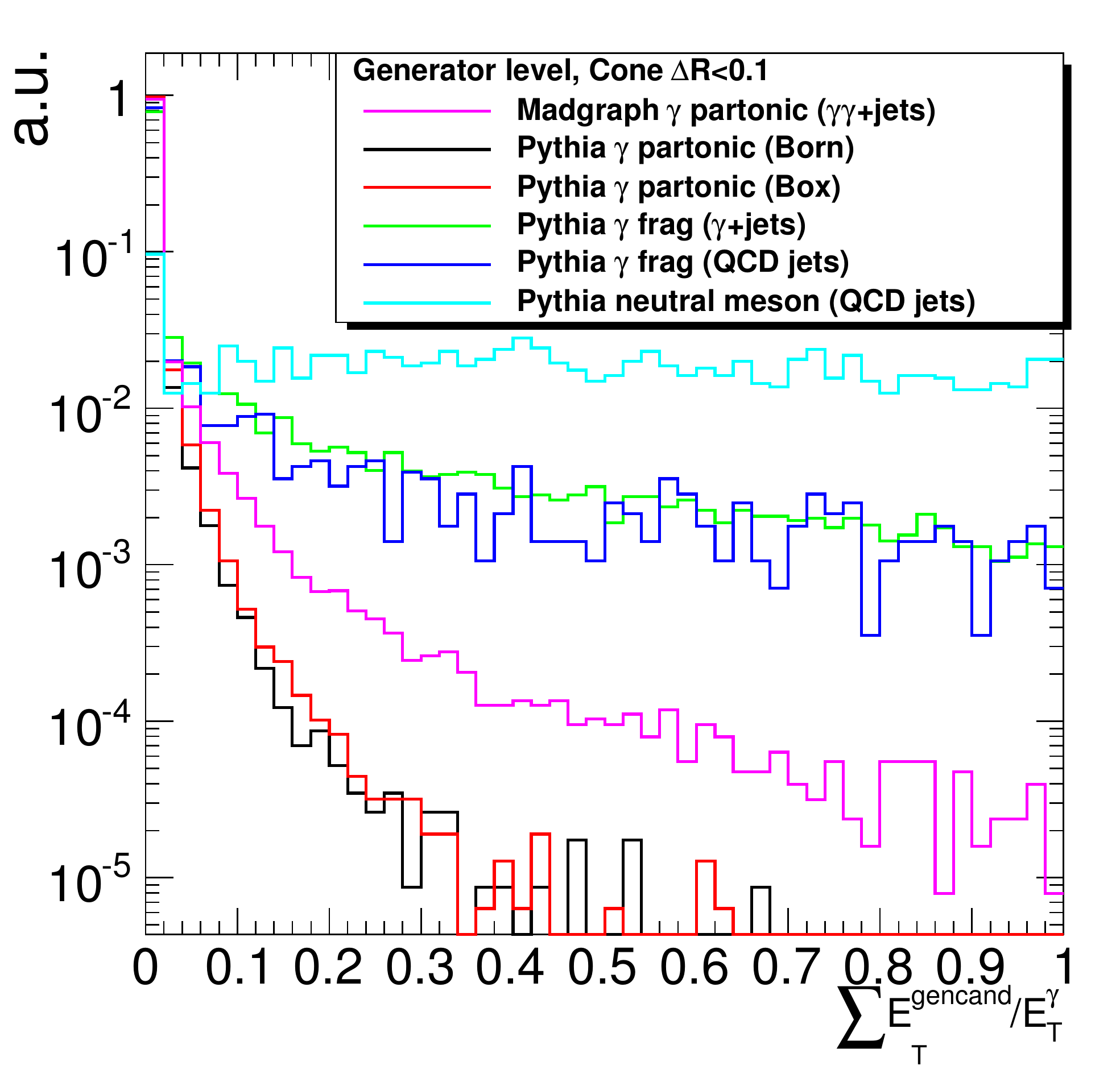}}
    \resizebox{7cm}{!}{\includegraphics{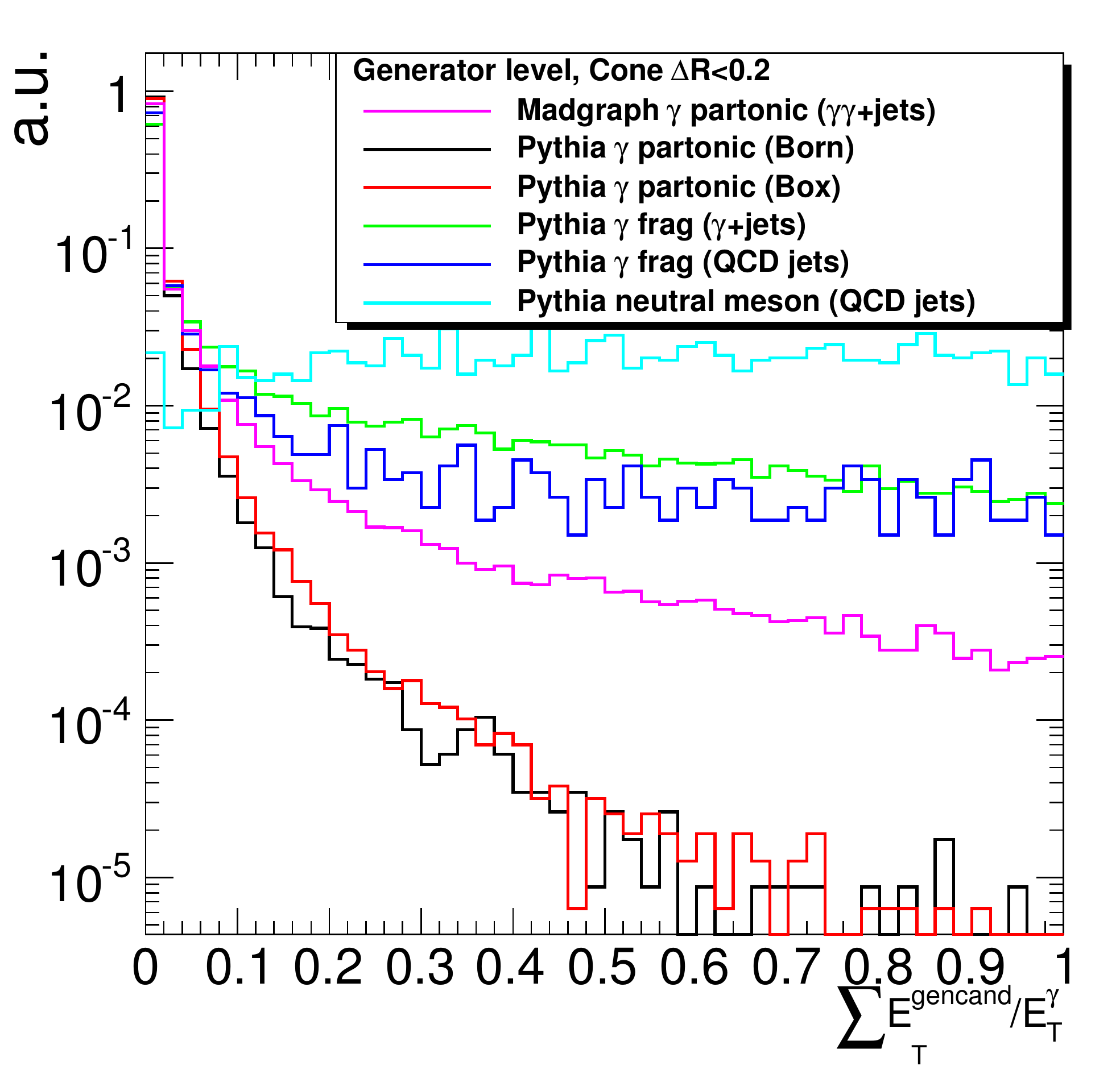}}
    \resizebox{7cm}{!}{\includegraphics{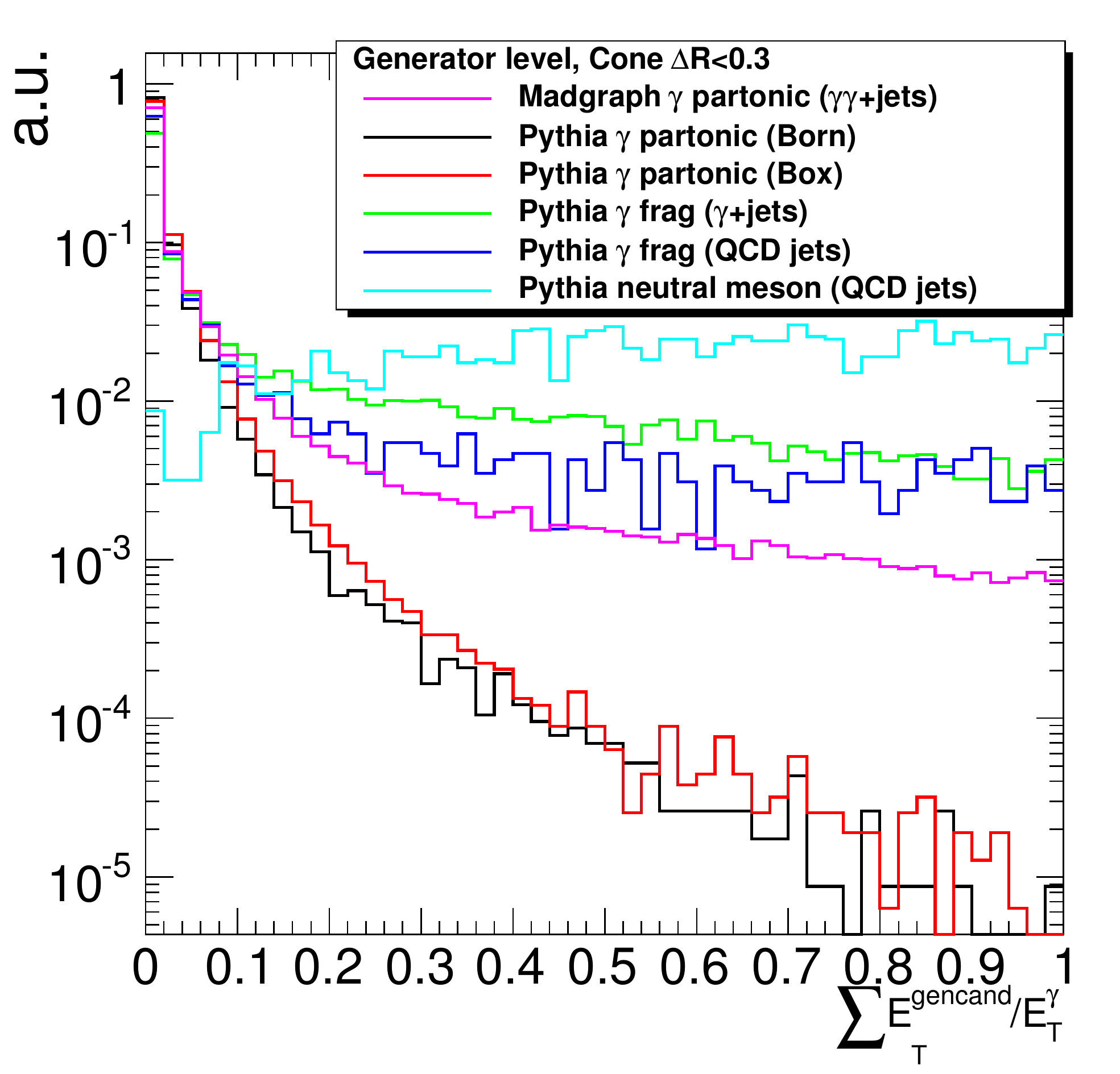}}   
    \resizebox{7cm}{!}{\includegraphics{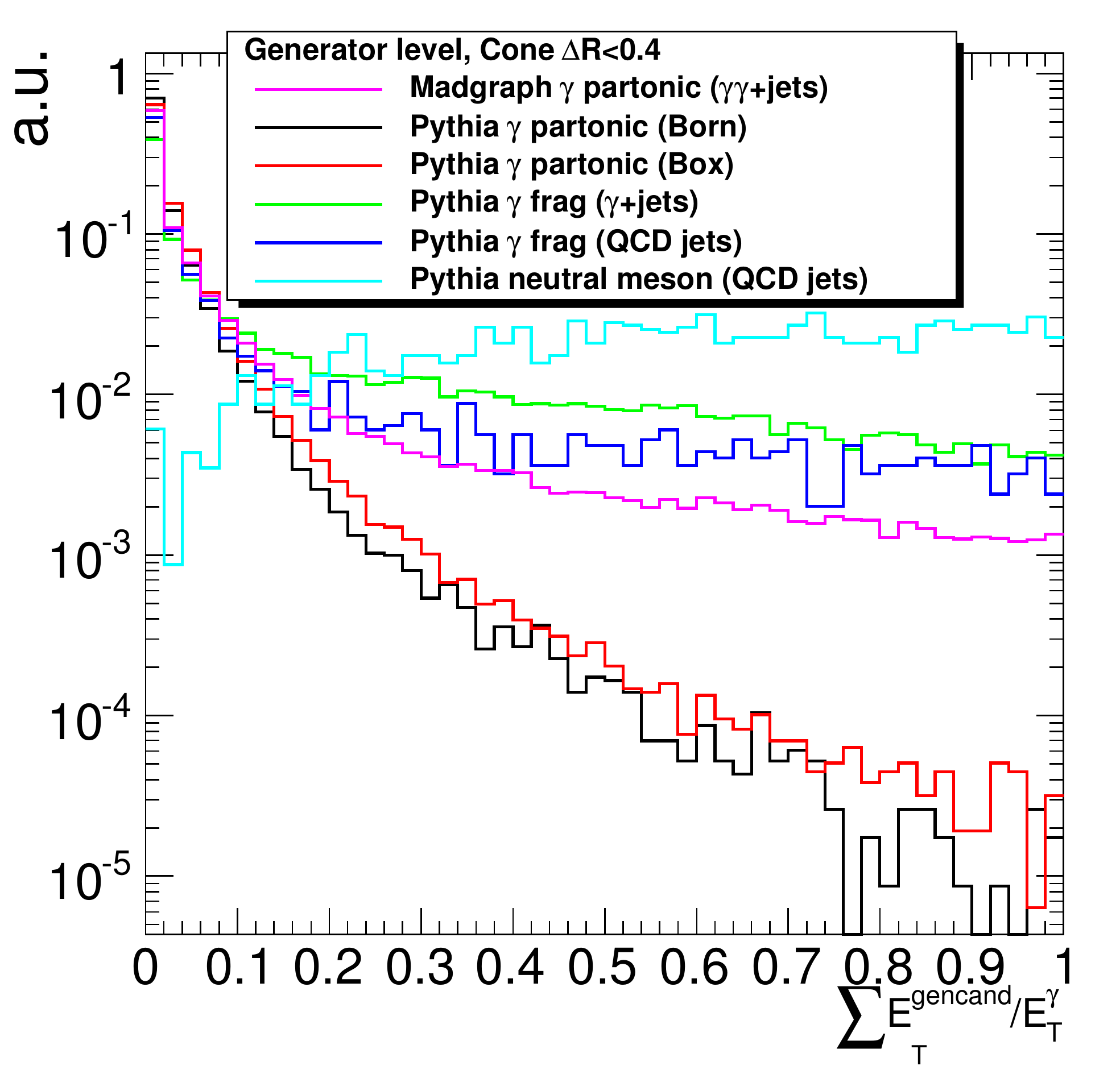}}
    \caption{Isolation sum normalized to the photon energy computed in cones of size $\Delta R<0.1$ (top left), 
$\Delta R<0.2$ (top right), $\Delta R<0.3$ (bottom left), $\Delta R<0.4$ (bottom right),  for neutral mesons, 
fragmentation photons and partonic photons.}
    \label{fig:IsolationProfiles}
  \end{center}
\end{figure}

\subsubsection{Impact of hollow cones on the fragmentation contribution}

In NLO generators, the products of the quark fragmentation are along the fragmentation photon direction. In 
parton-shower generators we have seen that this is not necessarily true. In NLO generators, the ``hollow'' or ``crown'' 
isolation, where the energy sum has to be below a fixed threshold in a region $R_1 < \Delta R < R_2$ while in the 
region $R<R_1$ any arbitrary amount of energy is admitted, has been shown to enhance the fragmentation contribution 
with respect to the usual ``solid'' isolation. This ``hollow'' isolation is interesting also because this criterion is 
closer than the ``solid'' cone to what is used experimentally (it allows to exclude from the isolation sum the energy 
deposited by the photon itself). The first two lines of Tables~\ref{table:FrixionePurityBiased} 
and \ref{table:FrixionePurityUnbiased} show that in general the fragmentation fraction does not increase significantly
when moving from solid to hollow cone isolation, for the PYTHIA samples.



\subsubsection{Impact of Frixione isolation on the fragmentation contribution}

Tables~\ref{table:FrixionePurityBiased} and \ref{table:FrixionePurityUnbiased} report the fragmentation fraction inside 
the Pythia two-prompt sample for different Frixione isolation criteria. Eight isolation cones were used : 
$\Delta R<0.05$, $\Delta R<0.1$, $\Delta R<0.15$, $\Delta R<0.2$, $\Delta R<0.25$, $\Delta R<0.3$, $\Delta R<0.35$, 
$\Delta R<0.4$. The results are almost identical if instead four cones are used (0.1, 0.2, 0.3, 0.4), as found at NLO. 
The tables show that in both the electromagnetically-enriched samples and non-enriched samples, the discrete Frixione 
isolation with the usual functional form $f(R)$ does not reduce the fragmentation contribution with respect to the 
standard isolation criterion (with a cone $\Delta R < 0.4$) except when the parameter $\epsilon$ is at its smallest 
value, $\epsilon=0.05$, for which modest reductions of between 5 and 8\% can be achieved. The cause of this apparent
non-optimal behavior can be explained by 
the non-collinearity of the fragmentation debris around the fragmentation photon in PYTHIA. Frixione isolation is designed to 
apply tighter and tighter isolation criteria $E_T^{iso} < f(R) \rightarrow 0$ as $\Delta R \rightarrow 0$, assuming 
that most of the fragmentation debris are around $\Delta R \simeq 0$. As it is seemingly not the case in the 
parton-shower Monte-Carlo studied here, the criterion loses most of its discrimination power.

\begin{table}
\begin{center}
\caption{\label{table:FrixionePurityBiased}Fraction represented by the 1-fragmentation and 2-fragmentation 
contributions for various Frixione isolation criteria in Pythia two-prompt photon samples (with electromagnetic 
enrichment filter).}
\begin{tabular}{|c|c|c|c|}
\hline
\textbf{Criteria} & \textbf{1-frag fraction} & \textbf{2-frag fraction} & \textbf{1,2-frag fraction}\\
\hline
\hline
Solid $\Delta R<0.4$, $E_T^{iso}<5$ GeV & 0.335 & 0.157 & 0.492\\
\hline
Hollow $0.1<\Delta R<0.4$, $E_T^{iso}<4$ GeV & 0.337 & 0.168 & 0.505 \\ 
\hline
\hline
Frixione 8 cones, $E_T=20$ GeV, $\epsilon=0.05$, $n=1.0$ & 0.322 & 0.145 & 0.467\\
\hline
Frixione 8 cones, $E_T=20$ GeV, $\epsilon=0.05$, $n=0.2$ & 0.318 & 0.147 & 0.466\\
\hline
Frixione 8 cones, $E_T=20$ GeV, $\epsilon=1.0$, $n=0.2$ & 0.372 & 0.228 & 0.599\\
\hline
Frixione 8 cones, $E_T=20$ GeV, $\epsilon=1.0$, $n=0.1$ & 0.374 & 0.232 & 0.601\\
\hline
Frixione 8 cones, $E_T=20$ GeV, $\epsilon=1.0$, $n=1.0$ & 0.353 & 0.192 & 0.545\\
\hline
Frixione 8 cones, $E_T=20$ GeV, $\epsilon=1.0$, $n=0.5$ & 0.365 & 0.212 & 0.577\\
\hline
Frixione 8 cones, $E_T=20$ GeV, $\epsilon=0.5$, $n=1.0$ & 0.343 & 0.176 & 0.518\\
\hline
\end{tabular}
\end{center}
\end{table}

\begin{table}
\begin{center}
\caption{\label{table:FrixionePurityUnbiased}Fraction represented by the 1-fragmentation contribution for various 
Frixione isolation criteria in Pythia two-prompt photon samples (without enrichment filter).}
\begin{tabular}{|c|c|c|c|}
\hline
\textbf{Criteria} & \textbf{1-frag fraction}\\
\hline
\hline
Solid $\Delta R<0.4$, $E_T^{iso}<5$ GeV & 0.455 \\
\hline
Hollow $0.1<\Delta R<0.4$, $E_T^{iso}<4$ GeV & 0.458 \\ 
\hline
\hline
Frixione 8 cones, $E_T=20$ GeV, $\epsilon=0.05$, $n=1.0$ &  0.420\\
\hline
Frixione 8 cones, $E_T=20$ GeV, $\epsilon=0.05$, $n=0.2$ & 0.419\\
\hline
Frixione 8 cones, $E_T=20$ GeV, $\epsilon=1.0$, $n=0.2$ & 0.514\\
\hline
Frixione 8 cones, $E_T=20$ GeV, $\epsilon=1.0$, $n=0.1$ &0.519 \\
\hline
Frixione 8 cones, $E_T=20$ GeV, $\epsilon=1.0$, $n=1.0$ & 0.489\\
\hline
Frixione 8 cones, $E_T=20$ GeV, $\epsilon=1.0$, $n=0.5$ & 0.503\\
\hline
Frixione 8 cones, $E_T=20$ GeV, $\epsilon=0.5$, $n=1.0$ & 0.465\\
\hline
\end{tabular}
\end{center}
\end{table}

The previous study suggests that the previous working points studied with the Frixione functional form $f(R)$ might
not be optimal for the rejection of fragmentation debris. In figure~\ref{FracFragVsDirectGammaEff} we compare the performance of three different sets of criteria: 
1) non-Frixione isolation in a single cone $\Delta R<0.4$, 2) optimized working points for the parameters in the 
Frixione functional form (four cones 0.1, 0.2, 0.3, 0.4 were used to make the algorithm converge faster), 
3) re-optimized 'Frixione' isolation criteria on cones $\Delta R<0.1,0.2,0.3,0.4$ without using the explicit functional 
form (we no longer constrain the events to satisfy $E_T^{iso}<f(R)$ and let $f(R)$ free). In the second case, an 
optimization procedure is performed scanning over the parameters $\epsilon$ and $n$ to find the best working points 
(corresponding to a maximum efficiency for a given $s/b$). In the last case, an optimization code is used to find the 
best selection criteria to be applied on $E_T^{iso}$ for each $\Delta R$ cone. The optimization takes as input the 
target value of $s/b$ (partonic signal over fragmentation background ratio), then relaxes and tightens each cut 
separately with an iterative procedure to find the best signal efficiency for this $s/b$ target. The procedure was 
performed to find the working points corresponding to the $s/b$ obtained with the first Frixione criterion.

Figure~\ref{FracFragVsDirectGammaEff} shows that the optimized working points for the Frixione functional form perform 
slightly better than the standard isolation for a given photon efficiency, and that optimization using no functional 
form in turn performs slightly better than the Frixione functional form; for the same value of single-photon efficiency,
lower values of fragmentation fraction are attainable. It 
should be noted that this optimisation leads to a looser cut on the first cone, 
$\Delta R < 0.1$, than the usual functional form. Nevertheless, to obtain reductions in the fragmentation fraction
of more than 10\%,  increasingly significant reductions in single photon efficiency are required, since
the fragmentation reduction becomes nearly flat.

All in all, with the definition of the isolation in a cone of $\Delta R$ used here, which is a usual way of defining 
isolation at the experimental level (where one has however to remove the footprint of the photon from the isolation sum 
and to cope with pile-up), rejecting fragmentation photons can be done only at a cost of a lowered signal efficiency. 
 With this optimization procedure it was found 
that to decrease the fragmentation fraction by 10\%, a signal loss of about 60\% has to be achieved, leading to 
extremely tight cuts probably not applicable in experimental analysis.

\begin{figure}[hbt]
  \begin{center}
    \resizebox{7cm}{!}{\includegraphics{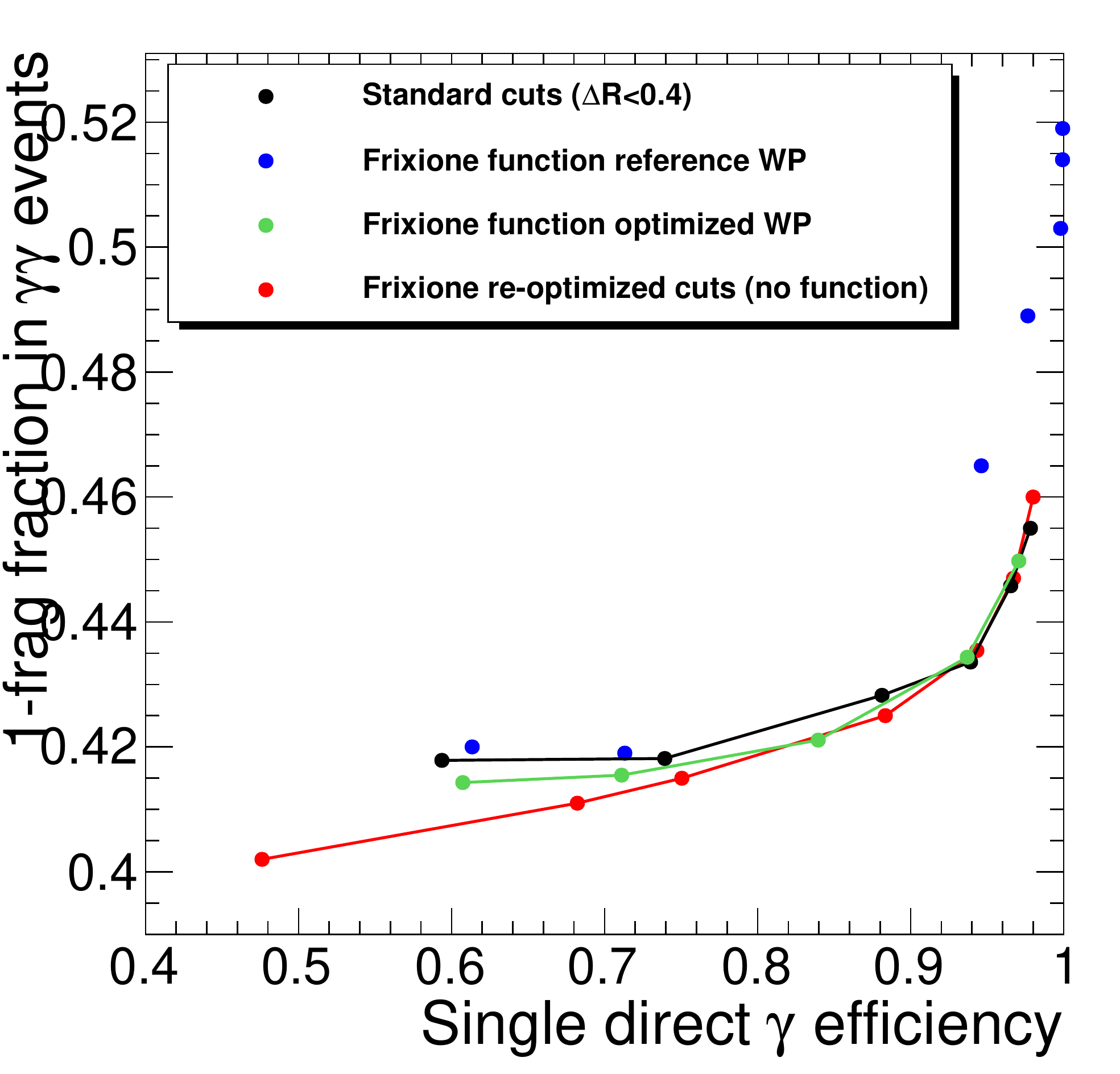}}
    \caption{The fraction of the 1-fragmentation contribution vs single $\gamma$ efficiency for various sets of criteria. 
Blue : tested working points reported on table~\ref{table:FrixionePurityUnbiased}. Black : selection criteria on 
isolation in $\Delta R<0.4$. Green : Optimized Frixione isolation using the usual functional form. 
Red : Re-optimized isolation criteria on cones $\Delta R<0.1,0.2,0.3,0.4$ without using the functional form.}   
 \label{FracFragVsDirectGammaEff}
  \end{center}
\end{figure}

\subsection*{CONCLUSIONS}

Firstly for the NLO cross sections, it was found that only
one of the Frixione isolation criteria suggested in~\cite{Binoth:2010ra}
actually performs better at removing the fragmentation contribution in the inclusive case than that
of the standard cone, although potentially too tight to use experimentally.
However, it is useful to see that the results are independent of the number
of cones used. This is also the case for the di-photon cross section where it compares well to the continuous criteria. 
A more promising result in the inclusive case is that the generalized version of Frixione isolation, with small 
values of $\epsilon$, do significantly reduce the fragmentation fraction without applying too tight a cut in the 
smallest cone. In addition it is confirmed that the `saddle point' can not be
found when altering the scale choice for the inclusive cross-section, suggesting
more corrections needed at NNLO. Finally when moving to the photon with
associated jet cross section, from the inclusive cross section, it is reported
that the jet algorithm (or size) used at NLO makes little difference to the
cross section.

In the study at the parton-shower level it was shown that the isolation profile of fragmentation photons is no 
longer collinear anymore, whereas the modelization of the fragmentation function in NLO generators leads the quark/gluon 
debris to be collinear to the photon. Furthermore, the hadronization process of the quark/gluon that emitted the 
photon leads to a $\Delta R$ profile which is no longer peaked at zero (but close to zero). In parton-shower 
programs isolation still has increasing discriminating power when going lower in $\Delta R$. 
However, it was found that a 10\% decrease in fragmentation fraction in di-photon events with respect to standard isolation leads to a drop in single photon 
signal efficiency to approximately 60\% of the initial value. The usual functional form for Frixione isolation was 
shown to be not completely optimal for suppressing the fragmentation contribution while preserving high signal 
efficiency. This can be mitigated by re-optimizing the cuts for each $\Delta R$ of the discrete Frixione prescription, which 
allows a looser cut in the innermost cone. Further studies using other fragmentation modelizations in 
parton-shower programs like SHERPA~\cite{Hoeche:2009xc} (LO matrix-element where photons and jets in the shower are 
matched to matrix element level) or POWHEG~\cite{D'Errico:2011sd} (NLO matrix-element with consistent fragmentation 
photon matching) would need to be investigated.

In conclusion the results from the two studies show there are differences and similarities at the two levels. 
Regarding the fragmentation fraction, this is far more reduced at the NLO level than at the parton-shower level. 
However, the two levels show agreement that the results from Frixione isolation are independant of the number of cones 
used and that similar shape cuts can be obtained by retuning cuts at the parton-shower level and by using the 
generalized prescription at NLO.

\subsection*{ACKNOWLEDGEMENTS} 
Thanks to the whole of the Les Houches photon subgroup, especially to 
J.P. Guillet and E. Pilon for their expert input.


\subsection*{APPENDIX: Selection details}

All of the studies are carried out for $pp$ collisions at $\sqrt{S} = 7$ TeV. For simplicity the the inclusive studies are carried out for the region where the photon lies in $|\eta|<0.6$.
When calculating the cone isolation around the photon a cone of 0.4 is used with the requirement that the energy in the cone is 
less than $4\rm{GeV}$. The renormalization scale 
$\mu$ and initial state factorization scale $M$ are set to the photon $p_T$, unless stated otherwise, and the CTEQ6.6 PDF\cite{Nadolsky:2008zw} is used and the photon 
fragmentation functions are BFG set II \cite{Bourhis:1997yu}. For the generalized Frixione isolation case, the cones used are: R = 0.4, 0.35, 0.3, 0.25, 0.2, 0.15 and 0.1.

In the NLO di-photon studies, the photons have: 
$p_{T}^{\gamma \, 1} \leq 25$ GeV, $p_{T}^{\gamma \, 2} \geq 22$ GeV, in the rapidity range $| \eta^{\gamma} | \leq 2.5$ for both photons, and a separation  
$\Delta R_{\gamma \gamma} \geq 0.4$ is required between the two photons. 
The mass range considered is $40$ GeV $\leq m_{\gamma \gamma} \leq 300$ GeV.  
In this case the scales $\mu$ and $M$ are chosen equal to
$\mbox{min} \, \{p_{T}^{\gamma \, 1}, p_{T}^{\gamma \, 2}\}$. 

In the parton shower di-photon studies, the photons are selected with: $M_{\gamma\gamma}>80$ GeV, $p_T>$21,20 GeV, $|\eta|<$2.5 and $E_{T}^{iso}<5$ GeV.

}

\section[Event-by-event pileup subtraction using jet areas]
{EVENT-BY-EVENT PILEUP SUBTRACTION USING JET AREAS \protect\footnote{Contributed by: M.~Cacciari, G.~P.~Salam and G.~Soyez}}
{\graphicspath{{pusub/}}

\newcommand{\ie}{{\it i.e.}\ }
\newcommand{\eg}{{\it e.g.}\ }
\newcommand{\rhoest}{\ensuremath{\rho_{\rm est}}\xspace}
\newcommand{\Dpt}{\ensuremath{\Delta p_t}\xspace}

\title{Event-by-event pileup subtraction using jet areas}

\author{Matteo Cacciari,$^{1,2}$ Gavin P. Salam$^{3,4,1}$ and Gregory Soyez$^{5}$}
\institute{\normalsize
  $^1$LPTHE, UPMC Univ.~Paris 6 and CNRS UMR 7589, Paris, France\\
  \normalsize
  $^2$Universit\'e Paris Diderot, Paris, France\\
  \normalsize
  $^3$CERN, Physics Department, Theory Unit, Geneva, Switzerland\\
  \normalsize
  $^4$Department of Physics, Princeton University, Princeton, NJ 08544,USA\\
  \normalsize
  $^5$Institut de Physique Th\'eorique, CEA Saclay, France}


\begin{abstract}
  In these proceedings, we compare the efficiency of several
  jet-area-based subtraction methods to correct for pile-up
  contamination at hadronic colliders. We study the dependence on
  various variables like the $p_t$ and rapidity of the jets, the
  number of pile-up vertices or the Monte-Carlo generator
  variations. We conclude that estimations of the pile-up density
  using a median computed over grid-cell patches, including a
  rescaling to correct for the rapidity dependence, perform
  particularly well, though alternative methods are possible.
\end{abstract}

\subsection{Introduction}\label{sec:pusub_intro}

With the LHC running at larger and larger luminosities, hard $pp$
interactions are accompanied by an increasing number of pile-up (PU)
collisions: from a few PU events per bunch crossing in spring 2011,
operation with $\sim 20$ PU events is now routine. Considering only
in-time PU, this would lead to an extra transverse momentum of $\sim 750$
GeV deposited in the event, and a jet of a typical radius $R=0.5$
would see its transverse momentum shifted by $\sim 10$ GeV. In order
to obtain a good energy resolution for the jets it is therefore
mandatory to correct for this contamination.

In these proceedings, we review several methods --- both existing
methods and new refinements --- to subtract the contamination due to
PU and provide a systematic study of their efficiency.

It is important to note already now that PU has not only the effect to
shift the momentum of the jets: it also smears their momentum. Indeed,
the number of PU vertices varies from one collision to the next
(following a Poisson distribution varying with the beam conditions),
all PU interactions, \ie minimum bias collisions, do not lead to the
same energy deposit, and finally, the energy produced in a minimum
bias collision is not deposited uniformly across the
detector. Altogether, on top of an average shift, PU will add two
sources of resolution smearing: {\em event-to-event} and {\em
  in-event} fluctuations corresponding respectively to variations of
the PU activity from one event to another and from one point to
another in a single event.

Here we shall primarily study {\it in-time PU}, that is the effects
coming from multiple $pp$ interactions that occur in the same bunch
crossing as the hard interaction one triggers on. Because of the
response time inherent to each detector this would come with a second
effect, {\it out-of-time PU}, corresponding to the PU activity in the
few bunch crossings preceding the one with the hard
interaction. Since these heavily depend on the details of each
individual detector --- and even varies from one sub-detector to another
--- it goes beyond the scope of this theoretical study. However, as we
shall discuss in further detail later on, the PU subtraction methods
proposed here do not make any assumption about a distinction between
in-time and out-of-time PU and thus should be robust enough in more
complex cases.

\subsection{Subtraction method(s)}\label{sec:pusub_methods}

We are interested in the situation where a {\em hard event} is
contaminated by a {\em background} coming from additional pileup
interactions. A reconstructed jet in that {\em full event} (hard
event + background), which we shall call a {\em full jet}, differs
from the {\em hard jet} in the original hard event because of the
presence of the background. By {\em background subtraction}, we mean
correcting the full jet in such as to recover the momentum of the
original hard jet, \ie subtract the pileup contamination from the
jet's momentum.

\subsubsection{Background effects}\label{sec:bkg_effects}

Our starting point is to realise \cite{Cacciari:2007fd} that a uniform
background affects the momentum of a jet in two ways: it
shifts its momentum because of the background particles clustered with
the jet, and it modifies the way the hard particles themselves are
clustered because the background particles are not infinitely soft.

This means that the reconstructed momentum has the form\footnote{This
  can be defined for the 4-momentum of the jet but we shall only
  discuss its transverse momentum for simplicity.}
\begin{equation}\label{eq:effects}
p_{t,full} = p_{t, \rm hard} + \rho A \pm \sigma \sqrt{A} + \Delta p_t^{BR}
\end{equation}
where $p_t$ denotes the transverse momentum of the reconstructed jet,
$p_{t, \rm hard}$ the momentum of the original hard jet (in the
absence of PU), $A$ the jet area, $\rho$ the background density per
unit area within a given event, $\sigma$ the fluctuations of that
background (per unit area) from place to place within the event, and
$\Delta p_t^{BR}$ the back-reaction describing the effect of the
background particles on the clustering of the hard ones.

If the background has a positional dependence (\eg depends on
rapidity) then $\rho$ and $\sigma$ will depend on the position of the
jet one tries to subtract.

Eq.~(\ref{eq:effects}) characterises the fact that the background has
the effects of shifting the transverse momentum of the jet and to
degrade its resolution. The shift comes from the ``$\rho$'' term in
(\ref{eq:effects}) and from potential back-reaction systematic
effects. Using the anti-$k_t$ jet algorithm the shift due to
back-reaction is negligible\footnote{For the $k_t$ or Cambridge/Aachen
  algorithms, it is usually negative and can be of the order of a
  GeV.}. Resolution smearing effects come from various sources: the
fluctuations of the background from within an event, \ie the
``$\sigma$'' term in (\ref{eq:effects}), fluctuations of the
background from one event to another, that is the fact that $\rho$ is
not the same in every event, and the fluctuations in the
back-reaction.

\subsubsection{Central subtraction formula}\label{sec:pusub_formula}

From (\ref{eq:effects}), the natural way to subtract the background
contamination is to define the {\em subtracted jet} as
\cite{Cacciari:2007fd}
\begin{equation}\label{eq:subtract}
p_{t,\rm sub} = p_t - \rhoest A
\end{equation}
where \rhoest is the estimated value for the background density per
unit area. 

To apply this subtraction we need to compute the jet area and find an
estimation \rhoest for the background density per unit area. 
The jet areas are readily available using FastJet, so we just need to
focus on \rhoest. The main goal of these proceedings is to investigate
various methods of obtaining \rhoest which are listed below. In all
cases, it is primordial to realise that the determination of \rhoest
is performed event-by-event, and even jet-by-jet when the positional
dependence of the background is taken into account.

As we shall see later on, the fact that $\rho$ is estimated for each
individual event is crucial: it corrects for the fluctuations of the
background from one event to another. If instead one uses an averaged
value for \rhoest (over many events), one would get an extra
resolution smearing due to the fluctuations of $\rho$ across different
events. Similarly, the jet area $A$ in (\ref{eq:subtract}) has to be
computed for each individual jet. Using an average area would lead to
an additional source of fluctuations of the form $\rho \sqrt{\langle
  A^2\rangle-\langle A\rangle^2}$.

\paragraph{Using {\em seen vertices}} Since experimentally it might be
possible --- within some level of accuracy that goes beyond the scope
of this discussion --- to count the number of pileup vertices using
charged track reconstruction, one appealing way to estimate the
background density in a given event would be to count these vertices
and subtract a pre-determined number for each of them:
\begin{equation}\label{eq:npuseen}
\rhoest^{(n{\rm PU})}(y) = f(y)\,n_{\rm PU,seen},
\end{equation}
where we have made explicit the fact that the proportionality constant
$f(y)$ can carry a rapidity dependence. $f(y)$ can be studied from
minimum bias collisions (see Section \ref{sec:rho_rapdep} below) and
can take into account the fact that only a fraction of the PU vertices
will be reconstructed.

\paragraph{Median subtraction} This technique divides the
rapidity-azimuthal angle plane in patches and estimates $\rho$ for
each event using
\begin{equation}\label{eq:median}
\rhoest^{\rm (global)} = \underset{i\in\, \rm patches}{\rm median}
  \left\{\frac{p_{t,i}}{A_i}\right\}
\end{equation}
This is motivated by the observation that many regions in the event
are populated just by the background. In these regions, $p_t/A$ is an
estimate of $\rho$ and the use of the median, rather than the average,
which ensures reduced bias from the hard jets.

This method was originally proposed in \cite{Cacciari:2007fd} using
jets (from a $k_t$ or Cambridge/Aachen clustering) as patches. Here,
we shall also test a new option where the $y-\phi$ plane is simply
subdivided into grid cells that we use as patches.

\paragraph{Using a local range} Eq. (\ref{eq:median}) provides a
unique, global, estimate of $\rho$ for the event but does not take
into account the positional-dependence of the background. One option,
assuming one wants to estimate $\rho$ at the location of a jet $j$, is
to limit the computation of the median to the jets in the vicinity of
$j$, that is\footnote{This option will only be considered in the case
  where jets are used as patches.}
\begin{equation}\label{eq:localrange}
\rhoest^{\rm (local)}(j) = \underset{\text{jets }i\in{\cal R}(j)}{\rm median}
  \left\{\frac{p_{t,i}}{A_i}\right\}
\end{equation}
where ${\cal R}(j)$ is a {\em local range} around $j$. A typical
example, that we shall study later on, is the case of a {\em strip
  range} where only the jets with $|y-y_j|<\Delta$ are included. This
option was already proven to be powerful in \cite{Cacciari:2010te}.

\paragraph{Using rescaling} Another option to correct for the rapidity
dependence of the background\footnote{The same technique should also
  work for the azimuthal-angle dependence of the underlying-event in
  heavy-ion collisions.} is to introduce a pre-computed
rapidity-reshaping function $f(y)$ (see Section \ref{sec:rho_rapdep})
and use
\begin{equation}\label{eq:rescaled}
\rhoest^{\rm (resc.)}(y) = f(y) \underset{i\in\, \rm patches}{\rm median}
  \left\{\frac{p_{t,i}}{A_if(y_i)}\right\}
\end{equation}
where now all patches (jets or grid cells) are included in the
computation of the median.

\subsection{Performance tests}\label{sec:performance-tests}

\subsubsection{Testing framework}\label{sec:testing-framework}

The remainder of these proceedings will be devoted to an in-depth
comparison of the subtraction methods proposed in Section
\ref{sec:pusub_methods}. Our testing framework will be very similar to
the one used in \cite{Cacciari:2010te}: we embed a hard event into a
pileup background (see again Section \ref{sec:pusub_methods}, we
reconstruct and subtract the jets in both the hard and full
events\footnote{One may argue whether or not one should subtract the
  jets in the hard event. We decided to do so to cover the case where
  the hard event contains Underlying Event which, as a relatively
  uniform background, will also be subtracted together with the
  pileup.}, for each jet in the hard event, we find the matching jet
in the full event and compute the shift
\begin{equation}\label{eq:deltapt}
\Dpt = p_t^{\rm full,sub} - p_t^{\rm hard,sub},
\end{equation}
\ie the difference between the reconstructed-and-subtracted jet with
and without pileup. A positive (resp. negative) \Dpt would mean that
the PU contamination has been underestimated (resp. overestimated).

Though in principle there is some genuine information in the complete
\Dpt distribution --- \eg it could be useful to deconvolute the extra
smearing brought by the pileup, see \eg \cite{Cacciari:2010te} and
\cite{Jacobs:2010wq} --- we shall focus on two simpler quantities: the
average shift $\langle\Dpt\rangle$ and the dispersion
$\sigma_{\Dpt}$. While the first one is a direct measure of how well
one succeeds at subtracting the pileup contamination on average, the
second quantifies the remaining effects on the resolution. One thus
wishes to have $\langle\Dpt\rangle$ close to 0 and $\sigma_{\Dpt}$ as
small as possible. Note that these two quantities can be studied as a
function of variables like the rapidity and transverse momentum of the
jets or the number of pileup interactions. In all cases, a flat
behaviour would indicate a robust subtraction method.

The robustness of our conclusions can be checked by varying many
ingredients:
\begin{itemize}
\item one can study various hard processes with the hope that the PU
  subtraction is not biased by the hard event. In what follows we
  shall study dijets with $p_t$ ranging from 50 GeV to 1 TeV, as well
  as fully hadronic $t\bar t$ events as a representative of busier
  final states.
\item The Monte-Carlo used to generate the hard event and PU can be
  varied. For the hard event, we have used Pythia 6.4.24
  \cite{Sjostrand:2006za} with the Perugia 2011 tune, Pythia 8.150
  with tune 4C \cite{Sjostrand:2007gs} and Herwig 6.5.10
  \cite{Corcella:2002jc} with the ATLAS tune and we have switched
  multiple interactions on (our default) or off. For the minimum bias
  sample used to generate PU, we have used Pythia 8, tune 4C, and
  checked that our conclusions remain unchanged when using Herwig++
  \cite{Gieseke:2011na} (tune LHC-UE7-2).
\end{itemize}

\paragraph{Additional details of the analysis} For the sake of
completeness, we list here the many other details of how the \Dpt
analysis has been conducted: we have considered particles with $|y|\le
5$ with no $p_t$ cut or detector effect; jets have been reconstructed
with the anti-$k_t$ algorithm with $R=0.5$ keeping jets with $|y|\le
4$; for area computations, we have used active areas with explicit
ghosts with ghosts placed\footnote{Note that we have used the ghost
  placement of FastJet 3 which differs slightly from the one in v2.4.}
up to $|y|=5$; for jet-based background estimations, we have used the
$k_t$ algorithm with $R=0.4$ though other options will be discussed
(and the 2 hardest jets in the set have been excluded from the median
computation to reduce the bias from the hard event); for grid-based
estimations the grid extends up to $|y|=5$ with cells of edge-size
0.56 (other sizes will be investigated); for estimations using a local
range, a strip range of half-width 1.5 has been used and we refer to
the Section \ref{sec:rho_rapdep} below for more information about the
rapidity rescaling. Jet reconstruction, area computation and
background estimation have all been carried out using FastJet (v3)
\cite{Cacciari:2005hq,Cacciari:2011ma}. Pile-up is generated as a
superposition of a Poisson-distributed number of minimum bias events
and we will vary the average number of pileup interactions. We shall
always assume $pp$ collisions with $\sqrt{s}= 7$ TeV. Finally, the
matching of a full jet to a hard jet is made by requiring that their
common constituents contribute for at least 50\% of the transverse
momentum of the hard jet. We shall not discuss matching efficiencies
here but they are extremely good: for a reconstructed (full) jet of 50
GeV and 20 PU events, the matching efficiency is 99.9\% and this
increases to 99.98\% for $p_t\ge 50$ GeV and 5 PU events and 99.995\%
for $p_t\ge 100$ GeV and 20 PU events.

\subsubsection{Minimum bias and rapidity shape}\label{sec:rho_rapdep}
\begin{figure}
\centerline{
\includegraphics[angle=270,width=10.0cm]{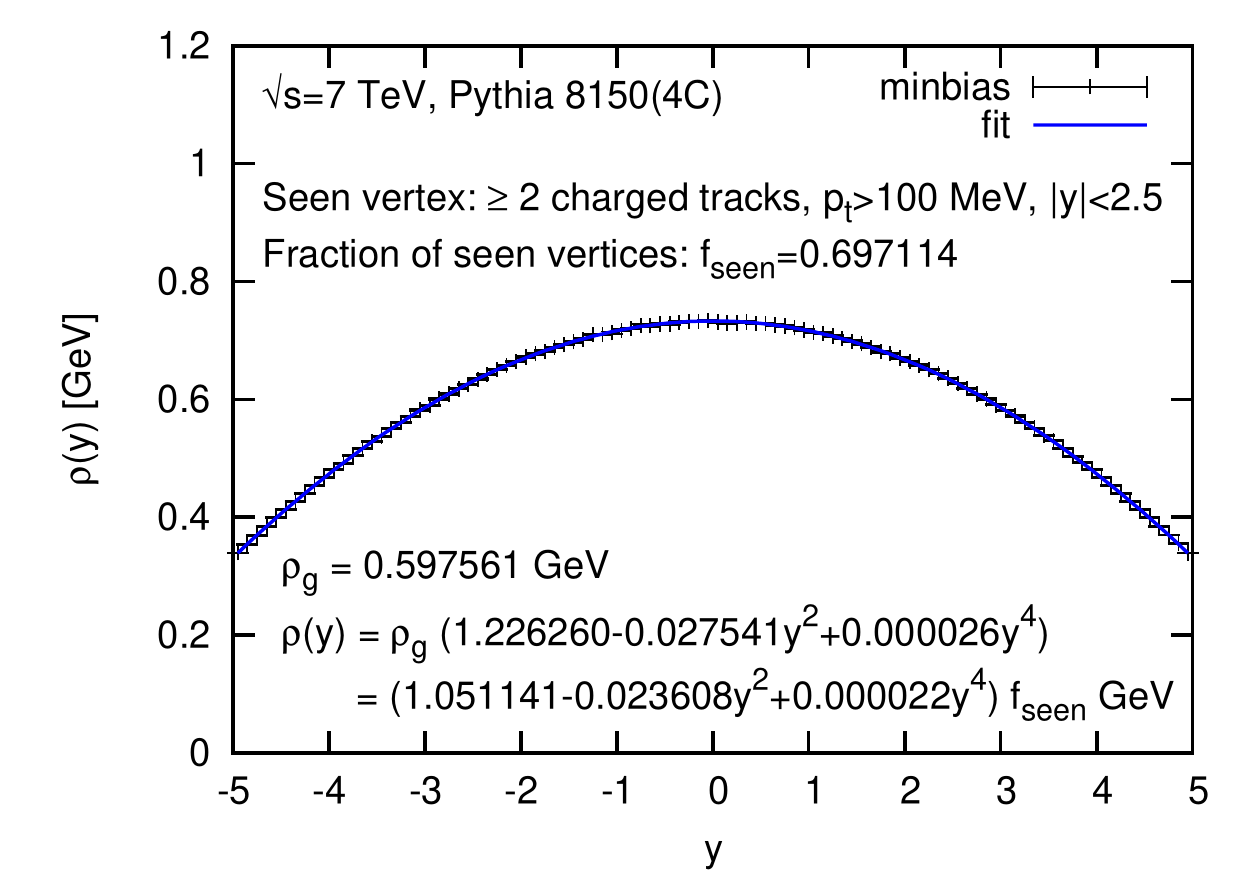}
}
\caption{Rapidity dependence of the transverse energy per unit area
  deposited in minimum bias events (obtained from Pythia 8, tune
  4C). The normalisation of the fit is such that $f_{\rm seen}$ is the
  fraction of {\em seen} minimum bias events \ie the fraction of
  events which have at least 2 charged tracks with $|y|\le 2.5$ and
  $p_t\ge 100$ MeV.}\label{fig:pusub_minbias_rapdep}
\end{figure}

Before discussing the performances of the subtraction methods
described in Section \ref{sec:pusub_methods}, there is still a
building block that has to be discussed, namely the rapidity
dependence of the background $f(y)$ that enters in
Eqs. (\ref{eq:npuseen}) and (\ref{eq:rescaled}). Letting aside the
question of in-time vs. out-of-time PU and non-linear effects in the
detectors, the shape $f(y)$ can be obtained directly from minimum
bias events. 

In our case, we have generated minimum bias events with Pythia 8 (tune
4C) and studied the rapidity dependence of the transverse momentum
deposited per unit area. The result is shown on
Fig. \ref{fig:pusub_minbias_rapdep} together with a quartic fit. If
$f(y)$ is used to rescale median-based estimates of $\rho$,
Eq. (\ref{eq:rescaled}), any global normalisation factor would cancel,
but in the case of Eq. (\ref{eq:npuseen}) \ie for the ``seen
vertices'' method, the normalisation has to match what we mean by a
{\em seen} PU vertex. In what follows, we shall define that as a
minimum bias interaction that has at least 2 charged tracks with
$|y|\le 2.5$ and $p_t\ge 100$ MeV, which corresponds to 69.7\% of the
events\footnote{This is a bit optimistic but does not affect in any
  way our discussion.}. In these conditions, we have found that the
rapidity dependence is well reproduced by
\begin{equation}\label{eq:mb_rapdep}
f(y) = 1.051141-0.023608\, y^2+0.000026\, y^4.
\end{equation}

\subsubsection{Generic performance and rapidity dependence}\label{sec:rapdep}

\begin{figure}
\centerline{
\includegraphics[angle=270,width=\textwidth]{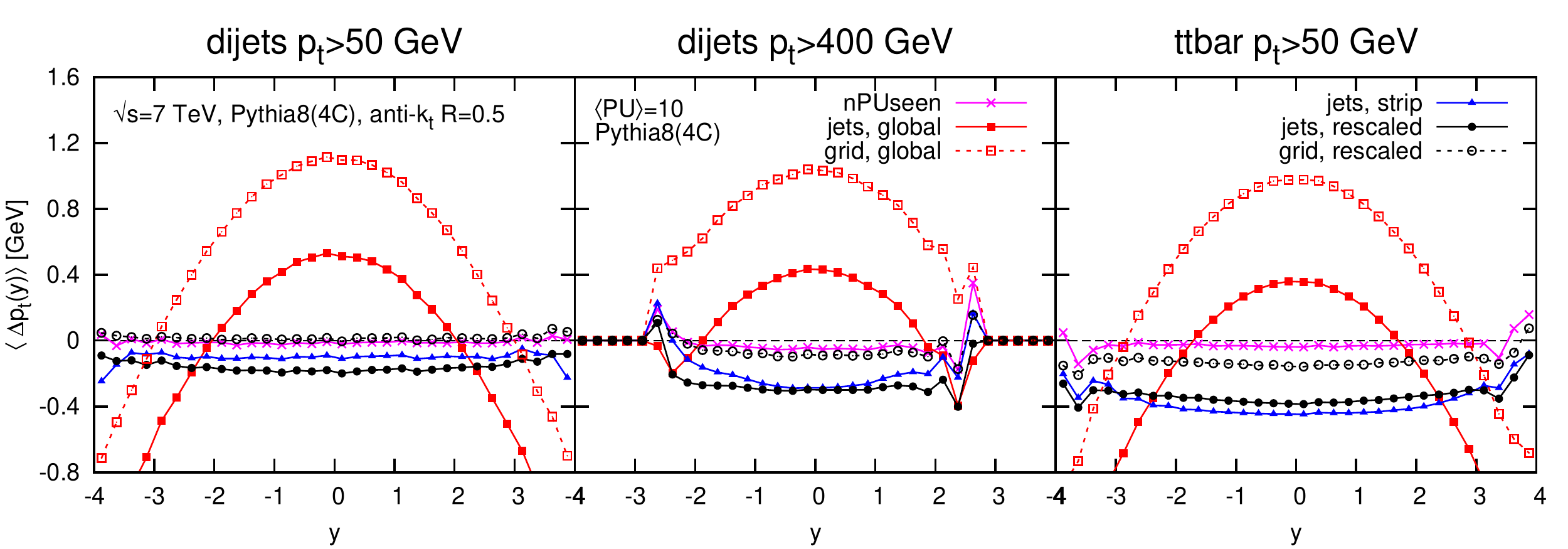}
}
\caption{Residual average shift as a function of the jet rapidity for
  all the considered subtraction methods. For the left (resp.  centre,
  right) plot, the hard event sample consists of dijets with $p_t\ge
  50$ GeV (resp. dijets with $p_t\ge 400$ GeV, and jets above $p_t\ge
  50$ GeV in $t\bar t$ events), generated with Pythia 8 (tune 4C) in
  all cases. The typical PU contamination (for unsubtracted jets) is
  around 5 GeV.}\label{fig:pusub_rapdep}
\end{figure}

Let us begin our performance benchmarks by the study of the rapidity
dependence of PU subtraction. First of all,
Fig. \ref{fig:pusub_rapdep} shows the residual average shift
($\langle\Dpt\rangle$) as a function of the rapidity of the hard
jet. These results are presented for different hard processes,
generated with Pythia 8 and assuming an average of 10 PU events per
hard interaction. Robustness w.r.t. that choice will be discussed in
the next Section but does not play any significant role for the
moment.

The first observation is that the subtraction based on the number of
seen PU vertices does a very good job in all 3 cases. Then, global
median-based (using jets or grid cells) estimations of $\rho$, \ie the
(red) square symbols, do a fair job on average but, as expected, fail
to correct for the rapidity dependence of the PU contamination. If one
now restricts the median to a rapidity strip around the jet, the
(blue) triangles, or if one uses rapidity rescaling, the (black)
circles, the residual shift is very close to 0, typically a few
hundreds of MeV, and flat in rapidity.

Note that the strip-range approach seems to have a small residual
rapidity dependence and overall offset for high-$p_t$ processes or
multi-jet situations. That last point, more clearly observed with some
Monte-Carlo generators like Pythia 6 than with others, may be due to
the fact that smaller ranges tend to be more affected by the presence
of the hard jets (see \eg Appendix A.2 of \cite{Cacciari:2010te}), an
effect which is reinforced for multi-jet events.
The fact that the residual shift seems a bit smaller for grid-based
estimates will be discussed more extensively in the next Section.

Next, we turn to the dispersion of \Dpt, a direct measure of the
impact of PU fluctuations on the $p_t$ resolution of the jets. Our
results are plotted in Fig. \ref{fig:pusub_resolution} as a function
of the rapidity of the hard jet (left panel), the number of PU
vertices (central panel) and the transverse momentum of the hard jet
(right panel). All subtraction methods have been included as well as
the dispersion one would observe if no subtraction were performed.

The results show a clear trend: first, a subtraction based on the
number of seen PU vertices bring an improvement compared to not doing
any subtraction; second, median-based estimations of $\rho$ give a
more significant improvement; and third, all median-based approaches
perform similarly well.

The reason why median-based estimations of $\rho$ outperform the
estimation based on the number of seen PU vertices is simply because
minimum bias events do not all yield the same energy deposit and this
leads to an additional source of fluctuations in the ``seen
vertices'' estimation compared to all median-based ones.
This is the main motivation for using an event-by-event determination
of $\rho$ based on the energy deposited in the event. This motivation
is further strengthened by the fact that additional issues like vertex
resolution or out-of-time PU would affect both $\langle\Dpt\rangle$
and $\sigma_{\Dpt}$ if estimated simply from the number of seen
vertices while median-based approaches are more robust.

\begin{figure}
\centerline{
\includegraphics[angle=270,width=\textwidth]{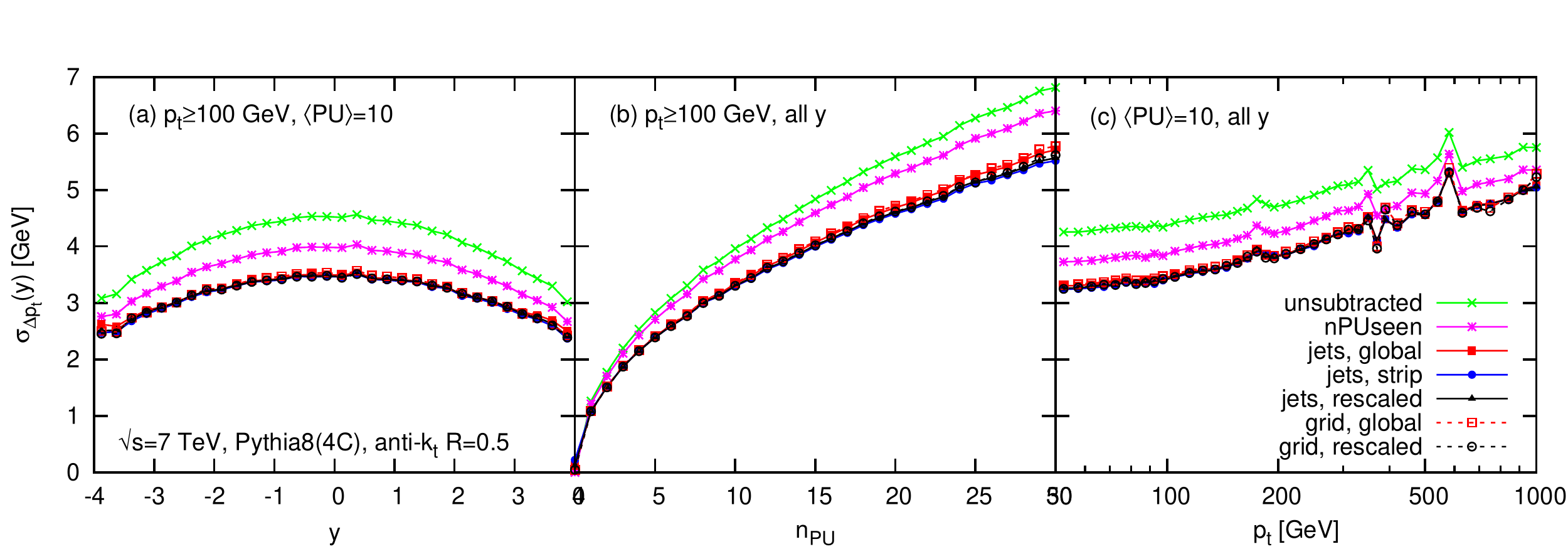}
}
\caption{Dispersion $\sigma_{\Dpt}$. Each curve corresponds to a
  different subtraction method and the results are presented as a
  function of different kinematic variables: left, as a function of
  the rapidity of the hard jet for a sample of jets with $p_t\ge 100$
  GeV and assuming an average of 10 PU events; centre: as a function
  of the number of PU events for a sample of jets with $p_t\ge 100$;
  right: as a function of the $p_t$ of the hard jet, assuming an
  average of 10 PU events}\label{fig:pusub_resolution}
\end{figure}

Note finally that even though local ranges and rapidity rescaling do
correct for the rapidity dependence of the PU on average, the
dispersion still depends on rapidity. The increase with the number of
PU vertices is in agreement with the expected $\sqrt{n_{\rm PU}}$
behaviour and the increase with the $p_t$ of the hard process can be
associated with {\em back-reaction}, see \cite{Cacciari:2010te}. These
numbers can also be compared to the typical detector resolutions which
would be $\sim$10 GeV for 100 GeV jets and $\sim$20 GeV at $p_t
=400$~GeV \cite{ATLAS-resolution,CMS-resolution}.

\subsubsection{Robustness and Monte-Carlo dependence}\label{sec:robustness}

\begin{figure}
\centerline{
\includegraphics[angle=270,width=\textwidth]{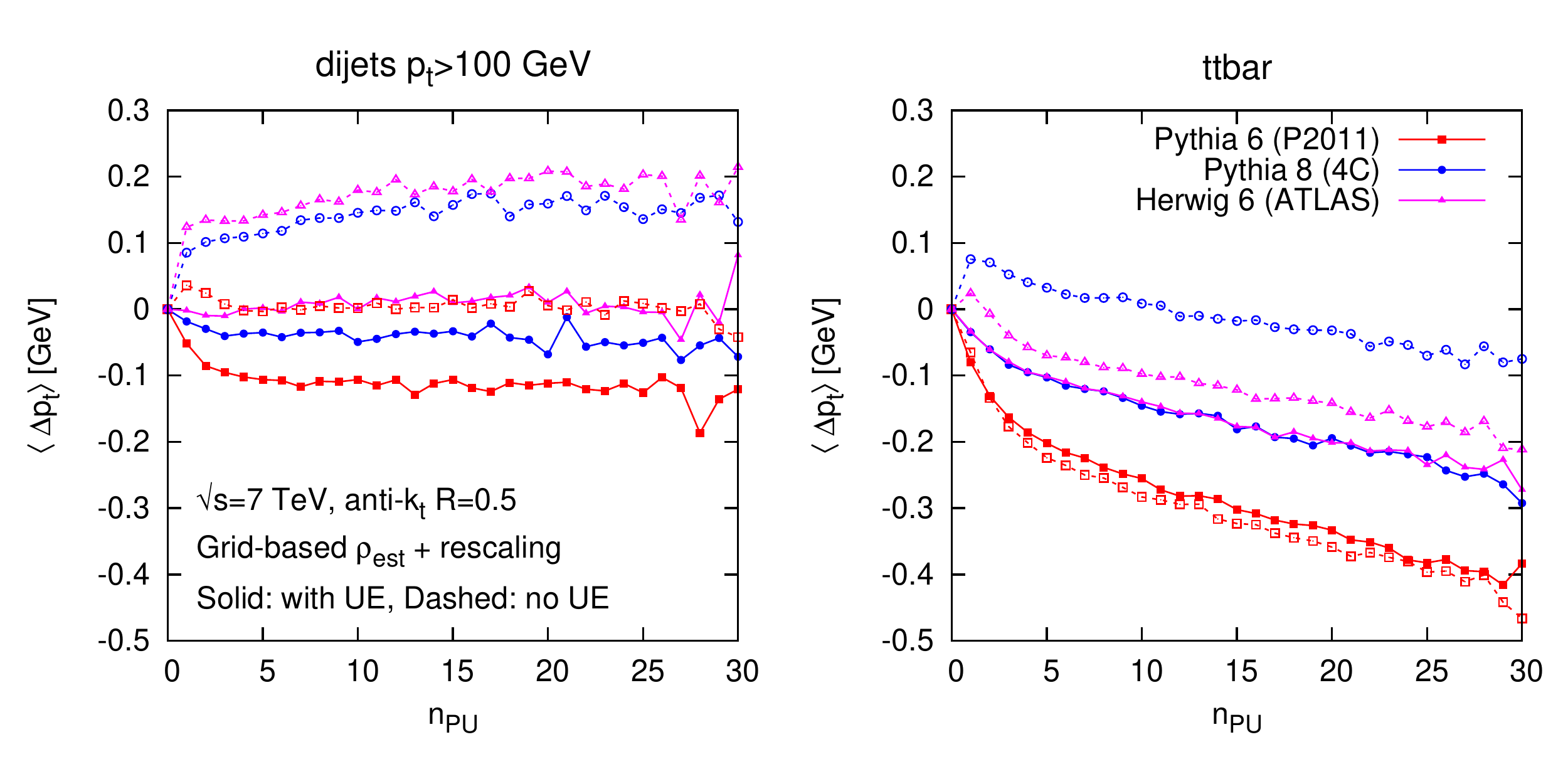}
}
\caption{Dependence of the average $p_t$ shift as a function of the
  number of PU vertices for various Monte-Carlo generators. For the
  left plot, the hard sample is made of dijets with $p_t\ge 100$ GeV
  while for the right plot, we have used a hadronic $t\bar t$
  sample. For each generator, we have considered both the case with
  the Underlying Event switched on (filled symbols) and off (open
  symbols). All results have been obtained using a grid-based median
  estimation of $\rho$ using rapidity rescaling.
}\label{fig:pusub_allmc}
\end{figure}

The last series of results we want to present addresses the
stability and robustness of the median-based estimation of the PU
density per unit area. 

To do that, the first thing we shall discuss is the Monte-Carlo
dependence of our results. In Fig. \ref{fig:pusub_allmc} we compare
the different Monte-Carlo predictions for the $\langle\Dpt\rangle$
dependence on the number of PU vertices in the case of a grid-based
median estimate of $\rho$ with rapidity rescaling. For each of the
three considered Monte-Carlos, we have repeated the analysis with and
without Underlying Event (UE) in the hard event. 
The first observation is that all the results span a range of 300-400
MeV in \Dpt and have a similar dependence on the number of PU
vertices. The dependence on $n_{\rm PU}$ is flat for dijet events but
shows a small decrease for the  busier $t\bar t$  events. The 300-400 MeV shift splits
into a 100-200 MeV effect when changing the generator, which is likely
due to the small but non-zero effect of the hard event on the median
computation, and a 100-200 MeV effect coming from the switching on/off
of the UE.

This question of subtracting the UE deserves a discussion: since the
UE is also a soft background which is relatively uniform, it
contributes to the median estimate and, therefore, one expects the UE,
or at least a part of it, to be subtracted together with the
PU. Precisely for that reason, when we compute \Dpt, our subtraction
procedure is not applied only on the ``full jet'' (hard jet+PU) but
also on the hard jet, see Eq. (\ref{eq:deltapt}). The 100-200 MeV
negative shift observed in Fig. \ref{fig:pusub_allmc} thus means that,
when switching on the UE, one subtracts a bit more of the UE in the
full event (with PU) than in the hard event alone (without PU).
This could be due to the fact (see \cite{Cacciari:2009dp} for details)
that for sparse events, as is typically the case with UE but no PU,
the median tends to slightly underestimate the ``real'' $\rho$, \eg if
half of the event is empty, the median estimate would be 0. This is in
agreement with the fact that for $t\bar t$ events, where the hard
event is busier, switching on the UE tends to have a smaller effect.
Note finally that as far as the size of the effect is concerned, this
100-200 GeV shift has to be compared with the $\sim$1 GeV
contamination of the UE in the hard jets.

\begin{figure}
\centerline{
\includegraphics[angle=270,width=\textwidth]{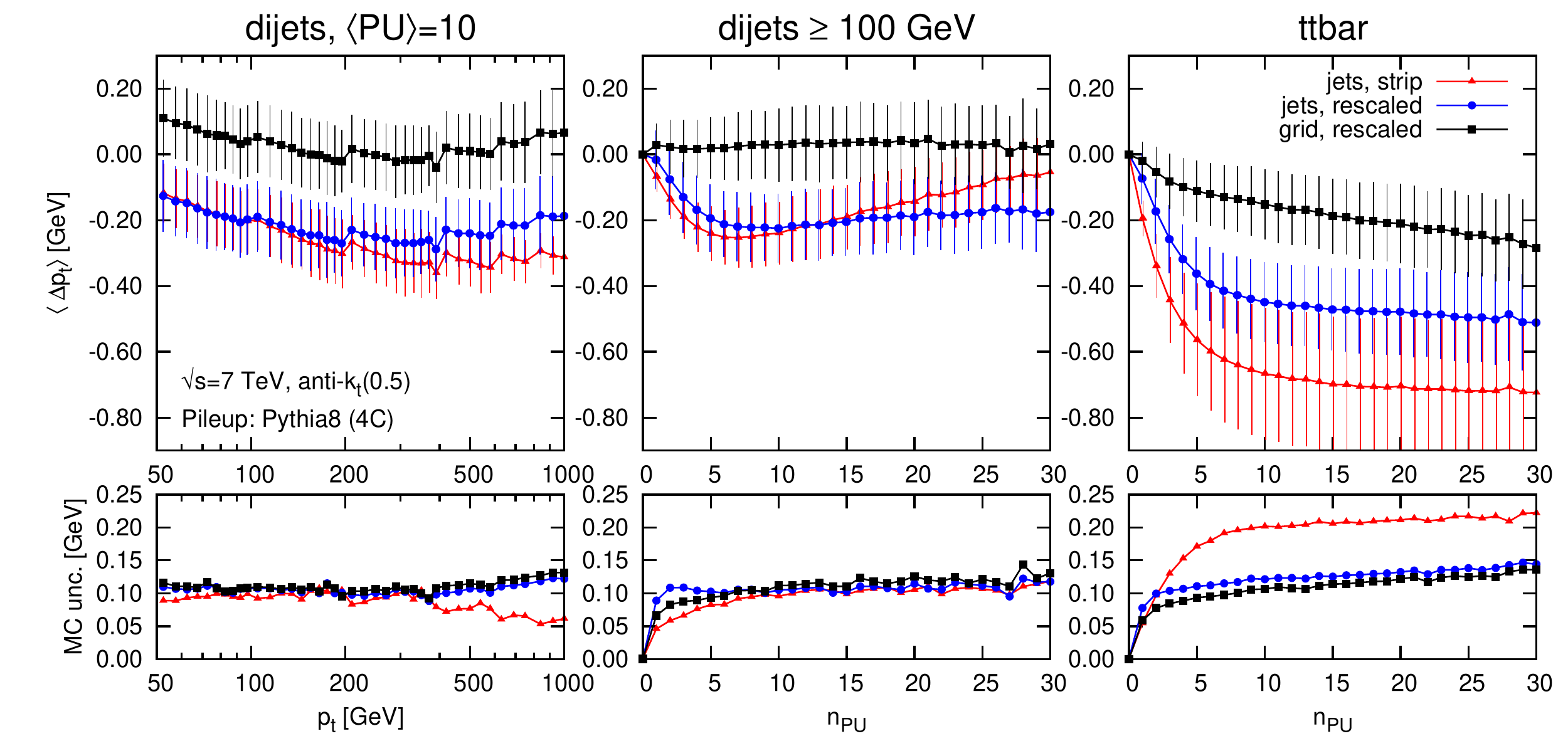}
}
\caption{Average residual shift after PU subtraction. $\langle
  \Dpt\rangle$ is plotted as a function of the $p_t$ of the jet for an
  average of 10 PU events (left panel), or as a function of the number
  of PU vertices for dijets with $p_t\ge 100$ GeV (central panel) and
  for $t\bar t$ events (right panel). In all cases, we compare 3
  methods: the rapidity-strip range, (red) triangles, the jet-based
  approach with $y$-rescaling, (blue) circles, and the grid-based
  approach with $y$ rescaling, (black) squares. Each curve is the
  result of averaging over the various Monte-Carlo generator options
  and the dispersion between them is represented both as error bars on
  the top row and directly on the bottom
  row.}\label{fig:pusub_summary}
\end{figure}

Finally, we wish to compare the robustness of our various subtraction
methods for various processes \ie hard events and PU conditions. In
order to avoid multiplying the number of plots, we shall treat the
Monte-Carlo (including the switching on/off of the UE) as an error
estimate. That is, an average measure and an uncertainty will be
extracted by taking the average and dispersion of the 6 Monte-Carlo
setups. The results of this combination are presented on
Fig. \ref{fig:pusub_summary} for various situations and subtraction
methods. For example, the 6 curves from the left plot of
Fig. \ref{fig:pusub_allmc} have been combined into the (black) squares
of the central panel in Fig. \ref{fig:pusub_summary}.

Two pieces of information can be extracted from these results.
First of all, for dijets, the quality of PU subtraction is, to a large
extent, flat as a function of the $p_t$ of the jets and the number of
PU vertices. When moving to multi-jet situations, we observe an
additional residual shift in the 100-300 MeV range, extending to
$\sim$500 MeV for the rapidity-strip-range method. This slightly
increased sensitivity of the rapidity-strip-range method also depends
on the Monte-Carlo. While in all other cases, our estimates vary by
$\sim$ 100 MeV when changing the details of the generator, for
multi-jet events and the rapidity-strip-range approach this is
increased to $\sim$200 MeV.

Overall, the quality of the subtraction is globally very good. Methods
involving rapidity rescaling tends to perform a bit better than the
estimate using a rapidity strip range, mainly a consequence of the
latter's greater sensitivity to multi-jet events. In comparing
grid-based to jet-based estimations of $\rho$, one sees that the
former gives slightly better results, though the differences remain
small.

Since the grid-based approach is considerably faster than the
jet-based one, as it does not require an additional clustering of the
event\footnote{Note that the clustering of the main event still needs
  to include the computation of jet areas since they are needed in
  Eq.~(\ref{eq:subtract}).}, the estimation of $\rho$ using a
grid-based median with rapidity rescaling comes out as a very good
default for PU subtraction. One should however keep in mind
local-range approaches for the case where the rapidity rescaling
function cannot easily be obtained.

\subsection{PU v. UE subtraction: an analysis on $Z$+jet events}\label{sec:more}

\begin{figure}
\centerline{
\includegraphics[angle=270,width=\textwidth]{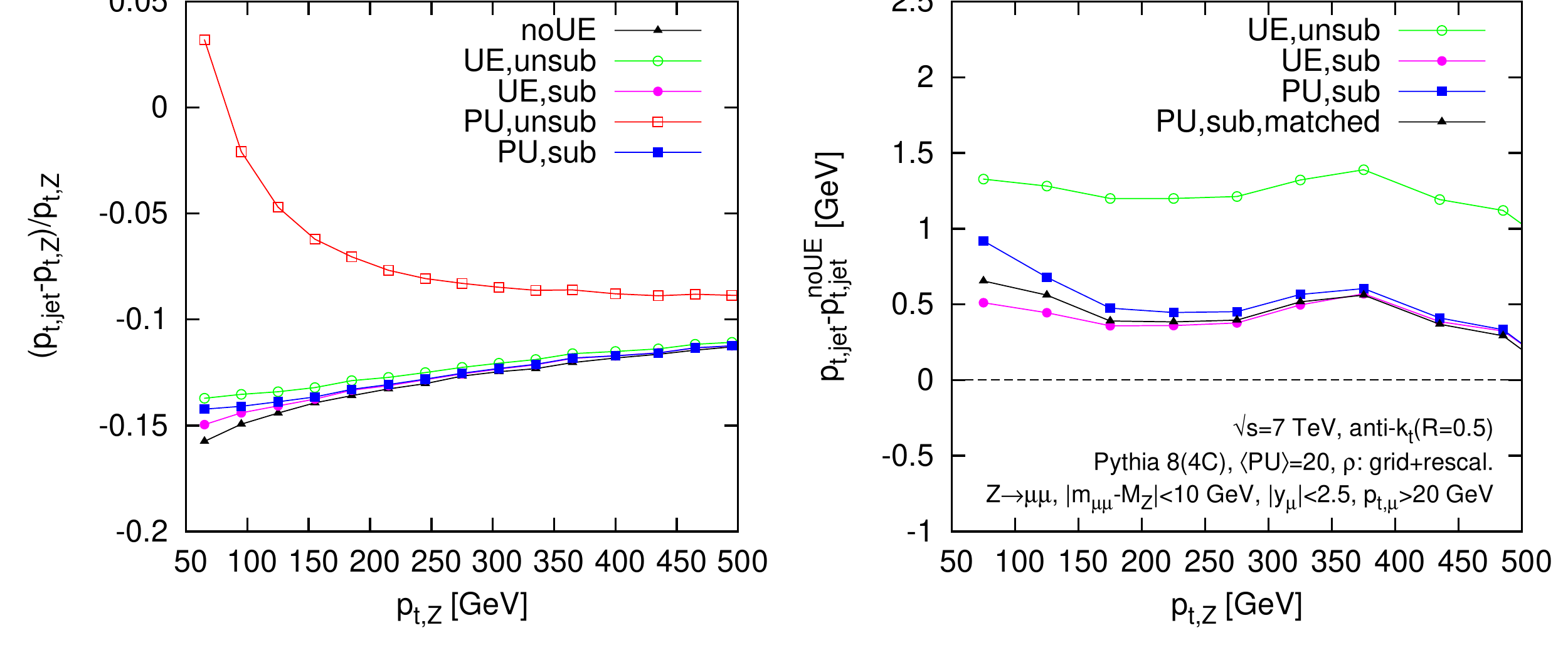}
}
\caption{Left: relative difference between the reconstructed jet and
  the reconstructed $Z$ boson transverse momenta. Right: at a given
  $p_t$ of the reconstructed $Z$ boson, difference between the
  reconstructed $p_t$ of the jet and the ideal $p_t$ with no UE or PU,
  {i.e.} $p_t$ shift w.r.t. the ``noUE'' curve, the (black) triangles,
  on the left panel. See the text for the details of the
  analysis.}\label{fig:zjet}
\end{figure}

To give further insight on the question of what fraction of the
Underlying Event gets subtracted together with the pileup, we have
performed an additional study of $Z+$jet events. We look at events
where the $Z$ boson decays into a pair of muons.
We have considered 5 different situations: events without PU or
UE, events with UE but no PU subtracted or not, and events with both
UE and PU again subtracted or not. 
Except for the study of events without UE, this analysis could also be
carried out directly on data.

Practically, we impose that both muons have a transverse momentum of
at least 20 GeV and have $|y|\le 2.5$, and we require that their
reconstructed invariant mass is within 10 GeV of the nominal $Z$
mass. As previously, jets are reconstructed using the anti-$k_t$ jet
algorithm and the pileup subtraction is performed using the
grid-based-median approach with rapidity rescaling and a grid size of
0.55. All events have been generated with Pythia 8 (tune 4C) and we
have assumed an average PU multiplicity of 20 events.

In Fig. \ref{fig:zjet}, we have plotted the ratio $p_{t,\rm
  jet}/p_{t,Z}-1$, with $p_{t,\rm jet}$ the transverse momentum of the
leading jet, for the various situations under considerations. Compared
to the ideal situation with no PU and no UE, the (black) triangles,
one clearly sees the expected effect of switching on the UE, the empty
(green) circles, or adding PU, the empty (red) squares: the UE and PU
add to the jet $\sim$1.2 and 13 GeV respectively. 

We now turn to the cases where the soft background is subtracted, \ie
the filled (blue) squares and (magenta) circles, for the cases with and
without PU respectively. There are two main observations:
\begin{itemize}
\item with or without PU, the UE is never fully subtracted: from the
  original 1-1.5 GeV shift, we do subtract about 800 MeV to be left
  with a 0-500 MeV effect from the UE. That effect becomes smaller and
  smaller when going to large $p_t$.
\item in the presence of PU, the subtraction produces results very
  close to the corresponding results without PU and where only the UE
  is subtracted. 
  This nearly perfect agreement at large $p_{t,\rm jet}$ slightly
  degrades into an additional offset of a few hundreds of MeV when
  going to smaller scales.
  This comes about for the following reason: the non-zero $p_t$
  resolution induced by pileup (even after subtraction) means that in
  events in which the two hardest jets have similar $p_t$, the one
  that is hardest in the event with pileup may not correspond to the
  one that is hardest in the event without pileup.
  This introduces a positive bias on the hardest jet $p_t$ (a similar
  bias would be present in real data even without pileup, simply due
  to detector resolution).
  The ``matched'' curve in Fig.~\ref{fig:zjet} (right) shows that if,
  in a given hard event supplemented with pileup, we explicitly use
  the jet that is closest to the hardest jet in that same event
  without pileup, then the offset disappears, confirming its origin as
  due to resolution-related jet mismatching.
\end{itemize}


\subsection{Conclusions and discussion}

In these proceedings, we have investigated several methods to correct
for the pile-up contamination to jets. They are all based on the
observation that the average PU contribution to a jet is on average
proportional to its area, which directly leads to
eq.~(\ref{eq:subtract}). The various methods then differ by the method
used to estimate the PU activity per unit area, $\rho$. The
subtraction efficiency has been studied by embedding hard events into
PU backgrounds and investigating how jet reconstruction was affected by
measuring the remaining $p_t$ shift after subtraction ($\langle\Delta
p_t\rangle$) as well as the impact on resolution ($\sigma_{\Delta
p_t}$).

There are 3 broad approaches to the estimation of $\rho$: (a) using
an average contamination per PU vertex, the {\em seen vertices}
approach, (b) using an event-by-event estimation and, the {\em median}
approach with jets or grid cells as patches, and (c) using an
event-by-event and jet-by-jet method, the {\em local range} or {\em
  rescaling} approaches.

The first important message is that, though all methods give a very
good overall subtraction ($\langle\Delta p_t\rangle\approx 0$),
event-by-event methods should be preferred because their smaller PU
impact on the $p_t$ resolution (see
Fig.~\ref{fig:pusub_resolution}). This is mostly because the ``seen
vertices'' method has an additional smearing coming from the
fluctuations between different minimum bias collisions. This does not
happen in event-by event methods that are only affected by
point-to-point fluctuations in an event. Note also that event-by-event
methods are very likely more robust than methods based on identifying
secondary vertices when effects like vertex identification and
out-of-time PU are taken into account.

The next observation is that event-by-event and jet-by-jet methods
have the additional advantage that they correct for positional-dependence of
the background like its rapidity dependence (see
Fig.~\ref{fig:pusub_rapdep}). The median approach using a local range
(with jets as patches) or rapidity rescaling (using jets or grid cells
as patches) all give an average offset in the 0-300 MeV range,
independently of the rapidity of the jet, its $p_t$ or the number of
PU vertices, see Fig.~\ref{fig:pusub_summary} and are thus very
suitable methods for PU subtraction at the LHC. Pushing the analysis a
bit further one may argue that the local-range method has a slightly
larger offset when applied to situations with large jet multiplicity like
$t\bar t$ events (the right panel of Fig.~\ref{fig:pusub_summary})
though this argument seems to depend on the Monte-Carlo used to
generate the hard-event sample. Also, since it avoids clustering the
event a second time, the grid-based method has the advantage of being
faster than the jet-based approach. 

At the end of the day, {\em we can recommend the median-based
  subtraction method with rapidity rescaling and using grid cells as
  patches} as a powerful default PU subtraction method at the LHC. But
one should keep in mind that the use of jets instead of grid cells
also does a very good job and that local-ranges can be a good
alternative to rapidity rescaling if the rescaling function cannot be
computed.
Also, though we have not discussed that in detail, a grid cell size of
0.55 is a good default as is the use of $k_t$ jets with $R=0.4$.

To conclude, let us make a few general remarks.
First, our suggested method involves relatively few assumptions,
which helps ensure its robustness.
Effects like in-time v.\ out-of-time PU or detector response should not
have a big impact.
Many of the studies performed here can be repeated with ``real
data'' rather than Monte-Carlo simulations. The best example is
certainly the $Z$+jet study of Section~\ref{sec:more} which could be
done using data samples with different PU activity from 2010 and 2011.
Also, the rapidity rescaling function can likely be obtained from
minimum bias collision data and the embedding of a hard event into
pure PU events could help quantifying the remaining ${\cal
  O}$(100~MeV) bias. Experimentally, it would also be interesting to
investigate hybrid techniques where one would discard the charged
tracks that do not point to the primary vertex and apply the
subtraction technique described here to the rest of the event. This
would have the advantage to further reduce fluctuation effects
(roughly by a factor $\sim \sqrt{1/(1-f_{\rm chg})}\approx 1.6$, where
$f_{\rm chg}\approx 0.61$ is the fraction of charged particles in an
event).
Finally, all the facilities to compute jet areas and background
estimation --- including jets or grid-cells as patches, local ranges
and rescaling functions --- are readily available from FastJet (v3.0.0
onward) using \eg the {\small \tt GridMedianBackgroundEstimator} or {\small\tt
  Subtractor} tools.

\subsection*{Acknowledgements}

We thank Andy Buckley, Suzanne Gascon-Shotkin, Paolo Francavilla,
Peter Loch, Emily Nurse, and Mark Stockton for very stimulating
discussion during and after the workshop.
This work was supported in part by grants ANR-09-BLAN-0060,
ANR-10-CEXC-009-01 and PITN-GA-2010-264564.


}

\part[MC TUNING AND OUTPUT FORMATS]{MC TUNING AND OUTPUT FORMATS}
\label{part:mc}
\section[Tune killing: quantitative comparisons of MC generators and tunes]
{TUNE KILLING: QUANTITATIVE COMPARISONS OF MC GENERATORS AND TUNES \protect\footnote{Contributed by: A.~Buckley, G.~Hesketh, H.~Hoeth, F.~Krauss, E.~Nurse, S.~Pl\"atzer,  H.~Schulz}}
{\graphicspath{{leshouches-tunekill/}}

\parindent 0mm
\title{Tune killing: quantitative comparisons of MC generators and tunes}

\author{Andy~Buckley$^a$, Gavin~Hesketh$^b$, Hendrik~Hoeth$^c$, Frank~Krauss$^c$, Emily~Nurse$^b$, Simon~Pl\"atzer$^d$,  Holger~Schulz$^e$}
\institute{%
  $^a$PPE Group, School of Physics, University of Edinburgh, UK.\\
  $^b$HEP Group, Dept. of Physics and Astronomy, UCL, London, UK.\\
  $^c$IPPP, Durham University, UK.\\
  $^d$DESY, Hamburg, Germany.\\
  $^e$Physics Dept., Berlin Humboldt University, Germany.}


\begin{abstract}
  We summarise the implementation, status, and scope of the ``tune killing''
  project, which classifies MC generator codes and tunes according to their
  quality of data description across a range of LHC-relevant observables. The
  primary aim of the project is to provide sufficiently clear information about
  generator performance that the current large collection of available tunes may
  be objectively reduced to a more manageable standard set for common use by LHC
  experiments and phenomenologists. We make final recommendations as to which
  generators and tunes are in rude health, and those which are obvious
  candidates for retirement from active service.
\end{abstract}

\subsection{INTRODUCTION}

Popular MC generators are nowadays associated with a bewildering array of
standard parameter configurations, called ``tunes''. This proliferation of tunes
is due to the ongoing project to provide optimised descriptions of LEP, Tevatron
and LHC data: as new data and techniques have become available, new tunes have
been created, usually but not always with increasing quality of data
description. This process looks set to continue, and hence there is a need for
agreement on which tunes are of most common interest at a given time.

The PYTHIA6\,\cite{Sjostrand:2006za} event generator in particular has been the
\textit{de facto} testbed for tuning due to the wealth of community expertise
and its ubiquity of tuning parameters for physical processes. At the time of
writing there are 77 tunes available via the built-in \texttt{PYTUNE} routine,
and a further 10 or more presented by the ATLAS experiment alone (this counting
of ATLAS tunes includes equivalently weighted tunes for multiple PDFs, but not
systematic variation tunes, of which there are many more). With such a
profligacy of configuration options, it is difficult to objectively decide which
are to be preferred for LHC simulation without
manually cross-referencing hundreds of plots. It is hence not uncommon for
different experimental or phenomenological studies to use entirely disjoint MC
generator setups, making comparison difficult. Ideally we would have a much
smaller set of agreed-upon generator setups, but choosing such a privileged
subset requires clear information on which to base our preferences.

As a first step to addressing this issue, we present here a comparative study of
event generator codes and tunes across a range of observables, particularly
those of relevance for LHC physics. The study is based on analyses from the
Rivet~\cite{Buckley:2010ar} toolkit, and the resulting data descriptions are quantitatively scored
based on measures of deviation from the data values, including $\chi^2$ and
median/maximum bin-wise deviations (in units of combined experimental,
statistical, and theoretical uncertainties). The results are presented as a
series of Web pages, using colour coded tables which are hyperlinked to provide
the necessary information in a compact, hierarchical form.


\subsection{Analysis system}
\label{sec:tunekill-analysis}

The data analysed for this project was produced by individual runs of various
generator/tune configurations into the Rivet analysis system. A choice of Rivet
analyses was made, intended to cover a number of core QCD modelling aspects for
LHC physics: these are documented in Table~\ref{tab:tunekill-obs}.  
In some cases only the most relevant range in the distribution is included, as indicated in the Table.
The
generators and tunes used are documented in Table~\ref{tab:tunekill-tunes}.

\begin{table}[tp]
  \centering
  \begin{tabular}{llll}
    \toprule
    Observable & Rivet analysis & Ref. & Range \\
    \midrule
    \textbf{Underlying event} & & \\
    Transverse region $N_\text{ch}$ vs. $p_\perp^\text{lead}, p_\perp^\text{ch} > 500\,\text{MeV}$ & ATLAS$\_$2010$\_$S8894728 & \cite{Aad:2010fh} & $p_\perp^\text{lead} > 5\,\text{GeV}$ \\
    Transverse region $\sum p_\perp$ vs. $p_\perp^\text{lead}, p_\perp^\text{ch} > 500\,\text{MeV}$ & ATLAS$\_$2010$\_$S8894728 &  \cite{Aad:2010fh} & $p_\perp^\text{lead} > 5\,\text{GeV}$ \\
    Transverse region $\langle p_\perp \rangle$ vs $N_\text{ch}, p_\perp^\text{ch} > 500\,\text{MeV}$ & ATLAS$\_$2010$\_$S8894728 & \cite{Aad:2010fh} & \\
    \addlinespace
    \textbf{Jets} & & \\
    Toward region $N_\text{ch}$ vs $p_\perp^\text{lead}, p_\perp^\text{ch} > 500\,\text{MeV}$ & ATLAS$\_$2010$\_$S8894728 &  \cite{Aad:2010fh} & $p_\perp^\text{lead} > 5\,\text{GeV}$ \\
    Toward region $\sum p_\perp$ vs $p_\perp^\text{lead}, p_\perp^\text{ch} > 500\,\text{MeV}$ & ATLAS$\_$2010$\_$S8894728 &  \cite{Aad:2010fh} & $p_\perp^\text{lead} > 5\,\text{GeV}$ \\
    Jet shapes, $30 < p_\perp < 40\,\text{GeV}$, $|y| < 2.8$ & ATLAS$\_$2011$\_$S8924791 & \cite{Aad:2011kq} & \\
    Jet shapes, $310 < p_\perp < 400\,\text{GeV}$, $|y| < 2.8$ & ATLAS$\_$2011$\_$S8924791 & \cite{Aad:2011kq} & \\
    Dijet $\Delta\phi$, $110 < p_\perp < 160\,\text{GeV}$ & ATLAS$\_$2011$\_$S8971293 & \cite{daCosta:2011ni} & $3\pi/4 \to \pi$ \\
    Dijet $\Delta\phi$, $310 < p_\perp < 400\,\text{GeV}$ & ATLAS$\_$2011$\_$S8971293 &  \cite{daCosta:2011ni} & $3\pi/4 \to \pi$ \\
    Dijet mass, $0.3 < |y| < 0.8$, anti-$k_\perp$(0.4) & ATLAS$\_$2010$\_$S8817804 & \cite{:2010wv} & \\
    Transverse thrust, $90\,\text{GeV} < p_\perp^{\text{jet1}} < 125\,\text{GeV}$ & CMS$\_$2011$\_$S8957746 &  \cite{Khachatryan:2011dx} &\\
    \addlinespace
    \textbf{ISR/intrinsic-$k_\perp$} & & \\
    D\O{} $\phi^*$, $|y| < 1.0$ & D0$\_$2010$\_$S8821313 & \cite{Abazov:2010mk} & $\phi^* < 0.4$ \\
    D\O{} $\phi^*$, $1.0 < |y| < 2.0$ & D0$\_$2010$\_$S8821313 &  \cite{Abazov:2010mk} & $\phi^* < 0.4$ \\
    \addlinespace
    \textbf{Fragmentation} & & \\
    $N_\text{ch}$, $\pi^+/\pi^-$, $K^+/K^-$ at LEP &  DELPHI$\_$1996$\_$S3430090 & \cite{Abreu:1996na} & \\
    $\rho/\pi$, $K/\pi$, $\Sigma^{\pm,+,-,0}/\pi$, $p/\pi$, $\Lambda/\pi$ & PDG$\_$HADRON$\_$MULTIPLICITIES & \cite{Amsler:2008zzb} & \\
    Inclusive $x_p$, thrust (+ major \& minor) & DELPHI$\_$1996$\_$S3430090 & \cite{Abreu:1996na} & \\
    $B$ fragmentation & DELPHI$\_$2002$\_$069$\_$CONF$\_$603 & \cite{delphi-b-quark} & \\
    \bottomrule
  \end{tabular}
  \caption{Observables used in the tune killing exercise.}
  \label{tab:tunekill-obs}
\end{table}

\begin{table}[tp]
  \centering
  \begin{tabular}{ll}
    \toprule
    Generator and version & Tunes \\
    \midrule
    Sherpa 1.3.1\cite{Gleisberg:2008ta}    & Default (CTEQ6.6) \\
    Herwig++ 2.5.2\cite{Gieseke:2011na}    & LHC-UE-EE-3 series (LO$**$ and CTEQ6L1) \\
    Pythia 8.150\cite{Sjostrand:2007gs}    & 4C \\
    PYTHIA 6.425\cite{Sjostrand:2006za}    & D6T, DW\cite{CDFtuneDW}, Z2, AMBT1\cite{Atlas:ambt1}, AUET2B (LO$**$ and CTEQ6L1)\cite{Atlas:mc11},  \\
                                           & Perugia\,2010\cite{Skands:2009zm}, Perugia\,2011\cite{Skands:2009zm}, prof-$Q^2$\cite{Buckley:2009bj} \\
    AlpGen\cite{mangano:2003} + PYTHIA 6.425 (*)   & Same tunes as PYTHIA6. \\
                                           & Perugia\,2011 using matched ME/PS $\Lambda_\text{QCD}$.\cite{Cooper:2011gk} \\
    HERWIG 6.510\cite{Corcella:2002jc} + JIMMY 4.31\cite{Butterworth:1996zw} & AUET2 LO$**$\cite{Atlas:auet2} \\
    AlpGen + HERWIG 6.5 + JIMMY 4.31 (*)   & Same as for HERWIG+JIMMY. \\
    \bottomrule
  \end{tabular}
  \caption{Generators and tunes used in the tune killing exercise.\qquad (*)~Jet and $Z$ boson $\phi^*$ observables only.}
  \label{tab:tunekill-tunes}
\end{table}

Note that not all observables are suitable for all generators. For example,
AlpGen has not been used for LEP fragmentation, although in a future iteration
we will extend the AlpGen coverage to include underlying event observables,
where the hard jets could interfere with those from the multiple parton interaction (MPI) mechanism. Several
observables, notably hard photon physics, minimum bias observables, and
fragmentation/strangeness from RHIC and LHC have not yet been included: this is
envisaged as a future extension of the project.

A Python program was written to load the histogram files for each generator/tune
combination from a hierarchical directory structure, and to perform some basic
statistical characterisation on each bin, histogram, and semantic group of
histograms. At the histogram level, specifications are used to determine
which bins are to be considered in the statistical comparisons, and to add a
nominal ``theoretical uncertainty''. In this study a 10\%\ theoretical uncertainty was added to the underlying event and fragmentation observables and a 5\%\ theoretical uncertainty on the rest.
The combined uncertainty for each bin $b$
is then computed from the sum in quadrature of the reference data error, the MC
statistical error and the theoretical uncertainty, and is used to compute a
MC--data deviation for that bin, expressed in units of the total bin error,
 $\text{dev}_b = (\text{MC}_b - \text{data}_b)/\text{err}_b$.

For each active histogram, the system then reports the $\chi^2/N_\text{bin}$,
and the median, mean, and maximum bin-wise deviation. A total ``metric'' value
for each histogram is reported as the maximum bin-wise deviation if that is
greater than $10\sigma$, otherwise the greater of the median and mean
deviations. This hybrid treatment of the metric allows the system to flag up
histograms in which there are either widespread moderate deviations or a
small number of very discrepant bins which might be missed with a pure median or
mean deviation treatment. An HTML table and set of histograms are rendered by
the system for each observable, with a continuous colour coding scheme used to
highlight the relative quality of data description from ideal (green) to very
poor (red).

The histograms are grouped to collect together observables from different
sources which reflect related aspects of QCD modelling. The current groups are
``Underlying Event (UE)'', ``Dijets'', ``Multijets'', ``Jet shapes'', ``W and Z'',
``Fragmentation'', and ``$B$ fragmentation''. In these groups, the same
$\chi^2/N_\text{bin}$, and mean/median/maximum deviation statistics are
calculated as before. For visual compactness of classification we again use a
hybrid performance metric for each histogram group: again this is the maximum bin-wise deviation found in the
contained histograms if that is greater than $10\sigma$, otherwise the maximum
histogram-wise deviation metric in the group if that is greater than $5\sigma$,
otherwise the maximum of the median/mean bin-wise deviation. 

The Web pages generated to present this data in a compact way consist of a
single top level page containing a colour-coded table of tune performance
metrics for each histogram group. Each cell in the table is hyperlinked to a
more detailed table for that tune/group where the various $\chi^2/N$,
max/mean/median deviation and hybrid metric are presented, again colour-coded,
for each histogram in the group. The table rows are then hyperlinked to a plot
page showing explicitly the tune/generator behaviour for each histogram and
indicating the active range of the histograms where appropriate. These pages are
shown in Figures~\ref{fig:tunekill-screenshot-1} to~\ref{fig:tunekill-screenshot-3}. This form of
presentation allows a rapid assessment of generator/tune performance, while
still permitting detailed investigation of any flagged-up issues with a few
mouse clicks. The system is easily extensible to more observables, groups, and
different theory uncertainty / visual classification thresholds.

The classification colours for each performance figure are generated in HSB
colour space as a linear variation in deviation $x$ between green (120) and red
(0) in the Hue parameter, i.e. $H = 120(1 - \min(x/x_\text{bad}, 1.0))$, with
fixed Saturation and Brightness parameters. The visual threshold $x_\text{bad}$
was chosen to be different for each metric type: $5\sigma$ for maximum
deviations, $4\sigma$ for $\chi^2/N$, and $2\sigma$ for mean and median
deviations, and for the hybrid performance metrics. These thresholds were
iterated from initial suggestions to the point where distinctions could be made
between the models: similar iteration of the discriminating criteria are
envisaged while significant model/tune variations exist as the motivation of
this study is model discrimination rather than passing or failing a natural
performance figure.

\begin{figure}[tp]
  \centering
  \includegraphics[width=0.7\textwidth]{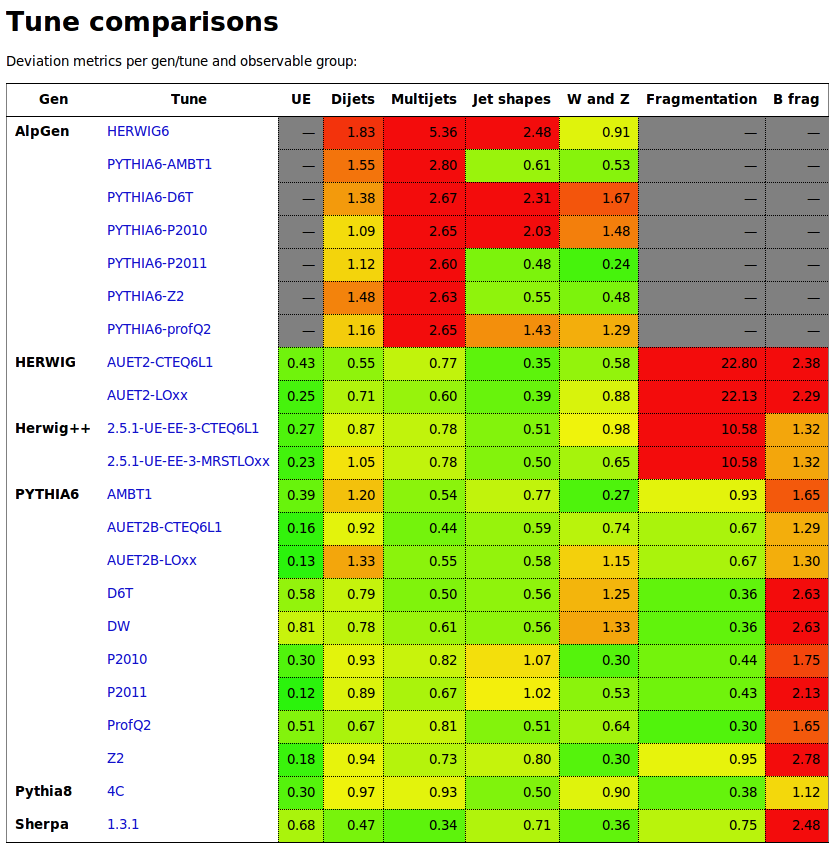}
  \caption{Screenshot of the top-level summary page produced by the tune comparison system.}
  \label{fig:tunekill-screenshot-1}
\end{figure}

\begin{figure}[tp]
  \centering
  \includegraphics[width=0.7\textwidth]{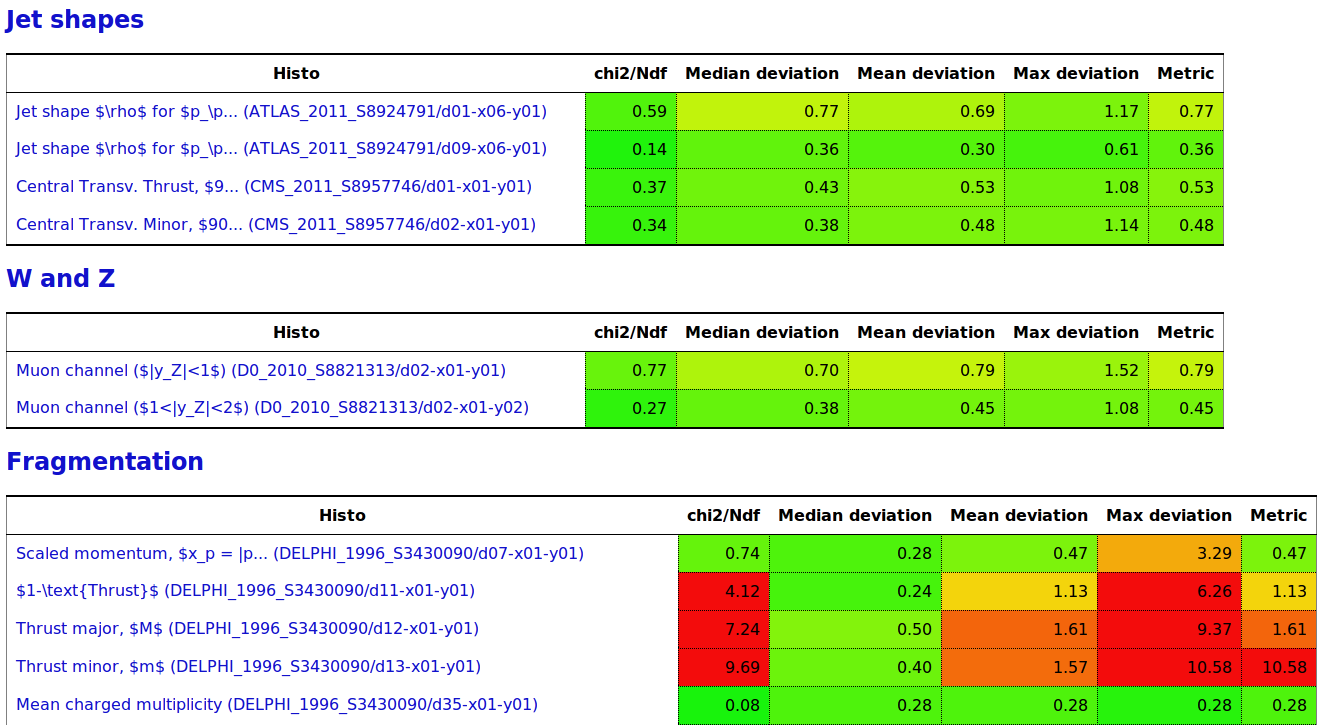}
  \caption{Screenshot of the mid-level performance metric page produced by the
    tune comparison system. This specific example is part of the performance
    metrics for the Herwig++ LHC-UE-EE-3 LO$**$ tune.}
  \label{fig:tunekill-screenshot-2}
\end{figure}

\begin{figure}[tp]
  \centering
  \includegraphics[width=0.7\textwidth]{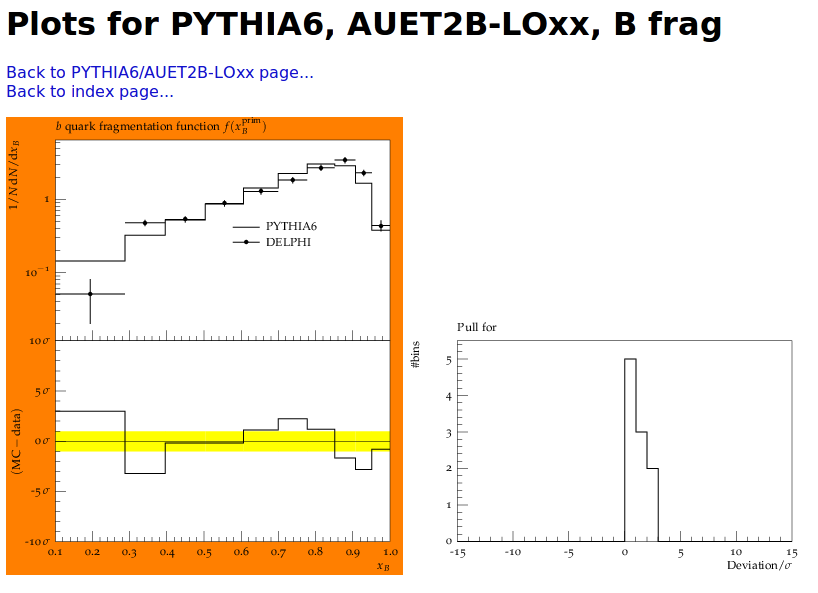}
  \caption{Screenshot of the observable plot page produced by the tune
    comparison system. This specific example shows the $B$ fragmentation
    performance of the PYTHIA6 AUET2B LO$**$ tune.}
  \label{fig:tunekill-screenshot-3}
\end{figure}

\subsection{Results}
\label{sec:tunekill-results}

As the central theme of this project has been to provide a comprehensible
visualisation of the relative performance of generators and tunes, and
hierarchical presentation via Web pages was key to achieving this, it would be
self-defeating to attempt to present the same information in this
summary. Additionally, the nature of tune comparison is that it evolves as new
data, tunes, and generator versions become available. Hence, for up-to-date
status information we refer the reader to the persistent ``tune killing'' web
page at \url{http://projects.hepforge.org/rivet/tunecmp/}.

However, it \emph{is} worth mentioning some of the most striking features of
generators which have been made more evident by this collating of data--MC
comparisons:
\begin{itemize}
\item The general quality of jet and W/Z data description is in fact better than
  expected: among PYTHIA tunes in particular there is sufficient variation in
  parton shower parameters that significant deviations in jet observables would
  reasonably be expected, but in fact the majority of tunes describe data
  fairly well.

\item The UE in particular has been a focus of tuning activity and
  this is evident in the consistency of UE data description. The worst
  performance in this group is from the DW tune of PYTHIA, but even this pre-LHC
  tune with the ``old'' PYTHIA MPI model achieves a deviation metric of less
  than $1\sigma$ on LHC UE observables.


  \item PYTHIA~D6T outperforms PYTHIA~DW -- an unexpected result since
  the MPI energy evolution of D6T is fixed to the default and disfavoured form
  $p_\perp^0(s) \sim (\sqrt{s}/1800\,\text{GeV})^{0.16}$, whereas the exponent
  in DW is closer to the tuned consensus of $\sim 0.25$. This may be a lucky
  behaviour at 7\,TeV, and hence care is needed with extrapolation of D6T to 8,
  10, or 14\,TeV, but it is clear that the PYTHIA $Q^2$-ordered parton shower is
  not yet dead on purely physics grounds. The best tune of this PYTHIA
  configuration, however, is Prof-$Q^2$, which in addition to general small
  improvements, is significantly better than DW or D6T at describing the vector
  boson $p_\perp$ distribution.

\item Pythia8 is generally seen to perform very well, and provides significant
  improvements over PYTHIA6 for jet shapes  and $B$ fragmentation. Tuning focus
  is accordingly beginning to shift towards Pythia8, also for minimum bias
  observables not yet considered here.

\item AlpGen interacts strongly with tunes on jet shape and vector boson data
  descriptions. In particular there appears to be little motivation to use
  AlpGen with the D6T or Perugia\,2010 tunes of PYTHIA6. AlpGen+HERWIG also has
  significant problems with jet shapes in particular, and the indication of this
  study is that AlpGen+PYTHIA~Perugia\,2011 is the most performant
  configuration, closely followed by AlpGen+PYTHIA~Z2. Notably, the
  Perugia\,2011 tune of PYTHIA was specifically developed to minimise ME/PS
  merging artefacts when used with AlpGen.

\item Both HERWIG and Herwig++ have problems describing LEP fragmentation data,
  but Herwig++ is a very significant improvement over its Fortran cousin. The
  identified hadron rates are in particular much improved, although $K^\pm$ and
  $\Sigma^0$ remain anomalous. However, a known problem with Herwig++ is the
  poor description of the LEP thrust distribution, which overshoots
  significantly in the multi-parton region.

\item AlpGen seems to have difficulty describing dijet azimuthal decorrelations,
  even when restricted to the 2/3 parton region of the plot. This is
  particularly surprising as AlpGen is intended to provide the multi-parton
  configurations needed to describe this observable.

\item $B$ fragmentation is in general quite poorly described. The best
  descriptions are by Pythia8, the AUET2B tunes of PYTHIA6, and Herwig++. Other
  generators and tunes are in decidedly dodgy shape for $B$-specific predictions
  at the LHC.
\end{itemize}

Insofar as it is within the scope of this project to make recommendations for
canonical generator and tune choices, we note that the Perugia\,2011, AUET2B,
and Z2 tunes of PYTHIA6 provide the best data descriptions currently available
with that generator and that the Prof-$Q^2$ tune is the best available
configuration using the $Q^2$-ordered PYTHIA parton shower. We hence recommend
these 4 PYTHIA tunes as the current minimal set of PYTHIA tunes for general use
at the LHC, particularly once an update of the ATLAS AUET2B tune has fixed the
tuning issue with the $Z$ $p_\perp$.

Among the other generators, where there is not such a proliferation of tunes, we
note again the apparent performance issues with AlpGen -- this is clearly in
need of further pursuit. However, to reduce the amount of comparison needed, we
note that Perugia\,2011 is the only PYTHIA6 tune now optimised for use with
AlpGen with avoidance of the worst effects of ME/PS coupling mismatches: hence
future studies can quite happily restrict themselves to this AlpGen+PYTHIA
configuration. As AlpGen+HERWIG has several problems with jet description,
HERWIG itself has serious problems with both light and $B$ fragmentation, and no
further tuning of the JIMMY MPI model is envisaged, the HERWIG generator cannot
be recommended for future use in any capacity where an alternative exists.

The ``new'' C++ generators Herwig++, Pythia8, and Sherpa all perform well, with
the exception of Sherpa's $B$ fragmentation and the Herwig++ light
fragmentation
. Pythia8 generally behaves well but some tuning or development may be
needed to improve inter-jet observables and the $Z$ $p_\perp$ spectrum. In
general, the C++ generators are in good health, and we anticipate further
improvements as the focus of tuning studies shifts to them.

\subsection{Outlook}
\label{sec:tunekill-outlook}

This project has put in place a system and a set of classification criteria
which have proven useful for summarising and investigating MC generator model
and tune predictivity for a variety of QCD phenomena. While we claim no mandate
to truly ``kill'' certain tunes or generators, and wish to emphasise that a poor
performance in a single observable type (in particular $B$ fragmentation)
certainly does not render that generator useless, the results from these
comparisons do provide strong arguments for deprecation of at least several
PYTHIA6 tunes and of the Fortran HERWIG generator in general.

It is the nature of a project like this that results are continually being
updated, and there are many natural avenues for extension which we wish to
pursue, in particular:
\begin{itemize}
\item Extra observables, e.g. minimum bias and $E_\perp$ flow, LHC and Tevatron
  photon physics, LHC $W$/$Z$ $p_\perp$ data, strangeness data from LHC and
  RHIC, explicit multijet observables, etc..
\item Extra generators and tunes, in particular
  POWHEG+PYTHIA/HERWIG/Pythia8/Herwig++, MadGraph+PYTHIA/Pythia8,
  MC@NLO+HERWIG/Herwig++. Comparison between Sherpa with the CTEQ6.6 and CTEQ6L1
  PDFs. New Pythia8 and PYTHIA6 tunes from ATLAS.
\end{itemize}

Greater automation of the data generation will be important, as finding
resources (human rather than CPU!) to produce and run combinatoric numbers of
generator/tune/PDF/observable combinations has been troublesome. We suggest that
this project can make use of the output of the CERN LPCC MCplots system (also
Rivet-based) for future extension. We also look forward to a forthcoming major
upgrade of the Rivet histogramming system which will greatly simplify the
treatment of multi-leg generators for which the $n$-parton samples must be
explicitly merged, e.g. AlpGen, MadGraph, etc.


}

\section[Compact ASCII output format for HepMC]
{COMPACT ASCII OUTPUT FORMAT FOR HEPMC \protect\footnote{Contributed by:  A.~Buckley, L.~Garren, G.~Hesketh, H.~Hoeth, L.~L{\"o}nnblad, E.~Nurse, S.~Pl\"atzer, G.~Salam, G.~Soyez}}
{\graphicspath{{CompactHepMC/}}

\parindent 0mm 
\title{Compact ASCII output format for HepMC}

\author{Andy~Buckley$^a$, Lynn~Garren$^b$, Gavin~Hesketh$^d$, Hendrik~Hoeth$^i$, Leif~Lönnblad$^c$, Emily~Nurse$^d$, Simon~Pl\"atzer$^j$, Gavin~Salam$^{efg}$, Gregory~Soyez$^k$}
\institute{$^a$PPE Group, School of Physics, University of Edinburgh, UK.\\
  $^b$Fermi National Accelerator Laboratory, Batavia, USA\\
  $^c$Department of Astronomy and Theoretical Physics, Lund University\\
  $^d$HEP Group, Dept. of Physics and Astronomy, UCL, London, UK.\\
  $^e$LPTHE, UPMC Univ. Paris 6 and CNRS, Paris, France.\\
  $^f$CERN, Physics Department, Theory Unit, Geneva, Switzerland\\
  $^g$Department of Physics, Princeton University, Princeton, NJ, USA.\\
  $^i$IPPP, Durham University, UK.\\
  $^j$DESY, Hamburg, Germany\\
  $^k$Institut de Physique Th\'eorique, CEA Saclay, France.}


\begin{abstract}
  We discuss the possibility of reducing the footprint of HepMC event
  files. Different compression options are discussed,
  and a suggestion for an update of the HepMC ASCII file format is
  presented.
\end{abstract}

\subsection{Introduction}

The HepMC~\cite{Dobbs:2001ck} event record has become the de-facto
standard for communicating events between event generators and
different kinds of analysis programs. HepMC also provides an
ASCII-based file format for storing and retrieving events to and from
disk, which has also become the standard. This file format is not at
all optimized for size, and although disk space today is fairly cheap,
there are still problems associated with handling very large files.

A typical minimum-bias 7 TeV LHC HepMC event occupies around 50~kB when
written on disk. More interesting events are usually bigger than this and one would typically want to store many
events to get anywhere near the statistics collected by any of the LHC
experiments; it is clear that such event files will become very
large and difficult to handle. Even with standard compression
algorithms such as \texttt{gzip} and \texttt{bzip2}, where these file
sizes can be reduced by a factor 3 or more, the problem is still
substantial.

One could imagine using a binary output format to reduce the event
size. Writing a 4 byte floating point number in an ASCII file
typically takes 10-12 characters, so here one could expect to reduce
file sizes up to a factor 3. However, standard compression
algorithms are rather good at identifying strings of numbers and
compressing them, so there is normally not much to be gained by using
a compressed binary format compared to a compressed ASCII one. In
addition one would lose the advantage of ASCII files that they are
(somewhat) readable to the human eye.

Instead the key to reducing file sizes is to remove redundant and
unnecessary information stored in the files. This could involve
completely reversible operations such as removing the information
about the momentum of an intermediate particle, as this can be
reconstructed from its decay products. It could also involve
irreversible operations such as reducing the precision on the momenta.
In the following we describe a number of such operations, which allows
us to reduce the file sizes by almost a factor 30.

\subsection{The Benchmarking procedure}
\label{sec:benchm-proc}

We started out by generating 1000 non-diffractive QCD events with
Pythia 6.425\cite{Sjostrand:2006za} using the
AGILe~\cite{Buckley:2010ar} interface. The resulting file size was
48~MB, which can be reduced to 16~MB or 13~MB using \texttt{gzip} or
\texttt{bzip2} respectively. We then investigated several ways of
reducing this size.

\paragraph{Removing irrelevant particles} The HepMC format contains
quite a lot of information about how the event was generated, such as
intermediate particles in the hard sub process, which may be
generator-dependent (and often unphysical) and is not relevant
when comparing to experimental data. In principle one could argue that
the only thing that should be written out is final-state stable
particles (with HepMC status code 1). However, there are circumstances
where information about intermediate unstable hadrons (status code 2)
is relevant. The AGILe event generator interface already includes
facilities for keeping only particle entries with status code 1 or 2.

\paragraph{Reconstructible information} Some information in the HepMC
file is redundant in the sense that it can be reconstructed from other
information in the file. Here are some examples.
\begin{itemize}\itemsep 0mm
\item Both energy, momentum and invariant mass of each particle is
  written out. Clearly, we can eg.\ reconstruct the energy given the
  three-momentum and mass\footnote{One could also reconstruct the mass
    given energy and momentum, but this typically gives large
    precision problems for small masses.}.
\item The three-momenta of decayed hadrons (status code 2) can be
  reconstructed from the sum of the momenta of the decay products.
\item The mass of a stable particle can be deduced from the particle
  ID.
\item The position of a vertex can be deduced from the previous vertex
  position and the life-time and momentum direction of the connecting
  particle.
\item Each particle in a HepMC event has a unique \textit{bar code},
  which is an otherwise arbitrary integer. No loss of information
  would result from renumbering the particles, simply inferring their
  bar code from the order in which they appear in the event.
\end{itemize}\vspace*{-5mm}

\paragraph{Precision} Clearly, having 8 byte floating point numbers is
not very relevant for many of the numbers in an event file. When
comparing with experimental data, there is no point in having much
larger precision than what is achievable in the experiment, and it  makes sense to match the the information in
the HepMC file to the precision of the actually measured variables in
the experiments. 

A possible example is to store masses and transverse momenta as integers in units of 0.1~MeV, azimuthal angles as integers in units of 0.00002$\times\pi$, pseudorapidities as integers in units of 0.00001 and vertex positions as integers 
in units of 0.001 mm.

\subsection{Benchmark Results}
\label{sec:benchmark-results}

We have investigated several of the options listed in the previous
section, and the resulting file sizes when applied to the benchmark
file is presented in table~\ref{tab:chepmc}. Firstly  we see
the size reduction using the standard format and simply reducing the number of particles,
keeping only those with status code 2 and 1 or only 1.
Next we present the same results, but using a compact format which keeps the structure of the HepMC ASCII file but applies all optimizations discussed above.
Finally, for reference, we present an aggressively
compacted \textit{Binary} format which uses the following optimizations for each particle:
  stores 1 float for transverse and 1 float for longitudinal momentum,
  a 3-byte integer for phi, and 1 byte for PDG IDs (rare PDG IDs are
  written out with 4 full bytes). This format loses the HepMC
structure of the event and in some sense this represents the target size,
below which it is difficult to go. 

\begin{table}
  \centering
  \begin{tabular}{|l||r|r|r|r|}
    \hline
    Format & Status codes & no comp. & \texttt{gzip} & \texttt{bzip2} \\
    & & (MB) & (MB) & (MB) \\
    \hline
    \hline
    Standard & All & 48 & 16 & 13 \\
              & 1 \& 2 & 43 & 15 & 13 \\
              & 1 & 17 & 6.0 & 4.8 \\
    \hline
    Compact   & All & 18 & 3.3 & 2.1 \\
              & 1 \& 2 & 13 & 2.9 & 1.9 \\
              & 1 & 4.0 & 1.9 & 1.6 \\
  \hline
    Compact binary & 1 & 1.8 & 1.7 & 1.7 \\  
  \hline
  \end{tabular}
  \caption{Size of the benchmark file after applying different
    compression methods.}
  \label{tab:chepmc}
\end{table}

It is clear that one does not gain much by using a binary format
provided one uses the
optimizations presented above together with \texttt{bzip2} compression
algorithm.

\subsection{Outlook}
\label{sec:outlook}

Given the results above, the work to include a more efficient file
format for the HepMC has begun. The suggestion is to keep the current
structure of the file format, but to add options to exclude all
particles except those with status code 1 (or 2). Furthermore options
for the representation and precision of momenta and vertex positions
will be included as well as options for excluding (simply replacing
with a single exclamation mark for easy parsing) information which can be
reconstructed. The new format will be included in a forthcoming HepMC
version during 2012.

In this report we have not looked carefully at the time it takes to
read the different formats. With the default HepMC format this can be
many times larger than the time taken for typical particle-level
analyses, while for minimal binary formats it is of the same order. We
defer detailed study of this question to future work.


}


\clearpage
\bibliography{nlmMC11}{}
\end{document}